\pdfoutput=1
\documentclass[12pt,a4paper]{article}

\usepackage{tikz}
\usetikzlibrary{backgrounds,fit,positioning}
\usetikzlibrary{shapes.geometric, arrows}
\usepackage[draft]{fixme}
\DeclareGraphicsExtensions{.jpg,.pdf,.png}

\tikzstyle{startstop} = [rectangle, rounded corners, minimum width=3cm, minimum height=1cm, text centered, text width=3cm, draw=black, fill=black!5!white, font=\footnotesize]
\tikzstyle{process} = [rectangle, minimum width=3cm, minimum height=1cm, text centered, text width=3cm, draw=black, fill=black!5!white, font=\footnotesize]
\tikzstyle{decision} = [diamond, aspect=3, minimum width=0.8cm, minimum height=0.3cm, text centered, text width=3cm, draw=black, fill=white!90!yellow, font=\footnotesize, yshift=-0.2cm]

\tikzstyle{optional_decision} = [diamond, dashed, aspect=3, minimum width=0.8cm, minimum height=0.3cm, text centered, text width=2.5cm, draw=black, fill=white!90!yellow, font=\footnotesize, yshift=-0.2cm]
\tikzstyle{process_description} = [rectangle, minimum width=3cm, minimum height=1cm, text centered, text width=3cm, font=\footnotesize]
\tikzstyle{process_cpu} = [rectangle, minimum width=3cm, minimum height=1cm, text centered, text width=3cm, draw=black, fill=white!85!blue, font=\footnotesize]

\tikzstyle{arrow} = [thick,->,>=stealth]
\tikzstyle{dashed_arrow} = [thick,->,>=stealth,dashed]

\usepackage{ifthen} 
\newboolean{pdflatex}
\setboolean{pdflatex}{true} 

\newboolean{articletitles}
\setboolean{articletitles}{true} 

\newboolean{uprightparticles}
\setboolean{uprightparticles}{false} 

\newboolean{inbibliography}
\setboolean{inbibliography}{false} 

\def\paperauthors{LHCb collaboration} 
\def\paperasciititle{The LHCb Upgrade} 
\def\papertitle{The \lhcb Upgrade I} 
\def\paperkeywords{{High Energy Physics}, {LHCb}} 
\def\papercopyright{\the\year\ CERN for the benefit of the LHCb collaboration} 
\def\paperlicence{CC BY 4.0 licence}
\def\paperlicenceurl{https://creativecommons.org/licenses/by/4.0/}

\usepackage[top=1in, bottom=1.25in, left=1in, right=1in]{geometry}

\usepackage{microtype}
\usepackage{lineno}  
\usepackage{xspace} 
\usepackage{caption} 

\usepackage{graphicx}  
\usepackage{color}
\usepackage{colortbl}
\graphicspath{{./figs/}} 

\usepackage{amsmath} 
\usepackage{amssymb}
\usepackage{amsfonts}
\usepackage{upgreek} 

\newcommand*\patchAmsMathEnvironmentForLineno[1]{%
\expandafter\let\csname old#1\expandafter\endcsname\csname #1\endcsname
\expandafter\let\csname oldend#1\expandafter\endcsname\csname
end#1\endcsname
 \renewenvironment{#1}%
   {\linenomath\csname old#1\endcsname}%
   {\csname oldend#1\endcsname\endlinenomath}%
}
\newcommand*\patchBothAmsMathEnvironmentsForLineno[1]{%
  \patchAmsMathEnvironmentForLineno{#1}%
  \patchAmsMathEnvironmentForLineno{#1*}%
}
\AtBeginDocument{%
\patchBothAmsMathEnvironmentsForLineno{equation}%
\patchBothAmsMathEnvironmentsForLineno{align}%
\patchBothAmsMathEnvironmentsForLineno{flalign}%
\patchBothAmsMathEnvironmentsForLineno{alignat}%
\patchBothAmsMathEnvironmentsForLineno{gather}%
\patchBothAmsMathEnvironmentsForLineno{multline}%
\patchBothAmsMathEnvironmentsForLineno{eqnarray}%
}

\usepackage{hyperref}
\usepackage{hyperxmp}
\hypersetup{
            pdfauthor={\paperauthors},
            pdftitle={\paperasciititle},
            pdfkeywords={\paperkeywords},
            pdfcopyright={Copyright (C) \papercopyright},
            pdflicenseurl={\paperlicenceurl}}

\usepackage[colorinlistoftodos,textsize=scriptsize]{todonotes}

\usepackage[bottom,flushmargin,hang,multiple]{footmisc}

\usepackage[all]{hypcap} 

\usepackage{xspace} 
\usepackage{upgreek}

\def\lhcb   {\mbox{LHCb}\xspace}

\def\cern {\mbox{CERN}\xspace}
\def\lhc    {\mbox{LHC}\xspace}

\def\velo   {VELO\xspace}
\def\rich   {RICH\xspace}
\def\richone {RICH1\xspace}
\def\richtwo {RICH2\xspace}

\def\hcal   {HCAL\xspace}
\def\MagUp {\mbox{\em Mag\kern -0.05em Up}\xspace}

\def\hltone {HLT1\xspace}
\def\hlttwo {HLT2\xspace}

\ifthenelse{\boolean{uprightparticles}}%
{
 
 \def\Pgamma      {\ensuremath{\upgamma}\xspace}

 \def\Peta        {\ensuremath{\upeta}\xspace}

 \def\Pmu         {\ensuremath{\upmu}\xspace}

 \def\Ppi         {\ensuremath{\uppi}\xspace}

 \def\Ptau        {\ensuremath{\uptau}\xspace}                 
                  
 \def\Pphi        {\ensuremath{\upphi}\xspace}

 \def\Ppsi        {\ensuremath{\uppsi}\xspace}

 \def\PDelta      {\ensuremath{\Delta}\xspace}                 
 \def\PXi         {\ensuremath{\Xi}\xspace}                 
 \def\PLambda     {\ensuremath{\Lambda}\xspace}                 
 \def\PSigma      {\ensuremath{\Sigma}\xspace}                 
 \def\POmega      {\ensuremath{\Omega}\xspace}                 
 \def\PUpsilon    {\ensuremath{\Upsilon}\xspace}
 \let\oldPi\Pi
 \def\PPi         {\ensuremath{\oldPi}\xspace}

 \def\PB      {\ensuremath{\mathrm{B}}\xspace}                 
                  
 \def\PD      {\ensuremath{\mathrm{D}}\xspace}

 \def\PJ      {\ensuremath{\mathrm{J}}\xspace}                 
 \def\PK      {\ensuremath{\mathrm{K}}\xspace}

 \def\PW      {\ensuremath{\mathrm{W}}\xspace}

 \def\PZ      {\ensuremath{\mathrm{Z}}\xspace}                 
                  
 \def\Pb      {\ensuremath{\mathrm{b}}\xspace}                 
 \def\Pc      {\ensuremath{\mathrm{c}}\xspace}                 
                  
 \def\Pe      {\ensuremath{\mathrm{e}}\xspace}

 \def\Pi      {\ensuremath{\mathrm{i}}\xspace}

 \def\Pp      {\ensuremath{\mathrm{p}}\xspace}

 \def\Ps      {\ensuremath{\mathrm{s}}\xspace}

 \def\Pv      {\ensuremath{\mathrm{v}}\xspace}

 \def\thebaroffset{0.0em}
}
{
 
 \def\Pgamma      {\ensuremath{\gamma}\xspace}

 \def\Peta        {\ensuremath{\eta}\xspace}

 \def\Pmu         {\ensuremath{\mu}\xspace}

 \def\Ppi         {\ensuremath{\pi}\xspace}

 \def\Ptau        {\ensuremath{\tau}\xspace}                 
                  
 \def\Pphi        {\ensuremath{\phi}\xspace}

 \def\Ppsi        {\ensuremath{\psi}\xspace}                 
                  
 \mathchardef\PDelta="7101
 \mathchardef\PXi="7104
 \mathchardef\PLambda="7103
 \mathchardef\PSigma="7106
 \mathchardef\POmega="710A
 \mathchardef\PUpsilon="7107
 \mathchardef\PPi="7105
                  
 \def\PB      {\ensuremath{B}\xspace}                 
                  
 \def\PD      {\ensuremath{D}\xspace}

 \def\PJ      {\ensuremath{J}\xspace}                 
 \def\PK      {\ensuremath{K}\xspace}

 \def\PW      {\ensuremath{W}\xspace}

 \def\PZ      {\ensuremath{Z}\xspace}                 
                  
 \def\Pb      {\ensuremath{b}\xspace}                 
 \def\Pc      {\ensuremath{c}\xspace}                 
                  
 \def\Pe      {\ensuremath{e}\xspace}

 \def\Pi      {\ensuremath{i}\xspace}

 \def\Pp      {\ensuremath{p}\xspace}

 \def\Ps      {\ensuremath{s}\xspace}

 \def\Pv      {\ensuremath{v}\xspace}

 \def\thebaroffset{0.18em}
}
\newcommand{\offsetoverline}[2][\thebaroffset]{\kern #1\overline{\kern -#1 #2}}%

\makeatletter
\ifcase \@ptsize \relax
  \newcommand{\miniscule}{\@setfontsize\miniscule{4}{5}}
\or
  \newcommand{\miniscule}{\@setfontsize\miniscule{5}{6}}
\or
  \newcommand{\miniscule}{\@setfontsize\miniscule{5}{6}}
\fi
\makeatother

\DeclareRobustCommand{\optbar}[1]{\shortstack{{\miniscule (\rule[.5ex]{1.25em}{.18mm})}
  \\ [-.7ex] $#1$}}

\def\epem       {{\ensuremath{\Pe^+\Pe^-}}\xspace}

\def\mumu       {{\ensuremath{\Pmu^+\Pmu^-}}\xspace}

\def\tauon      {{\ensuremath{\Ptau}}\xspace}

\def\g      {{\ensuremath{\Pgamma}}\xspace}

\def\W      {{\ensuremath{\PW}}\xspace}

\def\Z      {{\ensuremath{\PZ}}\xspace}

\def\squark    {{\ensuremath{\Ps}}\xspace}

\def\cquark    {{\ensuremath{\Pc}}\xspace}

\def\bquark    {{\ensuremath{\Pb}}\xspace}

\def\pion   {{\ensuremath{\Ppi}}\xspace}
\def\piz    {{\ensuremath{\pion^0}}\xspace}
\def\pip    {{\ensuremath{\pion^+}}\xspace}
\def\pim    {{\ensuremath{\pion^-}}\xspace}

\def\kaon    {{\ensuremath{\PK}}\xspace}

\def\KorKbar {\kern \thebaroffset\optbar{\kern -\thebaroffset \PK}{}\xspace}

\def\Kp      {{\ensuremath{\kaon^+}}\xspace}
\def\Km      {{\ensuremath{\kaon^-}}\xspace}
\def\Kpm     {{\ensuremath{\kaon^\pm}}\xspace}

\def\KS      {{\ensuremath{\kaon^0_{\mathrm{S}}}}\xspace}

\def\Kstarz  {{\ensuremath{\kaon^{*0}}}\xspace}

\newcommand{\phiz}{\ensuremath{\Pphi}\xspace}

\def\D       {{\ensuremath{\PD}}\xspace}

\def\DorDbar {\kern \thebaroffset\optbar{\kern -\thebaroffset \PD}\xspace}
\def\Dz      {{\ensuremath{\D^0}}\xspace}

\def\Dp      {{\ensuremath{\D^+}}\xspace}
\def\Dm      {{\ensuremath{\D^-}}\xspace}

\def\DpDm    {\ensuremath{\Dp {\kern -0.16em \Dm}}\xspace}

\def\B       {{\ensuremath{\PB}}\xspace}

\def\BorBbar {\kern \thebaroffset\optbar{\kern -\thebaroffset \PB}\xspace}
\def\Bz      {{\ensuremath{\B^0}}\xspace}

\def\Bd      {{\ensuremath{\B^0}}\xspace}

\def\BdorBdbar {\kern \thebaroffset\optbar{\kern -\thebaroffset \Bd}\xspace}

\def\Bs      {{\ensuremath{\B^0_\squark}}\xspace}

\def\BsorBsbar {\kern \thebaroffset\optbar{\kern -\thebaroffset \Bs}\xspace}

\def\Bcm     {{\ensuremath{\B_\cquark^-}}\xspace}

\def\Bds     {{\ensuremath{\B_{(\squark)}^0}}\xspace}

\def\jpsi     {{\ensuremath{{\PJ\mskip -3mu/\mskip -2mu\Ppsi}}}\xspace}

\def\Y#1S{\ensuremath{\PUpsilon{(#1S)}}\xspace}

\def\proton      {{\ensuremath{\Pp}}\xspace}

\def\Lz          {{\ensuremath{\PLambda}}\xspace}

\def\LorLbar     {\kern \thebaroffset\optbar{\kern -\thebaroffset \PLambda}\xspace}

\newcommand{\decay}[2]{\ensuremath{#1\!\to #2}\xspace} 

\def\to                 {\ensuremath{\rightarrow}\xspace}

\def\BdKstGam     {\decay{\Bd}{\Kstarz \g}}

\def\BdKstee  {\decay{\Bd}{\Kstarz\epem}}

\def\AT#1     {\ensuremath{A_{\mathrm{T}}^{#1}}\xspace}           

\def\C#1      {\ensuremath{\mathcal{C}_{#1}}\xspace}                       
\def\Cp#1     {\ensuremath{\mathcal{C}_{#1}^{'}}\xspace}                    
\def\Ceff#1   {\ensuremath{\mathcal{C}_{#1}^{\mathrm{(eff)}}}\xspace}        
\def\Cpeff#1  {\ensuremath{\mathcal{C}_{#1}^{'\mathrm{(eff)}}}\xspace}       
\def\Ope#1    {\ensuremath{\mathcal{O}_{#1}}\xspace}                       
\def\Opep#1   {\ensuremath{\mathcal{O}_{#1}^{'}}\xspace}                    

\newcommand{\nospaceunit}[1]{\ensuremath{\text{#1}}}       
\newcommand{\aunit}[1]{\ensuremath{\text{\,#1}}}       

\newcommand{\tev}{\aunit{Te\kern -0.1em V}\xspace}
\newcommand{\gev}{\aunit{Ge\kern -0.1em V}\xspace}
\newcommand{\mev}{\aunit{Me\kern -0.1em V}\xspace}
\newcommand{\kev}{\aunit{ke\kern -0.1em V}\xspace}
\newcommand{\ev}{\aunit{e\kern -0.1em V}\xspace}
 
\newcommand{\mevc}{\ensuremath{\aunit{Me\kern -0.1em V\!/}c}\xspace}
\newcommand{\gevc}{\ensuremath{\aunit{Ge\kern -0.1em V\!/}c}\xspace}
\newcommand{\mevcc}{\ensuremath{\aunit{Me\kern -0.1em V\!/}c^2}\xspace}
\newcommand{\gevcc}{\ensuremath{\aunit{Ge\kern -0.1em V\!/}c^2}\xspace}

\def\km   {\aunit{km}\xspace}
\def\m    {\aunit{m}\xspace}
\def\ma   {\ensuremath{\aunit{m}^2}\xspace}
\def\cm   {\aunit{cm}\xspace}
\def\cma  {\ensuremath{\aunit{cm}^2}\xspace}
\def\mm   {\aunit{mm}\xspace}
\def\mma  {\ensuremath{\aunit{mm}^2}\xspace}
\def\mum  {\ensuremath{\,\upmu\nospaceunit{m}}\xspace}

\def\nm   {\aunit{nm}\xspace}

\def\nb {\aunit{nb}\xspace}
\def\invnb {\ensuremath{\nb^{-1}}\xspace}

\def\fb   {\ensuremath{\aunit{fb}}\xspace}
\def\invfb   {\ensuremath{\fb^{-1}}\xspace}

\def\sec  {\ensuremath{\aunit{s}}\xspace}
\def\ms   {\ensuremath{\aunit{ms}}\xspace}
\def\mus  {\ensuremath{\,\upmu\nospaceunit{s}}\xspace}
\def\ns   {\ensuremath{\aunit{ns}}\xspace}
\def\ps   {\ensuremath{\aunit{ps}}\xspace}

\def\mhz  {\ensuremath{\aunit{MHz}}\xspace}
\def\khz  {\ensuremath{\aunit{kHz}}\xspace}
\def\hz   {\ensuremath{\aunit{Hz}}\xspace}

\def\degc {\ensuremath{^\circ}{\text{C}}\xspace}
\def\degk {\aunit{K}\xspace}

\def\Xrad {\ensuremath{X_0}\xspace}

\def\neutroneq {\ensuremath{n_\nospaceunit{eq}}\xspace}
\def\neqcmcm {\ensuremath{\neutroneq/\nospaceunit{cm}^2}\xspace}
\def\kRad {\aunit{kRad}\xspace}

\newcommand{\chisq}{\ensuremath{\chi^2}\xspace}

\def\gsim{{~\raise.15em\hbox{$>$}\kern-.85em
          \lower.35em\hbox{$\sim$}~}\xspace}
\def\lsim{{~\raise.15em\hbox{$<$}\kern-.85em
          \lower.35em\hbox{$\sim$}~}\xspace}

\def\pt         {\ensuremath{p_{\mathrm{T}}}\xspace}

\def\ptot       {\ensuremath{p}\xspace}
\def\et         {\ensuremath{E_{\mathrm{T}}}\xspace}

\def\degrees{\ensuremath{^{\circ}}\xspace}
\def\murad{\ensuremath{\,\upmu\nospaceunit{rad}}\xspace}
\def\mrad{\aunit{mrad}\xspace}
\def\rad{\aunit{rad}\xspace}

\newcommand{\lum} {\ensuremath{\mathcal{L}}\xspace}

\def\boole      {\mbox{\textsc{Boole}}\xspace}

\def\davinci    {\mbox{\textsc{DaVinci}}\xspace}
\def\dirac      {\mbox{\textsc{Dirac}}\xspace}
\def\evtgen     {\mbox{\textsc{EvtGen}}\xspace}

\def\fluka      {\mbox{\textsc{Fluka}}\xspace}

\def\gaudi      {\mbox{\textsc{Gaudi}}\xspace}
\def\gauss      {\mbox{\textsc{Gauss}}\xspace}
\def\geant      {\mbox{\textsc{Geant4}}\xspace}

\def\moore      {\mbox{\textsc{Moore}}\xspace}

\def\pythia     {\mbox{\textsc{Pythia}}\xspace}

\def\root       {\mbox{\textsc{Root}}\xspace}

\def\cpp        {\mbox{\textsc{C\raisebox{0.1em}{{\footnotesize{++}}}}}\xspace}

\def\git        {\mbox{\textsc{git}}\xspace}

\def\kbit          {\aunit{kbit}\xspace}

\def\mbit          {\aunit{Mbit}\xspace}
\def\mbps        {\aunit{Mbit/s}\xspace}
\def\mbytes     {\aunit{MB}\xspace}

\def\gbps        {\aunit{Gbit/s}\xspace}
\def\gbytes     {\aunit{GB}\xspace}
\def\gbyps      {\aunit{GB/s}\xspace}

\def\tbps        {\aunit{Tbit/s}\xspace}

\def\tbyps      {\aunit{TB/s}\xspace}

\def\tell1  {TELL1\xspace}
\def\ukl1   {UKL1\xspace}

\def\fpga   {FPGA\xspace}

\def\cfourften     {\ensuremath{\mathrm{ C_4 F_{10}}}\xspace}
\def\cffour        {\ensuremath{\mathrm{ CF_4}}\xspace}
\def\cotwo         {\ensuremath{\mathrm{ CO_2}}\xspace}

\newcommand{\phz}{\phantom{0}}

\newcommand{\lhcborcid}[1]{\href{https://orcid.org/#1}{\hspace*{0.1em}\raisebox{-0.45ex}{\includegraphics[width=1em]{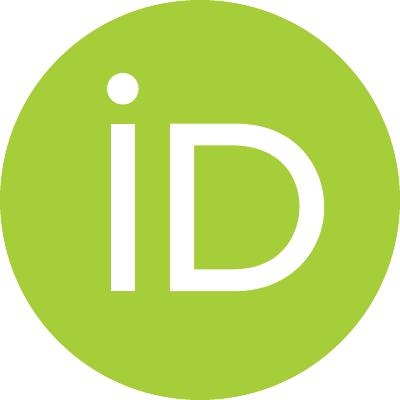}}}}

\usepackage{cite} 
\usepackage{mciteplus}

\usepackage{longtable} 
\usepackage{caption}
\usepackage{subcaption}
\usepackage{wrapfig}
\usepackage[multi-part-units=brackets,exponent-product={\cdot}, separate-uncertainty=true,range-phrase={\text{\ensuremath{-}}}]{siunitx}
\usepackage{cleveref}

\usepackage[normalem]{ulem}

\usepackage[acronym,nogroupskip]{glossaries}
\makeglossaries
\newcommand{\newAcr}[3]{\newacronym{#1}{#2}{#3}}
\newcommand{\Acr}[2][]{%
\ifthenelse{\equal{#1}{}}{{\gls*{#2}}}{
\ifthenelse{\equal{#1}{s}}{\acrshort*{#2}}{
\ifthenelse{\equal{#1}{m}}{\acrlong*{#2}}{
\ifthenelse{\equal{#1}{f}}{\acrfull*{#2}}{
\ifthenelse{\equal{#1}{p}}{\glspl*{#2}}{
\ifthenelse{\equal{#1}{c}}{\Gls*{#2}}{
\ifthenelse{\equal{#1}{P}}{\Glspl*{#2}}{}
}}}}}}%
}
\begin{document}

\renewcommand{\thefootnote}{\fnsymbol{footnote}}
\setcounter{footnote}{1}


\begin{titlepage}
\pagenumbering{roman}

\vspace*{-1.5cm}
\centerline{\large EUROPEAN ORGANIZATION FOR NUCLEAR RESEARCH (CERN)}
\vspace*{1.5cm}
\noindent
\begin{tabular*}{\linewidth}{lc@{\extracolsep{\fill}}r@{\extracolsep{0pt}}}
\ifthenelse{\boolean{pdflatex}}
{\vspace*{-1.5cm}\mbox{\!\!\!\includegraphics[width=.14\textwidth]{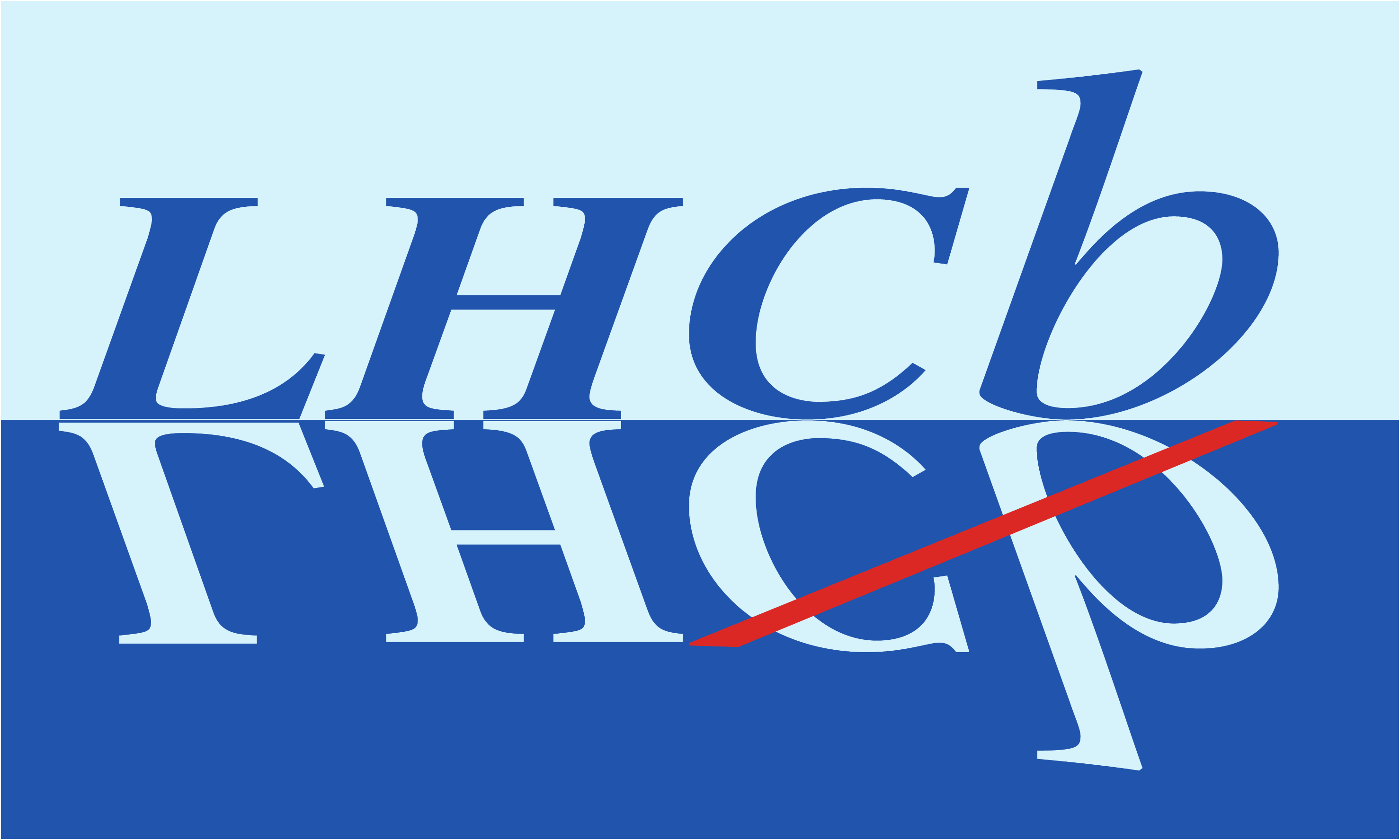}} & &}%
{\vspace*{-1.2cm}\mbox{\!\!\!\includegraphics[width=.12\textwidth]{figs/lhcb-logo.eps}} & &}%
\\
 & & LHCb-DP-2022-002 \\  
 & & May 23, 2024\\ 
 & & \\
\end{tabular*}

\vspace*{4.0cm}

{\normalfont\bfseries\boldmath\huge
\begin{center}
  \papertitle 
\end{center}
}

\vspace*{2.0cm}

\begin{center}
\paperauthors\footnote{Authors are listed at the end of this paper.}
\end{center}

\vspace{\fill}

\begin{abstract}
  \noindent
The LHCb upgrade represents a major change of the experiment. The
detectors have been almost completely renewed to allow running at an
instantaneous luminosity five times larger than that of the previous
running periods. Readout of all detectors into an all-software trigger
is central to the new design, facilitating the reconstruction of
events at the maximum LHC interaction rate, and
their selection in real time. The experiment’s tracking
system has been completely upgraded with a new pixel vertex detector,
a silicon tracker upstream of the dipole magnet and three
scintillating fibre tracking stations downstream of the magnet. The
whole photon detection system of the RICH detectors has been renewed
and the readout electronics of the calorimeter and muon systems have
been fully overhauled. The first stage of the all-software trigger is
implemented on a GPU farm. The output of the trigger provides a
combination of totally reconstructed physics objects, such as tracks
and vertices, ready for final
analysis, and of entire events which need further offline
reprocessing. This scheme required a complete revision of the
computing model and rewriting of the experiment's software.
\end{abstract}

\vspace*{2.0cm}

\begin{center}
  Published in JINST 19 (2024) P05065
\end{center}

\vspace{\fill}

{\footnotesize 
\centerline{\copyright~\papercopyright. \href{\paperlicenceurl}{\paperlicence}.}}
\vspace*{2mm}

\end{titlepage}


\newpage
\setcounter{page}{2}
\mbox{~}
%
%
%
%


\renewcommand{\thefootnote}{\arabic{footnote}}
\setcounter{footnote}{0}

\tableofcontents
\cleardoublepage


\pagestyle{plain} 
\setcounter{page}{1}
\pagenumbering{arabic}

%
\newAcr{lhc}{LHC}{Large Hadron Collider} 
\newAcr{lsone}{LS1}{LHC long shutdown 1}
\newAcr{lstwo}{LS2}{LHC long shutdown 2}
\newAcr{lsthree}{LS3}{LHC long shutdown 3}
\newAcr{prpr}{$pp$}{proton-proton}
\newAcr{heh}{HEH}{high-energy hadrons}
\newAcr{let}{LET}{linear energy transfer}
\newAcr{ip}{IP}{impact parameter}
\newAcr{sps}{SPS}{Super Proton Synchrotron}
\def\Lhc{\Acr{lhc}\xspace}
\def\runone {\mbox{Run 1}\xspace}
\def\runtwo{\mbox{Run 2}\xspace}
\def\runthree{\mbox{Run 3}\xspace}
\def\runfour{\mbox{Run 4}\xspace}
\def\runonetwo{\mbox{Run 1-2}\xspace}
\def\Lsone{\Acr{lsone}\xspace}
\def\Lstwo{\Acr{lstwo}\xspace}
\def\Lsthree{\Acr{lsthree}\xspace}
\def\Prpr{\Acr{prpr}\xspace}
\def\Ip{\Acr{ip}\xspace}
\def\Sps{\Acr{sps}\xspace}
%
\newAcr{velo}{VELO}{vertex locator}
\newAcr{ut}{UT}{upstream tracker}
\newAcr{sft}{SciFi Tracker}{scintillating fibre tracker}
\newAcr{ecal}{ECAL}{electromagnetic calorimeter}
\newAcr{hcal}{HCAL}{hadronic calorimeter}
\newAcr{rich}{RICH}{ring imaging Cherenkov detector}
\newAcr{plume}{PLUME}{probe for luminosity measurement}
\newAcr{bcm}{BCM}{beam conditions monitor system}
\newAcr{rms}{RMS}{radiation monitoring system}
\def\Velo{\Acr{velo}\xspace}
\def\Ut{\Acr{ut}\xspace}
\def\Scifi{\Acr{sft}\xspace}
\def\Ecal{\Acr{ecal}\xspace}
\def\Hcal{\Acr{hcal}\xspace}
\def\Rich{\Acr{rich}\xspace}
\def\Plume{\Acr{plume}\xspace}
\def\Bcm{\Acr{bcm}\xspace}
\def\Rms{\Acr{rms}\xspace}
%
\newAcr{l0}{L0}{level-0 trigger}
\newAcr{hlt}{HLT}{high level trigger}
\newAcr{hltone}{HLT1}{\Acr[m]{hlt} first stage}
\newAcr{hlttwo}{HLT2}{\Acr[m]{hlt} second stage}
\def\L0{\Acr{l0}\xspace}
\def\Hlt{\Acr{hlt}\xspace}
\def\Hltone{\Acr{hltone}\xspace}
\def\Hlttwo{\Acr{hlttwo}\xspace}
\newAcr{daq}{DAQ}{data acquisition system}
\newAcr{ecs}{ECS}{experiment control system}
\newAcr{hdr}{HDR}{high dynamic range}
\newAcr{scada}{SCADA}{supervisory control and data acquisition}
\newAcr{jcop}{JCOP}{joint control project}
\def\Scada{\Acr{scada}\xspace}
\def\Jcop{\Acr{jcop}\xspace}
%
\newAcr{dcs}{DCS}{detector control system}
\newAcr{fend}{FE}{front-end}
\newAcr{bend}{BE}{back-end}
\newAcr{tfc}{TFC}{timing and fast signal control}
\newAcr{asic}{ASIC}{application specific integrated circuit}
\newAcr{bxid}{BXID}{bunch-crossing identifier}
\newAcr{gwt}{GWT}{gigabit wireline transmitter}
\newAcr{gbt}{GBT}{gigabit transceiver}
\newAcr{gbtx}{GBTx}{gigabit transceiver}
\newAcr{gbtsca}{GBT-SCA}{gigabit transceiver slow control adapter}
\newAcr{sca}{SCA}{slow control adapter}
\newAcr{spi}{SPI}{serial peripheral interface}
\newAcr{vttx}{VTTx}{versatile twin transmitter}
\newAcr{vtrx}{VTRx}{versatile link transceiver}
\newAcr{vl}{VL}{versatile link}
\newAcr{tellfty}{TELL40}{PCIe40 board for data acquisition}
\newAcr{solfty}{SOL40}{PCIe40 board for controls}
\newAcr{pcie}{PCIe}{peripheral component interconnect express}
\newAcr{pciefty}{PCIe40}{PCIe generic back-end board}
\newAcr{feastmp} {FEASTMP}{radiation tolerant DC-DC converter} 
\newAcr{elmb}{ELMB}{embedded local monitoring board}
\newAcr{eb}{EB}{event builder}
\def\Daq{\Acr{daq}\xspace}
\def\Ecs{\Acr{ecs}\xspace}
\def\Dcs{\Acr{dcs}\xspace} 
\def\Fend{\Acr{fend}\xspace}
\def\Bend{\Acr{bend}\xspace}
\def\Tfc{\Acr{tfc}\xspace}
\def\Bxid{\Acr{bxid}\xspace}
\def\Gbt{\Acr{gbt}\xspace}
\def\Gwt{\Acr{gwt}\xspace}
\def\Gbtsca{\Acr{gbtsca}\xspace}
\def\Sca{\Acr{sca}\xspace}
\def\Vttx{\Acr{vttx}\xspace}
\def\Vtrx{\Acr{vtrx}\xspace}
\def\Feastmp{\Acr{feastmp}\xspace}
\def\Pcie{\Acr{pcie}\xspace}
\def\Eb{\Acr{eb}\xspace}
\def\Gbtx{\Acr{gbtx}\xspace}
\def\Asic{\Acr{asic}\xspace}
\def\Pciefourty{\Acr{pciefty}\xspace}
\def\Tellfourty{\Acr{tellfty}\xspace}
\def\Solfourty{\Acr{solfty}\xspace}
\newAcr{dss}{DSS}{detector safety system}
\newAcr{mss}{MSS}{magnet safety system}
\newAcr{ups}{UPS}{uninterruptible power supply}
\newAcr{pacl}{2PACL}{2-phase accumulator controlled loop}
\def\Novec{Novec\xspace} 
\def\aside{Side A\xspace}
\def\cside{Side C\xspace}
%
\newAcr{cmos}{CMOS}{complementary metal-oxide semiconductor}
\newAcr{lvcmos}{LVCMOS}{low-voltage complementary metal-oxide semiconductor}
\newAcr{fpga}{FPGA}{field-programmable gate array}
\newAcr{gwp}{GWP}{global warming potential}
\newAcr{pmt}{PMT}{photomultiplier tube}
\newAcr{mapmt}{MaPMT}{multi-anode photomultiplier tube}
\newAcr{pid}{PID}{particle identification}
\newAcr{qa}{QA}{quality assurance} %
\newAcr{hv}{HV}{high voltage}
\newAcr{lv}{LV}{low voltage}
\newAcr{feb}{FEB}{front-end board}
\newAcr{i2c}{I2C}{inter-integrated cicuit} 
\newAcr{jtag}{JTAG}{Joint Test Action Group industry standard}
\newAcr{gpio}{GPIO}{general purpose input/output}
\newAcr{mpo}{MPO}{multifibre push on connector}
\newAcr{lcc}{LC}{Lucent connector}
\newAcr{see}{SEE}{single event effect}
\newAcr{seu}{SEU}{single event upset}
\newAcr{sel}{SEL}{single event latchup}
\newAcr{pll}{PLL}{phase-locked loop}

\def\Fpga{\Acr{fpga}\xspace}
\def\Pmt{\Acr{pmt}\xspace}
\def\Mapmt{\Acr{mapmt}\xspace}
\def\Feb{\Acr{feb}\xspace}
\def\dcdc{DC-DC\xspace}
\def\Seu{\Acr{seu}\xspace}
\def\See{\Acr{see}\xspace}
\def\Sel{\Acr{sel}\xspace}
\def\Pid{\Acr{pid}\xspace}
\def\Qa{\Acr{qa}\xspace}
\def\Hv{\Acr{hv}\xspace}
\def\Lv{\Acr{lv}\xspace}
\def\I2c{\Acr{i2c}\xspace}
\def\elink{e-link\xspace}
\def\Pll{\Acr{pll}\xspace}
%
\def\pthundred{Pt100\xspace}
\def\ptthousand{Pt1000\xspace}
\newcommand{\sigmaIP}{\ensuremath{\sigma_\text{IP}}} 
\newcommand{\sigmaMS}{\ensuremath{\sigma_\text{MS}}}
\newAcr{velopix}{VeloPix}{VELO pixel chip}
\newAcr{spp}{SPP}{super-pixel packet}
\newAcr{neg}{NEG}{non-evaporable getter}
\newAcr{opb}{OPB}{opto- and power board}
\def\velopix{\Acr{velopix}\xspace}
\def\Vetra{\mbox{\textsc{Vetra}}\xspace}
\def\Storck{\mbox{\textsc{Storck}}\xspace}
\def\Titania{\mbox{\textsc{Titania}}\xspace}
\newAcr{salt} {SALT}{Silicon ASIC for LHCb Tracking}
\newAcr{pa}{PA}{pitch adapter}
\def\vera{VERA\xspace}
\def\susi{SUSI\xspace}
\def\dataflex{dataflex\xspace}
\newAcr{sipm}{SiPM}{silicon photomultiplier}
\def\Sipm{\Acr{sipm}\xspace}
\newAcr{scifi}{SciFi}{scintillating fibre} 
\def\dV {\ensuremath{\rm \Delta V}\xspace}
\newcommand{\dVeq}[1]{\ensuremath{\rm \Delta V=#1\,V}}
\def\vbd{\ensuremath{\rm{V_{BD}}}\xspace}
\def\vbdI{\ensuremath{\rm{V_{BD}^I}}\xspace}
\def\vbdAmp{\ensuremath{\rm{V_{BD}^{Amp}}}\xspace}
\def\vbdInt{\ensuremath{\rm{V_{BD}^{Int}}}\xspace}
\def\vbdG{\ensuremath{\rm{V_{BD}^G}}\xspace}
\def\vbias{\ensuremath{\rm{V_{bias}}}\xspace}
\newcommand\weightperweight[1]{#1\%~w/w}
\def\pox{$\rm{p_{DiXT}}$\xspace}
\newcommand{\poxeq}[1]{\ensuremath{\rm{p_{DiXT}=#1\,\%}}}
\def\pdx{$\rm{p_{DeXT}}$\xspace}
\newcommand{\pdxeq}[1]{\ensuremath{\rm{p_{DeXT}=#1\,\%}}}
\def\pap{$\rm{p_{AP}}$\xspace}
\newcommand{\papeq}[1]{\ensuremath{\rm{p_{AP}=#1\,\%}}}
\def\trec{$\rm{\tau_{rec}}$\xspace}
\newcommand{\treceq}[1]{\ensuremath{\rm{\tau_{rec}=#1\,ns}}}
\def\tlong{$\rm{\tau_{long}}$\xspace}
\newcommand{\tlongeq}[1]{\ensuremath{\rm{\tau_{long}=#1\,ns}}}
\def\tshort{$\rm{\tau_{short}}$\xspace}
\def\tfast{$\rm{\tau_{fast}}$\xspace}
\def\tap{$\rm{\tau_{AP}}$\xspace}
\newcommand{\tapeq}[1]{\ensuremath{\rm{\tau_{AP}=#1\,ns}}}
\def\tdx{$\rm{\tau_{DeXT}}$\xspace}
\newcommand{\tdxeq}[1]{\ensuremath{\rm{\tau_{DeXT}=#1\,ns}}}
\def\RQ{$\rm{R_Q}$\xspace}
\newcommand{\RQeq}[1]{\ensuremath{\rm{R_Q=#1\,k\Omega}}}
\def\Plum{$\rm{P_{lum}}$\xspace}
\newAcr{dcr}{DCR}{dark count rate}
\def\fdcr{\Acr{dcr}\xspace} 
\newAcr{niel}{NIEL}{non-ionising energy loss}
\newAcr{sey}{SEY}{secondary electron yield}
\newAcr{gfs}{GFS}{gas feed system}
\newAcr{smog}{SMOG}{system for measuring the overlap with gas}
\newAcr{stp}{STP}{standard temperature and pressure}
\newAcr{hpd}{HPD}{hybrid photon detector}
\newAcr{sin}{SIN}{signal-induced noise}
\newAcr{qe}{QE}{quantum efficiency}
\newAcr{ec}{EC}{elementary cell}
\newAcr{ecr}{EC-R}{R-type elementary cell}
\newAcr{ech}{EC-H}{H-type elementary cell}
\newAcr{pdmdb}{PDMDB}{photon detector module digital board}
\newAcr{pdm}{PDM}{photon detector module}
\newAcr{tcm}{TCM}{timing and control module}
\newAcr{dtm}{DTM}{data transmission module}
\newAcr{ams}{AMS}{Austria Micro Systems}
\newAcr{cfrp}{CFRP}{carbon fibre reinforced polymer}
\newAcr{coc}{CoC}{centre of curvature}
\def\Stp{\Acr{stp}\xspace}
\def\Sin{\Acr{sin}\xspace}
\def\Qe{\Acr{qe}\xspace}
\def\Ecr{\Acr{ecr}\xspace}
\def\Ech{\Acr{ech}\xspace}
\def\Pdm{\Acr{pdm}\xspace}
\def\Hpd{\Acr{hpd}\xspace}
\def\Ecel{\Acr{ec}\xspace}
\def\Pdmdb{\Acr{pdmdb}\xspace}
\def\Tcm{\Acr{tcm}\xspace}
\def\Dtm{\Acr{dtm}\xspace}
\newcommand{\ckvAngle}{\mbox{$\theta_{\text{C}}$}\xspace}
\newcommand{\ckvYield}{\mbox{$N_{\text{ph}}$}\xspace}
\newcommand{\ckvYieldOptimal}{\mbox{$N_{\text{ph}}^{\text{optimal}}$}\xspace}
\newcommand{\ckvYieldTypical}{\mbox{$N_{\text{ph}}^{\text{typical}}$}\xspace}
\newcommand{\ckvResTotal}{\mbox{$\Delta\ckvAngle$}\xspace}
\newcommand{\ckvResTracking}{\mbox{$C_{\text{tracking}}$}\xspace}
\newcommand{\ckvResSinglePhoton}{\mbox{$\sigma_{\theta}$}\xspace}
\newAcr{cw}{CW}{Cockroft Walton}
\newAcr{llt}{LLT}{low level trigger}
\newAcr{cccu}{3CU}{calorimeter control card unit}
\newAcr{ledtsb}{LEDTSB}{LED trigger signal board}
\def\Cccu{\Acr{cccu}\xspace}
\def\Cw{\Acr{cw}\xspace}
\def\Llt{\Acr{llt}\xspace}
\def\Ledtsb{\Acr{ledtsb}\xspace}
\newAcr{mwpc}{MWPC}{multi-wire proportional chambers}
\newAcr{gem}{GEM}{gas electron multiplier}
\newAcr{carioca}{CARIOCA}{CERN and Rio current amplifier}
\newAcr{dialog}{DIALOG}{diagnostic, time adjustment and logics}
\newAcr{ode}{ODE}{off-detector electronics}
\newAcr{node}{nODE}{new off-detector electronics}
\newAcr{nsync}{nSYNC}{new SYNC ASIC}
\newAcr{sb}{SB}{service board}
\newAcr{nsb}{nSB}{new service board}
\newAcr{nsbs}{nSBS}{new service board system}
\newAcr{npdm}{nPDM}{new pulse distribution module}
\newAcr{ncb}{nCB}{new custom backplane}
\newAcr{fec}{FEC}{forward error correction}
\newAcr{dco}{DCO}{digitally controlled oscillator}
\def\Mwpc{\Acr{mwpc}\xspace}
\def\Gem{\Acr{gem}\xspace}
\def\Carioca{\Acr{carioca}\xspace}
\def\Dialog{\Acr{dialog}\xspace}
\def\Ode{\Acr{ode}\xspace}
\def\Node{\Acr{node}\xspace}
\def\Nsync{\Acr{nsync}\xspace}
\def\Sb{\Acr{sb}\xspace}
\def\Nsb{\Acr{nsb}\xspace}
\def\Nsbs{\Acr{nsbs}\xspace}
\def\Npdm{\Acr{npdm}\xspace}
\def\Ncb{\Acr{ncb}\xspace}
\def\Dco{\Acr{dco}\xspace}
\newAcr{gpu}{GPU}{graphics processing unit}
\newAcr{gpgpu}{GPGPU}{general purpose computing on graphics processing unit}
\def\online{online\xspace}
\def\sodin{SODIN\xspace}
\def\Gpu{\Acr{gpu}\xspace}
\def\Gpgpu{\Acr{gpgpu}\xspace}
\newAcr{pv}{PV}{primary vertex}
\newAcr{gec}{GEC}{Global Event Cut}
\def\Pv{\Acr{pv}\xspace}
\def\Gec{\Acr{gec}\xspace}
\newAcr{vo}{VO}{virtual organization}
\newAcr{hpc}{HPC}{high performance computing}
\def\Vo{\Acr{vo}\xspace}
\def\Hpc{\Acr{hpc}\xspace}
%
\def\invsec{\ensuremath{\sec^{-1}}\xspace}
\def\invcma {\ensuremath{\cm^{-2}}\xspace}
\def\volt{\aunit{V}\xspace}
\def\kvolt{\aunit{kV}\xspace}
\def\Mvolt{\aunit{MV}\xspace}
\def\mvolt{\aunit{mV}\xspace}
\def\watt{\aunit{W}\xspace}
\def\kwatt{\aunit{kW}\xspace}
\def\Mwatt{\aunit{MW}\xspace}
\def\mwatt{\aunit{mW}\xspace}
\def\ohm {\aunit{$\Omega$}\xspace}
\def\kohm {\aunit{k$\Omega$}\xspace}
\def\Mohm {\aunit{M$\Omega$}\xspace}
\def\muamp{\ensuremath{\,\upmu\nospaceunit{A}}\xspace}
\def\muampcmcm{\ensuremath{\,\upmu\nospaceunit{A}}/\nospaceunit{cm}^2\xspace}
\def\namp {\aunit{nA}\xspace}
\def\mamp {\aunit{mA}\xspace}
\def\amp{\aunit{A}\xspace}
\def\bar{\aunit{bar}\xspace}
\def\mbar {\aunit{mbar}\xspace}
\def\MHz{\aunit{MHz}\xspace}
\def\MeV{\aunit{Me\kern -0.1em V}\xspace}
\def\Gy{\aunit{Gy}\xspace}
\def\cGy{\aunit{cGy}\xspace}
\def\kGy{\aunit{kGy}\xspace}
\def\MGy{\aunit{MGy}\xspace}
\def\litre {\aunit{l}\xspace} 
\def\mv{\ensuremath{\aunit{m}^3}\xspace}
\def\muvis{\ensuremath{\mu_{\rm vis}}}
\def\Mps{\aunit{million/s}\xspace}
%
\newAcr{ip8}{IP8}{interaction point 8}
\newAcr{dm}{DM}{dark matter}
\newAcr{dsp}{DSP}{digital signal processing}
\newAcr{dspor}{DSP}{digital signal processor}
%
\def\turbo{\mbox{\texttt{Turbo}}\xspace}
\def\FULL{\mbox{\texttt{FULL}}\xspace}
\def\turcal {\mbox{\texttt{TurCal}}\xspace}
\def\git{\mbox{\textsc{Git}}\xspace}
\def\gitlab{\mbox{\textsc{GitLab}}\xspace}
\def\github{\mbox{\textsc{GitHub}}\xspace}
\def\jenkins {\mbox{\textsc{Jenkins}}\xspace}
\def\tesla{\mbox{\textsc{Tesla}}\xspace}
\def\moore{\mbox{\textsc{Moore}}\xspace}
\def\root{\mbox{\textsc{Root}}\xspace}
\def\geant{\mbox{\textsc{Geant4}}\xspace}
\def\gauss{\mbox{\textsc{Gauss}}\xspace}
\def\gaussino{\mbox{\textsc{Gaussino}}\xspace}
\def\davinci {\mbox{\textsc{DaVinci}}\xspace}
\def\boole{\mbox{\textsc{Boole}}\xspace}
\def\smipp{\mbox{\textsc{SMI++}}\xspace}
\def\monet{\mbox{\textsc{Monet}}\xspace}
\def\jira {\mbox{\textsc{Jira}}\xspace}
\def\lhcbpr {\mbox{LHCb\textsc{PR}}\xspace}
\def\sprucing {\mbox{\texttt{sprucing}}\xspace}
\def\pythiaeight {\mbox{\textsc{Pythia8}}\xspace}
\def\redecay {\mbox{\textsc{ReDecay}}\xspace}
\def\lamarr {\mbox{\textsc{Lamarr}}\xspace}
\def\allen{\mbox{\textsc{Allen}}\xspace}
\def\ddforhep{\mbox{\textsc{DD4HEP}}\xspace}

%
\newcommand{\Trmk}[1]{#1\textsuperscript{\tiny TM}\xspace}
\newcommand{\ourfootnote}[1]{\footnote{\,#1}}
\def\mytimes{\times} 
\def\pHe{\ensuremath{\proton\text{He}}\xspace}
\def\pAr{\ensuremath{\proton\text{Ar}}\xspace}
\def\pp{\ensuremath{\proton\proton}\xspace}
\def\Ctot{C_{\rm tot}}


\section{Introduction}
\label{sec:introduction}
The \lhcb experiment~\cite{LHCb-DP-2008-001} is one of the four large
detectors at the \Lhc accelerator at CERN, and its primary purpose is
to search for new physics through studies of CP-violation and decays
of heavy-flavour hadrons. Although \lhcb was designed primarily for
precision measurements in heavy-flavour physics, the experiment has
demonstrated excellent capabilities in many other domains ranging from
electroweak physics to heavy ion and fixed target physics. The LHCb
Upgrade experiment has been designed with this wider physics programme
in mind as a general purpose experiment covering the forward region.
\lhcb has been successfully operated from 2010 to 2018 during the \Lhc
\runone (2010--2012) and \runtwo (2015--2018) data-taking periods with
excellent performance~\cite{LHCb-DP-2014-002}, collecting a total of
9\invfb of \Prpr data, about 30\invnb of lead-lead and \proton-lead
collisions and about 200\invnb of fixed target data.

Notwithstanding this considerable data set, the precision on many of
the key flavour physics observables studied and measured by \lhcb
remains statistically limited, as discussed in detail in
ref.~\cite{LHCb-PAPER-2012-031}, thus requiring significantly larger
data sets to probe the Standard Model at the level of precision
achieved by theoretical calculations and obtain the required
sensitivity to observe possible new physics effects.

While originally designed to take data at a maximum instantaneous
luminosity\break \mbox{$\lum = 2\times 10^{32}\invcma\invsec$} to keep the
average number of visible primary \Prpr interactions (\emph{pile-up})
close to unity~\cite{LHCbtechprop}, \lhcb has been successfully
operated for most of \runone and \runtwo at
$\lum\sim 4\times 10^{32}\invcma\invsec$, demonstrating the capability
to run and to produce excellent physics results at higher luminosity
and with a pile-up larger than initially foreseen.  A proposal for a
major upgrade to operate LHCb at substantially larger instantaneous
luminosity than \runonetwo was thus formalised in a Letter of
Intent~\cite{CERN-LHCC-2011-001} and detailed in a Framework
TDR~\cite{LHCb-TDR-012}.  The physics motivations for this upgrade
have been discussed in great detail in
refs.~\cite{LHCb-PAPER-2012-031,CERN-LHCC-2011-001,LHCb-TDR-012},
assuming an expected total luminosity of $\sim 50\invfb$ integrated by
the end of LHC \runfour.

\begin{figure}[h]
  \centering
  \includegraphics[width=0.48\linewidth]{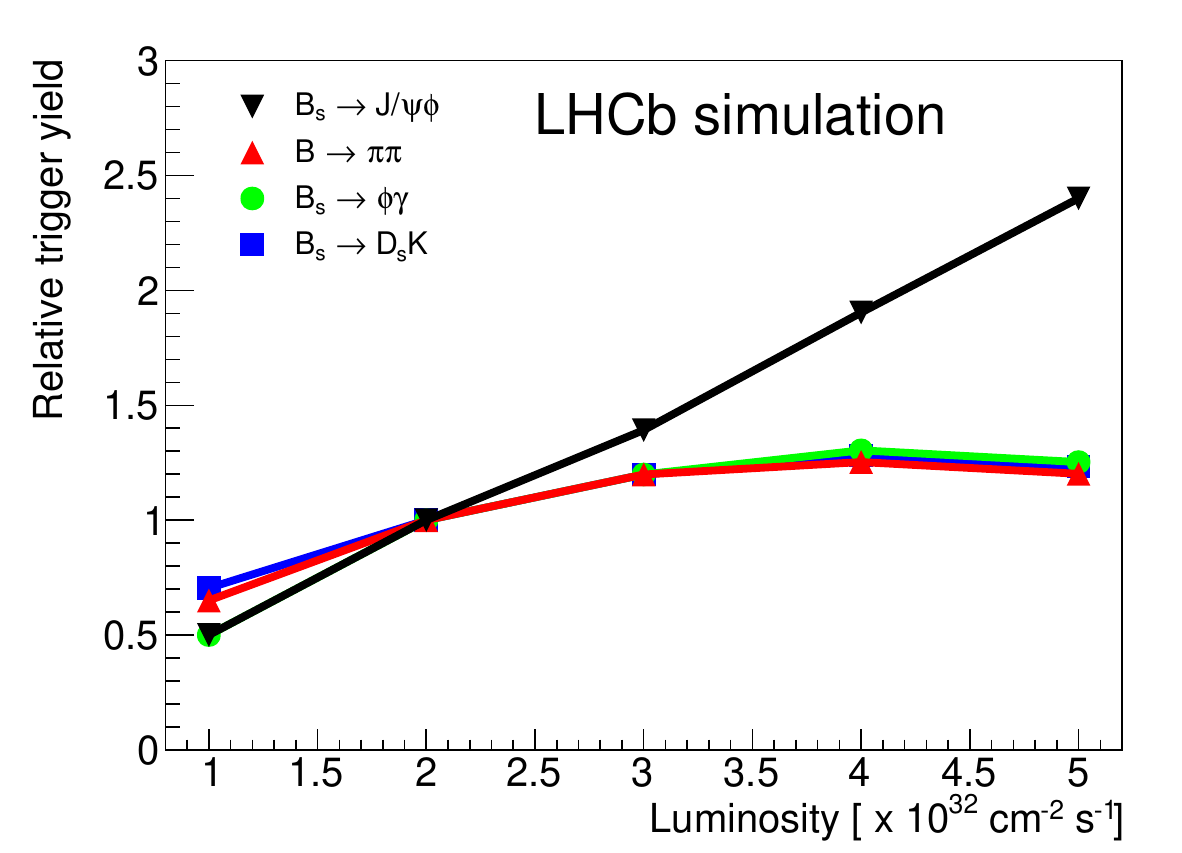}\hfill
  \includegraphics[width=0.48\linewidth]{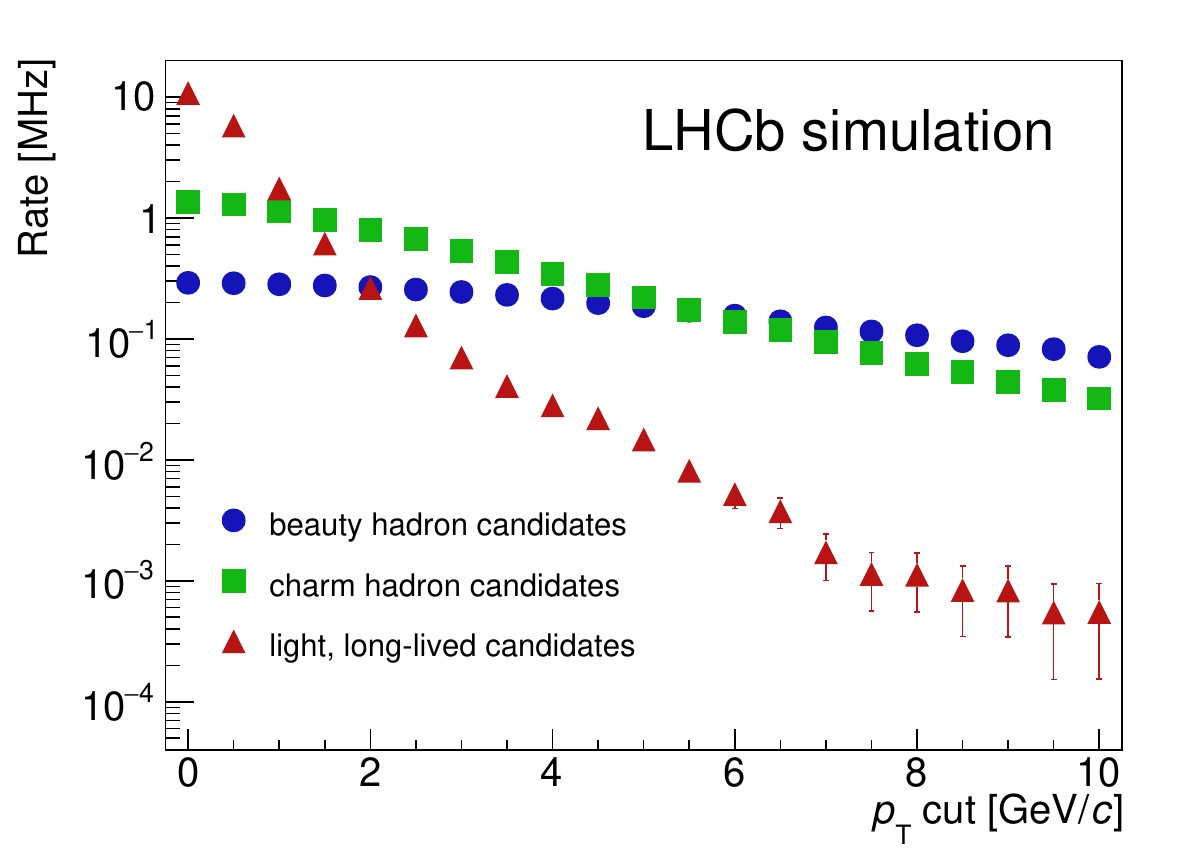}
  \caption{Left: relative trigger yields as a function of
    instantaneous luminosity, normalised to
    $\lum = 2\times 10^{32}\invcma\invsec$. Reproduced from~\cite{CERN-LHCC-2011-001}. CC BY 4.0. Right:~rate of decays
    reconstructed in the \lhcb acceptance as a function of the cut in
    \pt of the decaying particle, for decay time $\tau > 0.2\ps$. Reproduced from~\cite{Fitzpatrick:1670985}. CC BY 4.0.}
  \label{fig:l0limit}
\end{figure}

The LHCb \runonetwo system design would not allow a significant
increase in statistics, especially for fully hadronic final state
decays, the main limitation coming from the maximum allowed output
rate of the first trigger stage, the L0, implemented in
hardware~\cite{LHCb-DP-2012-004}.  The simple inclusive selection
criteria implemented in the L0 trigger stage, based essentially on
particle transverse momentum, would result in an effective loss of
efficiency with increasing luminosity, especially for the most
abundant processes with hadrons in the final state, and in the
saturation of the event yield, as clearly visible in the left panel of
figure~\ref{fig:l0limit}.

In addition, inclusive flavour physics signals have relatively large
cross sections and, at the upgrade luminosity, every event in the
\lhcb acceptance will contain on average two long-lived hadrons not
containing heavy
quarks~\cite{Fitzpatrick:1670985,LHCb-TDR-016}. Therefore, simple
cuts based on displaced vertices or on \pt would be either not
effective in rejecting background or, once enough purity is reached,
would amount to downscaling the signal as shown in the right panel of
figure~\ref{fig:l0limit}.  Profiting from a higher luminosity to
collect significantly more data is therefore only possible by removing
the L0 trigger stage and introducing selections that are more
discriminating than simple inclusive criteria. In particular a
full-software trigger discriminating signal channels based on the full
event reconstruction has been deemed essential for this strategy.

Based on these considerations the \lhcb upgrade has been designed to
run at a nominal instantaneous luminosity
$\lum = 2\times 10^{33}\invcma\invsec$ and to collect events at the
LHC crossing rate of 40\mhz. The events are discriminated by an
all-software trigger reconstructing in real time all events at the
visible interaction rate of $\sim 30\mhz$. By increasing the
instantaneous luminosity by a factor of five and improving the trigger
efficiency for most modes by a factor of two~\cite{LHCb-TDR-016}, the
annual yields in most channels will be an order of magnitude larger
than for the previous LHCb experiment. A total integrated luminosity
(including \runone and \runtwo) of around $50\invfb$ is expected by the
end of \runfour of the LHC.

The new trigger strategy, the higher luminosity and correspondingly
higher pile-up required a complete renewal of the \lhcb detectors and
readout electronics that are now able to read events at the 40\mhz LHC
bunch crossing rate and cope with the larger event multiplicity thanks
to a higher granularity. A full revision of the experiment's software
and of the data processing and computing strategy was also necessary
to deal with the expected large increase in data volume.

This paper describes the design and construction of the upgraded LHCb
experiment providing details on all the new subdetectors, on the
trigger and online systems and on the software and data processing
frameworks.

\section{The LHCb detector} 
\label{sec:infrastructure}
\subsection{Detector layout}

\begin{figure}[h]
  \centering
  \includegraphics[width=0.9\textwidth]{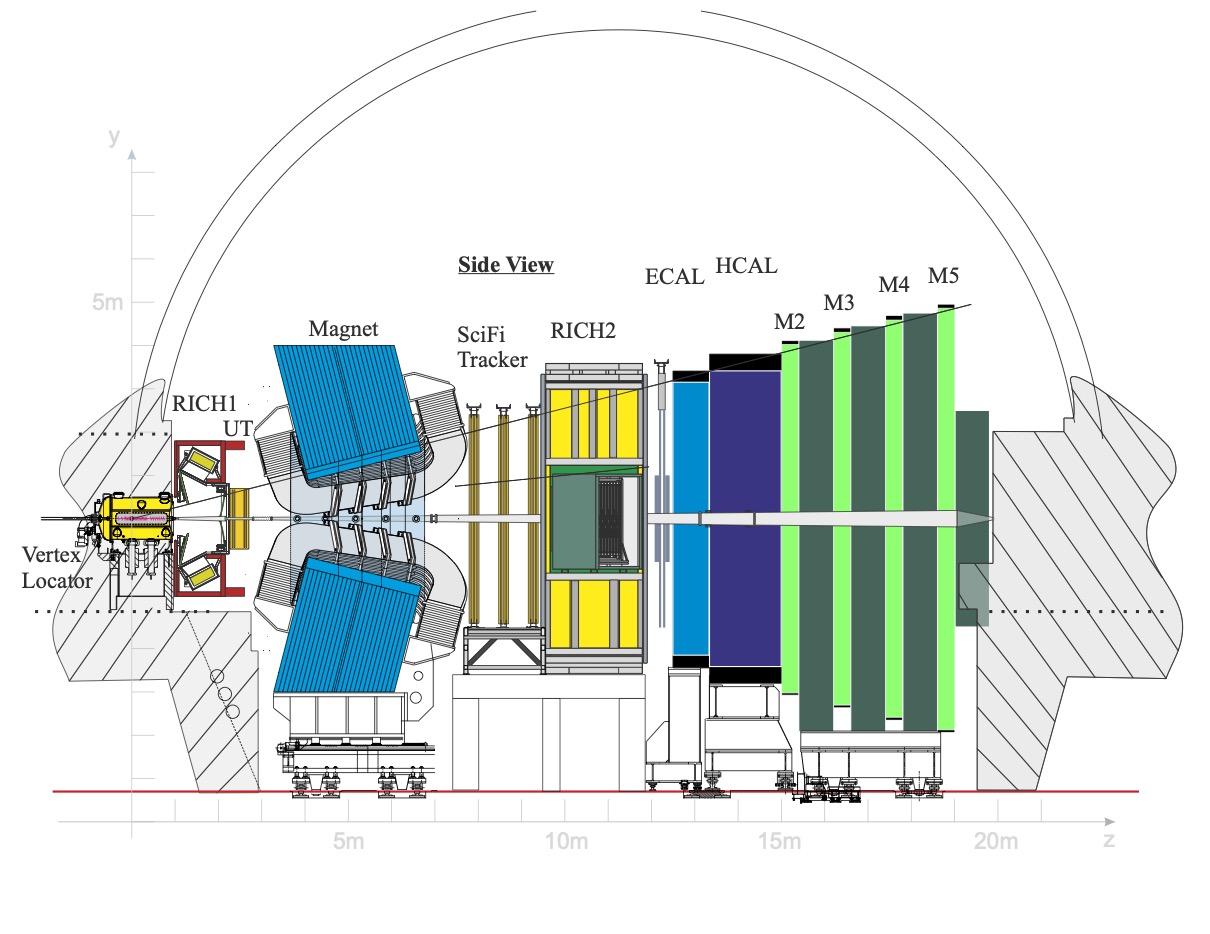}
  \caption{Layout of the upgraded \lhcb detector.}
  \label{fig:lhcb_layout}
\end{figure}

\lhcb is a single-arm forward spectrometer covering the pseudorapidity
range $2 < \eta < 5$, located at interaction point number 8 on the
\Lhc ring.  Figure~\ref{fig:lhcb_layout} shows the layout of the
upgraded detector.  The coordinate system used throughout this paper
has the origin at the nominal \Prpr interaction point, the $z$ axis
along the beam pointing towards the muon system, the $y$ axis pointing
vertically upward and the $x$ axis defining a right-handed system.
Most of the subdetector elements (with the notable exception of vertex
and Cherenkov detectors) are split into two mechanically independent
halves (the \emph{access side} or \emph{\aside} at $x>0$ and the
\emph{cryogenic side} or \emph{\cside} at $x<0$), which can be opened
for maintenance and to guarantee access to the beam pipe.

The particle tracking system comprises an array of pixel silicon
detectors surrounding the interaction region called \Velo, the
silicon-strip \Ut in front of the large-aperture dipole magnet, and
three \Scifi stations downstream of the magnet.\footnote{Upstream and
  downstream are intended in the direction of increasing $z$.}  All
three subsystems were designed to comply with the 40\mhz readout
architecture and to address the challenges associated with the
increased luminosity.  The upgraded \Velo, based on hybrid silicon
pixel detectors, is described in section~\ref{sec:velo}, and the \Ut
is described in section~\ref{sec:ut}.  The \Scifi, which replaces both
the straw-tube Outer Tracker and silicon-strip Inner Tracker systems
used in the downstream tracking stations in the original \lhcb
experiment, is described in section~\ref{sec:scifi}.

The \Pid is provided by two {\Acr[m]{rich}}s (\richone and \richtwo)
using C$_4$F$_{10}$ and CF$_4$ gases as radiators, a shashlik-type
\Ecal, an iron-scintillator tile sampling \Hcal, and four stations of
muon chambers (M2--5) interleaved with iron shielding.\footnote{The
  muon detector consisted of five stations of which the first (M1) has
  been removed --- see text. For historical reasons the remaining
  stations kept their original names.} The Scintillating Pad Detector
and Pre-Shower, which were part of the previous calorimeter system, as
well as the most upstream muon station, have been removed due to their
reduced role in the full software trigger compared to the former
hardware L0.  The upgraded \Acr[p]{rich} are described in
section~\ref{sec:rich}, the calorimeters are described in
section~\ref{subsec:calorimeters}, and the muon system is described in
section~\ref{subsec:muon}.

The \Daq comprises the \Fend and \Bend electronics connected by
long-distance optical links, the event-builder and the event-filter
farms, both described in section~\ref{sec:online}.

\subsection{Magnet}

The spectrometer's dipole magnet has been maintained unchanged with
respect to \runonetwo. It provides a vertical magnetic field with a
bending power of $\simeq 4$\aunit{Tm}.  It consists of two identical,
saddle-shaped coils, which are mounted symmetrically inside a
window-frame yoke.  To match the detector acceptance, the pole gap
increases both vertically and horizontally towards the downstream
tracking stations.  Detailed descriptions of the magnet design can be
found in refs.~\cite{MagnetTDR,Andre2002,Andre2004}.

Each coil is made from five triplets of aluminium pancakes and is
supported by cast aluminium clamps fixed to the yoke.

The initial magnetic field map was determined based on a set of
measurement campaigns (prior to \runone) complemented by
finite-element simulations.  Subsequent measurements for limited
regions inside the magnet, were carried out in 2011, 2014 and 2021,
and were used to apply corrections to the field map.

During data taking, the magnet polarity is reversed regularly (every
few weeks) to collect data sets of roughly equal size with the two
field configurations.

\subsection{Electronics architecture}
\label{sec:electronics_architecture}

\begin{figure}[h]
  \centering
  \includegraphics[width=\textwidth]{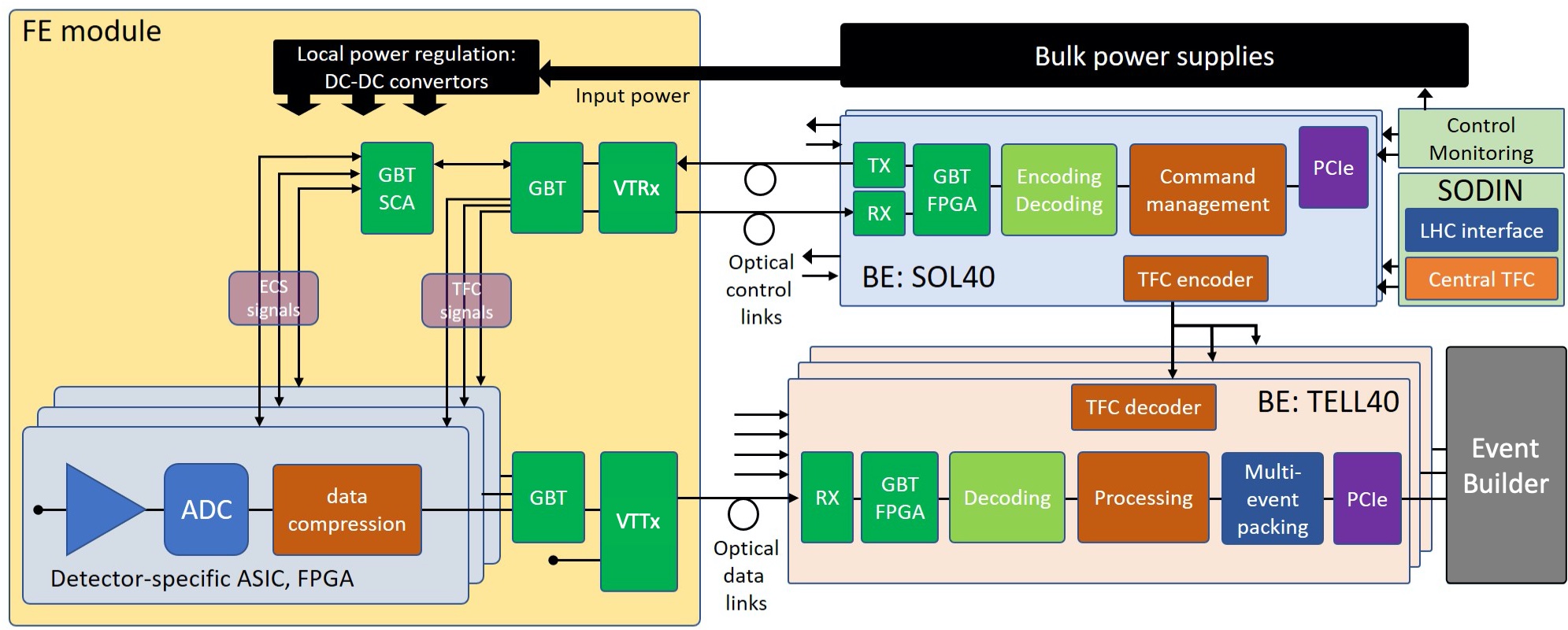}
  \caption{Electronics architecture of the upgraded \lhcb experiment. Reproduced from~\cite{Wyllie:2813379}. CC BY 4.0.}
  \label{fig:electronics_architecture}
\end{figure}

The architecture of LHCb Upgrade I is designed to transmit data
collected from every bunch crossing all the way to the event-builder
computing farm.  To implement this architecture, \lhcb maximised the
use of common building blocks to benefit from a unified
approach. Common developments for the \Lhc experiment upgrades, such
as radiation-tolerant optical links, have proven to be vital enabling
technologies for the new \lhcb detector.  The general architecture is
shown in figure~\ref{fig:electronics_architecture}. The \Fend
electronics amplify and shape the signals generated within the
particle detectors.

These signals are digitised and optically transmitted off the
detector.  All components of the \Fend electronics are located on or
close to the detector, and are therefore exposed to beam-induced
radiation.  Hence, they are all radiation tolerant by design and/or
qualified for the radiation environment in which they operate.  The
\Bend electronics are situated in a data centre on the surface and are
connected to the \Fend in the cavern by 250\m-long optical fibres. The
\Bend electronics preprocess and format the data for transmission to
the event builder. The data centre is a radiation-free environment and
commercial components have been used for the implementation of the
\Bend.  Clocks and fast, beam-synchronous commands are distributed by
a \Tfc system. The \Ecs configures and monitors the \Bend and
\Fend. The \Ecs implements also the slow controls like for example
\Hv, \Lv and temperature monitoring. The \Tfc and \Ecs systems are
described in sections~\ref{sec:tfc}
and~\ref{sec:experiment-control-system}.  \lhcb requires a much larger
bandwidth for data than for \Tfc and \Ecs signals. Even though the
radiation-tolerant optical links were conceived to provide data
transmission, \Tfc and \Ecs functionality in the same link, \lhcb has
separated these functions to maintain a strict independence between
data and controls. This has allowed a minimisation of the number of
links and construction of a modular system with clear boundaries
between functions. Hence, data from the detectors are transmitted on
dedicated unidirectional links whilst the \Tfc and \Ecs communications
to the \Fend electronics are merged onto a much smaller number of
separate bidirectional links.  The \Fend electronics are a mixture of
customised components for each subdetector and common components used
across all systems. The analog electronics connected to the detectors
have all been implemented as \Acr[p]{asic} and profit from the
intrinsic radiation tolerance of deep-submicron \Acr[s]{cmos}
technology. Data are digitised and compressed either in the \Asic or,
if the radiation levels allow, in commercially available
\Acr[p]{fpga}. All such \Fend digital electronics have been
implemented to resist \Acr[p]{see} by using techniques such as
triple-modular redundancy. Data compression was introduced into the
architecture to minimise the number of data links, and algorithms
ranging from simple zero-suppression to hit clustering have been
successfully implemented by different detectors. However, compression
offers little advantage when the channel occupancy is high and so has
not been used in some regions of the experiment. The relatively modest
radiation level in many parts of \lhcb has allowed the widespread use
of \Fpga{s} although only after careful qualification
procedures. These have brought many advantages such as shorter design
time and flexibility, as well as relaxing the demand for specialised
personpower and tools required for \Asic implementations. Efficient
data compression comes at the cost of variable latency, thus the data
transport system in \lhcb is asynchronous. To allow the proper
reconstruction of event fragments downstream in the system, data
packets are tagged as early as possible with a unique time-stamp based
on the \Bxid. Simplex data links, running at 4.8\gbps, are constructed
from the \Gbt serialiser/deserialiser \Asic~\cite{Moreira2009} and the
\Vttx opto-electrical converter~\cite{Amaral2009}. The only exception
is the \Velo, where the serialiser is embedded directly in the \Fend
\Asic. The duplex \Tfc and \Ecs links are constructed with the \Gbt,
\Vtrx and \Sca~\cite{Caratelli2015}. This standardisation across \lhcb
has had major benefits in easing both development and deployment, as
well as the sharing of experience across the wider \Lhc
community. Power distribution has followed a similar common approach
with the use of \Acr[s]{feastmp} \dcdc converters~\cite{Faccio2010}
for local power regulation. The specific implementations of the \Fend
architecture by each subdetector are described in subsequent sections.
The \Bend electronics consist of custom PCI-express modules mounted in
PC servers in the data centre. This module, known generically as
\Pciefourty, contains arrays of optical transmitters and receivers
connected to a powerful \Fpga. The \Pciefourty was conceived and
designed to fulfil the functionality required for both data
acquisition and controls. Hence, by the choice of the \Fpga firmware,
the module can be configured for either data acquisition or
controls. The role of the \Tellfourty is to decode and process data,
and then build multievent packets for transmission to the
event-builder. The \Solfourty is the \Ecs interface used to configure
the \Fend electronics and transmit \Tfc commands to the \Fend and
\Tellfourty{s}. The \Pciefourty also plays the role of interface to
the \Lhc machine timing when configured as a readout supervisor board
(\sodin) board. More details on the \Pciefourty and its functions are
given in section~\ref{sssec:pcie40}.

\subsection{Infrastructure}

A significant part of the \runonetwo \lhcb infrastructure has been
completely refurbished to comply with the upgraded detectors and
modernise old equipment.

\subsubsection{Power distribution}

The power supplies providing low and high voltages are housed in
electronics racks in counting rooms on \aside of the \lhcb cavern,
which is separated from the experimental area by a concrete shielding.
During \Lstwo, some optical fibres were removed to free space in the
long-distance cable trays and additional copper cables, with a total
length of $\sim 36\km$, were installed between the counting rooms and
the detector.

Most of the \lhcb equipment required for operating the experiment is
fed from the same electrical network (the so-called \emph{machine
  network}) used by the LHC\@.  Equipment such as lighting and
overhead cranes is fed from the separate, general services network. In
case of an outage, a change-over from one network to the other can be
performed.  Systems with particularly stringent up-time requirements,
such as the \Acr{dss} or the \Acr{mss}, are connected to an \Acr{ups}
network, with back-up provided by a diesel generator.  The electricity
consumption of the upgraded experiment is dominated by the dipole
magnet with 4.6\Mwatt of electric power and a yearly consumption of
$\sim20$\aunit{GWh}. This is followed by the surface data centre,
which draws $\sim2$\Mwatt, and the detectors' electronics and cooling
systems which consume $\sim200$\kwatt.  The electrical infrastructure
has been partially renewed for the upgrade. In particular, new power
distribution lines have been installed for the upgraded cooling system
(see section~\ref{ssec:infracooling}) and two new 18\aunit{kV}
high-voltage cells and two 3.15\aunit{MVA} tranformers were added for
the data centre at the LHCb site.

\subsubsection{Neutron shielding}
\label{subsubsec:neutronshielding}

As explained in section~\ref{sec:scifi}, the performance of the \Sipm
arrays used as photon detectors in the \Scifi is significantly
impacted by radiation damage.  The dominant contribution to the
fluence at the location of the \Sipm arrays comes from high-energy
neutrons produced in showers in the calorimeter.  During \Lstwo, a
dedicated neutron shielding has been installed upstream of the
calorimeter, taking the space formerly occupied by muon station M1 and
reusing its support structure.  The shielding, made from polyethylene
(C$_2$H$_4$) with a 5\% admixture of boron, has an inner region
($2\times 2\ma$) with a thickness of 300\mm and an outer region
($5\times 5\ma$) with a thickness of 100\mm. From simulations with
\textsc{FLUKA}~\cite{Ferrari:2005zk,BOHLEN2014211}, it is expected
to reduce the 1\mev neutron equivalent (\neutroneq) fluence at the
location of the \Scifi \Sipm{s} by a factor 2.2--3.0.  Polyethylene
was selected as material for the shielding since it efficiently
moderates fast neutrons by elastic scattering, while boron reduces the
activation due to the resulting thermal neutrons because of its high
cross-section for thermal neutron capture.

\subsubsection{Detector cooling}
\label{ssec:infracooling}

In order to further mitigate radiation effects, the \Scifi \Sipm
arrays will be kept at a temperature of $-40\degc$ using a monophase
liquid cooling system. Initially, the system has been operated with
the perfluorocarbon C$_{6}$F$_{14}$ as cooling fluid.  However, work
is underway to validate alternative fluids which possess similar
thermal properties and radiation tolerance as C$_{6}$F$_{14}$ but
feature a significantly lower global warming potential such as the
hydrofluoroether C$_4$F$_9$OCH$_3$ or the fluoroketone
C$_{6}$F$_{12}$O.\footnote{\Trmk{Novec 7100} and \Trmk{Novec
    649}. These fluids have \Acr{gwp} $\sim 1$ and $\sim 300$,
  respectively, compared to \Acr{gwp} $\sim~9300$ for
  C$_{6}$F$_{14}$.}

Both \Scifi and \Rich cooling plants have been constructed by CERN
which is also responsible for the demineralised water cooling system
used for the \Scifi electronics.

The \Velo and \Ut use evaporative CO$_{2}$ cooling for the thermal
management of their silicon sensors and front-end \Asic{s}.  Two
identical cooling plants based on the \Acr{pacl}
concept~\cite{Verlaat2018}, and each having a cooling capacity of
7\kwatt at $-30\degc$, have also been constructed by CERN.

The \Scifi and \Velo/\Ut cooling plants use a shared primary chiller,
with a capacity of 24\kwatt at $-56\degc$.  All cooling plants are
located in the \lhcb cavern \aside and are connected to the detector
via 50--80\m long transfer lines.  Primary cooling is provided by the
chilled water (at $\sim 6\degc$) and mixed water (at $\sim 14\degc$)
circuits.

\subsubsection{Data centre}
\label{sec:datacentre}

\begin{figure}[t]
  \centering
  \includegraphics[width=0.75\textwidth]{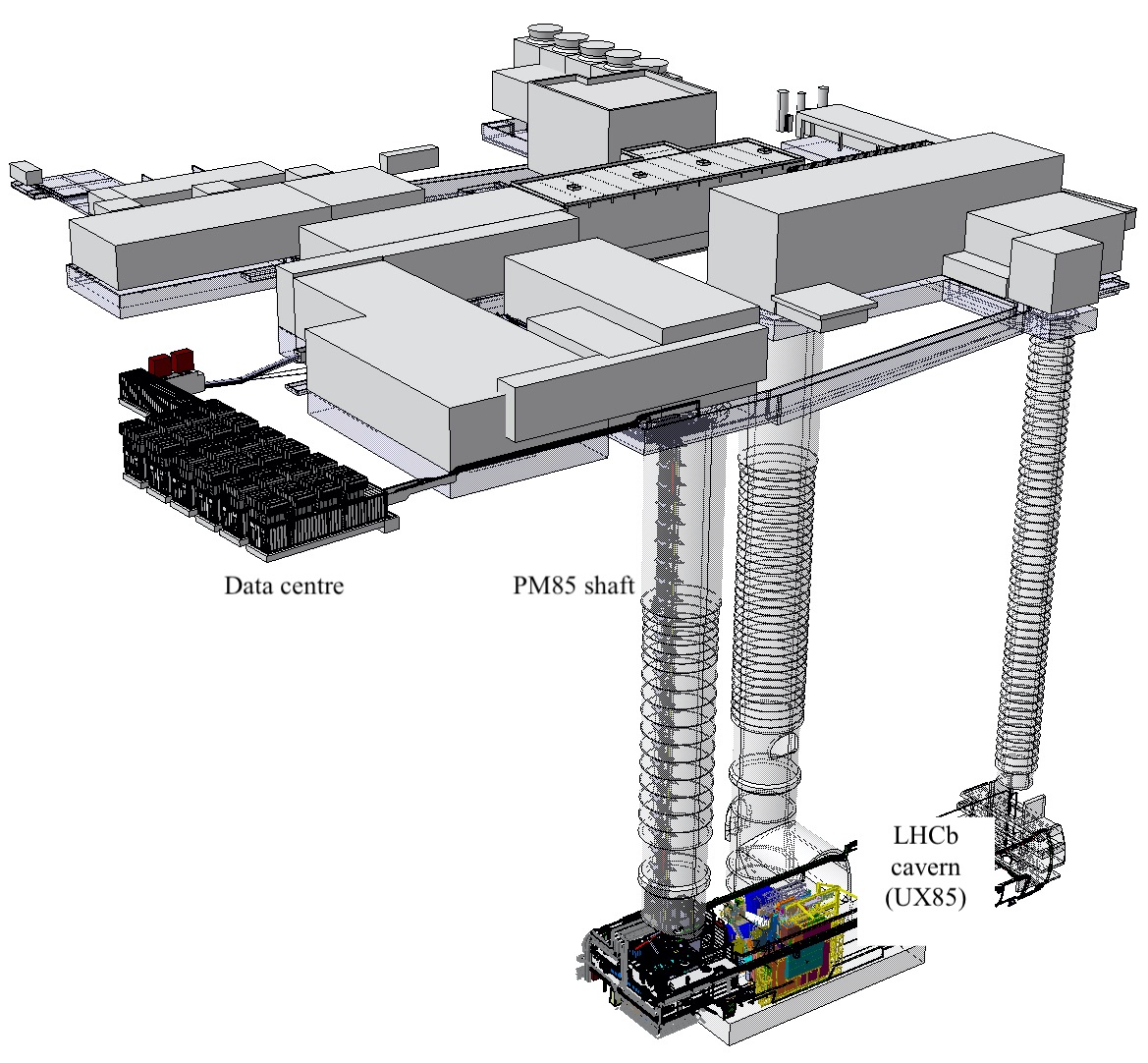}
  \caption{The readout system located in the modular data centre and
    the front-end electronics in the underground cavern are connected
    through long-distance optical fibres installed in the PM85 shaft.}
  \label{fig:cavern_and_dc}
\end{figure}

A new modular data centre has been built on the surface of the
experimental site to accommodate all the computing resources needed
for the upgraded readout system and event-filter farm.  The new data
centre comprises six 22-rack modules with a total power capacity of
$\sim$2\Mwatt of computing equipment.  The two central modules house
the event-builder servers, connected to the front-end electronics
underground via 190 thousand OM3 multimode optical fibres over a
length of $\sim 250\m$ (figure~\ref{fig:cavern_and_dc}).  The
remaining four modules host the servers of the event-filter farm.  The
data centre uses a highly efficient cooling system, based on a
combination of indirect free air cooling and evaporative water
cooling, with a power usage effectiveness smaller than~1.1.

\subsection{Beam pipe}
The vacuum beam pipe and its support structure have been optimised to
reduce background occupancy in the nearby tracking
detectors~\cite{LHCb-DP-2008-001,LHCb-TDR-009}.  The conical shape of
the beam pipe in \lhcb leads to unbalanced forces in the axial
direction due to the atmospheric pressure, which must be
counterbalanced with mechanical restraints.  In the aperture of the
dipole magnet a support system consisting originally of eight
stainless steel wires and rods, provided enough stiffness in all
transverse directions. The two wire systems were attached to aluminium
collars, connected to the beam pipe by graphite-reinforced
polyimide-based plastic\footnote{\Trmk{Vespel}.} rings.

During \runone this support system was identified as a significant
source of scattering in the experiment and was redesigned as part of
the upgrade programme for the \lhcb vacuum system.  The improved
support system was installed in 2014 during the \Lsone.

In the new support system, the volume of the collars has been reduced
and materials with longer radiation length have been selected for all
components.  Carbon fibre reinforced plastic tubes and synthetic ropes
have replaced the stainless steel rods and cables.  This led to a
remarkable increase of more than 90\% in material transparency for
these components, while retaining sufficient stiffness.  The aluminium
collars were redesigned and remade with beryllium, which has led to a
material transparency improvement of more than 85\%~\cite{Leduc2011}.
\footnote{Here, the material transparency is defined as $I = t/\Xrad$
  where \Xrad is the radiation length of the material and $t$ is the
  thickness of material seen by particles when traversing the various
  supporting elements.}  A range of innovative materials was selected:
aramid\footnote{Teijin \Trmk{Technora}.  } for the ropes; thermoset
carbon high-modulus fibres\footnote{Torayca \Trmk{M46J}.  } for the
tubes; and polybenzimidazole\footnote{\Trmk{Celazole PBI U-60}.} for
interface rings.  A thorough qualification process of these materials
has been carried out by the CERN vacuum group considering mechanical
strength, creep and radiation tolerance.

\subsection{Background and luminosity monitors}

A set of detectors has been developed to monitor the machine-induced
background conditions around the interaction point. To reduce
systematic uncertainties and facilitate the reconstruction, the
instantaneous luminosity delivered to \lhcb is kept constant
throughout a fill using a procedure known as \emph{luminosity
  levelling}~\cite{AlemanyFernandez2013}.\footnote{In the \Lhc jargon,
  a \emph{fill} is the full beam cycle from injection to beam
  dump. Experiments normally divide the data taking part of a fill in
  \emph{runs}, which correspond to samples of data taken at constant
  conditions.}  Every few seconds during nominal operation, \lhcb
publishes a measurement of the average number of visible \Prpr
interactions per beam-beam crossing, denoted by $\muvis$, which is
used by the \Lhc control system to adjust the offset of the two beams
in the direction perpendicular to the nominal crossing plane.  As the
number of visible \Prpr collisions per beam crossing follows a Poisson
distribution, the average of the distribution (i.e.\ $\muvis$) can be
determined by measuring the probability that a beam crossing has no
observable activity in the detector, $P_0 = \exp\left(-\muvis\right)$;
this is the \emph{logZero} method~\cite{Barsuk:2743098}.  During
\runone and \runtwo, the real-time luminosity measurement was based on
information available in the L0 hardware trigger.  For the upgraded
experiment, a dedicated luminosity subdetector, dubbed \Plume, has
been installed, see section~\ref{sec:plume}.

\begin{figure}[t]
  \centering
  \includegraphics[height=0.329\textwidth]{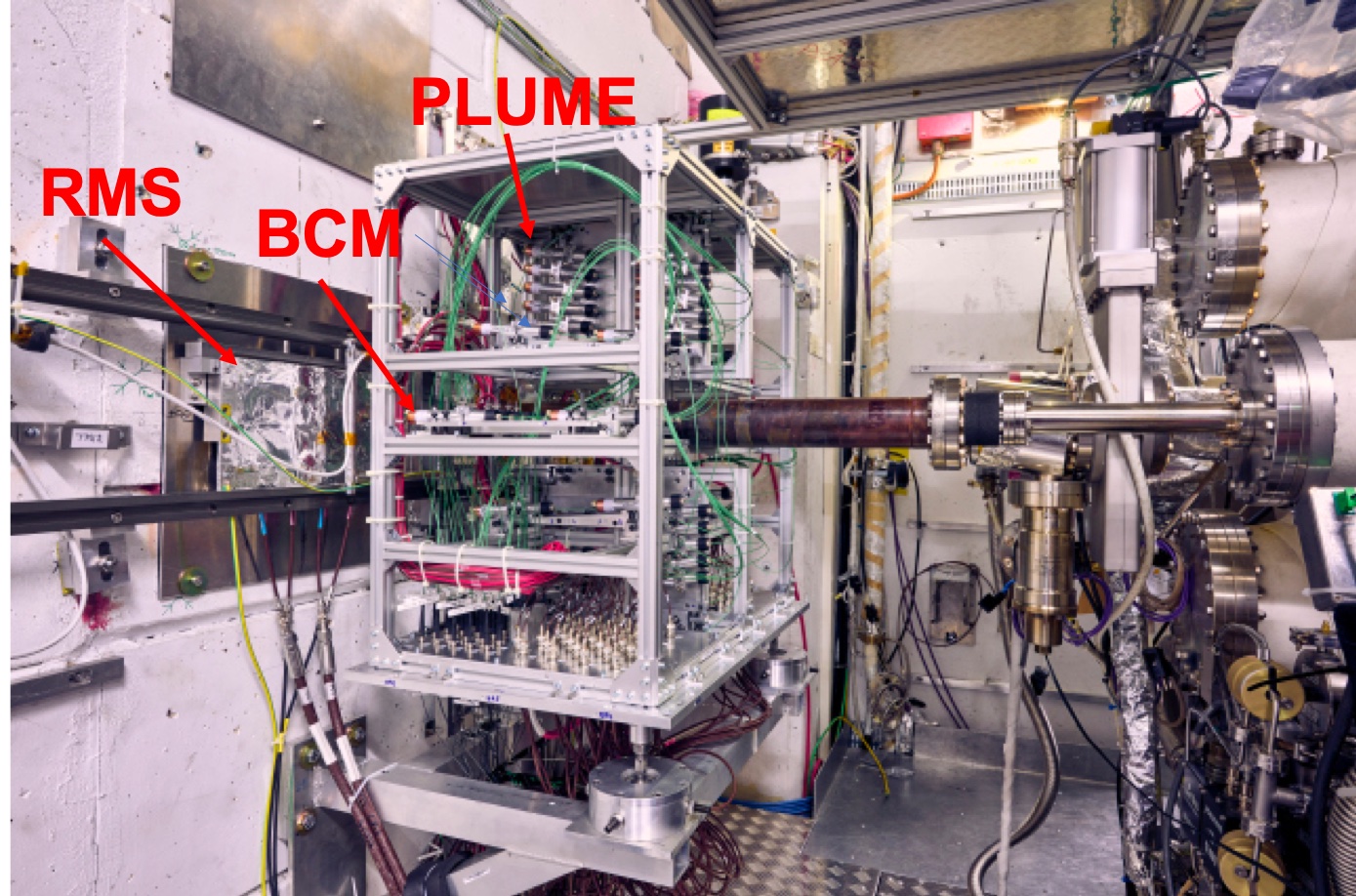}\hfill
  \includegraphics[height=0.33\textwidth]{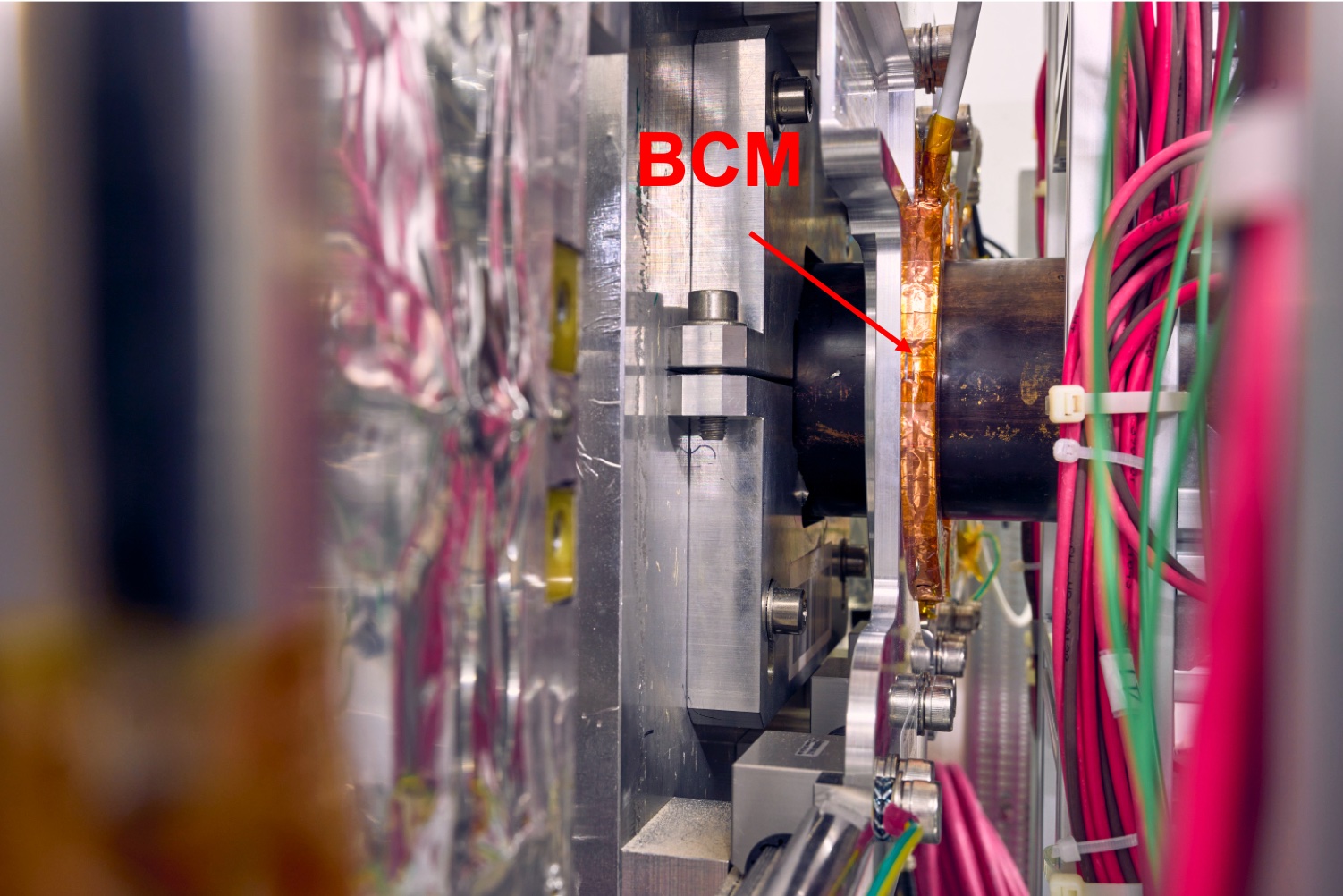}
  \caption{Left: \Rms (under the mylar protection foil at the left
    side) and \Plume (inside the scaffolding). The arrow indicates the
    position of \Bcm which is hidden behind the \Plume
    scaffolding. Right: upstream \Bcm detectors inside their
    kapton-insulated support surrounding the beam pipe.}
  \label{fig:plume_bcm_rms_view}
\end{figure}

Excursions in luminosity or elevated machine-induced background are
not only detrimental to the quality of the collected data but can also
cause damage to the detector.  For the safe operation of the
experiment, it is essential to quickly detect and react upon anomalous
beam conditions.  Such protection is ensured by the \Bcm, described in
section~\ref{sec:bcm}.  Additional monitoring of the beam environment
is provided by a metal foil detector called the \Rms, see
section~\ref{sec:rms}.  Figure~\ref{fig:plume_bcm_rms_view} shows a
picture of the \Rms, \Plume, and the upstream \Bcm station as
installed in the cavern.

\subsubsection[PLUME]{\Acr[s]{plume}}
\label{sec:plume}

\begin{figure}[h]
  \centering
  \includegraphics[width=0.9\textwidth]{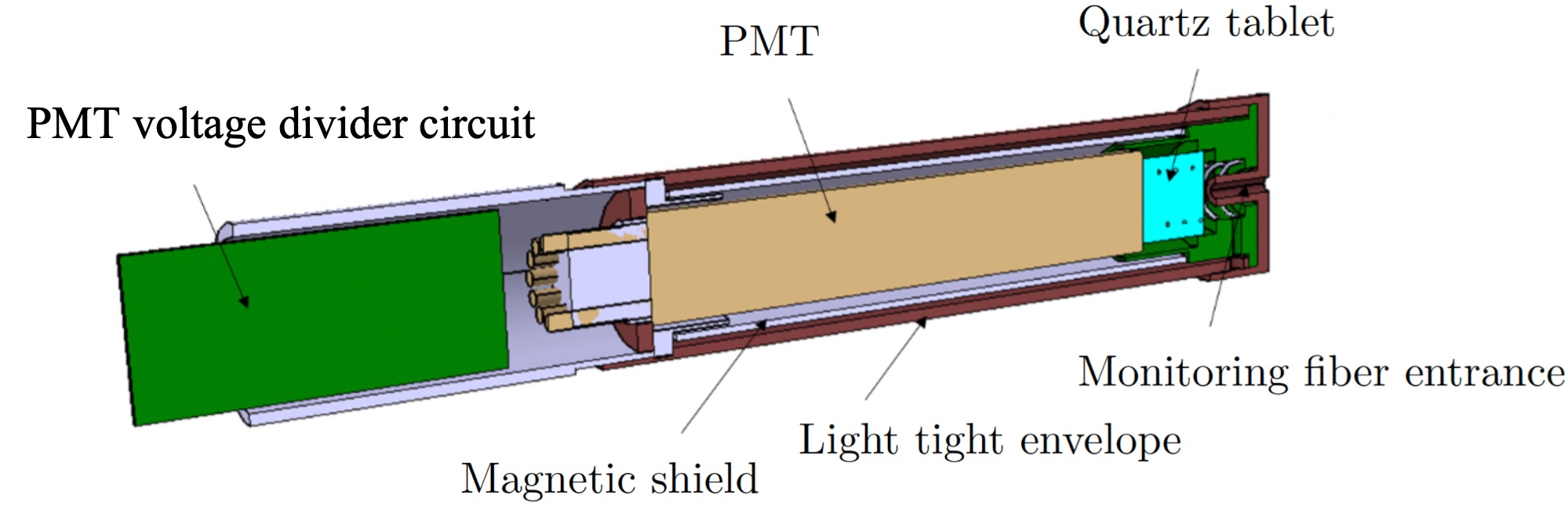}
  \caption{Schematic view of a \Plume elementary detection module. The
    module is 153\mm long and has a diameter of 24\mm. Reproduced from~\cite{PlumeTDR}. CC BY 4.0.}
  \label{fig:plume_elem}
\end{figure}

\Plume~\cite{PlumeTDR} is a luminometer measuring Cherenkov light
produced by charged particles crossing a quartz radiator.  Its basic
detection element, shown in figure~\ref{fig:plume_elem}, is a
\Pmt\footnote{Model R760 by \Trmk{Hamamatsu Photonics K.K.},
  Hamamatsu City, Japan.} with a 1.2\mm thick quartz entrance window
and a photocathode with a diameter of 10\mm. To increase the amount of
Cherenkov light, a 5\mm thick quartz tablet is placed in front of the
\Pmt window.\looseness=-1

The detector, located upstream of the \Prpr interaction region
(between $z = -1900$\mm and $z = -1680$\mm), is a hodoscope consisting
of two stations, each of which comprises 24 \Pmt modules arranged in a
star-shaped structure around the beam pipe.  The modules are placed at
radial distances between 157 and 276\mm with respect to the beam line
(corresponding to a pseudorapidity range $2.4 < \eta < 3.1$) and are
angled such that the \Pmt axes point to the nominal interaction point.
The \Fend, based on components developed for the upgraded \lhcb
calorimeters, and the LED monitoring system (described below) are
located at a distance of $\sim 20$\m from the \Pmt{s} to reduce the
level of radiation to which they are exposed.

\begin{figure}[t]
  \centering
  \includegraphics[height=0.348\textwidth]{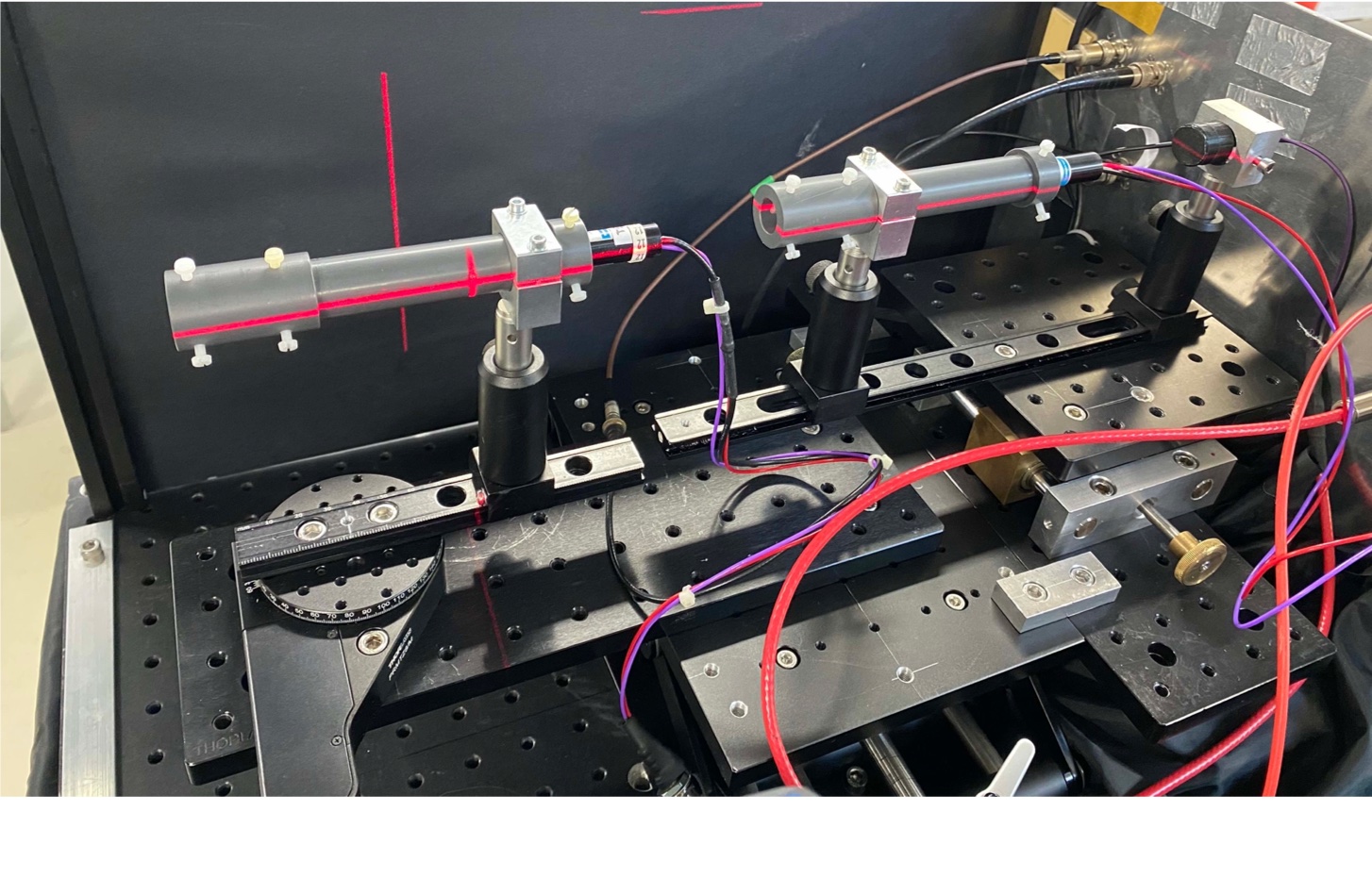}
  \includegraphics[height=0.35\textwidth]{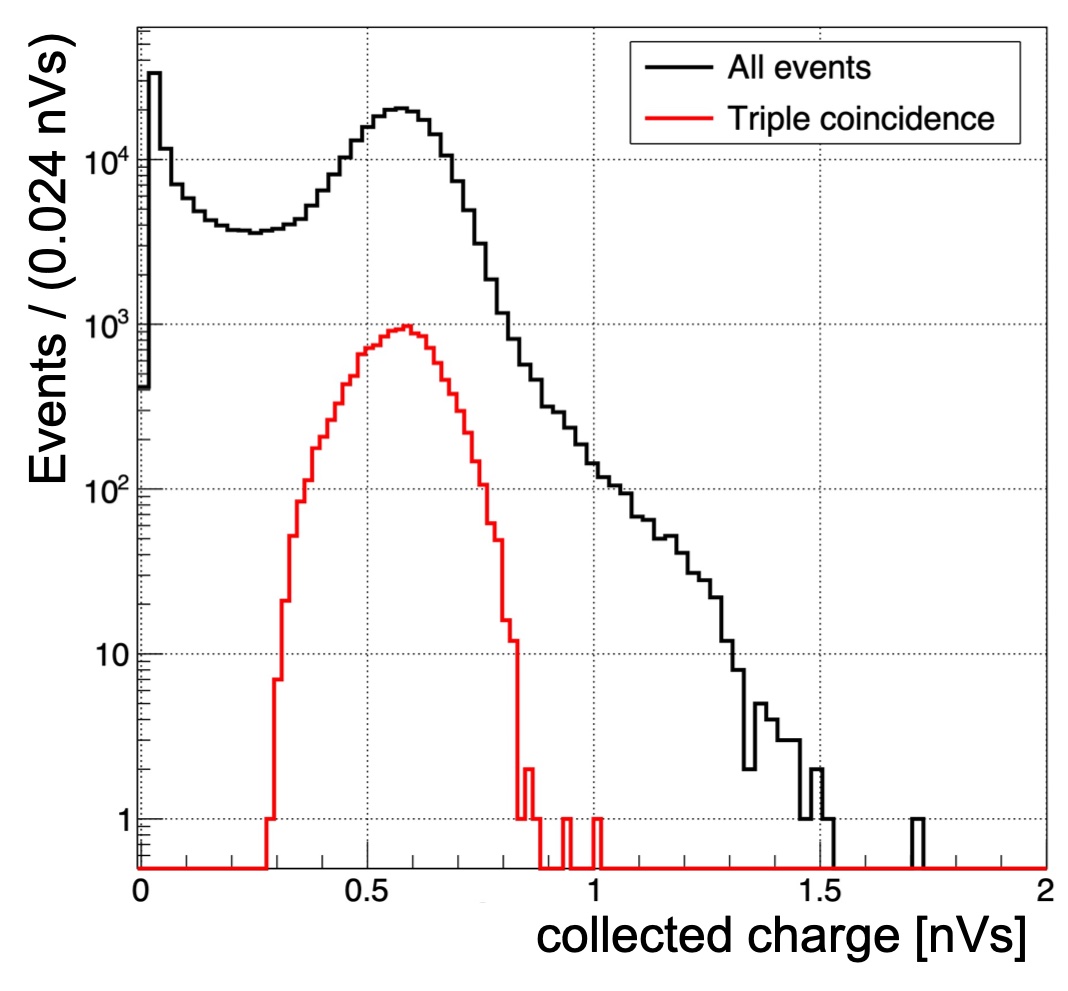}
  \caption{Left: \Plume test beam setup. Right: charge collected by
    the first \Pmt in the test beam for all events (black line) and
    events producing a simultaneous signal in the two \Pmt{s} and the
    trigger (red line). The scale is in nVs where 1\aunit{nVs} =
    20\aunit{pC}.}
  \label{fig:plume_TB}
\end{figure}

The detector design was optimised with simulations using
\pythia{8}~\cite{Sjostrand:2007gs} and
\geant~\cite{Agostinelli:2002hh,Allison:2006ve}, and was validated
using test beam measurements. A summary of the studies can be found in
refs.~\cite{Barsuk:2743098,PlumeTDR}.  Figure~\ref{fig:plume_TB}
(left) shows a test setup used in a 5.4\gev electron beam at
DESY.\footnote{Deutsches Elektronen-Synchrotron, Hamburg, Germany.} It
consists of two \Pmt modules placed one behind the other in the beam
line and a trigger scintillator at the rear.  An example of the charge
spectrum measured in the first \Pmt (operated at 1000\volt) is shown
in figure~\ref{fig:plume_TB} (right). For charged-particle tracks that
produced a signal in both \Pmt modules and the trigger scintillator,
the average collected charge was 12\aunit{pC} and a time resolution of
0.6\ns was found.

As is the case for the other \lhcb detectors, \Plume is fully
integrated in the \Ecs, \Daq and \Tfc systems.  A particular feature
of \Plume is that it needs to be read out even if the other \lhcb
detectors are off or the event-builder system is not available. Such
circumstances occur during injection and focusing of the \Lhc beams,
when feedback on the instantaneous luminosity is vital for the
optimisation of collisions at the \lhcb interaction point.

The online luminosity calculation is performed in the firmware of the
\Bend electronics.  For every pair of \Pmt modules, the \Plume
\Tellfourty counts the total number of bunch-crossings $N$ and the
number of those with a signal below threshold, $N_0$.  Using the
\emph{logZero} method, the average number of visible interactions per
bunch-crossing, $\muvis$, is given by~\cite{PlumeTDR}:
\begin{equation}
  \muvis = -\log P_0 = -\log \frac{N_0}{N} - \frac{1}{2}\left(\frac{1}{N_0}-\frac{1}{N}\right),
  \label{eq:l1}
\end{equation} 
where the second term accounts for second-order bias corrections to
the Poisson statistics.  In the 3\sec interval between counter resets,
the value of $\muvis$ is expected to be stable and deviations from the
Poisson statistics of eq.~\eqref{eq:l1} can be neglected. The
estimated statistical uncertainty on the average luminosity is given
by $5\%/\sqrt{n_{bb}}$, where $n_{bb}$ is the number of colliding
bunch pairs, and becomes negligible for $n_{bb}\gg 1$.

The systematic uncertainty depends, among other things, on the
stability of the \Pmt response which can change with time due to
variations in temperature or occupancy, or because of ageing. The
monitoring and calibration system, which was developed based on
experience from the \lhcb calorimeters~\cite{Guz:2196516} and the
ATLAS LUCID detector~\cite{Avoni:2018iuv}, is therefore an integral
part of \Plume. At regular time intervals (using suitable gaps in the
\Lhc filling scheme), the LED calibration system, located next to the
\Fend, sends light pulses over $\sim20$\m long quartz fibres to the
front face of each \Pmt. The stability of the LED light pulses is
monitored by PIN photodiodes located in the same rack as the
LEDs. Finally, the degradation in the transparency of the quartz
fibres due to radiation damage is monitored with dedicated fibres
looped back to the LED position and read out by \Pmt{s} placed next to
them. Based on the \Pmt response to the injected light, the high
voltage of the \Pmt{s} is adjusted in steps of $\Delta V = 0.5\volt$,
which corresponds to a $\sim2\%$ change in gain.  Tracks reconstructed
in upstream \Velo stations\footnote{These are \Velo stations placed
  upstream of the interaction point.} and passing through \Plume can
be used to cross-check the reliability of the calibration and
monitoring system.

\subsubsection[BCM]{\Acr[s]{bcm}}
\label{sec:bcm}

The \Acr[f]{bcm} comprises two stations, one upstream (at
$z = -2131$\mm) and one downstream (at $z = +2765$\mm) of the
interaction region.  A detailed description of the system can be found
in ref.~\cite{Ilgner2010}.  Each station consists of eight
poly-crystalline chemical vapour deposition (pCVD) diamond pad sensors
arranged symmetrically around the beam pipe.  For each of the diamond
sensors, the average current, integrated over periods of 40\mus, is
measured and a beam abort is requested through the \Lhc beam interlock
system if three adjacent diamond sensors exhibit a current above
threshold for two consecutive periods.  In addition, running sums over
32 consecutive measurements are computed and a dump of the \Lhc beams
is triggered if the average of the running sums in one station exceeds
a given threshold. For monitoring purposes the \Bend also calculates
average and maximum values over intervals of few seconds, which are
read by the \Ecs and used for calculating the normalised background
figures of merit, which are made available to the \Lhc control system.
The system has been operating successfully throughout \runone and
\runtwo.  During \Lstwo the diamond sensors were replaced, the support
structures rebuilt, and the \Bend have been upgraded to be compatible
with the new readout architecture.

\subsubsection[RMS]{\Acr[s]{rms}}
\label{sec:rms}

The upgraded \Acr[f]{rms} consists of four metal-foil detector modules
located upstream (at $z\sim -2200$\mm) of the nominal interaction
point, at a radial distance of $\sim 30$\cm from the beam line.  Each
module houses two five-layer stacks of copper foils, with the central
50\mum thick foil serving as the sensor.  The detector concept
exploits the phenomenon of secondary-electron emission at the metal
surface due to charged particles crossing the foil. The readout
electronics, located $\sim 80$\m away in the accessible part of the
\lhcb cavern, convert the resulting current to a frequency. The \Rms
is integrated in the \Ecs and the measurements are displayed in the
\lhcb control room.  A similar system was used during \runone and
\runtwo to monitor the charged particle fluence~\cite{Okhrimenko2011}.\looseness=-1

\section{Vertex locator}
\label{sec:velo}
\subsection{Overview}

The \Velo detects tracks of ionising particles coming from the beam
collision region and thereby measures the location of interaction
vertices, displaced decay vertices and the distances between them.
\Velo tracks seed the reconstruction algorithm of the \lhcb
spectrometer and provide discriminatory information for event
selection.  The \Velo has been redesigned~\cite{LHCb-TDR-013} to be
compatible with the luminosity increase and the trigger-less 40\MHz
readout requirement of the upgraded experiment. It must continue to
provide pattern recognition within an acceptable CPU budget, whilst
maintaining the highest track-finding efficiency.  The core technology
of the new \velo is pixelated hybrid silicon detectors, which are
arranged into modules and cooled by a silicon microchannel cooler.  Of
the mechanical structures, only the principal vacuum vessel and motion
services remain from the version that was in operation until 2018.  In
particular, the RF boxes, the enclosures that interface the detector
to the \Lhc beams, were entirely redesigned reducing both material and
the inner radius of the \velo along the beam line.  Furthermore, a new
structure, a \emph{storage cell}, is fitted immediately upstream of
the \Velo detector in the beam vacuum, see section~\ref{sec:smog}.  A
summary of the changes is shown in table~\ref{tab:mainChanges}.

\begin{table}[h]
  \centering
  \caption{Specifications of the upgraded \Velo compared to those of
    the original version.}
  \label{tab:mainChanges}
  \begin{tabular}{|l|c|c|}
    \hline
    & 2009--2018 & 2022 \\
    \hline
    RF box inner radius (minimum thickness) & $5.5\mm$ ($300\mum$) & $3.5\mm$ ($150\mum $) \\
    Inner radius of active silicon detector & $8.2\mm$  & $5.1\mm$ \\
    Total fluence (silicon tip)
    [\neqcmcm\!] & $4\times 10^{14}$ &  $\sim 8\times 10^{15}$ \\
    Sensor segmentation & $r-\phi$ strips & square pixels\\
    Total active area of Si detectors & $0.22\m^2$ & $0.12\m ^2$ \\
    Pitch (strip or pixel) &  37--97\mum  & $55\mum $\\
    Technology &  n-on-n & n-on-p\\
    Number of modules & 42 & 52 \\
    Total number of channels & 172~thousand  & 41 million \\
    Readout rate [\mhz] & 1, analogue & 40, zero suppressed \\
    Whole-\Velo data rate & 150\gbps & $\sim 2$\tbps\\
    Total power dissipation (in vacuum) & $800\watt$ & $\sim 2\kwatt$ \\
    \hline
  \end{tabular}
\end{table}

The combination of the pixel geometry, a smaller distance to the first
measured point and reduced material means the performance of the \Velo
is significantly improved. However, the closer proximity to the \Lhc
collisions and the step-change in design luminosity means the design
must prepare for hit rates and radiation doses that are an order of
magnitude higher than those experienced by the earlier \Velo.

\subsection{Design requirements}
\label{sec:velo:design}

The principal metric for a vertex detector design is impact parameter
resolution $\sigmaIP$, the precision with which the perpendicular
distance of a track to a point is measured.  This metric is a function
of: track transverse-momentum, \pt; the average axial distance of the
material before the second measurement, $r_1$; the distances from the
point to the first and second measurements, $\Delta_i$ ($i=1,2$); and
the position uncertainties of those measurements, $\sigma_i$.  In the
\Velo case, it can be approximated
as~\cite{papadelis2009characterisation},
\begin{equation}
  \sigmaIP^2 \approx \underbrace{\left(\frac{r_1}{\pt[\gevc]}\right)^{\!2}\!\left(0.0136 \gevc
               \sqrt{\frac{x}{X_0}}\left(1+0.038\ln\frac{x}{X_0}\right)\right)^{\!2}}_{\rm multiple~scattering}
               + \underbrace{\frac{\Delta_{2}^2\sigma_1^2 + \Delta_{1}^2\sigma_2^2}{\Delta_{12}^2}}_{\rm extrapolation}\label{eq:iperr}\\
\end{equation}
where $x/X_0$ is the fraction of radiation length traversed before the
second measurement.  The first term describes the degradation induced
by multiple scattering.  The second term is the extrapolation error,
which is dominated by detector geometry: pixel size and lever arm,
$\Delta_{12}$, between the first and second measured points.  The
upgraded \Velo design was optimised to achieve, within the nominal
\lhcb acceptance, a performance at least as good as that of its
predecessor \Velo, in terms of both $\sigmaIP$ and track-finding
efficiency, despite the increased instantaneous luminosity.

\subsubsection[LHC interface]{\Acr[s]{lhc} interface}

The \Velo performance, as described by eq.~(\ref{eq:iperr}), improves
by reducing the radius of the first pixel hits, though this must be
balanced against the limitations of proximity to the beam line.  The
minimal \Velo aperture allowed by the requirements of the \Lhc
collimation and protection depends on several factors: the maximum
expected separation of the counter-rotating beams; their transverse
sizes ($\beta$ functions and transverse emittances); the beam
direction relative to the longitudinal axis of the \Velo (crossing
angle); the mechanical accuracy and stability of the RF boxes.  A
reduced, but still conservative, radial clearance of 3.5\mm is
chosen~\cite{VELOapertureNote2012,VELOapertureNote2018} for the RF
boxes, the structures that directly interface with the \Lhc beam
environment. This allows the silicon sensors to be arranged such that
the radius of the closest active pixel edge is 5.1\mm from the beam
line. The decision takes into account the intended luminosity, beam
crossing schemes, luminosity levelling and the requirements of special
running scenarios such as {van der Meer}
scans~\cite{Balagura2020VanDM}.

The electromagnetic fields of the two \Lhc beams, pulsing at radio
frequencies, must also be taken into account. The principle of
electrical continuity is maintained for the \Velo upgrade with the
reimplementation of flexible wakefield suppressors at the entrance and
exit of the \Velo vacuum vessel.  The RF box shape was optimised to
reduce the beam impedance while maintaining a good impact parameter
resolution.  Simulation and measurements with a full-size mock-up of
the new RF boxes show the longitudinal and transverse impedance to
have good, broadband behaviour when the \Velo is closed with $3.5\mm$
inner radius~\cite{Popovic:2701362}.  In the open position,
simulation of the cavity predicts several resonance modes but the
total beam power loss due to the impedance presented by the whole
\Velo is a tolerable
$14\watt$~\cite{Popovic:2666879,Wanzenberg:2263336}.  To prevent
possible beam-stimulated electron emission, which can lead to
instabilities, the beam-facing surfaces of the RF boxes are coated
with a material with a low \Acr{sey}; a low activation temperature
\Acr{neg} is chosen~\cite{CHIGGIATO2006382}.

\subsubsection{Mechanics, vacuum and cooling}

The concept of separating primary (beam) and secondary (detector)
vacua is preserved for the \Velo upgrade.  The RF boxes must be
leak-tight with a tolerance on the pressure difference between vacuum
volumes of 10\mbar. The detector components inside the secondary
vacuum must be constructed of materials with minimal outgassing and
bespoke vacuum-tight solutions are needed to route high-speed data,
power and high voltages cables in and out of the vacuum.  The vacuum
vessel, which is integrated into the \Lhc beam pipe, remains from the
original vertex detector, as well as the large rectangular bellows and
detector supports, so that the total allowed detector length of about
1\m, the size and location of access ports and the mechanisms for the
horizontal and vertical movement of the detectors are unchanged.

\begin{figure}[h]
  \centering
  \includegraphics[width=0.48\linewidth]{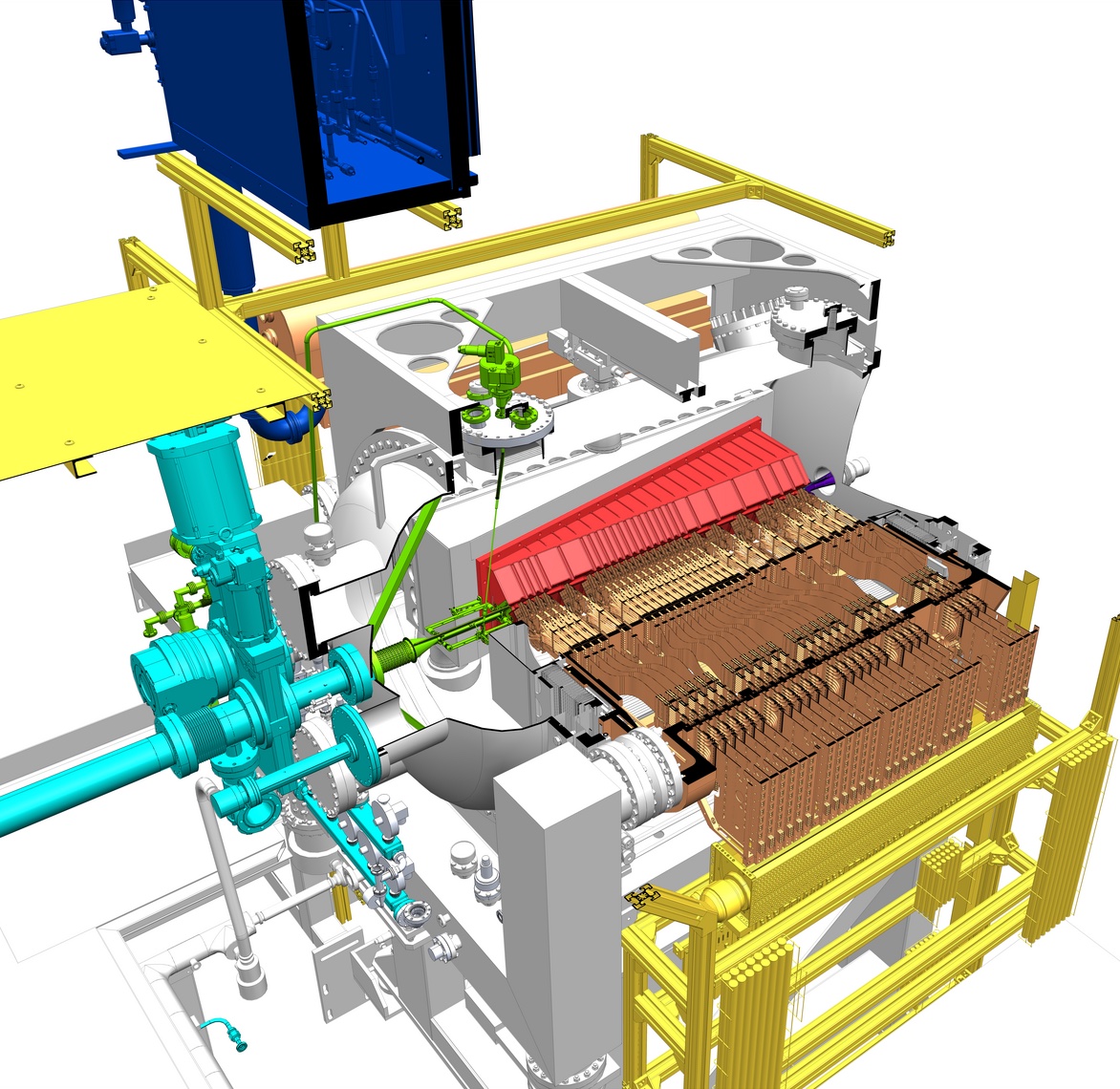}
  \includegraphics[width=0.48\textwidth]{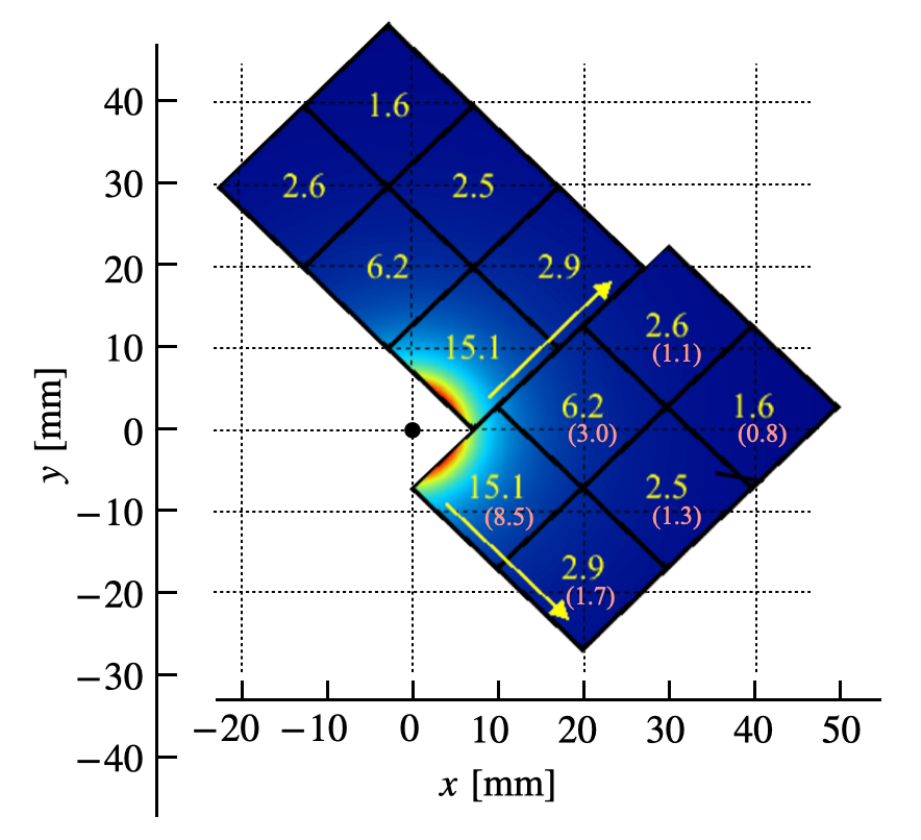}
  \caption{\label{Overview_changes} Left: a 3D view of the upgraded
    \Velo, with cut-out.  Some of the new items are highlighted, such
    as the \cside pixel modules and readout electronics (brown), the
    \aside RF box (red), the internal gas target system with a storage
    cell (green), the upstream beam pipe with a sector valve (cyan).
    Right: data rate per pixel \Asic in \gbps for the most active
    module.  The numbers in parenthesis are the number of traversing
    tracks per \Lhc bunch crossing for an average number of
    interactions per crossing equal to 7.6. Arrows indicate the
    readout direction.}
\end{figure}

The power dissipation of the \Fend \Asic{s} operating in the detector
vacuum must be removed by a cooling system.  Moreover the sensors must
be maintained at low temperatures (typically $<-20\degc$) for the
entire life of the detector, including shutdown periods. Bi-phase
\cotwo cooling is chosen, following the same principle as in the
predecessor \velo. However, the system is entirely redesigned.  In the
secondary vacuum, the \cotwo flows in microchannels within a silicon
cooler to which the active components are glued.  The risk of a
cooling system rupture in the secondary vacuum is mitigated by
additions to the mechanical design.  A tertiary vacuum volume, the
\emph{isolation} volume, is added to house the local distribution of
the \cotwo supply, including a fast-response bypass valve system.  The
preserved vacuum vessel is shown in figure~\ref{Overview_changes}
(left) contrasted against the parts that have been added or upgraded.\looseness=-1

\subsubsection{Detector geometry and layout}

\begin{figure}[h]
  \centering
  \includegraphics[width=0.99\textwidth]{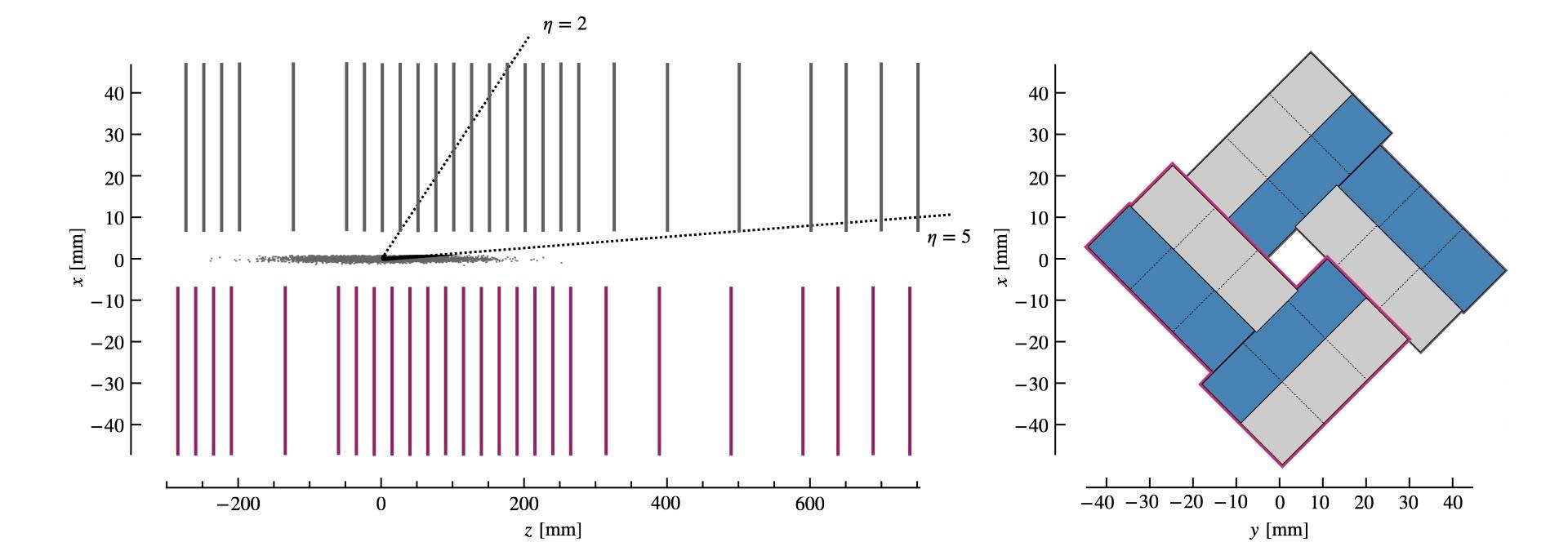}
  \caption{Left: schematic top view of the $z-x$ plane at $y = 0$
    (left) with an illustration of the $z$-extent of the luminous
    region and the nominal \lhcb pseudorapidity acceptance,
    $2<\eta<5$.  Right: sketch showing the nominal layout of the
    \Asic{s} around the $z$~axis in the closed \Velo
    configuration. Half the \Asic{s} are placed on the upstream module
    face (grey) and half on the downstream face (blue).  The modules
    on the \cside are highlighted in purple on both sketches.}
  \label{layout_from_PUB-2019-008}
\end{figure}

With the \lhcb acceptance unchanged, the optimised layout of \Velo is
similar to its predecessor.  Active elements and their services are
assembled into a series of identical modules, populated with pixelated
\Asic{s}, arranged perpendicular to the beam line.  The decision to
use identical modules throughout greatly simplifies the production
process and quality control.  The distribution of the modules must
cover the full pseudorapidity acceptance of \lhcb ($2<\eta<5$) and
ensure that most tracks from the interaction region traverse at least
four pixel sensors, for all azimuthal
directions~\cite{Bird:2256124}.  With the chosen sensor arrangement
shown in figure~\ref{layout_from_PUB-2019-008} (left), 52 modules are
necessary to satisfy these requirements, including the modules placed
upstream of the interaction region whose purpose is to improve the
unbiased measurement of primary vertices.

The modules are arranged into two movable halves, the \cside and
\aside.  Except for a shift, the distribution in $z$ (parallel to the
beam line) is identical for the two sides.  The minimal, nominal
spacing between modules is 25\mm and the \aside modules are displaced
in $z$ by $+12.5\mm$ relative to the \cside modules to ensure the two
sides overlap when closed to provide a complete azimuthal coverage.\looseness=-1

The rectangular pixel detectors are arranged in a rotated `L' shape,
as shown on figure~\ref{layout_from_PUB-2019-008} (right).  The
purpose of the $45\degrees$ rotation around the $z$~axis is to
minimise any risk of the detectors grazing the RF box during
installation.

\subsubsection{Expected particle fluxes and irradiation}

The most-occupied $2\cma$ \Asic will experience 8.5 charged particles
in every bunch crossing.  The \lhcb upgrade expects an average
bunch-crossing rate of 27\mhz, with a peak rate of 40\mhz. Particles
traverse detectors at relatively high angle and on average, given the
pixel size of $55 \mum \times 55 \mum$, 2.6~pixels will record the
passage of an ionising particle.  For the busiest \Asic, this implies
a peak pixel-hit rate of $\sim 900\Mps$.

Section~\ref{sec:velo:velopix} describes the dedicated \Asic developed
for the \velo upgrade, which has digital logic that groups hits into
\Acr[p]{spp} encoded by $30\,$bits.  The busiest \Asic records hits in
$\sim1.5$ \Acr[p]{spp} per traversing particle, giving a maximum
\Acr{spp} rate of $520\Mps$, or 15.1\gbps from the most central \Asic,
see figure~\ref{Overview_changes} (right).  The peak total data rate
out of the whole \Velo may reach 2.85\tbps and the readout scheme is
designed accordingly.  The power needed for such \Fend processing is
significant and performant on-detector cooling is vital.

The pixel \Asic and silicon sensors are designed to tolerate a high
and non uniform fluence, which ranges from $5\times 10^{12}$ to
$1.6\times 10^{14}\neqcmcm$ per $1\invfb$ of integrated luminosity
exposure.  With 50~\invfb, it is expected that some \Asic{s}
accumulate an integrated flux of $8\times 10^{15}\neqcmcm$.  With this
dose, leakage currents of around $200\muampcmcm$ ($\sim 7\namp$ per
pixel) are expected with 1000\volt of bias voltage at $-25\degc$.  In
terms of total ionising radiation dose, the \Asic{s} must remain fully
operational up to $4\MGy$.

\subsection{The pixel tile}
\label{sec:velo:detectors}

The \velo pixel \emph{tile} is composed of a pixelated, planar silicon
sensor and three pixelated \Asic chips. Known as
\velopix~\cite{VANBEUZEKOM201392}, these bespoke \Asic{s} provide
analogue signal processing and digitisation.  They are bonded to the
sensor by an array of solder bumps (SnPb) to form each of the four
tiles composing a module.

\subsubsection[VELOPIX]{\Acr[s]{velopix}}
\label{sec:velo:velopix}

The \velopix is an \Asic based on the Timepix3~\cite{Gromov:135262}
developed by the Medipix/Timepix consortia. It has an active matrix of
$256\times 256$ pixels, each $55\mum\times55\mum$ in size, giving a
sensitive area of $1.98\cma$. Each \Asic chip is thinned from
$700\mum$ down to $200\mum$ after fabrication. On three sides, the
distance between the edge of the pixel matrix and the physical edge of
the device is $30\mum$. On the fourth side the \Asic extends by
2.55\mm and contains common digital processing and \mbox{wire-bond~pads}.

The \Asic is fabricated in 130\nm \Acr{cmos} process,\footnote{By
  \Trmk{TSMC} Taiwan Semiconductor Manufacturing Company.}  a
technology which has proven radiation hardness above $\SI{4}{\MGy}$.
In addition, \velopix is designed with protection against \Acr{seu}
with the use of dual interlocked storage cells. The space for this
extra logic is obtained by removing some of the functionality present
on Timepix3 from the pixel cell, such as the fine-time measurement
(640\MHz clock).  The main commonalities with Timepix3 are the fast
analogue \Fend with a time walk $<25\ns$ and zero-suppressed readout
using a data-push scheme. Whenever a pixel-hit is recorded, it is
time-stamped, labelled with the pixel address and sent from the \Asic
immediately, without the need for a trigger signal.  One of the main
differences between Timepix3 and the \velopix is the hit rate
capability. Timepix3 is limited to a maximum hit rate of 80 million
hits/s while the \velopix can handle 900 million hits/s.  Several
modifications have been necessary to achieve this capability.  The
time-of-arrival information is removed so \velopix records only the
occurrence of the hit (\emph{binary} readout).  The time-stamp
granularity increases from 1.56 to 25\ns, the \Fend is optimised for a
negative input charge and the power budget of the \velopix is raised
to facilitate an increased throughput.  An additional data reduction
($\sim30\%$) comes from grouping $2\times4$ neighbouring pixels into a
\Acr{spp}, thereby removing duplication of the time stamp and address
fields.

With a 40\mhz acquisition rate, up to 20.48\gbps can flow from one
\Asic~\cite{poikela2015} via four, highly optimised serial links each
running at 5.12\gbps.  A custom serialiser, \Gwt, has been designed
for this purpose~\cite{Gromov:2015asa}.  This bandwidth is
significantly greater than the 15.1\gbps anticipated from the busiest
\Asic.  A notable achievement of the \velopix development is the power
consumption: 1.2\watt typical, 1.9\watt maximum, which is 65\% of the
original expectation. Table~\ref{tab:velopixboasts} lists the key
\velopix features.

\begin{table}[t]
  \centering\renewcommand\arraystretch{1.1}
  \caption{Summary of the \velopix capabilities.}
  \label{tab:velopixboasts}
  \begin{tabular}{|c|c|}
    \hline
    Technology & TSMC $130\nm$ CMOS \\
    Radiation hardness & $>\SI{4}{\MGy}$, \Acr{seu} tolerant \\
    Pixel size (analogue part) & $55\mum\times 55\mum$ ($55\mum\times 14.5\mum$) \\
    Peak rate per \Asic (per pixel) & $9\times 10^8\,{\rm hits}/\!\sec$ ($5\times 10^4\,{\rm hits}/\!\sec$) \\
    Maximum of charge distribution & $16\,000\,e^-$ \\
    Minimum threshold & $500\,e^-$ \\
    Timing resolution (range) & 25\ns (9~bits)\\
    Super-pixel data size & 30 bits\\
    Maximum data rate per \Asic & $20.48\,$\gbps \\
    Power consumption per \Asic & $\sim1.2\watt$ (spec. $3\watt$) \\
    \hline
  \end{tabular}
\end{table}

\subsubsection{Sensors}

The 208 silicon pixel sensors are each $200\mum$ thick and
$43.470\mm\times14.980\mm$ large, including $450\mum$ wide inactive
edges in which lie the guard rings.  They are
manufactured\footnote{\Trmk{Hamamatsu Photonics K.K.}, Hamamatsu,
  Shizuoka 435-8558, Japan.} using a float-zone p-bulk with n-type
implants insulated between pixels by p-stops.  The quoted bulk
resistivity is 3--8\kohm\cm.  The sensors are designed to provide
charge collection efficiency greater than $99\%$ and signals of at
least $6000~e^-$ after $\SI{4}{\MGy}$ and $1000 \volt$ applied bias
voltage. Key characteristics are listed in
table~\ref{tab:sensorspecs}.

Each sensor comprises $768\times256$ pixel implants, matching the
pixel arrays of three \Asic{s}.  The sensors are delivered with
under-bump metallisation.  On three sides the sensor dimensions are
larger than the \Asic{s} because of the guard ring. On one side the
\Asic{s} extend beyond the sensor, this is the periphery region with
the wire-bonding pads. The metallisation process deposits an
additional row of solder bumps on the innermost guard on one side of
the sensor and connects the guard ring to the ground row of the
\Asic. This serves to tie the guard ring to the \Asic grounds, on the
periphery side.\looseness=-1

\begin{figure}[t]
  \centering
  \includegraphics[width=0.99\linewidth]{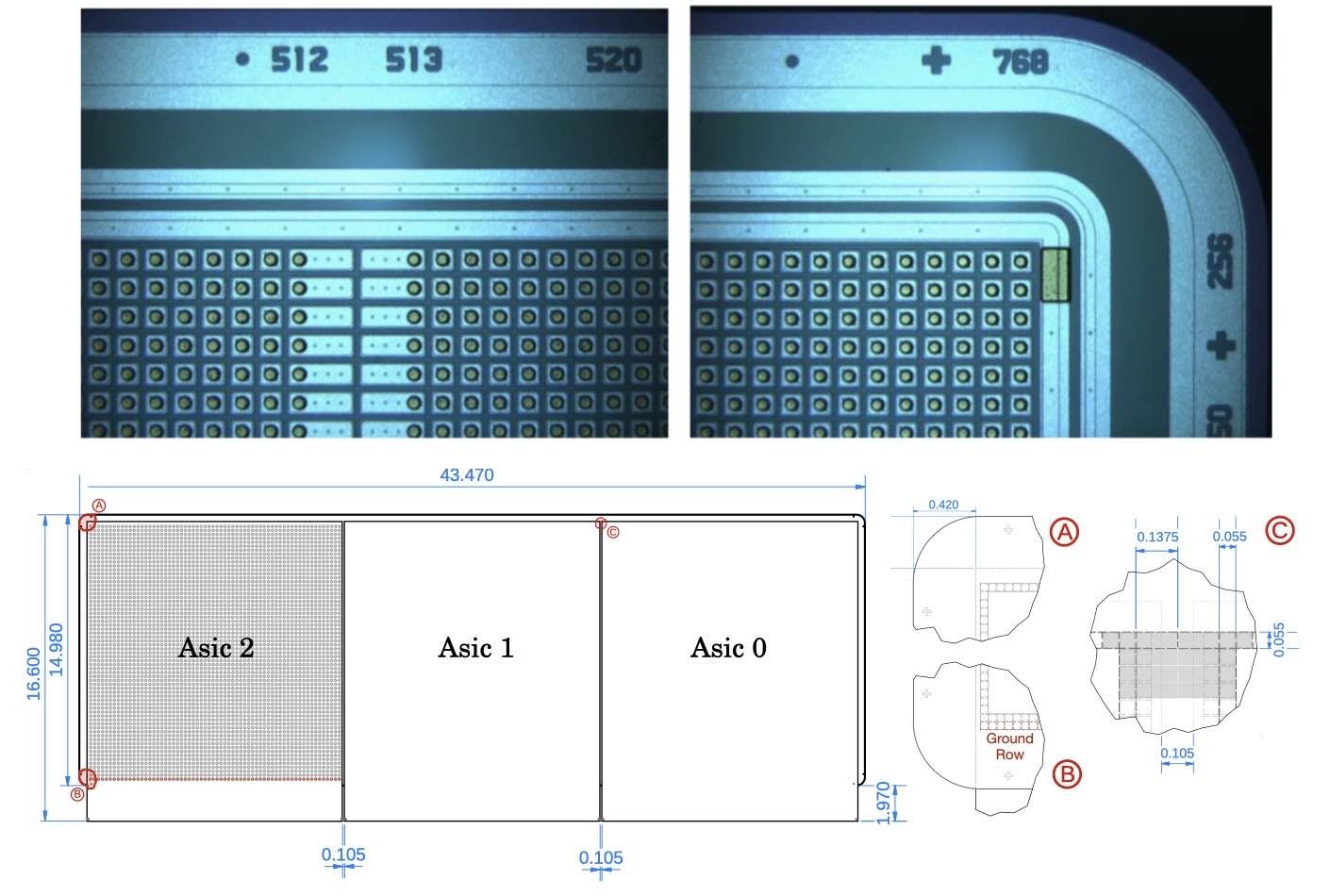}
  \caption{Top left: microscope image showing the elongated sensor
    pixels above the inter-\Asic region.  Top right:~image
    highlighting the ion-etched round corner of the sensor.  Bottom:
    schematic of the sensor tile, showing the overall dimensions of
    the sensor and \Asic.  The pixel layout is shown only under
    \Asic~2.  There are $256 \times 256$ active bonded pixels (only
    every fourth pixel is shown in the figure). An additional row of
    pixels identified in dark red provides a connection between the
    \Asic ground and the innermost guard ring of the sensor.  Three
    corners, encircled in red, are shown in detail on the left (A, B,
    C).}
  \label{SchematicWithPixels}
\end{figure}

For even distribution of the bias voltage to the backside, a $1\mum$
aluminium layer is applied on top of the sensor backside.  The corners
of the sensor are ion-etched to a curved shape to increase the
clearance to the RF box and thus minimise risk of contact damage
during insertion.  To provide sensitivity in the gap between \Asic{s},
sensor pixels that span the distance between two \Asic{s} are
elongated to $137.5\mum$.  These features are highlighted in
figure~\ref{SchematicWithPixels}~(top).

\begin{table}[h]
    \centering\renewcommand\arraystretch{1.1}
    \caption{\velo sensor specifications.}
    \label{tab:sensorspecs}
    \begin{tabular}{|c|c|}
      \hline
      Bulk material thickness & $200\mum$ n-on-p silicon \\
      Most probable unirradiated signal charge & $16\,000\, e^-$ \\
      Minimum end-of-life signal charge & $~6\,000\, e^-$ \\
      Maximum operational voltage &$1000\volt$  \\
      Required charge collection efficiency& $>99\%$\\
      \hline
    \end{tabular}
\end{table}

\subsubsection{Tile production and quality control}

After production by the manufacturer, the \velopix wafers, each
containing 91 \Asic chips, are individually quality-controlled using a
semiautomatic probe station\footnote{\Trmk{Karl Suss} PA200.} with 140
tungsten-rhenium needles.  A probe card\footnote{Manufactured by
  \Trmk{Technoprobe} S.p.A., 23870 Cernusco Lombardone, LC (Italy).}
routes signal from the \velopix to a SPIDR readout
board~\cite{Heijden_2017}.  Needle probes are placed in contact with
the \velopix \Asic{s} pads to provide power, control and readout
during testing. In a first round, the digital functionalities are
tested such as power-up, matrix readout behaviour, the calibration of
digital to analogue converters and the response to trigger and control
commands.  In a second round, the quality of the output links and the
behaviour of the analogue part of each \Asic are tested. The eye
diagram of each output link is verified on an oscilloscope while the
pixel noise and equalisation are checked.  The sensors and the \Asic
wafers are sent to a specialist firm\footnote{\Trmk{ADVACAM Oy},
  Tietotie 3, FI-02150 Espoo, Finland.} for \Asic wafer thinning to
$200\mum$, deposition of solder bumps on the \Asic pads, dicing and,
finally, tile production by bump-bonding three \Asic{s} and one
sensor.  Figure~\ref{SchematicWithPixels} (bottom) shows a schematic
of the bump-bonded tile and the pixel matrix layout.

The quality control programme continues by checking that the \Asic{s}
remain functional and that the sensor can withstand a biasing up to
$-1000 \volt$. The tiles are held on a vacuum jig that allows one to
position ten tiles with a $40\mum$ mechanical precision such that they
can be tested with the semiautomatic probe station.  While the \Asic
pads are connected to the probe card via the tungsten needles, the
tile sensor is biased to $-140 \volt$, the sensor depletion voltage
before irradiation. The pixel discriminator response is equalised and
the single pixel noise is measured. Analogue test pulses are fired on
individual pixels to detect cross-talk, which would indicate a short
in the solder bumps. Finally, after equalisation of the
discriminators, a strontium source is placed on top of the probe card
while a $1000\,e^-$ threshold is set on all pixels. Empty pixels
indicate missing bonds.  Across the whole tile production, no shorted
bumps were found.  Tiles with missing bumps were returned to the
bump-bonding firm for reworking.

\subsection{Microchannel cooling}
\label{sec:velo:cooling}

The silicon detectors operate in vacuum and the significant heat
generated must be dissipated by conduction.  There are 12 \Asic{s} per
module, each with a power budget of 3\watt. With the power consumption
of the \Fend electronics and the ohmic heating from the irradiated
sensors as well, the cooling is designed to extract up to $40\watt$
per module, which is $\sim2\kwatt$ from the whole \velo. Moreover, to
mitigate the effect of the radiation damage, the silicon must be
permanently cooled to below \SI{-20}{\degc}.  The chosen solution is
an evaporative cooling system with bi-phase \cotwo flowing through
microchannels within a silicon substrate approximately \SI{100}{\cma}
in area, herein referred to as a
\emph{cooler}~\cite{DEAGUIARFRANCISCO2022166874}.\looseness=-1

Carbon dioxide is attractive as a coolant because it is inert,
inexpensive and has a large latent heat of evaporation which is
exploited in a bi-phase system. By circulating the \cotwo in
microchannels inside the cooler on which heat-generating components
are glued, a thermal performance of 2--4$\degk\cma\watt^{-1}$ was
demonstrated during construction, where the range depends primarily on
the 50--$100\mum$ glue thickness between cooler and heat-generating
\Asic{s}.  The maximum temperature difference between the \cotwo exit
temperature and the sensor tip, which overhangs the cooler by
\SI{5}{\mm}, is \SI{6}{\degc} with nominal \Asic operation ($100\mum$
glue layer).

The \cotwo cooling system operates at typical pressures of $14\bar$ at
\SI{-30}{\degc} and up to $62\bar$ at room temperature.  For detector
safety, the integrity of the entire cooling network, including every
cooler, must be verified at three times the maximum operational
pressure i.e., $187\bar$.

\subsubsection{Microchannel cooler design and fabrication}

\begin{figure}[h]
  \centering
  \includegraphics[width=0.47\linewidth]{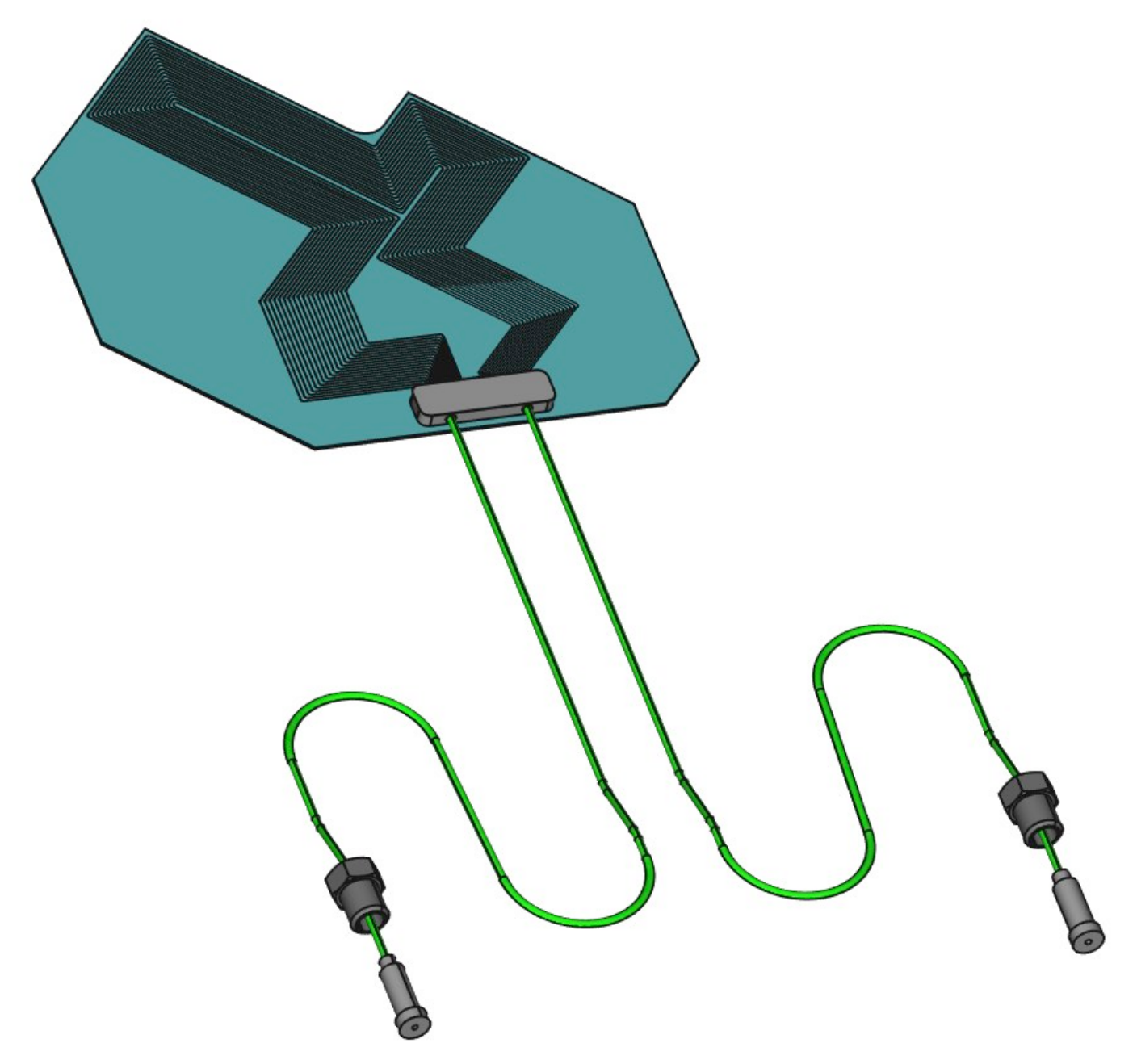}
  \includegraphics[width=0.52\linewidth]{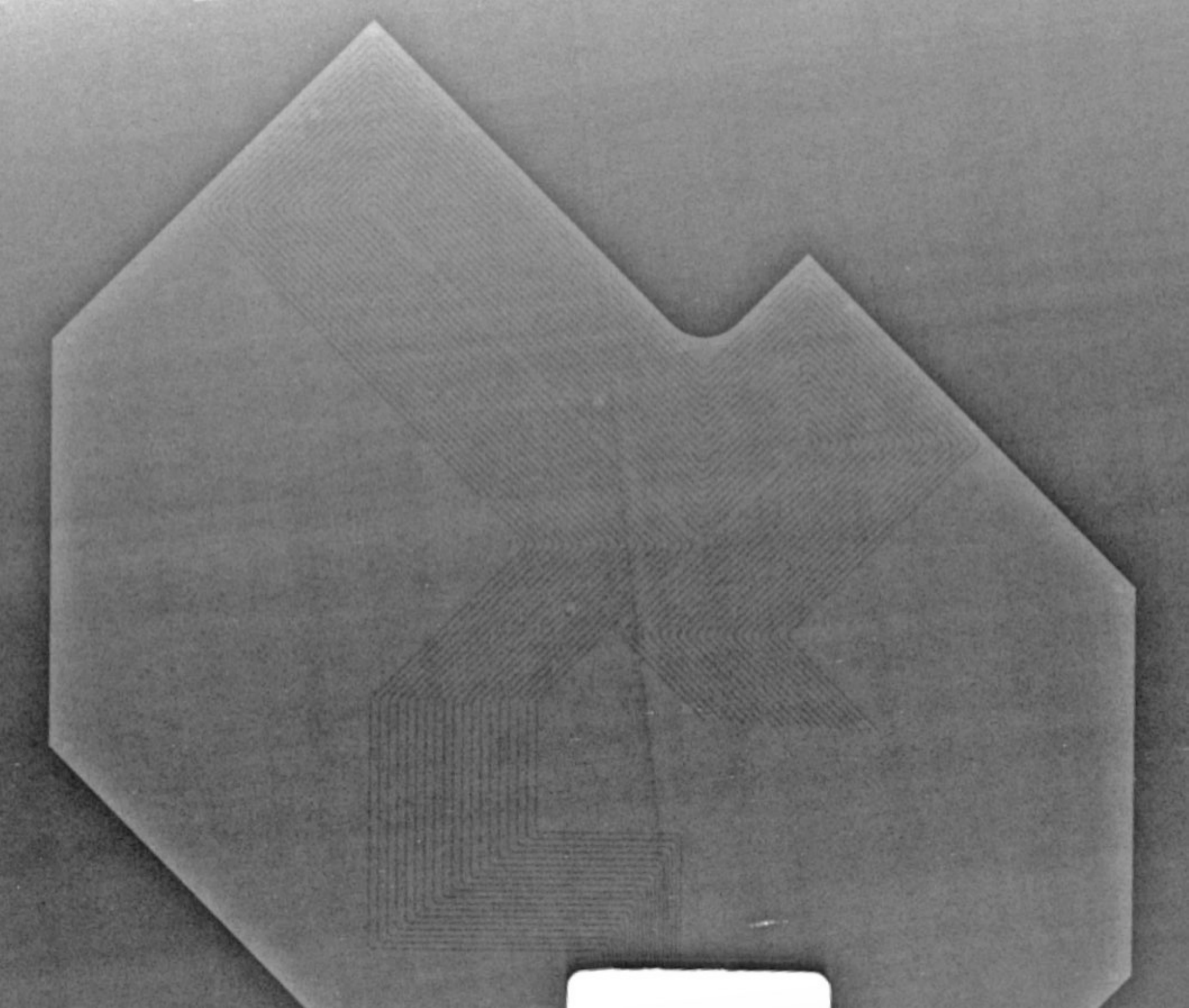}
   \caption{Left: illustration of the silicon microchannel coolers and fluidic connector. 
    Right: the parallel lines represent the etched microchannels, which can be seen in the X-ray image.
    \label{fig:cooling_illustration}}
\end{figure}

The 20 microchannels within each \SI{500}{\mum}-thick cooler range in
length from $271\mm$ to $332\mm$.  The channels stretch from the inlet
manifold, pass under the detector elements before returning to the
outlet manifold located adjacent to the inlet.  Each channel is
$200\mum$ wide and $120\mum$ deep. For the first $40\mm$ of every
channel, the width and depth is reduced to $60\mum\times 60\mum$ to
provide a uniform flow restriction. This ensures an even distribution
of coolant despite the varying heat load; the \cotwo boils in a
uniform manner as the cross section increases by a factor $\sim7$ at
the end of the restrictions.  An illustration of the composite device
is shown in figure~\ref{fig:cooling_illustration} beside an X-ray
image that reveals the microchannels inside the cooler.  The \cotwo
passes into and out of the silicon through slits machined in fluidic
connector that match the position and size of the microchannel
manifolds.  The fluidic connector is made of Invar36 whose thermal
expansion closely matches that of silicon.  Two vacuum-brazed,
$^1\!/\!_{16}\,$inch pipes with inner diameter of $0.57~(0.87)\mm$
service the inlet (outlet) of the connector, which is attached to the
silicon by a fluxless soldering assembly.

The microfabrication\footnote{CEA-Leti, 38054 Grenoble, France.} of
the silicon was performed using deep-reactive ion etching and direct
bonding techniques.  The process starts with double-side polished
silicon wafers in which the microchannel patterns are etched.  The
microchannels are closed by applying a second silicon wafer and using
hydrophilic bonding.  Afterwards, the second wafer is thinned to
$240\mum$ and the first wafer to $260\mum$ achieving a final thickness
of $500\mum$.  A soldering pad comprising three layers of
metallisation, Ti ($200\nm$), Ni ($350\nm$) and Au ($500\nm$) is
deposited around the microchannel manifold. Alignment marks are also
deposited during this step. Last, the inlet and outlet slits are
opened by ion-etching before plasma-dicing cuts the silicon into the
characteristic shape.

\subsubsection{Cooler assembly and quality control}

In order to attach the fluidic connector to the silicon in a reliable
and reproducible manner, a novel, fluxless soldering technique was
developed.  The technique involves many steps, including polishing,
cleaning (acetone, ethanol, ultrasound bath, plasma cleaning),
outgassing, and pretinning solder on each piece.  Pretinning is
performed in a reducing atmosphere of nitrogen with 3\% concentration
of formic acid vapour.  Subsequently, the opposing silicon and Invar
surfaces are aligned and the solder is reflowed in vacuum. During this
procedure, nitrogen at atmospheric pressure is reintroduced to
minimise the size of voids forming in the molten solder.  After
soldering, a rectangular piece of silicon is glued behind the
manifold, where it serves to reinforce the area where the blow-out
force is largest.  The full procedure is described in
ref.~\cite{DEAGUIARFRANCISCO2022166874}.

\begin{figure}[t]
  \centering
  \includegraphics[width=0.39\linewidth]{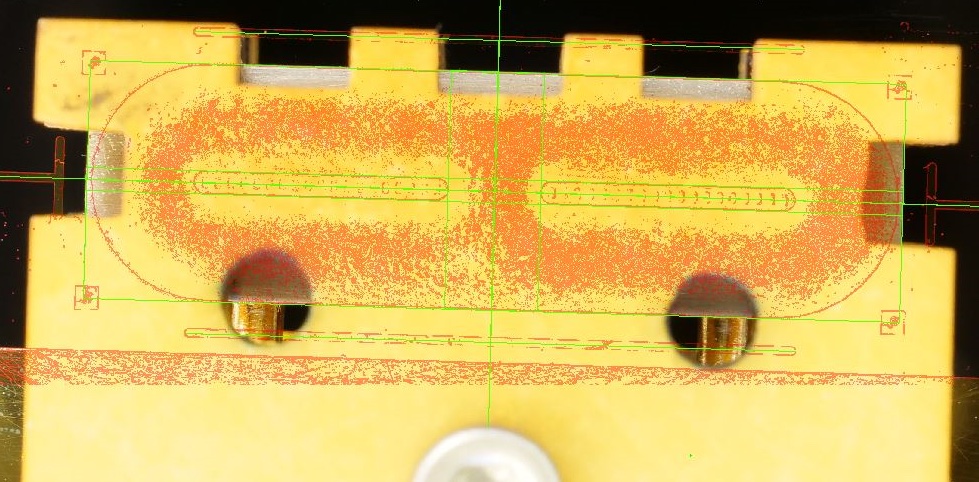}
  \includegraphics[width=0.59\linewidth]{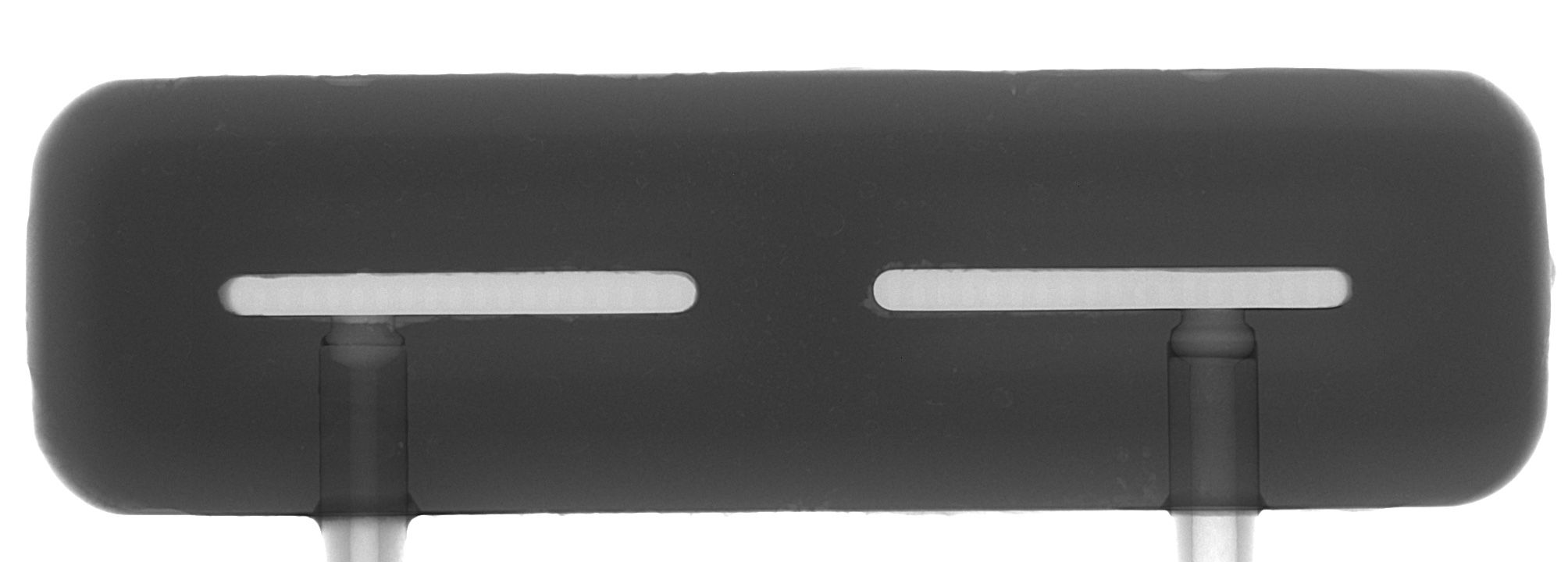}
  \caption{Left: image of a fluidic connector aligned in the
    horizontal plane with the corresponding solder layer on the
    silicon cooler, indicated with the red shadow. Right: X-ray of the
    solder joint attaching the fluidic connector to the microchannel
    cooler. No solder has entered the two inlet regions, and no large
    voids are seen in the solder layer. }\label{fig:mc}
\end{figure}

Due to the risks associated with this new technology, an exhaustive
quality control programme was developed.  An X-ray of the solder
joint, e.g.\ figure~\ref{fig:mc} (right), checks for voids in the
solder layer and verifies that no solder has entered the fluidic
pathway.  Each cooler is checked for leak tightness by helium
detection to below $10^{-9}\mbar\litre/\sec$.  Tests are done in two
ways: in ambient air with $60\bar$ helium inside the cooler and with
the cooler placed in a $1\bar$ helium atmosphere whilst pumping on the
microchannels.  The silicon surface is then checked for flatness with
a tolerance on planarity of $100\mum$.  To check for robustness, the
cooler is pressurised with nitrogen to $130\bar$ for 45 minutes (the
burst-disk release pressure of the final system) as well as a 15
minute, $187\bar$ stress test.  Finally, a visual inspection requires
the silicon surface to be clean of any residue and the shape of the
pipe loops to be within tolerance for the module construction.  Around
30 full-size prototypes were trialled before 81 installation-quality
coolers were produced with an acceptance yield of 87\%.

\subsection{The module}
\label{sec:velo:module}

The \Velo module brings together the silicon detectors, their cooling,
powering, readout and mechanical support into a single, repeating
unit. This section describes the design, construction, and quality
control of these objects.

\subsubsection{Anatomy}

\begin{figure}[t]
  \centering
  \includegraphics[width=0.94\linewidth]{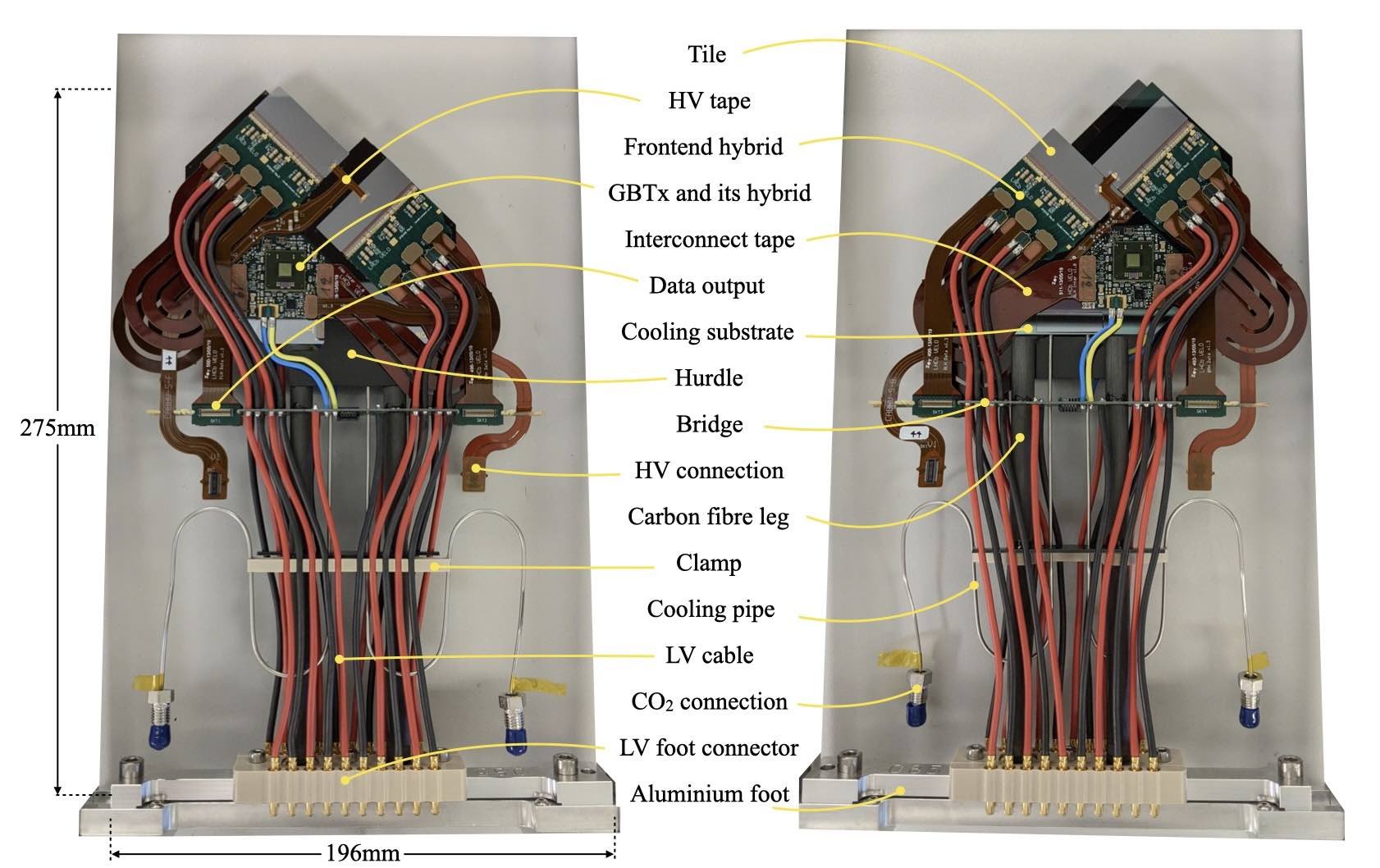}
  \caption{Photo of the (left) upstream and (right) downstream faces
    of a fully-assembled \Velo module.}
  \label{fig:velo:module_sides}
\end{figure}

A \Velo pixel module is a double-sided detector structure with a
microchannel cooler at its core.  The cooler is glued at the Invar
cooling connector to the \emph{mid-plate}, a carbon fibre plate which
sits atop two carbon fibre legs anchored on an aluminium foot.  On
each cooler face, a pair of tiles are glued at right angles such that
they approach the \lhcb beam line along their inner edge.  Along their
periphery, each triplet of \velopix \Asic{s} are wire-bonded to the
\Fend electronics \emph{hybrid} that transmits control signals and
routes their data out to a PCB flex cable.  The 2 \Fend hybrids (per
module side) are connected to a third hybrid housing a \Gbtx control
\Asic, see section~\ref{sec:velo:hybrids}.  Power is delivered through
an assembly of 20 silicone-coated copper cables and a PCB transition
bridge.  The bridge is used to change to thinner cables near the \Fend
in order to reduce the material budget close to the beam line. It also
provides mechanical support for the various cables.  The high voltage
(HV) for the silicon sensors is delivered by PCB flex cables which are
bonded to the top surface of the sensors.
Figure~\ref{fig:velo:module_sides} shows these components on both
sides of an assembled module.

\subsubsection{Assembly and quality control}
\label{sec:velo:Assembly and qc}

The module production was performed in clean rooms at two production
sites.  All steps followed preagreed instructions to ensure
consistency across the two sites.  Every process and qualification was
recorded in an online database, which automatically aggregated the
results to provide immediate feedback.  Following several years
developing the assembly process, 53 identical, installation-quality
modules were built over the course of 16 months.

\begin{figure}[t]
  \centering
  \includegraphics[width=0.3\linewidth]{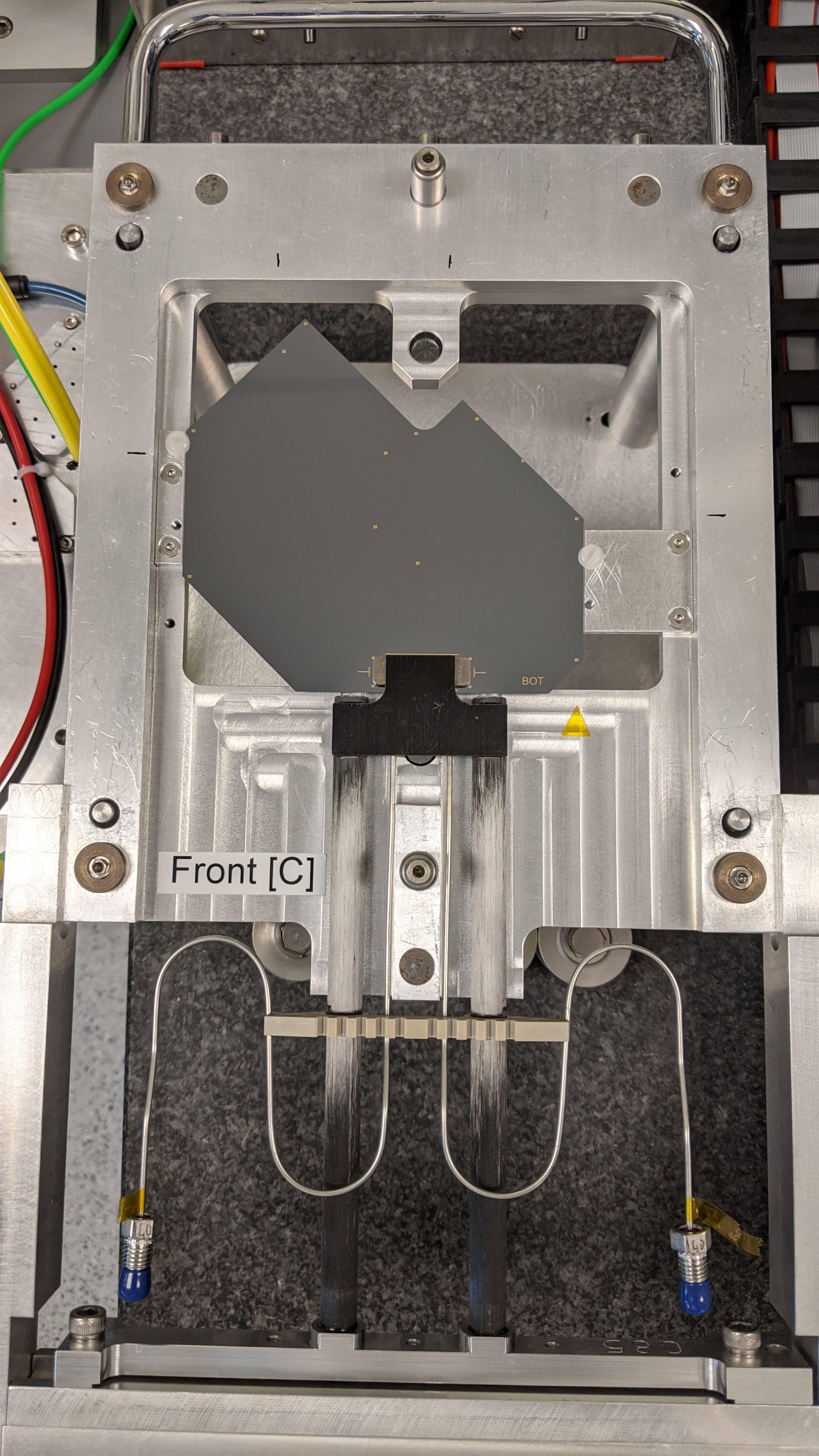}
  \includegraphics[width=0.3\linewidth]{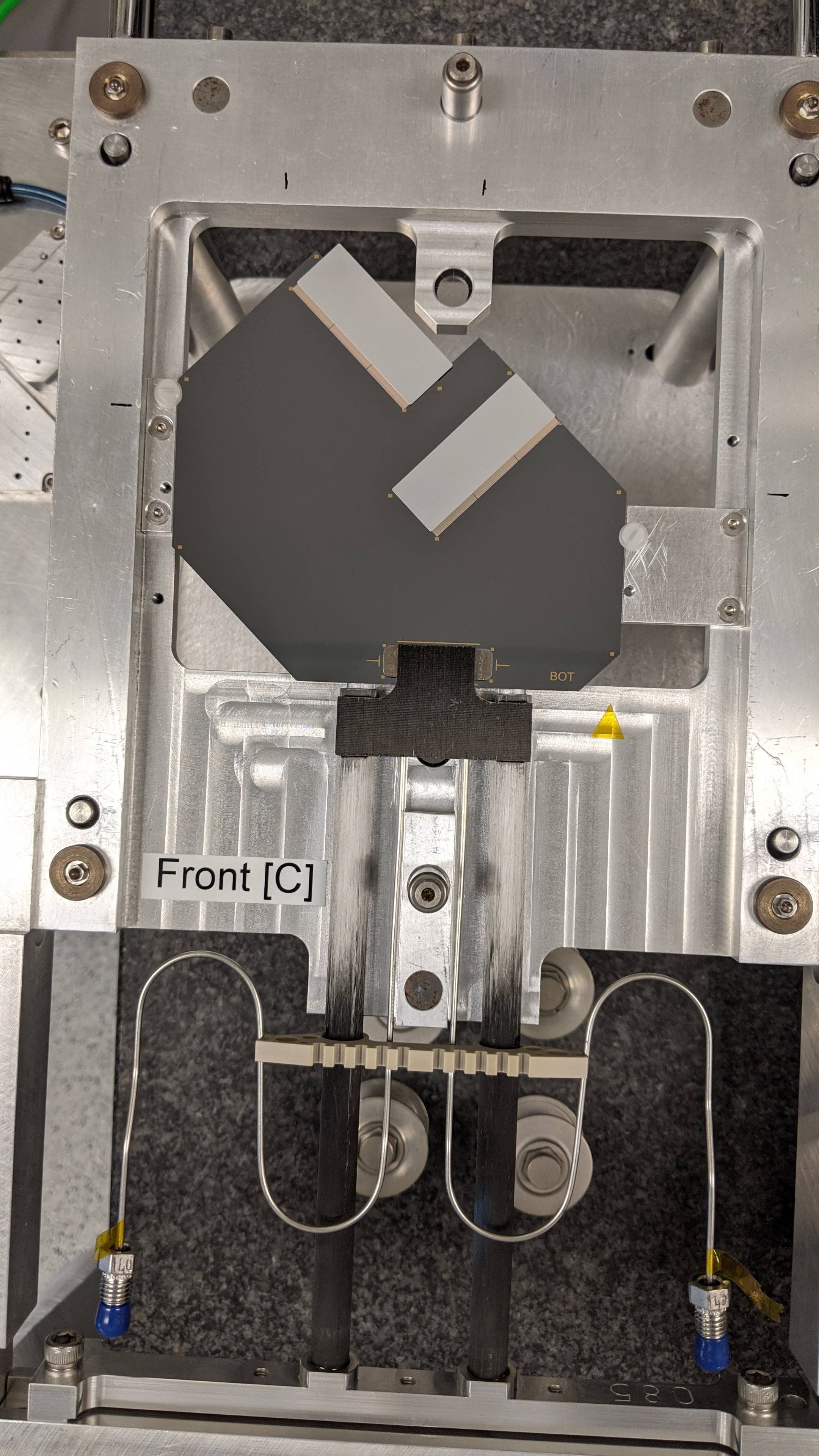}
  \includegraphics[width=0.3\linewidth]{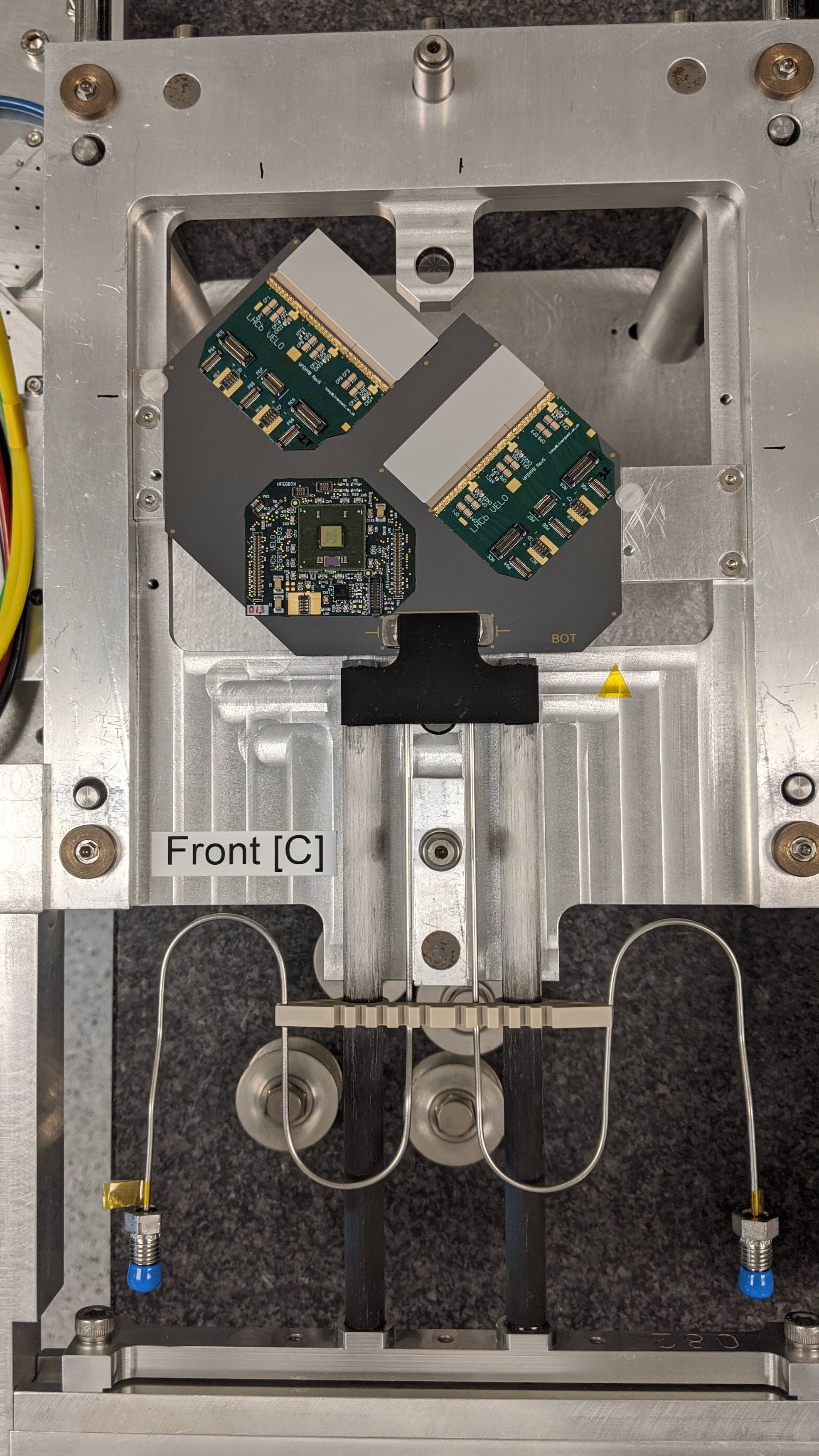}
  \caption{From left to right: bare module; module with tiles; then
    with hybrids too.}
  \label{fig:velo:module_stages}
\end{figure}

Different stages during the assembly of a module are shown in
figure~\ref{fig:velo:module_stages}.  First, two carbon fibre legs are
glued\footnote{Using \Trmk{Araldite} 2011.} to the aluminium foot, and
at the other end to the carbon fibre mid-plate. This rigid structure
is glued to the cooler at the Invar fluidic connector, using the same
glue.  The cooling pipes are clipped into a clamp attached to the
legs, which minimises the risk of transmitting stress to the cooler
from manual handling of the pipes.  The four tiles are
glued\footnote{Using \Trmk{Stycast} FT2850 and 23LV catalyst.} to the
microchannel cooler with precision-made jigs.  The placement of the
tiles is performed with an in-plane precision relative to the foot of
better than $30\mum $.  In addition, the thickness of the glue layer
is a critical parameter of the module quality, as it affects not only
the mechanical properties of the attachment, but also the cooling
performance of the module.  In order to meet all the requirements on
the glue interface, a uniform layer with thickness in the range of
50--100\mum is required. The placement of the \Fend and \Gbtx hybrids
is performed in a similar manner, although the placement tolerance is
relaxed to $100\mum $. A silicone adhesive\footnote{\Trmk{Loctite} SI
  5145.} is used for this process as its flexibility ensures minimal
stress on the cooler given the different coefficients of thermal
expansion on either side of the glue layer. The HV cables are
connected to the cooler with a fast-curing
glue\footnote{\Trmk{Araldite} 2012.} and manually bonded to the sensor
surface. Finally the module is placed in a wire-bonding machine, where
an automated programme carries out this process one side at a time,
for a total of 1680 bonds.  The last step consists in attaching all
interconnecting and power cables to the module.

Once built, the functionality of each module is verified in vacuum
with cooling close to \SI{-30}{\degc}, both before and after a thermal
cycle. The two-way communication with the \Gbtx chips and the 12
\velopix \Asic{s} is verified as well as the response of each \velopix
to fast trigger signals and the equalisation of the pixel matrix. A
bit error rate test of the 20 output links is performed and the sensor
leakage current is checked with a HV scan.

\subsubsection{Metrology}

\begin{figure}[t]
  \centering
  \includegraphics[width=0.32\linewidth]{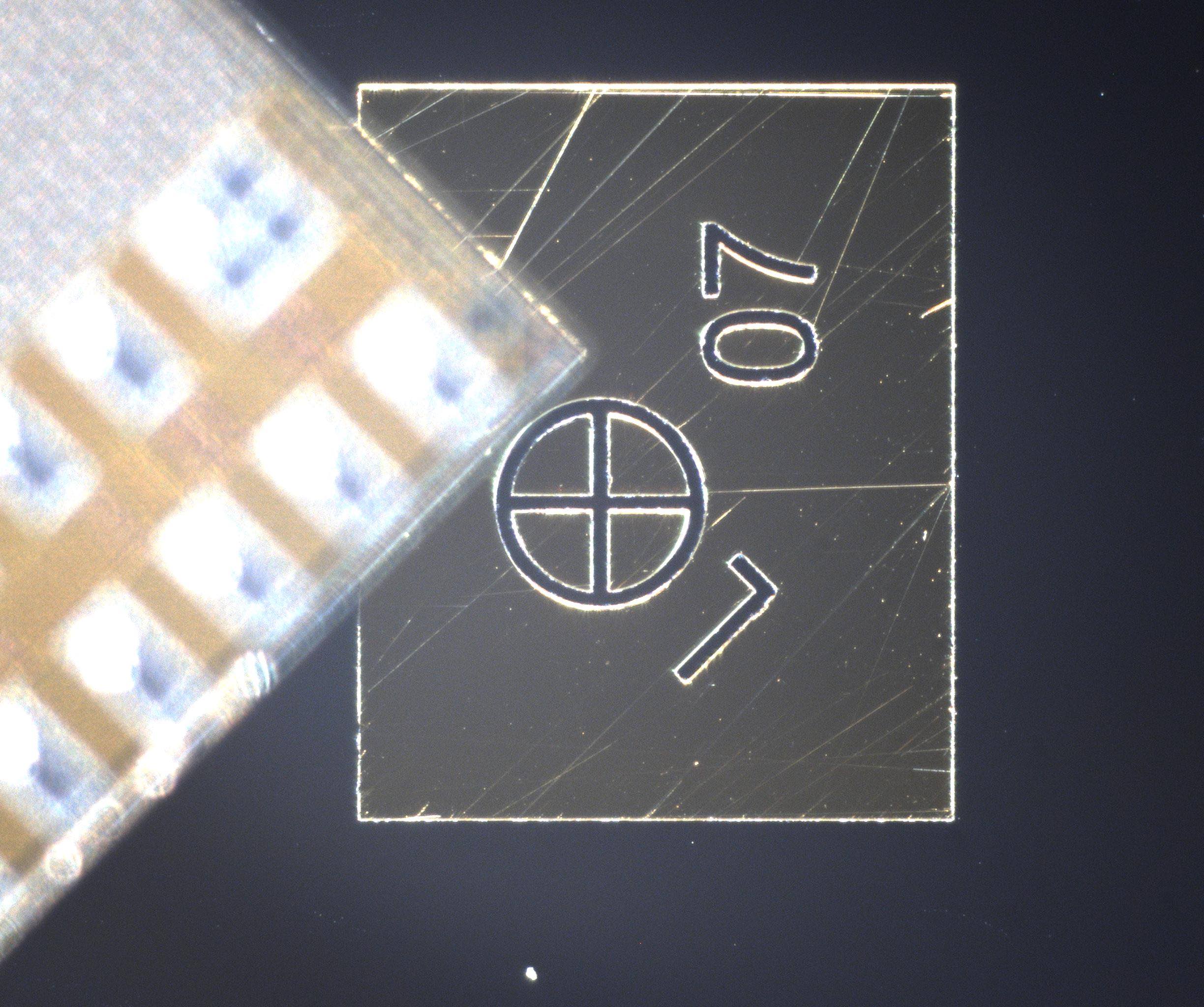}
  \includegraphics[width=0.32\linewidth]{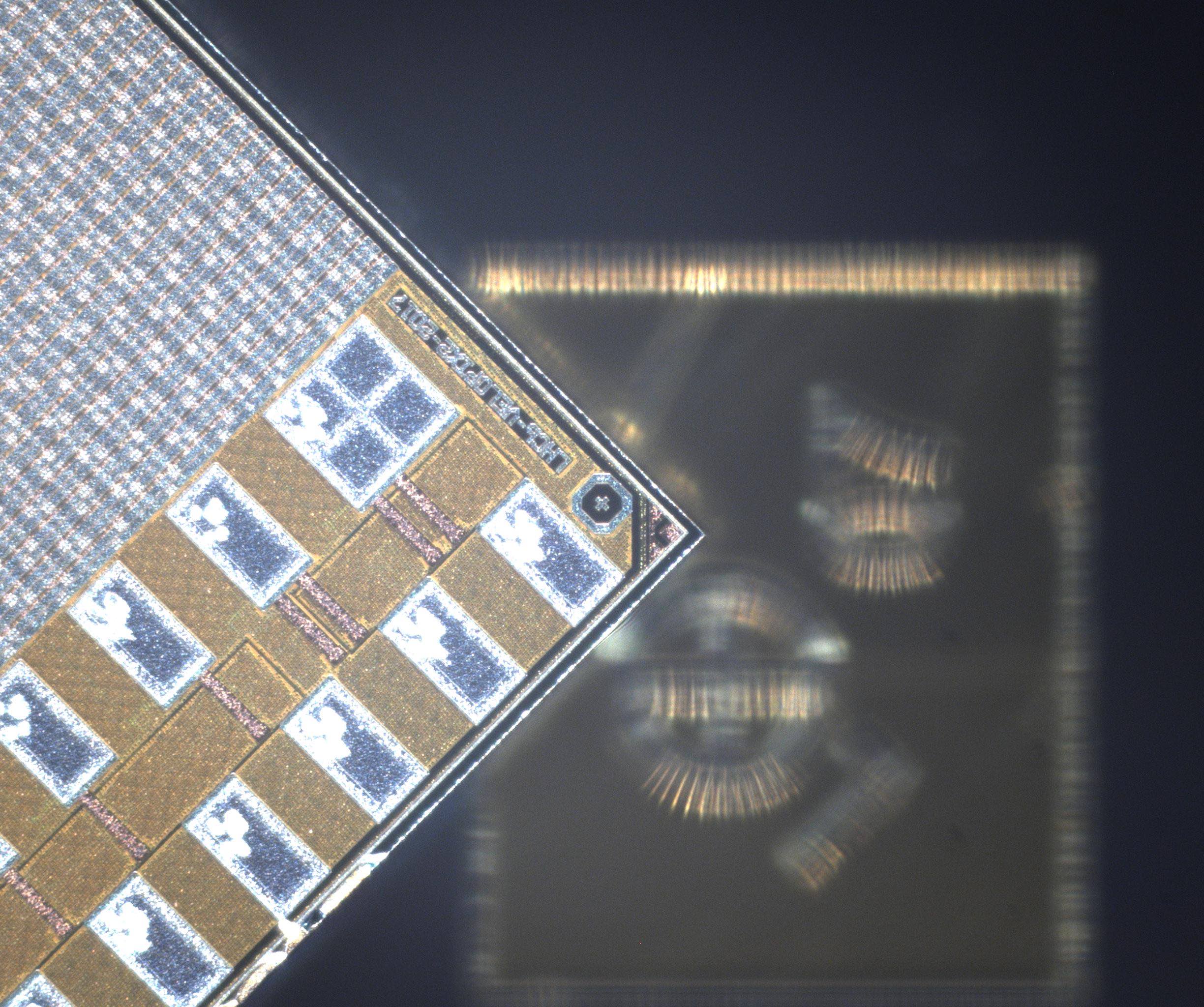}
  \includegraphics[width=0.32\linewidth]{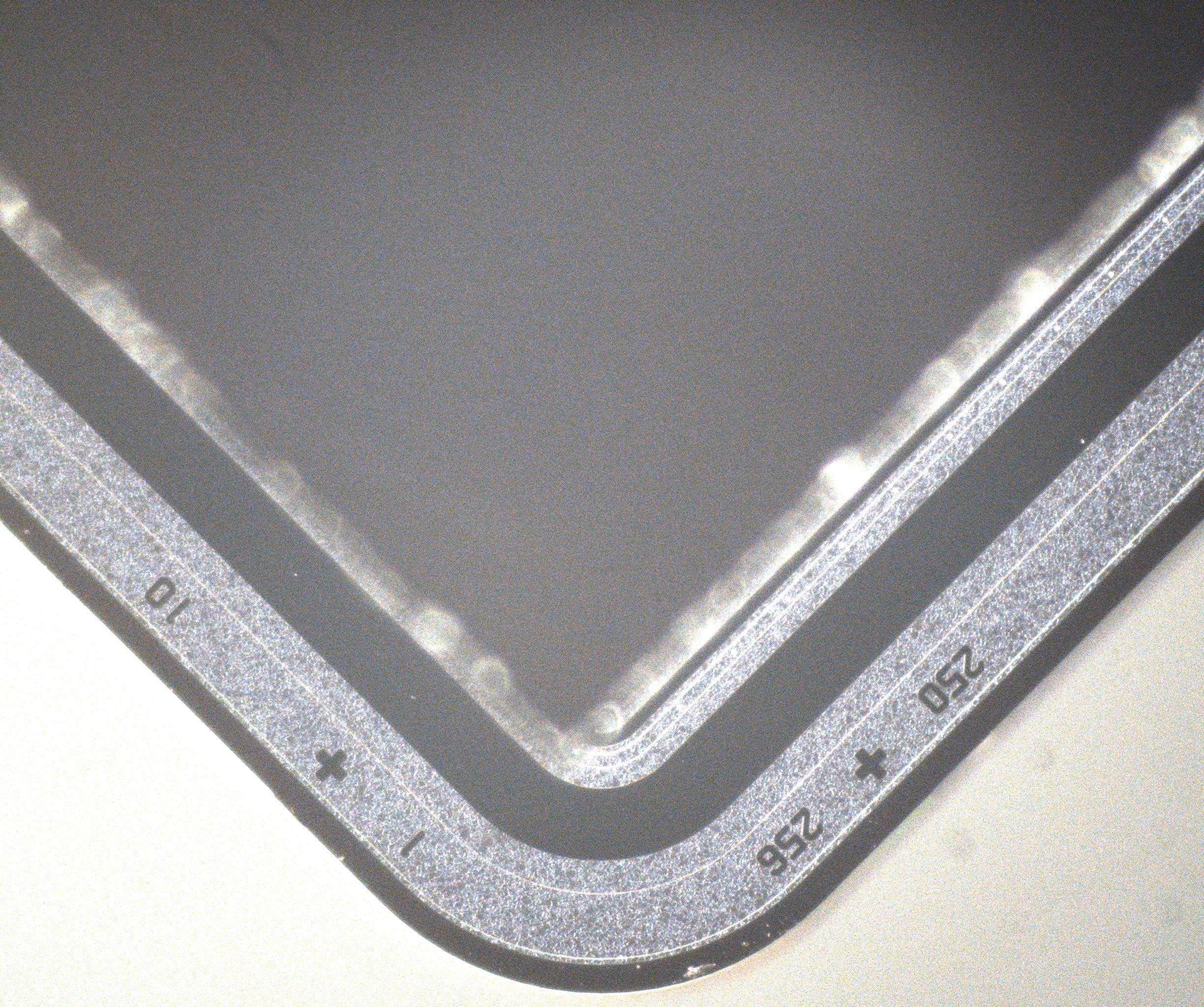}
  \caption{Cross-shaped markers used for module metrology. Left:~on
    the microchannel cooler; centre: on the \velopix \Asic; right:~on
    the \Asic-side of the sensor.}
  \label{fig:velomodulemetrologymarkers}
\end{figure}

Since the nominal distance of the module tips to the RF box is as
small as $0.8\mm$, direct knowledge of the position of the innermost
tiles is vital to confirm suitability for installation.  The relative
position of the sensors on the module is also an important input to
the track-based alignment algorithm.  The metrology is divided in two
parts: orientation and position in the plane of the sensors ($xy$
alignment); and position perpendicular to that plane, parallel the
beam axis ($z$ alignment).  The $xy$ alignment is based on
measurements of markers (see
figure~\ref{fig:velomodulemetrologymarkers}).  The absolute positions
of these markers are measured relative to the dowel pin on the module
foot.  These measurements extract deviations of the $x$ and $y$
positions and rotation of a sensor around the $z$~axis in the nominal
module frame.  In the $z$ direction, the relative position and
orientation of the sensors is affected by any curvature of the cooler
and variations in the thickness of the glue layer.

\begin{figure}[t]
  \centering
  \includegraphics[width=0.99\linewidth]{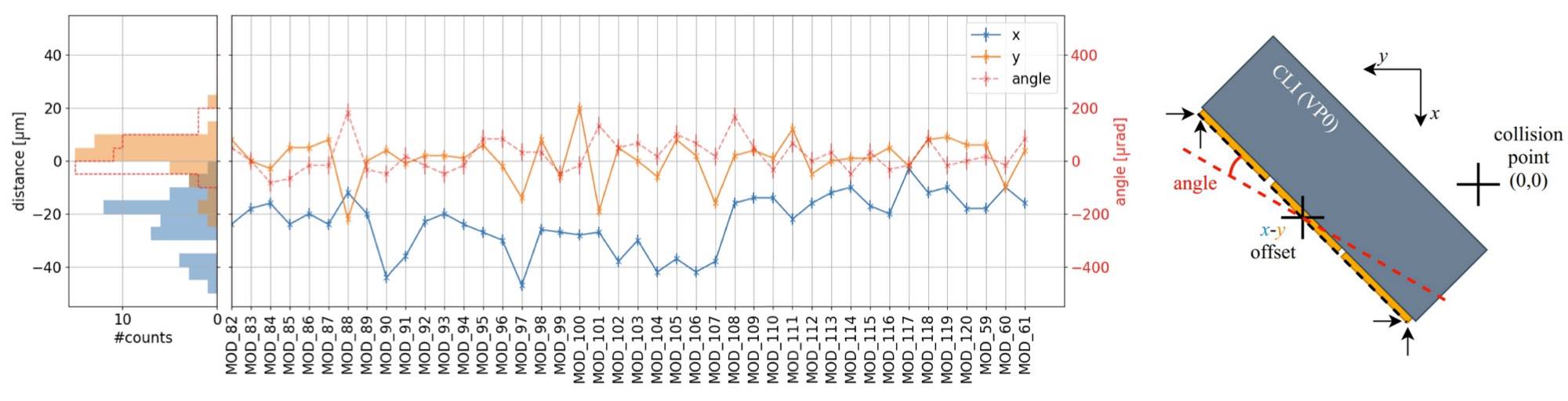}
  \caption{Tile metrology results showing $x$ and $y$ positions and
    angles for each module.}
  \label{fig:tilemetrology}
\end{figure}

Each production site developed its own method for metrology of the
modules, however both achieved a resolution of a few microns in all
coordinates. The results in $z$ reveal thicknesses of the glue between
tile and cooler within the range 30--110\mum, with a mean value of
$80\mum $. The results in $xy$, shown in
figure~\ref{fig:tilemetrology} show excellent tile positioning in the
$y$ direction, which is most critical in terms of clearance to the RF
foil.

\subsection{Electronics and readout chain}
\label{sec:velo:electronics}

The main role of the \velo electronics system is to transport data
from the \velopix \Asic{s} to the off-detector processing units.  The
system also delivers clock and control signals to the modules, as well
as low voltage to power the electronics and high voltage to bias the
sensors.

\subsubsection{System architecture}

\begin{figure}[t]
  \centering
  \includegraphics[width=0.9\textwidth]{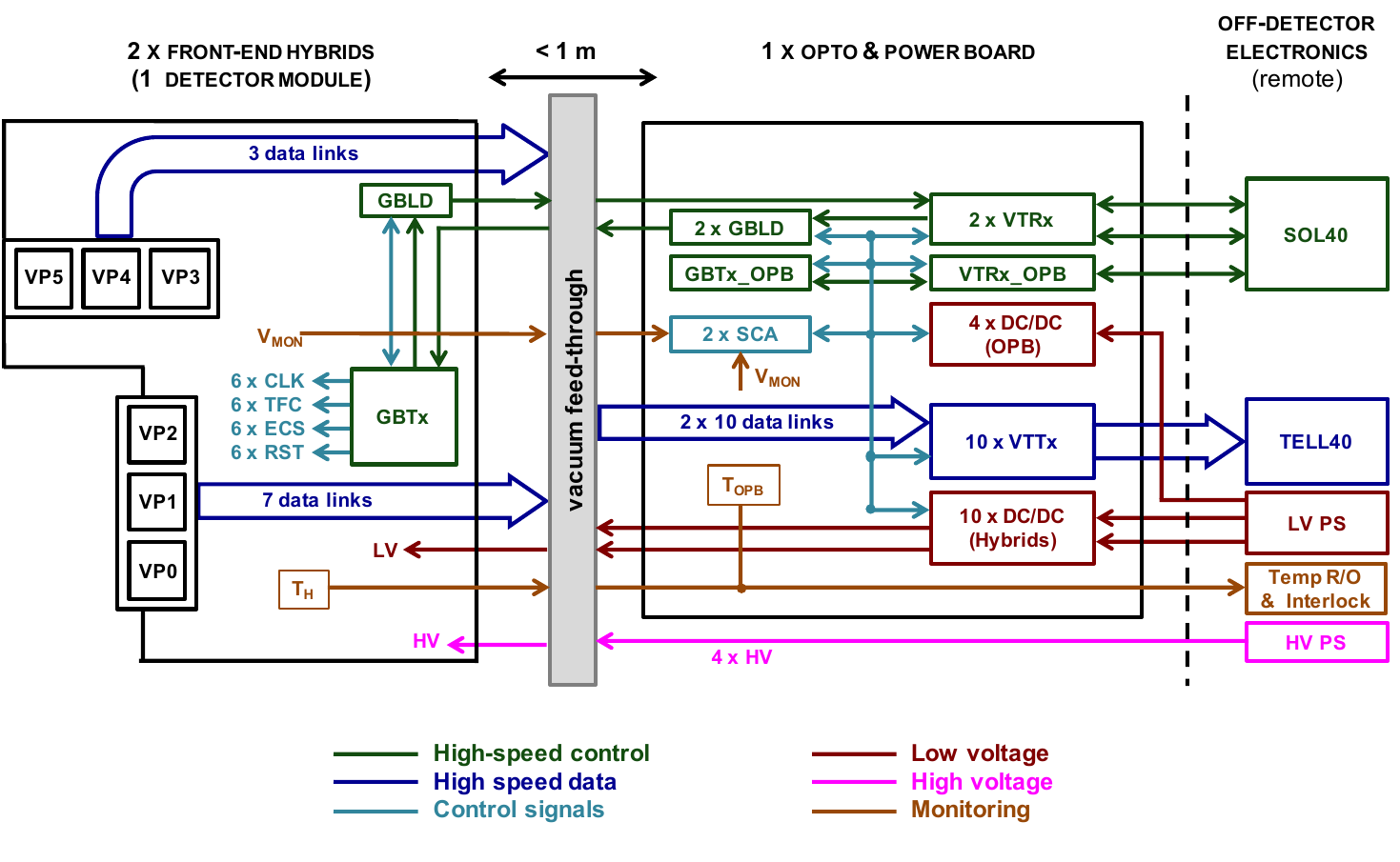}
  \caption{Block diagram showing the main parts of the \Velo
    electronics system.}
  \label{fig:velo_elec_overview}
\end{figure}

The electronic components of the system are located in three places:
on the detector module, immediately outside the \Velo vacuum vessel
and off detector, with dedicated cabling running between them.  An
overview of the system is shown in
figure~\ref{fig:velo_elec_overview}.

The \velopix \Asic~\cite{poikela2015} reads the analogue signals from
the sensor and sends binary hit data in \Acr{spp} (see
section~\ref{sec:velo:velopix}) over serial links at rates up to
5.12\gbps per link.  Serial data routing as well as distribution of
clock, control and power is managed on the \Fend by hybrid circuits
(section~\ref{sec:velo:hybrids}). The serial data from the hybrids is
transmitted out of the secondary vacuum on high-speed serial links,
through a vacuum feedthrough board to the \Acr{opb}, see
section~\ref{sec:velo:opb}, mounted on the exterior of the vacuum
vessel. The control signals to and from the \Fend are transmitted on
identical serial links.  The low and high voltage is supplied through
the same vacuum feedthrough board to separate cables.  The temperature
monitoring (section~\ref{sec:velo:tempmoni}) is routed through the
data flex cables and the \Acr{opb}.

Through optical links, the \Acr{opb} transmits the data to the
\Tellfourty data acquisition cards (section~\ref{sec:velo:daq}) whilst
receiving control signals from the \Solfourty readout supervisor.  The
\Tellfourty and \Solfourty boards are located in the \Daq server rooms
in the data centre, which requires over 300\m of optical fibre.  The
high and low voltage supplies are located in the electronics barracks
in the \lhcb cavern, requiring about $60\m$ of cable.

\subsubsection{Front-end circuits (hybrids)}
\label{sec:velo:hybrids}

\begin{figure}[t]
  \centering
  \includegraphics[width=0.99\linewidth]{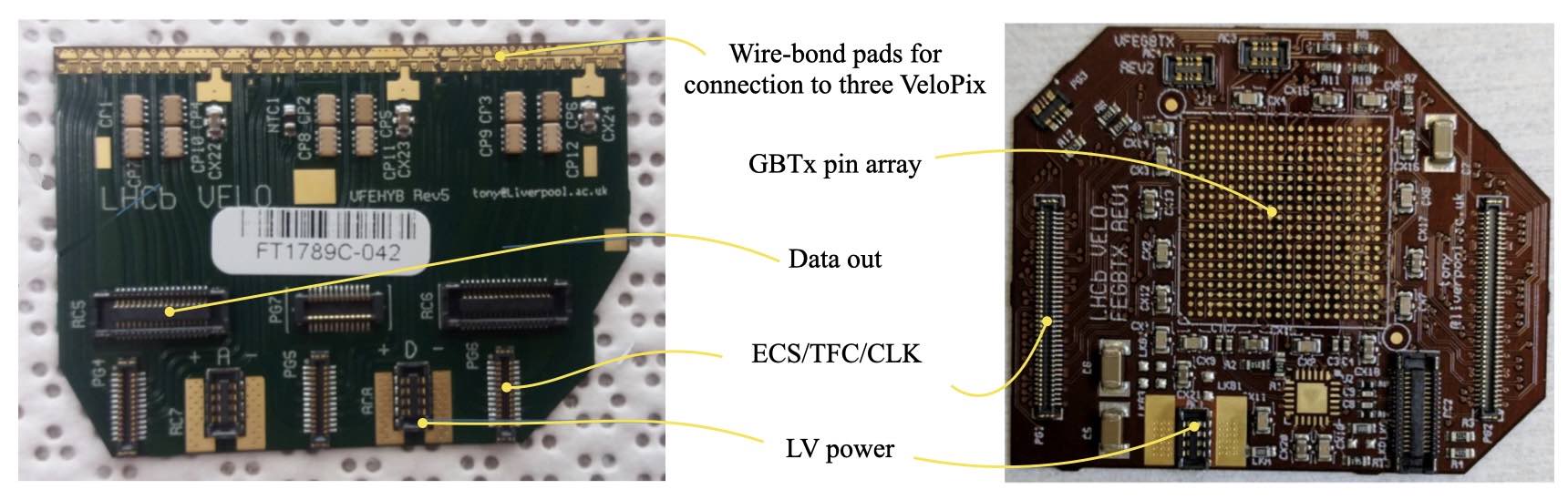}
  \caption{\label{hybrids_annotated}The \Fend (left) and \Gbtx
    (right) hybrids.}
\end{figure}

On the module, the distribution of \Asic power, clock and control
signals and the routing of outward-bound data is provided by
hybrids. These are four-layered, flexible printed circuits
interconnected with two-layered polyimide cables. The hybrids have a
total thickness of $390\mum$ and come in two types, as shown in
figure~\ref{hybrids_annotated}. The first type provides the \Fend
electronic interface to each \velopix where wire bonds are used to
connect the hybrid to the \Asic periphery. The second type houses the
\Gbtx chip and distributes timing signals and fast control
instructions to the \velopix \Asic{s} via the \Fend hybrids.  Slow
controls are routed through the \Gbtx hybrid as well as monitoring of
the bias voltage.\looseness=-1

On each module face there are two \Fend hybrids and one \Gbtx hybrid.
The \Fend electronics are packaged into these three pieces rather than
one larger hybrid circuit in an effort to minimise stress on the
module.  The necessary thickness of these hybrids is driven by the
cross section of copper necessary to transfer a suitable current to
the \Asic{s}, and leaves them relatively rigid.  As described in
section~\ref{sec:velo:Assembly and qc}, a sufficiently flexible glue
was chosen to attach the hybrids to the cooler and absorb contraction
differences when cooling to \SI{-30}{\degc}.

\subsubsection{High speed serial cables}
\label{sec:velo:links}

All module readout and control signals are routed inside the vacuum
using low-mass, high speed PCB flex cables~\cite{Bates:2017fjl}.
Polyimide microstrip technology is used on the module where low-mass
materials are of critical importance.  From the module to the vacuum
wall, four flex cables each carry up to seven 5.12\gbps serial links.
The dielectric used for these flex cables has to be radiation tolerant
and provide low dissipation loss for high-frequency signals. In
addition, high reliability and yield are required for the impedance
control.  The chosen material for this purpose is an all-polyimide
thick copper-clad laminate\footnote{DuPont \Trmk{Pyralux}
  AP-\emph{PLUS}.} offering high signal integrity, reliability and has
wide use in medical and aerospace applications.

\subsubsection{Vacuum feedthroughs}

\begin{figure}[t]
  \centering
  \includegraphics[width=0.99\textwidth]{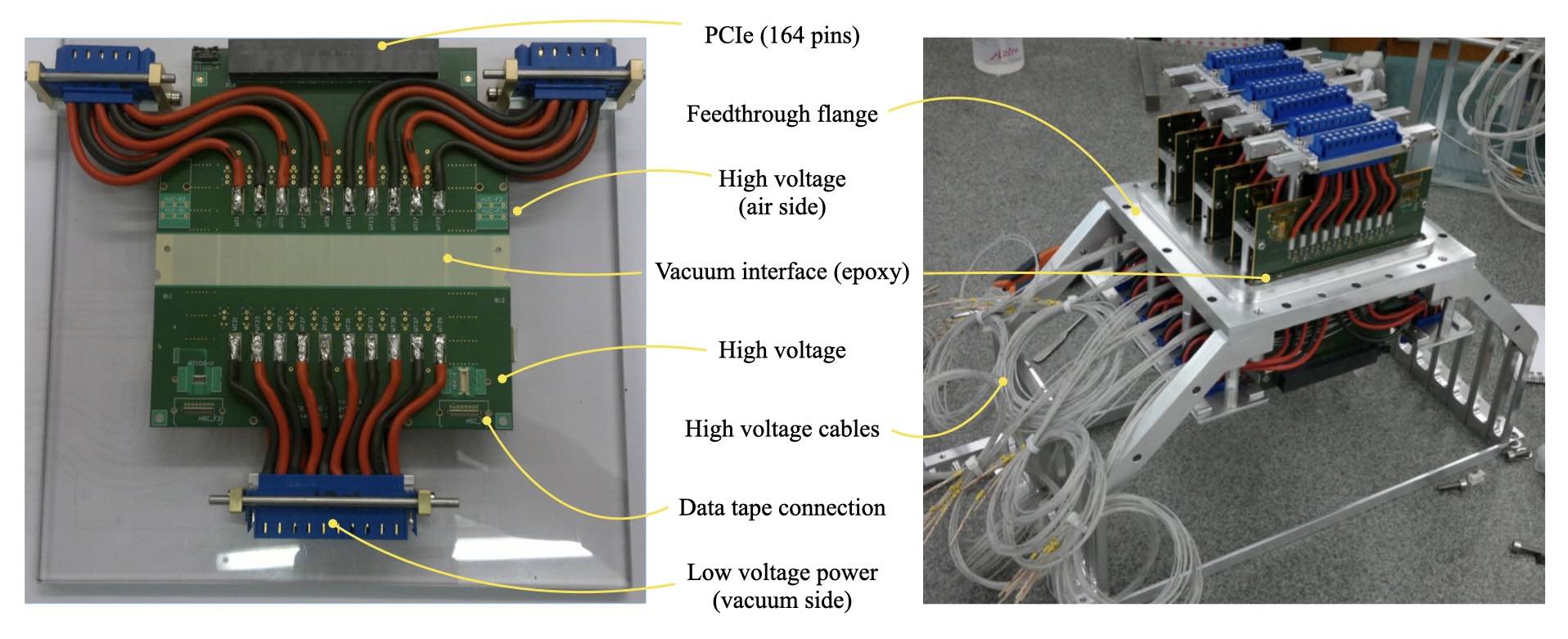}
  \caption{Left: custom-made vacuum feedthrough board. Right: a
    populated vacuum flange.}
  \label{fig:vfb}
\end{figure}

The routing of such a dense number of high speed signals through a
vacuum barrier represented a challenge which could not be solved by
any commercially available solution.  A bespoke solution was developed
using a PCB with edge metallisation glued into a vacuum flange with
epoxy.  The board is a 12-layer printed circuit ($2\mm$ thick) through
which the high speed data signals are transmitted.  The low voltage
power supply to the detector electronics ($\sim 2.5\amp$), the bias
voltage ($<1000\,$V) and the temperature data also pass through this
board.  Tests show that sealing with epoxy\footnote{\Trmk{Araldite}
  2011 (2020) on the air (vacuum) side.}  provides better leak
tightness than standard O-ring vacuum sealing.  The feedthrough boards
are grouped in sets of six or four, depending on position in the main
vacuum flanges, which are sealed by O-rings to the \velo vacuum
vessel.  An image of the vacuum feedthrough board is shown in
figure~\ref{fig:vfb} with an image of a flange populated with six
boards.

\subsubsection{Opto- and power board}
\label{sec:velo:opb}

The \Acr[f]{opb}~\cite{velo:opbnote} are the interface between the
detector modules and the off-detector electronics.  The \Acr{opb} has
14 DC/DC converters~\cite{Faccio2010,faccio:DCDC} that transform
the five input voltages to the ten supplies needed by the detector
module and the four voltages required by the \Acr{opb} itself. It
performs the electrical-to-optical conversion of the 20 high-speed
data links from each module and the bidirectional
electrical-to-optical conversion for the three control links, one for
each side of the detector module and one for the \Acr{opb} itself.
The \Acr{opb} hosts a \Gbtx
\Asic~\cite{Moreira:GBTx1,Moreira:GBTx2} that decodes the
high-speed control links and interfaces to two \Gbtsca
\Asic{s}~\cite{Caratelli2015}. These latter two devices provide ADC
channels that monitor the supplied and regulated voltages as well as
the received optical power.  They interface with \I2c buses that
control the GBLD laser drivers~\cite{Mazza:GBLD1,Mazza:GBLD2} on
the optical transceivers, and general-purpose I/O signals used for
local control.  The connections to the temperature sensors (NTC
thermistors) on the detector module and \Acr{opb} are available on the
front panel of the \Acr{opb}.

The \Acr{opb} is manufactured with eight metal layers and has an
overall size of $40\times15\cma$, plus a 10\mm protrusion which
connects with the \Pcie connector on the vacuum feedthrough board.
The layers that surround the high-speed electrical signals are made of
low-loss dielectric\footnote{\Trmk{Isola} I-Tera MT40.}  and the
remaining dielectric layers are made of standard FR4 glass fibre
laminate.  The \Acr[p]{opb} are vertically mounted on the exterior
wall of the \Velo vacuum vessel in a dedicated mechanical frame
integrated with the vacuum feedthrough flange. The frame acts as a
partial heat sink for the \Acr[p]{opb}.  Forced vertical air flow
through the crates provides additional cooling.

\subsubsection{Temperature and voltage monitoring}
\label{sec:velo:tempmoni}

Across each \Velo half there are 730 temperature sensors split between
monitoring and safety readout systems.  Within each cooling loop,
which contains two modules and one shared isolation valve, 50
temperatures are monitored (54 for the seventh module pair), which are
broken down into the following: 10 \pthundred probes attached to the
cooling pipes; 24 NTC sensors from the hybrids; 8 band gap
measurements implemented in the \velopix, one per tile; and 8 more
measurements, one per \Gbtx and four on the \Acr{opb} boards.  There
are a further 22 \pthundred sensors distributed over the RF box,
module support base, and in the isolation vacuum.  In addition, for
each of the 13 module pairs there are 10 low voltage readings and 4
independent readings of the high voltage, which gives 182 voltages to
be monitored.

\subsection{Mechanical design}
\label{sec:velo:mechanics}

The \Velo mechanical design retains the concept from the original
vertex detector of two movable halves, retracted from the beam line at
all times other than when stable beams are circulating.  Each \Velo
half moves independently in the horizontal direction from a $-29\mm$
retraction from the beam line to $+4\mm$ overclosure. The halves have
common vertical motion and may be moved $\pm4.7\mm$ above or below the
beam line.  The \Velo detector halves, their support structures and
the RF boxes are replaced to accommodate the requirements of the new
pixel detector.

\begin{figure}[t]
  \centering
  \includegraphics[width=0.99\textwidth]{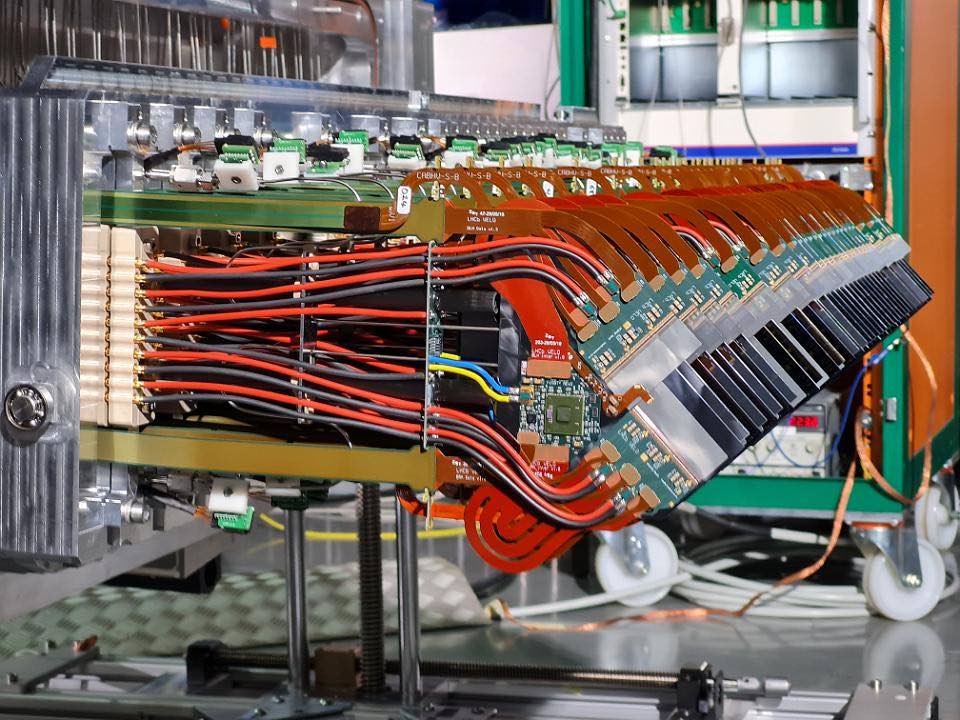}
  \caption{\label{halfassembly}The \cside half, with 26 modules, ready
    for the installation into the vacuum vessel.}
\end{figure}

\begin{figure}[t]
  \centering
  \includegraphics[width=0.96\textwidth]{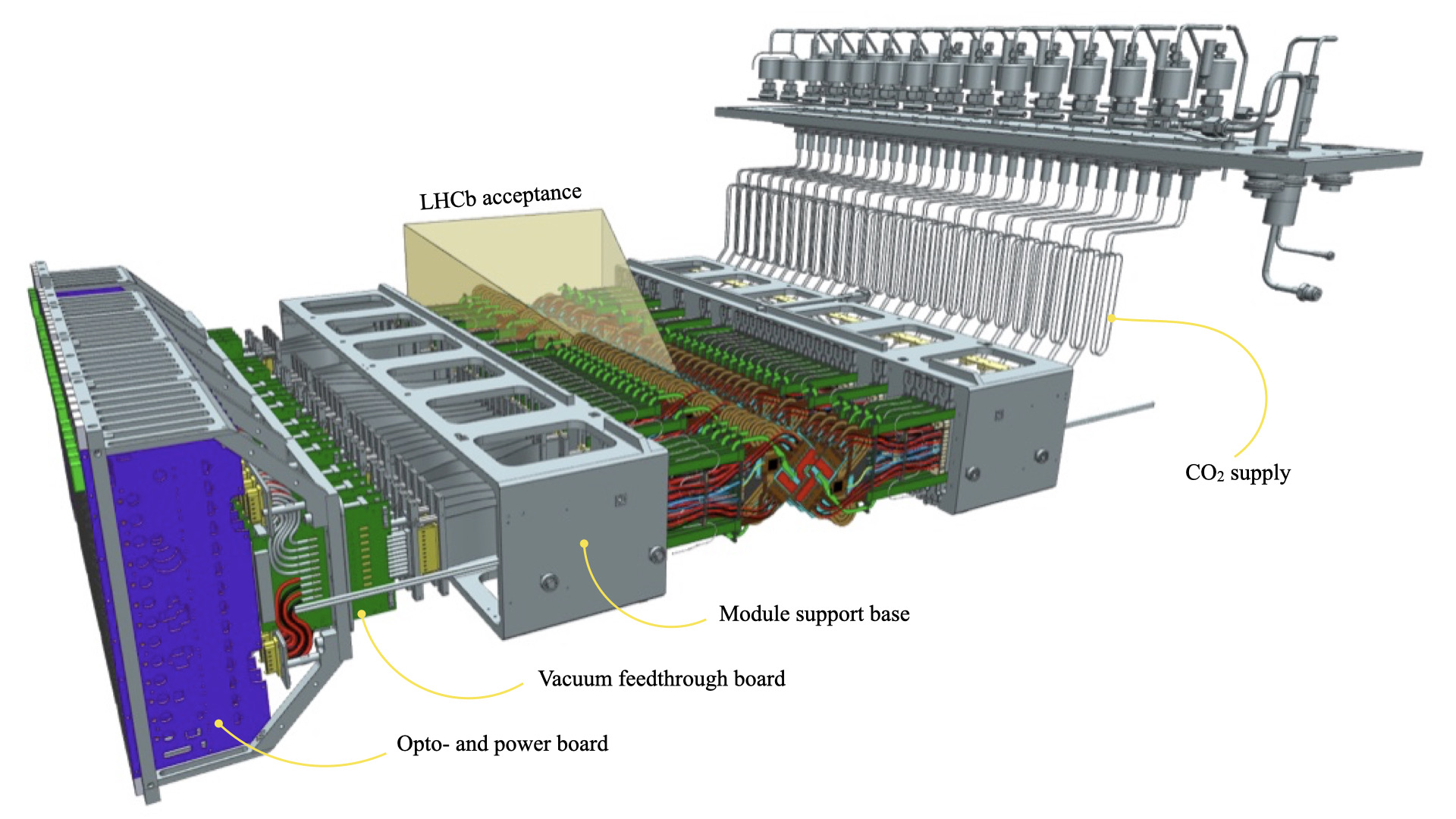}
  \caption{\label{CAD_whole_internal}Illustration of the \Velo halves
    showing modules on the module support bases and the \lhcb
    acceptance as a transparent pyramid. On the left, the flexible
    electronic cables are shown leading to the vacuum feedthrough
    boards and \Acr[s]{opb} boards in their custom frame. On the
    right, the flexible construction of long cooling loops is shown as
    well as the interface between the secondary and isolation vacua,
    in which sits an array of valves.}
\end{figure}

The central structure, onto which 26 modules are mounted, is the
aluminium module support base, visible in figure~\ref{halfassembly}.
Whereas the modules are designed to have minimal radiation length, the
bases are built for precision and rigidity. Any distortion of the
bases moves the module tip towards the RF box with a lever-arm equal
to the module height.  To avoid thermal distortions, they are
maintained to \SI{20}{\degc} (the manufacturing temperature) by
several adhesive heating pads.  Once installed, the bases are bolted
to the detector support which is, in turn, fixed to large, rectangular
bellows that provide a flexible barrier between the primary and
secondary vacua. All electronic and cooling services run from the
movable bases to the fixed \emph{detector hood}, which is the large
flange that seals the detector volume on the external wall of the
vacuum vessel. The $\sim 3\cm$ travel of the halves is absorbed by
flexible power and data cables running between the module foot and the
vacuum feedthrough.  These details are shown in
figure~\ref{CAD_whole_internal}.  For the \cotwo supply, an elongated
cooling loop, incorporated into every pipe running to/from each
module, absorbs the movement.  The cooling lines are connected to a
series of valves located in the tertiary vacuum, the isolation volume.

\subsubsection{RF boxes}

The RF boxes are the thin-walled corrugated enclosures that provide
the barrier between the primary (beam) vacuum and the secondary
(detector) vacuum and interface the \Velo detector halves to the \Lhc
beams.  They are made from aluminium, a light and electrically
conductive material.  Their complex shape accommodates overlaps
between sensors of opposing halves, while maintaining electromagnetic
effects to an acceptable level.  In order to compensate for the
reduced spatial resolution of the pixel detector, compared to that of
the innermost microstrips of the previous \Velo sensors, the distance
of approach to the beams was reduced from $8.2$ to $5.1\mm$.  The beam
aperture, as defined by the inner surface of the RF boxes, reduces
from $5.5$ to $3.5\mm$.  Special blocks of AlMg4.5Mn0.7 alloy were
forged to obtain a homogeneous material with small grain size and
without cavities.  The initial blocks had dimensions
\mbox{$1200\times300\times300\mm^3$} and were milled to the desired
shape with a 5-axis milling machine.

The RF box fabrication procedure included several steps, such as
verification of the block quality, rough milling of the outside and
inside shape, stress-relieving annealing, final milling to the nominal
$0.25\mm$ thickness with use of special moulds, supported by
wax-filling techniques, and interspersed thickness measurements to
achieve the desired thickness and geometry with the required precision
in a reproducible manner.

A vacuum test of each RF box was performed by closing the volume with
a flat flange and filling with helium at 1--5\mbar pressure inside a
large vacuum vessel. The leak rate was measured with a mass
spectrometer for different pressures to know the background level; no
leaks were detected.  The metrology was done with the RF box mounted
on its side.  RF box deformation studies at $\pm 10\mbar$ under- and
over-pressure showed a maximal wall displacement of $0.4\mm$ in the
central region (slot~22).

\begin{figure}[t]
  \centering
  \includegraphics[width=0.99\textwidth]{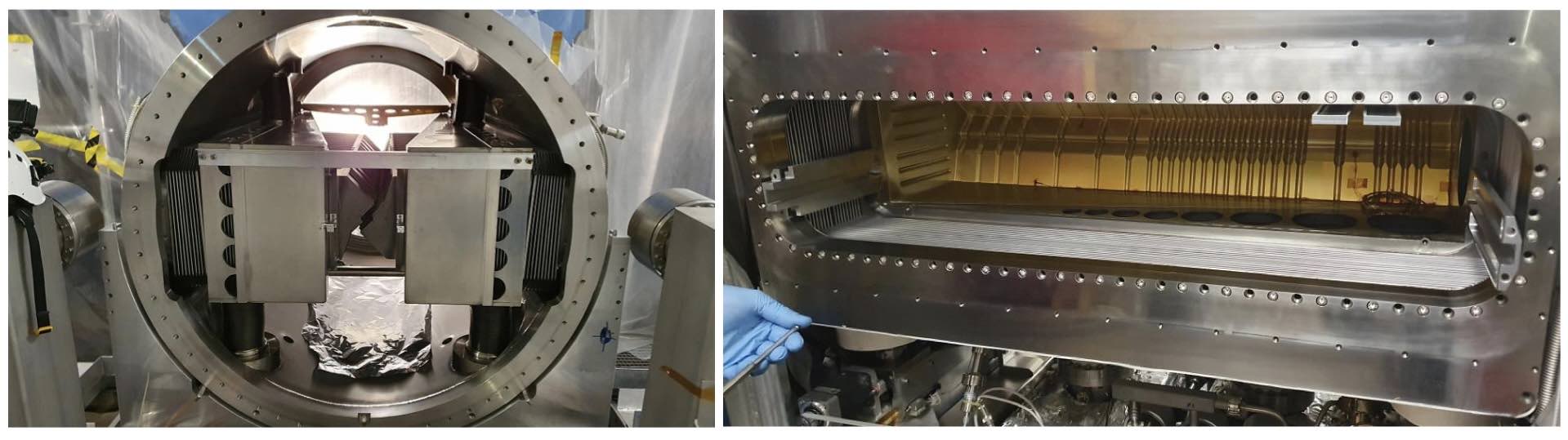}
  \caption{\label{RFboxInstallation}Left: the open \Velo vessel (seen
    from upstream) during the installation of the RF boxes.  Right:
    view inside the \aside RF box showing the module slot structure.}
\end{figure}

In total, two pairs of RF boxes (one for installation and one spare)
were fabricated.  The RF boxes to be installed were further thinned
down by chemical etching along the central spine within $\pm25\mm$ of
the beam line.  The etching proceeded in 20 minute steps of $0.02\mm$
using a painted mask on the aluminium to stop etching before the wall
could become locally too thin. The final result is a typical thickness
reduction from $0.30 \pm 0.07\mm$ to $0.25 \pm 0.10\mm$, with a
minimum intended thickness of $150\mum$.

To protect against electrical breakdown between the RF box and \Velo
sensors, the inside of the RF boxes is sprayed with
polyamide\footnote{\Trmk{Torlon} from Solvay Specialty Polymers.}
prepared in an aqueous solution, to about $10\mum$ thickness.  This
procedure included the attachment of \pthundred temperature probes on
the RF boxes.  Finally, the RF boxes were again cleaned before
applying a \Acr{neg} coating to the beam-facing surfaces.
Figure~\ref{RFboxInstallation} (left) shows the two RF boxes installed
in the \Velo vacuum vessel and the module side of one RF box with its
characteristic corrugated shape to accommodate modules.

\subsubsection{Wakefield suppressors}
\label{sec:Wakefield suppressors}

Wakefield suppressors at both ends of the \Velo vacuum vessel are used
to smooth the transition from the wide (50--56\mm) cylindrical beam
pipe to the narrower apertures of the storage cell and RF boxes.  The
wakefield suppressors must be flexible enough to accommodate the
motion of the \Velo halves, and provide a continuous electrical path
for the mirror currents of the \Lhc beams.  They are made from a
$50\mum$ thick, corrugated copper-beryllium sheet with a deposited
gold finish to reduce the secondary electron yield.  The upstream
connection attaches to the storage cell and is further described in
section~\ref{sec:smog}.  On the downstream end of the vacuum vessel,
the suppressor is clamped to the exit window on a $0.2\mm$ thick
gold-plated stainless-steel tension ring, as done previously.  The
attachment to the RF boxes is made by four keyholes in the wakefield
suppressor and matching mushroom features formed on the ends of the RF
boxes during their manufacturing.

\subsubsection{Vacuum safety system}

The risk of a microchannel rupture causing a sudden rise of pressure
in the secondary vacuum and damage to the RF boxes is mitigated by
continuous supervision of the pressure inside that volume.
Measurements of pressure over the range of 1~bar to $10^{-9}\mbar$ are
made using three different types of pressure sensors.  If a sudden
pressure rise is detected, the pneumatic shut-off valve system is
activated and coolant flow is diverted from all modules to bypasses.
This action immediately reduces the amount of the coolant injected in
each cooling loop and in the case of a microchannel rupture, would
minimise the amount of \cotwo released.  Each cooling loop serves two
microchannel coolers and contains two shut off valves (inlet and
outlet) and an additional, small, safety volume.  It is thermally
connected with the detector structure to ensure a temperature higher
than the coolant.  Consequently, in case of activation of the safety
system, coolant is trapped in between two shut off valves, will expand
and reduce in temperature following the typical \cotwo saturation
curve.  The 26 pneumatic bypass valves, two for each two-module
cooling loop, are installed in the isolation volume, an entirely new
feature of the \Velo mechanics.
\subsection[ECS and DAQ]{\Acr[s]{ecs} and \Acr[s]{daq}}
\label{sec:velo:daq}

The final component in the \Velo read-out chain are the \Pciefourty
cards, hosted in PCs in the data centre.  These cards are
hardware-wise identical for all subdetectors, including \velo, while
the firmware is partially subsystem-specific.  Detector data, control
and monitoring is performed within the \lhcb common and centrally
managed framework, based on several platforms, see
sections~\ref{sec:online},~\ref{sec:trigger} and~\ref{sec:software}.
Within these, a number of \Velo specific tools are developed and used.

\subsubsection{Firmware}

Cards with the \Solfourty firmware are used to control and monitor
thirteen \Velo detector modules, over a total of 39 bidirectional
optical fibre links.  This firmware, which is largely shared with the
other \lhcb subdetectors, includes the trigger and fast control
commands and the protocols for communication with and via the
\Gbtx~\cite{Moreira:GBTx1,Moreira:GBTx2} and
\Gbtsca~\cite{Caratelli2015} chips.  For the complete \Velo detector,
a total of four \Solfourty-flavoured \Pciefourty cards and 52
\Tellfourty-flavoured cards are used.

Cards with the \Tellfourty firmware are used for processing the twenty
high-speed \Gwt data serial links of a single module.  This firmware
is unique to the \Velo detector because of the different clock
frequency with which the \velopix serialiser protocol (\Gwt) works
compared to the \Gbt protocol used by all other subdetectors for their
data taking.  The firmware, described in detail in
ref.~\cite{Hennessy:2789034}, processes the data in the following
way.  First it performs the descrambling of the \Gwt serialiser output
frame, splitting it back into the \Acr{spp} format that was encoded on
the \velopix.  Time ordering of all hits within the last 512 clock
cycles is done.  This corresponds to the 9 bit resolution of the
timestamp inside the \Acr{spp}.  Any \Acr{spp} arriving outside the
512 clock cycle time window is dropped.  The overall \Lhc clock timing
information is added to each hit.  The firmware also performs pattern
recognition of cluster of pixels to recognise particle hits and
evaluate their coordinates and the cluster topology, dropping the raw
pixel data~\cite{Bassi:2825279,bassi2023fpgabased}. This
accelerates the track reconstruction process in both \Hltone and
\Hlttwo.

\subsubsection{Detector control}

Communication with the hardware components is done via the \lhcb \Ecs
software framework using both fast and slow control protocols.  The
project manages the configurable elements and monitors low-level
parameters of the hardware, including the \Daq system, the
\Acr[p]{opb}, the \Gbtx chips and the \velopix \Asic{s}. It also
collects, logs and displays information on the \Velo environment
(temperatures, voltages, pressures, etc.).  The calibration is handled
using fast readout, using the external DIM writer library to catch the
data points from the detector.  The control application sends the
calibration data, in a raw format, to a dedicated database.  The raw
data are analysed in \cpp calibration software (\Vetra).  Since the
calibration analysis is CPU-consuming, it is moved to a separate
machine.  When the control application receives back the calibration
result, it configures the detector accordingly.

\subsubsection{Calibration data processing}
\label{sec:velo:software}

The \Vetra software package, written in \gaudi framework, is dedicated
to handling standard \lhcb data, and special \Velo-only data collected
for specific purposes (e.g.\ IV scans).  The core part of \Vetra is
written in the \cpp and it can be used as a quasi-online monitoring
application to rapidly analyse raw data produced by the \Velo.  The
overall processing pipeline comprises data decoders, processing
algorithms and monitoring algorithms.  The \Velo is monitored with
this software tool and the result is fed back to the detector control
system where necessary.  For example, occupancies, dead-pixel maps,
charge-collection efficiency studies, alignment and other
calibrations.  This software is also used for track and vertex
reconstruction, providing beam position information to the control
algorithm that ensures safe closure of the \Velo halves.  A
significant part of the data handling, processing and visualisation
will be done outside \Vetra.  For this purpose, a database system,
called \Storck, and visualisation framework, \Titania, have been
created. \Storck is optimised to manage the calibration data files and
their child objects (e.g., calibration parameters, monitoring
histograms), whilst \Titania provides a means for rapid data
exploration and trending.  These tools are vital to the operation of
the subdetector and quality control of the \Velo data.

%
\section{Internal gas target} 
\label{sec:smog}
Fixed-target physics with \Lhc beams was pioneered in the \lhcb
experiment during \runtwo thanks to the availability of an internal
gas target.  A gas injection system, called
\Acr{smog}~\cite{SMOG_thesis}, originally conceived and implemented
for precise colliding-beams luminosity calibration, was used to inject
light noble gas into the \Velo vacuum vessel.  This produced a
temporary local pressure bump peaking at around $10^{-7}\mbar$ over
the length of the vessel (about 1\m) and decaying down to the \Lhc
background level ($\sim 10^{-9}\mbar$) over the 20\m \lhcb beam pipe
sections on each side of the interaction point.  The resulting
beam-gas interactions were used for precise imaging of the beam
profiles~\cite{LHCb-PAPER-2014-047}.  \Acr{smog} also gave the unique
opportunity to operate the \lhcb experiment in fixed target mode.
Gaseous targets of different nuclear size (He, Ne and Ar) were used in
combination with proton and lead beams at (nucleon-nucleon equivalent)
centre-of-mass energies of up to 115\gev, with negligible effect on
\Lhc operation.  Encouraged by first results and future prospects, an
upgrade of \Acr{smog} (also called SMOG2) was proposed and
implemented~\cite{LHCb-TDR-020}.

\begin{figure}[h]
  \centering
  \includegraphics[width=0.6\textwidth]{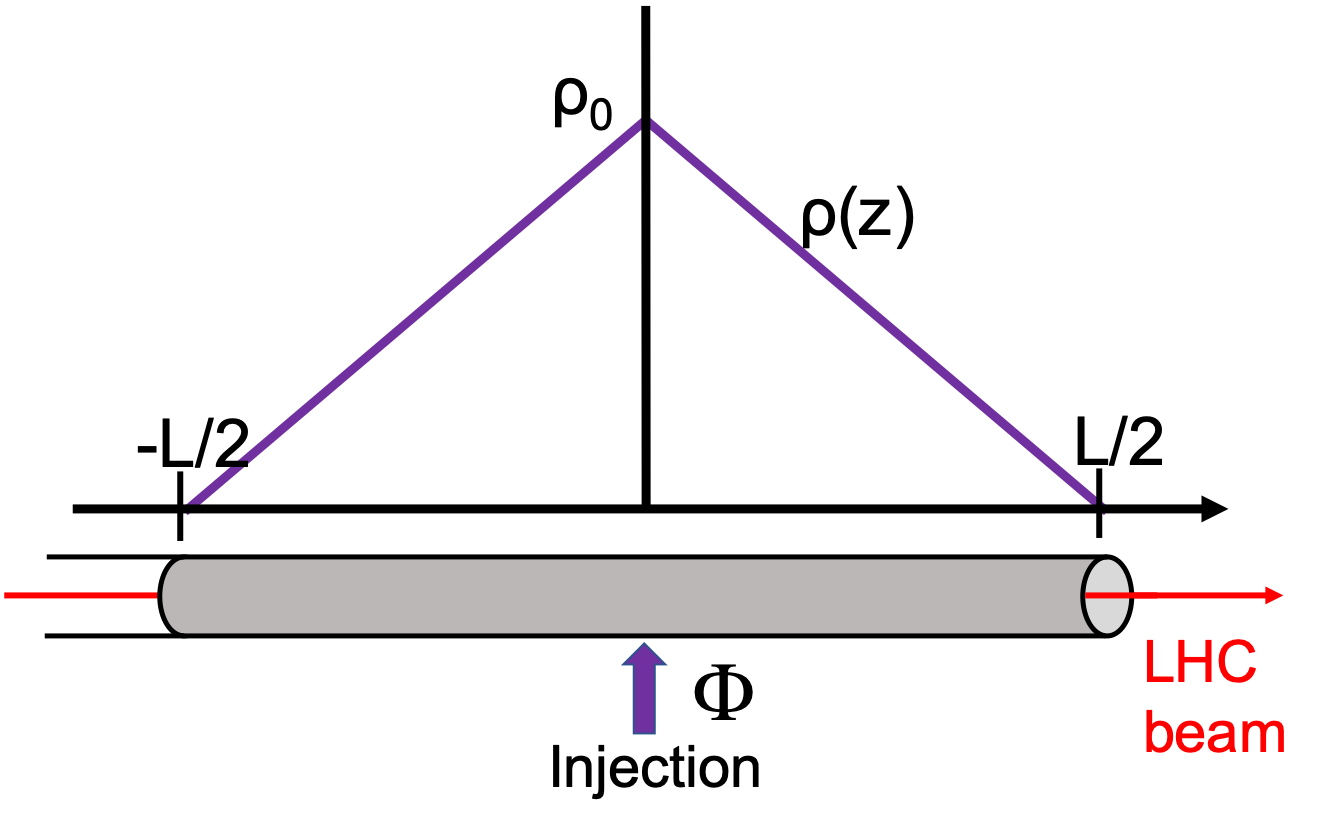}
  \caption{Sketch of a tubular storage cell of length $L$ and inner
    diameter $D$.  Gas is injected at the centre with flow rate
    $\Phi$, giving a triangular density distribution $\rho (z)$ with
    maximum $\rho_0$ at the centre. Reproduced with permission from~\cite{LHCb-TDR-020}.}
  \label{FigPrincStorageTube}
\end{figure}

The core idea of the \Acr{smog} upgrade is to inject the gas directly
into a so-called storage cell and benefit from the increased areal
density at an identical injected flux, as has been done in the past at
other accelerators~\cite{Ste03}.  The principle is sketched in
figure~\ref{FigPrincStorageTube}.  The open-ended cylindrical tube has
an inner diameter $D$ and a length $L$. Gas is injected via a
capillary at the storage cell centre at a flow rate $\Phi$ from a
\Acr{gfs}, resulting in an approximately triangular density
distribution $\rho(z)$ with maximum $\rho_0 = \Phi/\Ctot$ at the
centre ($z=0$). Here, $\Ctot$ is the total flow conductance of the
tube from the centre outwards and is given by the conductance of two
parallel tubular conductances of length $L/2$ in the molecular flow
regime~\cite{Ste03}, and thus amounts to (in $\litre \sec^{-1}$)
$\Ctot \approx 7.62 \sqrt{T/M}\,D^3\, (0.5\,L + 1.33\,D)^{-1}$, where
$L$ and $D$ are in \cm, the storage cell temperature $T$ in \degk, and
$M$ is the molecular mass number of the injected gas.  The areal
density seen by the beam is $\theta=\rho_0\, L/2$.

The \Velo upstream connection in the \Lhc vacuum between the \Velo
detector boxes and the beam pipe has been modified to accommodate the
integration of a storage cell composed of two cylindrical halves, each
attached to the upstream end of one of the \Velo detector halves and
moving together with them.  Thus, when the \Velo is brought into a
closed position, the two halves form an open-ended tube coaxial with
the \Lhc beam axis.  It is expected that the \Acr{smog} upgrade will
facilitate fixed-target runs with an effective gas areal density
higher by a factor of about 10 for He at the same flow rate, and even
higher gain factors for heavier gases.

The \Acr{smog} upgrade introduces other important improvements.
First, the determination of the target density (and beam-gas
luminosity) is significantly more precise because the target is
confined to the storage cell, whose conductance is well known and can
be combined with an accurate measurement of the injected gas flow rate
from the \Acr{gfs}. Second, it will be possible to select among
several gas species without intervention (including non-noble gases
such as H$_2$, D$_2$, O$_2$, etc.). Finally, the beam-gas interaction
region is much better defined and well separated from the beam-beam
collision region, which also opens the possibility to have concurrent
beam-gas and beam-beam collisions.\looseness=-1

The \Acr{smog} upgrade is composed principally of two systems: the
storage cell assembly, mounted inside the beam vacuum, and the
\Acr{gfs}, located on the ``balcony'', a platform near the detector
inside the experimental cavern.  Given its vicinity to the \Lhc beams,
the design of the storage cell assembly must fulfil several
requirements derived from aperture considerations, RF or impedance
related aspects, and dynamic vacuum phenomena, as already discussed
for the \Velo RF boxes (see section~\ref{sec:velo:mechanics}).\looseness=-1

\subsection{The storage cell system}

\begin{figure}[t]
  \centering
  \includegraphics[height=5.05cm]{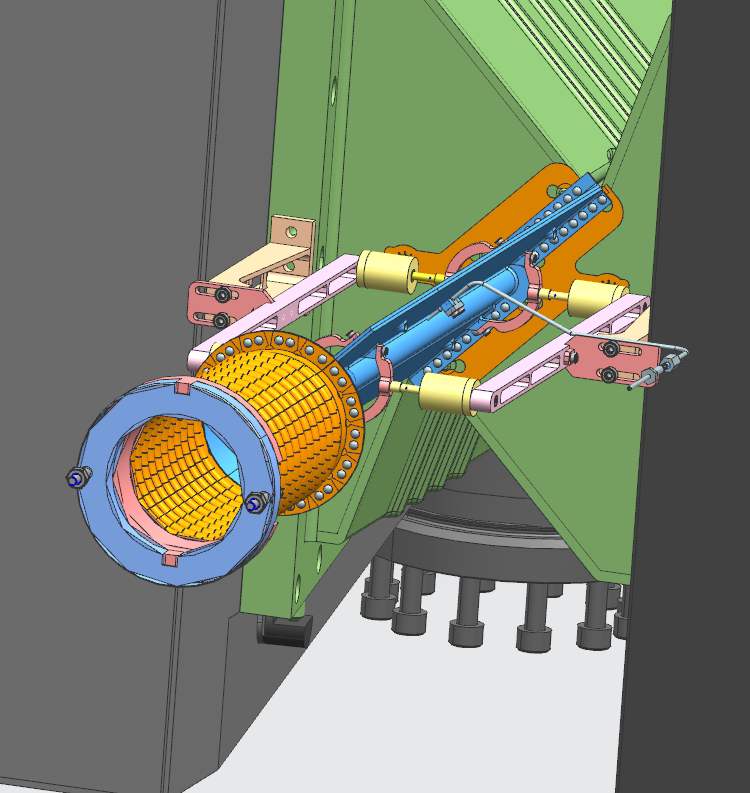}\hfill
  \includegraphics[height=5.05cm]{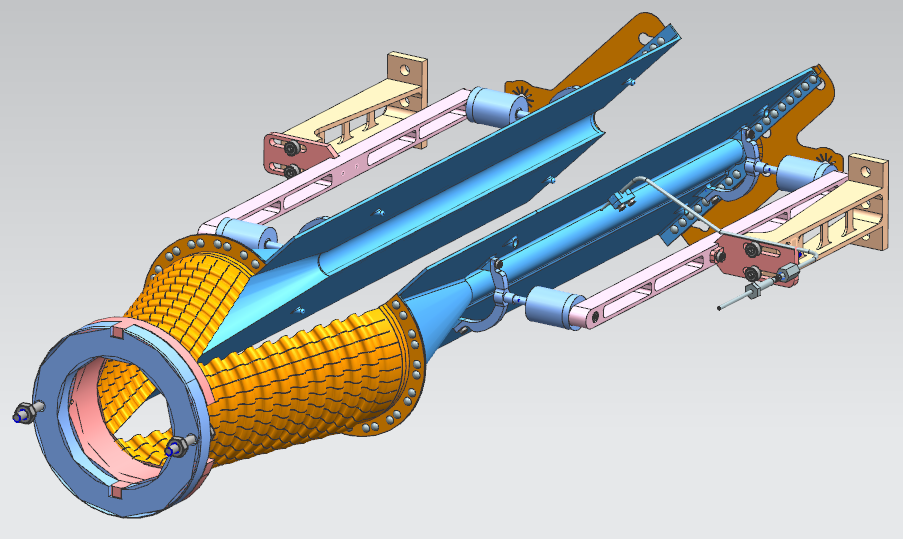}
  \caption{Left: view of the storage cell (blue) supported from the
    \Velo RF box flanges (in green) in the closed \Velo position. Two
    flexible wakefield suppressors (orange) provide the electrical
    continuity.  Right:~storage cell in the open position (without
    showing \Velo elements). Reproduced with permission from~\cite{LHCb-TDR-020}}
  \label{cell_overall2}
\end{figure}

The storage cell and its arrangement inside the \Velo vessel is
visible in figure~\ref{cell_overall2}.  The assembly fits into the
limited space available inside the existing \Velo vessel, upstream of
the \Velo detector.  In order to leave sufficient beam aperture for
beam operations (injection, energy ramp, squeeze, etc) the assembly is
split in two opposing halves each attached to its respective \Velo RF
box.  The assembly is composed of a flexible wakefield suppressor
split in two halves, two opposing storage cell shapes containing a
half cone, a half tube and side wings, a short wakefield suppressor
which connects to the \Velo detector box, and two arms to support the
storage cell halves from the \Velo RF box flanges.  The conical shape
allows for a smooth transition from the 56\mm diameter of the upstream
beam pipe to the 10\mm diameter of the storage cell tube.  The tubular
part is 20\cm long.  Gas is injected from the \Acr{gfs} into the tube
centre via a flexible line ended with a $0.8\mm$ inner diameter
stainless steel capillary pressed into a hole in the \cside half of
the storage cell.  All parts are sufficiently light to keep the
beam-induced background due to the material in the proximity of the
beams at a negligible level.

The cell opens and closes together with the \Velo detector boxes to
which it is mounted by two cantilevers rigidly attached to the flange
of the \Velo RF boxes.  Because the \Velo design foresees the
possibility to operate the detector at a slightly retracted position
(by $\sim 0.1\mm$) relative to the nominal closed position, the \cside
half of the storage cell is rigidly fixed to its cantilever, while the
\aside half is coupled via a spring system that allows to always reach
the final closed position, guided by the rigid half.  This spring
system allows to reach the storage cell nominal closure (thus,
sufficient gas tightness along the longitudinal slit) even if the
\Velo halves are not completely closed, within a range of up to 1\mm.
The minimum allowed aperture over the length of the storage cell is
imposed by the van der Meer scan configuration and amounts to
3\mm~\cite{VELOapertureNote2018}, well below the chosen radius of the
storage cell.

The surfaces surrounding the beam are made of electrically conductive
materials, in order to shield the chamber from the beam RF fields,
prevent excitation of RF modes and provide electrical continuity for
any position of the \Velo halves.  The cell structure is made from a
99.5\% pure Al block, milled with an accuracy of $\sim20\mum$.  The
cone, tube and wings are milled to a final thickness of 1.2, 0.2,
$1.2\mm$ respectively.  Before completion, the cell was heat-treated
to 290\degc for one hour to allow for stress release. All screws are
silver-coated and perforated in compliance with standards for high
vacuum systems.  Particular attention was given to the transitions to
the RF boxes (downstream side) and to the beam pipe (upstream
side). They are made from $0.075\mm$ thick Cu-Be foil.  The upstream
wakefield suppressor contains slits and corrugated strips which
ensures adequate flexibility to the wakefield suppressor.  It is
attached to the upstream support by screws and to the storage cell by
Al rivets.  The downstream wakefield suppressor is made of short
curved fingers.  It is connected on one end to the storage cell, in
the same manner as the upstream wakefield suppressor.  The fingers
press on the RF box, while the wakefield suppressor is locked to the
RF box on the mushroom shapes mentioned in section~\ref{sec:Wakefield
  suppressors}.

The storage cell is coated with amorphous carbon to present to the
beams a surface with a \Acr{sey} around
$\sim\! 1.0$~\cite{PhysRevSTAB.14.071001}.  This precaution avoids the
formation of a beam-induced electron cloud and the possible onset of
beam instabilities.  The coating is applied by sputtering.  First, a
$50\nm$ thick Ti adhesion layer is applied, then a $1-10\nm$ thick
layer of amorphous carbon.  From simulation studies it has been
concluded that such \Acr{sey} is largely sufficient to prevent
electron cloud build-up, even when taking into account that a higher
residual gas pressure can favour such phenomenon.

The storage cell is equipped with five 0.34\mm outer diameter K-type
thermocouples (with a precision of about $0.1\degk$) insulated with a
nickel-based super alloy\footnote{Special Metals Corporation
  \Trmk{Inconel}.} and terminated with a ceramic connector for use in
ultrahigh vacuum.  The temperature measurements are needed both for
determining the areal density $\theta$ ($\sqrt{T}$ dependence) and for
monitoring a possible temperature increase by beam-induced effects.

\begin{figure}[t]
  \centering
  \includegraphics[width=12.5cm]{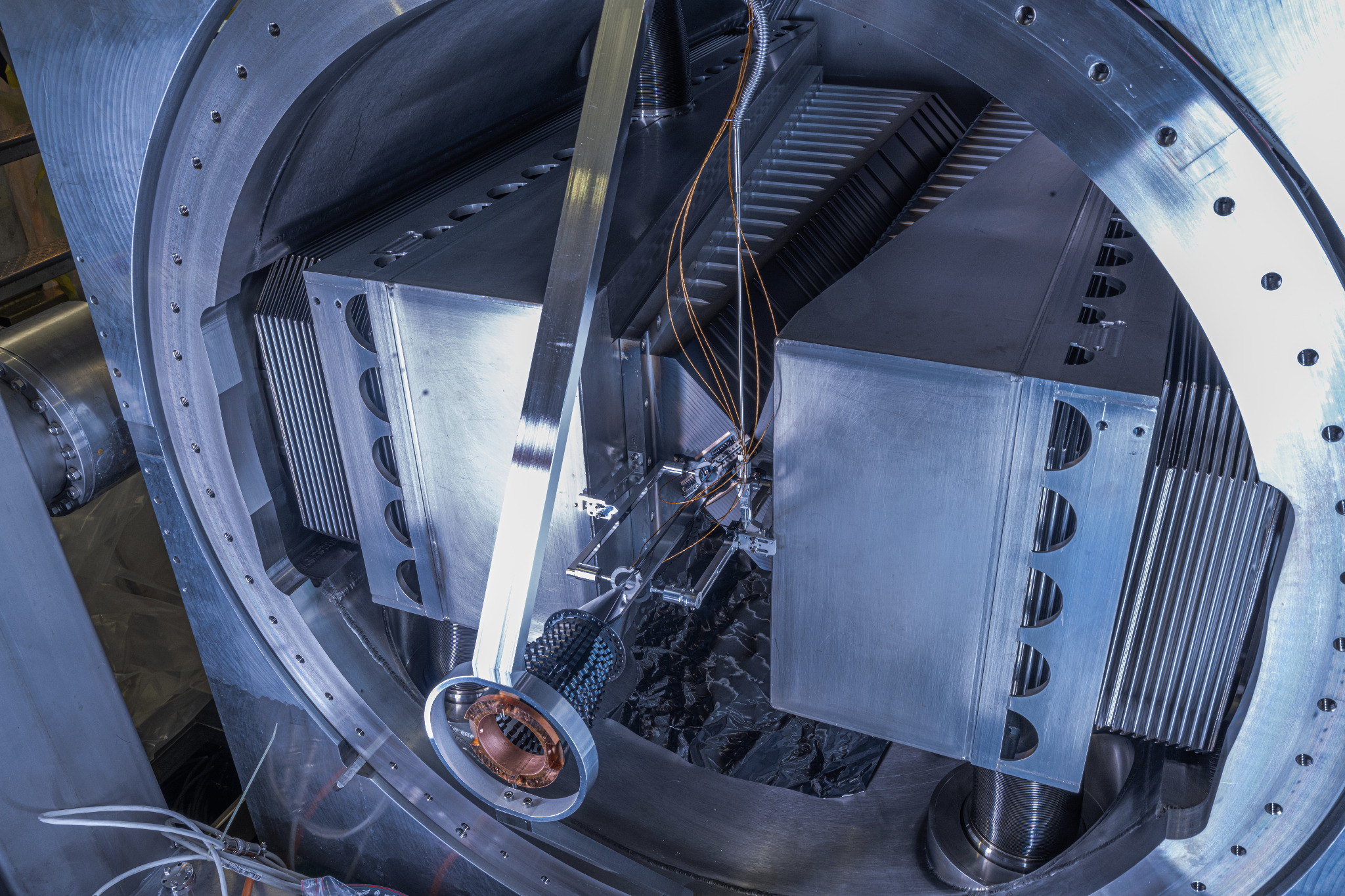}
  \caption{\label{photo}Picture of the \Acr[s]{smog} storage cell
    system installed into the \Velo vessel.}
\end{figure}

The electromagnetic compatibility of the \Acr[s]{smog}-\Velo assembly
was validated using frequency and time domain electromagnetic field
simulations that were benchmarked with RF measurements on a 1:1 scale
mockup with the wire
method~\cite{Popovic:2701362,Popovic:2666879}.  The wakefield
suppressor robustness has been demonstrated on a prototype with a
fatigue test (15\,000 open/close cycles), equivalent to about 15 years
of nominal operation in the experiment.  No sign of fatigue has been
observed.  The installation of the storage cell into the \Velo vessel
has been successfully completed in the summer of 2020, see
figure~\ref{photo}.  A detailed alignment survey of the storage cell
found no misalignment in excess of 0.25\mm.\looseness=-1

\subsection{Gas feed system}

\begin{figure}[t]
  \centering
  \includegraphics[width=12.5cm]{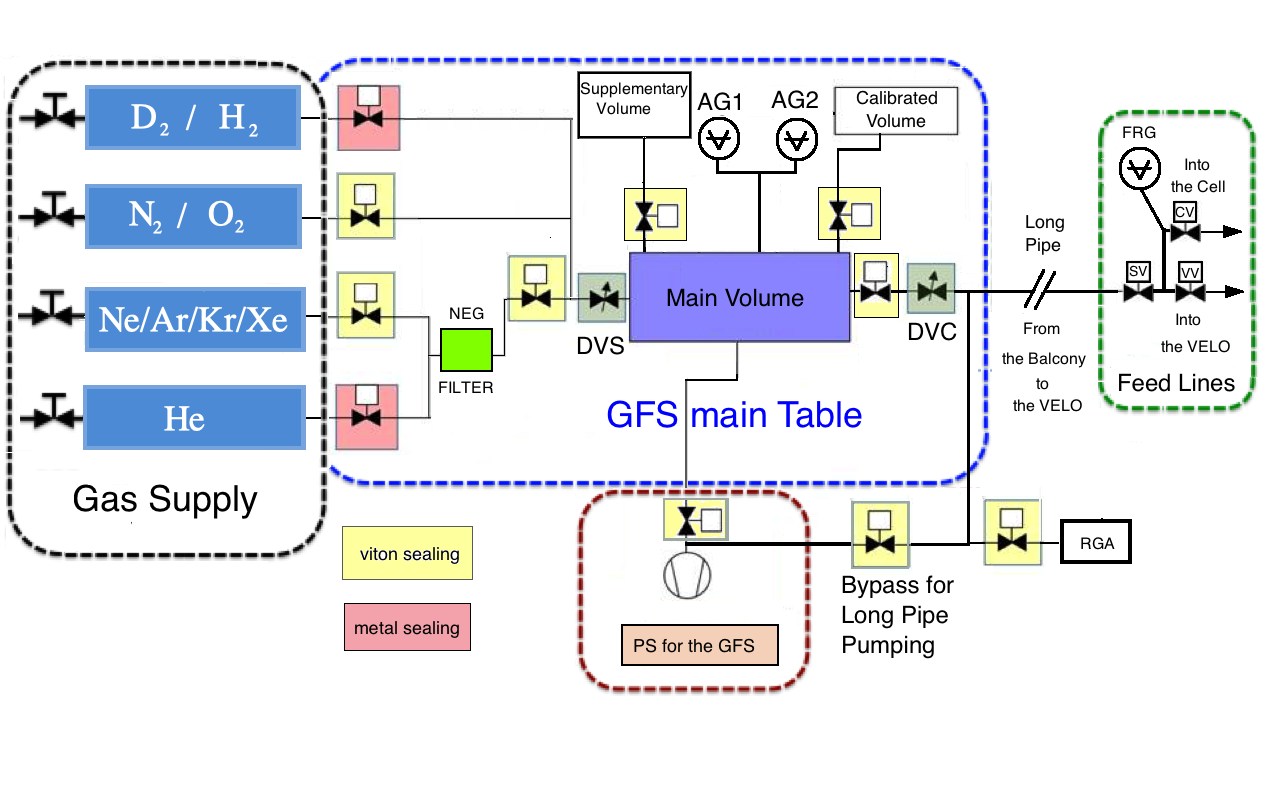}
  \caption{The four assembly groups of the \Acr[s]{smog} \Acr[s]{gfs}
    are the gas supply (with 4 reservoirs), the main table, the
    pumping station (PS) and the feed lines to the \Velo vacuum
    vessel.  The various components are described in the text. Reproduced with permission from~\cite{LHCb-TDR-020}.}
  \label{fig: GFS}
\end{figure}

The \Acr{gfs} allows one to choose the gas type to be injected among
those available in the four installed reservoirs.  The amount of
injected gas can be accurately set and measured in order to precisely
compute the target densities from the storage cell geometry and
temperature.  The \Acr{gfs} consists of four assembly groups, as shown
in figure~\ref{fig: GFS}, and is based on precise absolute
thermo-stabilised gauges that cover four decades of pressure reading:
absolute gauge 1 (AG1), which covers a pressure range from 1100 to
0.1\mbar, and absolute gauge 2 (AG2), which covers the range from 1.1
to $10^{-4}\mbar$.
The absolute gauges measure and monitor the pressure in the main
volume, and are therefore used in determining the stability of the
injected flow. The flow is obtained by setting a nominal pressure and
keeping it constant by means of a thermo-regulated valve (DVS).  After
stabilising the injection pressure, another thermo-regulated valve
(DVC) is set to the appropriate value depending on the gas type and
the gas flow rate chosen (typical values will be in the range
0.5--$8\times 10^{-5}\mbar\litre/\sec$).  The connection of the
\Acr{gfs} to the \Velo vessel is performed by a 10\mm inner diameter
bakeable stainless steel pipe with a length of 15\m terminated by two
feed lines (with valves CV and VV), one feeding directly into the
\Velo vacuum vessel and one into the storage cell centre.  A full
range gauge (FRG) monitors the pressure just upstream of these two
valves.  A rest gas analyser (RGA) is employed to analyse the
composition of the injected gas in the main volume.

Preliminary studies using the \textsc{Molflow} simulation
code~\cite{Kersevan:IPAC2019-TUPMP037} show that a total systematic
uncertainty of around 2--3\% in the determination of the target gas
areal density is achievable (slightly worse for light gases compared
to the heavier ones).

%
\section{Upstream tracker}
\label{sec:ut}
\subsection{Overview}
\label{sec:ut:overview}

The \Acr{ut} is located between the RICH1 detector and the dipole
magnet.  It is used for charged-particle tracking and is an integral
component of the first processing algorithm in the software
trigger~\cite{LHCb-TDR-016}: the \Acr{ut} hits are combined with the
\Velo tracks and, exploiting the magnetic field between the
interaction region and the \Acr{ut}, a first determination of the
track momentum $p$ with moderate precision ($\sim 15\%$) is obtained.
A momentum and charge estimate is performed only for tracks with \pt
$>0.2\gevc$ and this information is used to speed up the matching with
the \Acr{sft} hits. These two features provide significant improvement
of the speed of this matching algorithm.  Moreover, \Acr{ut} hit
information reduces the rate of fake tracks created by mismatched
\Velo and \Acr{sft} segments.  Lastly, it provides measurements for
particles decaying after the \Velo, e.g.\ long lived $K_S^0$ and
$\PLambda$ particles.

\subsubsection{Detector requirements}

The physics goals and environmental conditions dictate the following
requirements:
\begin{itemize}
\item Acceptance: in order to fulfil its function in the trigger
  algorithm and to be effective in suppressing fake tracks, the
  coverage of the \Acr{ut} detector in the nominal \lhcb acceptance
  should not have any gaps.
\item Single-hit efficiency: the detector should have a high enough
  hit efficiency to ensure that more than 99\% of the charged
  particles traversing the detector within the acceptance leave hits
  in at least three planes.
\item Hit purity: spurious hits due to noise or signal shape must be
  minimised. This is further specified in the \Fend \Asic section.
\item Radiation damage: the \Acr{ut} detector needs to maintain its
  performance up to an integrated luminosity of at least 50 \invfb,
  taking into account that the radiation dose has a strong radial
  dependence.  The sensitive elements near the beam pipe need to
  withstand fluences up to $4\times 10^{14}\neqcmcm$, while the
  maximum fluence is less than $2\times10^{13} \, \neqcmcm$ for most
  of the detector area, with the outermost sensors receiving less than
  $10^{12} \, \neqcmcm$.  In addition, the near detector electronics
  need to withstand a radiation level of the order of $\SI{1}{\kGy}$.
\item Occupancy: the charged particle density follows a similar radial
  trend as the radiation fluence.  The detector segmentation must be
  finer near the beam pipe in order to keep the occupancy below a few
  percent.
\item Material budget: the detector and its enclosure must be designed
  with the goal of obtaining a significant overall reduction of
  material in the forward region of the acceptance when compared to
  the \lhcb detector of \runone and \runtwo.
\end{itemize}

A silicon microstrip detector technology was chosen to fulfil these
requirements.

\subsubsection{Geometry overview}

\begin{figure}[h]
  \centering
  \includegraphics[width=0.7\textwidth]{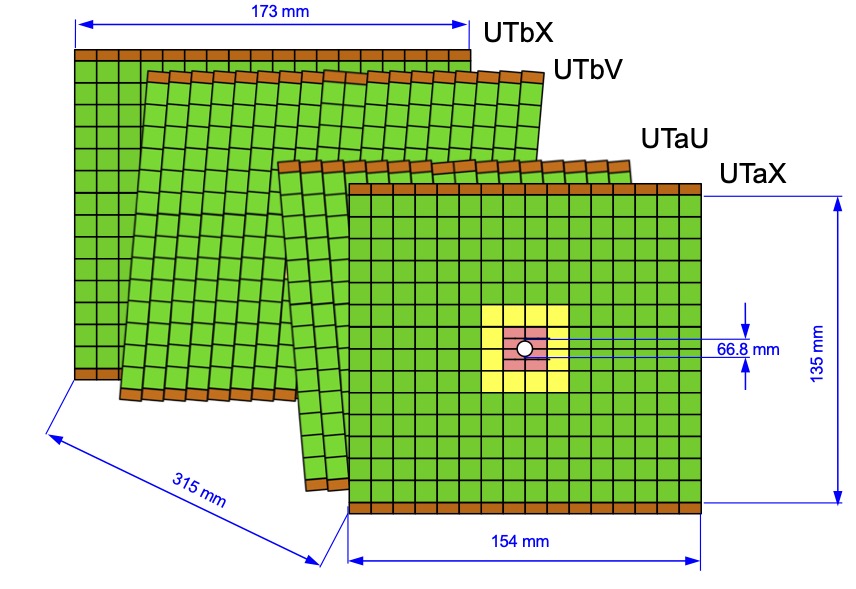}
  \caption{\looseness=-1\label{fig:ut:UT_planes}Drawing of the four \Acr{ut}
    silicon planes with indicative dimensions.  Different colours
    designate different types of sensors: Type-A (green), Type-B
    (yellow), Type-C and Type-D (pink), as described in the
    text. Reproduced from~\cite{LHCb-TDR-015}. CC BY 3.0.}
\end{figure}

The \Acr{ut} detector comprises four planes of silicon detectors
organised in two stations, as shown in figure~\ref{fig:ut:UT_planes}.
The circular hole in the middle provides clearance for the beam pipe.
The silicon strip pitches and lengths are matched to the expected
occupancy.  The silicon sensors, shown as coloured boxes in the image,
are described in section~\ref{sec:ut:sensors}.  They are arranged in
vertical units, called \emph{staves}, described in
section~\ref{sec:ut:Staves}.  The first station (labeled `a') is
composed of an $x$-measuring layer (UTaX) with vertical strips and a
stereo layer (UTaU) with strips inclined by $5\degrees$. Both layers
are made of 16 staves each.  The second station (`b') is similar, with
first a stereo layer (UTbV) with opposite inclination, and a layer
with vertical strips (UTbX).  Both layers contain 18 staves each.  The
two pairs of stations are symmetrically arranged around $z = 2485\mm$.
There is a gap of 205\mm between the nominal $z$ positions of UTaU and
UTbV, and of 55\mm between the UTaX and UTaU or UTbV and UTbX.\looseness=-1

To ensure full coverage in the vertical direction, the sensors are
arranged on both sides of the staves such as to obtain a vertical
overlap.  Similarly, a $z$-staggered arrangement of the staves with
horizontal overlaps facilitates a full horizontal coverage.  Moreover,
special sensors are utilised in the innermost area to maximise the
active area near the beam pipe.  Compared to its predecessor, the
Tracker Turicensis (TT) detector~\cite{LHCb-TDR-009} of \lhcb during
\runone and \runtwo, these improvements considerably reduce the gaps
in the acceptance.

In order to minimise the amount of material seen by particles, all the
components in the active volume of the detector have been thinned as
much as practical.

\subsection{Mechanics and near detector infrastructure}

\begin{figure}[h]
  \centering
  \includegraphics[width=\textwidth]{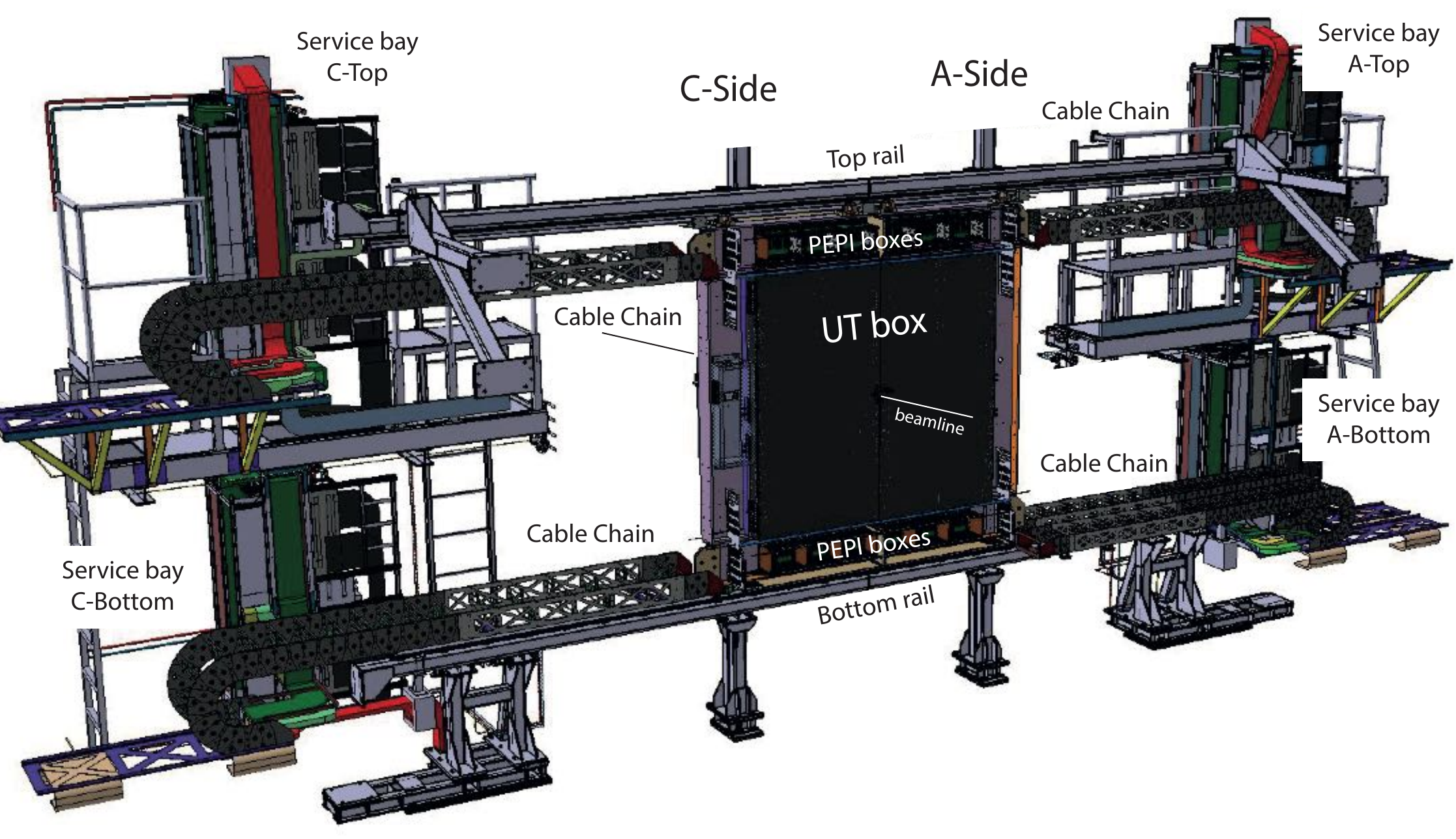}
  \caption{\label{fig:ut:UT-overview}A 3D view of the \Acr{ut}
    system.}
\end{figure}

An overview of the detector and its associated electronics and
mechanical services is shown in figure~\ref{fig:ut:UT-overview}.  The
staves are enclosed in a thermally insulating, light and air-tight box
that fits around the beam pipe.  The \Acr{ut} beam pipe section is
wrapped in a lightweight thermally insulating blanket.  A cooling
manifold distributes the \cotwo coolant to the staves.  Dry gas is
flushed through the box to prevent condensation on the stave
components.  These items are further described in
section~\ref{sec:ut:cooling}.  An off-detector electronics system,
described in section~\ref{sec:ut:readoutchain}, is mounted close to
the staves, just outside the detector box.  These electronics convert
the signals from the \Fend chips into optical signals and drive them
along optical cables.  The box is split into two halves and mounted on
rails, such that the detector can be opened for maintenance or during
beam pipe bake-out.  Four cable chains allow horizontal movement of
the two box halves without putting strain on electronic cables,
optical fibres and cooling fluid pipes.  Four service bays, on the
fixed racks, located on the \aside and \cside of the \Acr{ut}
detector, host the low voltage boards that power the near detector
electronics and the hybrids, and patch panels that distribute the high
voltage to the silicon detectors.

\subsection{Staves and modules}
\label{sec:ut:Staves}

\begin{figure}[h]
  \centering
  \includegraphics[width=.65\textwidth]{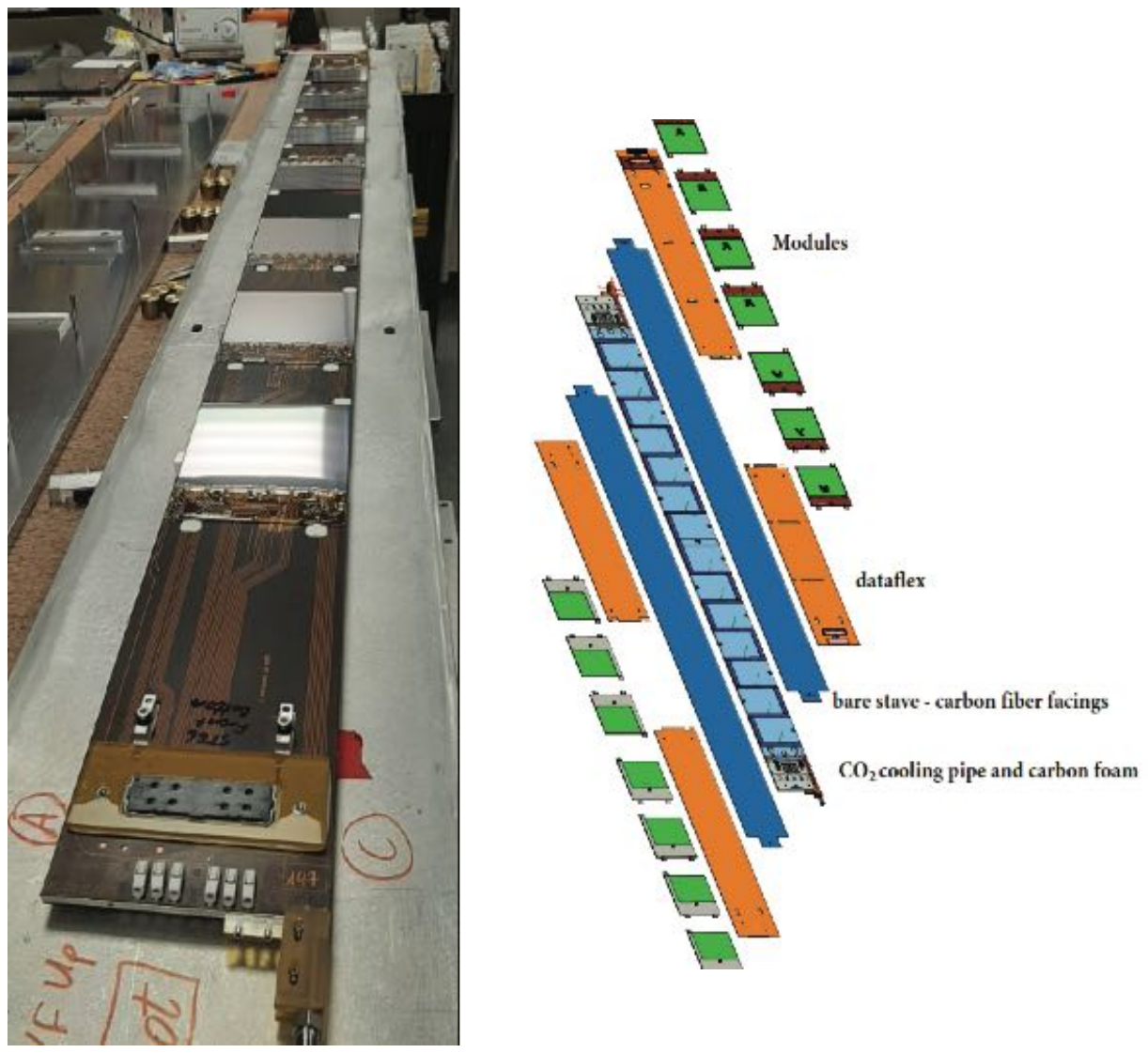}
  \caption{\label{fig:ut:full-stave2}Left: a completed stave.  At the
    bottom, the end of the cooling tube is visible with the high
    voltage and signal connections. In orange are the \dataflex cables
    and in brown the hybrids.  The reflective areas are the silicon
    detectors. Reproduced with permission from~\cite{Mountain:2807067}.  Right: an exploded view of an instrumented stave
    showing the individual components described in the text.}
\end{figure}

A \Acr{ut} stave provides the mechanical support for the sensors and
\Fend electronics, as well as active cooling.  An instrumented stave
is shown in figure~\ref{fig:ut:full-stave2} (left) and an exploded
view shows the individual components (right).  The bare staves are
designed to minimise the radiation length in the acceptance region of
the detector.  Each stave is approximately 10\cm wide, 1.4\m long and
is composed of a carbon fibre sandwich, about 3.9\mm thick, that
contains a 2.275\mm diameter titanium tube as part of the evaporative
\cotwo cooling.  The tube is embedded in a low density high thermal
conductivity carbon foam\footnote{\Trmk{Allcomp} K9.}  for good heat
conduction. Polymethacrylimide\footnote{\Trmk{Rohacell}.} structural
foam fills the remaining voids.  Four PCB flex cables, called
\emph{\dataflex}, are glued on the two faces of the bare stave in a
process involving stencils to ensure the use of a controlled and
optimal quantity of glue.  The \dataflex provides the electrical
connectivity from the stave outer edges to the \Fend electronics.
Three types of \dataflex cables are needed, Short, Medium and Long.
The \Fend readout electronics and silicon sensors are assembled in
\emph{modules}, themselves glued and bonded to the \dataflex cable.  A
\Acr{ut} module is composed of a silicon sensor (see
section~\ref{sec:ut:sensors}), supported on a boron nitride ceramic
stiffener, itself glued to a \emph{hybrid} flex circuit (see
section~\ref{sec:ut:hybrids}) which hosts the readout \Asic{s} (see
section~\ref{sec:ut:salt}).  The modules are glued to both sides of a
stave in a staggered manner such that sensors overlap in the vertical
direction.

In order to fulfil the occupancy requirements in different regions of
the detector, four sensor types (named A, B, C and D) are utilised
with different strip pitch and different strip lengths.  Hybrids can
host either 4 or 8 \Asic{s}, the former version being used for sensors
of Type-A and the latter version for all other sensor types.  The
modules are also of four types, A, B, C and D, named after the sensor
type that it hosts.  This, in turn, leads to three different stave
types (see figure~\ref{fig:ut:full-stave2}
and~\ref{fig:ut:flex2_and_groundgroups}):
\begin{itemize}
\item staves of variant A are used to cover most of the detector
  acceptance and host 14 sensors of Type-A; they have one Short and
  one Medium \dataflex cable on each face;
\item staves of variant B include 10 Type-A and 4 Type-B sensors
  served by one Short and one Medium \dataflex cable on each face;
\item staves of variant C are used for the region adjacent to the beam
  pipe and include 10 Type-A, 2 Type-B, 2 Type-C and 2 Type-D sensors;
  here, one Medium and one Long \dataflex cable are needed on each
  face.
\end{itemize}
Table~\ref{tab:ut:components} summarises the most salient features and
numbers of the \Acr{ut} staves and modules.

\begin{table}[h]
  \centering
  \caption{Summary of \Acr{ut} detector components.}
  \label{tab:ut:components}
  \begin{tabular}{ccll}
    \hline
    \multicolumn{4}{l}{Staves}\\
    \hline
    Variant  & Quantity & \dataflex cables & Module types \\
             &          &  per stave       & per stave \\\hline
    A       & 52       & $2 \times$Short,  $2 \times$Medium & $14 \times$A \\
    B       & 8        & $2 \times$Short,  $2 \times$Medium & $10 \times$A, $4\times$B \\
    C       & 8        & $2 \times$Medium, $2 \times$Long   & $10 \times$A, $2\times$B, $2\times$C, $2\times$D \\
    \hline
    \multicolumn{4}{l}{Modules}\\
    \hline
    Type  & Quantity & Sensor type  & Hybrid type \\
    \hline
    A    &      888 &      A       & 4-chip  \\
    B    &       48 &      B       & 8-chip  \\
    C    &       16 &      C       & 8-chip  \\
    D    &       16 &      D       & 8-chip  \\
    \hline
  \end{tabular}
\end{table}

\subsection{Silicon sensors}
\label{sec:ut:sensors}
\begin{figure}[h]
  \centering
  \includegraphics[width=0.48\textwidth]{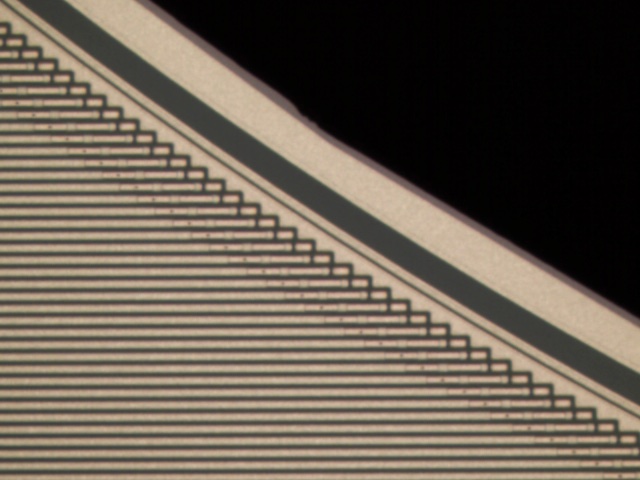}
  \includegraphics[width=0.48\textwidth]{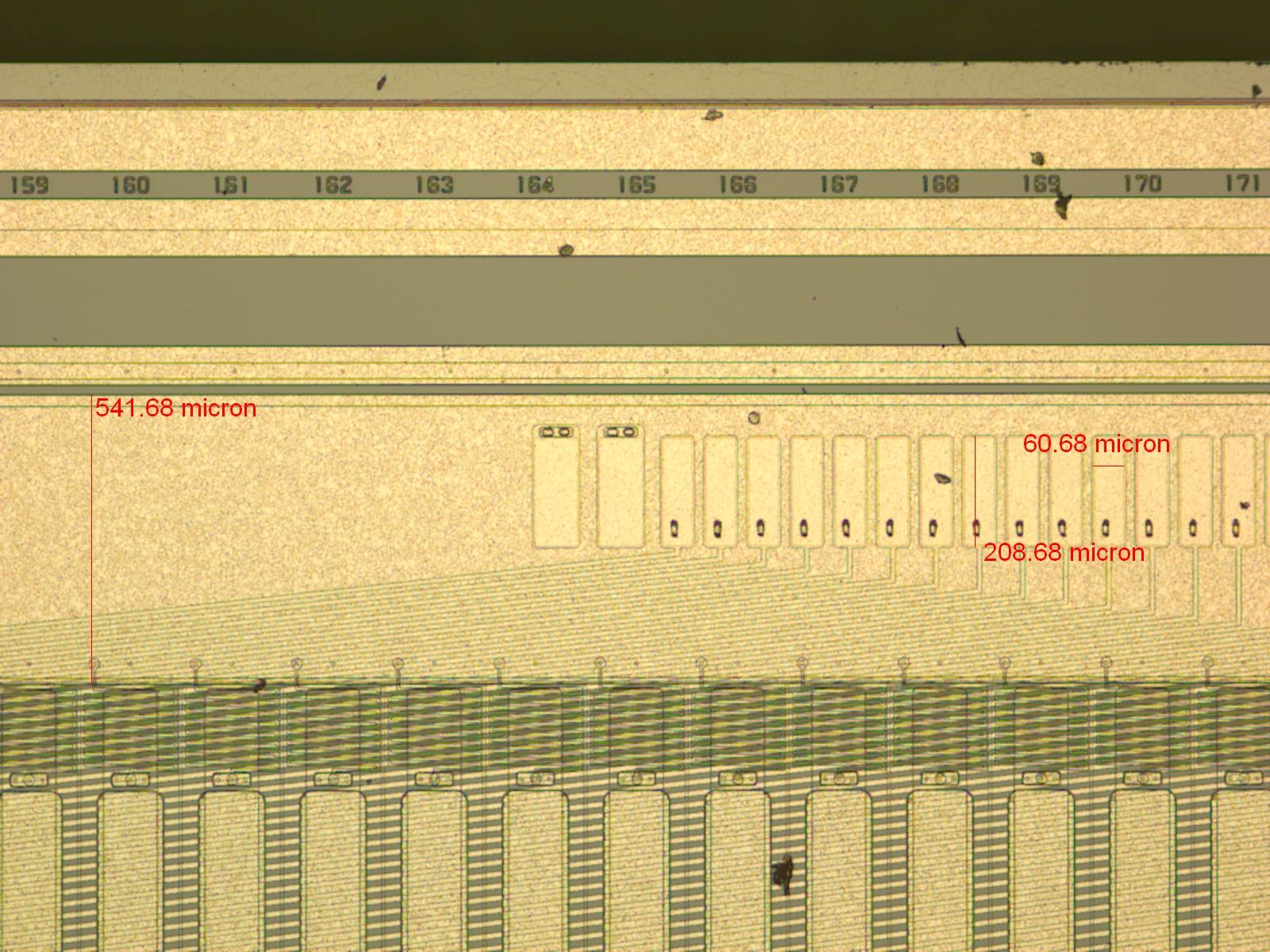}
  \caption{Design features of the \Ut silicon sensors (see text).
    Left: Type-D sensor cut-out region.  Right: embedded pitch
    adapter. Reproduced with permission from~\cite{Carli:2021z4}}
  \label{fig:ut:sensorfeatures}
\end{figure}

The silicon microstrip sensors of the \Ut need to cope with
occupancies and radiation fluences spanning different orders of
magnitudes.  For this reason, four different sensor designs are used,
with n-in-p for the central region and p-in-n in the outer region.
Strip isolation in the p-substrate sensors is achieved through p-stop
technology.  All sensors were produced by
Hamamatsu.\footnote{Hamamatsu Photonics K.K., Hamamatsu, Shizuoka
  435-8558, Japan.}  Type-A sensors constitute the majority of the
detector and cover the outermost parts of the instrumented area.
These n-substrate sensors have 512 p-type strips with a pitch of
187.5\mum.  All other sensor designs use p-substrates and have 1024
n-type strips with a twice smaller pitch of 93.5\mum.  Type-B sensors
are located at approximately 10\cm from the beam line.  The innermost
area is covered by Type-C and Type-D, which have half the strip
length.  Type-D is characterised by even shorter strips on part of the
sensor as its shape includes a circular cut-out that matches the beam
pipe.  In this way, a minimum distance to the beam line of 34\mm is
achieved, as seen in figure~\ref{fig:ut:sensorfeatures} (left).

The sensor design and arrangement in the \Acr{ut} detector keeps the
maximum occupancy below 1\%, being highest in the Type-D sensors.  The
signal-to-noise performance necessary for the efficiency requirements
has been demonstrated in the beam tests discussed in
section~\ref{sec:ut:testbeam}.  For all sensor types, high voltage is
brought to the sensor backplane via a silicon implant embedded along
the top-side edges of the sensor.  This simplifies the stave
construction by allowing high voltage contact to be made via wire
bonds on the same side as the connections between sensor strips and
readout electronics.  The pitch of the readout \Asic channels matches
the strip pitch on sensors of Type-B, -C, and -D.  The Type-A sensors
require a \Acr{pa}, which is achieved using additional metal layers on
the sensor surface, separated from the AC coupled metal strips and the
guard rings by a thin layer of silicon dioxide.  This embedded
\Acr{pa} design is shown in figure~\ref{fig:ut:sensorfeatures}
(right).  While prototype sensors showed signs of signal coupling to
these metal layers resulting in efficiency loss and spurious hits (see
section~\ref{sec:ut:testbeam}), this effect has been minimised in the
final design by locating the bonding pads above the guard ring of the
sensor.

The sensor thickness and resistivity have been chosen to ensure full
depletion of the sensors over the life of the detector.  Before
irradiation, full depletion is achieved applying between 200 and
300\volt bias voltage to the sensor.  It is expected that the most
irradiated p-substrate sensors (Type-D) will be fully depleted with
less than 500\volt at the end of the lifetime of the detector.

The main features of the four sensor designs are summarised in
table~\ref{tab:ut:sensors}.

\begin{table}[h]
  \centering
  \caption{\label{tab:ut:sensors}Main design parameters for the \Ut
    silicon sensors.  For Type-D the given length is for outside the
    cut-out region.}
  \begin{tabular}{ccrrrrrl}
    \hline
    Design & Implant/ & Length    & Width & Thickness & Pitch & Strips & Note\\
    type   & bulk     &      \mm  & \mm   & \mum      & \mum  &        & \\
    \hline
    A & p/n &  99.50& 97.50 & 320 & 187.5 & 512 & embedded \Acr[s]{pa}\\
    B & n/p &  99.50& 97.35 & 320 & 93.5 & 1024 \\
    C & n/p &  51.45& 97.35 & 250 & 93.5 & 1024 \\
    D & n/p &  51.45& 97.35 & 250 & 93.5 & 1024 & cut-out \\
    \hline
  \end{tabular}
\end{table}

\subsection[SALT readout chip]{\Acr[s]{salt} readout chip}
\label{sec:ut:salt}

\begin{figure}[b]
  \centering
  \includegraphics[width=0.95\textwidth,trim=40 0 40 0 ]{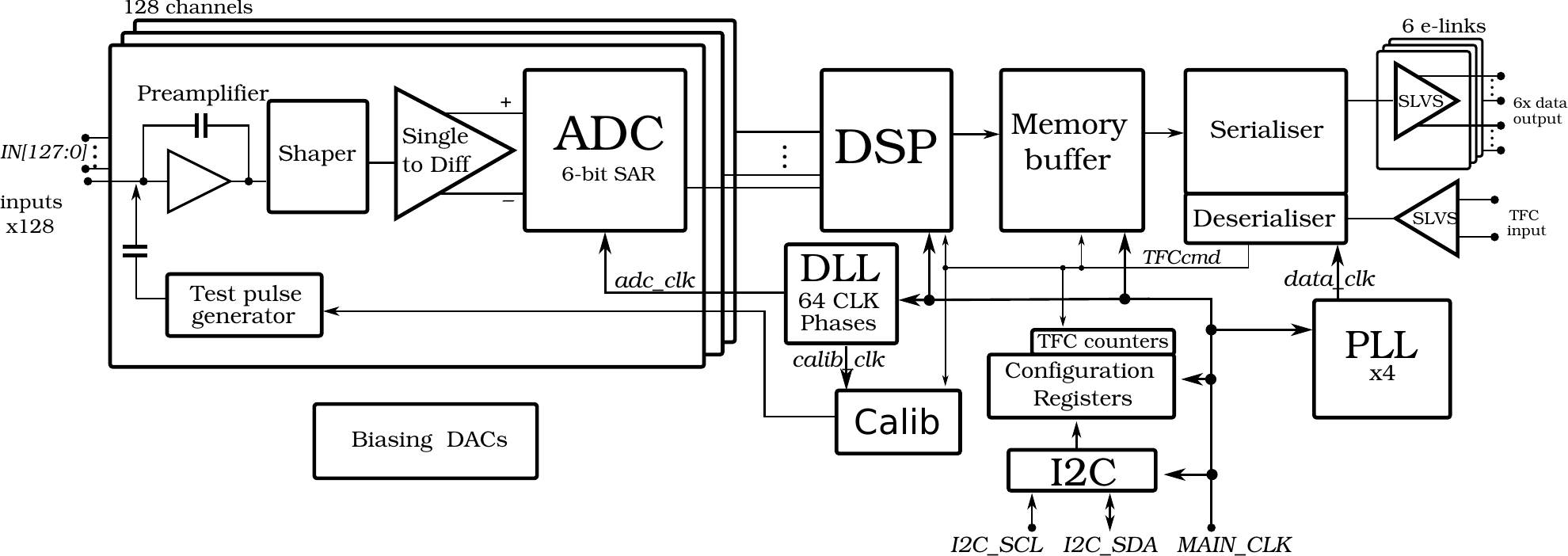}
  \caption{Block diagram of the 128-channel \Acr{salt} \Asic Reproduced from~\cite{SALT2}. CC BY 4.0.}
  \label{fig:ut:salt_block_diag}
\end{figure}

A 128-channel \Asic, called \Acr{salt}, was developed in a 130\nm CMOS
process\footnote{By \Trmk{TSMC}\, Taiwan Semiconductor Manufacturing
  Company.}  to read out silicon strip detectors of the \Acr{ut}.  A
block diagram is shown in figure~\ref{fig:ut:salt_block_diag}.  The
\Acr{salt} features an analog processor and a low-power (less than
1\mwatt/channel), fast (40\,MSps) 6-bit ADC per channel, followed by a
\Acr{dspor} block, a data formatting block and a serialiser block.

The analogue \Fend comprises a charge preamplifier with pole-zero
cancellation and a fast 3-stage shaper (with peaking time
$T_{\text{peak}}=25\ns$ and a fast recovery) required to distinguish between
the \Lhc bunch crossings at 40\MHz.  The \Fend is designed to work
with load capacitances in the range 1.6--12\,pF.  Each channel
contains an 8-bit trimming DAC for baseline equalisation.  In the last
stage of the analogue \Fend a single-to-differential block converts a
single-ended signal to a differential one.

A fully differential 6-bit successive-approximation-register (SAR) ADC
running at 40\MHz converts the analogue signal to the digital
domain~\cite{SALT-SAR-ADC}.  In order to achieve highest speed and
lowest power consumption (significantly below 1\mwatt) the SAR logic
is asynchronous and dynamic circuitry is used in the ADC logic and
comparator.  To synchronise ADC sampling instances with beam
collisions a dedicated ultra-low power ($< 1\mwatt$) DLL is used to
shift and align an external clock.  The ADC samples are signed 6-bit
numbers coded as two's complements.

The digital ADC output is processed by a \Acr{dspor} block, which
performs a pedestal subtraction, a mean common mode (MCM) subtraction
and a zero suppression.  For better testability the \Acr{dspor} can
also transmit raw ADC data or various combinations of partially
processed data.  In the \Acr{dspor} calculation, in each place where
subtraction is performed, a saturation arithmetic is used (giving
numbers inside the range from $-32$ to 31).  In the next step the data
packets are created and recorded in a local memory.

After the \Acr{dspor} the data are serialised with 320\mbps rate,
obtained by increasing the system clock frequency by a factor four and
double data rate (DDR) transmission.  An ultra-low power ($< 1\mwatt$)
\Pll is used to generate the 160\MHz clock from the 40\MHz system
clock.  The data are sent out by an SLVS interface.  The \Asic is
controlled via the \lhcb common protocol consisting of two interfaces:
the \Tfc and the \Ecs~\cite{LHCb-PUB-2011-011,LHCb-PUB-2012-017}.  The
\Tfc interface delivers the 40\MHz clock and other crucial information
and commands, synchronised with the experiment clock, while the \Ecs
serves to configure and monitor the \Asic and is realised through an
\I2c interface.  The main specifications of the \Acr{salt} \Asic are
shown in table~\ref{tab:ut:salt128_sum}.

\begin{table}[t]
\centering
\caption{Summary of the specifications of the \Acr{salt} \Asic.}
\label{tab:ut:salt128_sum}
\begin{tabular}{|l|l|}
  \hline
  Variable & Specification \\
  \hline
  Technology & TSMC CMOS 130\nm \\
  Channels per \Asic & 128 \\
  Input / Output pitch & $80\mum / 140\mum$ \\
  Total power dissipation & $<768\mwatt$ \\
  Radiation hardness & $\SI{0.3}{\MGy}$ \\
  \hline
  Sensor input capacitance & 1.6--12\,pF \\
  Noise & $\sim 1000\,e^-$@10\,pF $+ 50\, e^-$/pF \\
  Maximum cross-talk & Less than 5\% between channels \\
  Signal polarity & Both electron and hole collection\\
  Dynamic range & Input charge up to $\sim 30\,000\,e^-$ \\
  Linearity & Within 5\% over dynamic range\\
  Pulse shape and tail & $T_{\text{peak}} \sim 25\ns$, amplitude after $2\times T_{\text{peak}} <5$\% of peak\\
  Gain uniformity & Uniformity across channels within $\sim5$\%\\
  \hline
  ADC bits & 6 bits (5 bits for each polarity) \\
  ADC sampling rate & 40\MHz \\
  \Acr{dspor} functions & Pedestal and MCM subtraction, zero suppression \\
  Output formats & Non-zero suppressed, zero suppressed\\
  Calibration modes & Analogue test pulses, digital data loading\\
  Output serialiser & Three to five serial e-links, at 320\mbps\\
  Slow controls interface & \I2c \\
  Fast digital signals interface&  Differential, SLVS\\
  \hline
\end{tabular}
\end{table}

The 130\nm CMOS process is generally considered as intrinsically
radiation resistant against total ionising dose.  For this reason, and
given the main clock frequency and the expected charged particle
fluence, SEE protection in the \Acr{salt} is generally limited to
\Acr{seu} protection and monitoring.  In a few specific blocks working
with very small currents ($<1\muamp$), like 7-bit baseline DACs,
protection with NMOS enclosed layout transistors was applied.  To
improve the ADC robustness against SEE, the ADC is reset by the
sampling signal if the previous conversion is not yet completed.  The
triple-voting technique is applied to almost every flip-flop in the
whole digital part.  The combination logic is not triplicated, but
clock and reset signals are.  As a result, to triplicate the complete
clock and reset trees, there are three \Pll{s} and three input reset
synchronisers.  In addition, the configuration registers have built-in
self-correcting circuits that can correct the radiation-induced bit
flips (only one of the three copies) in each clock cycle.

Several \Asic design iterations were required to meet the \Acr{ut}
requirements.  The chip versions used in the \Acr{ut} are the
\Acr{salt}v3.5 for most of the 4-chip hybrids and the \Acr{salt}v3.9
for the 8-chip hybrids and some of the most exposed 4-chip hybrids.
More detailed information on the \Acr{salt} \Asic can be found in
ref.~\cite{SALT2}.

\subsection{Readout chain}
\label{sec:ut:readoutchain}

\subsubsection{Hybrids}
\label{sec:ut:hybrids}

\begin{figure}[h]
  \centering
  \includegraphics[width=0.49\textwidth]{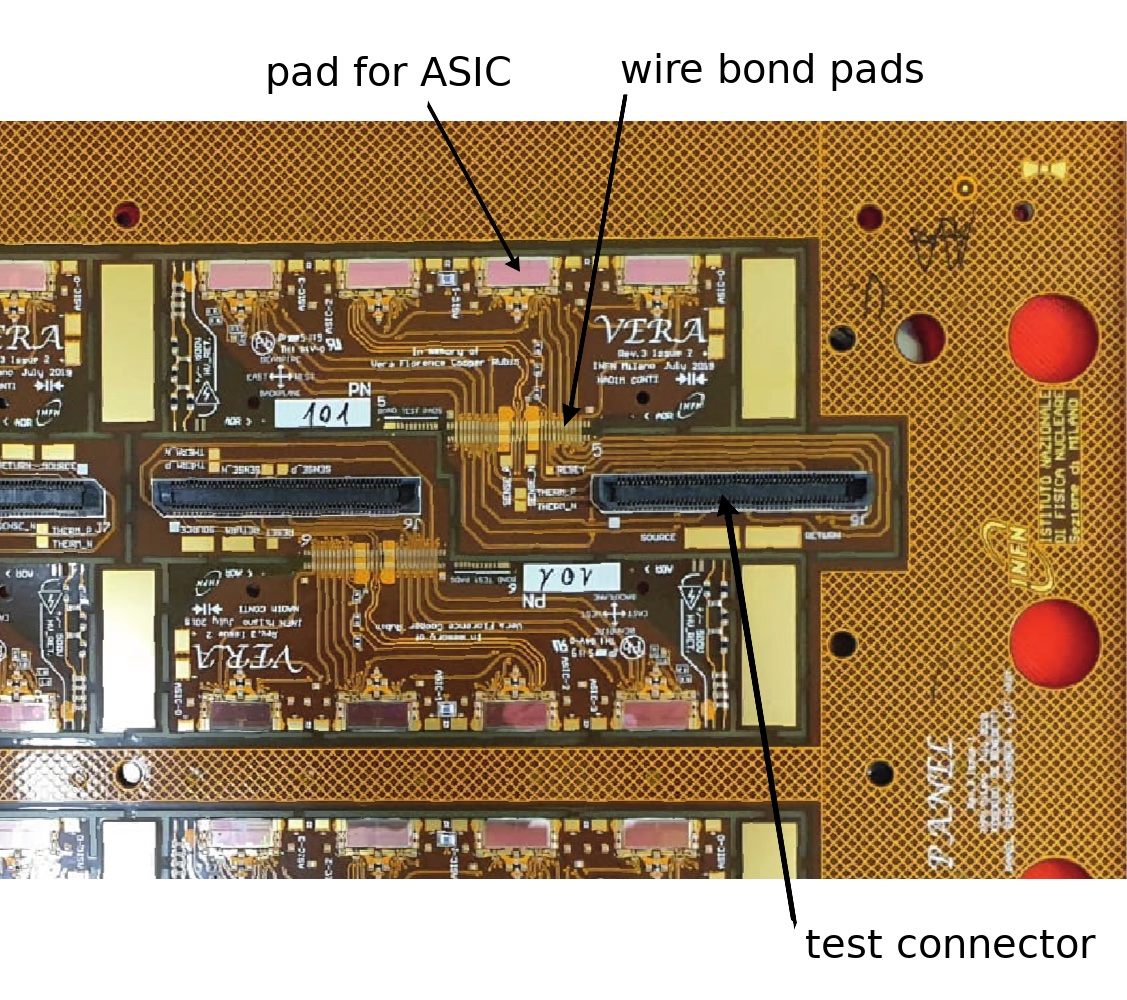}\hfill
  \includegraphics[width=0.452\textwidth]{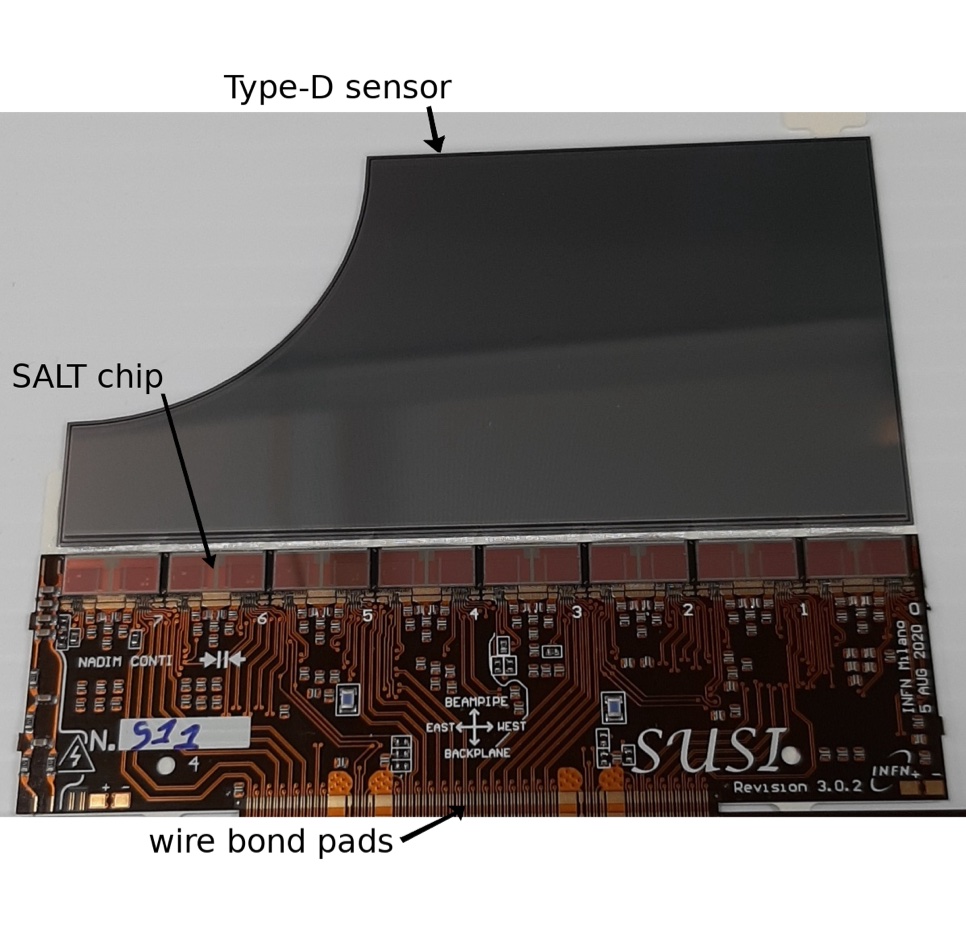}
  \caption{\label{fig:ut:hybrid2}Left: two fully visible \vera
    hybrids, not yet equipped with \Asic{s}, embedded in a carrier
    8-hybrid panel before cutting.  Right: \susi hybrid mounted in a
    final \Acr{ut} Type-D module.}
\end{figure}

The \Acr{ut} hybrids are the \Fend boards for the microstrip silicon
sensors.  Being mounted in the detector acceptance, they were designed
with mass minimisation in mind.  They are low-mass flex PCBs composed
of a stack of conductive and dielectric layers\footnote{\Trmk{DuPont}
  Pyralux: two FR~7013 and two AP~8535R sandwiching an adhesive
  LF~1500.}  with a total thickness of about 72\mum Cu, 176\mum
polyimide and 63\mum adhesive. Their function is to distribute \Tfc
and \I2c signals, and route e-links from the \Asic{s} to the the
\dataflex cables, while maintaining high signal integrity.  Hybrids
are designed to distribute low and high voltage (LV and HV) coming
from the \dataflex cable to the \Asic{s} and sensors with high
filtering performance.  The \Acr{ut} uses two types of hybrids: \vera
hybrids accommodate 4 \Acr{salt} chips and they are used in most of
the detector, while \susi hybrids host 8 chips each and instrument the
centre of the detector, where the strip density is double.  In total,
888~\vera and 80~\susi are needed to populate the full \Acr{ut}.
Hybrids have been produced in panels, as visible in
figure~\ref{fig:ut:hybrid2} (left), with \vera (\susi) panels
containing 8 (6) circuits each, fully instrumented with connectors and
alignment holes.  The connectors allowed a full electronic test to be
performed without the need of cutting the hybrid away from the panel.
Precut slits were added to ease the mounting on modules.  In
figure~\ref{fig:ut:hybrid2} (right) a picture of a \susi hybrid in a
final \Acr{ut} module is shown.\looseness=-1

\subsubsection{Passive PCB flex cables}

Two families of passive flex cables are used to transfer signals and
power from the peripheral electronics, located outside the acceptance,
to the hybrids: pigtails provide the connection between outside and
inside the detector box, \dataflex cables provide the connection
between pigtails and hybrids on the staves.

\begin{figure}[t]
  \centering \includegraphics[width=0.8\textwidth]{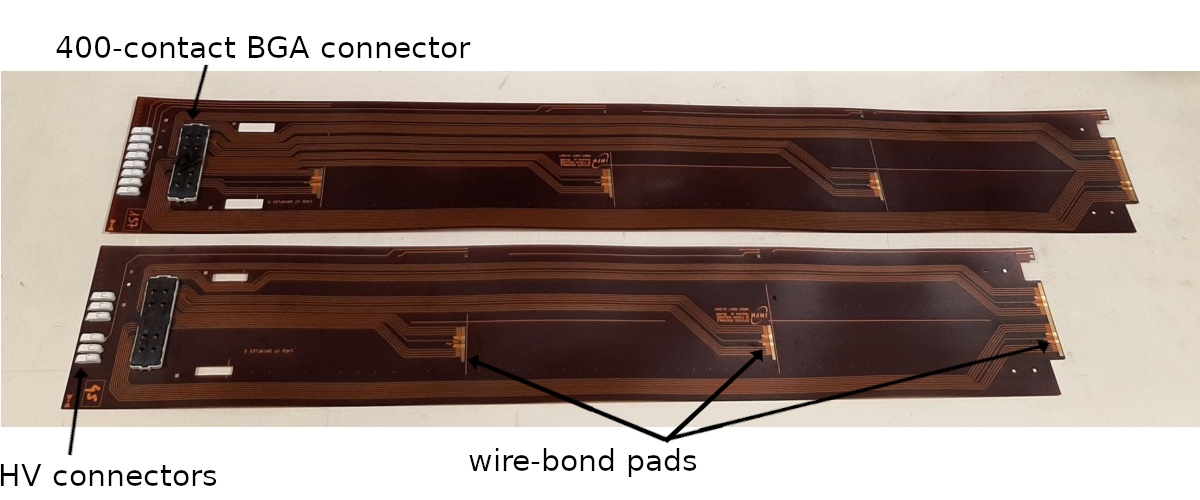}
  \caption{\label{fig:ut:flex1}Picture of one Short and one Medium
    \dataflex cable.}
\end{figure}

The \dataflex cables are flexible PCB passive boards glued directly on
the staves.  They are designed to provide high fidelity signal
integrity, be lightweight and maximise heat transfer from modules to
the underlying support, while minimising the voltage drops across the
board.  The sensor bias voltage lines are isolated from the low
voltage and data lines.  Each \dataflex cable connects to one pigtail
via a high-density 400-contact connector\footnote{\Trmk{MEG-Array} 400
  BGA connector from \Trmk{Amphenol}.}  and to the modules via wire
bonds.  \Hv connectors on the outer end connect directly to \Hv cables
that are fed through the detector box wall.  The \dataflex cables were
designed and produced in three different sises, Short, Medium and
Long, to match the various locations and types of modules along the
stave. Their respective lengths are about 604, 700 and 748\mm.  In
figure~\ref{fig:ut:flex1} a picture of two sample \dataflex cables is
displayed.  The stack\footnote{%
  \Trmk{DuPont} Pyralux: LF~0110, AP~8535, LF~0100, AP~9121, LF~0100,
  AP~8525 and LF~0110.}  includes about 225\mum of polyimide, 70\mum
of Cu and 100\mum of adhesive.  A pictorial representation of the
\dataflex cable usage in the \Acr{ut} is found in
figure~\ref{fig:ut:flex2_and_groundgroups} (left).

\begin{figure}[t]
  \centering
  \includegraphics[width=0.4\textwidth]{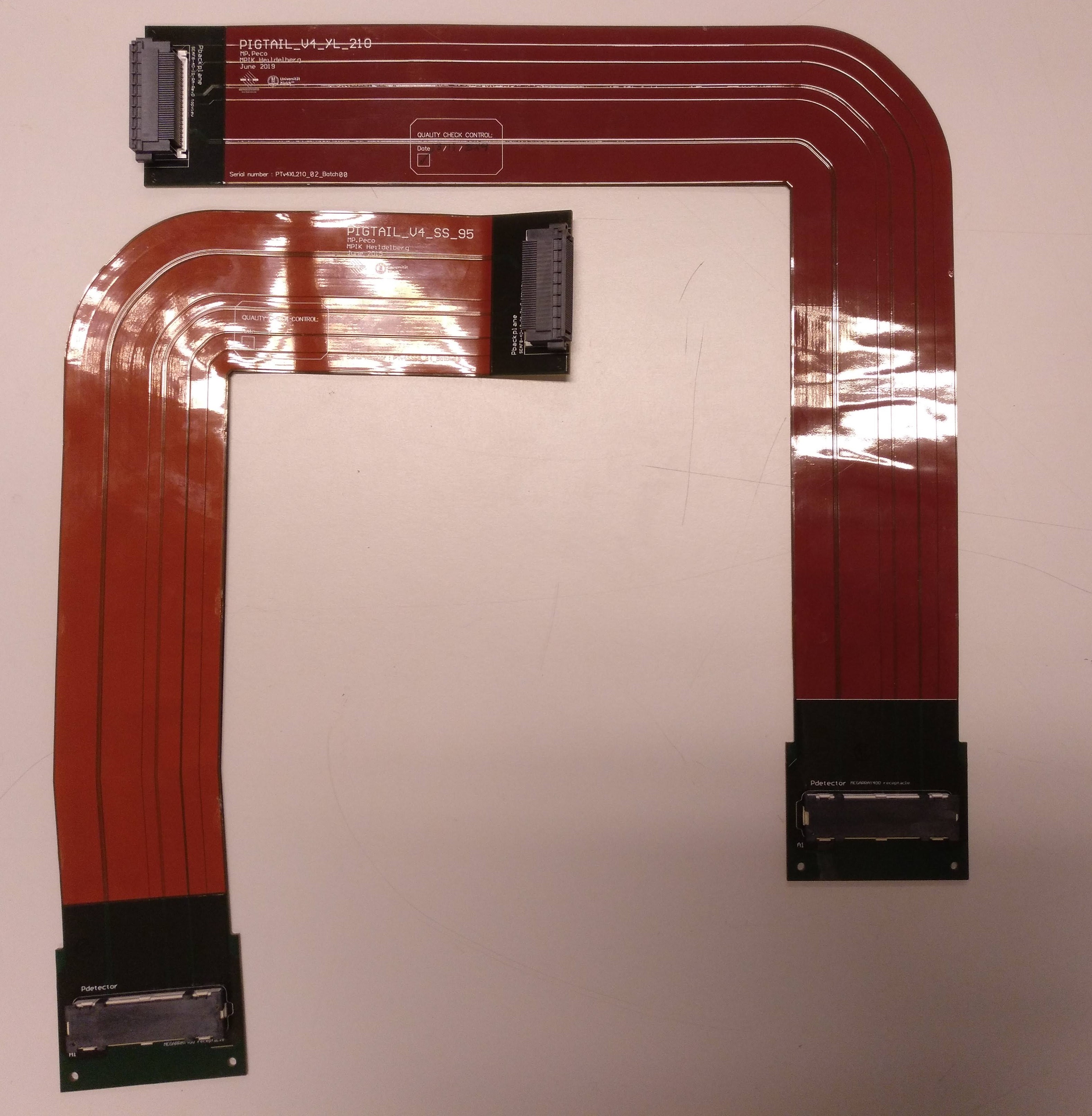}
  \includegraphics[width=0.315\textwidth]{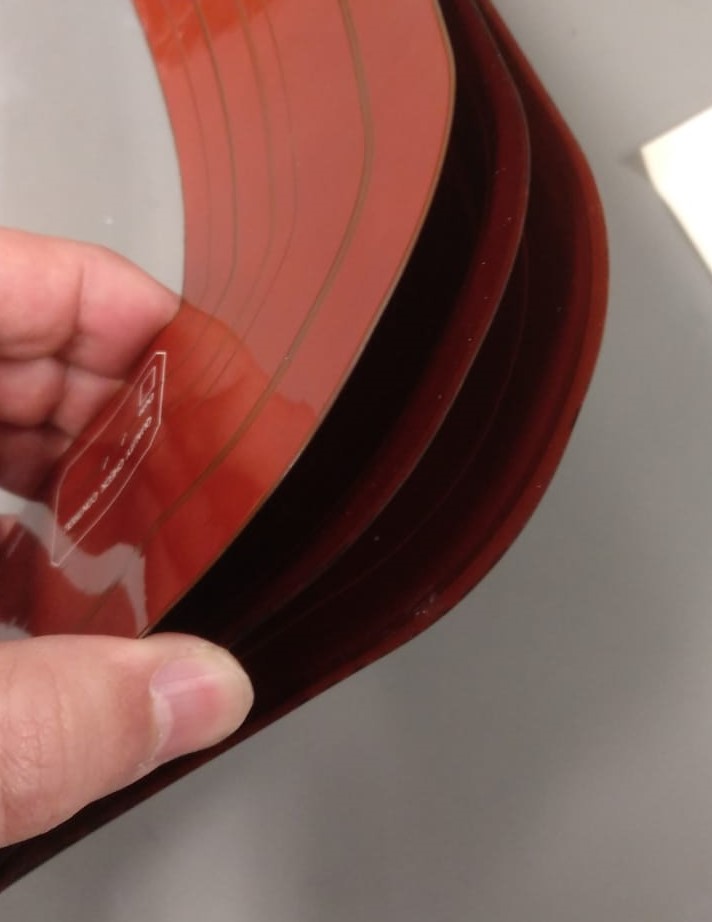}
  \includegraphics[width=0.41\textwidth, angle=90]{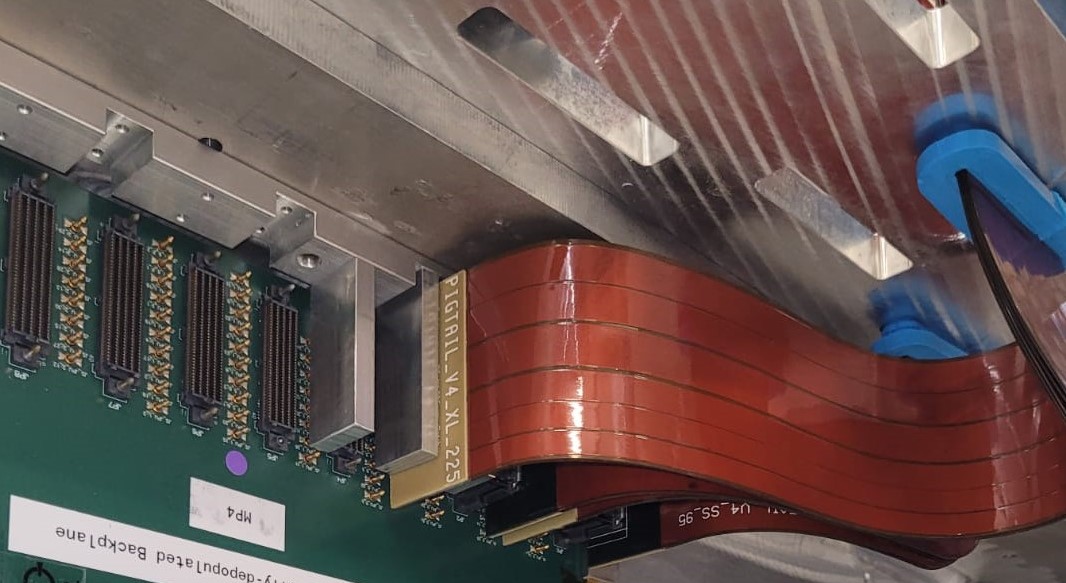}
  \caption{\label{fig:ut:pigtails}Left: pigtails in two different
    shapes; middle: pigtail section showing the 3 subcables; right:
    installed pigtails.  }
\end{figure}

Pigtails are flexible PCB boards that provide connection between
\dataflex cables (inside the detector box) and the peripheral
electronics (outside the box) discussed in section~\ref{sec:ut:PEPI}.
The latter connection is made via a high-speed high-density
connector.\footnote{SEAF8-40-1-S-10-2-RA connector from
  \Trmk{SAMTEC}.}  A complex design was developed to cope with all the
mechanical constraints and satisfy the electrical requirements related
to the low voltage power and the \Acr{salt} differential pair signals.
The 272 installed pigtails are identical from the electrical point of
view, despite the fact that \dataflex cables can host varying numbers
of chips.  To accommodate the required lengths and angles due to the
positions of the staves and the peripheral electronics, 11 different
pigtail shape variants have been designed.
Figure~\ref{fig:ut:pigtails} shows three pictures of the pigtails.  In
the leftmost picture two different variants are shown for
illustration.  The two 1\mm thick FR4 stiffeners used to protect the
fragile solder bumps under the connectors are also visible at both
ends.  The middle picture shows that each pigtail is composed of three
flex subcables to confer sufficient flexibility to the pigtail for
routing.  The rightmost picture shows an example of bending when
installed.  Each subcable contains three copper layers of 35\mum
thickness each, sandwiched between polyimide films and adhesive
(adding up to about 0.25\mum of polyimide and 75\mum of adhesive per
subcable).  The inner layer is used for the differential pair signals
and the outer layers for power lines.  This provides good signal
integrity and low resistance for the power traces.  The pigtails cross
the detector box walls through dedicated holes, which are sealed with
foam to provide sufficient gas and light tightness.

\subsubsection{Peripheral electronics}
\label{sec:ut:PEPI}

\begin{figure}[t]
  \centering
  \includegraphics[width=0.7\linewidth]{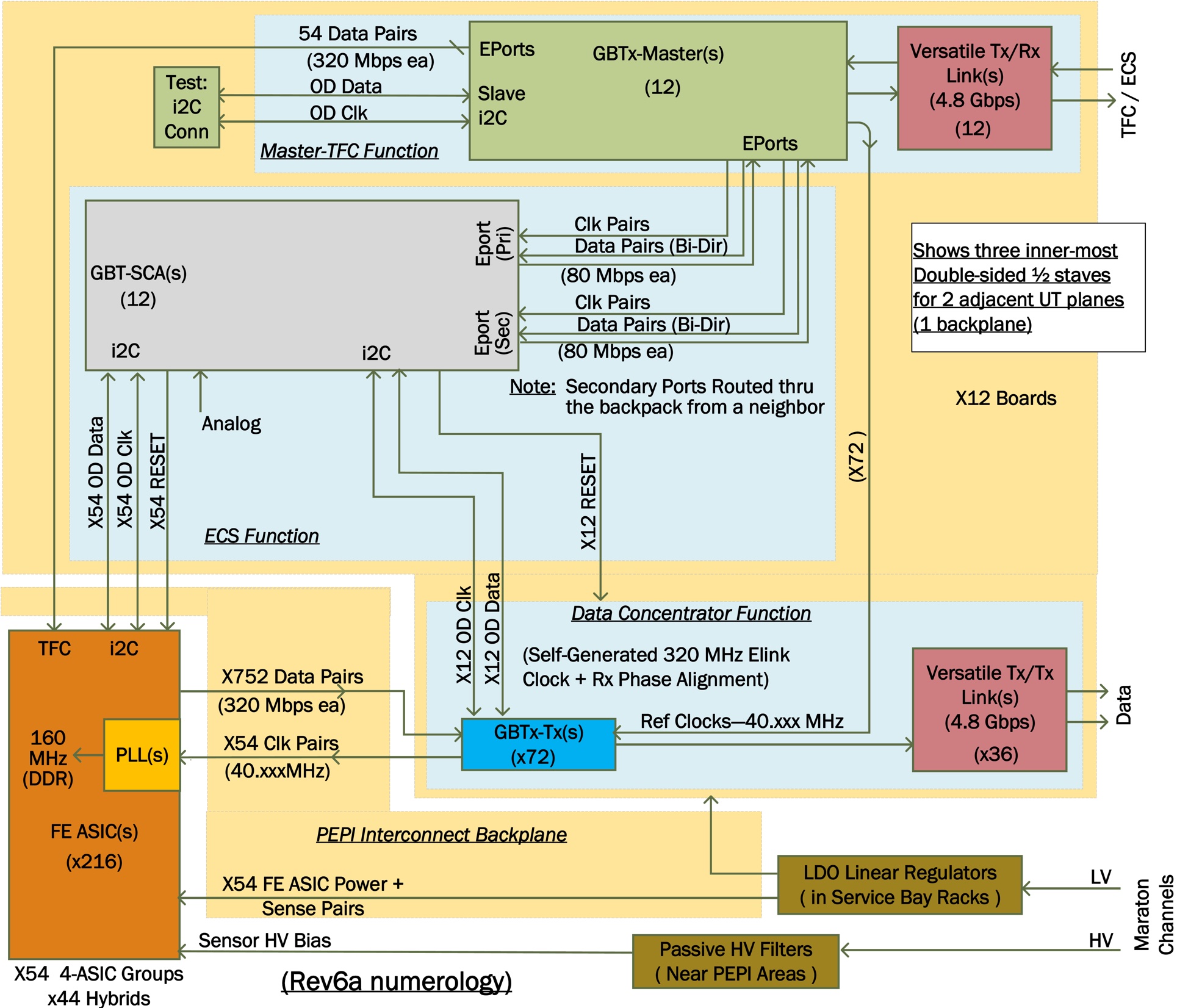}
  \caption{Schematic of LV, HV, data, fast and slow control signal
    distribution to \Acr{ut} \Fend electronics.}
  \label{fig:ut:data_flow}
\end{figure}

The periphery electronics processing interface (PEPI) units are
responsible for readout and control of the detector.  They connect to
the staves via the pigtails.  The full PEPI system comprises 24
backplanes, 24 pigtail power breakout boards (P2B2s), and 248 data and
control boards (DCBs).  The distribution of data, clock, and control
signals is shown in figure~\ref{fig:ut:data_flow}.  The \Gbtx, mounted
on the DCBs, implements bidirectional links between the detector and
the counting room with components that are radiation hard up to
$\SI{1}{\MGy}$~\cite{Moreira:GBTx1}.  Each PEPI unit houses 3
backplanes that route power, data, and control signals into and out of
the detector volume and three P2B2s that route additional power.  The
backplanes support up to 12~DCBs that connect to the counting room via
optical links.

\subsubsection{Backplanes and power breakout boards}

The routing of the backplanes balances the load carried by each DCB
and organises the data to minimise resources needed for event
reconstruction.  A fully utilised backplane supports 6 half-staves
read out by 12 DCBs.  Due to space constraints in the area around the
staves the size of the backplane is limited.  Matching the twelve
400-contact pigtail connectors leads to a very high density of traces
on the board.  The signal routing is achieved by a 28-layer board with
a PCB aspect ratio of 1:10 which is pushing the limits of
manufacturability.  There are two types of backplanes, referred to as
true and mirror, with different layouts to accommodate the geometry of
pigtails from the detector.  Each type also has three variants: full,
partial and depopulated.  The full backplane is used for the
high-occupancy region of the detector, while the depopulated one is
used for the outer region of the detector planes, and partial
backplanes are used everywhere else.  While the full backplane
transmits all data and control signals, on the partial and depopulated
backplanes some traces are unused and grounded.  The backplane routes
all of the 1.5\volt and 2.5\volt power to the DCBs and some 1.2\volt
power to the \Acr[p]{salt}.  However, the majority of \Acr{salt}
1.2\volt power lines are routed through the P2B2 in order to leave
room for data and control signals while maintaining the
manufacturability of the backplane.

\subsubsection{Data and control board}

Each DCB supports one master \Gbtx, one \Gbtsca, six data \Gbtx chips,
one \Vtrx~\cite{VTXx}, and two or three \Vttx chips.  The \Gbtx chip
receives 320\mbps sensor data from the e-ports connected to the \Fend
\Asic{s} and repackages it into 4.8\gbps data frames.  In addition it
controls data frames sent to and from the counting room via the
\Gbtsca~\cite{Caratelli2015}.  The \Gbtsca distributes slow control
and monitoring signals to the detector.  The \Vtrx incorporates a
laser driver and optical receiver to convert between optical signals
to and from the counting room and electrical signals to and from the
\Gbtx, while the \Vttx only transmits from the \Gbtx to the counting
room.  Communication between these chips is routed through the 16
signal and power layers of the DCB.  In order to preserve
high-frequency characteristics of the 4.8\gbps communication between
\Gbtx and \Vttx/\Vtrx chips, a special laminate
material\footnote{\Trmk{Isola Terragreen}.} is used in place of
standard FR4.

\subsubsection{Data and control signal distribution}

\begin{figure}[t]
  \centering \includegraphics[width=0.40\textwidth]{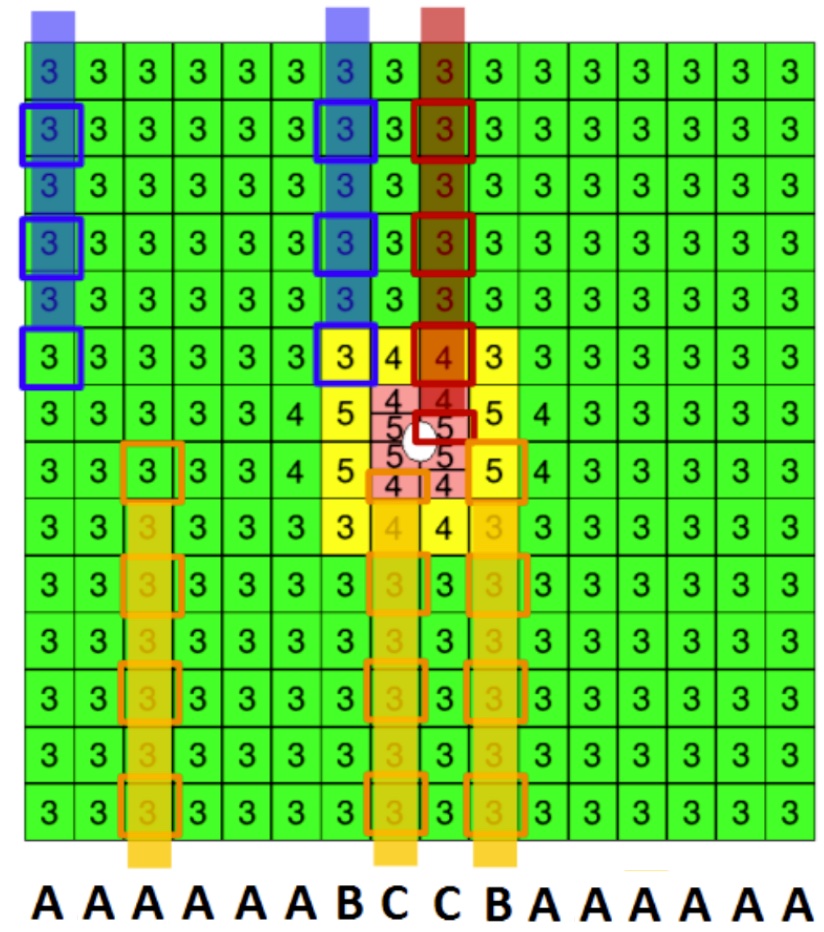}
  \includegraphics[width=0.55\linewidth]{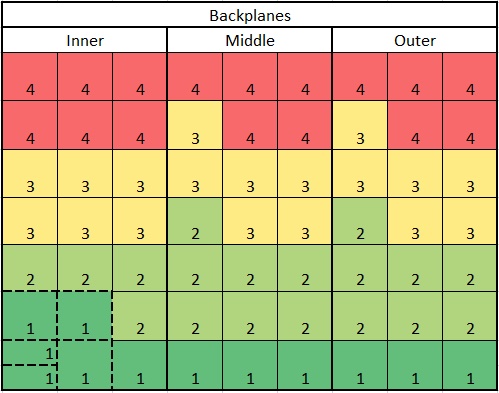}
  \caption{\label{fig:ut:flex2_and_groundgroups}Left: illustration of
    the usage of \dataflex cables on the different stave variants.
    Short cable shaded in blue, Medium in orange and Long in red.  The
    modules hosted by the cable are highlighted with a coloured
    contour.  The numbers 3, 4, 5 indicated the number of wire-bonded
    e-ports per chip on the given module. Right: arrangement of a
    single plane of the \Acr{ut} hybrids into power groups sharing a
    common \Lv output channel.  Only one quadrant of the plane is
    shown. Boxes represent hybrids and the numbers (1 to 4) the power
    group.}
\end{figure}

As shown in figure~\ref{fig:ut:data_flow}, between 6 and 12 e-ports of
each data \Gbtx are connected to the \Fend \Asic{s}.  Each data \Gbtx
receives data from either 2 or 4 \Asic{s} belonging to the same
4-\Asic group, depending on the number of \Acr{salt} e-ports and the
location in the detector.  The expected use of e-ports per \Acr{salt}
\Asic is shown by the numbers (3 to 5) in the boxes of
figure~\ref{fig:ut:flex2_and_groundgroups} (left).  Each pair of data
\Gbtx transmits data to the counting room via a \Vttx optical link.
The master \Gbtx receives the \Lhc clock from the counting room via a
\Vtrx optical link, and uses a \Pll to generate a local clock with the
same frequency (40\MHz).  The master \Gbtx clock serves as a reference
for the data \Gbtx{s} and \Gbtsca.  One data \Gbtx per DCB transmits
that clock directly to all \Fend read out by this DCB.  The clock
phase can be adjusted individually for each group of 4 \Acr{salt}
\Asic{s}.  \Tfc signals are received from the counting room via a
\Vtrx optical link by the master \Gbtx and transmitted directly to all
\Fend read out by the DCB.  \Ecs signals are transmitted from the
master \Gbtx to the \Gbtsca via a dedicated 80\mbps e-port.  The
\Gbtsca distributes slow control signals to both the data \Gbtx chips
and \Fend chips by the DCB.  The \Gbtsca also distributes a GPIO
signal that can reset the data \Gbtx and \Fend chips, and uses an ADC
to monitor the voltage of the DCB and thermistors on the DCB and \Fend
electronics.

\subsection{Low voltage power}

Low-voltage power for the \Acr{ut} hybrids and on-detector readout
chain electronics is sourced from a bank of 12 power
supplies\footnote{\Trmk{Wiener} MARATON HE LV.} providing 144 8V/50A
channels to four service bays situated around the \Acr{ut}
detector. The service bays house 268 8-channel low-voltage regulator
boards (LVRs) designed around the LHC4913 linear regulator chip. The
LVRs segment the \Lv power into individual 3.6\aunit{A} max derated
channels with full differential remote regulation to the detector
electronics at 1.26, 1.5 or 2.5\volt nominal potential at load,
depending on plug-in Current Control Mezzanine (CCM) boards.  The
individual channels also have the capability to be ganged in groups of
two to boost the derated output to 7.2A for high-draw loads such as
the \Gbtx chips and the 8-\Asic hybrid positions, with one channel
providing the voltage regulation and the other sharing the output
current with a specialised current-following CCM.

Groups of LVR channels served by a single power supply channel are
referred to as \emph{power groups} and these power groups have been
optimised at the load side to maintain power independence of different
backplanes. For the \Acr{salt} \Asic{s}, the non-negligible variations
in voltage drop at different positions along the stave flex cable
requires a \emph{horizontal} grouping of hybrids at similar voltage
drops rather than a \emph{vertical} grouping along the stave.  The
grouping implemented in the detector is illustrated in
figure~\ref{fig:ut:flex2_and_groundgroups} (right).

The LVR channels provide individual outputs referenced to shared LVR
ground which is isolated from the local mechanical interface so that
the floating power supply outputs remain floating through the LVR
boards. The grounding of electronics is done at the backplane. Between
the service bays and \Acr{ut} electronics, through approximately
8--10\m of cable, the LVRs must provide regulation across round-trip
voltage drops ranging from 150\mvolt to 500\mvolt while maintaining
smooth regulation performance, particularly at start-up.  Extensive
studies have been done to optimise the LVR regulation performance.
The overall bandwidth is limited to a few hundred \khz. Faster voltage
transients or oscillations are dealt with using passive decoupling
networks at the hybrids or DCB input.\looseness=-1

The \Acr{ut} LV power demands may be logically divided into \Acr{salt}
\Asic and DCB loads, with \Acr{salt} \Asic{s} requiring one 1.26\volt
LVR channel for \vera hybrids or a pair of ganged channels for the
\susi hybrids.  DCB boards require a pair of ganged channels at
1.5\volt for powering of the \Gbtx and \Gbtsca chips, and a single
2.5\volt channel for \Vttx and \Vtrx optical modules.  Due to the low
power demand for the latter, a single 2.5\volt channel serves a pair
of DCBs in the final \Acr{ut} system.

Monitoring and control for the LVRs is provided through a \Acr{spi} to
an on-board \Fpga\footnote{\Trmk{Actel} ProASIC3 \Fpga.} which
controls channel state and sequencing and provides state
information. A dedicated mezzanine board provides a \Gbtsca interface
to off-board electronics as well as monitoring analog outputs for
real-time diagnostic information about voltage and current levels in
the \Acr{ut} system.  The mezzanines are in turn interfaced to the
counting room via a control board in each LVR crate.

\subsection{Cooling}
\label{sec:ut:cooling}

Evaporative \cotwo cooling is provided using a titanium tube embedded
in the stave and running below the \Asic{s} in a serpentine shape.
The two-phase accumulator controlled loop is in common with the \lhcb
\Velo detector, see section~\ref{ssec:infracooling}.  The detector
cooling system has to extract the power dissipated by the read-out
chips, and keep the sensors at the target working-point temperature
(\SI{-5}{\degc}) to prevent thermal runaway in presence of radiation
damage.  The total detector power to be extracted is expected to be
about 4\kwatt, including 4192 \Asic{s} (about 0.8\watt/each), cables,
sensors and environment. A safety margin of $+25\%$ is also considered
in designing the system.

The core material of the box walls is a rigid polyetherimide-based
polymeric foam.\footnote{\Trmk{Airex R82.60}, from 3A Composites.}
\newcounter{airexfnc}\setcounter{airexfnc}{\value{footnote}} The foam
is sandwiched in two \Acr{cfrp} skins.  The walls parallel
(perpendicular) to the beam pipe are made of 24\mm (20\mm) thick core
and 1\mm (0.5\mm) thick skins.  A copper net (51\mum thick, 74\% open
area) covers the interior surfaces.  The Be beam pipe is wrapped in a
thermal blanket composed of 4 successive layers, a 0.2\mm PI PCB
heater jacket directly on the beam pipe, $\sim 0.1\mm$ PI tape, a 5\mm
thick thermal insulation\footnote{\Trmk{Pyrogel} XTF from Aspen
  Aerogels.}  and again $\sim 0.2\mm$ PI tape.  The interface between
the box and beam pipe blanket is made of rigid polymeric
foam,\footnotemark[\value{airexfnc}] EPDM foam and \Acr{cfrp}.

The detector box and beam pipe blanket were designed such that the
temperatures on the surfaces never decreases below \SI{13}{\degc},
i.e.\ stays above the cavern dew point.  Dry nitrogen, with a nominal
dew point of \SI{-50}{\degc}, is circulated through the detector box
via a supply line and a bubbler.  Hot air from the electronics is
forced to circulate near the pigtails outside the box to avoid
condensation.

Four manifolds, two per side, one above and one below the detector
box, distribute the cooling fluid to the detector, with one cooling
loop per stave.  Each loop is equipped with a restriction (0.2\mm
orifice) at the entrance in order to avoid cross-stave effects arising
from different heat loads.  The nominal flow per stave is about
0.76~g/s.  The pressure drop is driven by the inlet flow restrictions.
Nominally, the total pressure drop in the detector is about 5~bar.  On
the stave, the temperature is stable to about \SI{-0.5}{\degc}.  The
\cotwo inlet temperature can be set from ambient temperature to
\SI{-30}{\degc}.

\begin{figure}[h]
  \centering
  \includegraphics[width=0.9\textwidth]{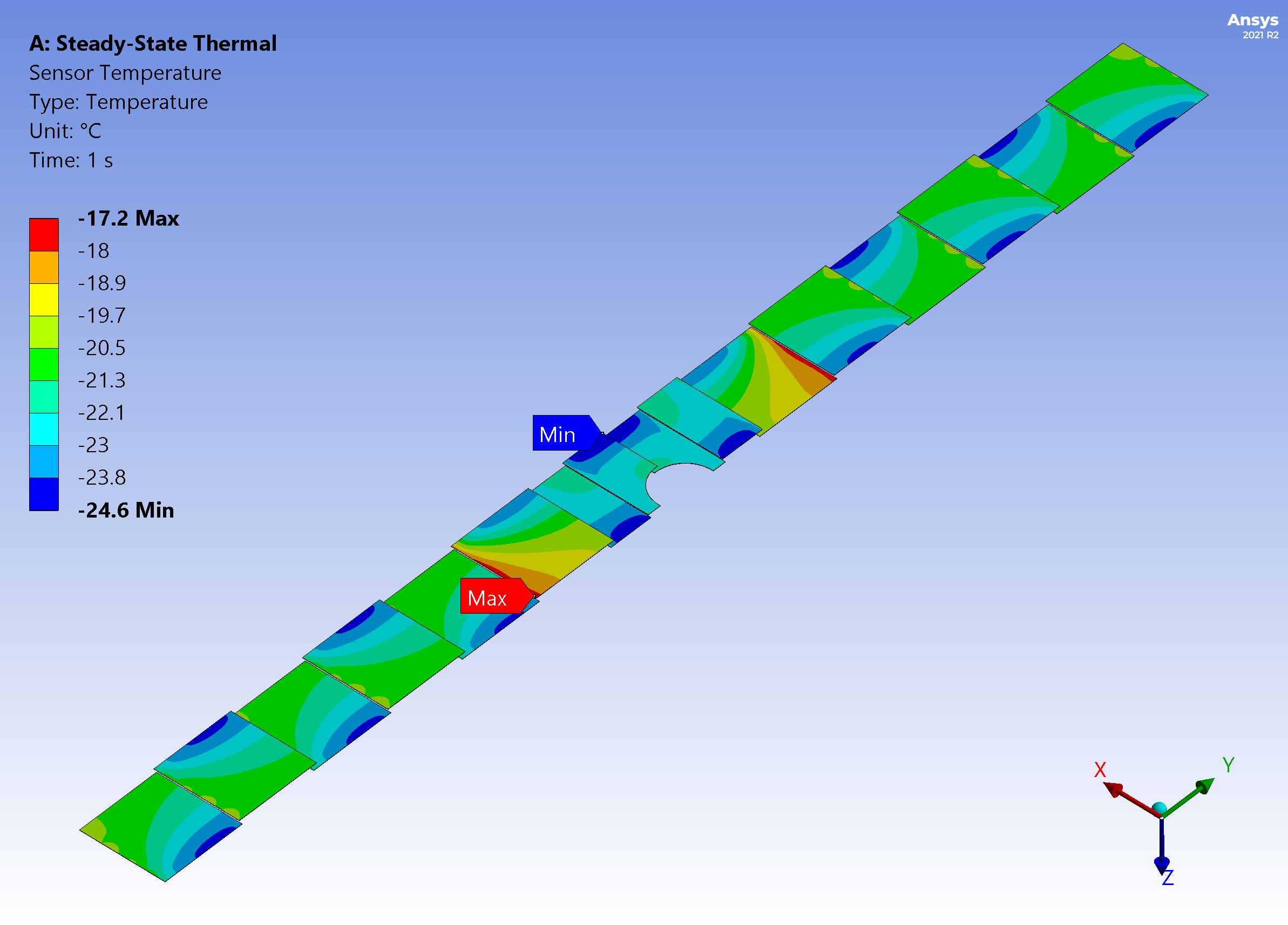}
  \caption{\label{fig:ut:cooling1} Simulation of the thermal behaviour
    of a central stave (variant C) at an inlet \cotwo temperature of
    \SI{-25}{\degc}.}
\end{figure}

The cooling requirements and properties of the detector have been
studied using thermal and mechanical finite element
models~\cite{Coelli_2017}.  A result of the simulated thermal profile
obtained on a central stave (variant C) is shown in
figure~\ref{fig:ut:cooling1} for an inlet \cotwo temperature of
\SI{-25}{\degc}. The \Asic power is assumed to be 0.8\watt per chip.
This shows that the temperature of the innermost sensor (the most
critically irradiated) can be kept well below the required temperature
of \SI{-5}{\degc}.

\subsection{Test beam results}
\label{sec:ut:testbeam}
 
To validate various aspects of the \Acr{ut} sensor design and readout,
a series of beam tests were carried out.  The test beam runs were
focused on R\&D and validation of three novel features of the sensor
design (see section~\ref{sec:ut:sensors}): (1) the double-metal region
where the $190\mum$ sensor strip pitch is reduced to $80\mum$ to match
the pitch of the custom \Acr{ut} \Asic input pads, (2) the top-side
biasing scheme, and (3) the circular cut-outs of the Type-D sensors.
Both unirradiated and irradiated sensors were tested, of full nominal
size as well as miniature sensors.  Sensor performance was measured,
in particular signal-to-noise ratio (S/N), for exposures up to twice
the maximum expected irradiation value over the lifetime of the
\Acr{ut} detector.

The key R\&D areas listed above were studied in
detail~\cite{NIM2016,LHCB-PUB-2016-007a}.  Early test beam results,
using the Beetle readout chip~\cite{BEETLE_REF}, showed that a S/N of
about 12 can be expected for the Type-A sensors.  The S/N of the
Type-D sensors was measured to be higher, about 17, owing to the lower
input capacitance of the shorter strips.  The studies also
demonstrated that the sensors with top-side biasing perform equally
well as those that are biased directly from the back of the sensor.
Studies of the Type-D sensors showed that they maintained excellent
signal efficiency all the way up to the edge of the circular cut-out.

Test beam runs with early prototypes of the Type-A sensors showed a
significant loss of signal efficiency in the \Acr{pa} region.
Although this region covers less than 1\% of the sensor's active area,
the loss in efficiency was deemed to be unsatisfactory.  Additional
studies of the test beam data, as well as CAD device
simulations,\footnote{Simulatios were performed with \Trmk{Synopsys}
  Sentaurus TCAD.} were performed to conclude that the loss of
efficiency was due to the capacitive coupling between the double-metal
layers, inducing charge pick-up on other strips~\cite{ARTUSO2018252}.
A new design was developed, with a thicker silicon dioxide insulating
layer below the \Acr{pa} and with most of the \Acr{pa} metal moved
outside the active area, and successfully tested.  The results proved
that the \Acr{pa} region exhibited no significant loss of hit
efficiency.

\begin{figure}[t]
  \centering
  \includegraphics[width=1\textwidth]{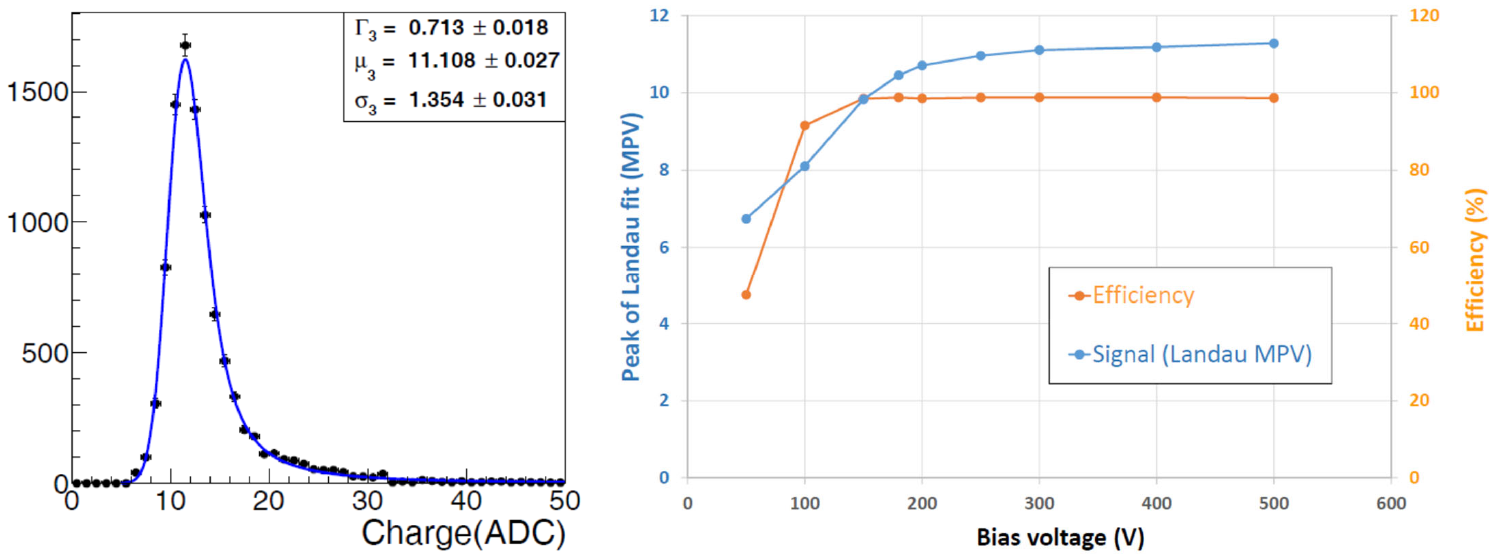}
  \caption{Test beam results for a Type-A unirradiated sensor for
    tracks with normal incidence.  Left: distribution of collected
    charge (in ADC counts) at 300\volt bias.  The data are fitted with
    a Landau distribution convoluted with a Gaussian function.  Right:
    most probable value of the Landau fit result and hit efficiency
    versus the applied bias voltage. Reproduced from ~\cite{LHCB-PUB-2019-009}. CC BY 4.0.}
  \label{fig:ut:FermilabTBResults}
\end{figure}

A final test beam run~\cite{LHCB-PUB-2019-009} was conducted which
successfully demonstrated the performance of a Type-A sensor with
\Acr{salt} v3.0 128-channel readout chips.  The most relevant results
are shown in figure~\ref{fig:ut:FermilabTBResults}.  The left panel
shows the distribution of collected charge, in ADC counts, for tracks
with normal incidence.  Fitting the distribution with a Landau
function convoluted with a Gaussian function, results in a most
probable value of 11.1 ADC counts, while the measured common-mode
subtracted noise was about 0.9 ADC counts.  Thus, a S/N of about 12 is
obtained.  An irradiated sensor was also tested and found to have a
S/N about 10\% lower, which is still large enough to meet the \Acr{ut}
requirements on signal efficiency and noise hit rejection at the end
of life of the detector.

\subsection{Slice test}
\label{sec:ut:slicetest}

A full electronics read-out test on a prototype stave, dubbed
\emph{slice test}, was set up to validate the detector design with
realistic power distribution and grounding.  To read out a complete
stave, two partially routed backplanes were used together with four
DCBs and associated power distribution break-out boards.  The \lhcb
MiniDAQ system was used to control the system, distribute the timing
signals and send trigger commands to the \Fend electronics, and to
process the data from the \Fend electronics.  The stave was cooled
using a bi-phase \cotwo cooling system (a reduced size version of the
one installed in the \lhcb cavern).  The stave was operated in a
light-tight box at temperatures between \num{-30} and 15\degc.

\begin{figure}[t]
  \centering
  \includegraphics[width=0.7\textwidth]{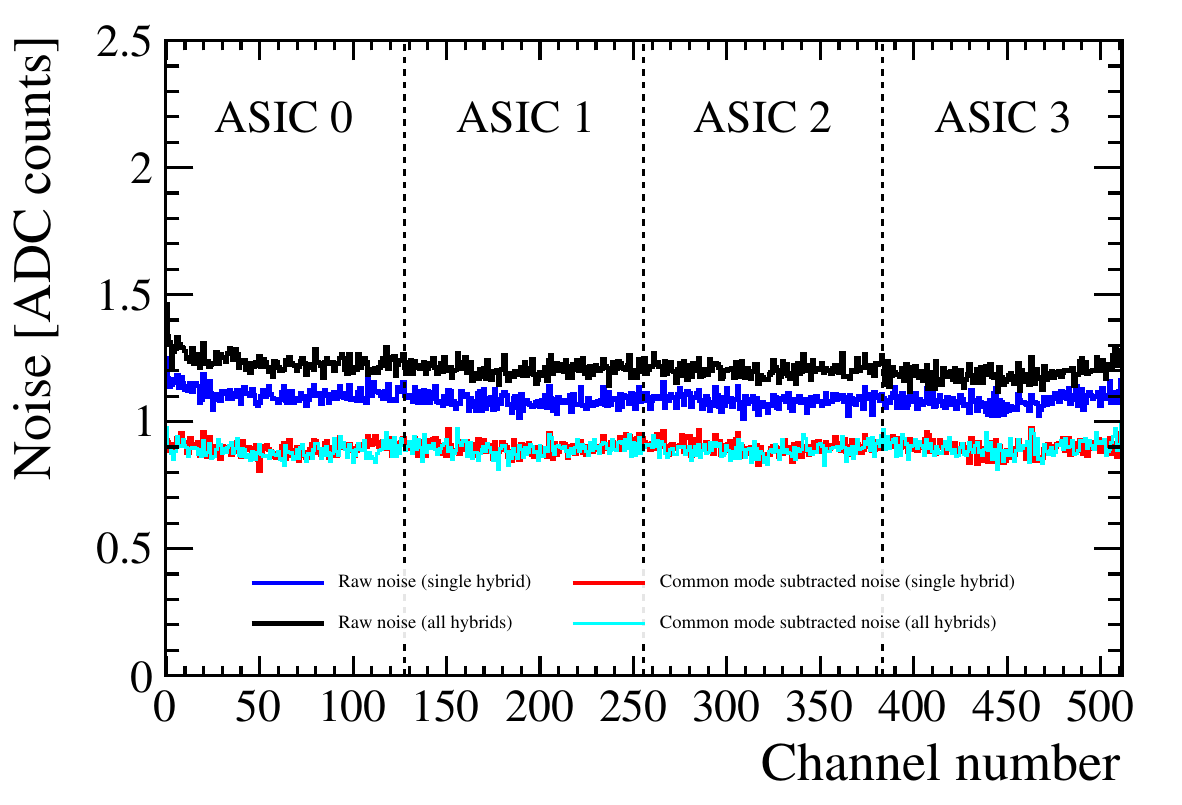}
  \caption{The raw and common mode subtracted noise measured in a
    single hybrid with only that hybrid powered and configured (blue
    and red curves, respectively) compared with the noise measured in
    the same hybrid when all hybrids in the stave were operating
    (black and cyan).}
  \label{fig:ut:SliceTestNoise}
\end{figure}

Data were collected with different numbers of hybrids powered and
configured to nominal settings, in order to study possible correlated
noise effects.  \Asic{s} on each hybrid were configured individually,
or all at the same time.  Finally, tests with all hybrids powered and
configured simultaneously were carried out.  An example result of the
noise before and after mean common mode suppression is shown in
figure~\ref{fig:ut:SliceTestNoise} for the innermost module on a
Medium \dataflex cable.  The raw and common mode subtracted noise were
around 1.1 and 0.9 ADC counts, respectively, when only that particular
single module was configured.  When all modules were operating, the
raw noise increased by around 10\% while the difference in the noise
after common mode subtraction was negligible.\looseness=-1

\subsection{Simulation and reconstruction software}
\label{sec:ut:software}

Unlike for the TT detector in the original \lhcb experiment, the
\Acr{ut} \Daq electronics does not cluster adjacent strip hits.  All
strip hits are saved with their ADC pulse heights.  In spite of
somewhat complicated \Acr{ut} raw bank organisation, driven by the
readout granularity and limited computing resources available in the
\Tellfourty boards, fast raw data processing is achieved via a smart
iterator with the knowledge of the raw data structure, pointing
directly to the raw bank in computer memory.  In the simplest
implementation, a strip hit in the raw bank is promoted directly to a
hit with a space location used by the \lhcb tracking software.  Since
only, 10\% of tracks passing a \Acr{ut} sensor are expected to light
up more than one strip, it is possible that this will be an optimal
decoding scheme in the first level software trigger.  A second version
of the iterator over the \Acr{ut} raw data has been developed, which
clusters on the fly while advancing over the raw bank.  It checks for
adjacent strip hits, by looking up the next element in the raw data,
and returns the highest pulse-height strip, while suppressing the
others. Detailed timing studies will be needed, once exact performance
of the \Acr{ut} \Asic{s} is known under running conditions, to
determine if such algorithm can run in the first level or must be
deferred to the second level of the software trigger.

\begin{figure}[t]
  \centering
  \includegraphics[width=0.49\linewidth]{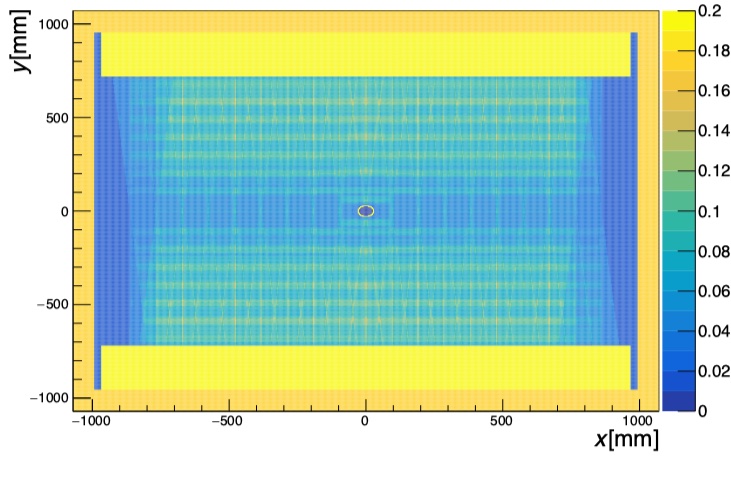}
  \includegraphics[width=0.47\linewidth]{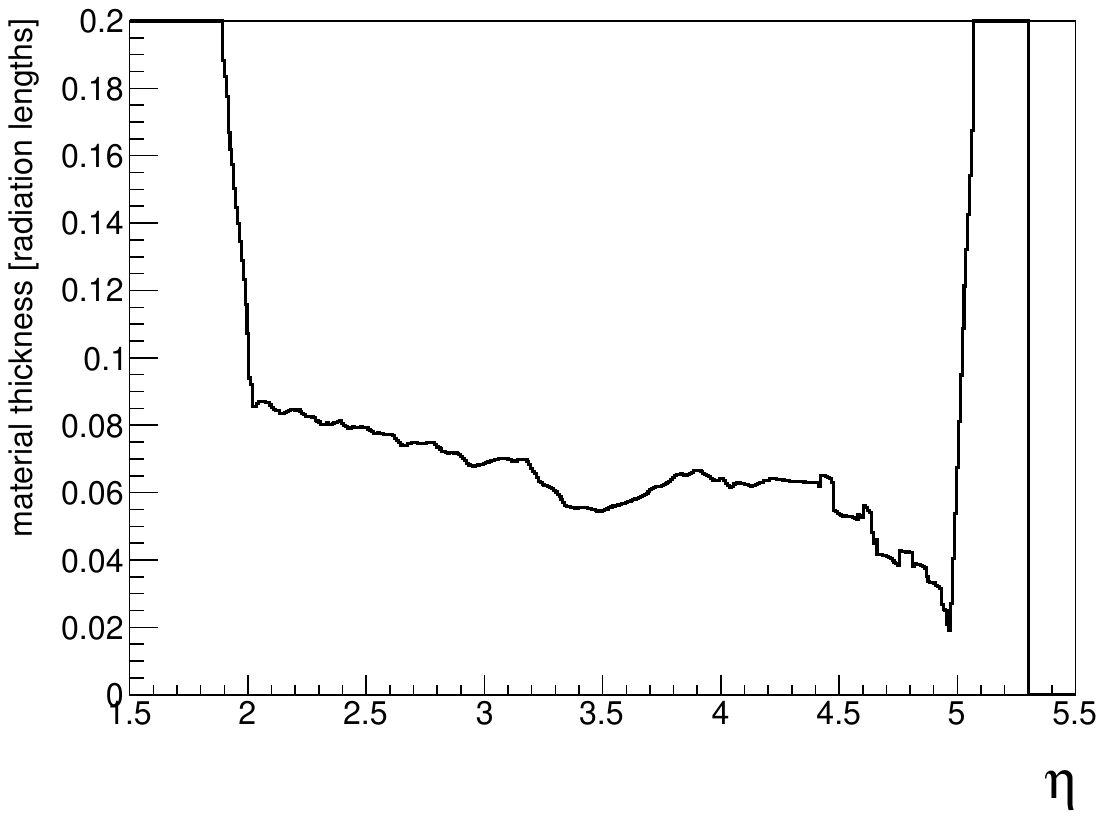}
  \caption{Material scan through the \Acr{ut} detector, with thickness
    given as a fraction of a radiation length.  Left: thickness map in
    $xy$ plane for normal track incidence.  Right: thickness as a
    function of pseudorapidity ($\eta$) as seen from the interaction
    point.  }
  \label{fig:ut:radlen}
\end{figure}

A \geant-level simulation software was developed for the \Acr{ut}
system.  The detector representation includes technical details of the
final design.  The hierarchy of geometrical volumes has been
structured to represent segmentation of the mechanical support and of
the final assembly with detector software alignment in mind.  For
example, layers were split into left and right half-layers to follow
the actual retractable C-frame design.  A map of the material in the
acceptance was obtained from simulation.  The \Acr{ut} material
thickness, expressed as a fraction of radiation length, is shown in
figure~\ref{fig:ut:radlen}.  Compared to its predecessor tracker (the
TT detector), the \Acr{ut} material in the region closer to the beam
pipe ($4.4<\eta<5$, where the track density is maximal) has been
reduced from about 15\% to 4\% of a radiation length.

\begin{figure}[t]
  \centering
  \includegraphics[width=0.9\linewidth]{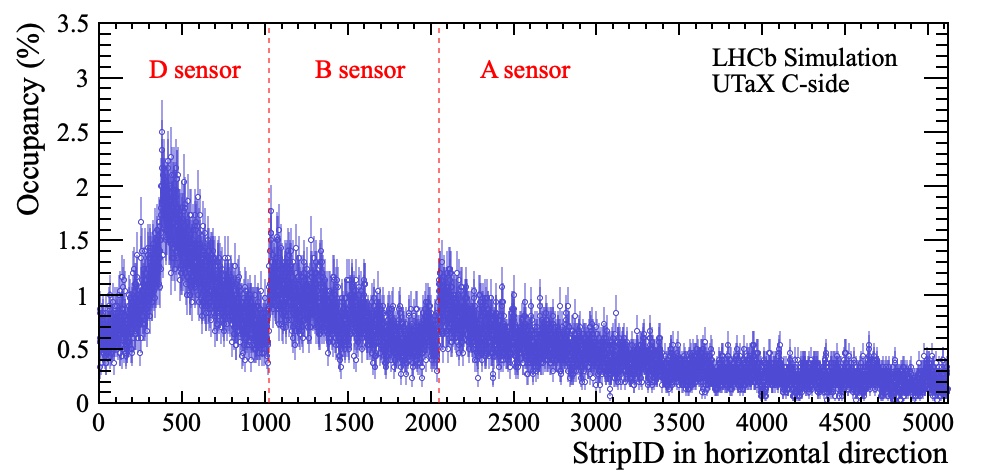}
  \caption{Expected \Acr{ut} strip occupancies for minimum bias events
    in the sensors near the detector midline ($y=0$).  }
  \label{fig:ut:occupancies}
\end{figure}

The expected strip occupancies in the \Acr{ut} near the detector $y=0$
mid line are illustrated in figure~\ref{fig:ut:occupancies}, where the
average fraction of minimum bias events leaving a hit in the strip is
shown as a function of strip number.  The plot focuses on the \cside
of the first \Acr{ut} layer (UTaX).  The occupancies are kept below
3\% even in the busiest region and are below 1\% in most of the
detector.

%
\section{Scintillating fibre tracker}
\label{sec:scifi}
\subsection{Overview}
\label{section:scifi:introduction}

The tracker, located downstream of the \lhcb dipole magnet, is
responsible for charged particle tracking and momentum measurement.  A
momentum resolution and track efficiency for \bquark- and
\cquark-hadrons comparable to the ones obtained in \runone and \runtwo
must be achieved, while working in an environment with higher particle
density.  To cover the nominal \lhcb acceptance it must have an area
of about $6\m \times 5\m$ in the $xy$ plane.

\subsubsection{Detector requirements}

The design of the tracker must take into account the following
requirements:
\begin{itemize}
\item Performance: the tracker should provide a single hit position
  resolution of better than \SI{100}{\mum} in the magnet bending plane
  and a single hit reconstruction efficiency better than 99\%.
\item Rigidity: the mechanical stability of the detector must
  guarantee that the positions of the detector elements are stable
  within a precision of 50 (300) \mum in $x$ ($z$); the detector
  elements should also be straight along their length within
  \SI{50}{\mum}.
\item Material budget: to limit further multiple scattering and
  secondary particle production, each of the 12 layers should not
  introduce more than 1\% of a radiation length.
\item Radiation hardness: the tracker should operate at the desired
  performance over the lifetime of the experiment, where 50\invfb of
  integrated luminosity is expected to be collected.
\item Granularity: the tracker must have an occupancy low enough so
  that the hit efficiency is not impacted with an instantaneous
  luminosity of
  $2 \times 10^{33}
  \cm^{-2}\sec^{-1}$~\cite{LHCb-DP-2012-001,LHCb-TDR-015}.
\end{itemize}
A tracker design based on \Acr{scifi} technology with \Acr{sipm}
readout was chosen to fulfil these requirements.

\subsubsection{Detector layout}

\begin{figure}[t]
  \centering
  \includegraphics[width=0.95\linewidth]{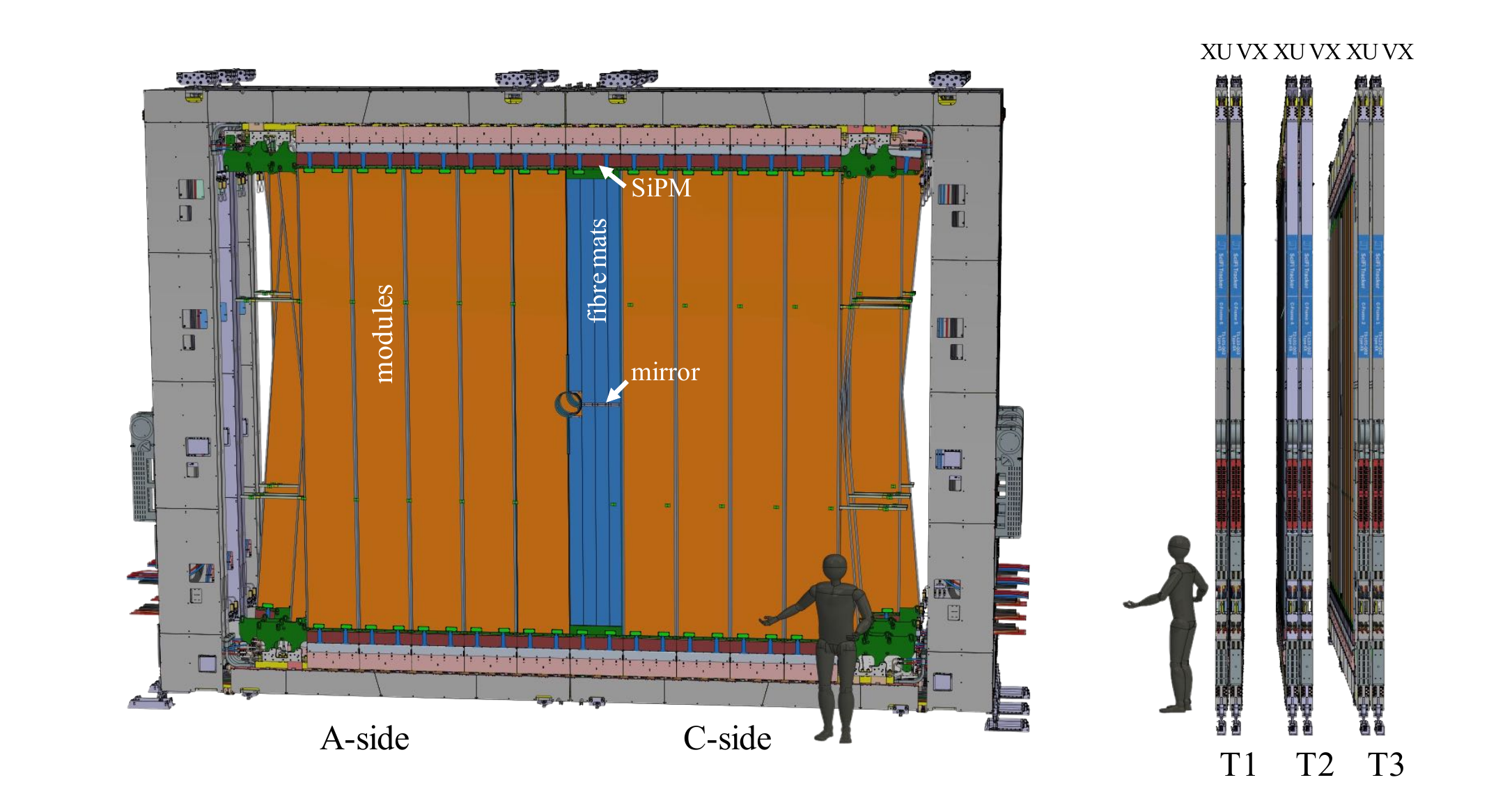}
  \caption{Front and side views of the 3D model of the \Acr{sft}
    detector.}
  \label{fig:scifistations}
\end{figure}

The \Acr{sft} acceptance ranges from approximately \SI{20}{\mm} from
the edge of the beam pipe to distances of $\pm$3186\mm and
$\pm$2425\mm in the horizontal and vertical directions, respectively,
with a single detector technology based on \SI{250}{\mum} diameter
plastic scintillating fibres arranged in multilayered fibre mats.  In
total there are 12 detection planes arranged in 3 stations (T1, T2,
T3) with 4 layers each in an $X-U-V-X$ configuration, as shown in
figure~\ref{fig:scifistations}. The $X$ layers have their fibres
oriented vertically and are used for determining the deflection of the
charged particle tracks caused by the magnetic
field~\cite{LHCb-TDR-015}.  The inner two stereo layers, $U$ and $V$,
have their fibres rotated by $\pm 5\degrees$ in the plane of the layer
for reconstructing the vertical position of the track hit.

Each station is constructed from four independently movable structures
referred here as C-Frames, with two C-Frames on each side of the beam
pipe.  The carriages of the C-Frames move along rails fixed to a
stainless steel bridge structure above the detector, supported by
stainless steel pillars.  To simplify production, each station is
built from identical \Acr{scifi} modules about $52\cm$ wide and
spanning the full height, except for a few modules near the beam pipe.
The T3 station is instrumented with six modules on each C-Frame. T1
and T2 stations have one less module on each side for instrumenting
the smaller acceptance at those locations due to the opening angle of
\lhcb.

\subsubsection{Detector technology}

The detector modules are constructed as a honeycomb and carbon-fibre
sandwich containing eight $\sim 2.4\m$ long and $\sim 13\cm$ wide
\Acr{scifi} mats made from six staggered layers of fibres.  A thin
mirror is glued to the fibre end to reflect additional light back to
the readout side~\cite{Joram:2134475}.  In the experiment, these
mirrors are located near the $y=0$ plane.  From there, four fibre mats
point upward and four downward, spanning a total height of almost
$5\m$.  Near $x=0$, a few special modules are used in order to take
into account the presence of the beam pipe.  Those modules have one
mat shortened to accommodate the beam pipe radius, as seen in the
centre cutaway module in figure~\ref{fig:scifistations}. A more
detailed description of the modules is found in
section~\ref{section:scifi:modules}.

The optical signal from the scintillating fibres are detected by
128-channel arrays of \Acr[p]{sipm} with a channel pitch of
\SI{250}{\mum}.  The \Acr[p]{sipm} are discussed in
section~\ref{section:scifi:sipms}.  At the readout end of each module,
16 \Acr{sipm} arrays are bonded to a 3D printed titanium alloy cooling
bar, which is aligned to the four fibre mats and housed in a cold-box,
described in section~\ref{section:scifi:services}.

\subsubsection{Detector irradiation}
\label{section:scifi:introduction:irradiation}

\begin{figure}[h]
  \centering
  \includegraphics[width=0.7\linewidth]{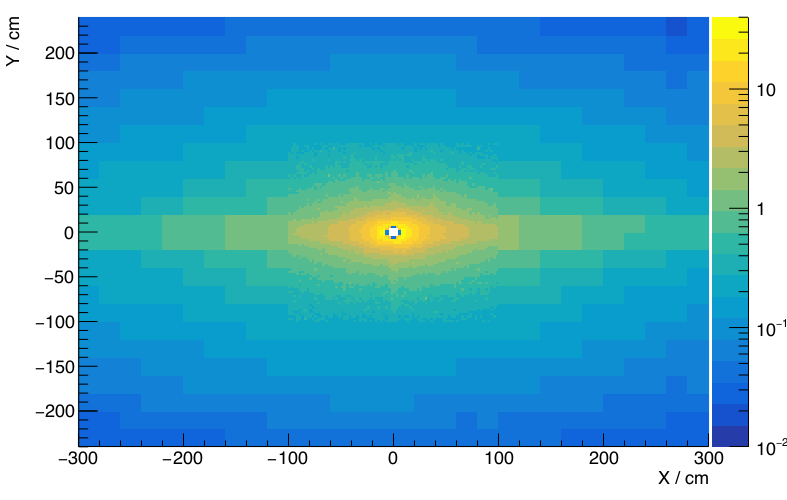}
  \caption{Map of the total expected ionising dose in \kGy for an
    integrated luminosity of 50\invfb at the T1 station of the
    \Acr{sft} from \fluka simulations of the \lhcb detector.}
  \label{fig:scifi:ionisingdose}
\end{figure}

Initially, the \Acr{sft} has a mean light yield about
18--20~photoelectrons for particles passing perpendicularly through
the detector plane near the mirrors.\footnote{The distribution is
  Poisson-like in its shape, due to variations in the total path of
  the charged particle through the scintillating fibre matrix,
  fluorescence processes, and saturation from large ionisation energy
  deposits.}  Irradiation will reduce the light yield. \fluka
simulations have shown that up to \SI{35}{\kGy} of ionising radiation
dose can be expected in the fibres in this region by the end of the
lifetime of the detector. The total received dose drops off sharply to
about \SI{50}{\Gy} at the readout end of the fibres, as seen in
figure~\ref{fig:scifi:ionisingdose}. Several irradiation campaigns
were performed to study the effect on light output and a signal loss
of about 40\% is expected for hits in the most irradiated regions
towards the end of the tracker's lifetime. This will still exceed
10~photoelectrons allowing for a single hit detection efficiency of
99\% in the sensitive regions, assuming low selection thresholds
(4~photoelectrons or greater), and a single hit position resolution
measured to be \SI{70}{\mum} which is needed for efficient tracking of
charged particles behind the magnet of \lhcb.

At the position of the \Acr[p]{sipm}, the expected total ionising dose
over the lifetime of the experiment is relatively low, \SI{50}{\Gy},
such that the main concern will be the \Acr{niel} which damages the
silicon crystal lattice.  A wall of borated polyethylene
(\weightperweight{5}) between the \richtwo and the \Ecal has been
installed to reduce the back-scatter of low energy neutrons from the
calorimeters by a factor of more than two in the region of the
\Acr[p]{sipm} (see also section~\ref{subsubsec:neutronshielding}).
The inner region of shielding installed $\pm$\SI{1}{\m} around the
beam pipe is \SI{30}{\cm} thick with the outer region having a
thickness of \SI{10}{\cm}.  \fluka simulations of the \lhcb Upgrade
show that a collected fluence of up to $6\times 10^{11}1\mev\neqcmcm$
is expected.  The \Acr{niel} damage will increase the rate of pixel
avalanches per channel initiated by thermal electrons, occurring
without incident light, and is commonly referred to as dark noise or
\fdcr.  The \Acr[p]{sipm} are to be cooled below \SI{-40}{\degc} as
the damage increases in order to suppress the rate of accidental
clusters and maintain the required detector performance.

A significant portion of the detector infrastructure is dedicated to
managing the radiation effects on the fibres and the \Acr[p]{sipm}, as
well as handling the large data rates generated by the increased
instantaneous luminosity of the \lhcb upgrade where, after zero
suppression, up to \SI{20}{\tbps} of data are sent to the \Daq from
the \Acr{sft} \Fend electronics.

\subsection{Scintillating fibres}
\label{section:scifi:fibres}

Blue-green emitting double-clad plastic scintillating fibres with a
\SI{250}{\mum} diameter were chosen for the \lhcb
\Scifi.\footnote{Fibres SCSF-78MJ by \Trmk{Kuraray}.}  The production
of the over \SI{11000}{\km} of fibre needed for the tracker was
delivered on time and of a very high and constant quality, which led
to negligible rejection rates.

The manufacturer states an intrinsic light yield of 8000 photons per
\MeV of ionisation energy deposited.  The decay time constant was
measured to be \SI{2.4}{\ns}, slightly faster than the \SI{2.8}{\ns}
declared by the manufacturer~\cite{scifi:kuraray, Borshchev_2017}.
The double-clad fibre achieves a minimum trapping efficiency of 5.3\%
per hemisphere through total internal refection with the two cladding
layers. The fibres have a mean nominal attenuation length of
approximately \SI{3.5}{\m}.  However, the transparency of the fibres
will degrade due to the received ionising dose. This effect has a
strong wavelength dependence and may result in losses that peak in the
UV/blue regions. The emission spectrum of the fibre has a maximum at
\SI{450}{\nm}. It has also been observed that the attenuation length
degrades by 1--2\% per year due to radical production resulting from
the interaction of oxygen with the polymer~\cite{Shinji:2658070,
  Shinji:2684438, Cavalcante_2018}.

\subsubsection{Fibre quality control} 

The fibre supply was delivered in 48 individual shipments (batches) in
weekly (\SI{100}{\km}) or bi-weekly (\SI{300}{\km}) intervals on
\SI{12.5}{\km} spools. After the arrival of each shipment, fibre
samples for quality control were prepared and the individual fibre
spools were processed on a fibre scanner developed in-house.  A
detailed description of the quality control process can be found in
ref.~\cite{Cavalcante_2018}.

The attenuation length of the fibres was measured for each batch both
by the manufacturer and upon reception by \lhcb, using light from a
photodiode and fitting the yield in a range between $1$ and $3\m$ from
the light source to avoid the contribution of a second exponential
with shorter attenuation~length.

The attenuation length was monitored over the full production time of
almost two years.  Averages over all spools (typically 24) of one
shipment were calculated.  Apart from an initial increase of the
attenuation length during the preseries and early main-series
production, the attenuation length was very stable throughout the full
production, with a mean of $3.5\pm 0.2\m$, well above the acceptance
limit~of~$3\m$.

The light yield was measured in a dedicated quality control setup
(described in ref.~\cite{LHCB-PUB-2015-012}) at $2.4\m$ from the
position of a $^{90}$Sr radioactive source (with magnetically selected
1.1\MeV electrons).  An improvement of the attenuation length observed
in the preseries production was also seen in the light yield.
Otherwise, the single-fibre light yield, extrapolated to zero distance
from the photon detector, was very stable around a mean of about
$6.8 \pm 0.5$~photoelectrons, and was always above the acceptance
limit of 5~photoelectrons.
\subsection{Fibre mats and modules}
\label{section:scifi:modules}

\subsubsection{Fibre mats}

As found in ref.~\cite{scifi-engineering-review} fibre mats are
produced by winding six layers of fibres on a threaded winding-wheel
with a diameter of approximately \SI{82}{\cm}.  The winding puts the
fibres in a regular hexagonal matrix with a horizontal fibre pitch of
\SI{275}{\mum}.  Titanium-dioxide loaded epoxy (\weightperweight{20})
is used to bond the fibres to each other and to provide some shielding
for cross-talk and for the diffusion of light escaping the cladding.
Thin black polyimide foils are bonded to the fibre mat on both sides
to provide mechanical stability as well as some amount of light
shielding.

\begin{figure}[h]
  \includegraphics[width=0.49\textwidth]{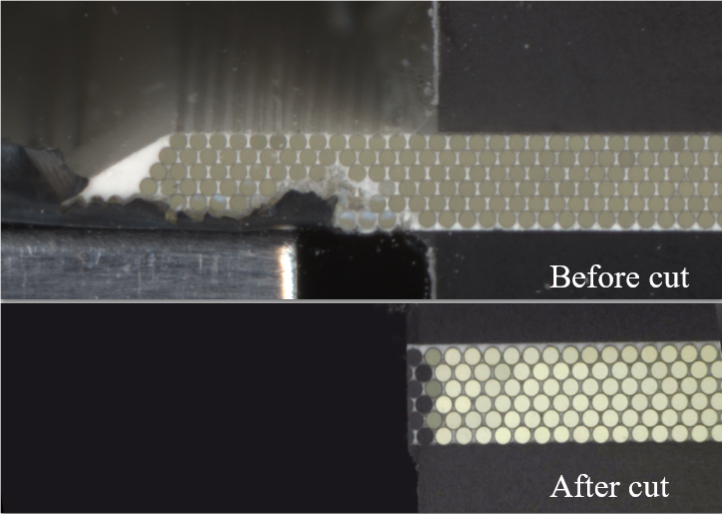}
  \hfill
  \includegraphics[width=0.465\textwidth]{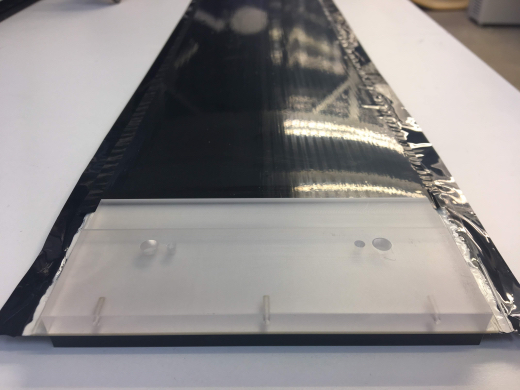}
  \caption{Left: view of a fibre mat with a microscope before and
    after the side fibres are cut away.  Right: a photo of a fibre mat
    with the polycarbonate end-pieces and \Acr{sipm} alignment holes.
    Reproduced from~\cite{scifi-engineering-review}. CC BY 4.0.}
  \label{fig:fibremats}
\end{figure}

Additionally, the fibre mats are cut precisely to the desired length
($2424\mm$) and width ($130.6\mm$) after having a polycarbonate
end-piece added with precision holes for aligning the \Acr[p]{sipm} to
the fibre mat at the readout end, as seen in
figure~\ref{fig:fibremats}.  The mirror end has two $2\mm$-thick
polycarbonate end-pieces for alignment and optical milling.  The foil
mirror\footnote{Enhanced Specular Reflector from \Trmk{3M} corp.} is
bonded with epoxy to them and the fibre mat.

\subsubsection{Modules and C-Frames}

\begin{figure}[t]
  \centering
  \includegraphics[width=\textwidth]{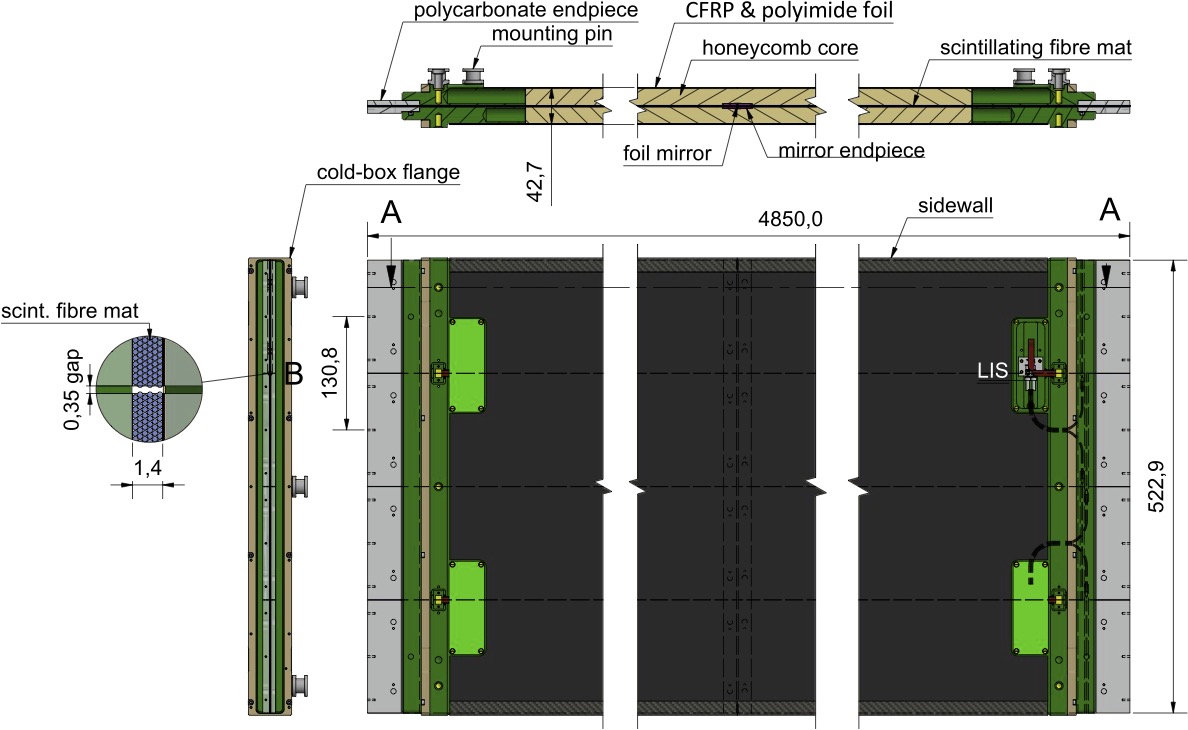}
  \caption{A cross-section and projections with dimensions (in \mm) of
    a scintillating fibre module and its components.  The outlines of
    the fibre mats and light injection system (LIS) are shown with
    dashed lines.  The nominal gap between fibre mats is shown in a
    zoomed inset on the left. Reproduced from~\cite{scifi-engineering-review}. CC BY 4.0.}
  \label{fig:SciFi:explodedview}
\end{figure}

Modules are built from eight fibre mats and two half-panels. Each
half-panel is made of \SI{19.7}{\mm} thick polyaramid honeycomb cores
laminated on the outer side with a single \SI{0.2}{\mm} carbon-fibre
reinforced polymer (CFRP) skin.  In order to minimise internal
stresses, which could deform the module, a symmetric design was
chosen, with the two half-panels sandwiching the fibre mats.  The
cross-section and design of a module is shown in
figure~\ref{fig:SciFi:explodedview}.  Each C-Frame contains two layers
of modules with 10\textemdash12 modules in total, depending on the
station.  A full description of the module production can be found in
ref.~\cite{scifi-engineering-review}.

Special modules contain a circular cut-out in the half-panels to
accommodate the radius of the beam pipe with a \SI{20}{\mm} safety
radial gap. The top (bottom) fibre mat at this location has been
shortened by 116~(116)\mm for $X$-modules, and 139 (93)\mm for stereo
modules to accommodate the cutout resulting in a rectangular hole in
the detector acceptance around the beam pipe. This results in three
types of modules: nominal, $X$-beam-pipe, and Stereo-beam-pipe.

The modules are fixed in position on the C-Frames by the top-centre
mounting pin. Five other mounting pins on the module constrain the
rotations. The modules in the $X$-layer are positioned with a pitch of
\SI{531}{\mm}, such that there is a gap of \SI{8}{\mm} between the
edge fibres of neighbouring modules. For the stereo modules the
horizontal pitch is \SI{532}{\mm}.  There is also a \SI{0.35}{\mm} gap
between neighbouring fibre mats within a module, as shown in the inset
of figure~\ref{fig:SciFi:explodedview}, though the inefficiency gap is
effectively slightly larger due to the likelihood of damaging the edge
fibre during the cutting of the fibre mats, as shown in
figure~\ref{fig:fibremats} (left bottom).  A \SI{2}{\mm} gap between
top and bottom fibre mats is also present.  The total geometric
inefficiency of one layer is approximately 1.7\% (within the nominal
acceptance).

\subsubsection{Material budget}

A module has a minimum estimated material thickness of 1.03\% of \Xrad
for perpendicular tracks.  The individual contributions are shown in
table~\ref{tab:scifi:X0}.  The carbon-fibre U-shaped foils that
enclose the sidewalls add an additional 0.145\% of \Xrad where they
overlap.  The region of the mirror end-piece of the fibre mat adds
\SI{4}{\mm} of polycarbonate, corresponding to an additional 1.15\% of
\Xrad.  In the acceptance, a particle passing through the 12 \Acr{sft}
layers traverses a minimum of 12.4\% of an \Xrad, or a bit more
depending on angle and exact location.

\begin{table}[t]
  \centering
  \caption{Material budget contributions from the scintillating fibre
    module components.  Densities and radiation lengths are taken from
    ref.~\cite{Zyla:2020zbs}.}
  \label{tab:scifi:X0}
  \begin{tabular}{lrrrrr} \\
    \hline
    Material             & Thickness              & Density  & Layers      & \Xrad & Fraction of \Xrad \\
                         & (\mum)                 & (kg/m$^3$) & per module  &  (\cm) & (\%)  \\
    \hline
    \emph{widespread}  &                         &         &        &            &            \\
    Fibre mat            & 1350                    & 1180    & 1      & 33.2       & 0.407      \\
    Honeycomb core       & 19700                   & 32      & 2      & 1300       & 0.303      \\
    CFRP skin            & 200                     & 1540    & 2      & 27.6       & 0.145      \\
    Glue                 & 260                     & 1160    & 2      & 36.1       & 0.144      \\
    Polyimide foil       & 25                      & 1410    & 4      & 35         & 0.029     \\
    \hline
    \emph{local}       &                         &         &        &            &            \\
    Sidewalls            & 200                     & 1540    & 2      & 27.6       & 0.145     \\
    Mirror polycarbonate & 2000                    & 1200    & 2      & 34.6       & 1.15      \\
    \hline
  \end{tabular}
\end{table}

\begin{figure}[t]
  \centering
  \includegraphics[trim=0cm 0.0cm 0cm 0cm,clip, width=0.99\textwidth]{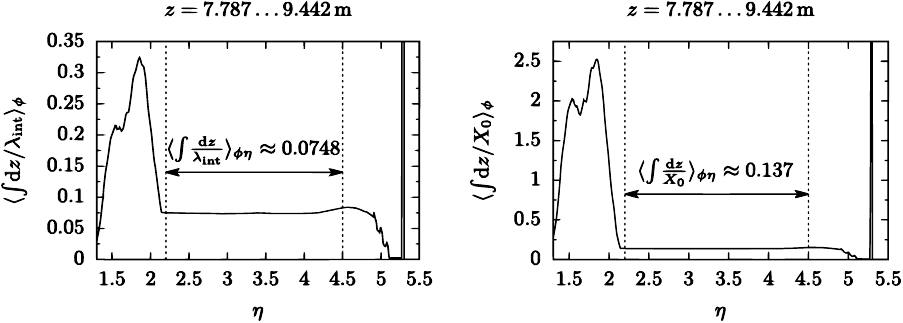}
  \caption{The traversed amount of material, averaged over azimuthal
    angle $\phi$, in units of (left) hadronic interaction length and
    (right) radiation length as a function of $\eta$ for a sample of
    simulated $\Bs\to\phi\phi$ decays.  The total average is also
    shown for $\eta$ between 2.2 and 4.5 (range limited to the
    acceptance of the \Acr{sft} shown with dashed lines).}
    \label{fig:scifi:radlengthsim}
\end{figure}

Results from particle tracking simulations using a sample of
$\Bs\to\phi\phi$ decays, shown in figure~\ref{fig:scifi:radlengthsim},
estimate the material budget for the three stations, averaged over
pseudorapidities in the range $2.2 < \eta < 4.5$, to be 7.48\% of a
hadronic interaction length, and 13.7\% of a radiation length.

\subsection{Silicon photomultiplier assemblies}
\label{section:scifi:sipms}

\begin{figure}[t]
  \centering
  \includegraphics[width=0.75\textwidth]{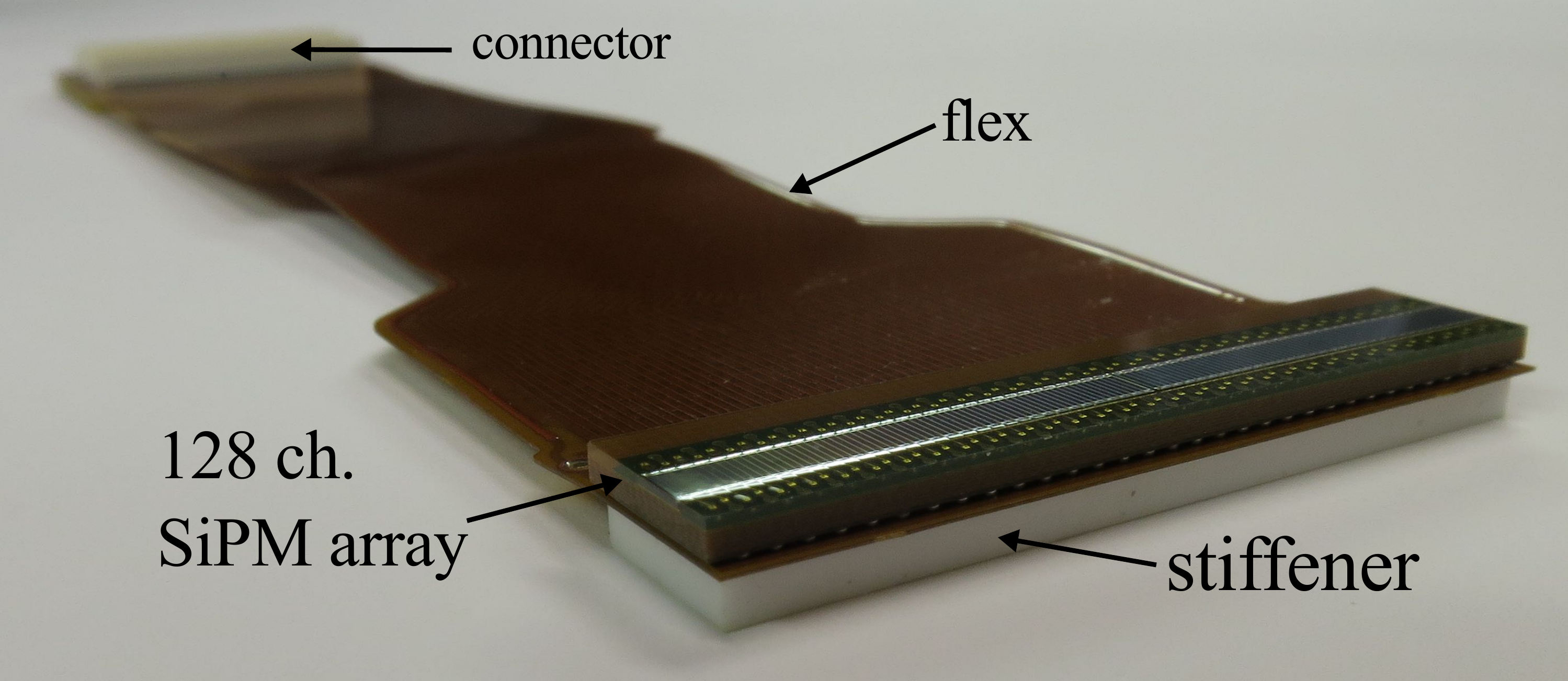}
   \caption{An H2017 \Acr{sipm} array bonded to a flex cable.
    The white stiffener is visible on the lower side of the flex cable.
    \label{fig:sipm_array} }
\end{figure}

The tracker will use a total of 524\,288 \Acr{sipm} channels
implemented as 4096 128-channel arrays to detect the light from the
scintillating fibres. The production version of the array,
manufactured by \Trmk{Hamamatsu}, is here referred to as
\emph{H2017}. Each channel comprises $4 \times 26$ pixels connected in
parallel with a pixel size of $57.5\mum\times 62.5\mum$ resulting in a
single channel with dimensions of $230\mum\times 1625\mum$. The
channel pitch is $250\mum$.  The array is composed of two 64-channel
dies wire-bonded to a PCB with a $220\mum$ gap between the dies.

A $100\mum$ thin transparent epoxy window protects the silicon and
bonding wires, such that the array can be pressed against the
scintillating fibres. The PCB is also instrumented with a \ptthousand
temperature sensor on the back to monitor the temperature. The 32.6\mm
wide PCB is bump bonded to a PCB flex cable which has been equipped
with a connector and a stiffener, glued to the back side, as seen in
figure~\ref{fig:sipm_array}.

\subsubsection{Bias and photon detection efficiency}

The pixel of the \Acr{sipm} is a reverse-biased photodiode operated in
Geiger mode. The breakdown voltage, \vbd, at which the avalanche
process can occur has been measured to have a mean value of
\SI{51.75}{\volt} and varies by up to \SI{\pm300}{\mvolt} across the
array~\cite{girard2018characterisation}.  The breakdown voltage for
every channel in every array has been measured and stored in a
database for use during operation.  The nominal operating voltage in
the \Acr{sft} is \SI{3.5}{\volt} above \vbd; details of the \Acr{sipm}
bias distribution system are reported in ref.~\cite{Joram:2147967}
The \vbd is linearly dependent on the temperature with a coefficient
of \SI{60\pm2}{\mvolt/\degk} requiring a stable cooling system, as
several parameters depend on the value of the bias over \vbd.  A
signal multiplication (gain) value of \SI{1.01\pm0.01E6}{\volt^{-1}}
was measured for these devices.

The photon detection efficiency (PDE) for the H2017 \Acr{sipm} was
measured with both current and pulse frequency methods described in
ref.~\cite{girard2018characterisation} and found to have a peak value
of \SI{43.5\pm 3.5}{\%} for nominal bias.

The avalanche is quenched as the bias voltage drops below \vbd due to
the increased current over a so-called quench resistor. The resistors
have a range of \SIrange{470}{570}{\kohm} in the H2017 devices with a
\SI{25}{\ohm} difference between odd and even channels. No simple
correlation was observed between the quench resistor value and \vbd.

\subsubsection{Cross-talk and correlated noise}

Infrared photons produced in the primary pixel avalanche can be
absorbed in neighbouring pixels causing additional pixels to fire,
either in time or with a short delay due to the depth and location of
the photon absorption.  This is referred to as direct and delayed
cross-talk. A reduction of the cross-talk over older \Acr{sipm}
designs has been achieved by the addition of trenches in the silicon
between pixels.  Trapped charge carriers in the silicon of a pixel
that has previously fired can result in a delayed pixel avalanche once
the bias in the pixel exceeds the breakdown voltage and the trapped
charge is freed.  This is called after-pulsing. Cross-talk and
after-pulsing are collectively referred to here as correlated noise.

\begin{figure}[h]
  \centering
  \includegraphics[trim=0.75cm 0.5cm 2.5cm 0.5cm,clip, width=0.42\textwidth]{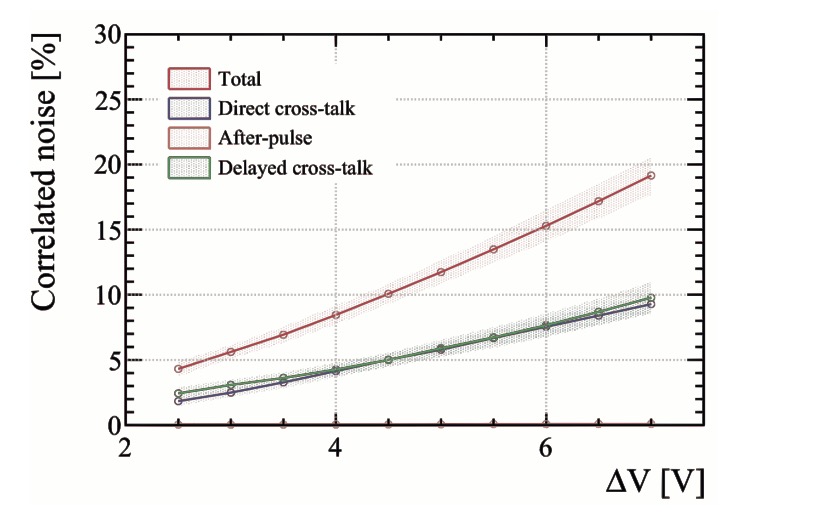}
  \hfill
  \includegraphics[width=0.46\textwidth]{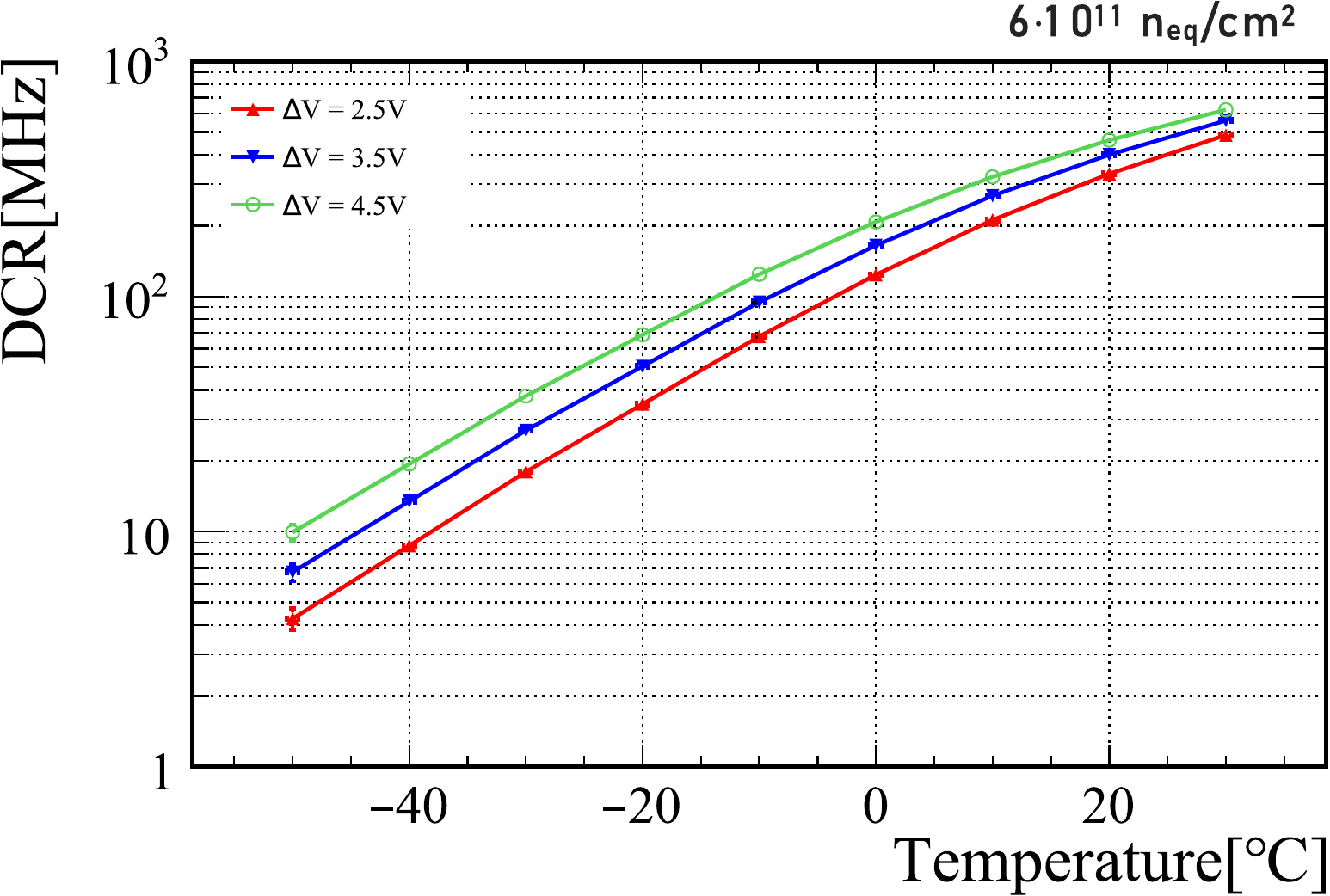}
  \caption{Left: correlated noise probabilities for an H2017 detector
    as a function of \dV.  Right:~dark-count rate for an irradiated
    \Acr{sipm} as a function of temperature for three
    overvoltages. Reprinted from~\cite{GRUBER2020162025}, Copyright
(2020), with permission from Elsevier.}
    \label{fig:scifi:sipm_results}
\end{figure}

The correlated noise probabilities were measured as a function of the
overvoltage~\cite{girard2018characterisation}.
The results are shown in figure~\ref{fig:scifi:sipm_results} (left).
Direct cross-talk has a probability of 3.3\% of occurring at nominal
overvoltage. Delayed cross-talk has a probability of 3.7\% and an
exponential decay time constant of \SI{17.7\pm0.4}{\ns} for the H2017
sensors.\footnote{The first batch that was delivered (about 10\% of
  the total), is slightly noisier, with direct and delayed cross-talk
  probabilities of 3.2\% and 5.6\%, respectively.}  After-pulsing is
negligible in these devices.

\subsubsection{Dark noise}

The photodetectors are expected to accumulate a total fluence of
$6\times 10^{11}1\mev\neqcmcm$ and \SI{50}{\Gy} of ionising radiation
over the lifetime of the experiment.  The ionising dose at this level
has been shown to not have significant impact on the \Acr{sipm}
performance.  The accumulated displacement damage, however, increases
the rate of single pixel avalanches caused by thermal excitation of
electrons, which increases the \fdcr by several orders of magnitude
over the lifetime of the detector.  A \fdcr of 14\MHz per channel is
expected towards the end of life of the detector, as seen in
figure~\ref{fig:scifi:sipm_results}~(right).\looseness=-1

Coupled together with cross-talk, single-pixel dark counts have some
significant probability to create signal amplitudes of two or more
pixel avalanches which appear similar to low amplitude signals caused
by particle tracks.  Channels with dark-noise amplitudes above a set
threshold will be mistaken for real signal clusters unless they are
rejected by other means, such as the clustering algorithm described in
section~\ref{section:scifi:clusterboard}.  In addition to screening
neutrons with the PE shielding, a further reduction of the \fdcr by a
factor of 100 is achieved by cooling the irradiated \Acr[p]{sipm} from
room temperature to \SI{-40}{\degc}, as it is described in
section~\ref{section:scifi:services}.

\subsection{Front-end electronics}
\label{section:scifi:electronics}

\begin{figure}[t]
  \centering
  \includegraphics[width=0.474\textwidth]{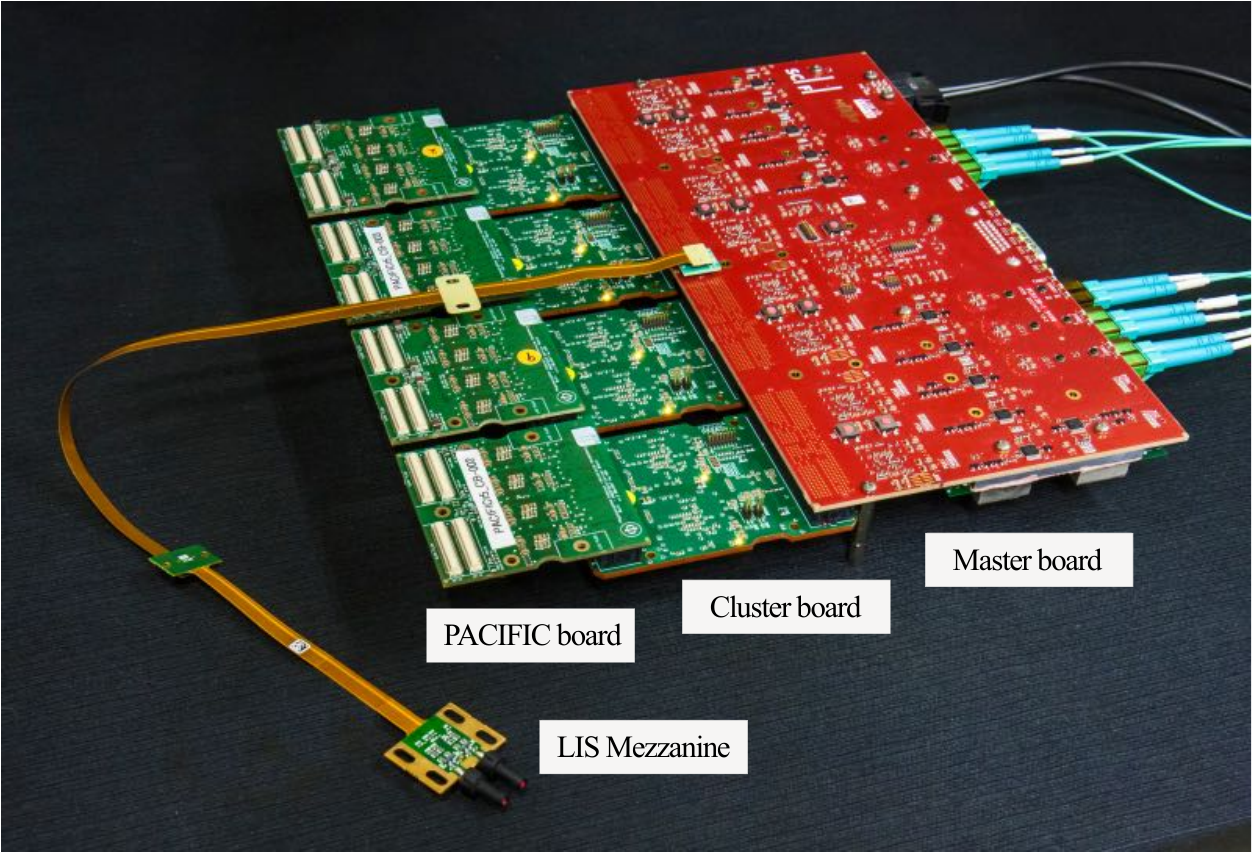}
  \hfill
  \includegraphics[width=0.38\textwidth]{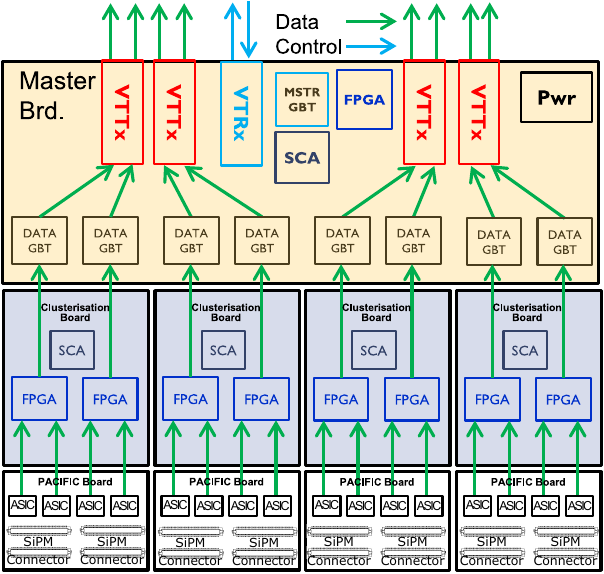}
  \caption{ Left: picture of assembled Master, Clusterisation and
    PACIFIC boards. Reprinted from~\cite{GRUBER2020162025},
Copyright (2020), with permission from Elsevier.  Right: corresponding schematics of signal data
    routing.  Reproduced from~\cite{scifi-engineering-review}. CC BY 4.0.}
  \label{fig:SciFi:elecArchi}
\end{figure}

The \Fend electronics consists of three types of boards: the PACIFIC
board, the Clusterisation board, and the Master board, shown in
figure~\ref{fig:SciFi:elecArchi} (left). The PACIFIC board performs
the digitisation of the analog signals received from the \Acr{sipm}.
The Clusterisation board performs zero-suppression and clustering of
the signals. The Master board serves multiple roles, such as the
distribution of control and clock signals, monitoring, as well as
distributing the low voltage and \Acr{sipm} bias to the other boards.\looseness=-1

Figure~\ref{fig:SciFi:elecArchi} (right) shows the layout and data
path of one set of \Fend electronics.  Each end of a \Acr{scifi}
module requires two sets of electronics, each one consisting of 4
PACIFIC boards, 4 Clusterisation boards and one Master board, to
digitise and transmit data from 16 \Acr[p]{sipm}.  The boards are
fixed to a thick aluminium chassis which is cooled by chilled
demineralised water.

\subsubsection{Data flow summary}

\Acr{sipm} avalanche pulses are processed and digitised in the PACIFIC
\Asic (described in more detail in
section~\ref{section:scifi:pacific}) with a system of three
hierarchical threshold comparators with four possible results, which
are encoded in a 2-bit output word.  Four channels are serialised
together and transferred from the PACIFIC to a \Trmk{Microsemi Igloo2}
\Fpga for clusterisation at a rate of \SI{320}{\mbps}.  The data
output after serialisation of the PACIFIC has a total rate of
\SI{10.24}{\gbps} delivered for each \Acr{sipm} array.  The \Fpga{s}
are programmed to perform a cluster search algorithm per \Acr{sipm}
array, as discussed in section~\ref{section:scifi:clusterboard}, and
to calculate the position for each found cluster.

Clustered data from each bunch crossing are labelled with an event
header and transferred to the data serialiser where the \Acr{sft}
\Fend electronics follows the architecture described in
section~\ref{sec:electronics_architecture}.  Eight \Gbtx chips
serialise the eight data streams on each Master board (\SI{4.8}{\gbps}
per link) and the \Vttx transmitters~\cite{Amaral2009} push the data
to the \Daq. In total, the \Acr{sft} transmits approximately
\SI{20}{\tbps} to the \Bend over 4096 data links.  A bi-directional
\Vtrx is used for the \Acr{ecs} and the \Tfc commands for each set of
electronics.  The \Gbtx, \Vttx, and \Vtrx are described further in
detail in section~\ref{sec:online}.

\subsubsection[PACIFIC ASIC]{PACIFIC \Acr[s]{asic}}
\label{section:scifi:pacific}

\begin{figure}[t]
  \centering
  \includegraphics[width=0.79\textwidth]{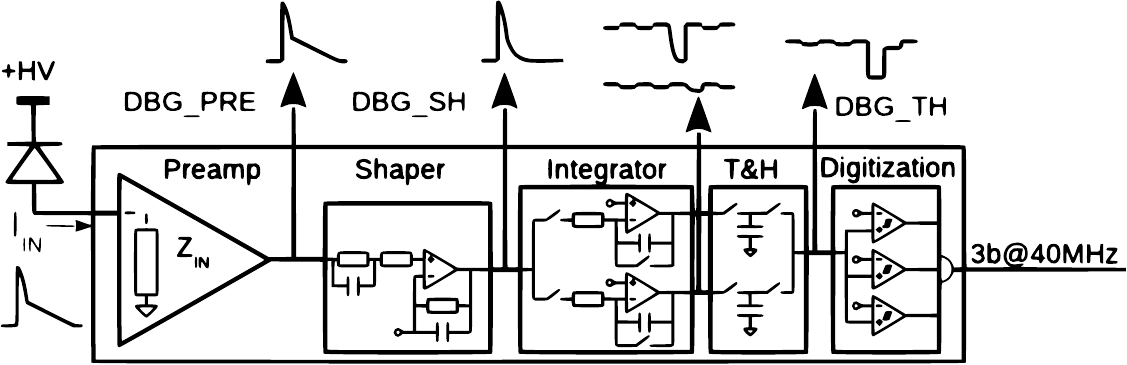}
  \caption{PACIFICr5q Channel block diagram. Reproduced with permission from~\cite{Leverington:2015rrn}.\label{fig:pacific_block}}
\end{figure}

The PACIFIC is a 64-channel \Asic with current mode input and digital
output developed to read out \Acr[p]{sipm}.  It is implemented in a
\SI{130}{\nm} CMOS process.\footnote{By \Trmk{TSMC}\,Taiwan
  Semiconductor Manufacturing Company.}  Each of the 64 channels
contains an analog processing, digitisation, slow control, and digital
output synchronised with the \SI{40}{\MHz} bunch clock of the \Lhc.
Several versions of the PACIFIC \Asic were developed. The version used
in the \Acr{sft} is the PACIFICr5q.  A detailed description of the
\Asic can be found in ref.~\cite{Comerma2020}.

A simplified representation of the analog processing is shown in
figure~\ref{fig:pacific_block}.  It consists of a preamplifier, a
shaper, and an integrator. The interleaved, double-gated integrator
operates at \SI{20}{\MHz} to avoid dead time as one integrator is in
reset while the other collects the signal.  The two integrator outputs
are merged by a track-and-hold to provide a continuous measurement.

The output voltage of the track-and-hold is digitised using three
comparators acting like a nonlinear flash ADC.  The result of the
three comparators is encoded into a two bit datum.  Four channels are
serialised together.  The design of the PACIFICr5q is complemented by
auxiliary blocks such as voltage and current references, charge
injection, control DACs, and power-on-reset circuitry.

An additional feature allows for the voltage present on the \Acr{sipm}
anode to be fine tuned using an internal configuration over a range
from 100 to 700\mvolt, to account for the variations in \vbd between
\Acr[p]{sipm}, sharing a common external bias. Four \Acr[p]{sipm}
($8~\times~64$ channels) share a single HV bias channel.
\Acr[p]{sipm} have been selected and grouped such that the variation
between the \Acr[p]{sipm} is within the tuneable range of the PACIFIC.

\subsubsection{PACIFIC analog circuit simulation}
\label{sec:scifi:PACIFIC analog circuit simulation}

\begin{figure}[b]
  \centering
  \includegraphics[width=0.48\linewidth]{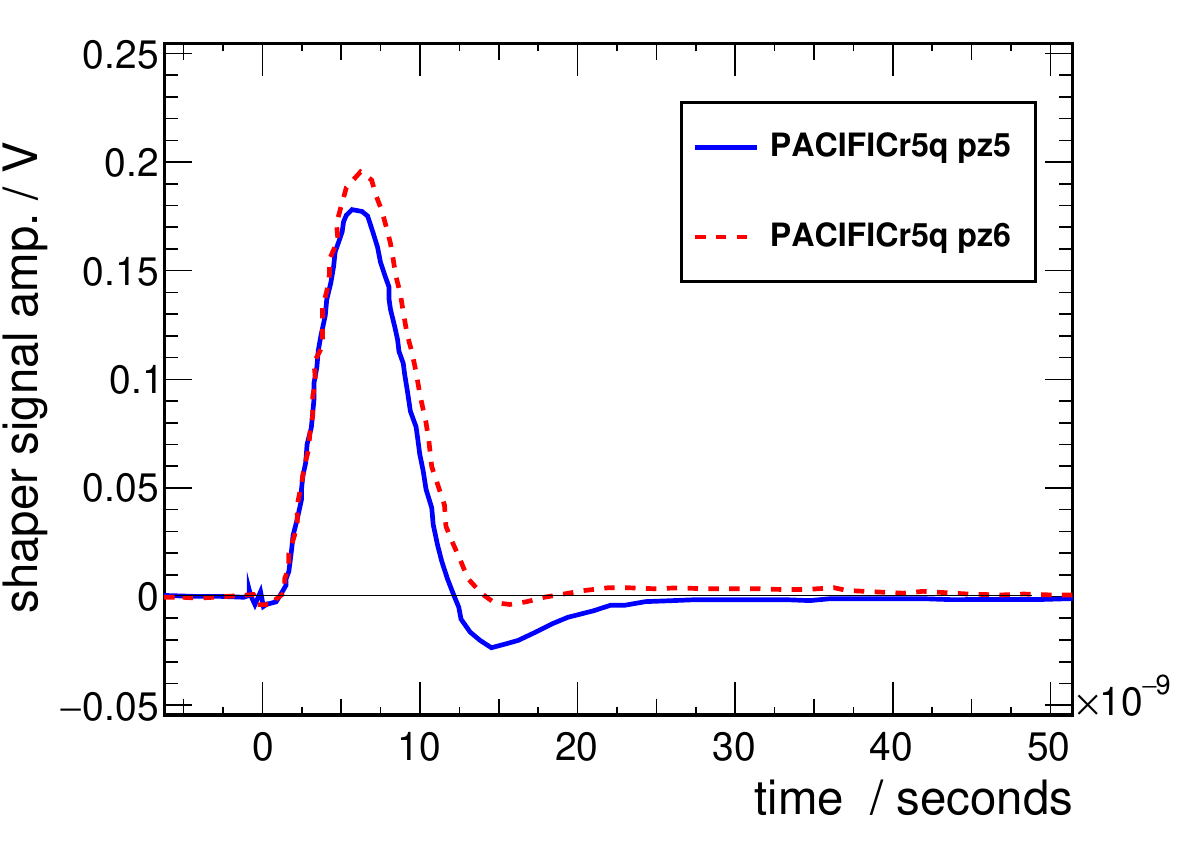}
  \hfill
  \includegraphics[width=0.475\linewidth]{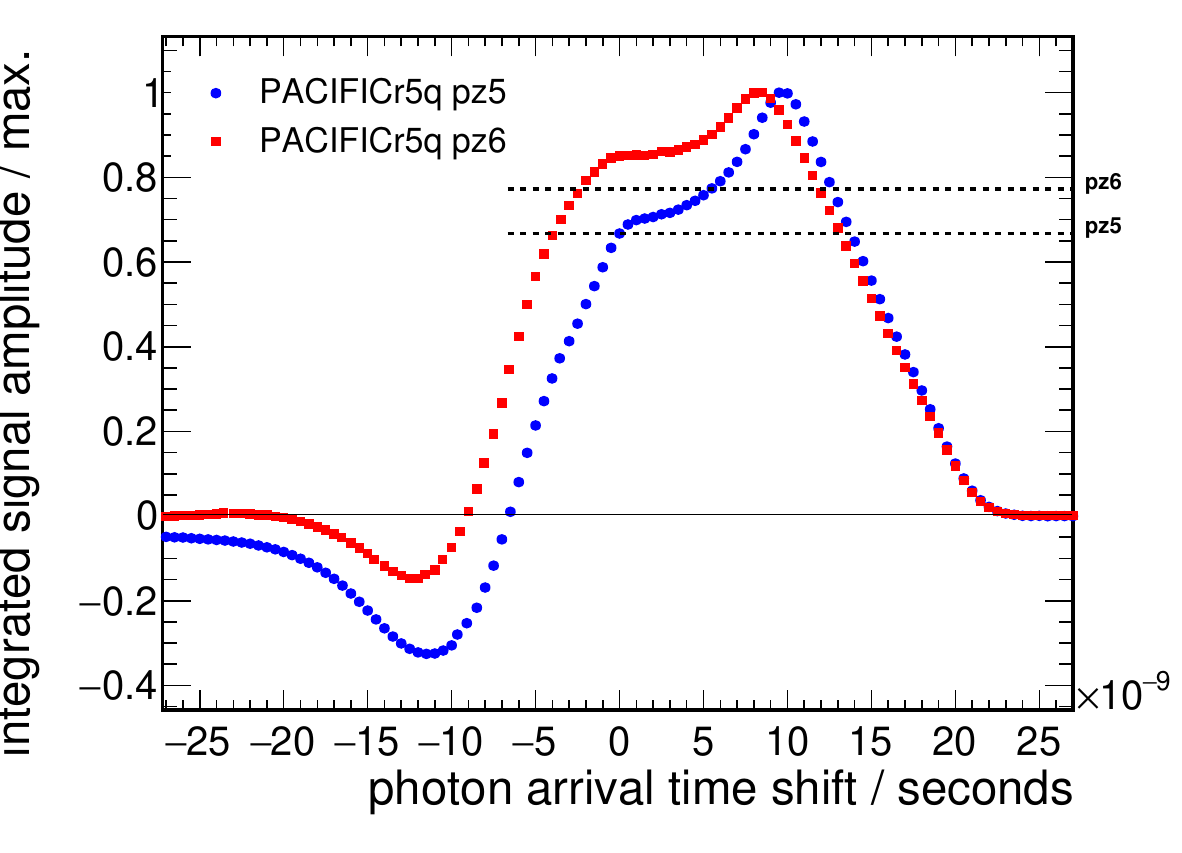}
  \caption{Left: the simulated shaper amplitude as a function of time
    for a single 10~photoelectron signal at \SI{4}{\volt} above
    breakdown. Right: the track-and-hold output values as a function
    of signal arrival time. The dashed lines indicate the nominal
    threshold value with respect to the maximum value for two settings
    (pz5 and pz6, see text).  \label{fig:scifi:pacificsim}}
\end{figure}

A 10~photoelectron \Acr{sipm} signal with an arrival time of
\SI{0}{\ns} at the input has been simulated at \SI{4}{\volt} above the
breakdown voltage.  The signal amplitude over time after the shaper
circuit is shown in figure~\ref{fig:scifi:pacificsim} (left) for two
separate \emph{pole-zero} shaper settings, pz5 and pz6.  The signal
has a positive component for \SI{10}{\ns} with a small undershoot
afterwards.  The pz5 setting is intended to create a larger undershoot
compared to pz6.  The integrated charge of the shaper signal that
falls in one bunch crossing period from 0 to \SI{25}{\ns} is measured
in one integrator.  The data points in
figure~\ref{fig:scifi:pacificsim} (right) are the track-and-hold
values (sampled output of the integrator) for separate signals for a
range of arrival times.  Data points with a negative arrival time have
been partially integrated in the previous bunch crossing by the other
integrator.  The threshold setting relative to the maximum value is
indicated by dashed lines in figure~\ref{fig:scifi:pacificsim}
(right).  This will ensure a relatively flat efficiency for separate
signals with the same number of photoelectrons occurring at different
times with respect to the clock phase of the integrator.  Dark-noise
signals will arrive randomly in time and be added on top of any signal
pulse and charge spillover across bunch crossing windows.

\subsubsection{Threshold calibration}

\begin{figure}[t]
  \centering
  \begin{subfigure}[c]{0.49\textwidth}
    \includegraphics[trim=0.2cm 0.2cm 1cm 0.5cm,clip, width=\linewidth]{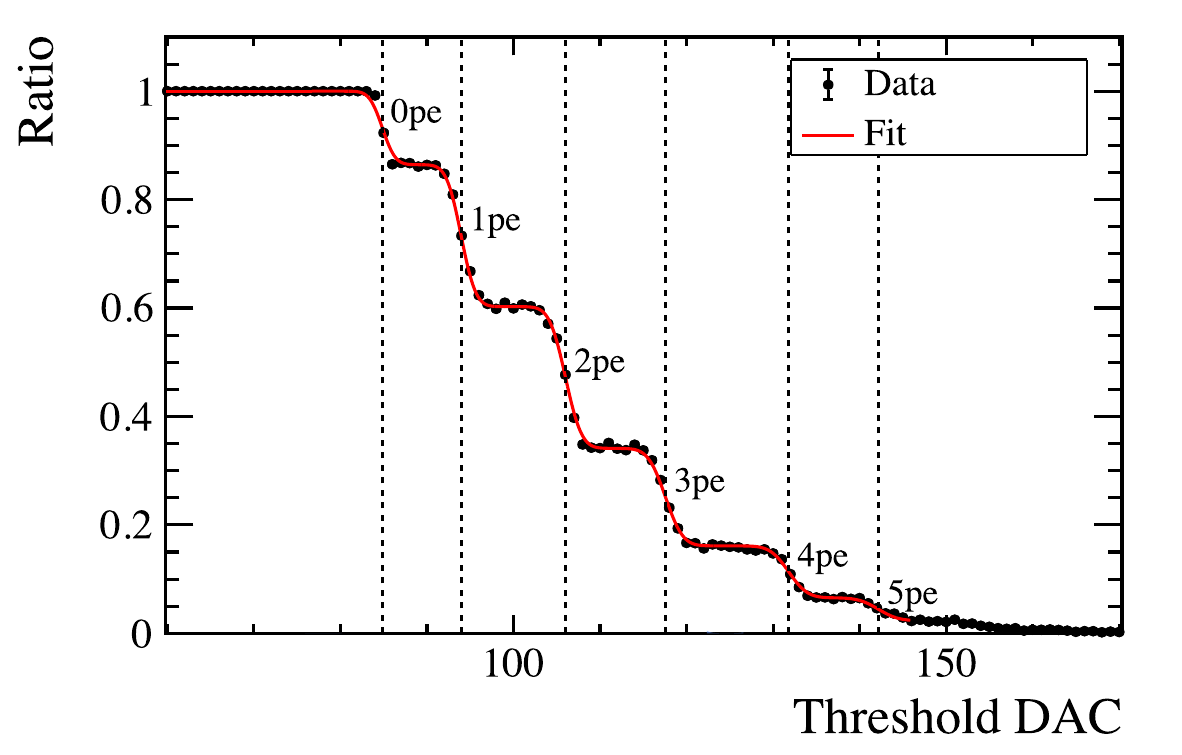}
  \end{subfigure}
  \begin{subfigure}[c]{0.47\textwidth}
    \includegraphics[width=\linewidth]{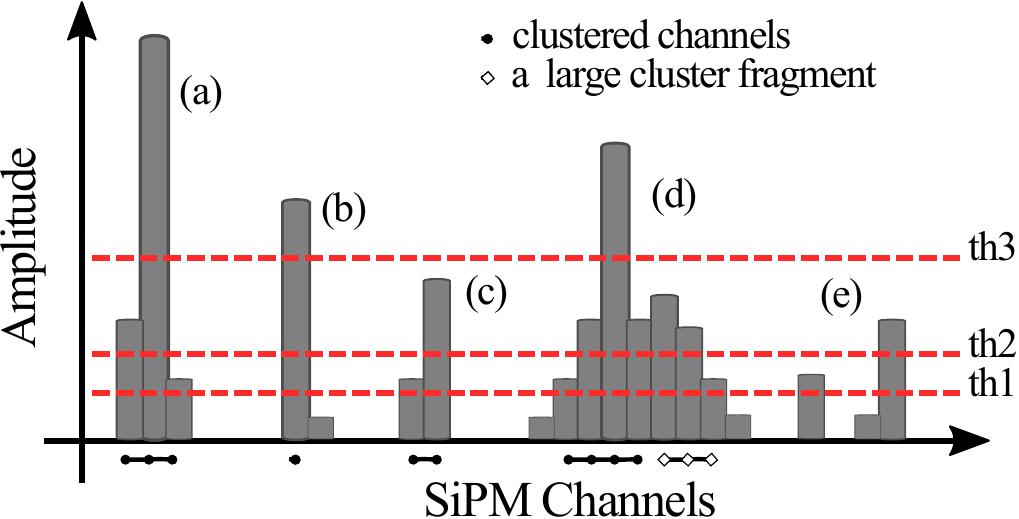}
  \end{subfigure}
  \caption{Left: an example threshold calibration for one comparator
    of one channel, showing the ratio (number of data above threshold
    to the total number of data) as a function of the threshold value.
    The red curve is the result of a fit.  The vertical dashed lines
    through the steps highlight the discrete photoelectron
    amplitudes. Reproduced with permission from~\cite{Witola2019}.
    Right: a diagram of the clustering algorithm. }
  \label{fig:scurvefit}
\end{figure}

The setting of the thresholds has a large impact on the single hit
efficiency and dark-count rate.  The calibration of the comparators
which are used to set the signal thresholds consists of two parts: (a)
determining the ratio of events over threshold for each DAC setting
for each of the three comparators of each channel
($3\times524$~thousand) and (b) an offline fit to the data to extract
the calibration constants.  The threshold DAC scan is performed under
pulsed illumination provided by the light injection system (LIS).  An
example of the results for one threshold is shown in
figure~\ref{fig:scurvefit} (left).  The fit to the data is based on an
analytical description of the Poisson-like \Acr{sipm} spectrum
described in ref.~\cite{VINOGRADOV2012247}, which includes
contributions from cross-talk.  It not only allows one to determine
the threshold DAC values corresponding to the discrete photoelectron
amplitudes, nominally 1.5, 2.5, and 4.5~photoelectrons, but also gives
access to other important parameters such as the mean light intensity
of the LIS. A more detailed discussion of the calibration procedure
can be found in ref.~\cite{Witola2019}.

\subsubsection[Clusterisation FPGA]{Clusterisation \Acr[s]{fpga}}
\label{section:scifi:clusterboard}

The PACIFIC 2-bit output data are processed by the Clusterisation
board.  The \Fpga clustering algorithm groups neighbouring channels
from the same \Acr{sipm} array into clusters, calculating an 8-bit
cluster position which reduces the data volume from \SI{10.24}{\gbps}
per \Acr{sipm} to less than \SI{4.8}{\gbps}.  The combination of
PACIFIC thresholds and channel selection in the \Fpga suppresses the
rate of accidental clusters from dark noise while maintaining a high
track hit efficiency.

The clustering algorithm is best explained by the example shown in
figure~\ref{fig:scurvefit} (right), where the three hierarchical
thresholds th1, th2 and th3 of the comparator are visualised.  A
cluster is formed when the sum of the weights of two or more
neighbouring channels exceeds the weight of th2, such as clusters (a)
and (c) in the figure. A weight of 1, 2, and 6 is attributed to
channels which exceed th1, th2, or th3 respectively.  Exceptionally, a
cluster is also formed when a single channel has an amplitude greater
than th3, such as cluster (b) in the same figure. The channels around
(e) will not form clusters.\looseness=-1

The 8-bit barycentre of a cluster is calculated from a weighted
average of all participating channels in the cluster, rounding to a
half channel position.  This digital half channel precision is enough
to provide a position reconstruction resolution better than
\SI{100}{\mum}.

A maximum of four channels can be included in a single cluster before
it is flagged as \emph{large} in a ninth bit and combined with
subsequent cluster fragments, such as (d) in
figure~\ref{fig:scurvefit} (right).  An unweighted geometric
barycentre is determined for the large clusters. The maximum number of
clusters per event that can be sent by a clusterisation \Fpga is
limited. In the high occupancy region, a flexible data format is used,
which sends a maximum of 16 clusters per \Acr{sipm} per across
additional bunch crossings when needed. In the rest of the detector a
fixed data format is used which has a limit of 10 clusters, sent
synchronously with each bunch crossing.

An \Fpga may fail due to the passage of ionising particles over time.
Irradiation tests performed at the CHARM facility at CERN indicated
that the chosen \Fpga{s} remain re-programmable up to \SI{23}{\Gy}
corresponding to approximately 10\invfb of integrated luminosity at
\lhcb.  In total, the boards received up to \SI{300}{\Gy} of ionising
dose and $3\times 10^{12}1\mev\neqcmcm$. The observed speed
degradation indicates that the loss will be lower than 5\% during the
detector lifetime and should not affect the operation of the
detector. Increased power consumption was not observed to any
significant level.  During the irradiation tests three \Fpga{s} (out
of 26) ceased to respond and had to be power-cycled to make it
function again.  No \Fpga was permanently damaged.

\subsubsection{Master boards}
\label{section:scifi:masterboard}

The cluster data from the \Fpga is serialised by the \Gbtx
\Asic{s}~\cite{Moreira2009} on the Master board and shipped over
optical fibres. The \Fend architecture is such that the cluster data
of each \Acr{sipm} are sent over a single fibre to the \Tellfourty
\Daq.  There are four \Vttx (8 links) on each Master board while the
boards are controlled by using the \Ecs through the \Vtrx
connector~\cite{EDMS-ID-1140665}.

A housekeeping \Fpga on the Master board provides slow and fast
control to the light-injection system, the voltage-level shifting
necessary to drive the monitor LEDs connected to the \Gbtx status
outputs, and the power-up reset signals to the data \Gbtx chips.

The Master boards are powered by an external \SI{+8}{\volt} power
supply. The \Feastmp are used to power the various other components on
the \Fend boards.  Each \Fend box
contains 2 Master, 8 PACIFIC and 8 Clusterisation boards.  It requires
approximately \SI{200}{\watt} of power and is cooled by circulated
chilled water.

\subsubsection{Light injection system}
\label{section:scifi:LIS}

For calibration and commissioning purposes a light injection system
(LIS) is implemented at the ends of the \Acr{scifi} modules, as it can
be seen in figures~\ref{fig:SciFi:explodedview}
and~\ref{fig:scifi:coldbox}.  The system consists of an external
GBLD-based light driver and a scratched optical fibre which injects
light into the scintillating fibres.  Fast and slow control signals,
as well as power, are transferred through a PCB flex cable connected
to the Master board.  The pulsed injection is controlled by two GBLD
laser driver chips on the LIS mezzanine via \I2c such that a
distribution from one to five photoelectrons is observed by each
\Acr{sipm} channel.  This range was chosen such that the three DAC
threshold values of the PACIFIC can be set to correspond to the
desired 1.5, 2.5, and 4.5~photoelectrons.

\subsection{Mechanics and alignment}
\label{section:scifi:mechanics}

The \Acr{sft} C-Frames are hung from the rails of the former \lhcb
Outer Tracker bridge, downstream of the magnet.  The C-Frames are made
from extruded aluminium profiles.  The majority of the service cables
and cooling pipes are distributed along these profiles.  Additional
aluminium covers are fixed to the outside of the profiles to provide
electromagnetic shielding and additional stiffness to the profiles.
The total mechanical structure when completely assembled with cables,
pipes, detector modules and electronics weighs slightly less than
\SI{1.5}{tonnes}.

A system of adjustment screws on the table, bridge, and carriages at
the top and bottom of each C-Frame allow for adjustment in three
spatial dimensions plus rotations with a precision of about
\SI{100}{\mum}.  A threaded drive mechanism which is fixed at the
bottom of the C-Frame allows for a controlled movement to the final
run-position when the C-Frames are closing around the beryllium beam
pipe.\looseness=-1

\subsubsection{Survey}

Each \Acr{scifi} module has four laser tracker survey points, two at
each end. There are four more on the beams of the C-Frames which are
used during installation and alignment of the detector.  The location
of these targets are known to an accuracy better than \SI{0.2}{\mm}
with respect to their nominal design positions. Additionally, the
curvature of all modules has been measured by photogrammetry with
reflective targets on the surface of all the modules after
installation on the C-Frames.  The data from each half of a module was
fit to a single plane, where the standard deviation from the plane is
of the order of \SI{0.1}{\mm}, approximately the resolution of the
measurement.  A single plane fit to the entire C-Frame layer has
maximum deviations of \SI{1.5}{\mm}, typically localised at the edges
or corners of the plane~\cite{Sainvitu2021}.\looseness=-1

\subsubsection{Real-time 3D position monitoring system}

An online 3D metrology system has been developed to permanently
monitor the evolution of the position of one detection plane in each
of the three \Acr{sft} stations while they are exposed to the magnetic
field and other slowly varying experimental conditions.  The system
relies on triangulation measurements by $24$ BCAM
cameras~\cite{Hashemi:684119} of passive reflective targets placed
on the detector surface. The camera positions (orientations) are known
at the level of a few \mum (\murad).  A sequential image acquisition
cycle of all cameras provides one determination of the detector
position approximately every minute.  The intrinsic accuracy of
relative movement measurements is better than 100\mum, and averaging
over cycles, for up to one hour, can lead to an improvement of the
precision to better than 20\mum.  These measurement are used to study
the geometric evolution of the \Acr{sft} detector, and to validate the
alignment constants obtained from the offline tracking alignment
algorithm.

\subsection{Sensor cooling}
\label{section:scifi:services}

\begin{figure}[h]
  \centering
  \includegraphics[trim=0cm 0cm 0cm 1cm,clip,width=0.99\textwidth]{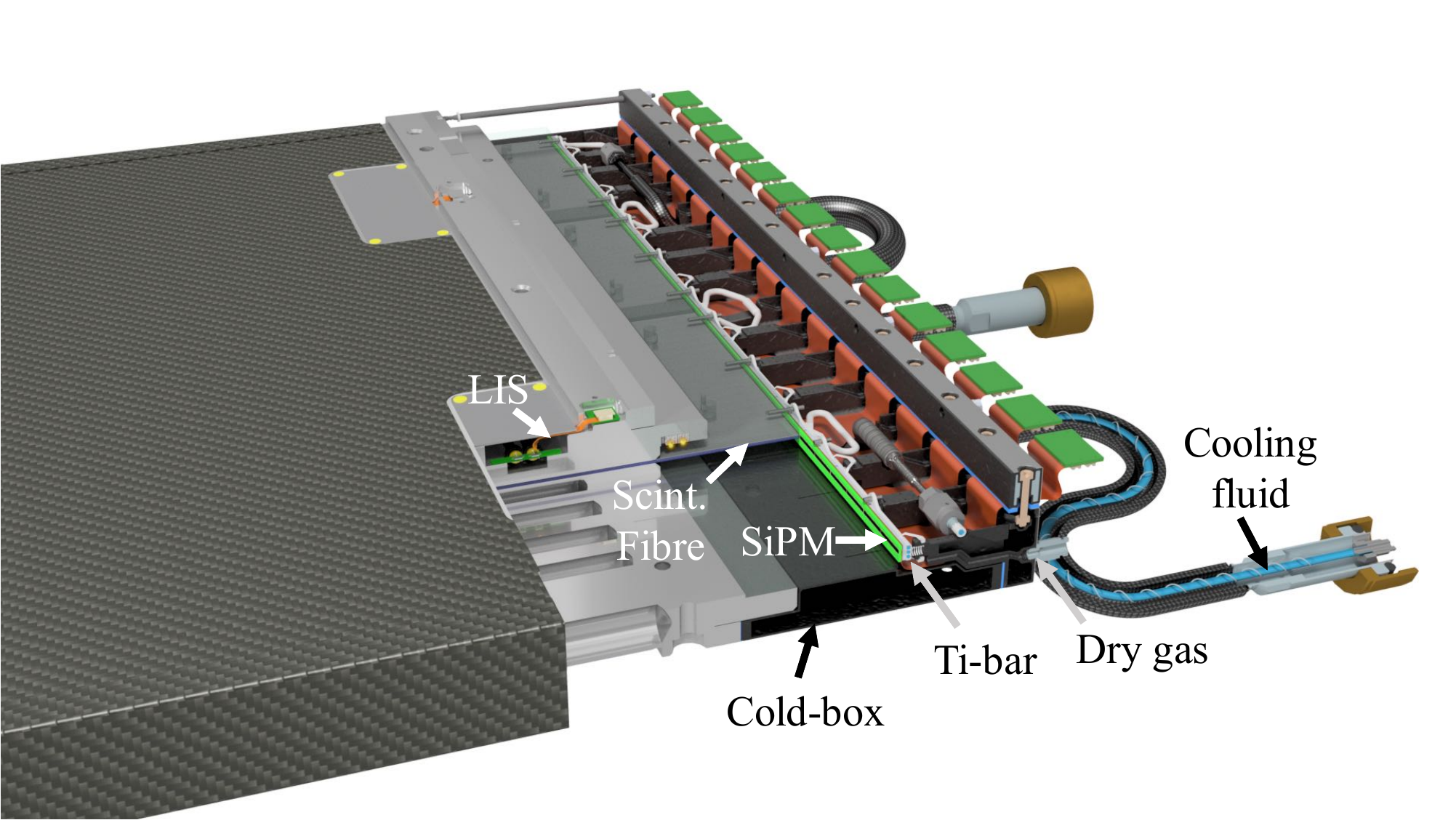}
  \caption{A cutaway view of the cold-box fixed to the fibre module.}
  \label{fig:scifi:coldbox}
\end{figure}

The \Acr[p]{sipm} are coupled to the \Acr{scifi} modules inside a
cold-box.  The details can be seen in figure~\ref{fig:scifi:coldbox}.
There is a gap of $0.48\mm$ between each of the 16 neighbouring
\Acr{sipm} sensors on a module.  They are glued to a 3D-printed
titanium alloy cooling bar and pressed against the fibre ends with
springs.  The whole assembly is housed in a multipart 3D-printed
polyamide shell containing an expanded polyurethane foam layer.  A
\SI{0.12}{\mm} thick tin-plated copper foil is glued to the shell to
improve thermal uniformity on the outer surface.  The cooling liquid
is supplied and returned via vacuum-insulated stainless steel pipes.

The \Acr{sipm} cooling circuit uses a single phase thermal transfer
fluid.\footnote{Both fluoroketone and C$_{6}$F$_{14}$ are possible.}
The delivered coolant temperature can be adjusted between +30 and
\SI{-50}{\degc}, depending on the desired cooling performance needed.
The fluid is circulated through vacuum insulated lines to the detector
by a dedicated plant located in the shielded underground area of \Lhc
Point 8.  To remove any risk of frost and condensation building up on
the cold-bar or \Acr[p]{sipm}, every cold box is individually supplied
with dry air at a dew point of \SI{-70}{\degc}.  Thin heater wires are
wrapped around the cold-box of each module, as well as transfer
bellows on the modules to maintain a stable outer temperature above
the cavern dew point.  The dry air flow rate out of each box is
monitored individually.  Four power supplies\footnote{\Trmk{Wiener}
  MARATON.} are needed for the detector which provide up to
\SI{50}{\watt} of heating power per cold-box.

\subsection[ECS and DAQ]{\Acr[s]{ecs} and \Acr[s]{daq}}
\label{section:scifi:ecs}

The \Acr{sft} control system is implemented in the \lhcb \Ecs platform
described in section~\ref{sec:experiment-control-system}.  The \Ecs
mainly includes the controls of the high-voltage power supplies for
the \Acr[p]{sipm}, the low-voltage power supplies (electronics and
heating wires), the \Fend and \Bend electronics (including \Daq).
Additionally, the \Ecs provides monitoring of the voltages,
temperatures and heating powers, of the water cooling system, of the
\Acr{sipm} cooling system, including the integrated vacuum system, of
the dry gas system for controlling humidity inside the cold-box, and
of the BCAM position monitoring system.\looseness=-1

The \Acr{sft} \Daq uses multiple data formats.  Two formats are used
for clustered data in order to cope with the varying occupancy across
the detector, as described earlier. In addition to the two basic data
formats and some \Tfc related outputs, a special data format provides
a nonzero-suppressed data mode which allows for the output of the raw
2-bit data along with the clustered data. This requires sending only a
fraction of the channels per each bunch crossing, due to the larger
amount of information. Detailed documentation can be found in
refs.~\cite{DelBuono2021} and~\cite{DelBuono2020}.

\subsection{Simulation and reconstruction software}
\label{section:scifi:simrecon}

To study the tracking performance a detailed \Acr{sft} simulation was
performed.  It consists of three main parts.

First, the energy deposition for each particle that traverses a fibre
mat is computed in the \gauss application~\cite{Clemencic_2011}.  In
\boole, where the signal created from the energy deposited is
simulated and digitised, the number of photons produced in each
channel at the track hit location is calculated assuming 8000 photons
produced per \MeV of energy deposition.  The survival probability for
these photons to reach the \Acr{sipm} is dependent on the properties
of the mirrors and the total integrated ionising dose along the fibre.
The fibre properties can be adjusted to take into account ageing and
exposure to radiation.

\begin{figure}[h]
  \centering
  \includegraphics[width=\linewidth]{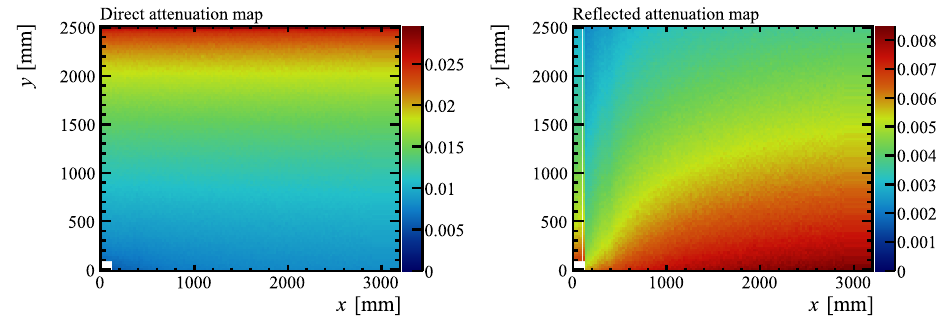}
  \caption{The survival probability (attenuation) map of direct and
    reflected photons in the \Acr{sft} after 50\invfb from
    simulation. Reproduced with permission from~\cite{Beranek:2673602}.}
  \label{fig:scifisurvivalmap}
\end{figure}

A second standalone \geant-based simulation was developed to propagate
optical photons in a single irradiated fibre in order to produce the
survival probability map, shown in figure~\ref{fig:scifisurvivalmap},
that can be used in the \boole simulation of the detector.  The total
photon signal is divided amongst neighbouring channels based on the
crossing angle of the track and a slight smearing to account for the
cluster widths observed in test beam data.  The photon propagation
simulation and the full detector simulation have been tuned to data
obtained from experimental measurements of irradiated fibres and test
beam data.  A detailed description is given in
ref.~\cite{Beranek:2673602}.

In the third part, the \Acr{sipm} signal and PACIFIC chip digitisation
are simulated.  A signal in each channel is obtained based on the
quantum efficiency and other properties of the \Acr{sipm}, the
thresholds of the PACIFIC comparators, and the electronics response
function of the \Asic, as it can be seen in
figure~\ref{fig:scifi:pacificsim} (right).  The dark-noise avalanches
of the \Acr{sipm} are also (optionally) simulated at this stage and
added to the analog signal of each channel before digitisation.  The
simulated digitisation of the thresholds outputs are then passed to
the clusterisation output and the position of a cluster can be
calculated and encoded. At this stage the output of the simulation
corresponds to what is obtained on detector electronics output.  The
simulations are used for the development of the analysis and
monitoring chain, as well as testing various scenarios such as the
impact of threshold settings on the production of dark-noise clusters,
single hit efficiency, and tracking performance.

\subsubsection{Spillover clusters}

A spillover signal in a given bunch crossing is a signal due to a
particle associated to an interaction from a previous or following
bunch crossing.  There are two main sources of spillover signals. The
first is associated with real hits from particles from previous or
following bunch crossings that end up inside the current integration
window due to flight times and timing offsets.  The second is
associated with the extended shape of the analog electronics pulse,
which will result in a (positive or negative) charge contribution to
the PACIFIC integrators in multiple bunch crossings.
\begin{table}[h]
  \centering
  \caption{The average number of clusters per event occurring in the
    current bunch crossing for different PACIFICr5q models based on a
    sample of $B_s\rightarrow \phi\phi$ events generated at the given
    delay from the current~crossing.}
  \label{tab:spillover}
  \begin{tabular}{lcccc}
    \hline
    & $-50$\ns    & $-25$\ns    & 0\ns      & $+25$\ns   \\
    \hline
    PACIFICr5q  pz5   & 76        & 350       & 4082      & 34 \\
    PACIFICr5q  pz6   & 91        & 503       & 4021      & 25 \\
    \hline
  \end{tabular}
\end{table}
To mitigate the first type of spillover, an undershoot at the end of
the pulse was added and tuned to reduce the effect arising from $-50$
and $-25$\ns crossings.  The average number of clusters observed in
the current (0\ns) bunch crossing from events generated in the bunch
crossings ($-50$ to $+25$\ns) are displayed in
table~\ref{tab:spillover} for two different models of the PACIFICr5q
settings (see section~\ref{sec:scifi:PACIFIC analog circuit
  simulation}).

\subsubsection{Dark-noise cluster rates}
Initial estimates indicate that the tracking algorithms performance is
degraded if the rate of dark clusters per \Acr{sipm} increases above
\SI{2}{\MHz}, or approximately 200 dark-noise clusters across the
entire tracker per bunch crossing.
\begin{figure}[t]
  \centering
  \includegraphics[width=0.7\linewidth]{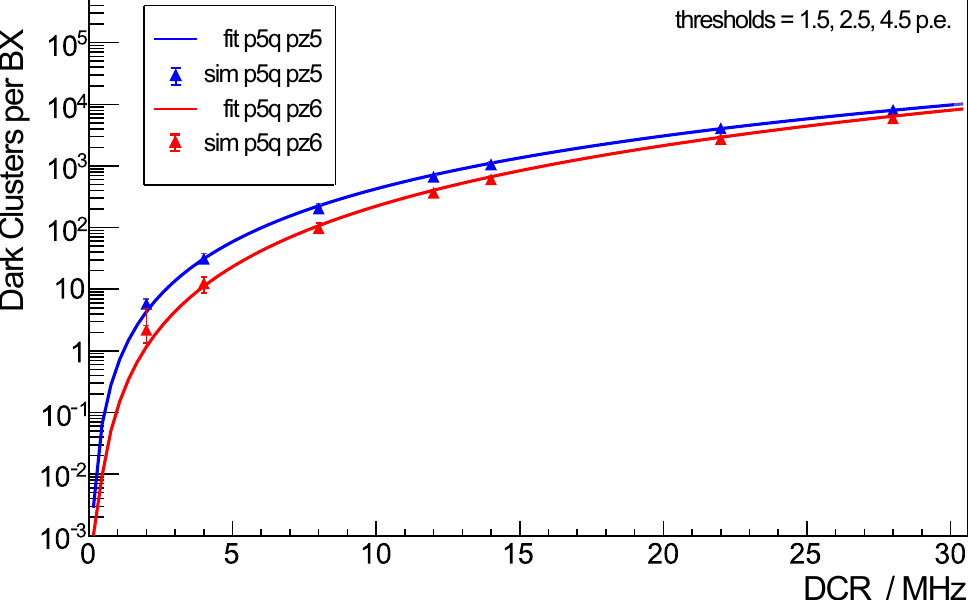}
  \caption{Simulated number of thermal-noise clusters per 25\ns clock
    cycle as a function of the \fdcr.  The curves stem from a power
    law fit to the data points.  The comparison is made for two
    different models of the PACIFIC settings (in blue the pz5
    settings, in red the pz6).  The three threshold values used are
    1.5, 2.5 and 4.5 photoelectrons.  }
  \label{fig:val_PACcomp}
\end{figure}
However, this rate is dependent upon the single channel \fdcr and the
thresholds set in the PACIFIC.  Ideally, the thresholds are set as low
as possible to maximise the efficiency, while accepting a tolerable
amount of dark-noise clusters that can be removed in the track finding
algorithms.  A comparison of the number of thermal-noise clusters per
bunch crossing for two different electronics response configurations
is shown in figure~\ref{fig:val_PACcomp}.  The data points are
generated from a detailed model of the \fdcr, PACIFIC response
function, \Acr{sipm} cross-talk, and the clustering algorithm.  A
simple power law function, overlaid in the figure, describes well the
data.

\subsection{Test beam results}
\label{section:scifi:testbeam}

A slice test of two \Acr{sft} modules coupled to a complete set of
nearly final \Fend readout electronics was performed in 2018, with a
preliminary standalone version of the \lhcb \SI{40}{\MHz} \Pciefourty
readout~\cite{Cachemiche2015}.  The system was tested for its single
particle hit reconstruction efficiency and position reconstruction
resolution using a beam from the \Sps at CERN's North Area consisting
of 180\gevc pions, protons and/or muons.  The \Acr[p]{sipm} were not
irradiated and kept at room temperature with chilled water.

To ensure that the reconstructed particle tracks are of good quality
and to provide a precise hit location, the TimePix3 telescope in the
H8A area was operated synchronously with the \Pciefourty readout.  The
telescope provides a fine time stamp that divides the \SI{25}{\ns}
long clock cycles into 96 intervals of about 0.25\ns.\footnote{Due to
  the multiple \Pll{s} used to generate this fine time binning, the
  intervals are not all equally long.}  This fine time stamp is
required as the arrival of the particles from the \Sps are not
synchronised in time with the \Daq system, unlike the particles that
will be generated at the \Lhc.\looseness=-1

For every trigger and high-quality track from the telescope a search
is made in the \Acr{sft} data for a corresponding signal cluster (one
or more neighbouring channels with a signal).  The relative position
of the found cluster is compared to the telescope track position in
the module to determine the position resolution of the fibre modules,
as well as the single hit efficiency.

\subsubsection{Single-hit efficiency}

\begin{figure}[t]
  \centering
  \includegraphics[width=0.525\linewidth]{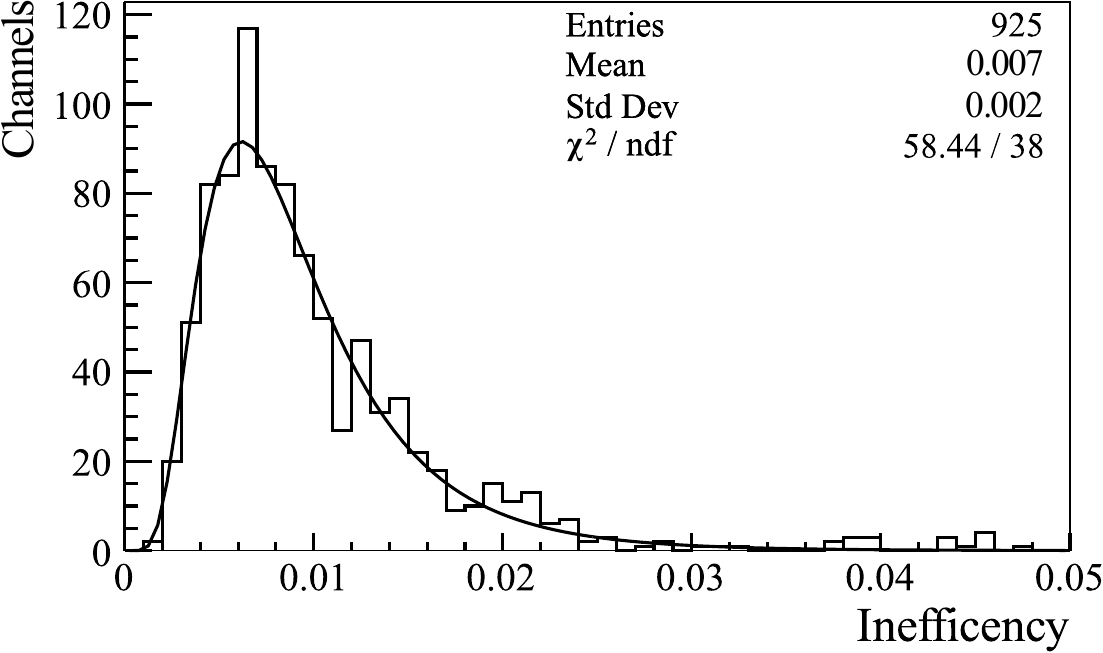}\hfill
  \includegraphics[width=0.44\linewidth]{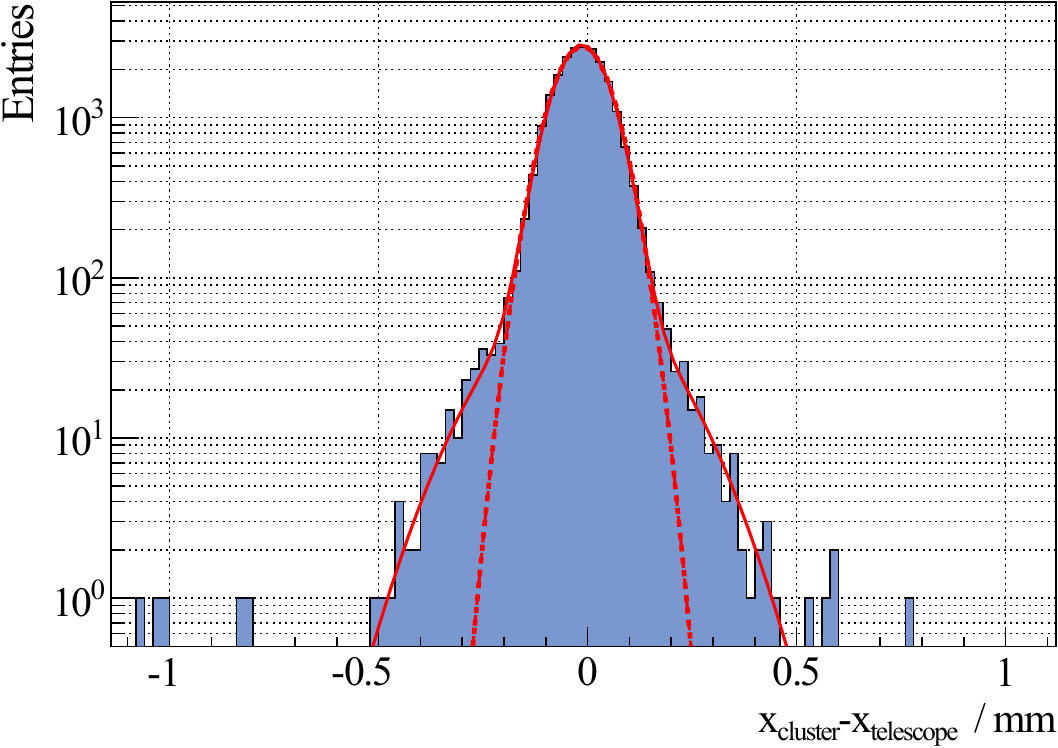}
  \caption{Left: the single hit inefficiency of the five most
    efficient fine time bins of all channels measured at several
    positions across the module. The distribution is fit to a
    log-normal distribution. The mean and standard deviation of the
    log-normal distribution is also shown. Right:~an example hit
    position residual distribution fitted with a single (dashed curve)
    and double (solid curve) Gaussian function. Gap regions have been
    excluded.\label{fig:scifiresidual} }
\end{figure}

A log-normal distribution describes well the measured inefficiency
distribution observed in the test beam as can be seen in
figure~\ref{fig:scifiresidual} (left).  From these data a mean hit
efficiency for the sensitive regions of the \Acr{sft} (excluding gaps)
is estimated to be $0.993\pm0.002$ (quoting the standard deviation of
the log-normal as an estimator for the expected spread).

\subsubsection{Hit position resolution}

The hit position residual is defined as the difference between the
cluster position and the position of the telescope track extrapolated
to the fibre plane, after having applied alignment corrections using
Millepede~\cite{millipede}.  An example hit position residual
distribution, merging the data from three beam spot positions, at the
centre and outer edges of one \SI{32.6}{\mm} wide \Acr{sipm} arrays,
is shown in figure~\ref{fig:scifiresidual} (right).  As the telescope
resolution contributes insignificantly, the width of the residual
distribution can be taken as the resolution of the \Acr{sft} module.
The distributions of the hit position residuals measured across the
modules in the test beam campaign indicate a single hit position
resolution of $64\pm$\SI{16}{\mum} for perpendicular tracks.

\section{RICH}
\label{sec:rich}
Charged hadron discrimination, namely the separation between pions,
kaons and protons, is a crucial aspect of the \lhcb physics
programme. Hadron \Pid is provided in \lhcb by the \rich system in the
2.6--100\gevc momentum range, and plays a central role in the
measurements performed in \lhcb with \runonetwo
data~\cite{LHCb-DP-2012-003,Calabrese:2022eju}. The \rich system
allows to: distinguish between final states of otherwise identical
topologies, e.g.\ $\Bds \to \pip \pim, \Kp \pim, \Kp \Km$ decay modes;
heavily reduce the combinatorial background in decay modes involving
hadrons in the final state, such as $\Bs \to \phi \phi$, where
$\phi \to \Kp \Km$, that would be prohibitively large without \Pid
requirements; perform the flavour tagging of a \Bds meson at the
production vertex, relying on charged kaon identification from the
$\bquark \to \cquark \to \squark$ decay chain. The information
provided by the \rich system is also used to suppress the
combinatorial background at the \Acr[s]{hlttwo} level (see
section~\ref{ssec:LHCbDAQ}).

\begin{figure}[t]
  \centering
  \includegraphics[width=0.27\linewidth]{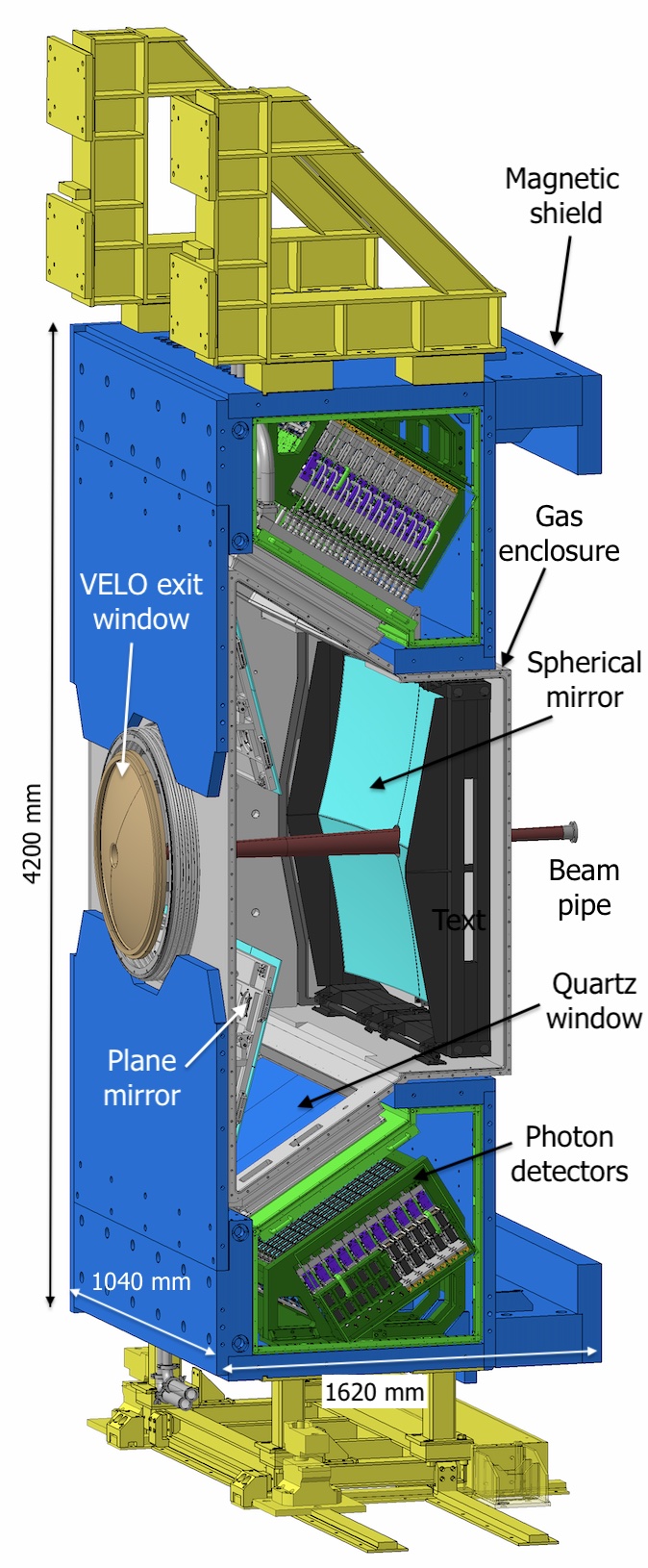}\hfill
  \includegraphics[width=0.57\linewidth]{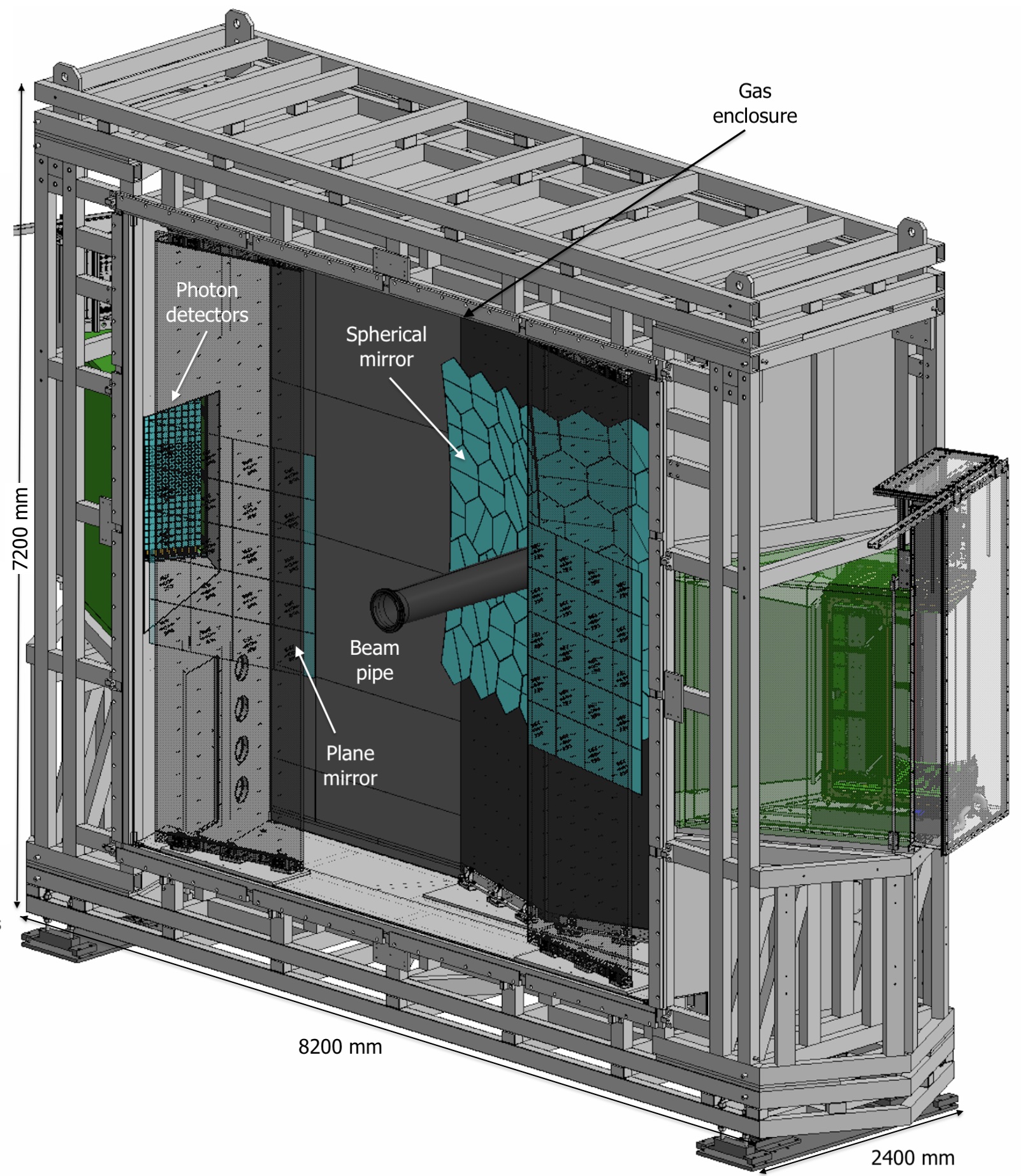}
  \caption{Schematic view of the (left) \richone and (right) \richtwo
    detectors. Reproduced from ~\cite{Okamura_2022}. \textcopyright\ 2022
IOP Publishing Ltd and Sissa Medialab. All rights reserved.}
  \label{fig:RICH:layout}
\end{figure}

The overall layout and concept of the \rich system remains unchanged
with respect to \runonetwo \lhcb~\cite{LHCb-DP-2008-001}, although
critical modifications were needed to allow the system to operate at
the higher design luminosity while maintaining a performance
comparable to that of \runone and
\runtwo~\cite{DAmbrosio:2013ijj,LHCb-TDR-014}. It consists of two
detectors, \richone and \richtwo, as shown in
figure~\ref{fig:RICH:layout}.  \richone covers an angular acceptance
from 25 to 300\mrad in the magnet bending plane and from 25 to
250\mrad in the vertical direction.  The photon detector planes are
located above and below the beam pipe, where the residual magnetic
field is minimal.  \richtwo is located downstream the dipole magnet,
covering an angular acceptance from 15 to 120\mrad in the magnet
bending plane and 15 to 100\mrad in the vertical direction.  The
photon detector planes are located on the sides.  In both detectors
the Cherenkov photons produced inside fluorocarbon gaseous radiators
are reflected outside the \lhcb acceptance by means of a system of
spherical and planar mirrors, focusing the ring images on the photon
detector planes.\looseness=-1

\richone is located upstream of the dipole magnet and employs a
\cfourften gas radiator with a refractive index $n=1.0014$ for
Cherenkov radiation of wavelength $\lambda = 400\nm$ at \Stp, allowing
to provide \Pid in the momentum range between 2.6 and 60 \gevc. The
average path length of particles inside the radiator is approximately
110\cm. \richtwo is designed to provide \Pid for higher momentum
particles, between 15 and 100\gevc, with a \cffour gas radiator with
$n=1.0005$ for Cherenkov radiation of $\lambda=400\nm$ at \Stp, and an
average track path of 167\cm.

In order to read the detectors out at 40\mhz rate, the full photon
detection chain was replaced in both \richone and \richtwo detectors,
since the former \Hpd~\cite{ALEMIA200048} had embedded \Fend
electronics limited to a 1\mhz output rate. The \Acr[p]{hpd} have been
replaced with \Acr[p]{mapmt} equipped with new \Fend electronics. The
upgraded photon detection modules are described in
section~\ref{subsec:richPhotonDetectionChain}.

One of the key parameters driving the performance of the \rich system
is the efficiency of the pattern recognition algorithm, optimal for
detector occupancies not exceeding 30\% as determined from experience
in \runone and \runtwo operations.\footnote{The occupancy is defined
  as the number of fired channels over the total number of channels in
  a given region at 40\mhz readout rate.} With the five-fold increase
in the instantaneous luminosity, a redesign of the \richone optics was
necessary to reduce the peak occupancy, as described in
section~\ref{subsec:richone}. The optical system and mechanical
envelope of \richtwo was left unchanged, but redesigned support
structures to house the new photon detectors were required, as
reported in section~\ref{subsec:richtwo}.\looseness=-1

Monitoring and control systems and the calibration procedure have also
been updated, to cope with the changes in the photon detection chain
and the readout infrastructure, as described in
sections~\ref{subsec:richMonitoringAndControls}
and~\ref{paragraph:richCalibration}. The upgraded \rich system has
been designed to improve the single photon resolution and to keep the
excellent \Pid performance provided in \runone and \runtwo in the more
challenging conditions of \runthree. Its expected performance is
discussed in section~\ref{subsec:richPerformance}.

\subsection{The upgraded photon detection chain}
\label{subsec:richPhotonDetectionChain}

The design of the upgraded photon detection chain has been optimised
in order to cope with the highly nonuniform occupancy expected in the
\rich system, ranging from about 30\% in the central region of
\richone down to 5\% in the peripheral region of \richtwo, if the
previous photon detectors were kept unchanged. The largest hit rates
correspond to Cherenkov photons associated to the large number of
tracks produced at high pseudorapidity. Given the observed occupancy
distribution, the detector geometry and channel granularity have been
optimised taking into account existing overall mechanical constraints
and the number of readout channels, which have significant impact on
the cost. As described in the following, the photon detection planes
are subdivided into two regions having different granularity, with the
aim of keeping the optimal performance while maximising the cost
savings. In addition, in order to ensure stable operations of the
upgraded \rich detectors, an evaluation of the photon detection chain
performance under high radiation fields has been performed as
described in section~\ref{subsec:richIrradiation}.

\subsubsection{Photon detectors}
\label{paragraph:richMaPMTs}

\begin{figure}[t]
  \centering
  \includegraphics[width=0.5\linewidth]{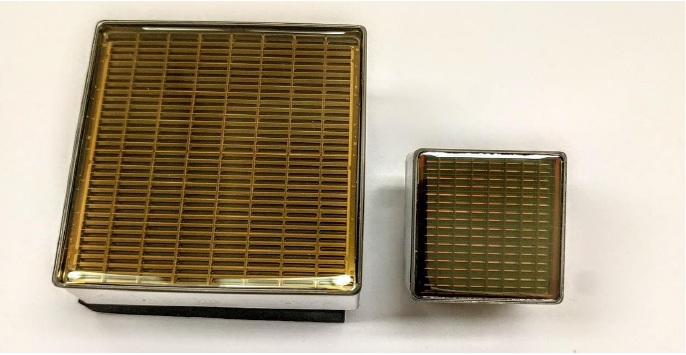}\hfill
  \includegraphics[width=0.5\linewidth]{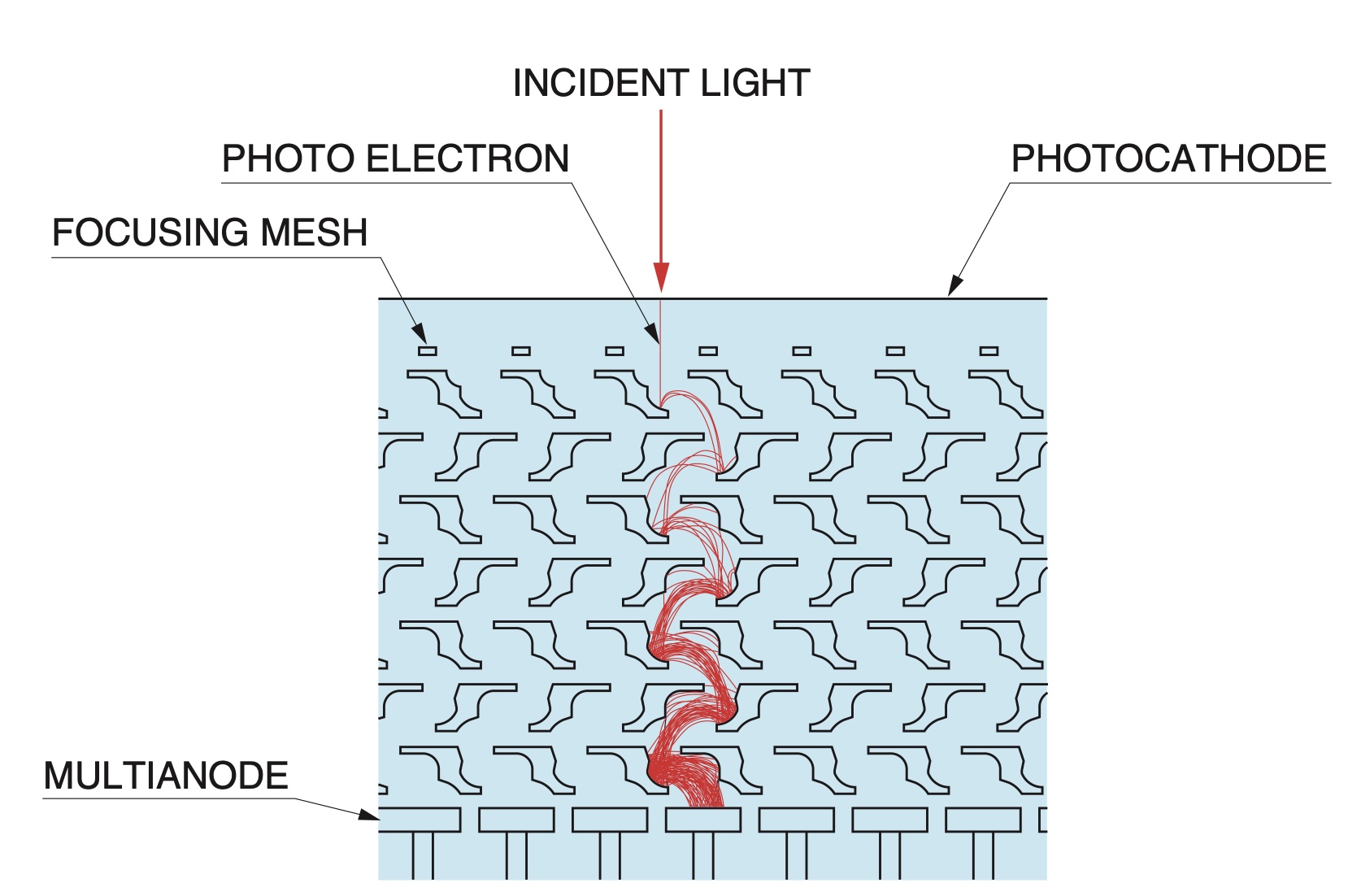}
  \caption{Left: the \Mapmt{s} selected for the upgraded \rich
    detectors with the 2-inch model on the left and the 1-inch model
    on the right. Right:~scheme of the internal structure of the
    \Mapmt{}. Reproduced from~\cite{LHCb:2021okg}. CC BY 4.0.}
  \label{fig:RICH:mapmt1}
\end{figure}

The main parameters driving the choice of photon detectors have been
good spatial resolution on a large active area, high detection
efficiency in the wavelength range 200--600\nm, and very low
background noise to allow single photon detection despite the high
occupancy foreseen.  The \Mapmt{s} had already been considered during
the first construction of the \rich detectors, but were rejected
mostly due to the limited fill factor ($\sim40$\%) of such devices at
the time.\footnote{The fill factor is the fraction of active area with
  respect to the total detector area.} The latest models available on
the market are instead characterised by a fill factor exceeding 80\%
and could be adopted as the upgraded \rich photon detectors. The
selected \Mapmt models both consist of a matrix of $8\times 8$ anodes.
\richone and the central region of \richtwo are equipped with 1-inch
\Mapmt modules\footnotemark{} with a pixel size of
$2.88\times2.88\mma$, ideal for the high occupancy areas of the \rich
system. The outer region of \richtwo has been equipped with a 2-inch
device\footnotemark[\value{footnote}]\footnotetext{\Trmk{Hamamatsu}
  R13742 (1-inch) and R13743 (2-inch). The R13742 and R13743 models
  are custom variants of the commercial models R11265 and R12699,
  respectively. The difference between the custom and commercial units
  stands on the requested technical specifications described in the
  text.}, with a pixel size of $6\times6$\mma. The decision to install
detectors with a coarser granularity in the peripheral regions of
\richtwo allowed a significant reduction in the number of \Mapmt units
and readout channels, while having a negligible impact on the overall
\rich performance as demonstrated by simulation studies. The \Mapmt{s}
installed in the upgraded \rich detectors, together with the schematic
view of their internal structure, are shown in
figure~\ref{fig:RICH:mapmt1}.\looseness=-1

\begin{figure}[h]
  \centering
  \includegraphics[width=0.5\linewidth]{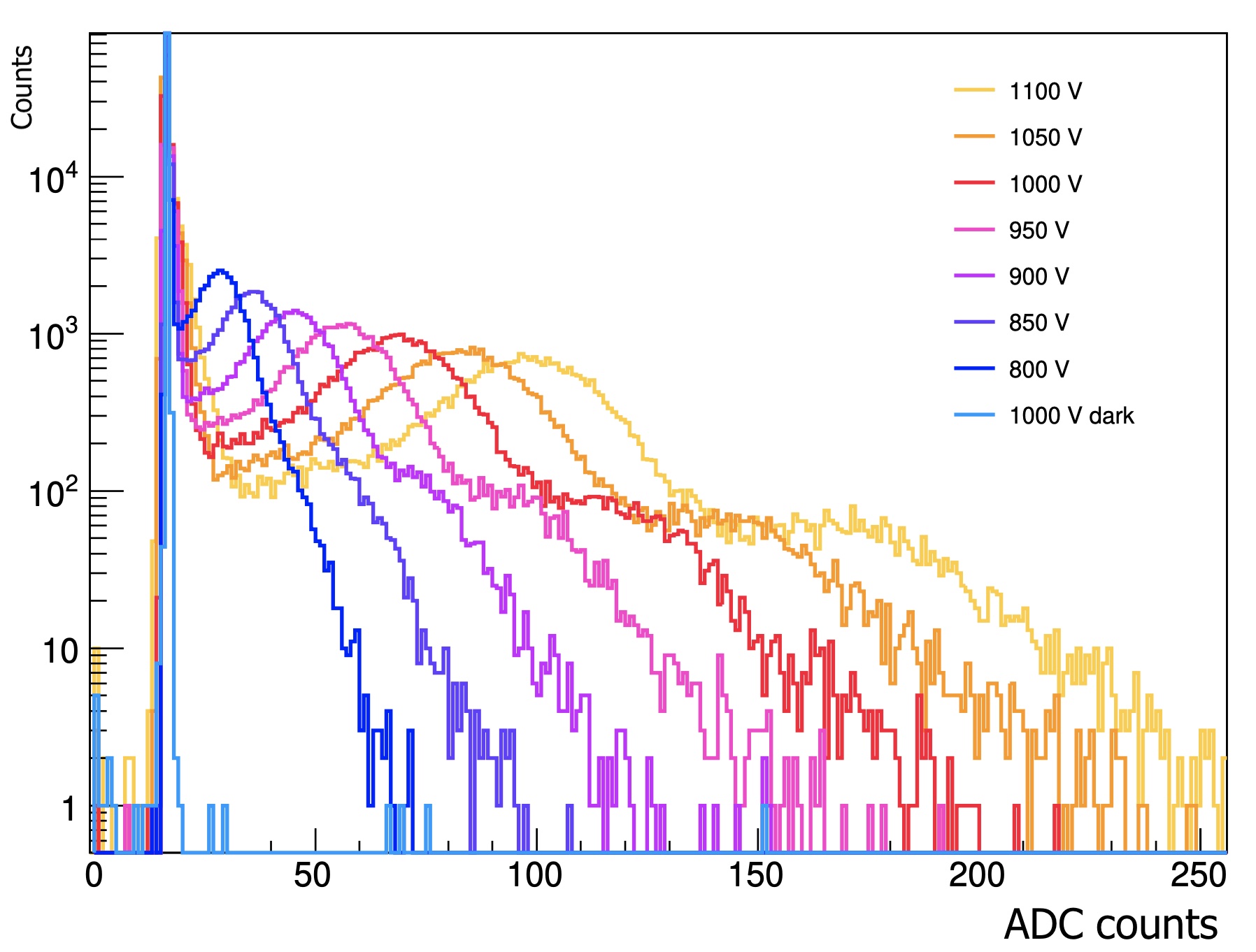}\hfill
  \includegraphics[width=0.44\linewidth]{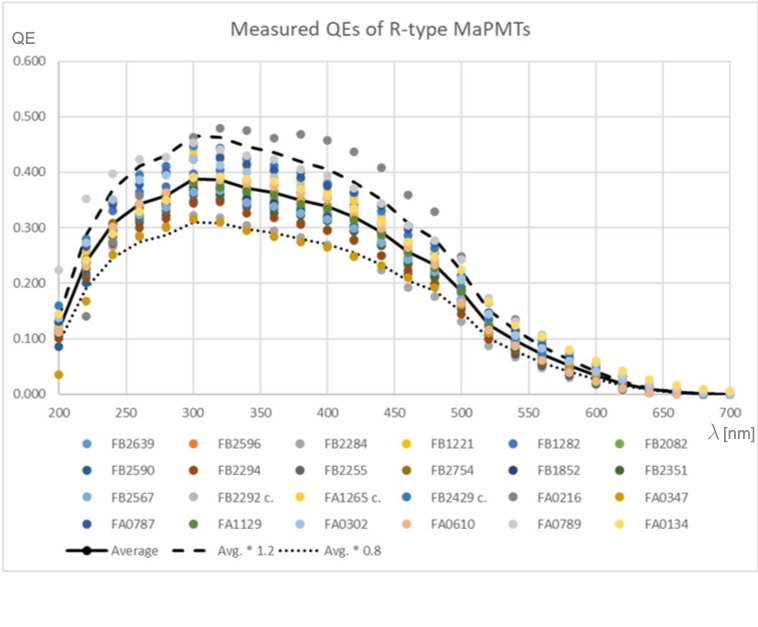}
  \caption{Left: typical signal amplitude spectra for a pixel as a
    function of the \Hv value. Reprinted from~\cite{GIUGLIANO2023168436},
Copyright (2023), with permission from Elsevier. Right:~\Qe curves for a batch of 1-inch
    \Mapmt{s} from the production: the ultra bi-alkali photocathode
    allows to reach excellent \Qe values.}
  \label{fig:RICH:mapmt2}
\end{figure}

A total of 1888 (768) 1-inch \Mapmt{s} are installed in \richone
(\richtwo) and 384 2-inch \Mapmt{s} are installed in \richtwo. Over
3500 units, including spares, have been purchased by the \rich
collaboration and quality-assured to verify the requested technical
specifications, among which are a gain larger than $10^6$ and a
dark-count rate less than $<2.5\ \mathrm{kHz/cm^2}$. A full set of \Qa
tests was implemented to qualify the whole \Mapmt{} production. Two of
the typical \Qa parameter scans, the signal amplitude as a function of
the \Hv and the \Qe as a function of the wavelength, are shown in
figure~\ref{fig:RICH:mapmt2}.

\subsubsection{Front-end electronics and elementary cell}
\label{paragraph:richEC}

\begin{figure}[t]
  \centering
  \includegraphics[width=0.45\linewidth]{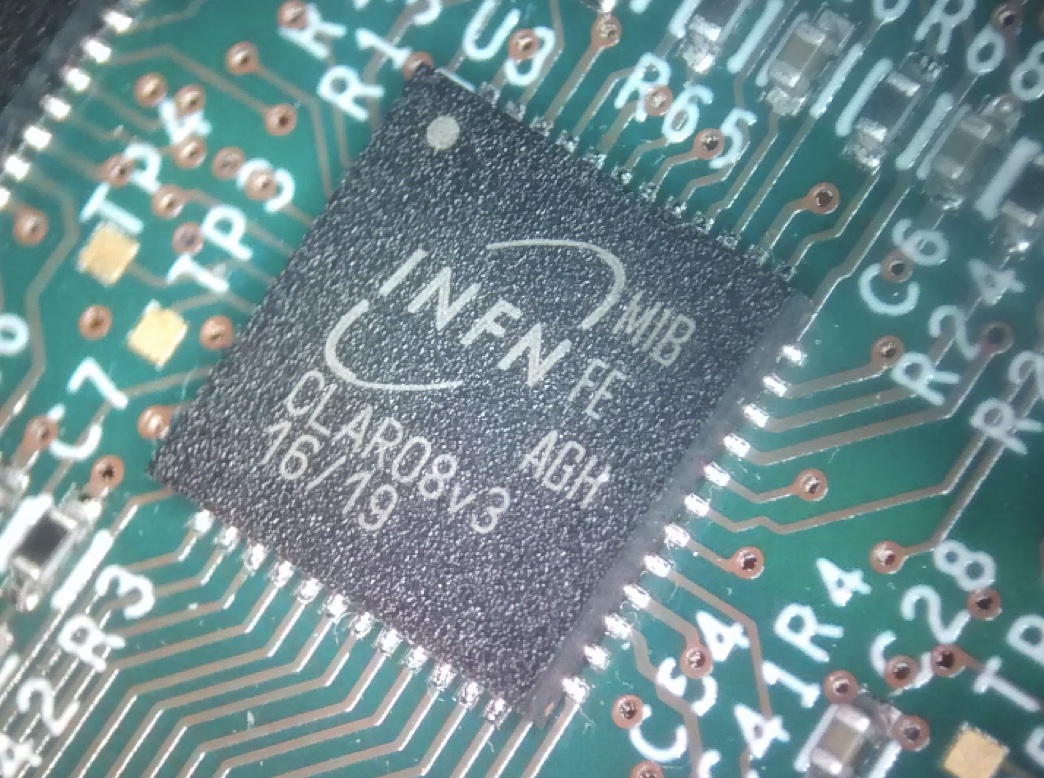}\hfill
  \includegraphics[width=0.52\linewidth]{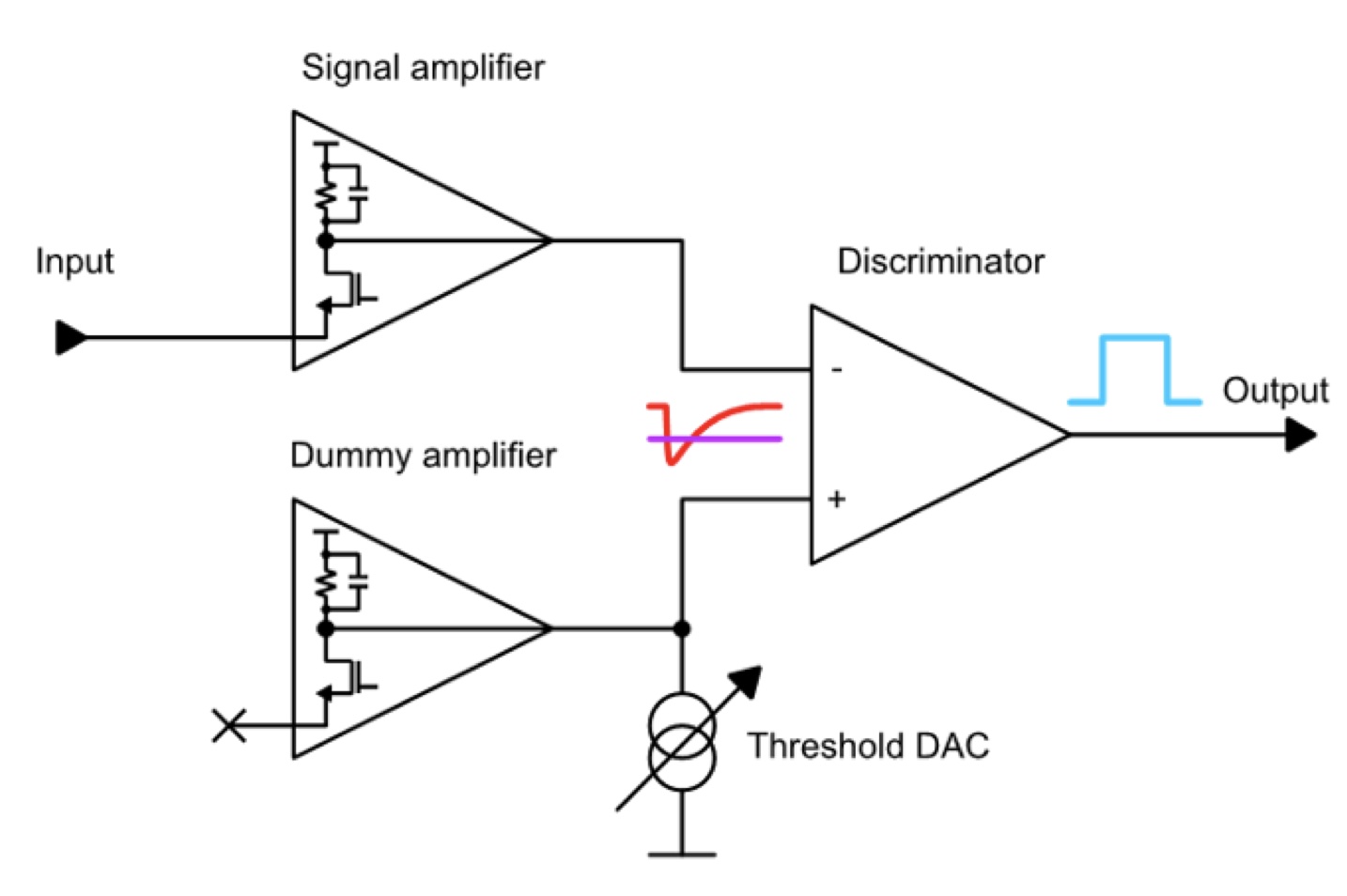}
  \caption{Left: CLARO \Asic with its packaging. Reproduced from~\cite{Baszczyk:2017fiz}. \textcopyright\ 2017 IOP Publishing Ltd
and Sissa Medialab. All rights reserved. Right:~block
    schematic of a CLARO channel. The purpose of the dummy amplifier
    is to give each channel a differential structure, improving the
    power supply rejection ratio and allowing DC-coupled input to the
    discriminator. Reproduced from\cite{Gotti_2017}. \textcopyright\ 2022 IOP Publishing Ltd and Sissa
Medialab. All rights reserved.}
  \label{fig:RICH:claroChip}
\end{figure}

The average hit rate in the high occupancy regions of \richone can
exceed 10$^7$ hits/s per pixel on average. Furthermore, the estimated
total integrated dose over the detector lifetime, in the regions
closer to the beam pipe is estimated to be about 2\kGy for
\richone~\cite{8115435}.  Radiation-hard fast readout electronics
is therefore needed, with low power consumption to minimise heating.
This motivated the design of a custom 8-channel front-end \Asic named
CLARO~\cite{Baszczyk:2017fiz}. CLARO, shown in
figure~\ref{fig:RICH:claroChip}, is designed in 350\nm CMOS \Acr{ams}
technology, with the exception of configuration registers as will be
discussed in section~\ref{subsec:richIrradiation}.

Each CLARO channel is composed of an analogue transimpedance amplifier
followed by a discriminator.  Converted and discriminated input
current signals trigger asynchronous digital pulses at the output. The
output signals have a voltage swing of 2.5\volt, and a variable length
allowing time-over-threshold measurements.

A 128-bit register allows CLARO single channel configuration. In
particular, to allow for channel-by-channel gain differences, input
signals can be attenuated by factors 1, 1/2, 1/4, and 1/8. Individual
thresholds can also be set, with the possibility to cancel the
discriminator offset. Thresholds are calibrated with test signals
injected at the input through a dedicated test capacitor.  A more
detailed description of the CLARO design and its functionalities can
be found in ref.~\cite{Baszczyk:2017fiz}.

The CLARO number of channels matches the $8\times 8$~pixel modularity
of the \Mapmt{s} and allows placing the \Asic as close as possible to
the \Mapmt{} anodes, minimising the parasitic capacitance at the input
and the susceptibility to electromagnetic interference noise.

\begin{figure}[t]
  \centering
  \includegraphics[width=0.5\linewidth]{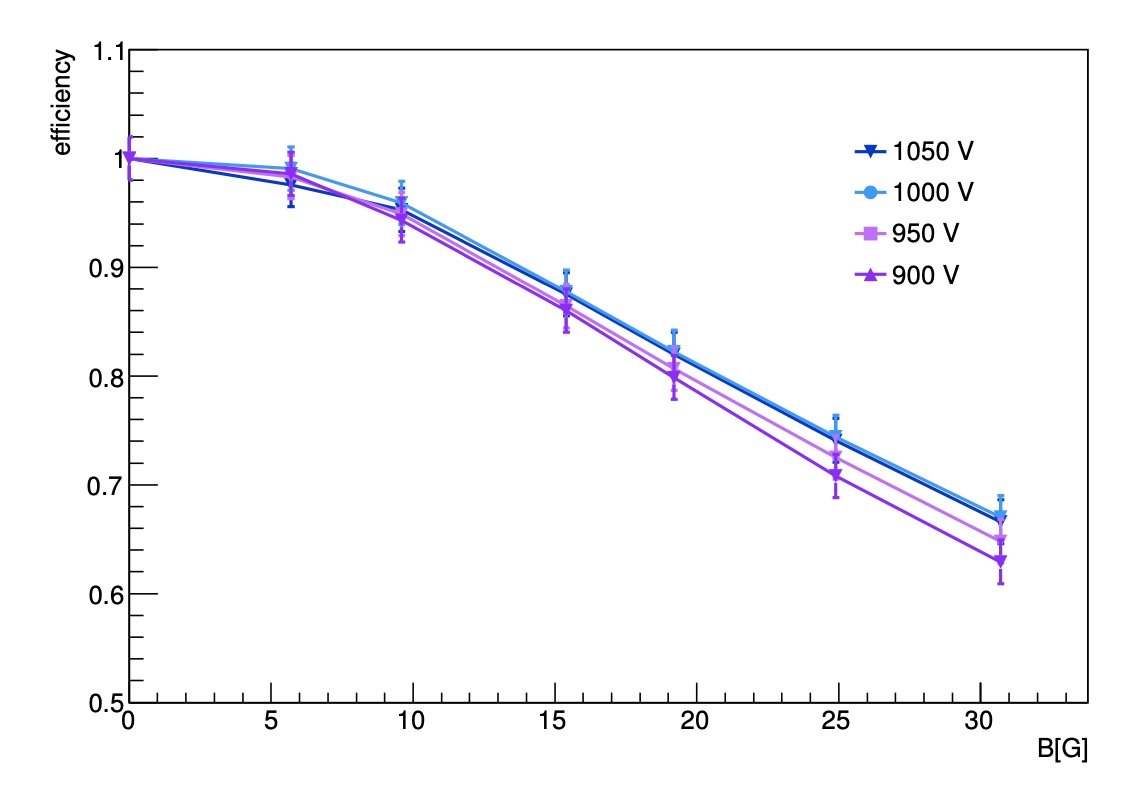}%
  \includegraphics[width=0.5\linewidth]{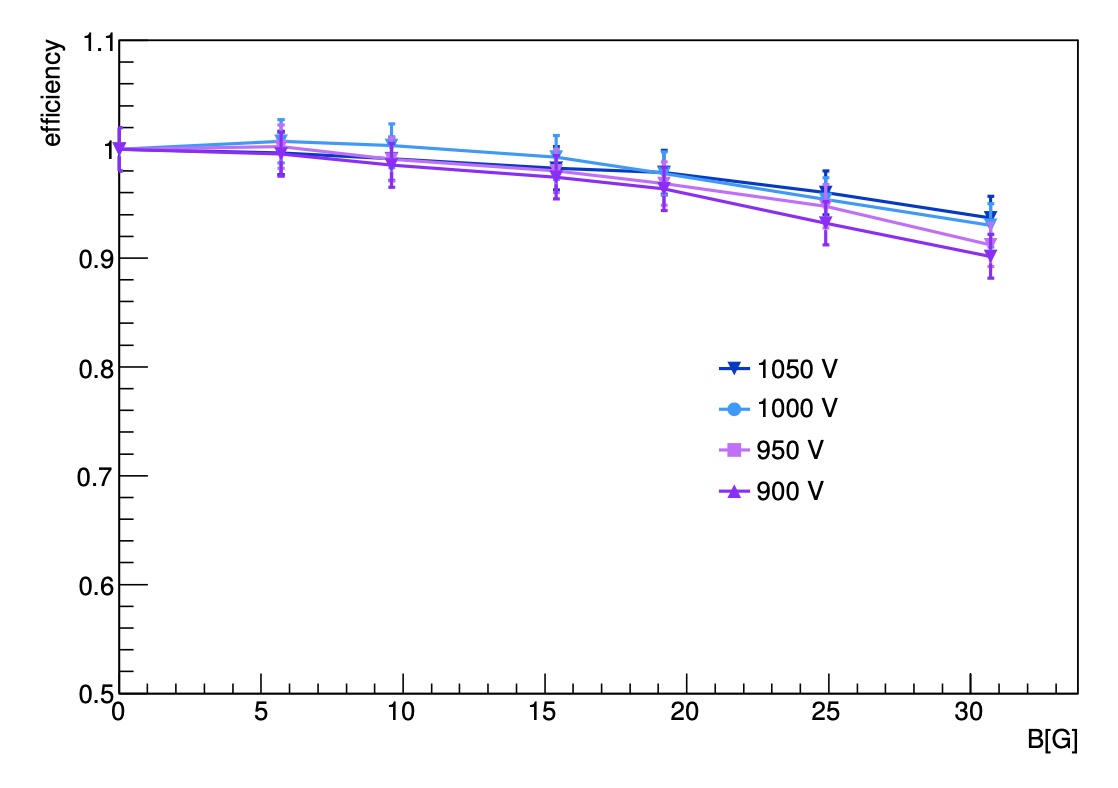}
  \caption{Counting efficiency as a function of the longitudinal
    magnetic field for an edge pixel, at different values of \Hv, for
    an \Ecr (left) without and (right) with the magnetic shield.}
  \label{fig:RICH:MS}
\end{figure}

The readout system was arranged in compact units named
\Acr[p]{ec}. Two types of \Ecel{s}, adapted to the different \Mapmt{}
models, are used: the \Ecr and the \Ech.
\begin{figure}[t]
  \centering
  \includegraphics[width=\linewidth]{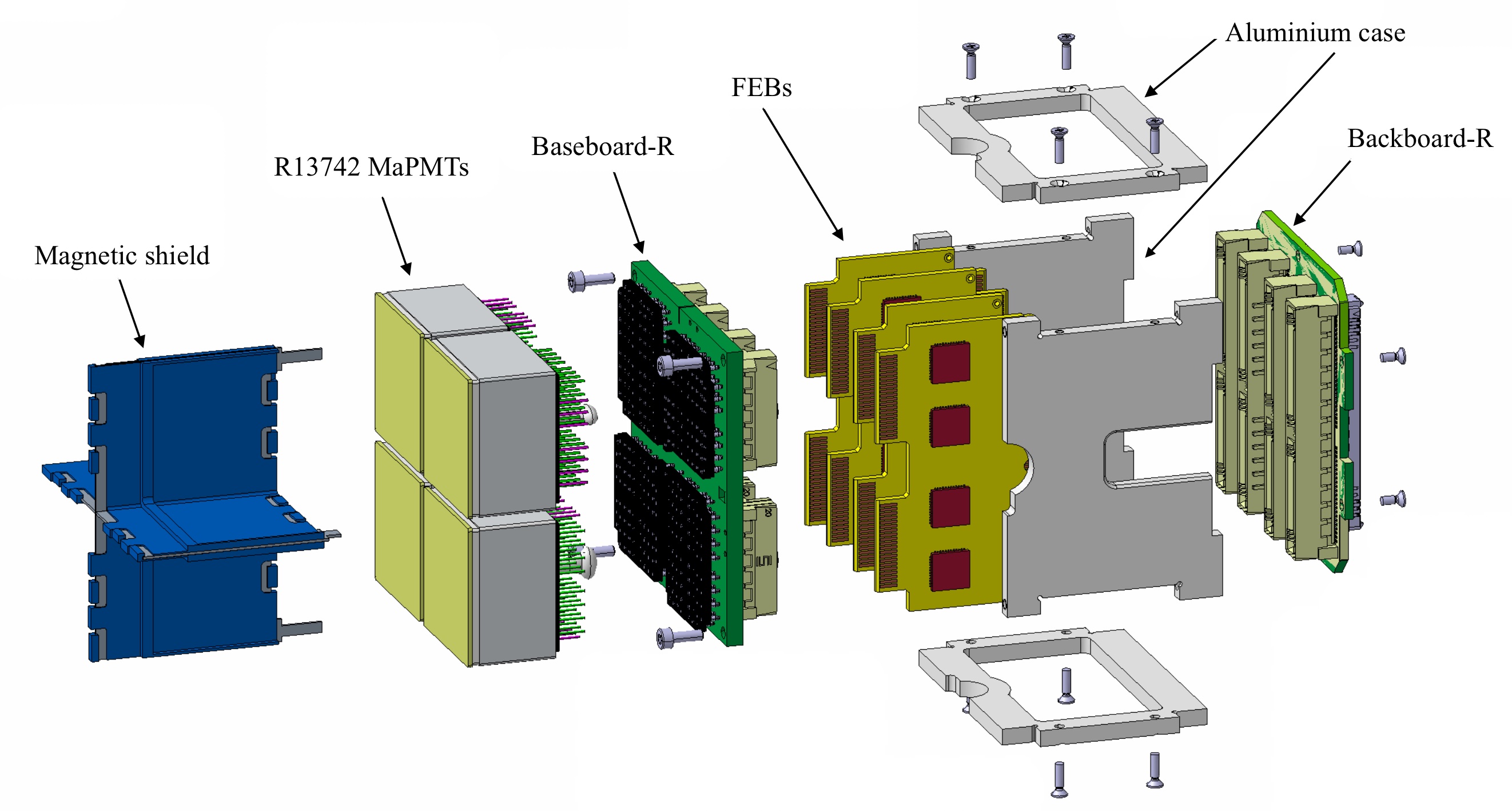}
  \caption{Exploded view of the \Acr[m]{ecr}. Reproduced from~\cite{LHCb:2021okg}. CC BY 4.0.}
  \label{fig:RICH:ECRexploded}
\end{figure}
A view of the \Ecr is shown in figure~\ref{fig:RICH:ECRexploded}.  It
reads out four 1-inch \Mapmt{s}, for a total of 256 pixels in
approximately $2\times 2$ square inches.  The \Mapmt{s} are plugged
into a baseboard, which hosts four 3\aunit{M$\Omega$} resistive
dividers in parallel, to bias the dynodes of each \Mapmt{}.  The last
two dynodes of each chain can be powered by dedicated supply lines in
high-occupancy regions, where the drawn current is higher and can
induce nonlinear effects in \Mapmt{} gain.  A magnetic shield is
placed in front of the \Mapmt{s} in the \richone \Acr[p]{ecr} where,
even inside the magnetic shield, the stray magnetic field from the
\lhcb magnet is up to about 2\aunit{mT}. The shield is cross-shaped
and made of a 500\mum thick mu-metal. It deflects the field lines,
attenuating the magnetic field that reaches the \Mapmt{} by a factor
of approximately 20, down to a value where its effect on the
performance of the \Mapmt{s} becomes negligible, as shown~in~figure~\ref{fig:RICH:MS}.\looseness=1

The baseboard propagates the anode signals to four \Acr[p]{feb},
hosting eight CLARO \Asic{s} each (four on each face of the
board). The \Feb{s} are in turn connected to a backboard routing the
output signals to the \Acr[p]{pdmdb}, described in
section~\ref{paragraph:richPDMDB}, through two high-density
connectors. The CLARO power supply and control signals are generated
on the \Pdmdb{s} and are routed through the backboard as well. A
3.0\mm thick and 40.5\mm long aluminium case serves as a mechanical
support structure for the electronic components and allows thermal
transfer by conduction, with the heat dissipation from the voltage
dividers enhanced by copper layers inside the baseboard. Temperature
monitoring is also ensured by temperature probes. There are 472
{\Ecr}s in \richone and 192 in the central region~of~\richtwo.\looseness=1

\begin{figure}[t]
  \centering
  \includegraphics[width=\linewidth]{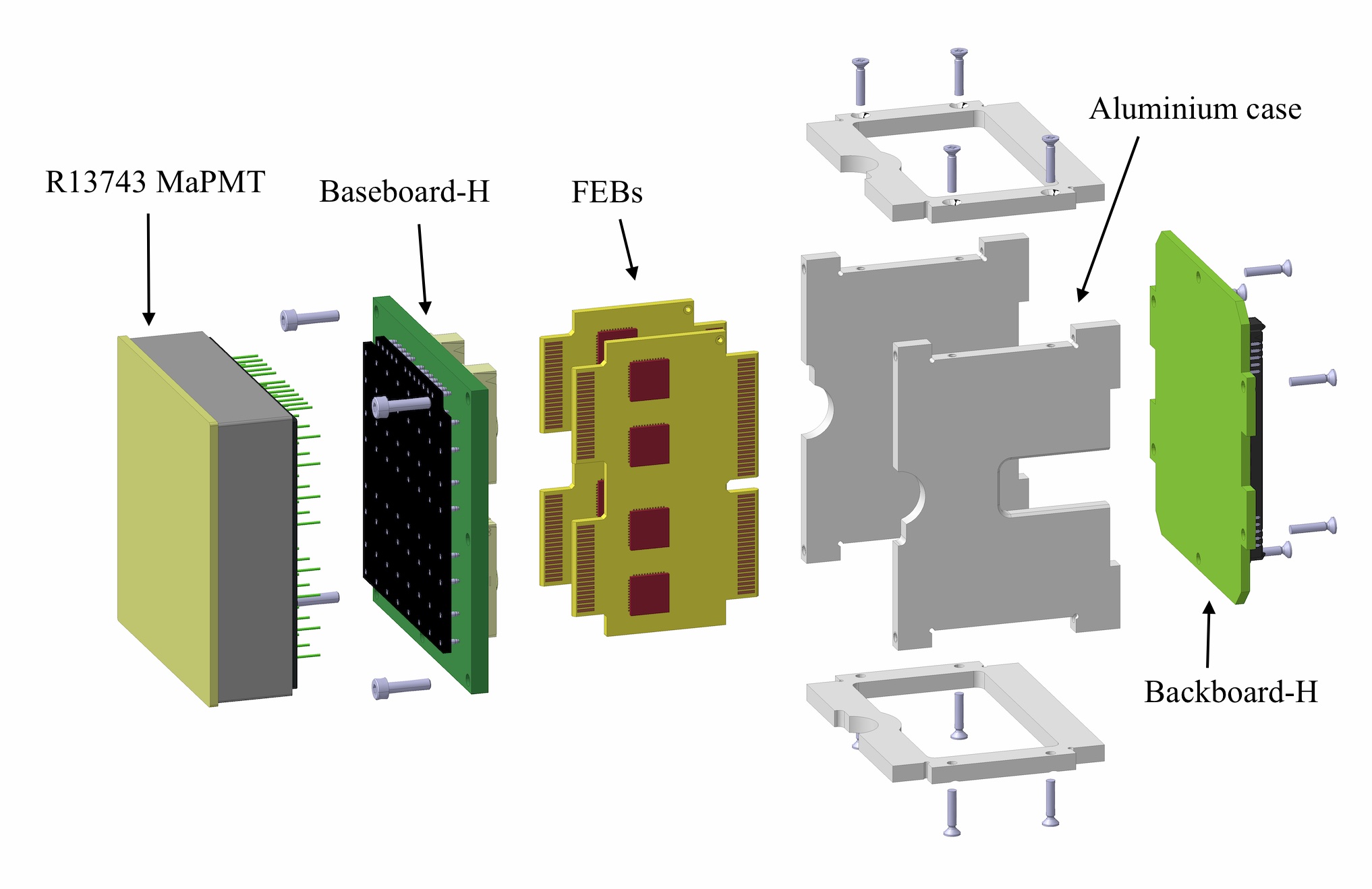}
  \caption{Schematic view of the \Acr[m]{ech}. Reproduced from~\cite{LHCb:2021okg}. CC BY 4.0.}
  \label{fig:RICH:ECHexploded}
\end{figure}

The \Ech, shown in figure~\ref{fig:RICH:ECHexploded}, reads out a
single 2-inch \Mapmt. Accordingly, it consists of a single
2.5\aunit{M$\Omega$} voltage divider and two \Feb{s} with half the
CLARO channels disabled. There are 384 {\Ech}s in the peripheral
region of \richtwo.

\subsubsection{Photon detector module digital boards}
\label{paragraph:richPDMDB}

The \Pdmdb is required to transport the digitised photon detector
signals away from the high-radiation region of the detector without
introducing dead time and ensuring the interface with the \lhcb \Ecs.

An \Fpga-based approach is adopted as a flexible way to capture and
format the data and to interface between the different electrical
signalling standards of the front-end \Asic{s} and \Gbt chipset. A
comprehensive set of measurements at a number of irradiation
facilities, reported in section~\ref{subsec:richIrradiation}, has
demonstrated that the chosen \Fpga\footnote{Xilinx Kintex-7.} is
sufficiently tolerant to the effects of radiation in the \rich
environment, provided certain mitigating design features are
incorporated. Nevertheless, a modular design, with the radiation-hard
components on pluggable modules, allows these parts to be reused in
case it becomes necessary in the future to replace the \Fpga{s}.

Two variants of the \Pdmdb are used, corresponding to the different
granularity of the photon detector planes. A pair of back-to-back
\Pdmdb-Rs is coupled to a group of four \Ecr{s} and a single \Pdmdb-H
is coupled to a group of four \Ech{s}. The assembly of four \Ecel{s}
and one or two \Pdmdb{s} is called a \Pdm. Each \Pdmdb hosts one \Tcm
and up to three \Acr[p]{dtm}, implemented as pluggable mezzanine
boards, following the concept outlined above.

\begin{figure}[t]
  \centering
  \includegraphics[width=0.48\linewidth]{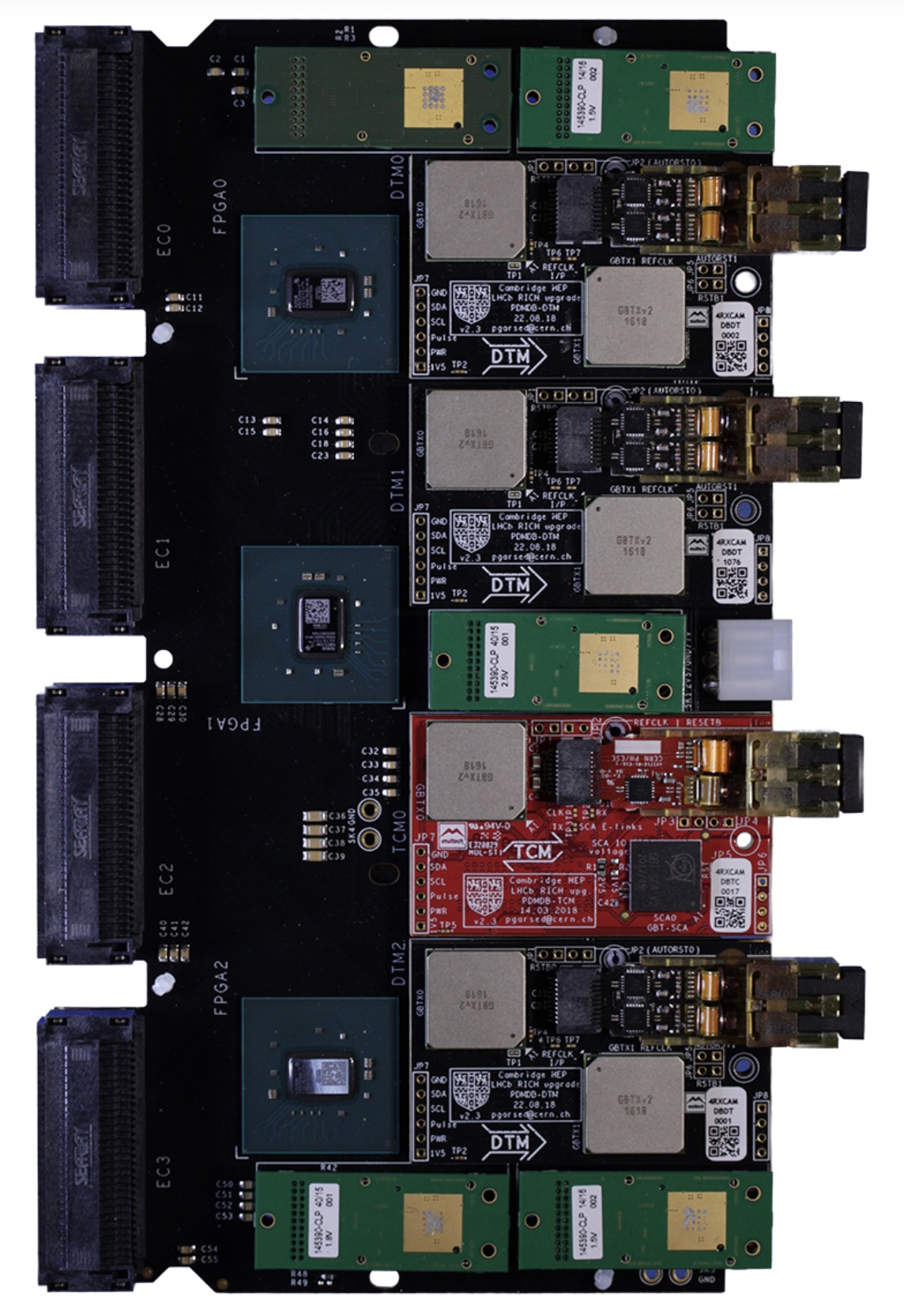}\hfill
  \includegraphics[width=0.42\linewidth]{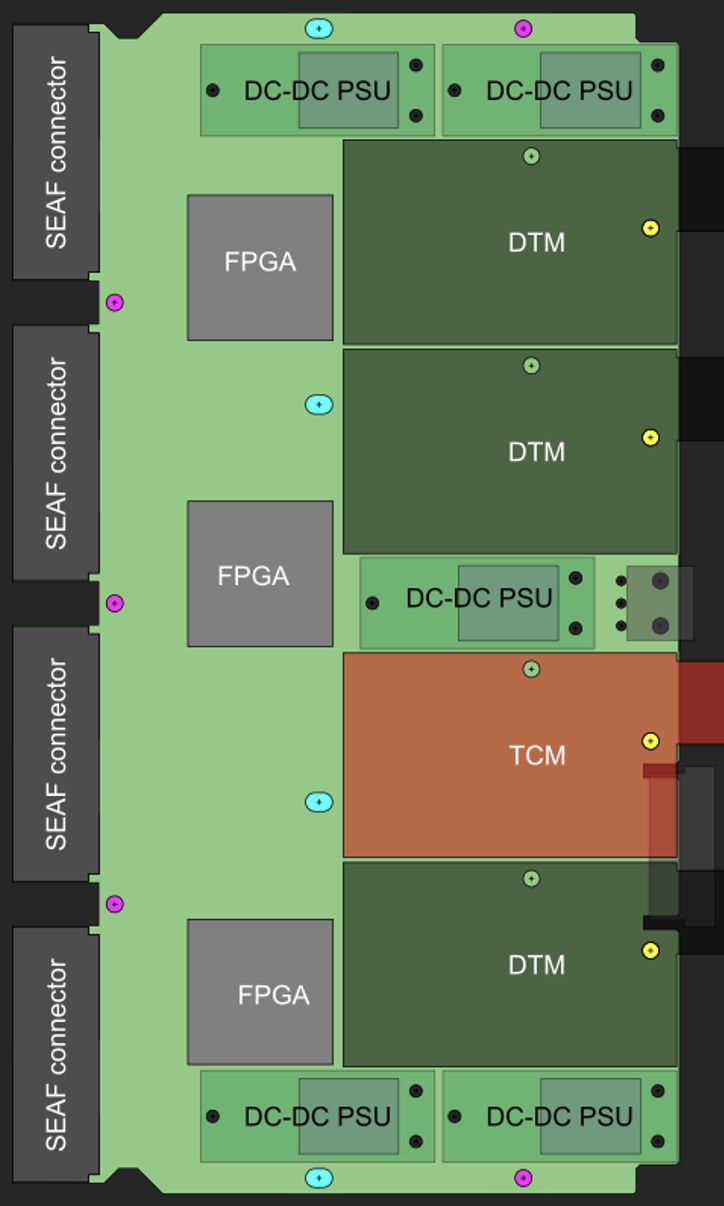}
  \caption{Left: picture of a fully populated \Pdmdb-R
    board. Right:~schematic view of the \Pdmdb-R board main
    components. The \Pdmdb-H differs by having one less \Fpga and \Dtm
    with respect to the \Pdmdb-R.}
  \label{fig:RICH:PDMDBphoto}
\end{figure}

The \Tcm is a $3\times 6\cma$ module that provides an interface for
the fast- and slow-control data exchanged between a \Pdm and the \lhcb
\Ecs. The physical link is implemented using a \Vtrx and \Gbtx{s}
operating in bidirectional forward-error-correction mode. The initial
configuration of the \Tcm is programmed into its e-fuses to ensure
proper operation at power-on.  Configuration protocols provided by the
\Tcm include \I2c to program the \Dtm \Gbtx, \Acr[s]{jtag} to program
the \Fpga{s}, \Acr{spi} to configure the CLARO \Asic{s}, ADCs for
temperature and voltage monitoring for the \Pdm, DACs to generate the
voltage level for CLARO test pulse generation and \Acr[p]{gpio} for
local resets and digital control.

The \Dtm is a $3\times6$\cma plug-in module that provides the
high-speed data transmission interface for the \Pdm. There are three
(two) \Dtm{s} on each \Pdmdb-R (\Pdmdb-H). The physical uplink is
implemented by the \Vttx dual optical transmitter with each channel
connected to a \Gbtx \Asic, each operating in wide-bus transmission
mode. The \Gbtx{s} are configured through their \I2c configuration
port. The two \Gbtx{s} and the \Fpga are connected to a dedicated \Tcm
\I2c bus master.\looseness=-1

The \Pdmdb motherboard acts as a bridge for the signals between the
\Ecel{s} and the \Tcm and \Dtm{s}. The board also incorporates local
power regulation for the \Fpga{s} as well as for the active components
on the \Ecel{s}, \Tcm and \Dtm{s} using CERN \Feastmp \dcdc
converters. The only active components on the motherboard apart from
the \dcdc converters are the \Fpga{s}. These receive the 2.5\volt
\Acr[s]{lvcmos} digital outputs of the CLARO \Asic{s}. No
zero-suppression is applied, thus the \Fpga{s} effectively sample the
CLARO data at 40\mhz and transport the sampled data transparently to
the up-links with constant latency.

\subsubsection{Photon detector columns}
\label{paragraph:richColumns}

The \Ecel{s} and \Pdmdb{s} are arranged into two types of \Pdm: the
\Pdm-R, composed of four {\Ecr}s and two back-to-back \Pdmdb-Rs,
installed in the whole \richone and in the central region of \richtwo;
the \Pdm-H, composed of four {\Ech}s and one \Pdmdb-H, installed in
the peripheral regions of \richtwo.\looseness=-1

\begin{figure}[t]
  \centering
  \includegraphics[width=0.42\linewidth]{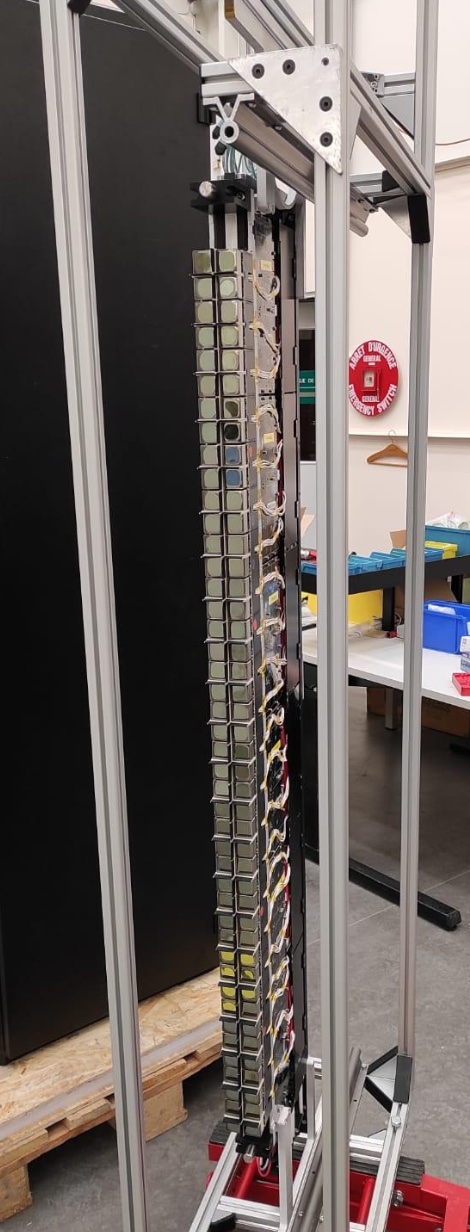}\hfill
  \includegraphics[width=0.54\linewidth]{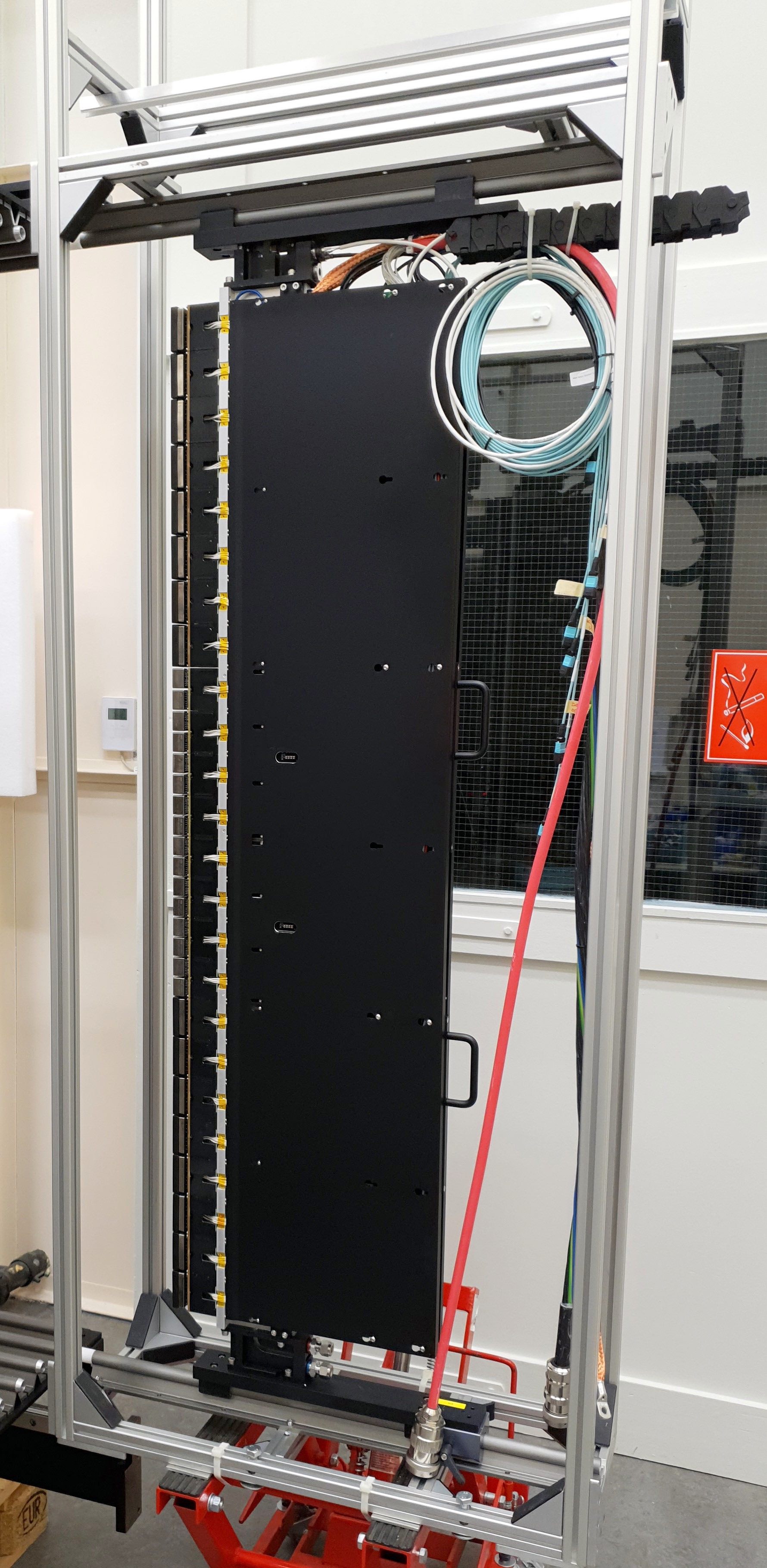}
  \caption{Pictures of (left) a completed \richone column (front view)
    and (right)~a \richtwo column (side view, \Ecel{s} on the left),
    with the photon detector chain and the complete set of services.}
  \label{fig:RICH:columns}
\end{figure}

For both \richone and \richtwo detectors, six {\Pdm}s are assembled on
a T-shaped aluminium structural element, referred hereafter as T-bar,
to build one \rich column, including the distribution of services and
the cooling circuit. In order to minimise the production of specific
mechanical components for \richone and \richtwo, the T-bar is kept
identical between the two detectors. The T-bar provides a precise
reference for the positioning of the \Ecel{s} and \Pdmdb{s}, at the
level of 0.2\mm. The overall length of a column is approximately 1.6\m
with a width of 55\mm and a depth of 40\cm. Fully populated \richone
and \richtwo columns are displayed in figure~\ref{fig:RICH:columns}.

The \Ecel{s} are fixed at the front of the T-bar base and plugged into
the corresponding connectors of \Pdmdb connectors which are mounted on
the T-bar sides. The active components on the \Pdmdb are a significant
source of heat and are therefore cooled by means of an aluminium plate
that provides a thermal coupling with the T-bar. The thermal exchange
is favoured by using commercial thermal pads placed on the active
elements of the \Pdmdb{s}.

\richone columns contain 22 \Ecel{s}, instead of the 24 corresponding
to six {\Pdm}s, since the two \Ecel{s} at the upper and lower end of
each array are not mounted to facilitate the installation, handling
and maintenance, while maintaining the complete acceptance. For the
same reason, only four {\Pdm}s (16 \Ecel{s}) are installed in the
outer column of each \richone photon detector plane. \richtwo columns
are fully populated with 24 \Ecel{s} and have four \Pdm-Hs interleaved
by two \Pdm-Rs.

The cooling of the photon detection chain is achieved by circulating a
fluoroketone coolant\footnote{\Trmk{Novec 649}.} in two 6\mm diameter
ducts deep-drilled into the spine of the T-bar. The cooling keeps the
temperature at the \Mapmt{} surface well below 30\degc.

The instrumented columns require the distribution of services such as
power supply cables, data, \Tfc and \Ecs optical fibres, and
monitoring devices for the slow controls and \Acr{dss}. These services
run on both sides of the T-bar. The \Lv supply is provided by two
power supply channels\footnote{\Trmk{Wiener} MARATON.} for \richone
columns, while one channel is used for the \richtwo columns. Dedicated
distribution boards, that are located at one end of the column,
provide the 2.5\volt supply by means of \dcdc regulators. Each
\Mapmt{} is supplied by a high voltage (900\volt) and an intermediate
voltage (90\volt) to power the last dynode to mitigate possible
nonlinear effects on the gain within the \Mapmt{s}. The \Hv supplies
are provided by common floating ground A1538DN CAEN boards. The \Hv
supply has a common floating ground for the complete column.  Control
signals of the \Pdmdb{s} and data are transmitted through
long-distance optical fibres with twelve-fibre ribbon with
\Acr[s]{mpo}-to-\Acr[s]{lcc} connectors\footnote{\Acr[s]{mpo},
  \Acr[m]{mpo} and \Acr[s]{lcc}, \Acr[m]{lcc} are standard connectors
  for optical fibres.} used to fan-out the optical links to the
individual connectors on the \Pdmdb{s}. To monitor the overall
temperature, a total of 168 (112) temperature probes are installed on
\richone (\richtwo) columns. These temperature sensors are monitored
through the \Ecs as described in section~\ref{paragraph:richDCS}. For
safety purposes and in order to cope with eventual network
disruptions, a thermo-switch (normally closed), is mounted directly on
the T-bar and will issue a \Acr{dss} alarm if the temperature exceeds
35\degc.

At each end of the column, the T-bar is fixed to a trolley composed of
an interface plate and two open cylindrical bearings, made of low
friction and electrically insulating polymer, that slide on
cylindrical rails. In this way it is possible to easily extract the
columns from their operational position for maintenance. The alignment
of the columns inside the corresponding rack can be affected by
mechanical tolerances and differential thermal dilation, that are
compensated by a small degree of freedom of the bearing at the top-end
of each column. This clearance is recovered by preloaded washer
springs acting between the floating trolley and the T-bar end.

\subsubsection{Irradiation campaigns and mitigations}
\label{subsec:richIrradiation}

According to \fluka simulations, at the \rich photon detection system
location during the whole upgrade phase (corresponding to about
50\invfb of integrated luminosity) a total ionising dose of 200\kRad
and a fluence of $3\times10^{12}$ 1\mev\neqcmcm and $1\times10^{12}$
\Acr{heh} per square \cm are expected, where the estimations include a
safety factor of two. Several irradiation campaigns have been carried
out in order to assess the impact of radiation on photon detection
chain components, in particular on \Mapmt{s}, CLARO \Asic{s} and
\Pdmdb \Fpga{s}, taking into account additional safety factors with
respect to the expected dose. Irradiation tests on other components,
such as passive elements, cables and mechanical components, were also
performed, showing no radiation-induced degradation.

Radiation effects on \Mapmt{s} elements have been carefully
studied. Radiation-induced effects on photocathode sensitivity and
secondary emission ratios were found to be negligible. Radiation
damage on optical entrance windows was also investigated.  Multiple
samples of \Mapmt{} windows made of borosilicate and UV-transmitting
glass were irradiated at different particle fluences. UV glass windows
were found to suffer substantially smaller degradation than
borosilicate ones and were therefore chosen for the installed
\Mapmt{s}.

Radiation hardness tests of a CLARO prototype have been performed with
neutrons, X-rays and protons, as described in
refs.~\cite{Fiorini:2014uqa,Andreotti:2015eia}.  Further tests
with ion, proton and mixed-field high-energy beams where performed on
the first two full versions of CLARO, which showed soft \Sel events at
values of \Acr{let} of about 20\mev/\aunit{mg}/\cma.  As a result of
these tests a third version of CLARO was produced, where configuration
registers were resynthesised using cells radiation-hard by
design~\cite{6131335, 6937402}, which exhibited a higher
\Acr{let} threshold for \See with respect to \Acr{ams} standard cells.
Tests on this CLARO version~\cite{8115435} have confirmed that the
design modification was effective, with threshold for \Seu and \Sel
having increased by a factor of three with respect to the previous
versions.

\Fpga{s} are sensitive to \Seu which may flip configuration bits
disrupting the correct operation of controlled devices and of the
\Fpga itself.  Therefore, the \Pdmdb \Fpga{s} have been tested with
various species of ionising particles. In particular, to emulate as
closely as possible the \lhc environment, the \Fpga{s} have been
tested under a mixed neutron and \Acr{heh} irradiation fields at the
CHARM facility~\cite{Thornton:2149417} at \cern. The \Seu
cross-section within the configuration memory has been measured by
counting single- and multi-bit errors arising while emulating a fixed
pattern through the \Fpga logic. The estimated \Seu cross-section over
the full 19\mbit \Fpga configuration memory (CRAM) has been determined
to be $(1.02 \pm 0.37) \times 10^{-7}$
\cma/device~\cite{vladThesis}. No \Sel events were observed in the
tests at CHARM while a \Sel threshold of approximately
15\mev\cma/\aunit{mg} has been found when irradiating the \Fpga{s}
with Kr, Ni and Ar ions. This threshold is considered acceptable but
risk mitigation actions have been put in place as described below.  In
order to further decrease the risk of logic failures, the \Pdmdb
firmware was designed in order to keep minimum complexity reducing the
CRAM usage to about 30\kbit. As much data processing as possible was
shifted to the \Bend \Daq boards, resulting in a worst-case upper
limit of approximately 28 logic failures per
hour.\footnote{\looseness=-1The number is relative to all the \Fpga{s}
  used in \richone and \richtwo when considering the worst case
  scenario irradiation level of RICH1 everywhere, a safety factor of
  four, and averaged over an operational time corresponding to
  50\invfb.}  The \Pdmdb output data are presented to the \Tellfourty
processing logic after a constant delay that simply adds to the
optical fibre propagation delay. As a result, any synchronous
processing required by the data transmission protocol can be safely
performed in the \Tellfourty, therefore saving substantial logic
resources in the \Pdmdb \Fpga{s} and reducing significantly the
probability of radiation-induced upsets. Furthermore, the most
critical parts of the \Fpga logic are protected using an extended
triple modular redundancy technique that also allows selective partial
reconfiguration of the \Fpga without disrupting the logic
operation. Finally, a fast recovery procedure is implemented in the
\Bend electronics and, for redundancy, in the slow control system.

\subsection[RICH1 optical and mechanical systems]{\richone optical and mechanical systems}
\label{subsec:richone}

The upgraded \richone detector, located upstream of the \lhcb dipole
magnet between the \Velo and the \Ut, underwent major design changes
and a subsequent rebuilding.  Nevertheless, the fluorocarbon
C$_4$F$_{10}$ gas radiator was retained and the angular acceptance was
kept unchanged.

A schematic (CAD model) of the \richone detector is shown in
figure~\ref{fig:RICH:layout}. \richone is aligned to the LHCb
coordinate axes and occupies the region beyond the \Velo exit window
947.5\mm$\le z \le 2245$\mm and $\pm$920\mm in $x$. The $z$ axis
follows the beam line which is inclined at 3.6\mrad to the horizontal.

\begin{figure}[t]
  \centering
  \includegraphics[width=0.45\linewidth]{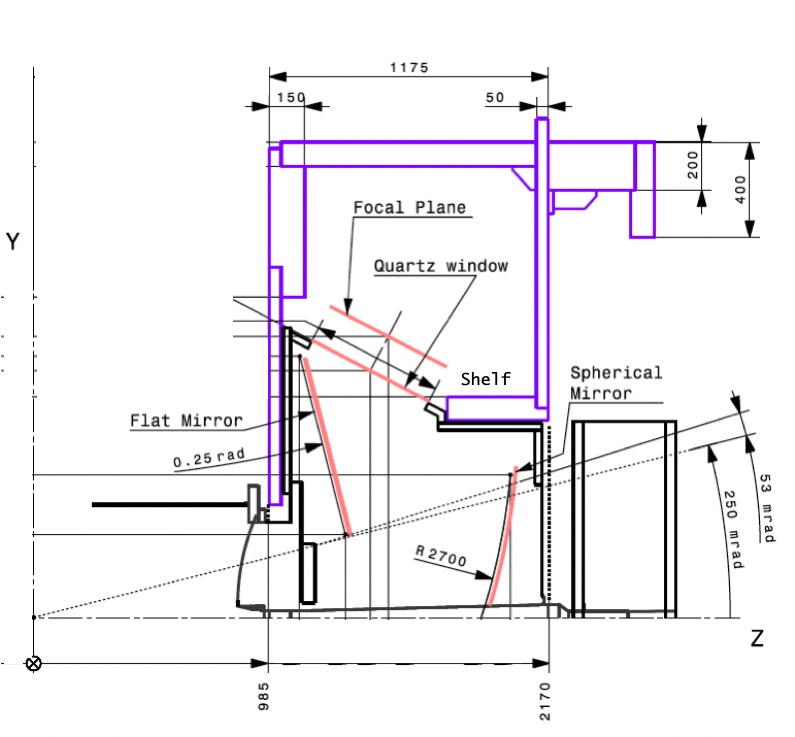}\hfill
  \includegraphics[width=0.45\linewidth]{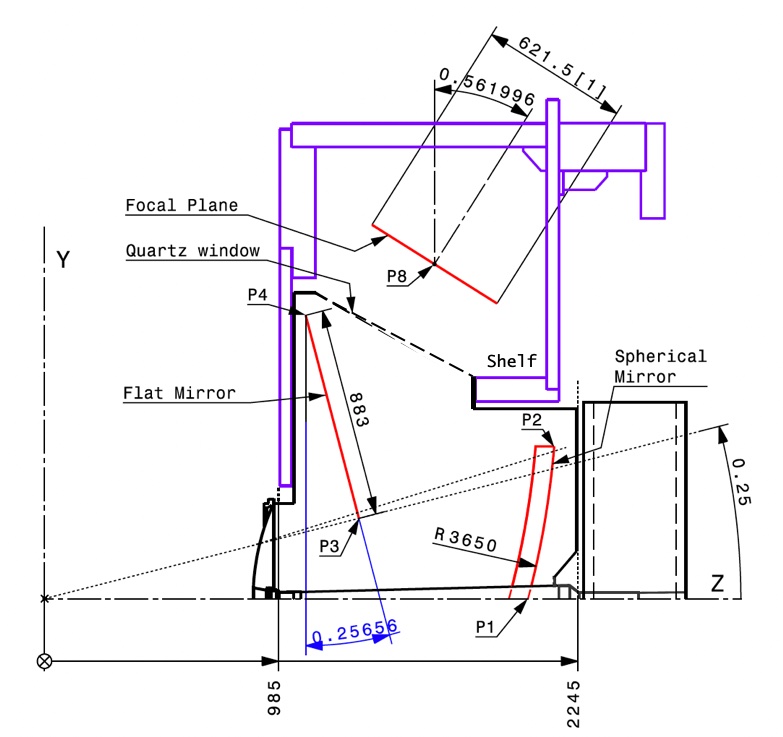}
  \caption{The optical geometries of (left) the original and (right)
    the upgraded \richone.}
  \label{fig:RICH1optics}
\end{figure}

The optical layout of the upgraded \richone has been modified to
reduce the larger hit occupancy expected in the central region of the
detector. The occupancy has been halved by increasing the focal length
of the spherical mirrors by a factor of approximately $\sqrt{2}$,
which also improves the Cherenkov angle resolution by reducing mirror
aberrations.  A comparison of the previous and upgraded optical layout
is shown in figure~\ref{fig:RICH1optics}. As a consequence of the
increased focal length, the photon detector planes have been moved
parallel to and outwards from the beam line by approximately 270\mm
(225\mm in $y$, 150\mm in $z$).
To minimise the material budget within the acceptance, lightweight
carbon-fibre spherical mirrors are used and all other components of
the optical system are located outside the acceptance. The average
\richone material budget inside the LHCb acceptance is $\sim4.8$\%
\Xrad. Planar (flat) mirrors reflect the image from the tilted
spherical mirrors onto the photon detector planes.\looseness=-1

The arrays of \Mapmt{s}, described in
section~\ref{paragraph:richColumns}, are located at the upper and
lower focal planes, and each array occupies an active area of
$605 \times 1199\mma$. The \Mapmt{s} are shielded from the
60\aunit{mT} fringe field of the \lhcb dipole by magnetic shielding
boxes made of ARMCO$^{\text{\tiny\textregistered}}$ iron, placed above
and below the beam line outside the LHCb acceptance. These shields are
retained from the original \richone detector, however with the
so-called \emph{shelves} cut off by 70\mm to ensure photon
acceptance. To allow \Mapmt{} column extraction and insertion for
maintenance and installation, additional apertures of
$666 \times 462\mma$ and $810\times 56\mma$ were machined on the sides
of each shielding box and covered with removable 10\mm thick plates.
Inside the shielding boxes, the \Mapmt{s} have additional local
mu-metal shielding as described in section~\ref{paragraph:richEC} and
are able to work efficiently in fields of 3\aunit{mT}. To guide the
new design, the magnetic field was simulated with the
\textsc{Opera}/\textsc{Tosca} software and later measured at a
position displaced by about 10\cm from the nominal \Mapmt{}
plane. These studies confirmed that the \Mapmt{s} will operate in a
magnetic field in the range 0.6-2.2\aunit{mT}, with the axial field
below 1\aunit{mT} for all \Mapmt{s}.

\subsubsection{Gas enclosure}
\label{paragraph:richOneGasEnclosure}

\begin{figure}[t]
  \centering
  \includegraphics[height=0.595\linewidth]{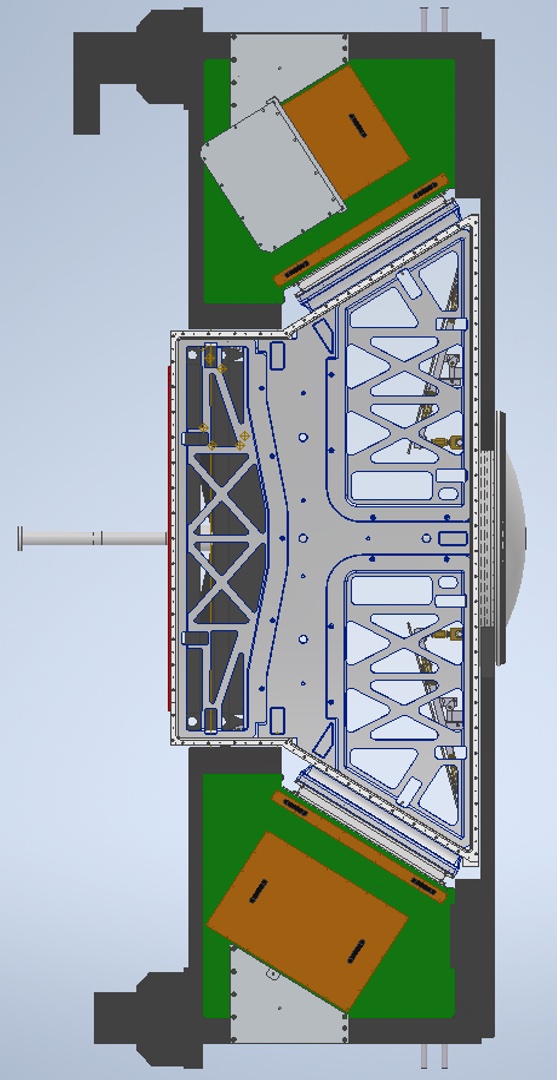}\hspace{2cm}
  \includegraphics[height=0.60\linewidth]{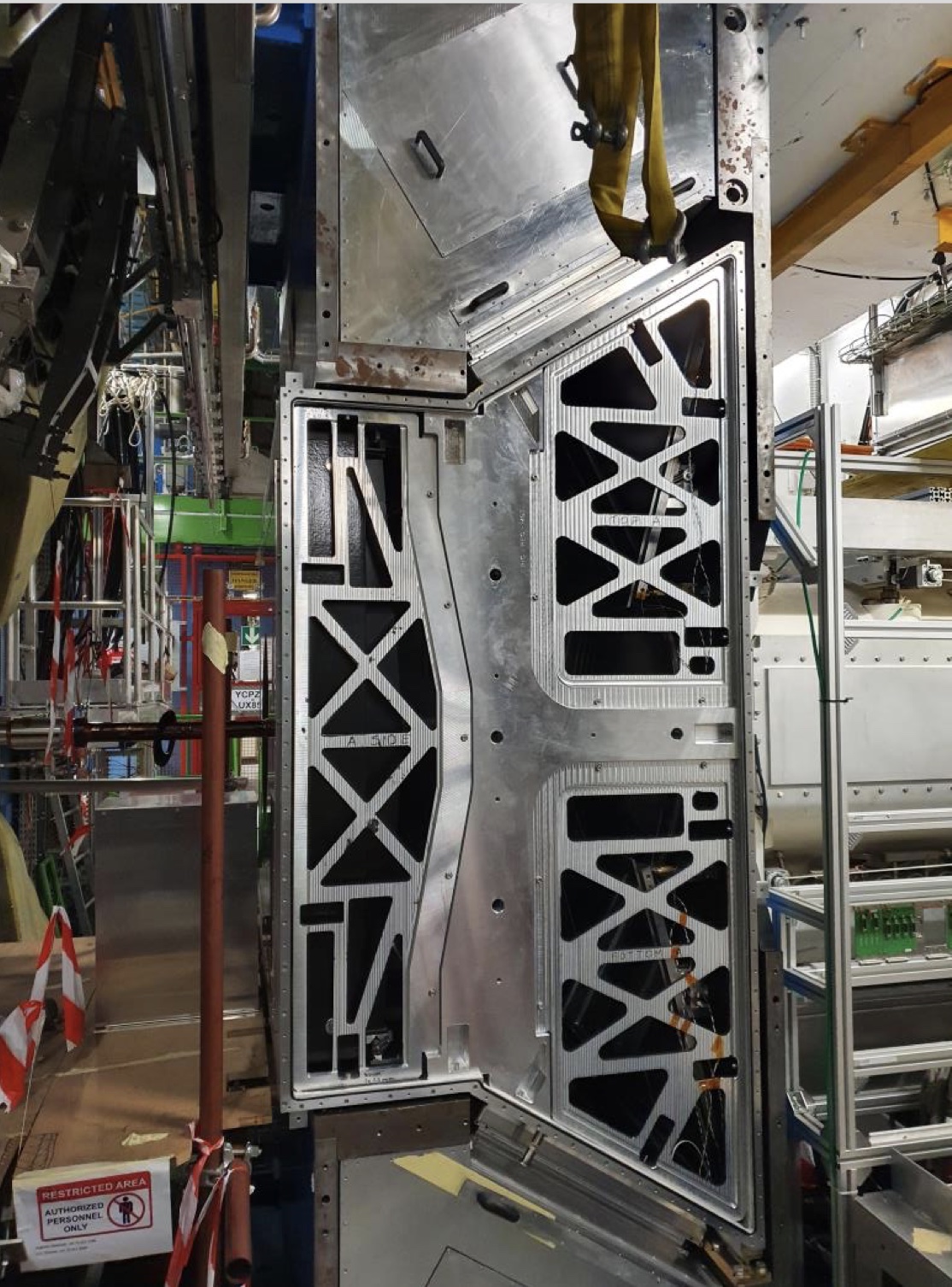}
  \caption{Left: side view CAD layout of the \richone gas enclosure.
    Right:~photo of the \richone gas enclosure after its installation
    in the \lhcb cavern.}
  \label{fig:RICH1-gas-enclosure}
\end{figure}

The purpose of the gas enclosure is to contain the C$_4$F$_{10}$ gas
radiator, to provide an optical bench for all optical components, and
to ensure gas and light tightness. A schematic and a picture of the
gas enclosure are shown in figure~\ref{fig:RICH1-gas-enclosure}. The
C$_4$F$_{10}$ radiator gas pressure follows the atmospheric pressure
within $\pm 3$\mbar. The total gas volume is approximately 3.8\mv.

The enclosure is machined from 30\mm thick aluminium alloy tooling
plates. The six sides are bolted and epoxy-sealed at their edges, and
internally sealed with flexible silicone sealant\footnote{Bluestar
  CAF4 Silicone Sealant.} to ensure leak tightness. The side faces of
the gas enclosure are open to allow access for installation of mirrors
and to the beam pipe. The structure has removable stiffening
hatch-plates at the side apertures to prevent deflections of the
structure when the side panels are removed, following loading with the
mirrors, or when under ambient operational pressure. During normal
operation, the sides are sealed by 15\mm aluminium panels. The maximum
deflection of the superstructure is limited everywhere to 150\mum.\looseness=-1

The upstream and downstream faces of the gas enclosure have apertures
to allow passage of particles with minimum scattering within the LHCb
acceptance. The upstream face attaches to 300\mum thick stainless
steel bellows, the so-called \emph{\Velo seal}, which provides a
gas-tight, mechanically compliant (longitudinal $\pm10$\mm and
transverse $\pm1$\mm) seal to the downstream face of the \Velo vacuum
tank

The downstream face is closed by a low-mass (16.2\mm thick, estimated
0.7\% \Xrad) exit window manufactured from a sandwich of two 0.6\mm
thick carbon fibre skins filled with 15\mm of foam.\footnote{Airex
  R82.80.}  The window is sealed to a flange (a fixed fin machined in
the beryllium beam pipe) using a 1\mm thick opaque moulded silicone
diaphragm,\footnote{Dow Corning Sylgard 186, with 5\% black pigment
  added.} as used in the original
\richone~\cite{LHCb-DP-2008-001}. The deflection of the window due to
the gas enclosure pressure differential is approximately
$\pm0.8$\mm. All removable aperture covers (\Velo seal, exit window,
quartz window and side doors) are sealed with 4.5--5.8\mm diameter EPDM
O-rings.\footnote{Ethylene Propylene Diene Monomer (M-class) rubber.}
The gas enclosure is supported by the lower magnetic shield through
mounts that allow its alignment to the nominal beam line. Unloaded
with optical components but including the quartz windows, described
below, the overall weight of the gas enclosure is approximately
1130\aunit{kg}.

Square apertures above and below the beamline allow Cherenkov light to
reach the \Mapmt{s}, located behind. The apertures are sealed with
polished fused silica windows 8\mm thick, of dimensions
655$\times$475\mma. The windows are each fabricated from three
equal-size panes, glued together along one edge and then glued into an
additional frame.  The six quartz panes were individually coated with
an antireflective coating with approximately a quarter wavelength of
MgF$_2$, which provides a gain in transmitted light at 270\nm of an
additional 3.5\%.  The transmission has been measured on test samples
over the range of interest of 270\nm to 500\nm to be better than 95\%.

\subsubsection{Mirrors}
\label{paragraph:richOneOptics}

\begin{figure}[t]
  \centering
  \includegraphics[width=0.609\linewidth]{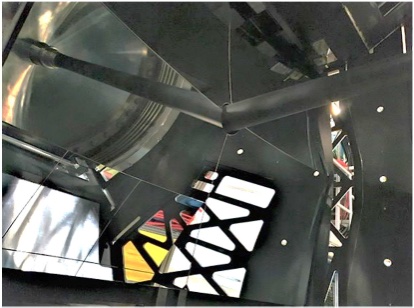}
  \includegraphics[width=0.342\linewidth]{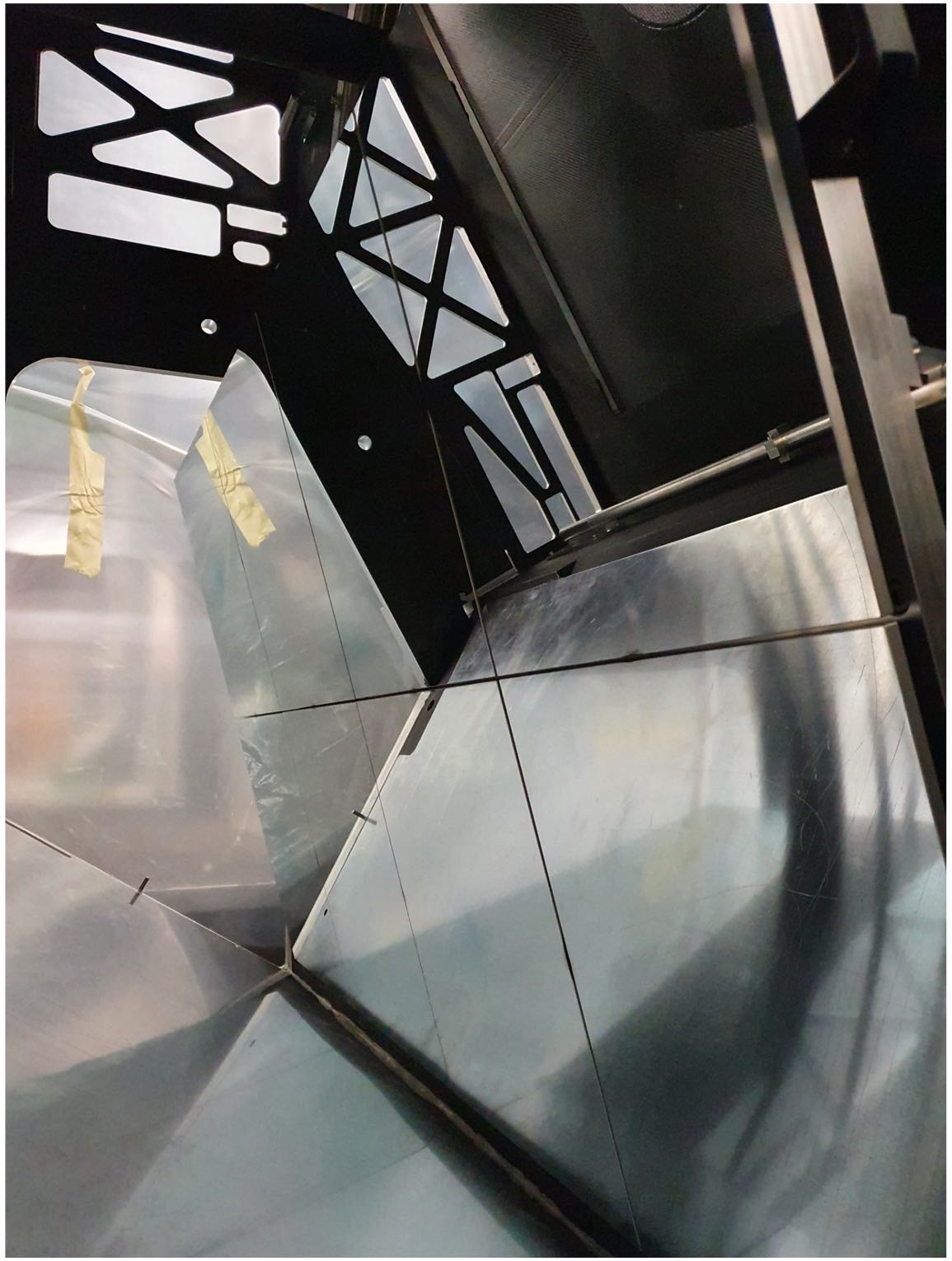}
  \caption{Pictures of (left) the spherical and (right) the bottom
    flat mirror assemblies.}
  \label{fig:RICH:richMirrors}
\end{figure}

Four tilted \Acr{cfrp} spherical mirrors and 16 glass planar mirrors
are used to focus the Cherenkov photons onto the photon detector
planes, positioned outside the \lhcb detector acceptance. Pictures of
the mirror assemblies are shown in figure~\ref{fig:RICH:richMirrors}.

Each spherical mirror has a width of 740\mm, a height of 650\mm, a
thickness of 33\mm and a radius of curvature of 3650\mm. Each mirror
has a weight of 2.9\aunit{kg}. The mirrors are arranged into four
quadrants centred around the beam line. The inner corner of each
mirror has a quarter-circle cutout in order to accommodate the beam
pipe with a clearance of approximately 12\mm. In order to allow a
mirror alignment with a precision of the order of tenths of mrad, each
mirror is supported at the outer three corners by means of spherical
rod end adjusters, bolted to a \Acr{cfrp} frame made of a 2-inch
square tubular structure. The \Acr{cfrp} frame is divided along the
$y$ axis into two C-shaped halves, each one supporting two mirrors
positioned in the vertical direction, with a mirror-to-mirror
separation of 3\mm to allow for alignment and as a clearance for
deformations. The top (bottom) pair of mirrors are aligned to point to
the same top (bottom) centre-of-curvature. Each C-shaped half has a
weight of approximately 10\aunit{kg}, including the weight of the
mirrors, and is bolted to V-shaped blocks sitting on a cylindrical
load rail positioned at the floor of the gas enclosure.

The planar mirrors have a width of 370\mm, height of 440\mm, thickness
of 8\mm and a radius of curvature larger than 60\m. Each mirror has a
weight of 3.3\aunit{kg}. The 16 mirrors are arranged in two sets of
eight mirrors each, positioned outside of the detector acceptance
above and below the beam line, with a mirror-to-mirror separation of
3\mm as in the case of spherical mirrors. The tilt of the plane is
257\aunit{mrad} to the vertical. Each mirror is bonded at its centre
to a polycarbonate mount, bolted into machined pockets on four rigid
1-inch thick aluminium support frames, each frame supporting four
mirrors. In addition, a polycarbonate ring centred on each mirror is
bonded over a larger area than the polycarbonate mounts and secured to
the support frame to retain the mirrors in case of failure of the
polycarbonate mounts.

Each support frame with its mirrors weighs approximately 43\aunit{kg}
and is bolted to V-shaped blocks sitting on a rail bolted to the front
panel of the gas enclosure.

The mirror quality is characterised by the diameter $D_0$ of the
circle which contains 95\% of the light intensity from a point source
placed at the mirror \Acr{coc} imaged at the \Acr{coc} of the
mirror. The $D_0$ for all planar and spherical mirrors was found to be
better than the specification of 2.5\mm. The spherical and flat
mirrors were individually aluminised with 10\nm chromium (adherence
layer) and 100\nm aluminium (reflective layer). Additional enhancement
in reflectivity and protection against oxidisation were ensured by
coating the mirrors with 70\nm SiO$_2$ and 60\nm HfO$_2$. The
reflectivity of all mirrors is everywhere $>90$\% in the wavelength
range 260\nm to 500\nm, peaking around 95\%.

There are three stages to the mirror alignment process: prealignment
on the optical rig before installation, survey in situ, and alignment
with data.  The prealignment on the optical rig is crucial to the
process as the \Acr{coc} of the mirrors are inaccessible when the
mirrors are installed in \richone. At this stage the upper and lower
pairs of the two spherical mirrors are aligned to a common
\Acr{coc}. For the planar mirrors, the top and bottom set of mirrors
are aligned to form a single plane parallel to both support frames
which will point to the corresponding photon detector plane.  Survey
points on the spherical mirror frame and on the flat mirror backing
plates then reference the \Acr[p]{coc} and the flat mirror tilts
respectively.

\subsubsection{Photon detector region}
\label{paragraph:richOnePmtEnclosure}

The \Mapmt{} columns make use of common cooling and electronics, but
have custom support mechanics and services. Eleven \richone columns
are arranged side by side to form an 11$\times$22 array of elementary
cells. One such array is placed below the beam pipe, with a second
above. Both arrays are horizontal in the plane perpendicular to the
beam pipe, and are tilted with an angle of 562\mrad with respect to
the vertical towards the interaction point.

\begin{figure}[t]
  \centering
  \includegraphics[height=0.4\textheight]{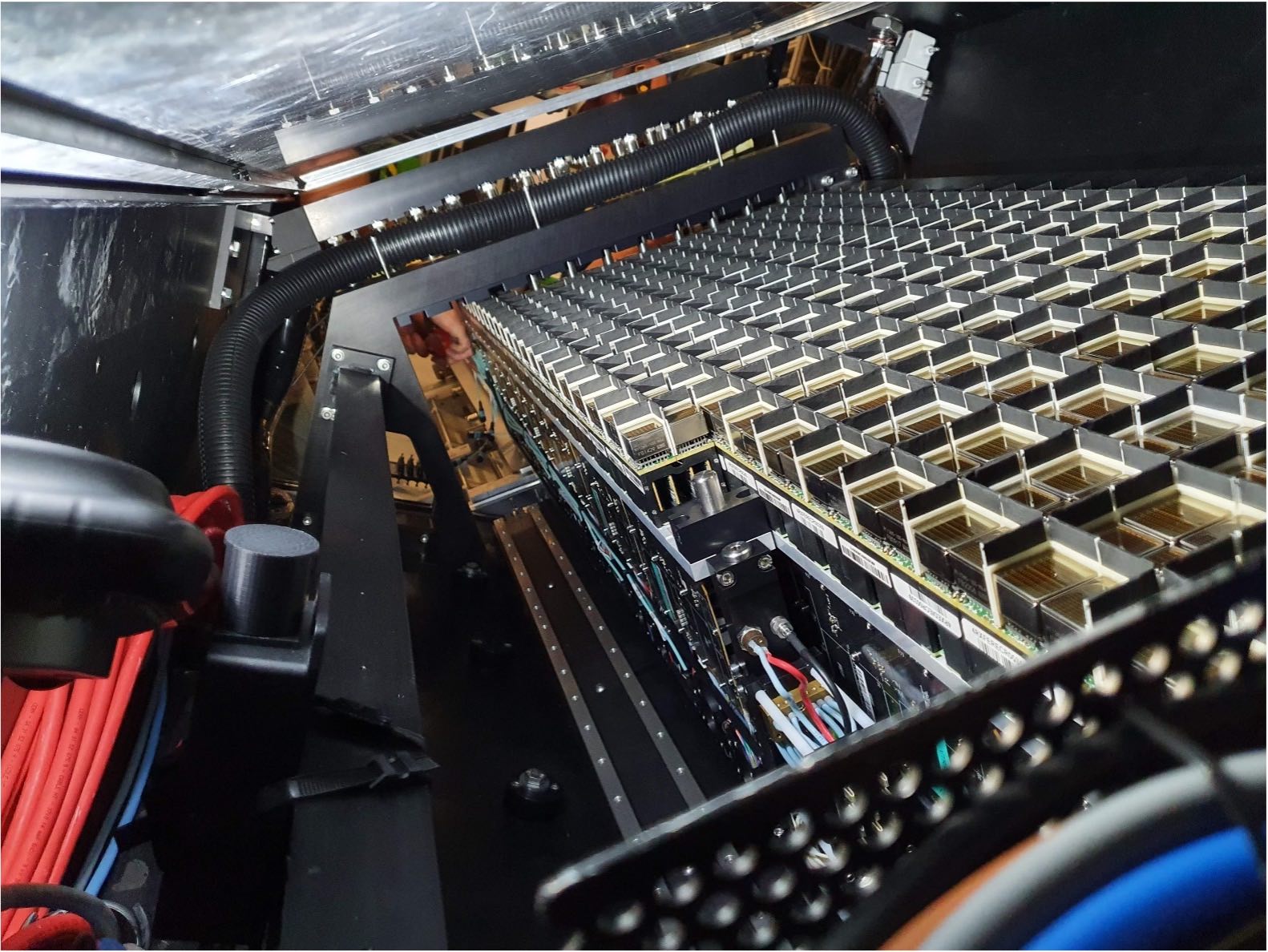}
  \caption{Picture of \richone columns while being inserted into their
    support structure, the lower \Mapmt{} chassis. The rails and
    alignment structures are visible on the left side. The chassis is
    mounted to the soft-iron magnetic shielding that surrounds the
    \Mapmt{} region. The push connector for the copper-carried
    services is visible as well in the left-most column. Reproduced from~\cite{Okamura_2022}. \textcopyright\ 2017 IOP Publishing Ltd
and Sissa Medialab.}
  \label{fig:RICH1-chassis}
\end{figure}

\richone columns are supported from the face opposite the \Mapmt{s}
and held in place on rails allowing for easy removal for
maintenance. Pivoting around the rails is prevented by precision
alignment pins at both ends of the columns on the same face as the
\Mapmt{s}. The rails are held at the correct angle and aligned on the
\Mapmt{} chassis, a mechanical support structure which acts as an
optical bench for the columns. At each end, the chassis has precision
slots to match the alignment pins mounted on the columns. A picture of
the lower chassis hosting the corresponding columns is shown in
figure~\ref{fig:RICH1-chassis}.

Columns are inserted and removed along the rails from one side of the
chassis. At the far end, services carried on copper cables
(low-voltage, high-voltage and monitoring cables) are connected to the
column through a push connector. Services that require manual
disconnection (cooling and data fibres) are attached to the column at
the extraction side. This allows the column to be removed for
maintenance by accessing only one side of the detector.

\subsection[RICH2 photon detector planes]{\richtwo photon detector planes}
\label{subsec:richtwo}

\richtwo is located downstream of the dipole magnet, covering an
angular acceptance of $\pm 15 \text{--} \pm 120$\mrad in the
horizontal direction and $\pm 15 \text{--} \pm 100$\mrad in the
vertical direction. It extends between 9500 and 11832\mm along the
$z$-axis. The photon detector planes are located on the \lhcb detector
sides.

The superstructure including the large entry and exit windows, the two
magnetic shields located on each side and the optical system (mirrors
and their supports) inside the enclosure and the quartz windows are
retained without any modification~\cite{LHCb-DP-2008-001}. These
components have been demonstrated to sustain the occupancy and
radiation levels expected in \lhcb upgrade running conditions.

The main upgrade of \richtwo concerns the installation of the new
photon detector planes. The integration of the new photon detection
chain and its services in \richtwo requires a new mechanical structure
to host the photon detector planes.
\begin{figure}[b]
  \centering
  \includegraphics[width=0.49\linewidth, trim={0 2cm 0 0}, clip=true]{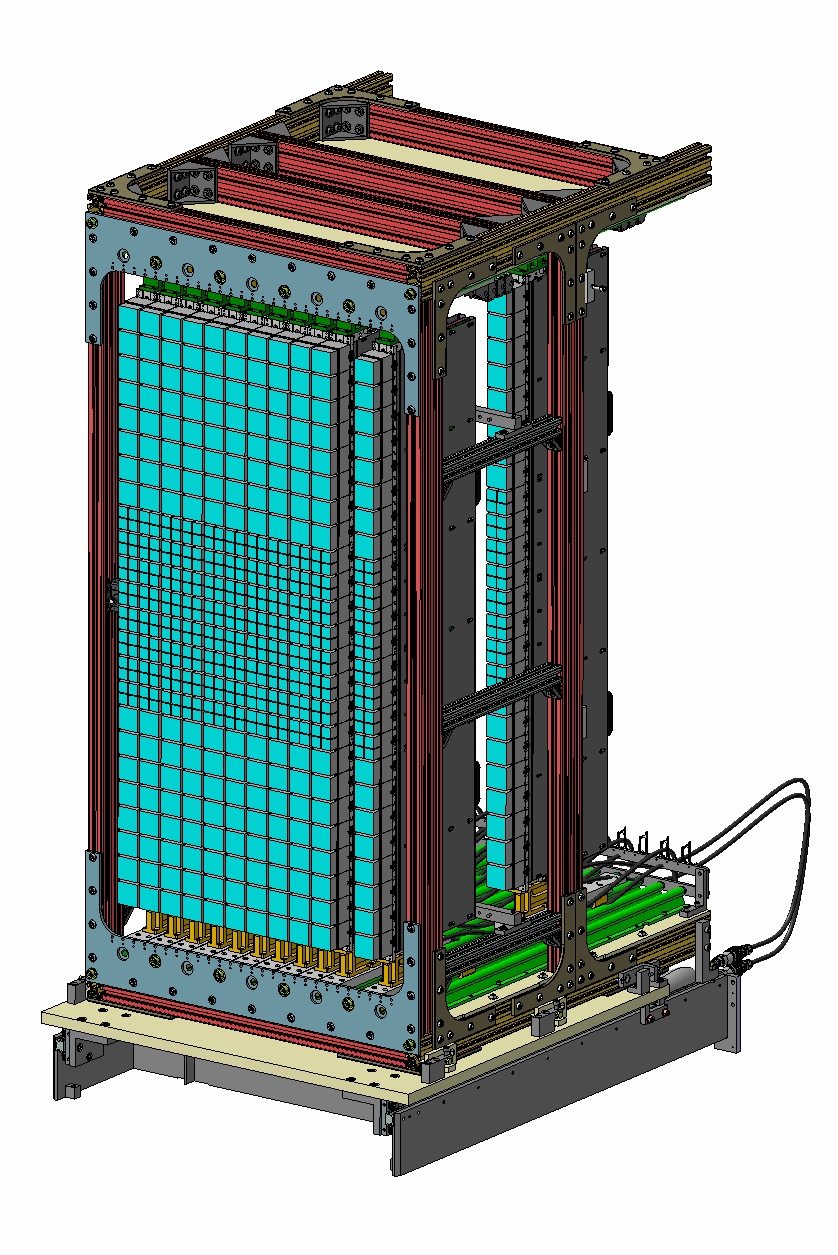}
  \includegraphics[width=0.49\linewidth]{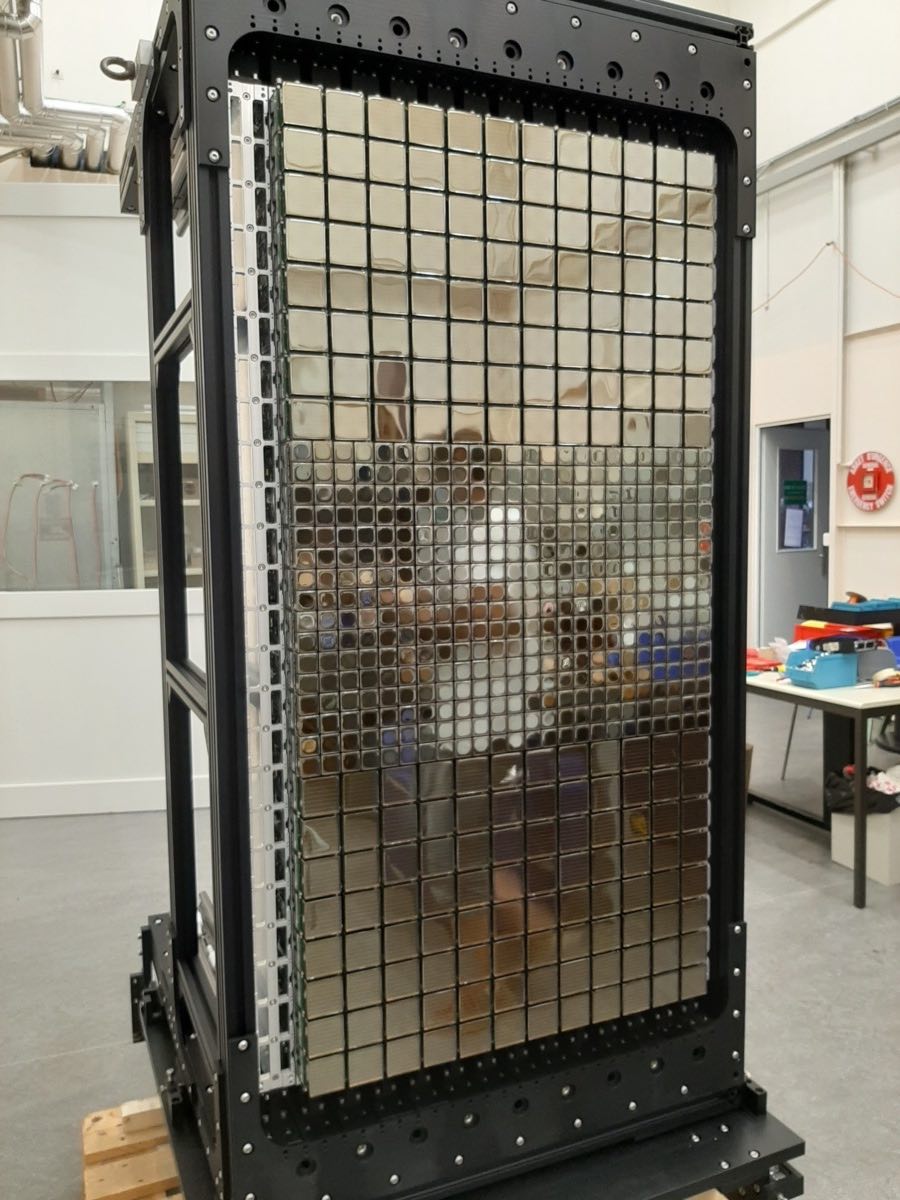}
  \caption{Fully assembled and commissioned \richtwo photon detector
    array. Left: CAD view; right: photograph taken in the assembly
    area. Reproduced from~\cite{Okamura_2022}. \textcopyright\ 2017 IOP Publishing Ltd
and Sissa Medialab.}
  \label{fig:RICH:richTwoArray}
\end{figure}
The structure holding the columns is made of aluminium profiles,
allowing to place up to fourteen \richtwo columns side by side with a
clearance of 1\mm between them. This arrangement of the columns form
the array of the photon detectors. Twelve columns were found to be
enough to cover the detector acceptance and were installed, with two
empty slots at the periphery, as shown in
figure~\ref{fig:RICH:richTwoArray}, for a total of 288 \Ecel{s}.

Two such structures, or \emph{racks}, are installed into the magnetic
shields on the \lhcb \aside and \cside. The arrays are placed
vertically and parallel to the \richtwo structure which is not exactly
perpendicular to the \lhc beam. Both arrays are tilted by an angle on
the horizontal plane of 1.065\rad with respect to the \lhcb $x$ axis,
with the first column towards the interaction point being the farthest
from the beam pipe.

Each rack is installed on a trolley, mounted on rails, which allows
the movement of the complete photon detector system perpendicular to
the focal plane for installation, maintenance and dismounting of the
rack. Furthermore, anchor points between the rack and the trolley,
allow to adjust the position of the racks.  Inside each rack, top and
bottom rails allow to slide each column individually in the direction
perpendicular to the focal plane. The rails are made of hard anodised
aluminium and fastened on base plates.  On the upper side of each
rack, cable chains connected to all columns route the electric cables
and optical fibres to the patch panel located above the rack. Thanks
to this design, each column can be extracted fully independently,
allowing to continue the operations on the other columns. This
arrangement greatly simplifies installation and maintenance with
respect to \runonetwo setup.\looseness=-1

Two cooling manifolds are located under the trolley. The first
distributes the cooling fluid to each column in parallel while the
second collects back the fluid. The manifolds are connected to the two
transfer lines via custom-made feedthroughs. Each column is connected
with two polyurethane hoses and double shut-off couplers to the
manifolds. The couplers offer the possibility to disconnect a column
avoiding purging the fluid or closing any valve.

A patch panel installed at the top of the photon detector enclosure
provides the interface where all services are connected to the
columns. It provides the connections to the safety ground, \Lv, \Hv,
\Acr{dss}, \Tfc, \Ecs and data transmission lines. The interface has
been designed to provide an easy way to disconnect the columns and to
ensure a light- and gas-tight enclosure. Nitrogen is flushed
permanently to ensure a dry atmosphere in the enclosure, minimising
the risk of condensation and therefore maintaining a good dielectric
environment.

\begin{figure}[t]
  \centering
  \includegraphics[width=0.528\linewidth]{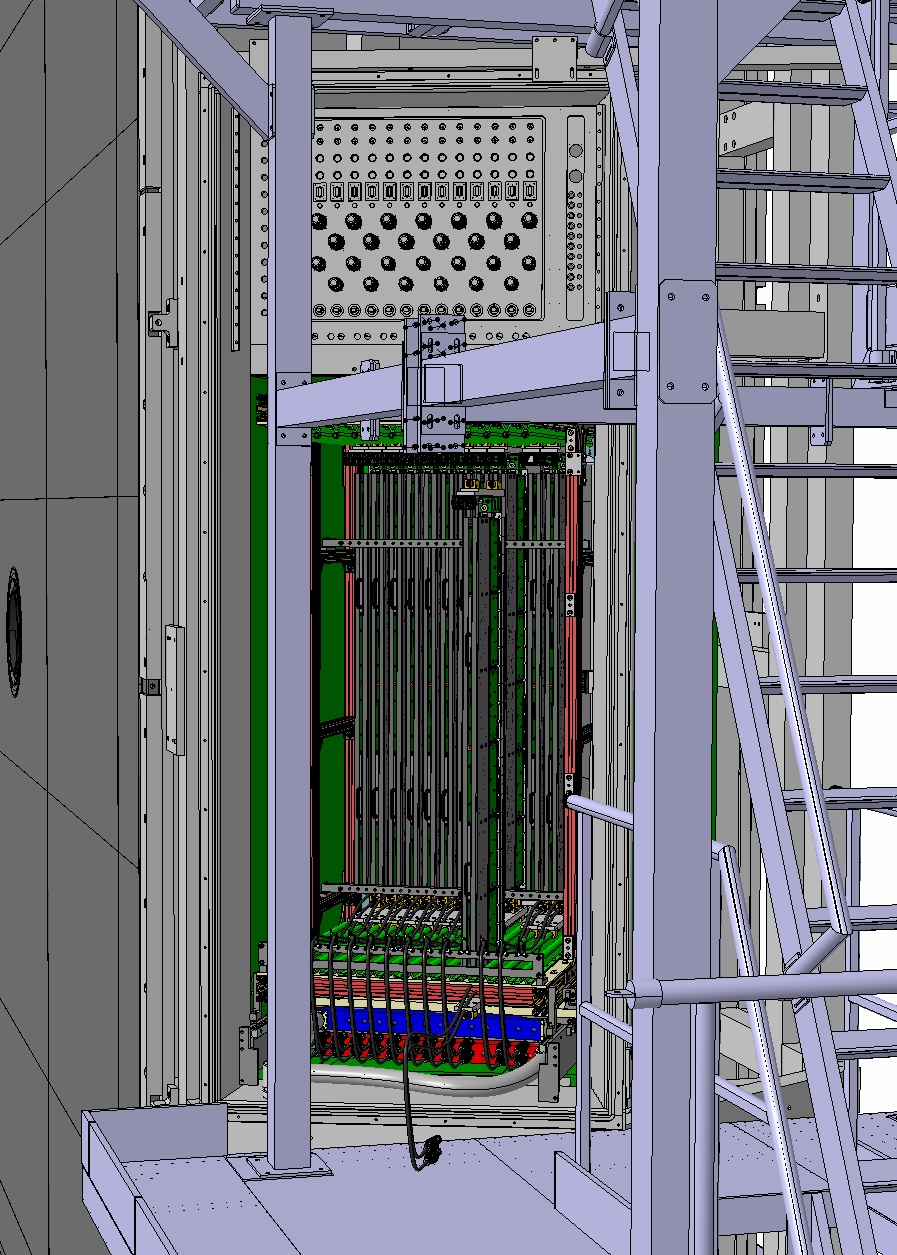}\hfill
  \includegraphics[width=0.415\linewidth]{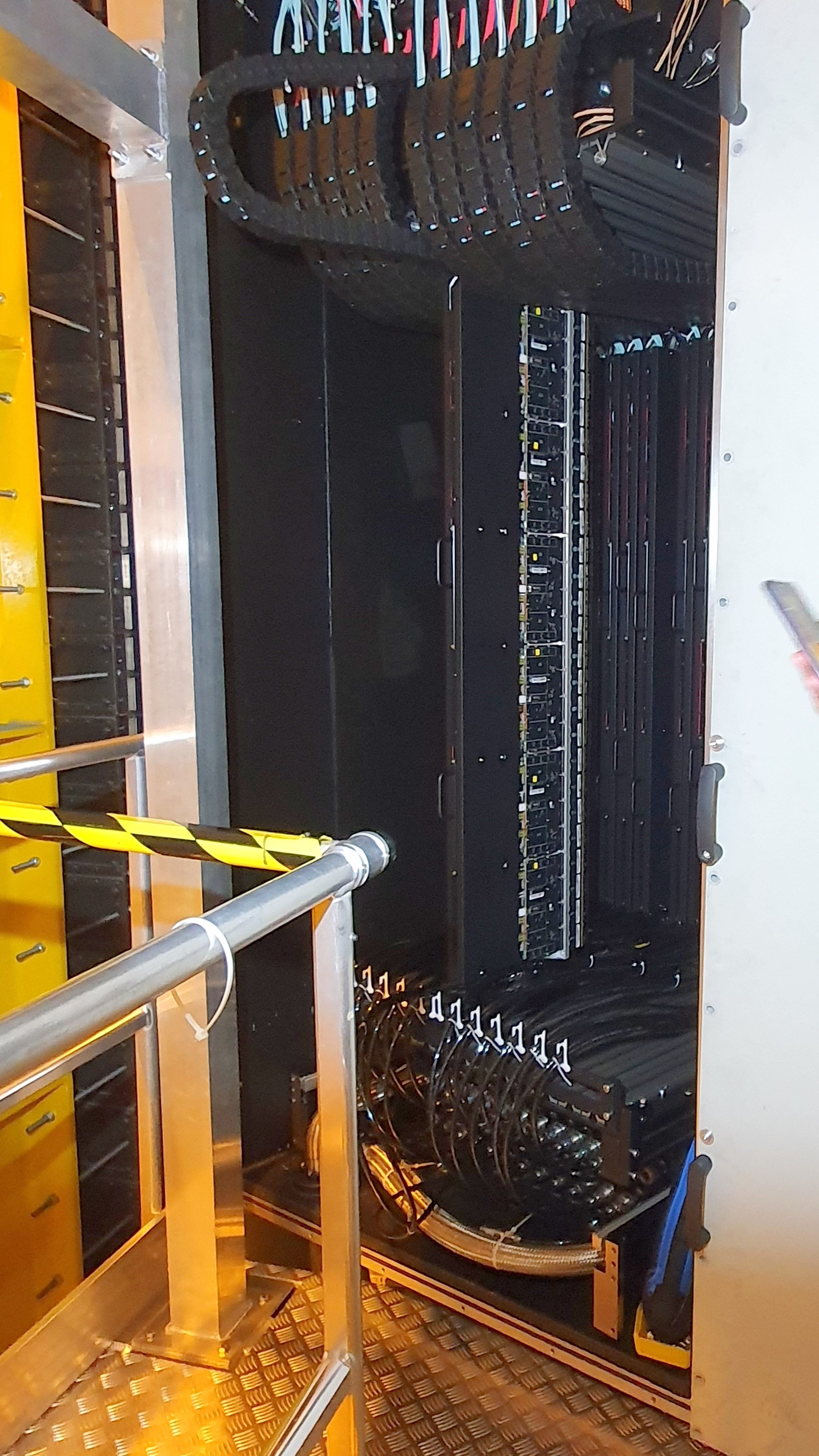}
  \caption{\richtwo photon detection system inside its
    enclosure. Left: CAD view; right: photograph.}
  \label{fig:RICH:richTwoEnclosure}
\end{figure}

\subsection{Monitoring, controls and data acquisition}
\label{subsec:richMonitoringAndControls}

Monitoring, controls and data flow of the \rich system are fully
integrated in the LHCb online structure, with \Tfc and \Ecs
controlling timing and detector control and configuration, and optical
links transporting data from \Fend to \Bend electronics.

\subsubsection{Detector control system}
\label{paragraph:richDCS}

The \rich \Dcs is a subset of the \Ecs that controls \Lv power
supplies and monitors detector safety and environmental parameters for
both the photon detection system and the gas radiators. A variety of
sensors are interfaced using an
\Acr{elmb}~\cite{Boterenbrood:690030}, while a large number of
temperature sensors mounted on baseboards, \Feb{p} and backboards are
read out using the ADC of the \Sca readout chip with a resolution of
approximately 0.5\degc. As these temperature readings are only
available when the detectors are operational, a separate \pthundred
sensor mounted on every photon detector column, read out via the
\Acr{elmb} in a 4-wire configuration, provides additional information
when the detectors are switched off. In addition, two temperature
sensors are installed at the input and output of each cooling
manifold. A further check on the circulation of the cooling fluid is
performed by using two pressure sensors per cooling manifold.
Temperature and humidity in the photon detector enclosure are
monitored with dedicated sensors. A safe switch-on procedure is
implemented, where automatic configuration of the \Sca readout chip at
power-up allows the read out of the temperature sensors without any
action of the operator. If the automatic configuration fails, the
detectors are switched back off.

When the detector is operational, parameters from sensors allow to
constantly monitor the electronics, the detector environment and the
condition of the cooling system to ensure safe operation. These
parameters are used by the \Ecs to check for possible signs of
abnormal conditions and take automatic actions.  Possible actions
include switching off the \Lv system and in extreme conditions also
the \Mapmt{} \Hv system. Finally a smaller number of sensors are
connected to the \Acr{dss} running on a PLC system with many
redundancies, ensuring an additional layer of detector safety.

The \Dcs also collects information about the temperature and pressure
of the Cherenkov gas radiators, as any change in the gas density
affects directly the refractive index. Both \rich radiator gases are
left free to follow the atmospheric pressure variations. Their
temperature is the same as the \lhcb cavern temperature, which is kept
stable at a value of $20.0 \pm 0.5$\degc. Temperature and pressure are
recorded in a database and are extracted by the \lhcb event
reconstruction software to calculate the correct refractive index.

\subsubsection[DAQ controls, monitoring and data flow]{\Acr[s]{daq} controls, monitoring and data flow}
\label{paragraph:richDAQ}

The \rich control system operates on a minimal set of devices,
composed by one \Pdmdb, one \Solfourty and one \Tellfourty, and
replicates the commands to up to thousands of \Fend and about a
hundred \Bend devices.  The \Solfourty provides, via each of its 48
bidirectional optical links, the reference 40\mhz clock and \Tfc
commands to the \Pdmdb{s} through their \Tcm{s}. Uploaded commands are
decoded by the combination of a \Gbtx and an \Sca readout chip while
the \Fend data is sent by the \Pdmdb to the \Tellfourty, via
unidirectional optical links, through \Dtm{s}, depending on the
variant of the \Pdmdb.\looseness=-1

The relevant \Pdmdb registers are periodically monitored through the
\Gbt server and differences between writing and readings in any of
such registers will raise errors.

\Tfc commands are used at power on and configuration time to enable
communication buses, initialise temperature sensors, load the firmware
on the \Pdmdb \Fpga{s} and set thresholds on the CLARO
discriminators. A stateless implementation of the \Pdmdb firmware,
where data are transported transparently towards the links to the
\Bend, was chosen in order to minimise the impact of \See due to
radiation. This approach required to move to the \Tellfourty the
association of event data with the corresponding \Bxid, needed for
event building.

Since the pixel occupancy in the RICH detector varies by orders of
magnitude between \richone and \richtwo, their centre and periphery,
and the highest occupancy region corresponds to a relatively small
fraction of the acceptance, the \Tellfourty input bandwidth had to be
carefully optimised to minimise the number of needed boards to save on
costs.

Where possible, the bandwidth was kept under control by applying a
simple lossless compression algorithm.  Then, after a Monte Carlo
simulation of the hit occupancy, two different configurations of
\Tellfourty have been implemented: where the bandwidth could be
balanced, 48 input links were used, while 24 where used elsewhere,
allowing the number of needed boards to be reduced by about 30\%.
Each \Tellfourty merges and compresses \Bxid aligned data from all the
connected input links into packets that are transferred to the host
\Eb server via \Pcie, up to a maximum bandwidth of 102\gbps. Although
the \Eb network allows for a maximum average bandwidth of 90\gbps,
after optimisation none of the \rich \Eb servers exceeded 70\gbps,
being limited by the maximum instantaneous bandwidth rather than its
average.

\subsection{Calibration of photon detectors and front-end electronics}
\label{paragraph:richCalibration}

The single photon detection efficiency is a crucial parameter of the
\rich system, and is mainly driven by intrinsic properties of the
\Mapmt{s}, such as the photocathode quantum efficiency, the collection
efficiency at the first dynode and the single photon gain. Additional
contributions to the detection efficiencies arise from the anode
signal digitisation, provided by the corresponding CLARO channel by
means of a programmable threshold.

Dedicated calibration procedures are used in order to minimise the
inefficiency arising from the threshold setting, while ensuring the
suppression of noise hits due to the \Mapmt{} and \Fend electronics
pedestals. In addition, the same procedures allow to monitor the
single photon gain variation with time and ageing of each \Mapmt{}
channel, and the stability of each CLARO channel.

\begin{figure}[t]
  \centering
  \includegraphics[height=5.4cm]{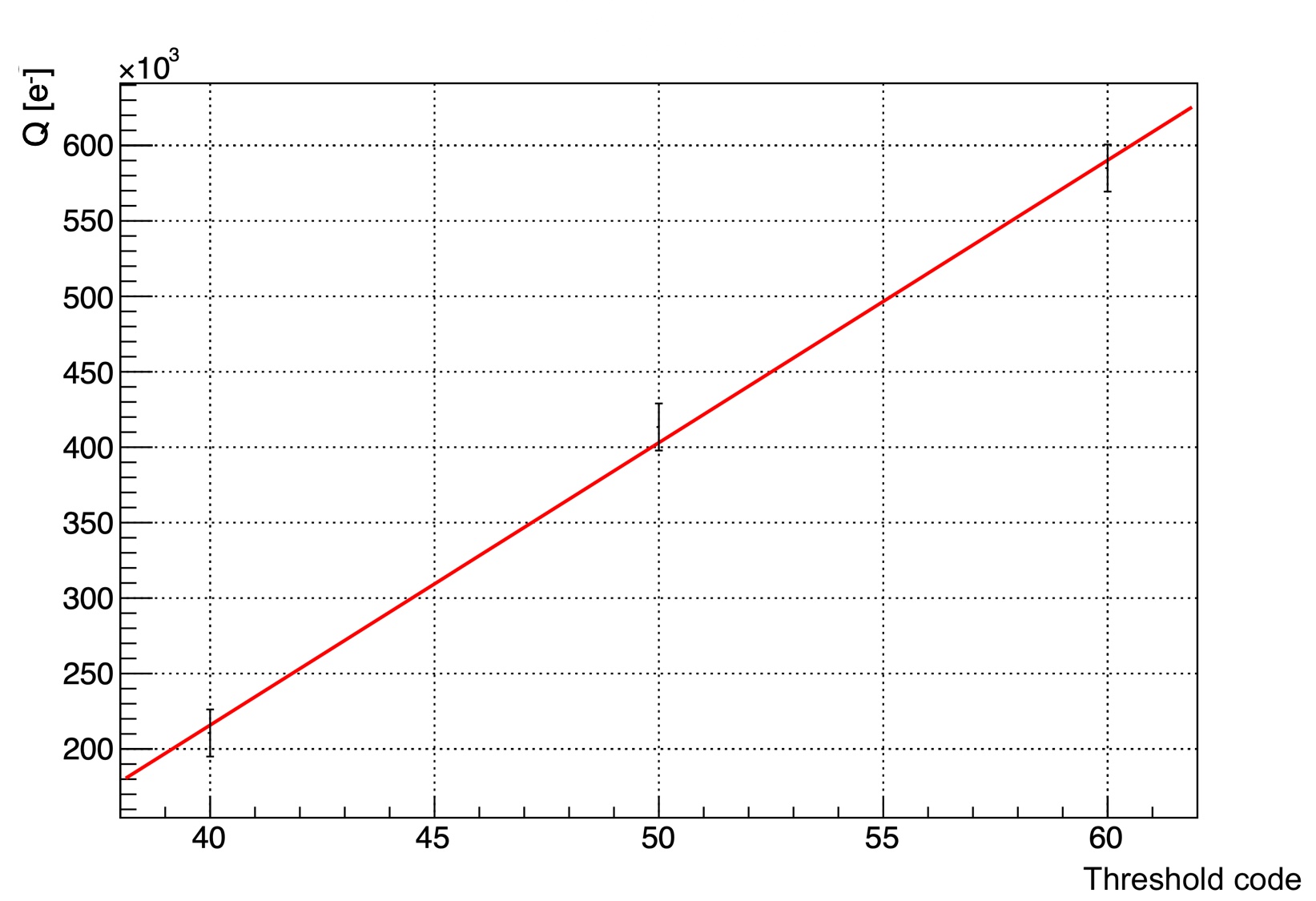}\hfill
  \includegraphics[height=5.2cm]{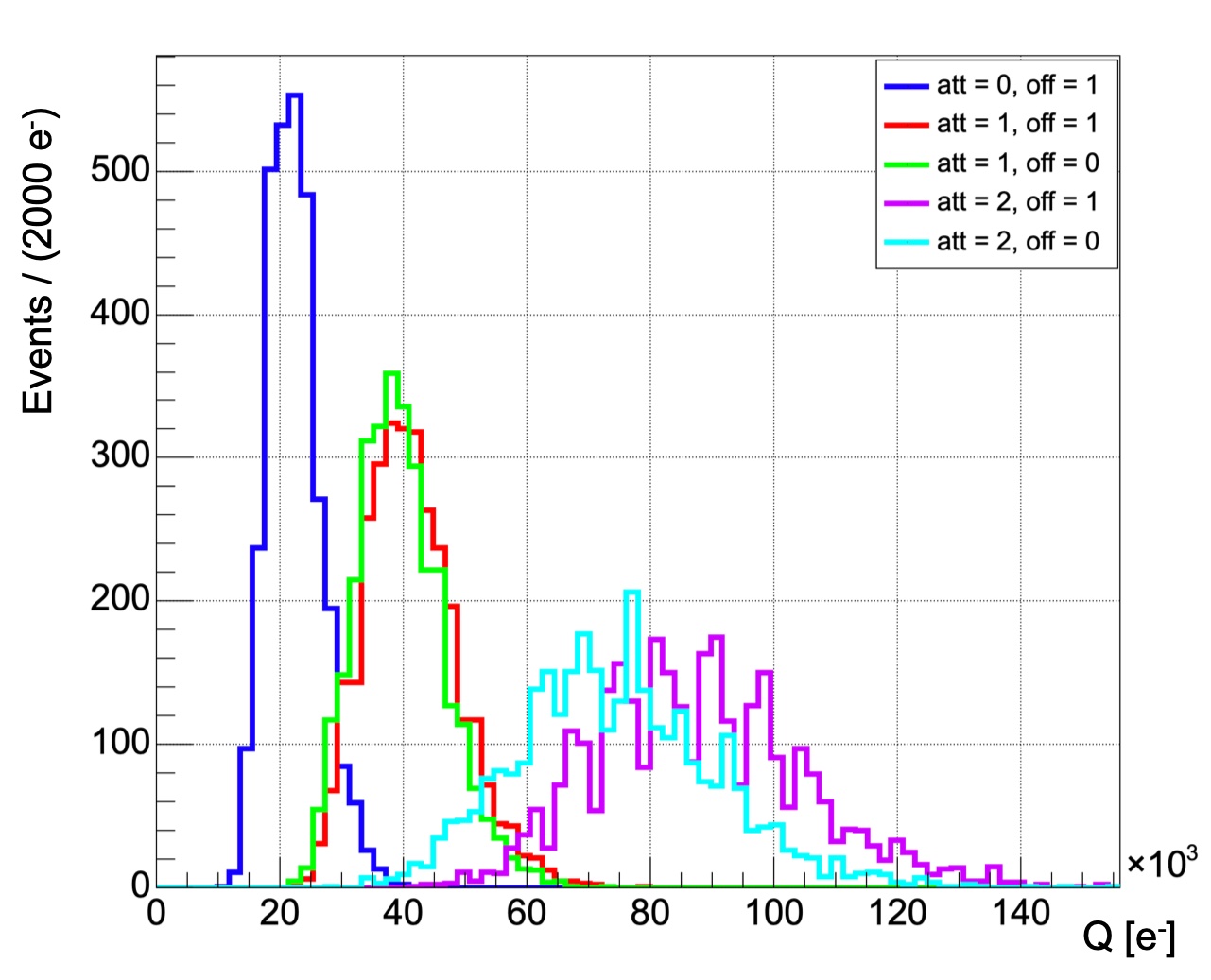}
  \caption{Typical output of the DAC scans procedure. On the left, the
    calibration of a single CLARO channel with offset bit enabled and
    no attenuation is shown. The charge (in units of electron charge)
    corresponding to a threshold DAC code (th) is determined by the
    linear relation $Q=Q_0 + Q_{\text{th}}\cdot\text{th}$. On the
    right, the distribution of the charges (in units of electron
    charge) corresponding to one threshold step ($Q_{\text{th}}$) for
    a \richtwo column is shown. The linearity of the threshold setting
    as a function of the injected charge is found to be excellent for
    all attenuation and offset values.}
  \label{fig:RICH:DACscans}
\end{figure}

Each CLARO channel is calibrated by using DAC scans, where a known
charge is injected at the input in 256 steps of $15.6\times 10^3$
electrons worth of charge each, for different CLARO settings. It
allows to determine the conversion between a threshold DAC code and
the corresponding absolute charge. An example of the output of DAC
scans performed on a \richtwo column is reported in
figure~\ref{fig:RICH:DACscans}.\looseness=-1

\begin{figure}[t]
  \centering
  \includegraphics[width=0.54\linewidth]{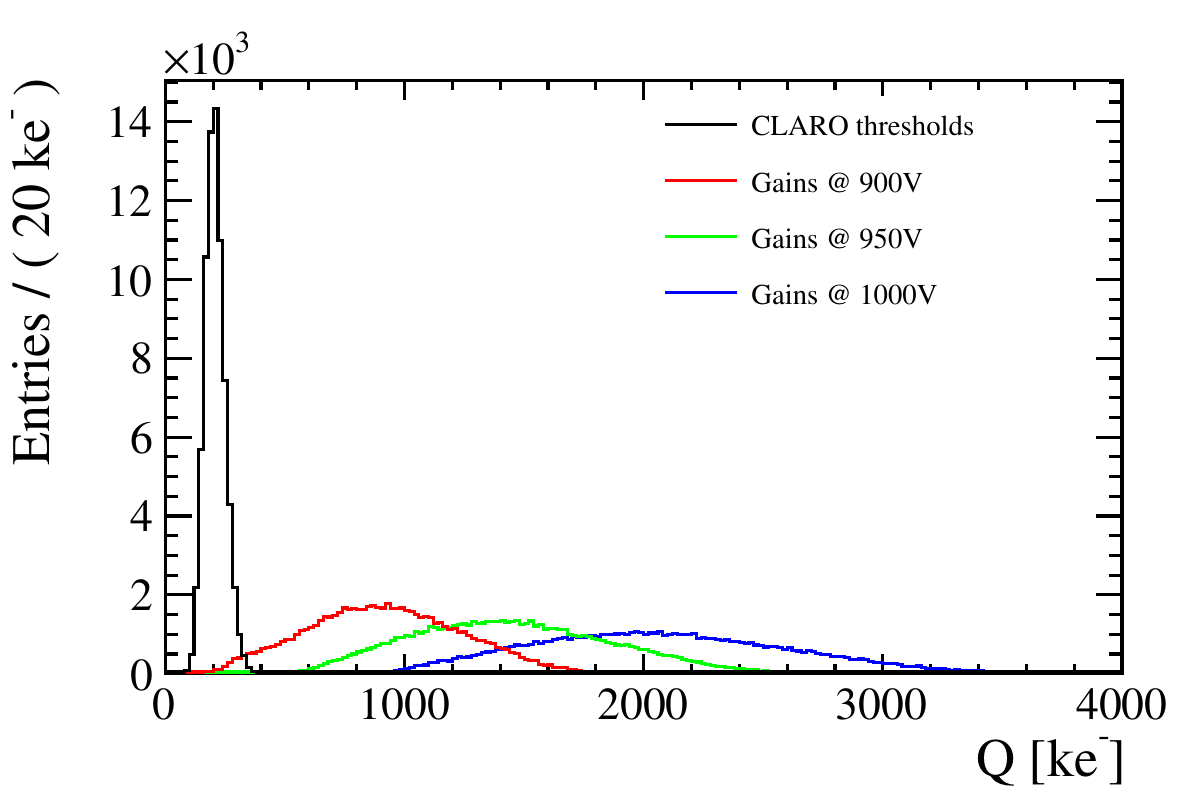}\hfill
  \includegraphics[width=0.44\linewidth]{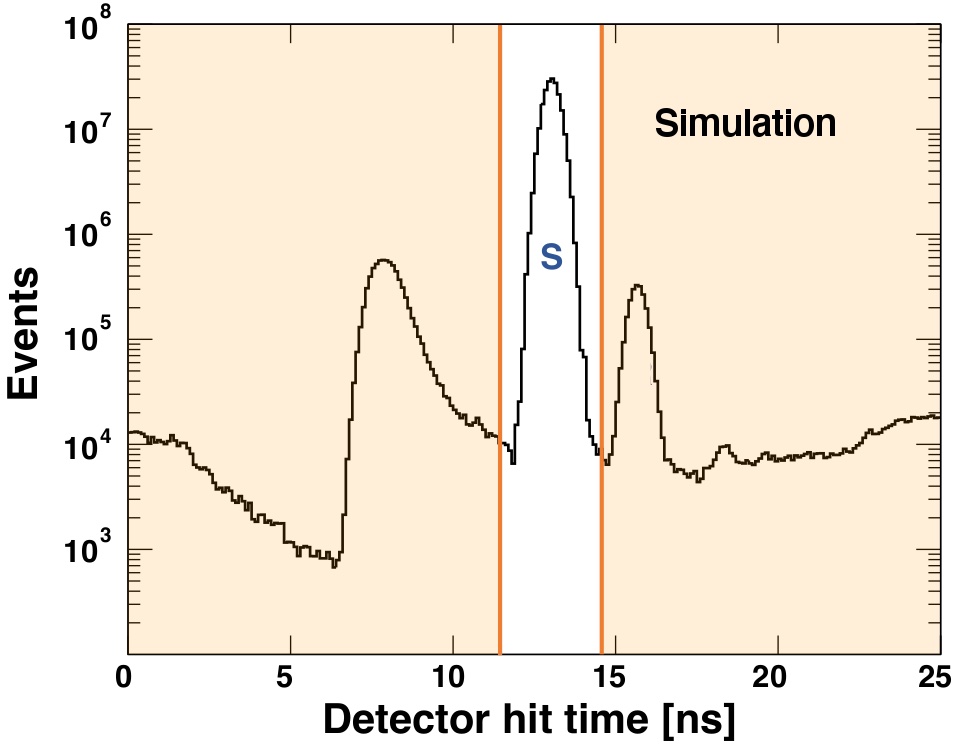}
  \caption{Left: distribution of \richtwo thresholds of the CLARO
    comparator converted into absolute charge (black). The mean and
    standard deviations of the distribution are
    $(207.58 \pm 0.16)\times 10^3$ electrons and
    $(39.64 \pm 0.10)\times 10^3$ electrons. The threshold settings
    can be compared to the pixel gain at 900\volt (red), 950\volt
    (green) and 1000\volt (blue). Reprinted from ~\cite{GIUGLIANO2023168436}, Copyright (2023), with permission from Elsevier. Right: \richone simulated photon
    detector hit time distribution showing the signal peak (S) and a
    possible time gate in the front-end electronics. Reproduced from ~\cite{LHCb:2021okg}. CC BY 4.0.}
  \label{fig:RICH:richTwoWorkingPoints}
\end{figure}

Threshold scans are performed in order to find the set of front-end
working points that maximise the single photon efficiency. The photon
detection chain is illuminated at very low light intensity (single
photon regime), and the threshold of the CLARO comparator is
decremented in unit steps. It is found that the best operational
points correspond to threshold settings that are five steps above the
pedestal. The distribution of the threshold settings for all the
\richtwo channels, converted to absolute charge as determined from DAC
scans, is compared to the single photon peak distributions of the
corresponding \Mapmt anodes at different \Hv values in
figure~\ref{fig:RICH:richTwoWorkingPoints}. Threshold scans, providing
the integral pulse height spectrum, are also used to estimate the
single photon peak position for each channel, allowing to implement
the monitoring of gain variations with \Mapmt{} ageing.

\subsubsection{Time alignment}
\label{paragraph:richTiming}

The prompt Cherenkov radiation and focusing mirror optics lead to the
nearly simultaneous time-of-arrival to the detection plane of photons
from a track in the \rich detector. This unique feature allows the
application of a time gate at the \Fend electronics to exclude
out-of-time background hits from the output data whilst accepting the
photon signals within a narrow time interval.
Figure~\ref{fig:RICH:richTwoWorkingPoints} shows the distribution of
photon hit times in \richone from a simulation. The signal peak spans
approximately 2\ns due to the spread of primary interactions in \lhcb,
which dictates the minimal width for the \Fend time gate. In practice,
the combination of CLARO time walk, channel-to-channel variations,
\Mapmt{} transit time spread and digital sampling rate at the \Fend
electronics require a time gate whose width must be set to 3.125\ns or
doubled to 6.250\ns if needed. In addition to the background from the
beam interactions in figure~\ref{fig:RICH:richTwoWorkingPoints}, the
time gate can exclude sensor noise, such as \Mapmt{} cross-talk and
afterpulses~\cite{LHCb:2021okg}.  The achieved background reduction
significantly improves the \Pid performance using the \rich pattern
recognition algorithms.

Gate generation and the time alignment procedure are implemented in
the \Pdmdb \Fpga firmware. The time gating logic exploits the
deserialiser embedded in every input-output logic block of the \Fpga
which can operate at gigabit rates. The deserialiser samples the CLARO
signals using both edges of the 160\mhz clock and shifts the sampled
data at 320\mbps into an 8-bit shift register. This byte can be
checked against specific signal patterns using a lookup table, which
is a readily available memory resource with a small logic footprint in
the general purpose logic of the \Fpga. If the CLARO signal pattern
matches one of the configured lookup table patterns, a hit is
registered on the 40\mhz system clock edge. The programmable lookup
table allows flexibility between data-taking modes such as different
time gate widths, edge detection and basic spillover checks.\looseness=-1

The time gate is applied at a fixed latency with respect to the \lhcb
clock. The \Fpga receives the 40\mhz system clock and 160\mhz sampling
clock from the \Gbt, where the clock phases can be adjusted over the
25\ns range in fine steps of 49\ps. This allows the position of the
time gate to be fine-tuned with respect to the signal time-of-arrival
in the \rich detector.

Gate generation and the time alignment procedure are implemented in the \Pdmdb \Fpga firmware. The time gating logic exploits the deserialiser embedded in every input-output logic block of the 
\Fpga which can operate at gigabit rates. The deserialiser samples the CLARO signals using both edges of the 160\mhz clock and shifts the sampled data at 320\mbps into an 8-bit shift register. This byte can be checked against specific signal patterns using a lookup table, which is a readily available memory resource with a small logic footprint in the general purpose logic of the \Fpga. If the CLARO signal pattern matches one of the configured lookup table patterns, a hit is registered on the 40\mhz system clock edge. The programmable lookup table allows flexibility between data-taking modes such as different time gate widths, edge detection and basic spillover checks.

The time gate is applied at a fixed latency with respect to the \lhcb clock. The \Fpga receives the 40\mhz system clock and 160\mhz sampling clock from the \Gbt, where the clock phases can be adjusted over the 25\ns range in fine steps of 49\ps. This allows the position of the time gate to be fine-tuned with respect to the signal time-of-arrival in the \rich detector.
\subsection{Expected performance}
\label{subsec:richPerformance}

The performance of the \rich detectors is evaluated using the \lhcb
simulation framework.  The main parameters used to evaluate the
performance are the Cherenkov angle resolution \ckvResTotal and the
photoelectron yield.  The Cherenkov angle resolution is estimated
starting from the single-photon resolution, \ckvResSinglePhoton, which
can be split into roughly independent contributions:
\begin{itemize}
\item \emph{chromatic}, due to the chromatic dispersion of the
  radiators which leads to a dependence of the Cherenkov angle on the
  photon energy;
\item \emph{emission point}, due to the tilting of the spherical
  mirrors which leads to a smearing of the Cherenkov angle depending
  on the point of emission of the photons along the track;
\item \emph{pixel error}, due to the finite size of the \Mapmt pixels;
\end{itemize}

The single photon resolution is investigated from simulated events,
using a simplified Cherenkov angle \ckvAngle reconstruction, assuming
that the photon is emitted from the track at the middle of the
radiator length and that it hits the centre of the relevant detector
pixel. The reconstructed Cherenkov angle is then compared with the
expected value from the simulation.  The photoelectron yield,
\ckvYield, i.e.\ the number of Cherenkov photons emitted by a track
detected on a Cherenkov ring by the \Mapmt{s}, is studied using high
momentum tracks with $\beta \sim 1$ where the Cherenkov angle as well
as the number of emitted photons are maximal.  For these tracks, the
total Cherenkov angle resolution is given by:
\begin{equation}
  \ckvResTotal = \frac{\ckvResSinglePhoton}{\sqrt{\ckvYield}} \oplus \ckvResTracking,
  \label{eq:rich:ckvResolution}
\end{equation}
where the constant factor \ckvResTracking, added in quadrature, is the
contribution from track reconstruction uncertainties. This includes
the uncertainties associated to multiple scattering, to the track
curvature inside the \rich radiator volumes, and to the intrinsic
resolution of the tracking detectors. As such \ckvResTracking is a
function of momentum, and it takes an average value of 0.35\mrad while
asymptotically tending to approximately 0.15\mrad at high momentum as
determined from simulation studies~\cite{LHCb-TDR-014}.

\subsubsection{Simulation setup and typical output}
\label{paragraph:richPerformance:setup}

The simulation used to obtain the results presented in this section
includes the up to date information on the \rich geometry and on the
properties of its optical system and photon detection chain. The
properties of the individual photon detector channels, such as the
\Mapmt{} gain and noise, are accounted for; in addition, background
sources such as cross-talk, afterpulses and scintillation photons
produced by charged particles in the \cffour radiator are included as
well~\cite{Blake:2015wia}. In the simulation, the values related to
the mentioned aspects of the individual channels are based on the data
acquired during quality assurance procedures with a 900\volt bias
voltage for the \Mapmt{s}.  The study is performed using a sample of
10\,000 events containing a \ensuremath{\Bs \to \phi \phi} decay as
typical signal events. The simulated data are obtained with the
standard upgrade configuration, corresponding to an instantaneous
luminosity of $\lum =2 \times 10^{33}\invcma\invsec$. The \Pid
performance, after the application of the reconstruction algorithms,
is reported for tracks in the momentum range of 2--100\gevc and \pt
larger than 0.5\gevc. Only the tracks that traverse the full \lhcb
tracking system acceptance are used.\looseness=-1

\begin{figure}[p]
  \centering
  \includegraphics[width=0.414\linewidth]{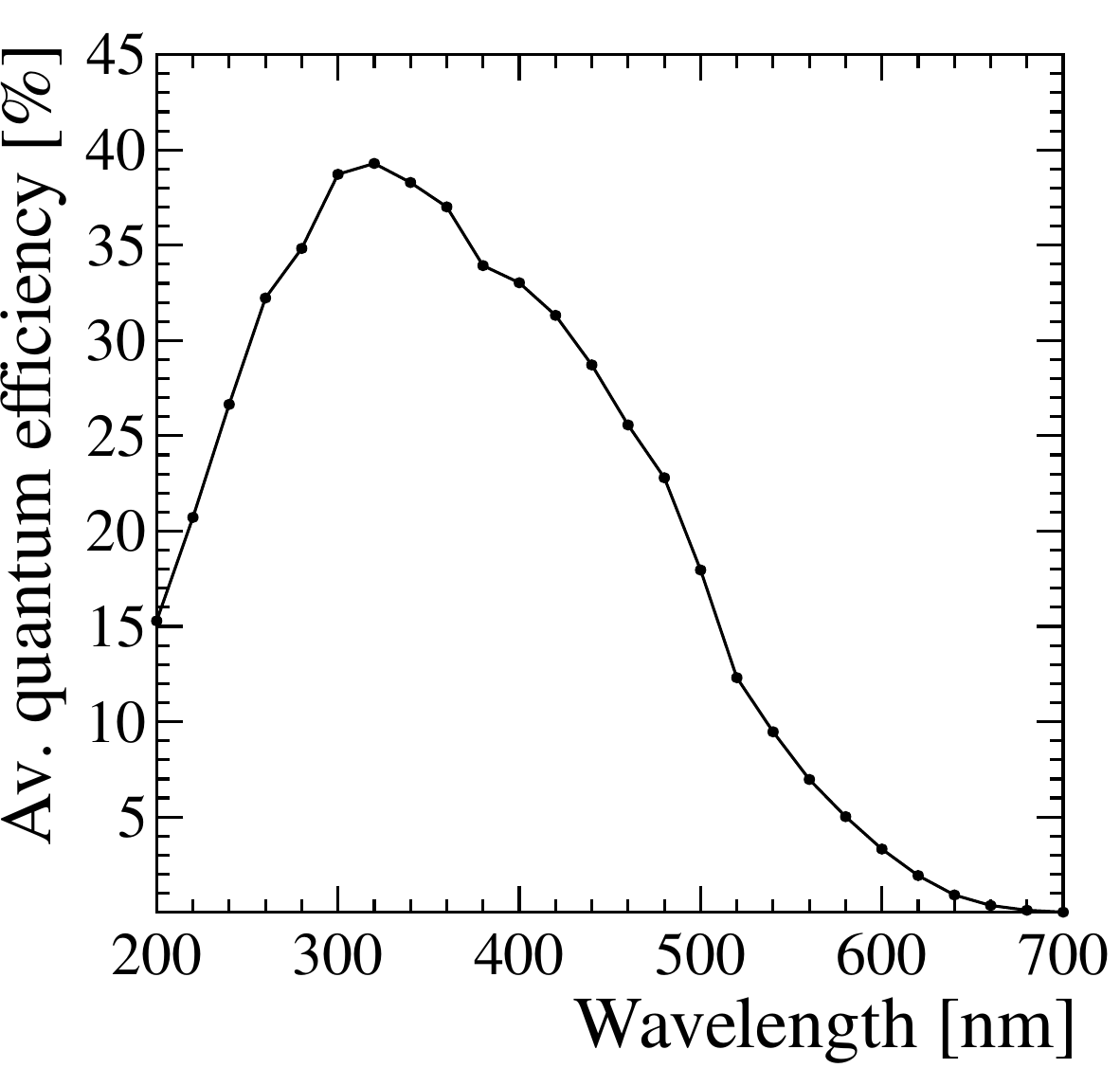}\hfill
  \includegraphics[width=0.576\linewidth]{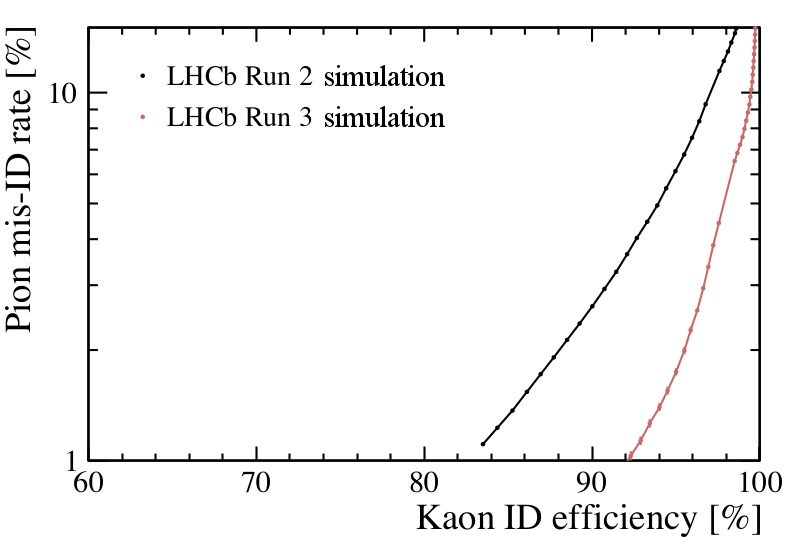}
  \caption{Left: average quantum efficiency of the \Mapmt{s} used in
    the \rich detectors. Right:~a typical \Pid performance of the kaon
    identification obtained from the \lhcb software for the
    configuration described in the text (red). A corresponding curve
    for the \runtwo conditions (prepared using the simulation with
    \lhcb \runtwo geometry and luminosity, as reported in
    ref.~\cite{Easo:2017zxn}) is shown for reference (black).}
  \label{fig:rich:qe}
\end{figure}

\begin{figure}[p]
  \centering
  \includegraphics[width=0.45\linewidth]{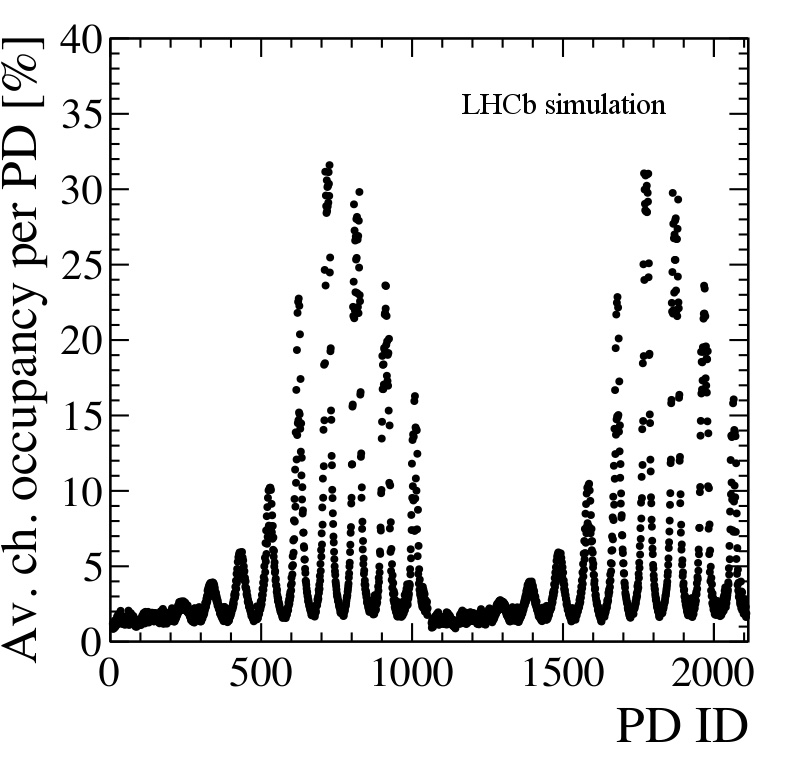}\hfill
  \includegraphics[width=0.45\linewidth]{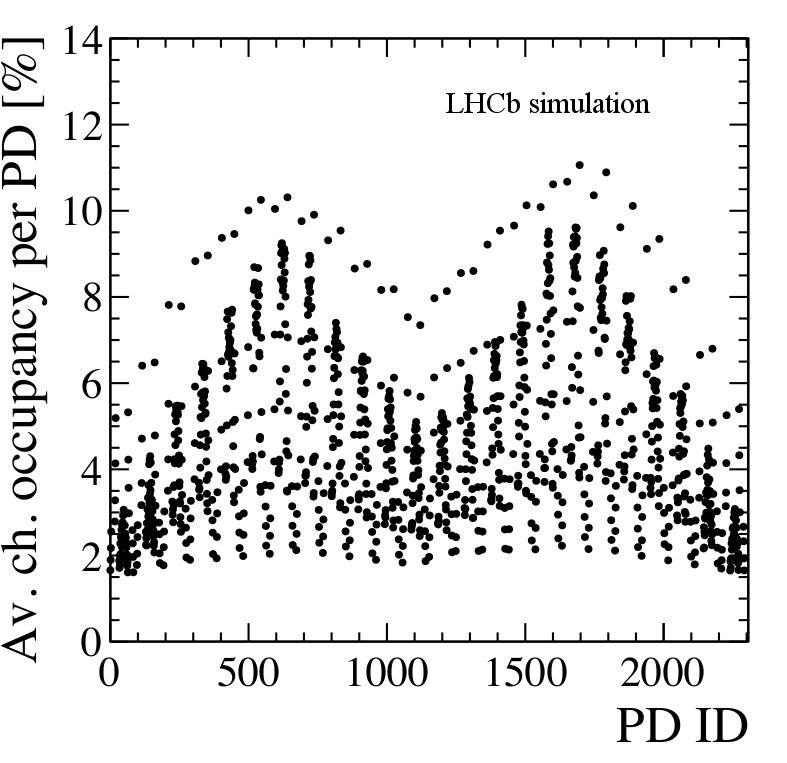}
  \caption{Average expected occupancy per channel for different
    \Mapmt{s} in the (left) \richone and (right) \richtwo detector.}
  \label{fig:rich:occupancy}
\end{figure}

The average \Mapmt quantum efficiency curve, which is used in the
simulation, is shown in figure~\ref{fig:rich:qe}, together with a
typical \Pid performance curve, representing the probability to
misidentify a pion as a kaon versus the probability to correctly
identify the particle as a kaon.  The expected hit occupancy in the
\Mapmt{} as a function of the \Mapmt identifier is reported in
figure~\ref{fig:rich:occupancy}.

\subsubsection{Performance studies}
\label{paragraph:richPerformance:performance}

\begin{table}[p]
  \centering
  \caption{Simulated performance of the upgraded \rich detectors. For
    \richtwo, the values are given for the inner detector regions
    populated with the 1-inch \Mapmt{s}.}
  \begin{tabular}{cccccccc}
    \hline
    & \multicolumn{2}{c}{\bf Photoelectron yield} & \multicolumn{5}{c}{\bf Cherenkov angle resolution [mrad]} \\
    & \ckvYieldOptimal & \ckvYieldTypical & chromatic & emission point & pixel & \ckvResSinglePhoton & \ckvResTotal \\ 
    \hline
    \richone & 63 & 59 & 0.52 & 0.36 & 0.50 & 0.81 & 0.18 \\ 
    \richtwo & 34 & 30 & 0.34 & 0.32 & 0.22 & 0.52 & 0.17 \\ 
    \hline
  \end{tabular} 
  \label{tab:rich:performance}
\end{table}

A single particle simulation is used to evaluate the best achievable
photoelectron yield (\ckvYieldOptimal) and the Cherenkov angle
resolution. The simulation is configured to provide 80\gev muons,
which ensures that the tracks are saturated and are not significantly
curved by the magnetic field to minimise the uncertainty arising from
the tracking system. In addition, the acceptance region where the
\rich performance is expected to be optimal is used: the polar angle
of the tracks is required to be in the 90--180\mrad and 40--90\mrad
for \richone and \richtwo, respectively. The results are summarised in
table~\ref{tab:rich:performance}. Consistent results are found when
using $\Bs \to \phi \phi$ events provided that the selected tracks
fulfil the constraints described above for the single particle
simulation.  The photoelectron yield decreases when the optimal track
requirements are relaxed. This is shown in
table~\ref{tab:rich:performance} where a typical photoelectron yield,
\ckvYieldTypical, is reported. The typical photoelectron yield values
are lower than the optimal ones, mainly due to the limitations in the
acceptance due to the beam pipe region. As it can be extracted from
table~\ref{tab:rich:performance}, the contributions to the total
Cherenkov angle resolution from the \rich and tracking systems at high
momentum are of similar magnitude, and slightly dominated by the
\ckvResTracking term. The \ckvResTracking contribution used here is
the asymptotic limit of approximately 0.15\mrad at high momentum,
dominated by the intrinsic resolution of the tracking detectors.

\section{Calorimeters}
\label{subsec:calorimeters}
To cope with the new LHCb readout scheme, the \Fend and readout
electronics of the electromagnetic and hadronic calorimeters have been
entirely redesigned and replaced. Moreover, two subdetectors of the
previous calorimeter system, namely the Scintillating Pad Detector
(SPD) and the PreShower (PS)~\cite{LHCb-TDR-002}, have been removed,
given their reduced role in the new LHCb all-software trigger.

The layout of both the \Acr[f]{ecal} and \Acr[f]{hcal} remains
unchanged for the upgrade. A complete description of the geometry and
technological aspects can be found in ref.~\cite{LHCb-TDR-002}.  To
minimise the required modifications, the \Ecal and \Hcal calorimeter
modules, their photomultiplier tubes, \Cw bases and coaxial cables
were also maintained unmodified. However, to keep the same average
anode current of the phototubes at the higher luminosity, their high
voltage has been reduced implying an increased gain of the
amplifier-integrator in the \Fend cards. This modification is
described in section~\ref{sssec:calo_electronics_FEB}.

The \Fend electronics boards have been fully redesigned to comply with
the 40\mhz readout frequency, but their number and format have been
chosen such that they are compatible with the existing crates and
racks, as described in section~\ref{calo_electronics_subsection}.

The decision to keep the calorimeter modules, their \Pmt{s} and \Cw
bases, assumes that they can operate at radiation levels corresponding
to the foreseen integrated luminosity. This is discussed in
section~\ref{sec:Calo_detector_radissues}.

\subsection{General detector structure}

The LHCb calorimeter system presents a classical structure of an
electromagnetic calorimeter followed by a hadronic calorimeter. The
most demanding performance constraint concerns the identification of
electrons and photons with an optimal energy resolution requiring the
full containment of the showers from high energy particles. For this
reason, the thickness of the \Ecal was chosen to be 25 radiation
lengths~\cite{Barsuk:691508}. On the other hand, the trigger
requirements on the \Hcal resolution do not impose a stringent
hadronic shower containment condition, thus its thickness is limited
to 5.6 interaction lengths~\cite{Djeliadine:691688}, due to space
limitations.

\begin{figure}[t]
  \centering
  \includegraphics[width=0.98\linewidth]{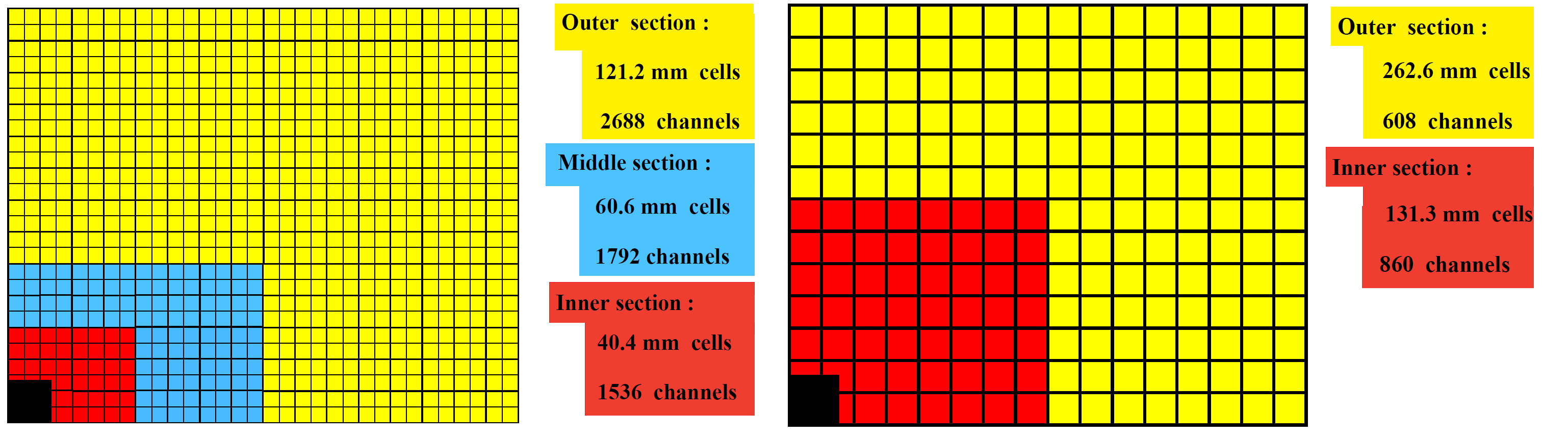}
  \caption{Lateral segmentation of (left) the \Ecal and (right) the
    \Hcal. One quarter of the detector front face is shown. Reproduced from ~\cite{LHCb-TDR-002}. CC BY 3.0.}
  \label{fig:Calo_segmentation}
\end{figure}

To account for different hit densities across the calorimeter surface,
the \Ecal is segmented laterally in three regions referred to as
inner, middle and outer, with increasing dimensions going from the
beam pipe outwards, as shown in
figure~\ref{fig:Calo_segmentation}. The \Hcal is segmented in two
regions with a larger granularity with respect to the \Ecal, given the
typical spread of hadronic showers.
The regions are segmented in cells of transverse dimensions roughly
projective with respect to the interaction point. Their dimensions are
optimised to provide uniform measurements of the \emph{transverse
  energy}, $\et = E_c\sin{\theta}$, where $E_c$ is the energy measured
by a cell and $\theta$ is the angle between the vector pointing to the
centre of the cell from the interaction point. This quantity is
particularly useful for hadron selection at trigger level.

The two calorimeters share the same basic detection principle:
scintillation light from plastic scintillator modules is transmitted
to the \Pmt{s} by wavelength-shifting fibres. Fibre bundles from
calorimeter modules are then fed to the \Pmt{s}. In order to have a
constant energy scale across the calorimeter surface the gain of the
\Ecal and \Hcal \Pmt{s} is set proportionally to the distance from the
beam pipe of the corresponding modules. Since the light yield of an
\Hcal module is a factor 30 less than \Ecal modules, the \Hcal
phototubes operate at higher gain.

\subsubsection{The electromagnetic calorimeter}
\label{calo_ecal_subsection} 

The \Ecal front surface is located at about 12.5\m from the
interaction point.  The square cell sizes for the inner, middle and
outer regions are 121.2\mm, 60.6\mm, 40.4\mm side for outer, middle
and inner regions, respectively, and scale with the distance from the
beam-pipe in order to make the particle rate per cell roughly uniform.
The outer dimensions of the \Ecal match projectively those of the
tracking system, $\theta_{x} < 300$\mrad and $\theta_{y} < 250$\mrad,
while the inner angular acceptance of \Ecal is limited to
$\theta_{x,y} >$ 25\mrad around the beam pipe, where $\theta_{x}$ and
$\theta_{y}$ are the polar angles in the $xz$ and $yz$ planes in the
LHCb reference frame.

\begin{figure}[t]
  \centering
  \includegraphics[width=0.8\linewidth]{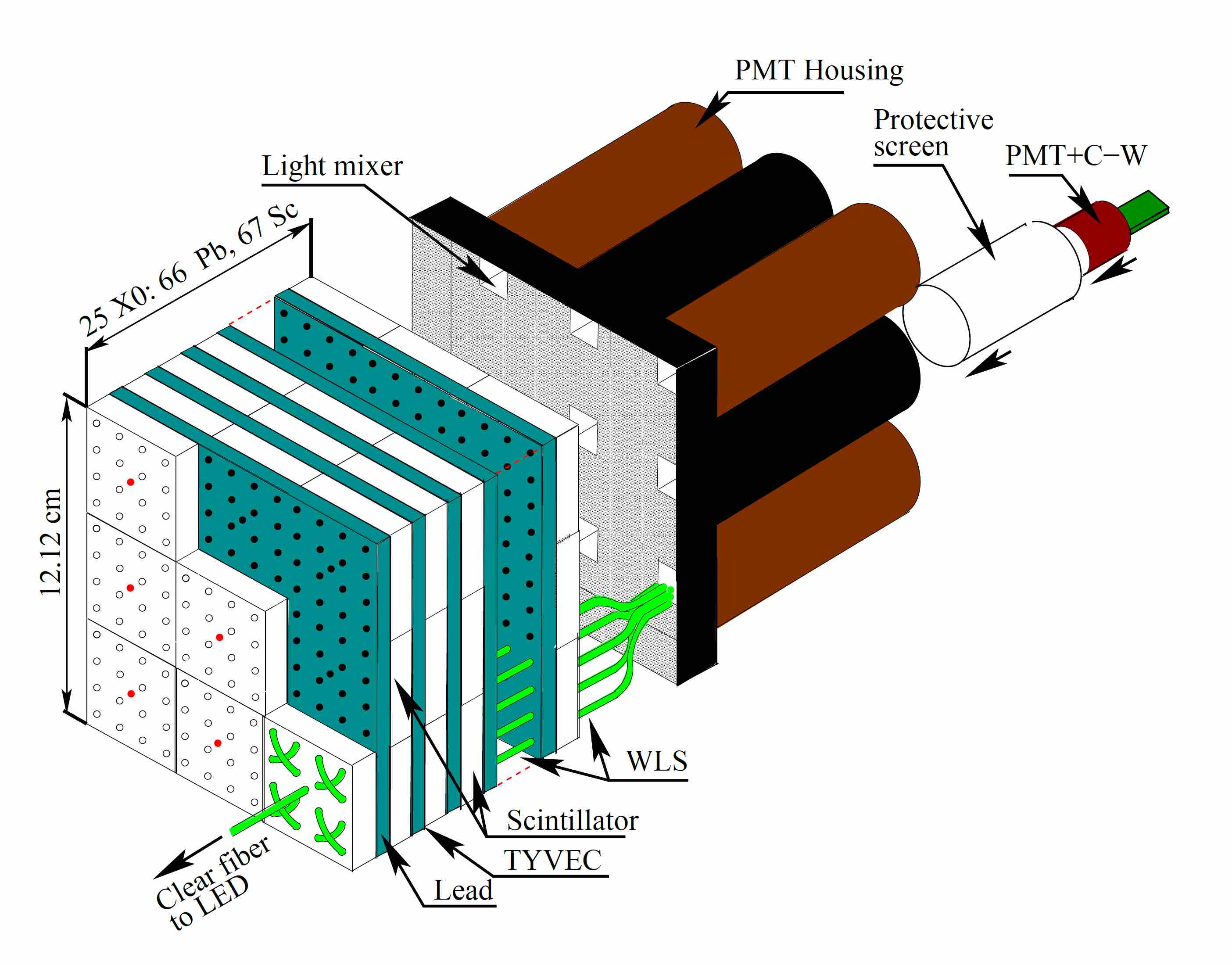}
  \caption{Schematic of an \Ecal cell. Reproduced from ~\cite{LHCb-DP-2020-001}. CC BY 4.0.}
  \label{fig:Calo_ECAL_cell}
\end{figure}

The \Ecal cells have a \emph{shashlik} structure, as shown in
figure~\ref{fig:Calo_ECAL_cell}, with alternated scintillator (4\mm)
and lead (2\mm) layers. The scintillation light readout is performed
by dedicated phototubes \Pmt{s}.\footnote{Hamamatsu R7899-20.}  The
total number of cells is 6016.
The energy resolution of a given cell, measured with a test electron
beam, is parametrised as~\cite{LHCb-TDR-002}:
\begin{equation}
  \frac{\sigma\left( E \right)}{E}=\frac{\left( 9.0 \pm 0.5 \right)\%}{\sqrt{E}}\oplus \left( 0.8 \pm 0.2 \right)\% \oplus \frac{0.003}{E \sin \theta}
\end{equation}
where $E$ is the particle energy in \gev, $\theta$ is the angle
between the beam axis and the line from the LHCb interaction point to
the centre of the \Ecal cell. The second contribution is a constant
term taking into account mis-calibrations, nonlinearities, energy
leakage out of the cell and other effects, while the third term is due
to the noise of the electronics which is evaluated on average to 1.2
ADC counts~\cite{LHCb-DP-2008-001}.

\subsubsection{The hadronic calorimeter}

\begin{figure}[t]
  \centering
  \includegraphics[width=0.75\linewidth]{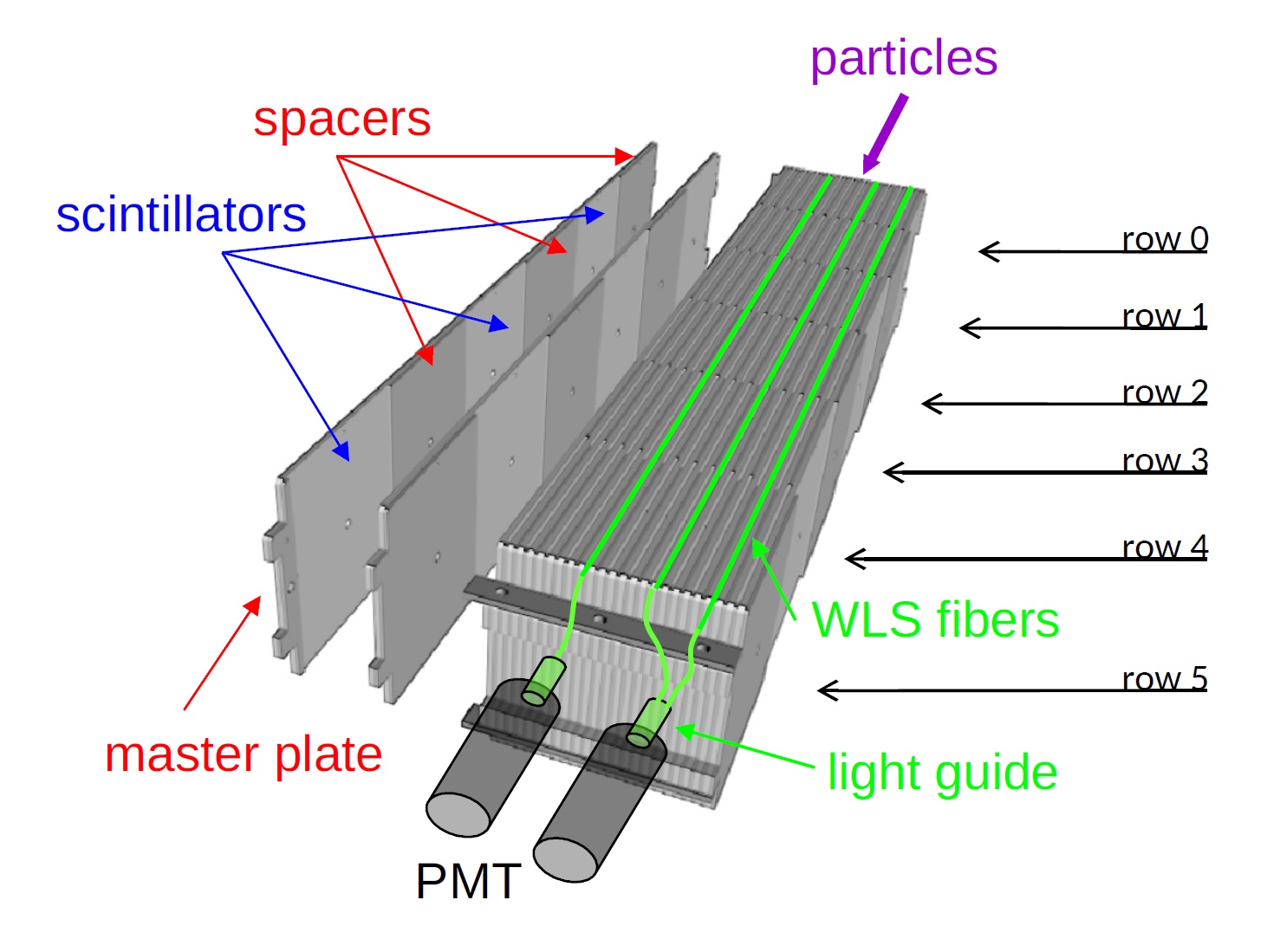}
  \caption{Schematic of an \Hcal cell. Reproduced from ~\cite{LHCb-DP-2020-001}. CC BY 4.0.}
  \label{fig:Calo_HCAL_cell}
\end{figure}

The \Hcal is a sampling tile calorimeter with a thickness of 5.6
interaction lengths.  The sampling structure consists of staggered
iron and plastic scintillator tiles mounted parallel to the beam axis
(figure~\ref{fig:Calo_HCAL_cell}) to enhance the light collection.
The same \Pmt{} type as in \Ecal is used for the readout. The \Hcal
has a total of 1488 cells arranged in an inner and an outer region,
segmented in square cells with sides of 131.3\mm and 262.6\mm,
respectively. The energy resolution, as measured in beam tests with
pions, is parametrised as follows:
\begin{equation}
  \frac{\sigma\left( E \right)}{E}=\frac{\left( 67 \pm 5 \right)\%}{\sqrt{E}}\oplus \left( 9 \pm 2 \right)\%
\end{equation}
where $E$ is the deposited energy in \gev~\cite{LHCb-DP-2020-001}.

\subsubsection{Radiation effects and ageing}
\label{sec:Calo_detector_radissues}

The effects of radiation on the calorimeter system components was
assessed in a series of measurement campaigns at different irradiation
facilities. The radiation resistance of calorimeter components such as
scintillators, wavelength shifter fibres, \Pmt{s} and \Cw bases has
been studied to extrapolate their lifetime over the foreseen
integrated luminosity of the upgraded LHCb. First irradiation tests
performed on an \Ecal module prototype, indicated that the performance
of the innermost cells remains satisfactory up to $\sim 25$\kGy at the
position of the shower maximum, corresponding to about 20\invfb of
integrated luminosity~\cite{LHCb-TDR-002}. Tests conducted at CERN PS
IRRAD facility with 24\gev protons and on \Ecal inner modules, placed
in the LHC tunnel during \runone, confirmed this result. These
measurements indicated that the innermost region of \Ecal will have to
be replaced during the \Lsthree.  Radiation tolerance of \Pmt{s} and
\Cw bases was assessed with irradiation tests conducted at a 50\gev
proton beam.\footnote{The beam is provided by the IHEP U-70
  accelerator in Protvino, Russia.} The \Cw bases remained operational
up to doses of 15--20\kGy corresponding to 30--40\invfb of integrated
luminosity in the central \Ecal cells.  The replacement of \Cw bases
can be easily performed during annual shutdown periods. Replacement of
about 500 \Cw bases is estimated to be needed over the full upgrade
programme.  The transmittance loss of the \Pmt{} window in the
wavelength peak range did not exceed 5\%, ensuring that the radiation
tolerance of the \Pmt{} entrance window will maintain sufficient
transparency during the full upgrade lifespan.\looseness=-1

As far as the \Hcal is concerned, only the tile modules will suffer
radiation damage effects, since the \Pmt{s} are installed behind the
calorimeter and shielded by the iron.  The radiation damage of the
\Hcal modules has been studied during \runone, using calibration data
obtained with a $^{137}$Cs source. The relative light yield of the
scintillator tile rows (see figure~\ref{fig:Calo_HCAL_cell}) was
measured with respect to the rearmost tile row, which receives the
smallest dose and is not expected to suffer from significant radiation
damage.  A reduction of the light yield at the level of 15\% was
observed for the central \Hcal cells after collecting 3.4\invfb of
luminosity. This result indicates that the \Hcal innermost cells
cannot survive the full lifetime of the upgrade. Since it is not
possible to replace these cells in an easy way during the LHC runs,
they have been removed and replaced by slabs of tungsten absorber, to
mitigate the low energy particle background on the innermost regions
of M2 and M3 muon stations (see section~\ref{ssec:muonHRmit}).  The
impact of this reduction in acceptance on the physics programme was
estimated to be negligible as the information from the \Hcal innermost
cells will not affect significantly the software trigger analysis.

\subsection{Electronics}
\label{calo_electronics_subsection}

The two LHCb calorimeters share the same electronics which consists
of:
\begin{itemize}
\item a \Acr[f]{feb} (described in
  section~\ref{sssec:calo_electronics_FEB}), where the \Pmt signals
  are amplified, shaped and digitised and then, after proper
  formatting, shipped to the back-end electronics;
\item a \Cccu board (detailed in
  section~\ref{ssec:calo_electronics_3CU}) to distribute clocks and
  \Ecs commands to the \Feb{s};
\item a set of calibration and monitoring boards
  (section~\ref{ssec:calo_electronics_monitoring}).
\end{itemize}
The \Feb{s} provide the digitised analog data in the form of \et
measurements. They also perform simple data preprocessing, to send to
the software trigger some precalculated quantities such as simplified
energy clusters, calibrated energy measurements and global quantities
such as total energy and hit multiplicity. This preprocessing is
identified for simplicity as \Llt, in analogy with the L0 hardware
stage of the previous trigger scheme~\cite{LHCb-DP-2012-004} and is
discussed in section~\ref{sssec:calo_electronics_FEB} and
section~\ref{sssec:calo_electronics_LLT}. The information provided by
the \Llt is sent to the \Hlt through the \Tellfourty for a possible
use in event reconstruction or selection (see
section~\ref{ssec:hlt1:selections}).

In order to keep the same average anode current of the \Pmt{s} and
minimise the ageing effects at the higher luminosity, the high
voltage, and consequently the gain of the photomultipliers, must be
reduced by a factor of five. Therefore, a partial gain compensation of
a factor 2.5 was implemented into the \Fend electronics allowing also
to double the \et dynamic range of the calorimeter system. The gain
compensation can be tuned within a factor ranging from 0.5 to 2.0, by
configuring the analog electronics which allows large flexibility in
adapting to physics requirements. The equivalent input noise of the
preamplifier has also been decreased in the \Fend analog design in
order to maintain the same performances in spite of the larger gain of
the electronics.  Finally, the signal transmission protocol had to be
adapted to the new 40\mhz readout scheme and to the new back-end
electronics. As a consequence, the \Feb{s}, \Cccu and monitoring
boards have been fully redesigned, although maintaining the same
format and number of the previous system to minimise the cost of the
infrastructure.\looseness=-1

\subsubsection{The front end board}
\label{sssec:calo_electronics_FEB}

The analog signals from the \Pmt{s} are clipped inside the \Cw base to
keep them within the 25\ns \Lhc bunch crossing window, and then
transmitted to the \Feb through a 12\m long 50\ohm coaxial cable.
Since the new \Feb amplifier has a five times higher gain than the
previous \Feb, more stringent requirements in terms of noise
($\lesssim$1 ADC counts) were imposed, so that passive termination at
the \Feb level was not possible. Since each \Feb receives the signals
from 32 \Pmt{s} and an actively terminated input stage is required, an
\Asic solution (the ICECAL \Asic~\cite{Picatoste_2012}) has been
implemented for the analog signal processing stage of the \Feb (see
section~\ref{sssec:calo_electronics_icecal}).
\begin{table}[t]
  \centering
  \caption{Summary of the requirements for the calorimeter analog
    \Fend.}
  \label{table:Calo_ElecRequirements}
  \begin{tabular}{|c|c|}
    \hline
    Parameter                & Requirements                             \\
    \hline
    Energy range             & $0 \leq \et \leq 10\rm{GeV} $ (\Ecal)   \\
    Calibration/Resolution   & 4\aunit{fC}/5\MeV \et per ADC count               \\
    Dynamic range            & 4096-256 = 3840 counts: 12 bits            \\
    Noise                    & $ \lesssim 1 $ ADC counts (ENC $ < 4 $\aunit{fC})  \\
    Termination              & $ 50\pm 5\ohm$                       \\
    Baseline shift prevention & Dynamic pedestal subtraction            \\
    Max. peak current        & 4--5\mamp over $ 50\ohm$               \\
    Spill-over residue level &  $ \pm 1\% $                             \\ 
    Integrator peak plateau  & $ < 1\%$ variation in $\pm 2\ns$             \\
    Linearity                & $ < 1\% $                                \\
    Cross-talk               & $ < 0.5\% $                              \\
    Timing                   & Individual (per channel)                 \\
    \hline 
  \end{tabular} 	 
\end{table}
Table~\ref{table:Calo_ElecRequirements} summarises the main
requirements for the analog \Fend of the calorimeter system. Except
for the \Pmt{} current and noise, the other requirements are similar
to the ones for the previous calorimeter
\Fend~\cite{LHCb-DP-2008-001,Beigbeder-Beau:691705,LHCb-TDR-002}.

The calorimeter electronics~\cite{LHCb-DP-2008-001,LHCb-TDR-002} is
based on 246 \Feb{s}, 192 for the \Ecal and 54 for the
\Hcal~\cite{Beigbeder-Beau:691705}.  This number includes \Feb{s}
needed for the measurement of signals of photodiodes of the LED
monitoring system (4 \Feb{s} for each subdetector, see
section~\ref{ssec:calo_electronics_monitoring}).  Each board is
connected to 32 \Pmt{} outputs.  The region of the calorimeter which
is covered by a \Feb is a rectangle of $4 \times 8$ cells.  Each \Feb
provides digitised data in the form of 32 transverse energy
measurements to the \Bend electronics, and preprocessed information
based on the ADC data of the board and on data received from
neighbouring boards, to be used by the trigger farm.
\begin{figure}[t]
  \centering
  \includegraphics[width=0.8\linewidth]{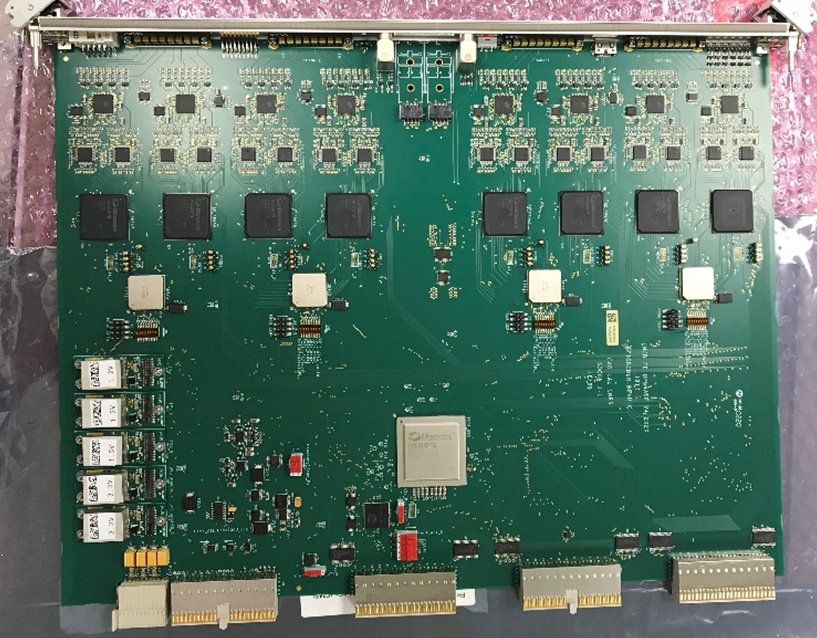}
  \caption{Picture of the \Feb.}
  \label{fig:Calo_FEB_picture}
\end{figure}
The digital section of the \Feb is based on a 12 bit ADC.  Each \Feb
handles 32 channels, thus the requested output bandwidth is
$12 \times 32\times 40\mbps$. This load is distributed over four
optical links driven by \Gbt{} chips in wide bus mode.\looseness=-1

The location of the front-end electronics is unchanged for the
upgrade. The 18 necessary 9U crates will be gathered in racks located
on the calorimeter gantries, 14 on the \Ecal platform and 4 on the
\Hcal one. The clock, fast and slow control for the front-end
electronics will be ensured by 18 \Cccu{s} plugged in the central slot
of each crate. The \Cccu boards are powered directly by the backplane
of the crate as for the \Feb and are connected to the \Bend
electronics with bi-directional optical links.\looseness=-1

There are 6 major sections in the \Feb:
\begin{itemize}
\item 4 \Fend{m} blocks consisting of two ICECAL chips, four
  dual-ADCs, two \Fpga{s},\footnote{Microsemi Igloo2 family type
    M2GL025-1FG484.} called \Fend{m} \Fpga{s} in the following, and a
  \Gbtx component (driving 4 \Vttx{} emitters), producing \et
  measurements for the data stream and calibrated \et values for the
  \Llt processing;
\item the trigger and sequencer \Fpga (TrigSeq
  \Fpga\footnote{Microsemi Igloo2 family type M2GL150-1FC1152.}),
  which is used to perform the \Llt calculations but also sends
  additional global information (like e.g.\ \Bxid) over the optical
  links;
\item the \Gbtsca that converts the \elink{s} from the \Gbtx of the
  \Cccu in the fast and slow control signals;
\item the block of (de-)serialiser for the exchange of the \Llt data
  between different boards in the same or neighbouring crates;
\item the block containing the \dcdc converters and the protection
  delatchers;
\item the \Vttx{} transmitters that receive the data from the four
  \Gbtx and send them to the back-end electronics.
\end{itemize}
The \Gbtsca \Asic is used to distribute control and monitoring signals
to the \Feb and perform monitoring operations of detector
environmental parameters. It provides various user-configurable
interfaces (\I2c, \Acr{spi}, \Acr{gpio}) used by the \Cccu and the
\Feb{s}.  All registers storing configurations and permanent
information are protected with the triple voting technique.

\subsubsection[The ICECAL ASIC]{The ICECAL \Acr[s]{asic}}
\label{sssec:calo_electronics_icecal}

\begin{figure}[t]
  \centering
  \includegraphics[width=0.98\linewidth]{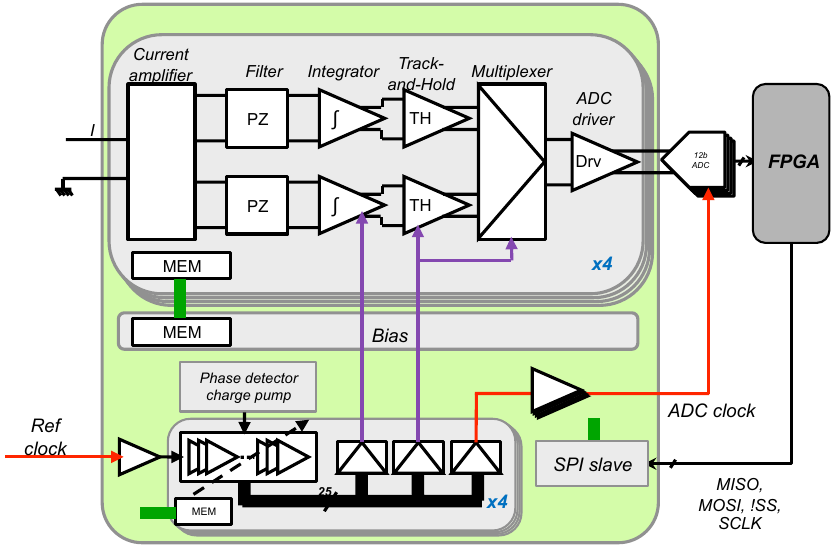}
  \caption{Block diagram of the ICECAL \Asic. Reproduced from~\cite{Picatoste_2015}. \textcopyright\ 2015 IOP Publishing Ltd and
Sissa Medialab srl. All rights reserved.}
  \label{fig:Calo_analog_schematic}
\end{figure}

The ICECAL is a four-channel fully differential amplifier implemented
in SiGe BiCMOS 0.35\mum technology.  The input stage consists of a
current amplifier featuring an active line termination in order to
avoid resistor noise. After the preamplifier stage, the signal is sent
to two interleaved processing lines running at 20\mhz synchronous with
the 40\mhz global clock. Each processing line shapes the signal with a
pole zero compensation, integrates the signal, stores the integrated
signal in a track-and-hold (T/H) module and finally sends it to the
ADC driver through a multiplexer.  A block diagram of the ICECAL ASIC
is shown in figure~\ref{fig:Calo_analog_schematic}.  The charge
integration is performed in the two processing lines by two switched
fully differential amplifiers. In a first half-clock cycle, the first
amplifier integrates the main part of the \Pmt{} signal, which is
sampled by the T/H stage while the other amplifier is reset. In the
second half cycle, the second amplifier integrates the tail of the
signal which is sampled by the T/H stage, while the first amplifier is
reset. The fully differential architecture largely eliminates the
switching noise intrinsic in this scheme. The samples are then
presented through the multiplexer to the ADC driver.
The ICECAL receives its clock from the \Gbtx of the \Fend module it
belongs to. It is configured through the TrigSeq \Fpga with the
\Acr{spi} protocol. Among the most important configuration parameters
there are the pole zero compensation parameters, the gain of the
integrators and the clock phase to be used to integrate the \Pmt{}
signals.

The four ICECAL channels are connected to two 12-bit dual
ADCs.\footnote{AD9238 from Analog Devices.} The ADCs require a clock
to properly sample the ICECAL output. The clock is produced by the
ICECAL and directly injected into the ADC. The clock phase of each of
the 32 channels of the \Feb can be adjusted independently.

\subsubsection[FEB digital processing section]{\Acr[s]{feb} digital processing section}

As described in section~\ref{sssec:calo_electronics_FEB}, the 32 \Feb
input channels are grouped in four front-end blocks, each one
consisting of 2 ICECAL, 4 ADCs, 2 \Fpga{s} and one \Gbtx.  The
front-end \Fpga processing is divided into three distinct functional
stages.

The first stage processes the input ADC data, which needs to be
resynchronised (each ADC channel has its own clock) and processed to
remove the low frequency noise and to subtract the pedestal. A tunable
latency is also introduced after the data synchronisation in order to
correct for coarse bunch crossing misalignment between channels. The
data are then sent to the \Gbtx for serialisation.

In the second stage, the \Llt relevant quantities are calculated by
first applying a conversion factor from 12-bit encoding to words of 10
bits. The data is then sent at 80\mhz to the TrigSeq \Fpga for further
processing or to the neighbouring \Feb{s} for cluster calculations of
the \Feb border regions of the calorimeter (see
section~\ref{sssec:calo_electronics_LLT}). The low frequency noise
subtraction is implemented by subtracting signals previous to the
current one, with the assumption that occurrence of consecutive hits
in the same channel is extremely rare.

In the third stage, an \Acr{spi} interface is implemented for the
configuration and monitoring of the \Fpga. Several FIFOs are used to
inject digital patterns, store digital processing results, and store
\Llt calculations.

The data flow of the board can be tested by spy functionalities using
either pattern or signal injection. Patterns of 12-bit words to
simulate the ADC input can be injected very early in the
processing. Each channel can be configured to receive the standard ADC
input or in test pattern mode.  The test sequence is controlled by the
TrigSeq \Fpga so that the patterns can be synchronised between all the
\Fpga{s} that are in the test pattern mode.  Test sequences can run in
single trains of 1024 clock cycles, in loops or in single steps
started by a calibration sequence.  Additionally, charge injection
into the inputs of the amplifiers through small capacitors can also be
performed. Charge injection can be made on all channels simultaneously
or on any combination of individual channels.  Charge injection of
pulse pattern tests are fully configurable through the TrigSeq \Fpga.

\subsubsection{Data preprocessing}
\label{sssec:calo_electronics_LLT}

In order to facilitate the event reconstruction in the calorimeters by
the full-software LHCb trigger, and to ease the electron, photon and
hadron identification, a simple data preprocessing, which would be
otherwise time-consuming in the software trigger, has been implemented
in the \Feb{s}. This is historically indicated as \Acr[f]{llt}, in
analogy with the previous L0 hardware stage of the \runonetwo LHCb
trigger scheme, although this preprocessing does not provide any
triggering mechanism by itself.

The output of the \Llt calculations is fed to the trigger farm
concurrently to the raw ADC data. About 7\% of the bandwidth allowed
by the four optical links of the \Feb{s} is used by this stream
without any loss for the ADC data stream.

Electron and hadron candidates are defined as \emph{clusters}, defined
as sums of signals from $2 \times 2$ cells in \Ecal and \Hcal
respectively. Since the Scintillator Pad Detector and the Pre-Shower
systems have been dismantled for the LHCb upgrade, there is no way to
separate electron and photon candidates at the early \Feb stage, and
the identification is deferred to the software trigger
algorithms. Therefore, for this reason, in the context of the \Llt,
electron candidates indicate both electrons and photons.\looseness=-1

Clusters are built either within a single \Feb, if the $2 \times 2$
cells are contained in the 32 \Feb channels, or using neighbour
\Feb{s} in a crate or \Feb{s} belonging to neighbour crates. The
communication between \Feb{s} is ensured by 280\mhz serial links.

The first steps of the computations needed to obtain the electron and
hadron candidates in the \Llt are realised in the TrigSeq \Fpga. The
processing consists in a rough calibration of the energy deposited in
the calorimeter cells (performed in the front-end \Fpga) and in the
computation of the \et of the $2 \times 2$ clusters in each \Feb. In
addition to the cluster transverse energy, a few more quantities are
evaluated:
\begin{itemize}
\item the maximum transverse energy measured from the clusters built
  from $2 \times 2$ cells;
\item the address of the cluster giving the largest transverse energy
  as measured in the previous calculation;
\item the total transverse energy from the contributions of the 32
  channels handled by the \Feb,
\item the number of cells on the region covered by the \Feb for which
  the measured transverse energy is larger than a programmed threshold
  (hit multiplicity).
\end{itemize}
The results are added to the raw data on the optical links in order to
be further processed in the event building farm or to be possibly used
as electron, photon or hadron seeds in the software trigger. It is
also planned to use the total calorimeter transverse energy as a fast
luminosity counter for the experiment.\looseness=-1

\subsubsection[Monitoring and control: 3CU boards]{Monitoring and control: \Acr[s]{cccu} boards}
\label{ssec:calo_electronics_3CU}

\begin{figure}[t]
  \centering
  \includegraphics[width=0.8\linewidth]{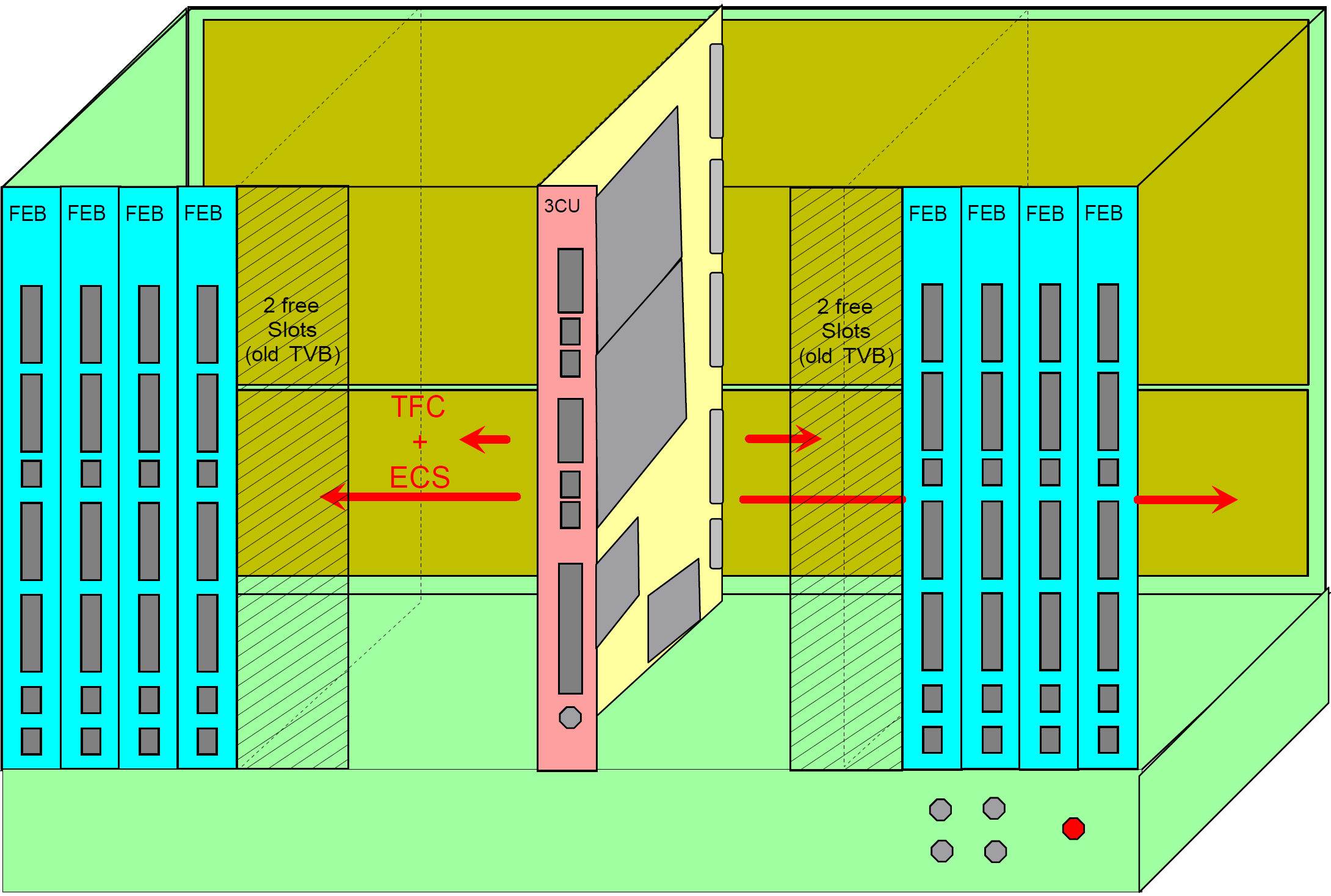}
  \caption{Schematic of the calorimeter crate.}
  \label{fig:Calo_crate}
\end{figure}

In each calorimeter \Fend crate (figure~\ref{fig:Calo_crate}), the
central slot is reserved to the \Acr[f]{cccu} board. The main role of
the \Cccu boards is to distribute the signals from the LHCb control
system to the \Feb{s} contained in the crate.  The crates are standard
9U VME-like crates with two custom backplanes. The lower one (3U
backplane) provides the power supplies, the \Tfc commands and the
clock distribution. The upper one (6U backplane) is reserved to the
exchange of signals between the boards and with the other crates.
The main role of the \Cccu is to receive the \Gbt{} frame through the
optical link and to extract the information which is needed by the
\Feb{s} inside a given crate: the 40\mhz clock, the \Tfc and \Ecs
commands. The processing of the \Tfc commands and the control of the
\Cccu board is performed by a dedicated \Fpga.\footnote{Microsemi
  IGLOO2 family.}

The \Cccu board contains a \Vtrx and a \Gbtx chip that can be used to
implement multipurpose high speed bidirectional optical
links. Logically, the link provides three distinct data paths for
timing and trigger control, \Daq and slow control information. In
practice, the three logical paths do not need to be physically
separated and are merged on a single optical link. The \Gbtx component
of the \Cccu is configured to drive 17 \Gbtsca{s}: the on-board
\Gbtsca and the \Gbtsca{s} of the 16 \Feb{s} that can populate a
crate.  As mentioned in section~\ref{sssec:calo_electronics_FEB}, the
\Feb{s} are protected by delatchers which detect any current increase
that could be due, for example, to a \Sel. If such a situation occurs,
the current is switched off for a few \ms, and then switched on again.
The \Cccu{s} monitor the \Sel{s} and control the
delatching. Additionally, the \Cccu \Fpga can enable \Feb firmware
reloading through the \Ecs if needed.

\subsubsection[LED monitoring, high voltage and HCAL calibration systems]{LED monitoring, high voltage and \hcal calibration systems}
\label{ssec:calo_electronics_monitoring}

The LED monitoring and \Hv systems of LHCb \Ecal and \Hcal are
described in details in~\cite{LHCb-DP-2008-001}.  The \Hv needed to
bias the photocathodes and the ten dynodes of each phototube is
generated by the \Cw generator boards, and is controlled by a single
analog voltage in a range of 0--5\volt applied to the control input of
each \Cw base and generated by a dedicated DAC.

The monitoring of the \Pmt{} gain during data taking is performed by
measuring the response to an LED flash of constant magnitude injected
into the \Pmt{} entrance window.  The LED flash magnitude is
adjustable by applying a control voltage (0--5\volt) to the inputs of
each LED driver.

The control voltages of \Cw bases and LED drivers are produced in the
common \Hv-LED DAC board~\cite{Konopl:2006}, which is interfaced to
the \Ecs.  Every \Hv-LED DAC board provides 200 outputs for \Pmt{} \Hv
control and 16 outputs for LED control.  The 6016 \Pmt{s} of \Ecal and
1488 \Pmt{s} of \Hcal are served by 40 \Hv-LED DAC boards, 32 for
\Ecal and 8 for \Hcal.

In addition to the LED flash magnitude, the LED monitoring system
controls also LED flash timing, which is performed by a dedicated
\Ledtsb~\cite{Konopl:2008}.  Each \Ledtsb board provides 64 LED
flash delays configurable via \Ecs.  A total of 8 \Ledtsb boards are
used for \Ecal and 2 boards for \Hcal.

In order to account for possible instabilities, the LED flash
magnitude is independently monitored using PIN
photodiodes.\footnote{Hamamatsu S1223-01.} A fraction of the LED light
is sent to these photodiodes and the corresponding signals are
digitised by dedicated standard \Feb{s} (see
section~\ref{sssec:calo_electronics_FEB}).  Since the \Hcal is placed
behind the \Ecal, it is difficult to calibrate its energy scale on
physics events.  Therefore, the absolute \Hcal calibration is
performed using a $\sim10$\aunit{mCi} $^{137}$Cs radioactive source
that can be moved across every cell~\cite{rustem:2003}.  The gain
is obtained by measuring the anode current of each \Pmt{} under source
irradiation.  The \Hcal calibration is performed during \Lhc technical
stops in dedicated data taking runs.  The readout of the \Pmt{} anode
currents during the \Hcal calibration is performed via \Ecs by means
of dedicated boards, named CsCalib, reading out a quarter of \Hcal
each.\looseness=-1

\begin{figure}[t]
  \centering
  \includegraphics[width=0.7\textwidth]{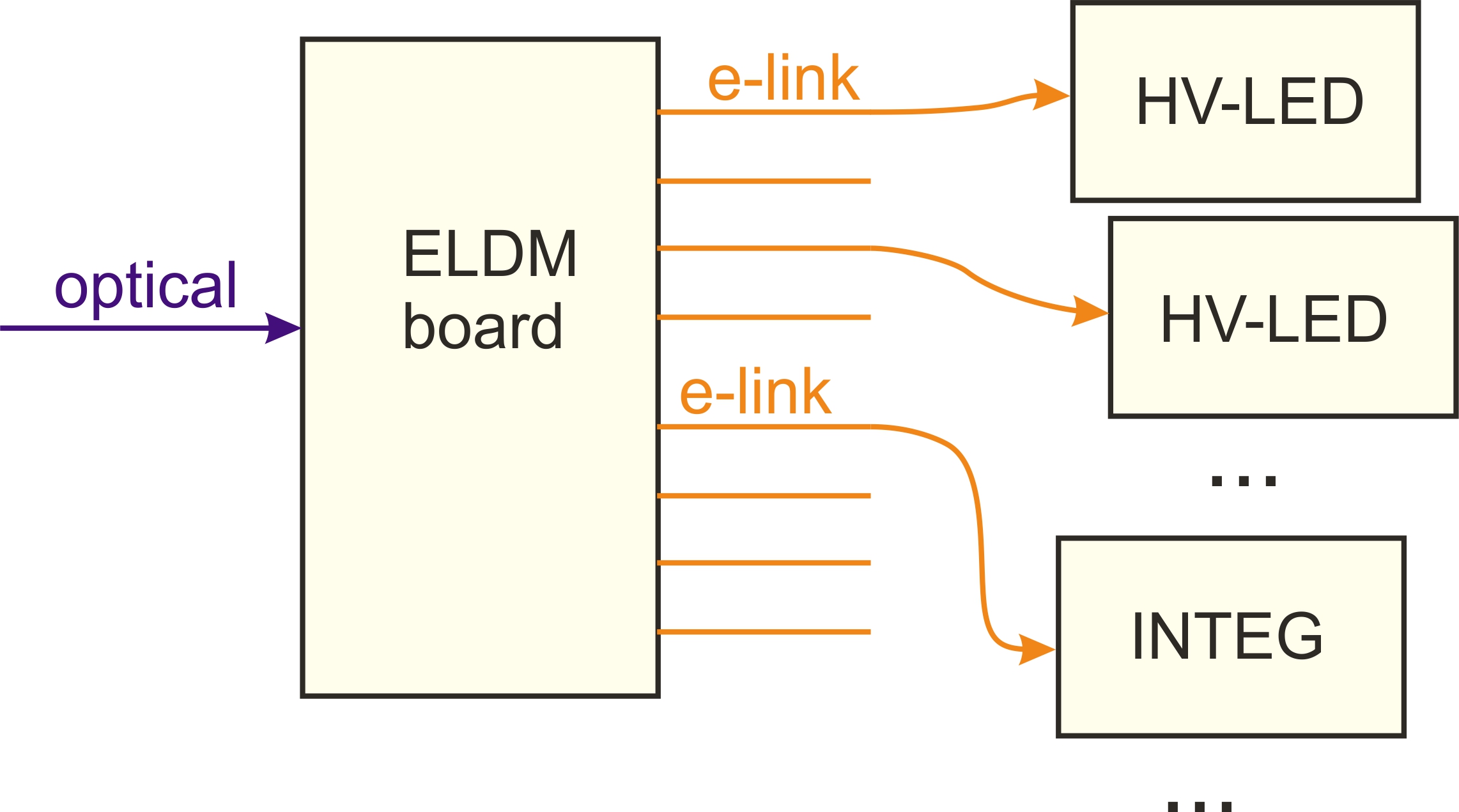}
  \caption{Control board connection scheme with the new \Gbt{}
    protocol.}
  \label{fig:Calo_SPECS_GBT_layouts}
\end{figure}

The three types of control boards described above have similar
features in their architecture.  Namely, they consist of a motherboard
bearing the elements executing the main function of the board (DACs
for the \Hv-LED DAC boards, delay chips for \Ledtsb{s}, ADC for the
CsCalib boards), on which additional mezzanine boards are installed to
implement data transmission and board configuration functions.  The
upgrade of these boards was dictated by the phasing out of the SPECS
bus~\cite{SPECS:2003}, used in the previous boards, and its
replacement by the universal \Gbt{} protocol.  The upgrade allowed
also to replace the obsolete \Fpga{s} of the configuration mezzanine
with a new radiation tolerant \Fpga.\footnote{Microsemi IGLOO2
  series.}

The transition from SPECS to \Gbt{} required also the introduction of
an additional board, the \elink~distribution module (ELDM) (see
figure~\ref{fig:Calo_SPECS_GBT_layouts}).  This board, featuring a
\Vtrx module and a \Gbtx chip, distributes the information from the
optical duplex \Gbt{} line to several (up to 10) control boards
connected via copper \elink{s}.

\subsection{Test beam results}

To check the upgraded \Feb functionality in more realistic conditions
a test beam campaign was organised, mainly at the CERN T4-H8 beam
line. An \Ecal module equipped with photomultiplier tubes and their
bases was exposed to a beam of electrons with energy ranging from 20
to 120\gev. The signal was triggered using two scintillators and sent
to a \Feb prototype and, in parallel, to a charge integrator and a
TDC, which allowed to verify ADC linearity and provided a precise time
stamp.  The tests were carried out with prototypes of the electronics
at different development stages in 2012, 2015 and 2018. They allowed
to study the analog electronics in realistic detector conditions
(final design \Pmt{s}, bases, signal cables, etc.) and to verify that
the performance, from the point of view of energy resolution,
linearity, noise and signal shaping was within specifications. A test
with the final version of the \Feb and a simplified version of LHCb
\Daq system was made to test the whole acquisition and configuration
chain.

\begin{figure}[h]
  \centering
  \includegraphics[height=0.4\linewidth]{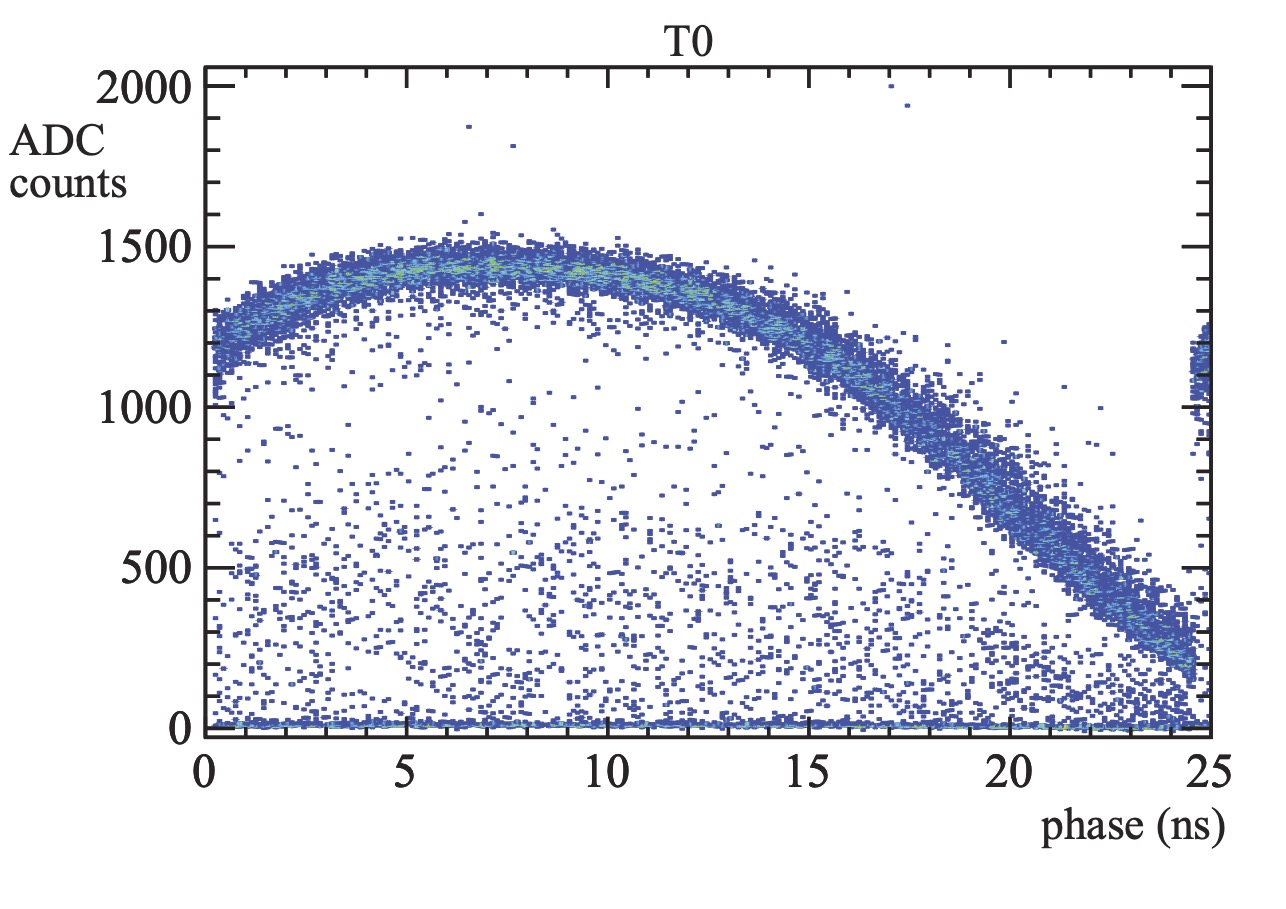}
  \includegraphics[height=0.4\linewidth]{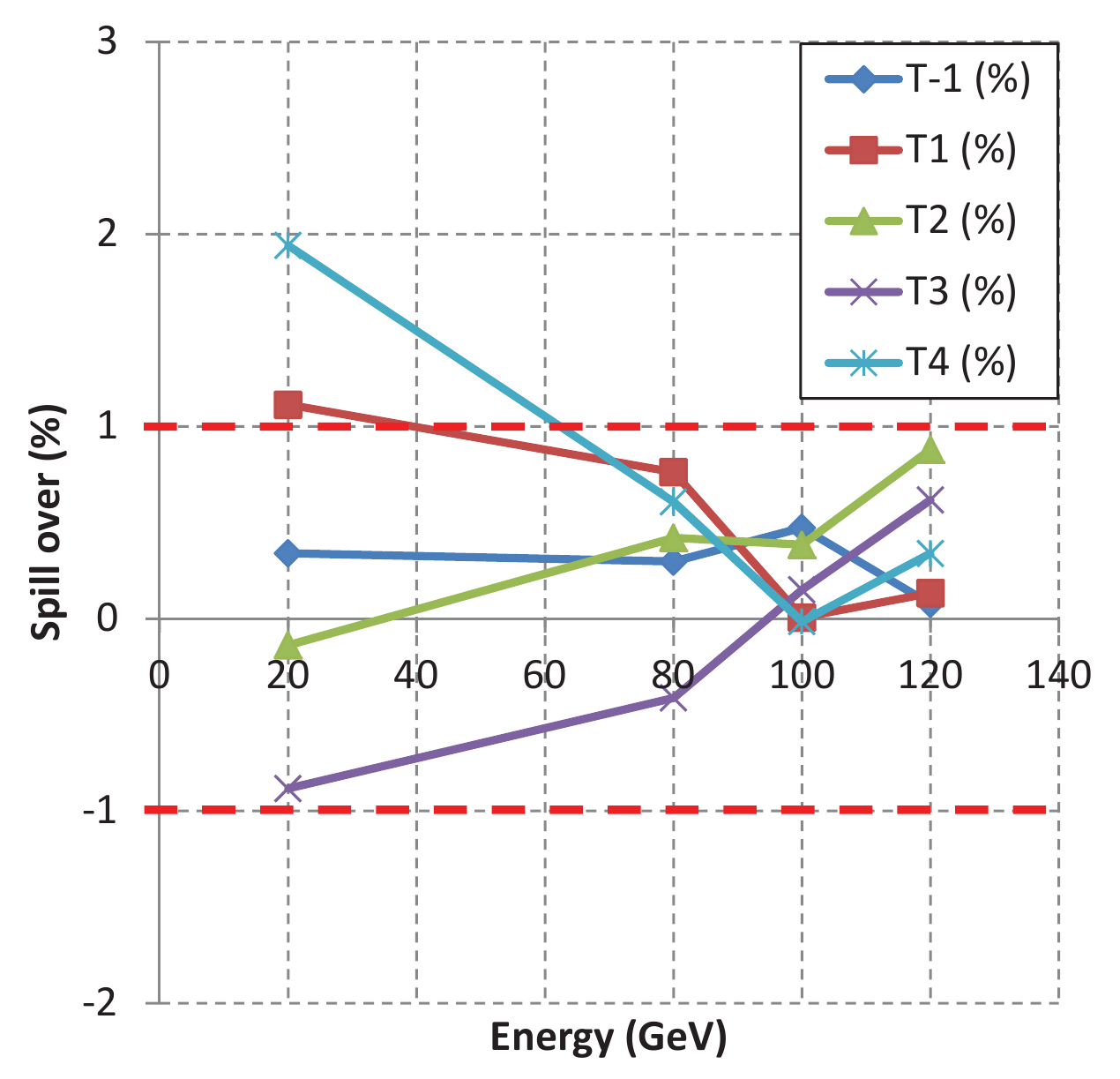}
  \caption{\Feb test beam results. Left: integrated charge (ADC
    counts) as a function of hit time phase with respect to the
    internal 40\mhz clock. Right: spillover measured in different
    clock cycles as a function of the beam energy.}
  \label{fig:Calo_TestBeam_int_spillover}
\end{figure}

The electron energy range, from 20 to 120\gev, allowed a detailed
characterisation of the \Fend amplifier.  The electron beam was
obtained from the main H8 hadron beam by inserting a lead target into
the beam line. Magnetic optics downstream the converter were used to
select monochromatic electrons.  With a 6\mm thick lead absorber, the
electron beam had a pion contamination larger than 60\%. Changing to a
12\mm thick absorber improved the electron purity. However, increasing
the energy of the beam resulted in a decreased purity and made it
difficult to discriminate between low energy electron events, muons
and noise.

The CERN \Sps beam is not synchronised with the 40\mhz LHC clock,
hence hits are recorded by the \Feb under test with a random phase
with respect to the internal clock.  Therefore, the relative phase
between each event and the local clock was measured with a TDC. To
measure the ICECAL integrator stability, signals were sampled in
consecutive 25\ns windows and their time phase with respect to the
40\mhz clock was measured. A selection to separate electrons from
pions was applied although it was complicated by the low beam
purity. The measured ADC values as a function of the hit relative
phase within the 25\ns window are shown in
figure~\ref{fig:Calo_TestBeam_int_spillover} (left). A plateau of
$\sim 4$\ns is clearly visible, with a stability better than 1\%
showing that if properly time-aligned, the ICECAL is able to integrate
the full signal charge within the 25\ns \Lhc bunch-crossing.

By choosing events where the signal is fully integrated within 25\ns
(\emph{T0 cycle}), it is possible to look into previous and following
clock cycles to measure the residual spillover charge not integrated
within the main clock cycle.  The result of this test is shown in
figure~\ref{fig:Calo_TestBeam_int_spillover} (right), where the
spillover is measured in the previous (\emph{T-1}) and following
(\emph{T1-T4}) clock cycles with respect to T0, at different beam
energies. The measured spillover is within $\pm 1\%$ as required, with
a maximal deviation observed at low energy, where however the
separation between electrons and pions was very difficult. Note that
the spillover can be positive as well as negative due to the signal
shaping used to reduce the pulse width to equal or less than 25\ns
inside the ICECAL ASIC.

\begin{figure}[h]
  \centering
  \includegraphics[width=0.46\linewidth]{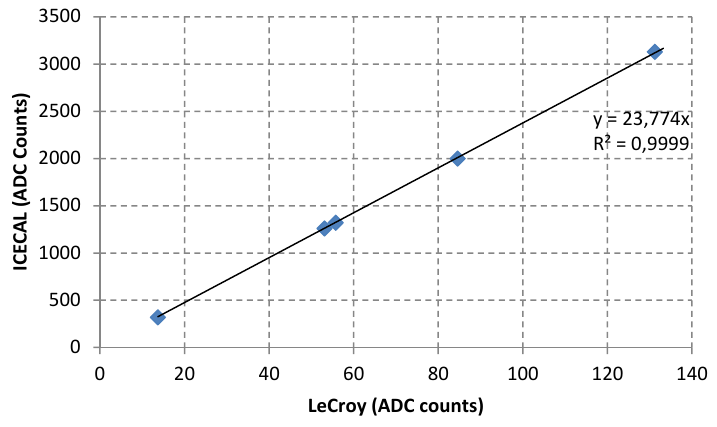}
  \includegraphics[width=0.46\linewidth]{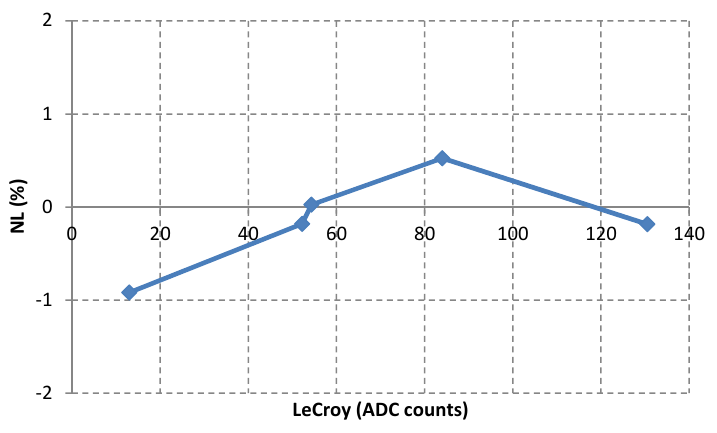}
  \caption{Left: energy values measured with the \Feb prototype at a
    beam test with respect to the reference charge integrator
    values. Right: the nonlinearity deviation is shown to be less than
    1\%.}
  \label{fig:Calo_TestBeam_linearity}
\end{figure}

The key parameters of the analog circuit were checked and showed a
behaviour within specifications. Linearity was checked to be better
than 1\% by comparing the charge integrator and the \Feb prototype
readings for different electron energies. The results are shown in
figure~\ref{fig:Calo_TestBeam_linearity}. As shown before, the
integrator stability was better than 1\% in 4\ns. The spillover was
also tested for different \Pmt{s}, \Pmt{} biasing voltage, \Pmt{}
bases and signal cables and was found to be stably below~1\%.\looseness=-1

\begin{figure}[t]
  \centering
  \includegraphics[width=0.88\linewidth]{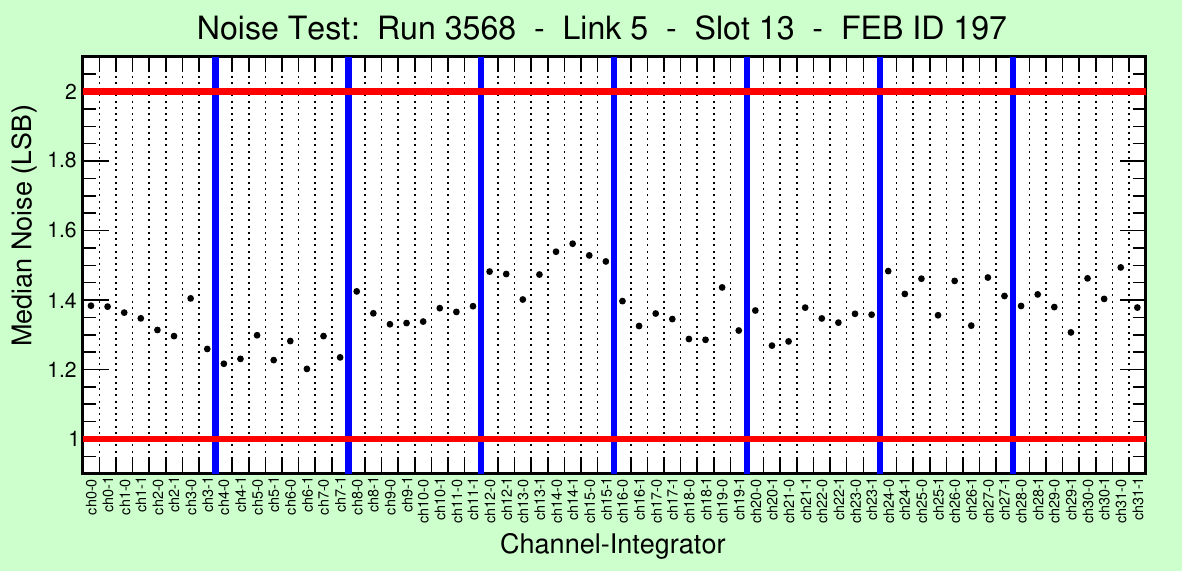}
  \caption{Noise of all channels of a \Feb, measured in laboratory
    conditions.}
  \label{fig:Calo_TestBeam_noise}
\end{figure}

The noise was checked in different conditions both in the laboratory
and at the test beam. Results with and without pedestal and low
frequency noise subtraction were also obtained.  Test bench
measurements without \Pmt{} biasing yielded a noise level of 1.3 and
1.6 ADC counts before and after pedestal subtraction, respectively,
slightly above specifications. In the beam tests, the noise behaviour
varied significantly, ranging from 1.6 to 3.4 ADC counts, probably due
to setup changes and grounding imperfections.  To obtain a more
realistic estimation, a measurements was performed directly in LHCb,
connecting to a channel at the crate level with proper grounding.
After pedestal subtraction, the noise was measured to be about 1.4
LSB, in agreement with test bench results as shown in
figure~\ref{fig:Calo_TestBeam_noise}, a level that was considered
acceptable.

\section{Muon system}
\label{subsec:muon}
\subsection{Overview}
\label{subsec:muonoverview}

The LHCb muon
detector~\cite{LHCb-DP-2008-001,LHCb-TDR-004,muonbib:TDR1,muonbib:TDR2}
has been successfully operated during \lhc \runone and \runtwo with an
excellent performance~\cite{LHCb-DP-2012-002,LHCb-DP-2013-001}.

The LHCb muon system is composed of four stations M2 to M5 comprising
1104 \Mwpc for a total area of 385\ma.  The LHCb \runonetwo muon
system additionally included a station M1 located upstream of the
calorimeters and comprising 12 \Gem{s} in the innermost region and 264
MWPCs. Station M1 was utilised in the hardware L0 trigger and thus is
no longer needed in the upgraded system.  Its supporting structure has
been maintained to host the \Scifi neutron shielding as described in
section~\ref{subsubsec:neutronshielding}.  Each station is composed of
two mechanically independent halves, \aside and \cside.  The four
stations M2 to M5, located downstream of the calorimeter system, are
equipped with \Mwpc{s} and interleaved with 80\cm thick iron absorbers
to filter low energy particles. Each station is divided into four
regions, R1 to R4, of increasing area moving from the central beam
axis outwards. The area and the segmentation of the four regions scale
in such a way to uniformly distribute the particle flux and the
channel occupancy across each station.  The \Mwpc{s} are made up of
four independent layers (or \emph{gaps}), each consisting of anode
wires between two cathode planes, to achieve a high efficiency and a
high redundancy.  The \Fend electronics host an
amplifier-shaper-discriminator stage implemented in a dedicated \Asic
as well as a digital section that allows time alignment of the signals
and logical combinations of readout channels in so-called
\emph{logical channels}.  The \Fend electronics was designed to be
radiation tolerant up to 100\kGy which is expected to be adequate also
at the new running conditions. Thus, the current \Fend electronics is
kept unchanged.  Digitised signals from the \Fend electronics were
originally sent at 40\mhz rate to the L0 trigger but recorded for
further processing only at a maximum rate of 1\mhz. To comply with the
new LHCb readout scheme a complete overhaul of the readout electronics
has been carried out representing the main upgrade of the muon
system~\cite{LHCb-TDR-014}.

The monitoring and control electronics have also been completely
redesigned to comply with the new 40\mhz readout rate and the new
experiment's \Daq and control systems. Despite the significant changes
required, the new electronics have been designed to be
backward-compatible with the original architecture in order to
minimise the cost and allowing to reuse of the original crates,
cabling, and power supplies. The new electronics is presented in
section~\ref{ssec:muonELT}, while section~\ref{ssec:muonT40fw}
describes the muon-system-specific processing implemented in the
back-end electronics system. Section~\ref{ssec:muonECS} presents the
new \Ecs software developed to monitor and control the muon system.

The high particle rates expected in the upgraded muon system required
the introduction of specific strategies to mitigate the otherwise
unacceptable level of induced inefficiency.  These are discussed in
section~\ref{ssec:muonHRmit}.

The \Mwpc{s} have been operated for eight years and are expected to be
left in place for the whole lifetime of the experiment. To secure the
smooth operation of the muon detector, the number of needed spares has
been estimated based on the operational experience during \runone and
\runtwo. In addition to the ones already available from past
productions, 54 new chambers have been built (30 for M5R2 and M5R4
regions and 24 for M2R3, M2R4, M3R4, and M4R2 regions). Studies on
expected long term operation of the chambers have been performed and
are summarised in section~\ref{ssec:muonLTO}.
\subsection{Electronics}
\label{ssec:muonELT}

The muon system electronics is designed to convert, format and
transmit downstream the analog signals extracted from the detector. As
introduced in section~\ref{subsec:muonoverview}, the \Fend electronics
is fully compliant with the upgraded experiment running conditions and
will be therefore maintained.  The \Mwpc signals are digitised by the
front-end CARDIAC boards~\cite{muonbib:CARDIAC} which host two kind
of chips, two \Acr[p]{carioca}~\cite{muonbib:CARIOCA} and a
\Dialog~\cite{muonbib:DIALOG}, both implemented in a IBM 250\nm
technology combined with a specific layout technique to be radiation
tolerant up to 100\kGy. The \Carioca is an eight-channel \Asic that
implements a current mode amplifier, a three stage discriminator and a
LVDS output driver. In each CARDIAC one \Dialog chip receives sixteen
input channels from two \Carioca boards. Each input channel can be
configured with a programmable delay in steps of about 1.6\ns realised
by a voltage-controlled delay line. Each channel can be propagated or
disabled by means of a masking circuit. In order to optimise the
number of readout channels, the \Carioca outputs are combined within
the CARDIAC into \emph{logical channels} by means of selectable logic
functions suitably chosen according to the position of the
corresponding detector channels. The \Dialog provides also individual
threshold settings for the \Carioca discriminators and test pulses.
The \Dialog{s} are fully accessible and configurable via an \I2c
interface.

\begin{figure}[t]
  \centering
  \includegraphics[width=0.45\linewidth]{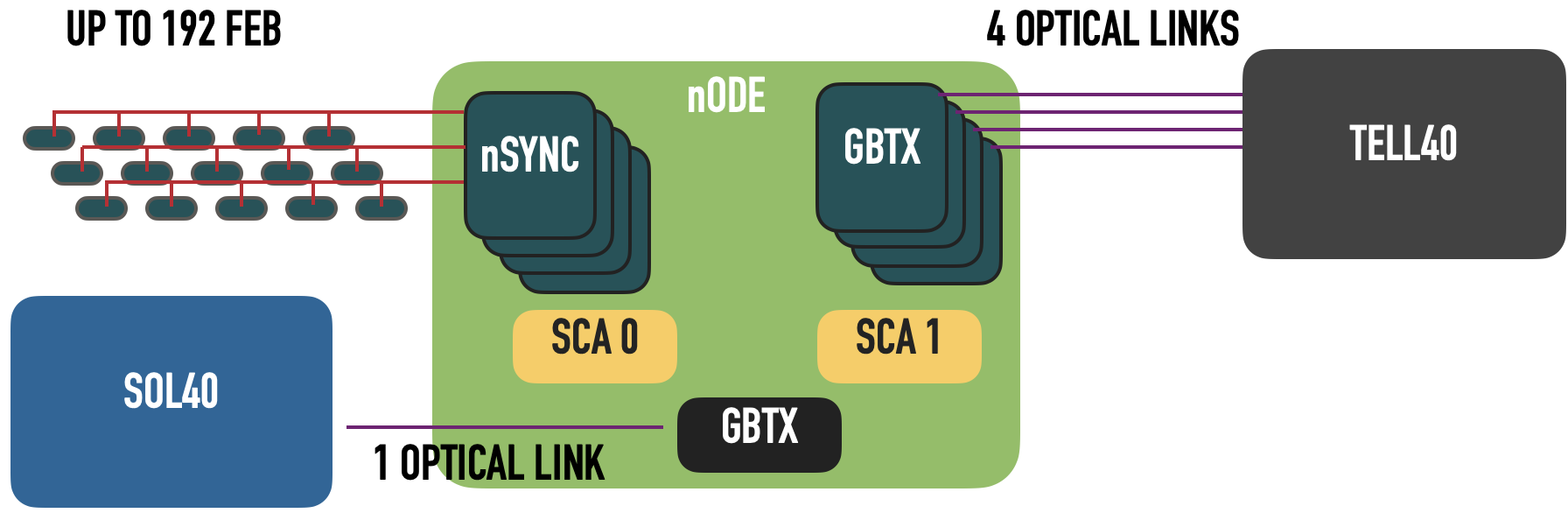}
  \hspace{0.05\linewidth}
  \includegraphics[width=0.45\linewidth]{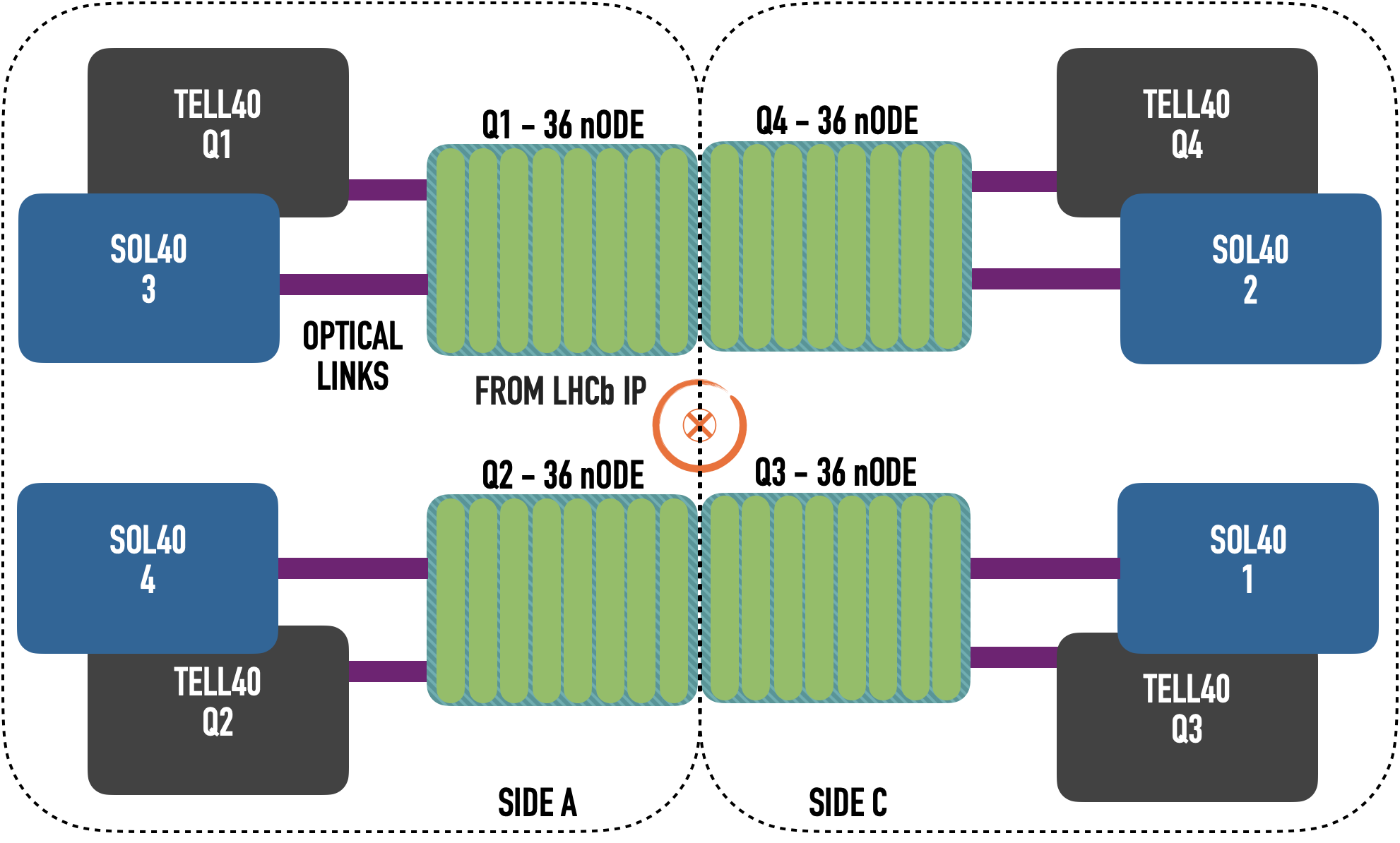}
  \caption{\Node communication scheme.}
  \label{fig:nODE}
\end{figure}

During operation in \runonetwo, signals from CARDIAC boards, which are
in some cases further combined by the Intermediate Boards, were
received by the \Acr{ode} boards which provided a time stamp through a
4-bit TDC, formatted and shipped the data to the \Bend
electronics. The TDC was implemented in a dedicated \Asic named SYNC
which also allowed coarse grained time alignment of signals. Data was
shipped by the \Ode to the L0 trigger at a rate of 40\mhz and, upon
L0 trigger positive decision, to the \Daq boards at a maximum rate of
1\mhz. This scheme can no longer be maintained in the upgraded readout
system and the \Ode has been redesigned and upgraded into the \Node
boards. The TDC has also been fully redesigned and upgraded into the
\Acr{nsync}. In the \Node boards, signals are synchronised with the
master LHCb clock, compressed, formatted and sent to the downstream
\Tellfourty readout boards via high-speed optical links. The \Node
boards are also connected with the \Solfourty boards to interface and
distribute the \Tfc and \Ecs information.  Figure~\ref{fig:nODE} shows
the communication scheme and the architecture of the new \Acr[m]{ode}
system comprising 144 \Node{s}. Each quadrant will be served by 36
\Node{s}, installed in five crates.

\begin{figure}[t]
  \centering
  \includegraphics[width=0.45\linewidth]{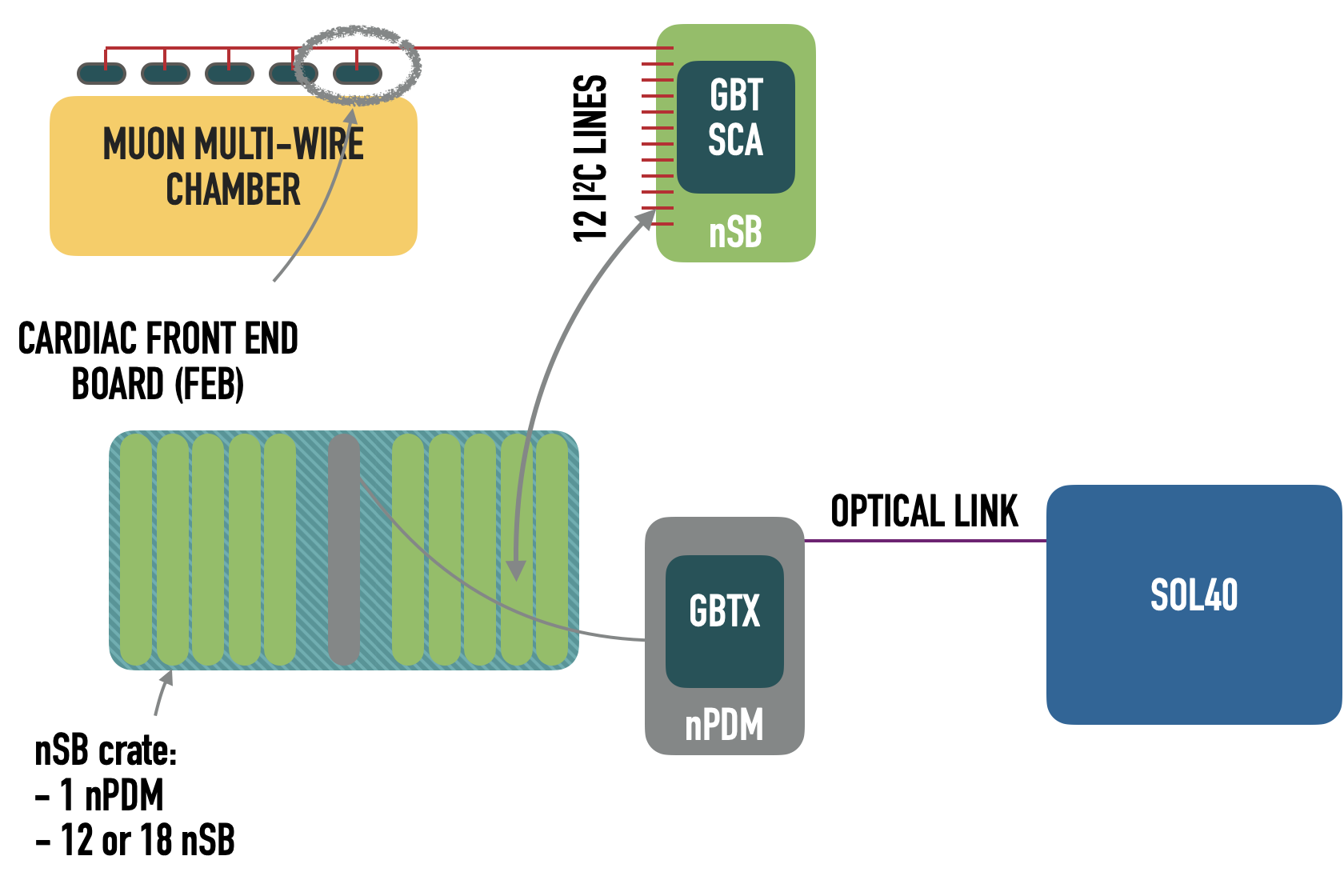}
  \hspace{0.05\linewidth}
  \includegraphics[width=0.45\linewidth]{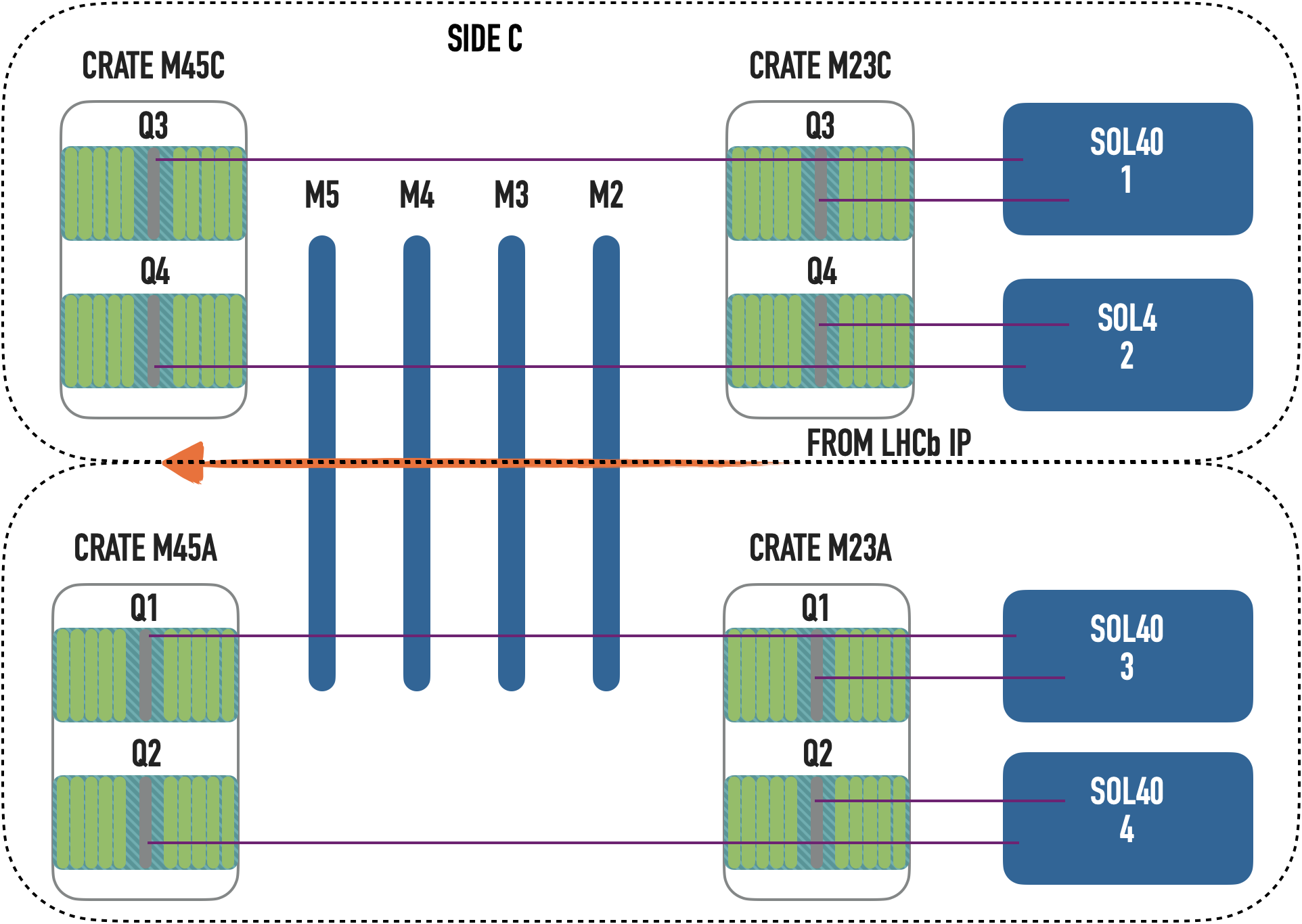}
  \caption{\Fend electronics communication scheme.}
  \label{fig:nSB}
\end{figure}

The muon control and monitoring system~\cite{muonbib:nSBS},
formerly provided by the \Acr[p]{sb}, has also been fully redesigned
to comply with the new \Ecs and \Tfc systems and has been implemented
in the \Nsbs. The \Nsbs is comprised of the \Acr[p]{npdm}, the
\Acr[p]{nsb}, and the \Acr[p]{ncb}.  Figure~\ref{fig:nSB} shows the
communication scheme and the architecture of the \Nsbs. Eight crates
comprising 12 or 18 \Nsb{s} and one \Npdm are installed for a total of
8 \Npdm and 120 \Nsb{s}.

\subsubsection[The nODE]{The \Acr[s]{node}}
\label{subsubsec:nODE}

The \Node conceptual design is similar to the previous \Ode board and
is intended to be backward-compatible with the previous architecture
in terms of board format, crate occupancy and cabling.

The \Node receives up to 192 input logical channels which are
processed by the four onboard radiation-tolerant \Nsync chips which
provide clock synchronisation, bunch-crossing alignment, time
measurements, histogram capability and buffering (see
section~\ref{subsubsec:nSYNC}).  In order to optimise the output
bandwidth and minimise the number of long-distance optical links the
\Node transmits data frames with packed nonzero-suppressed hit maps of
the corresponding input channels and zero-suppressed time stamps from
the \Nsync{s}, thus transferring the decoding part to the \Bend
\Tellfourty boards. Details of data formatting are discussed in
section~\ref{subsubsec:nSYNC}.

\begin{figure}[t]
  \centering
  \includegraphics[width=0.8\linewidth]{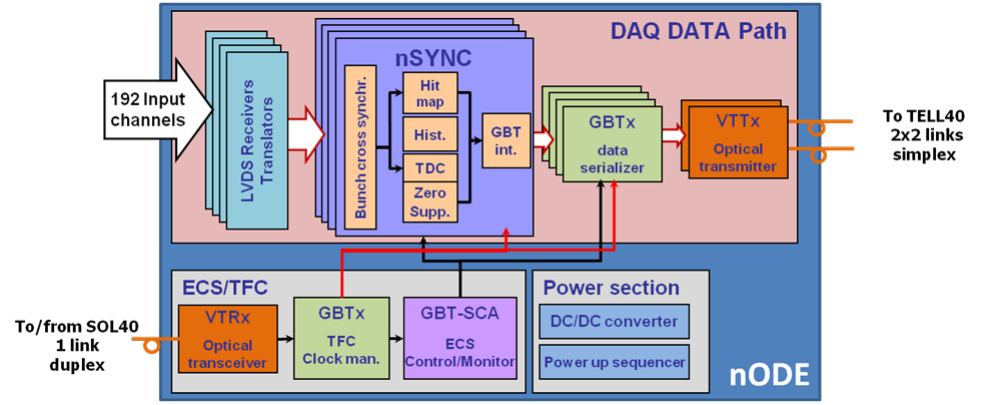}
  \caption{\Node block diagram. Reproduced from~\cite{Cardini_2014}. \textcopyright\ CERN 2014 for the benefit of the LHCb
collaboration. CC BY 3.0.}
  \label{fig:nODE_BD}
\end{figure}

\begin{figure}[t]
  \centering
  \includegraphics[width=0.5\linewidth]{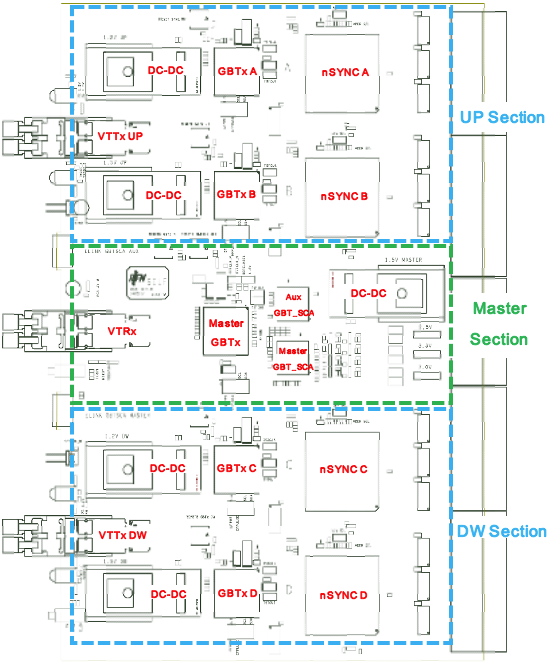}
  \caption{\Node board layout.}
  \label{fig:nODE_Layout}
\end{figure}

The logical block diagram of the \Node is shown in
figure~\ref{fig:nODE_BD}. Board functionalities can be grouped in
three main logical blocks: \Tfc and \Ecs management block with related
distribution sections, \Fend and data electronics block, and power
converter and distribution block.  The board layout, shown in
figure~\ref{fig:nODE_Layout}, is divided in three functional sections:
a master section implementing management and distribution of \Tfc and
\Ecs information and two slave sections, UP and DW, implementing \Fend
and data functionalities.  Each stage includes dedicated power
management circuits.

The master section is based on one \Gbtx chip, two \Gbtsca chips and
one \Vtrx optical transceiver. The \Gbtx is configured in transceiver
forward error correction mode and generates the reference clock
synchronous with the master LHCb clock, which is then distributed to
slave sections. The \Ecs/\Tfc path relies on two \Gbtsca{s} to control
the slave \Gbtx{s} and \Nsync{s}.  The \Tfc information of the \Node
implements a subset of seven of the standard \Tfc commands of the LHCb
experiment.  The encoded \Tfc command is replicated four times in the
data frame and distributed to each \Nsync via seven
\elink{s}~\cite{Bonacini:1235849}.  The \Ecs interface is used to
configure the electronics components of the board, to monitor their
status and to download the time histograms.

The slave sections are comprised of a total of up to 192 input
channels, four \Nsync chips, four \Gbtx chips and two \Vttx{} optical
transmitters. All such elements are arranged into two identical
sections, UP and DW. Each section is further subdivided into two
identical data paths comprising one \Nsync with 48 input channels, one
\Gbtx and one channel of the \Vttx{} module.  The slave \Gbtx{s} are
configured as simple transmitters in widebus mode with fixed header
and fixed frame scheme. The \Nsync data frame is received by the \Gbtx
via fourteen input \elink{s} at 320\mbps. The 112 bits data field of
\Gbtx data frame is composed of a Header Field of 12 bits encoding the
\Bxid, followed by four dummy bits and by the \Nsync data frame.

\subsubsection{The nSYNC}
\label{subsubsec:nSYNC}

\begin{figure}[t]
  \centering
  \includegraphics[width=0.60\linewidth]{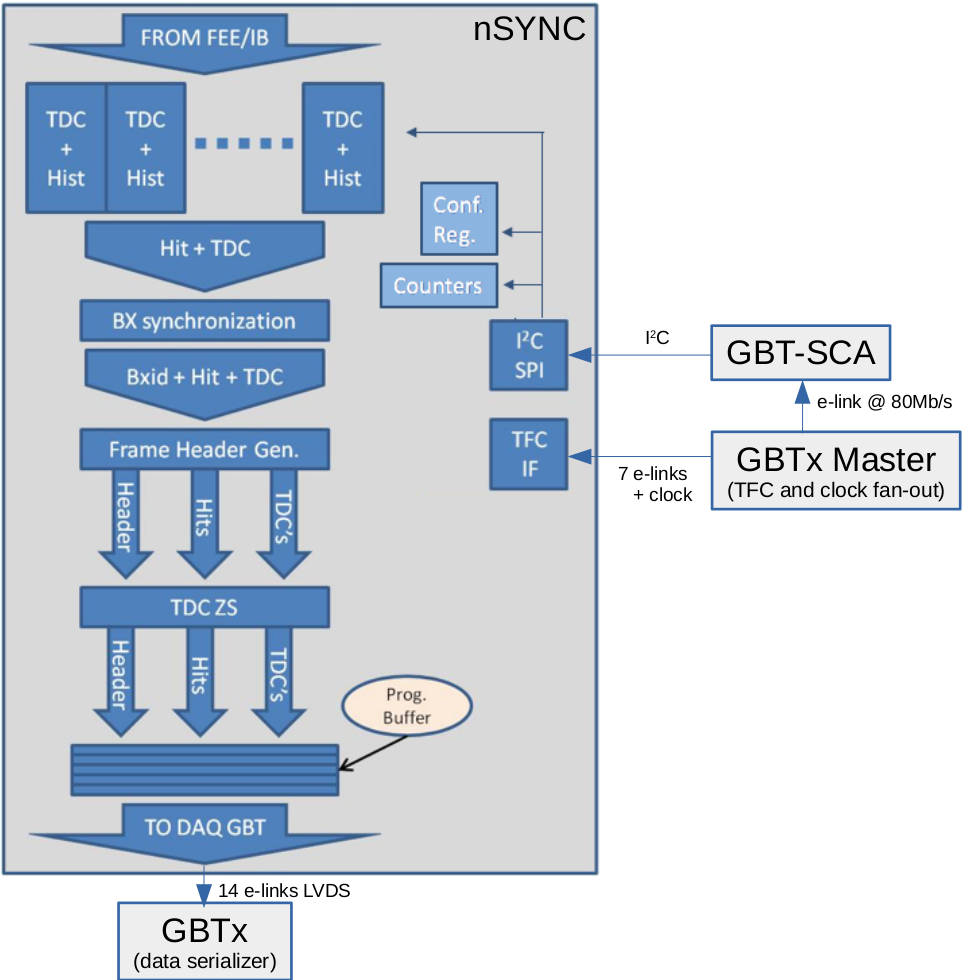}
  \caption{Schematic view of the \Nsync architecture and its interface
    with the GBT chipset. Reprinted
from~\cite{muonbib:CADEDDU2019378}, Copyright (2019), with permission from Elsevier.}
  \label{fig:nSYNCdiagram}
\end{figure}

The main building block of the new readout electronics of the LHCb
muon system is the \Nsync chip~\cite{muonbib:CADEDDU2019378}, a
radiation tolerant custom \Asic developed in UMC 130\nm technology as
an evolution of the SYNC chip~\cite{muonbib:5571049}, used in the
old \Ode boards. The \Nsync architecture is composed of several
functional blocks, schematically shown in
figure~\ref{fig:nSYNCdiagram}.  The main purpose of the \Nsync is to
integrate all the required functionalities for the upgrade of the
readout system, such as clock synchronisation, bunch crossing
alignment, hit map production, time measurements, histogram capability
and buffers. The \Nsync handles also the zero-suppression algorithm
for the time measurements and the interfaces to the \Daq and \Tfc/\Ecs
systems.  The \Nsync receives the digital signals coming from the muon
chambers, through 48 LVDS input channels, and synchronises them with
respect to the \Bxid. Concurrently, the phase of the arriving signal
is measured by a TDC (one for each channel) with respect to the LHC
40\mhz master clock. This information is crucial for the time
alignment of the whole muon detector. The \Bxid information, the hit
maps and the TDC counts are then combined to build a frame that, after
the TDC data zero-suppression, is transmitted to the \Daq through the
GBT output interface.

The TDC is composed of a fully digital patented
\Dco~\cite{muonbib:7348815}, which produces a clock signal based on
an input digital word. The TDC works with a nominal resolution of
1.56\ns. This is obtained by dividing the 25\ns \Lhc clock cycle into
16 slices, allowing to encode the TDC time stamp in 4-bit words. Lower
(8 slices) or higher (32 slices) resolution can also be configured.
Incoming signals trigger the \Dco clock, whose periods are then
counted until both the counter and the \Dco are stopped by the arrival
of the rising edge of the master clock. The phase measurement
corresponds to the counter output and is stored in an output buffer.
Since the \Dco is composed of a digital delay chain, a systematic
error is continuously accumulated during the measurements.  Therefore
a dithering system is implemented in order to add or remove a unit to
the digital input word of the \Dco, thus inverting the systematic
error and preventing its accumulation.  At nominal resolution, the
output data of each TDC consist of a binary flag (corresponding to the
hit/no-hit information) and a 4-bit-wide word with the measured phase,
if a hit is present. Each \Nsync channel is also equipped with a
histogram facility, comprised of 16 counters, in order to monitor the
measured phases.

The TDC output data are sent in parallel through a pipeline in order
to align all the hits belonging to the same bunch crossing. This
allows to uniquely associate the 12 bit-wide \Bxid number to all the
data. Finally the extended output frame is built by combining a
header, the 48 bit-wide hit map and the TDC data. While the hit map is
always sent nonzero-suppressed, the TDC data are instead subject to a
zero-suppression algorithm: only the first nonempty block of TDC data
are added to the frame, up to a maximum of 12 TDC measurements at
nominal resolution. The TDC data address decoding is therefore
deferred to the \Bend electronics, allowing an optimisation of the
output bandwidth. The last eight bits of the frame are dedicated to
the Hamming code, a feature that can be disabled to increase the TDC
occupancy.

Output communication between \Nsync and \Gbtx chips is ensured by 14
LVDS links at a transmission rate of 320\mbps. An intermediate
asynchronous FIFO is implemented to interface the two clock
domains. The high frequency clock is generated internally using a \Pll
with the 40\mhz master clock as reference. Moreover, an 8-steps
programmable output pipeline allows the correct alignment from the
receiver side. The 40\mhz clock and all other synchronous fast
commands sent by the \Tfc system are received through the \Gbtx master
chip. Asynchronous slow commands, like configuration commands, are
received from the \Gbtsca chip through a \I2c bus, as explained in
section~\ref{subsubsec:nODE}.\looseness=-1

The radiation level at the muon readout crate location is not
critical, being about 40 times smaller than in the detector
region. For 10 years of LHCb upgrade operation the expected total
ionising dose is 130\Gy, with an expected fluence of $2\times10^{12}$
1\mev \neqcmcm~\cite{muonbib:Brundu:2020ixi}.  Nevertheless, to
ensure proper \Asic operation, two radiation hardness techniques have
been implemented in the \Nsync design to mitigate the \Seu rate:
triple modular redundancy to protect the most critical registers, such
as the first input stage of \Tfc commands and the configuration
registers, and the Hamming code with the corresponding error detection
and correction logic, to protect all the internal counters, buffers
and output FIFO.  The radiation hardness of the \Nsync has been
verified with several tests, in particular using a 60\mev proton beam
and X-ray
irradiation~\cite{muonbib:Brundu:2018vdm,muonbib:Brundu:2316037},
up to $2\times10^{12}$ 1\MeV \neqcmcm and a total ionising dose of
1.3\kGy.  The \Nsync showed an excellent performance under radiation
with no failure or \Seu behaviour after an accumulated total ionising
dose ten times larger than the one expected for 10 years of \lhcb
upgrade operations. The chip current consumption and the internal
clock jitter are expected to increase by less than 5\% for the same
time period. The \Seu cross section per bit has been also measured to
be $(0.53\pm 0.04)\times 10^{-13}\cma$, corresponding to an expected
\Seu rate of less than 0.1 events per day for the whole muon detector
for the most important registers, thus having a negligible impact on
the overall muon detector performance and efficiency.

\subsubsection{The new service board system}

\begin{figure}[t]
  \centering
  \includegraphics[width=0.8\linewidth]{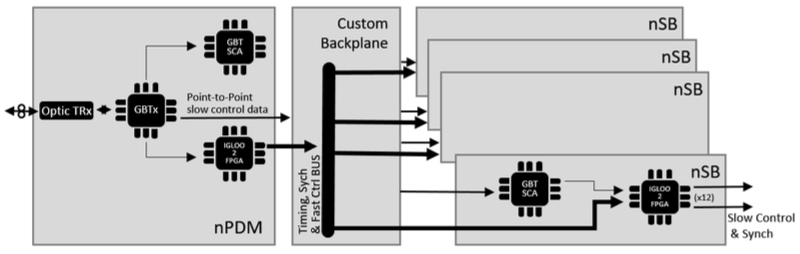}
  \caption{\Nsbs block diagram.}
  \label{fig:nSBS_BD}
\end{figure}

\begin{figure}[t]
  \centering
  \includegraphics[width=0.45\linewidth]{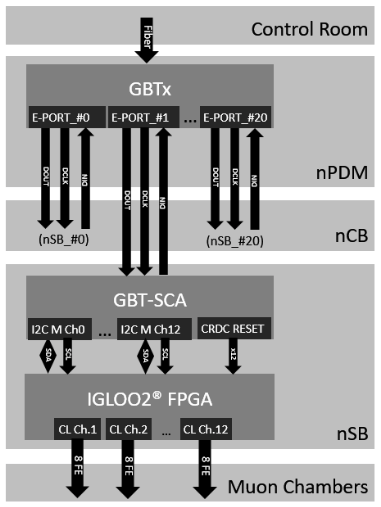}
  \hfill
  \includegraphics[width=0.45\linewidth]{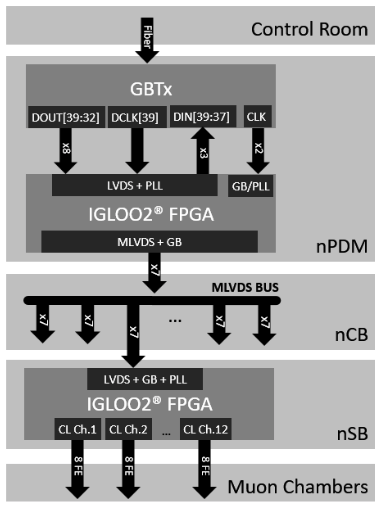}
  \caption{Left: functional representation of the \Nsbs \Ecs
    path. Right: functional representation of the \Nsbs \Tfc
    path. Reproduced from~\cite{muonbib:nSBS}, with permission from Springer Nature.}
  \label{fig:nSBS_TFC}
\end{figure}

The main purpose of the \Nsbs is the management and distribution of
\Tfc and \Ecs information to the CARDIAC \Fend boards. The \Nsbs
consists of a crate hosting a master board, called the \Acr[f]{npdm},
and up to 20 slave boards, called \Acr[m]{nsb}{s} (\Nsb), which
replaces the previous system. The \Npdm communicates with the \Nsb{s}
via \elink{s} on a custom backplane, called \Acr[f]{ncb}. Each \Nsb is
interfaced with up to 96 CARDIAC boards via 12 serial links. The block
diagram of the \Nsbs is shown in figure~\ref{fig:nSBS_BD}.
Control signals from the \Ecs and \Tfc are routed to two separate
paths in order to reduce the complexity and the costs of the whole
system. The functional representation of each path is shown in
figure~\ref{fig:nSBS_TFC}.

The \Npdm main components are a \Gbtx chip, a \Gbtsca chip, an IGLOO2
\Fpga and a \Vtrx optical transceiver. A \Gbtsca and an IGLOO2 \Fpga
constitute also the main components of each \Nsb.  The \Npdm \Gbtx is
controlled through the \Vtrx by the interface board \Solfourty. The
\Gbtx is configured in transceiver mode with forward error correction
and it generates the reference clock for the whole \Nsbs, synchronous
with the master \lhcb clock.  The \Tfc interface relies on eight
\elink{s} at 40\mbps from the same group and propagates the \Tfc
commands to the \Npdm \Fpga, which in turn distributes them to the
\Nsb \Fpga via backplane lines.  The local \Npdm \Ecs interface uses
the EC field of the frame to communicate with the \Npdm \Sca which
controls and configures the \Npdm \Fpga.  Twenty more \elink{s} from
the remaining groups of the \Npdm \Gbtx working at 80\mbps, routed
through the backplane, are used to propagate directly the \Ecs
information to the \Nsb \Sca{s} of up to 20 \Nsb{s}.  The \Nsb \Sca{s}
are used to configure and control the \Nsb \Fpga which in turn
distributes all the timing and control information to the \Fend
CARDIAC boards via \I2c links.

\subsection{The muon readout board specific processing}
\label{ssec:muonT40fw}

As discussed in section~\ref{sssec:pcie40}, the \Tellfourty readout
boards are provided with a generic firmware framework where
subdetector specific firmware can be plugged in. The specific muon
system firmware scheme is driven by bandwidth considerations and aims
at optimising the number of long distance optical links and
\Tellfourty boards.

The maximum theoretical bandwidth for one PCIe link is 64\gbps. The
PCIe protocol encapsulation and the DMA processing limit in fact the
bandwidth to 54\gbps but a conservative limit of 50\gbps was imposed
to each of the two output PCIe links of \Tellfourty
boards~\cite{muonbib:ReadoutProtocol}.  In the \Gbt wide bus mode
96 bits of data can be transmitted per each link at the maximum
trigger rate of 40\mhz, resulting in an input rate of 3.84\gbps per
link. Assuming that all the input data are transferred unmodified to
the output of the \Tellfourty board using its generic output data
format, this would limit the maximum number of links per board to 22.
In addition, imposing a 70\% limit on the \fpga resources of the
\Tellfourty, the maximum number of links that can be handled at
maximum input rate is limited to 32.  However, as a low occupancy is
expected, in particular in the outer regions and in the most
downstream stations of the muon detector, a specific data processing
block with a zero-suppression algorithm has been implemented in the
muon \Tellfourty, in order to minimise their number and optimise the
number of long distance optical links.

To estimate the expected output bandwidth several minimum bias events
acquired in \runtwo at $\lum = 3.7\times 10^{32}\invcma\invsec$ have
been superimposed.  The output bandwidth per event obtained using the
output data format described in section~\ref{subsubsec:nSYNC} is
reported in table~\ref{table:nonlin}. The average rate is well below
the maximum allowed of 50\gbps.

\begin{table}[h]
  \centering
  \caption{Maximum output bandwidth (\gbps) per \Pcie interface in the
    muon system stations at two different luminosity values when
    zero-suppression is applied; for comparison, also the output rate
    with no zero-suppression is reported.}
  \begin{tabular}{c c c c c}
    \hline
    Station &  \# \Tellfourty & output rate (\gbps)   & output rate (\gbps) &output rate (\gbps)\\
            &&$\lum=2\times 10^{33}\invcma\invsec$ & $\lum=4\times 10^{33}\invcma\invsec$ & no zero-suppression \\ 
    \hline
    M2 & 10 & 22 & 27 & 54\\
    M3 & 4 & 24  & 33 & 61\\ 
    M4 & 4 & 13 & 18 &35 \\
    M5 & 4 & 18 & 25 &42 \\
    \hline
  \end{tabular}
  \label{table:nonlin}
\end{table}

The number of readout boards per muon station and links per board are
listed in table~\ref{TELL40xstation}.

\begin{table}[h]
  \centering
  \caption{Number of \Tellfourty per station and of input links per
    board.}
  \begin{tabular}{l c c c c}
    \hline
    station & M2 & M3 & M4 & M5 \\
    \hline
    \# \Tellfourty & 10 & 4 & 4 & 4 \\ 
    input links/\Tellfourty & 28 & 32 & 18 & 22\\
    \hline
  \end{tabular}
  \label{TELL40xstation}
\end{table}

The muon specific data processing block consists mainly of a
zero-suppression algorithm which decodes TDC data when available and
formats the output data taking care of the different number of optical
links connected to different \Tellfourty{s}.  The zero-suppression
procedure is fully configurable via \Ecs, both at \Nsync level and in
the \Tellfourty.  This implementation allows different options to be
used for boards covering regions of the detector with different
occupancy levels if needed, although the baseline is to use the same
settings everywhere.
 
\subsection{Monitoring and Control}
\label{ssec:muonECS}

\begin{figure}[t]
  \centering
  \includegraphics[width=0.75\linewidth]{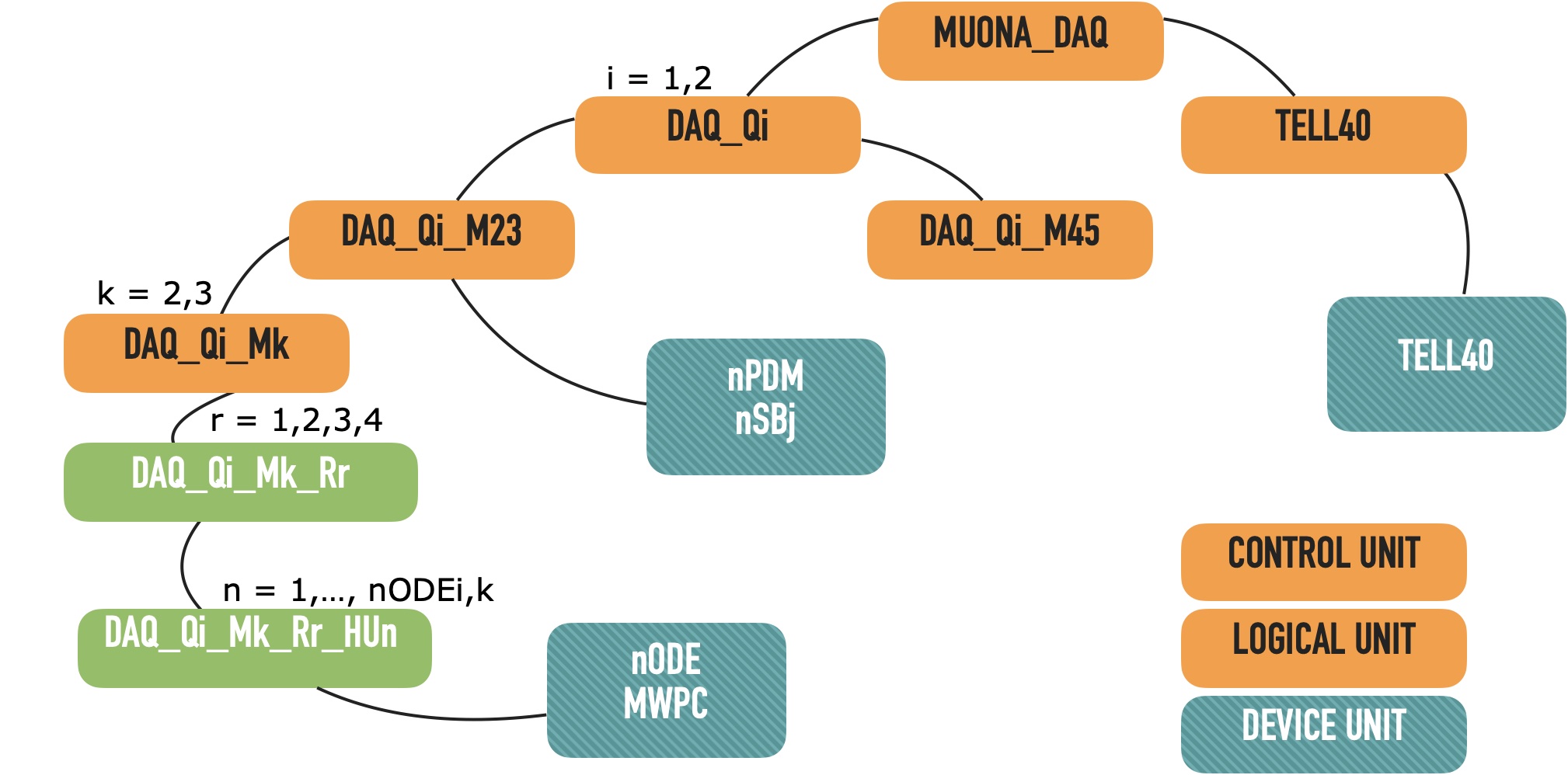}
  \caption{Finite state machine hierarchy scheme (example for the
    \aside).}
  \label{fig:nFSM}
\end{figure}

Control and configuration of the muon system is performed within the
LHCb \Ecs system framework
(section~\ref{sec:experiment-control-system}).  Muon-specific software
to monitor and control the muon system consists of a finite state
machine organised in the hierarchical structure reported in
figure~\ref{fig:nFSM}, through which all the system can be configured
and operated.

Moving down the tree from the top-level node, the system is
partitioned into two nodes, \aside and \cside, serving the two
mechanically and functionally independent halves of the muon
detector. From each node it is possible to access specific subelements
of the \Daq, high voltage, and slow-control \Ecs sector. The \Daq node
is organised in quadrants, stations and regions coherently with the
detector layout and is further subdivided in \Fend electronics, \Nsbs
and \Acr[m]{ode} domains. The slow-control \Ecs sector is further
subdivided to separately control and monitor low voltage, gas and
cooling systems.

\subsection{Mitigation of the high rate inefficiency}
\label{ssec:muonHRmit}

\begin{figure}[t]
  \centering
  \includegraphics[width=0.6\linewidth]{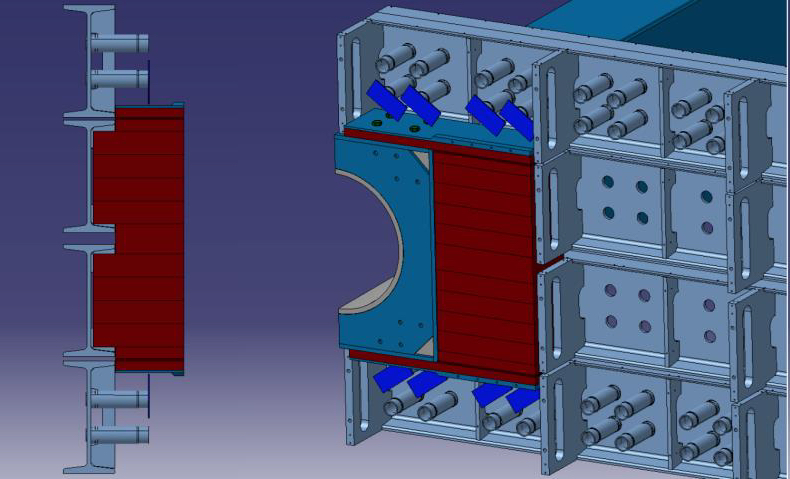}
  \caption{Mechanical drawing of the tungsten shielding around the
    beam pipe.}
  \label{fig:tungsten_plug}
\end{figure}

Consequences of the high rates expected at the upgrade running
conditions on the performance of the muon system have been extensively
studied~\cite{LHCb-TDR-014}.  From the analysis of data taken at
$\lum= 1\times 10^{33}\invcma\invsec$ during a test run in
2012~\cite{Pinci:1546731}, it has been demonstrated that no space
charge effects are expected at this luminosity. The only expected
degradation of detector efficiency due to the increased rate is caused
by the dead time induced by the \Fend electronics.  The particle flux
in the innermost region of station M2 is expected to be very high,
resulting in a very unevenly distributed efficiency drop of about
$7\%$ on average in the region closest to the beam pipe.  To reduce
the inefficiencies an additional shielding has been installed around
the beam-pipe before M2. In particular, as shown in
figure~\ref{fig:tungsten_plug}, a tungsten \emph{plug} shielding has
been inserted in place of the removed innermost cells of the \Hcal
(see section~\ref{sec:Calo_detector_radissues}).  A rate reduction of
about $30\%$ in region R1 of M2 has been estimated by simulation with
a consequent maximum rate estimated to be about 600\khz/\cm$^2$.
The high rate will induce inefficiencies in the \Fend electronics
resulting from both the CARIOCA dead-time, in the range of 70\ns to
100\ns, and the DIALOG signal formation time, which was successfully
reduced from 28\ns at the beginning of \runone to 12\ns at the end of
\runtwo.  The CARIOCA component is estimated to introduce the dominant
contribution to the inefficiency, amounting to $\sim$8\% in M2R1 and
$\sim$4\% in M3R1 and M2R2, respectively.

\begin{figure}[h]
  \centering
  \includegraphics[height=0.30\linewidth]{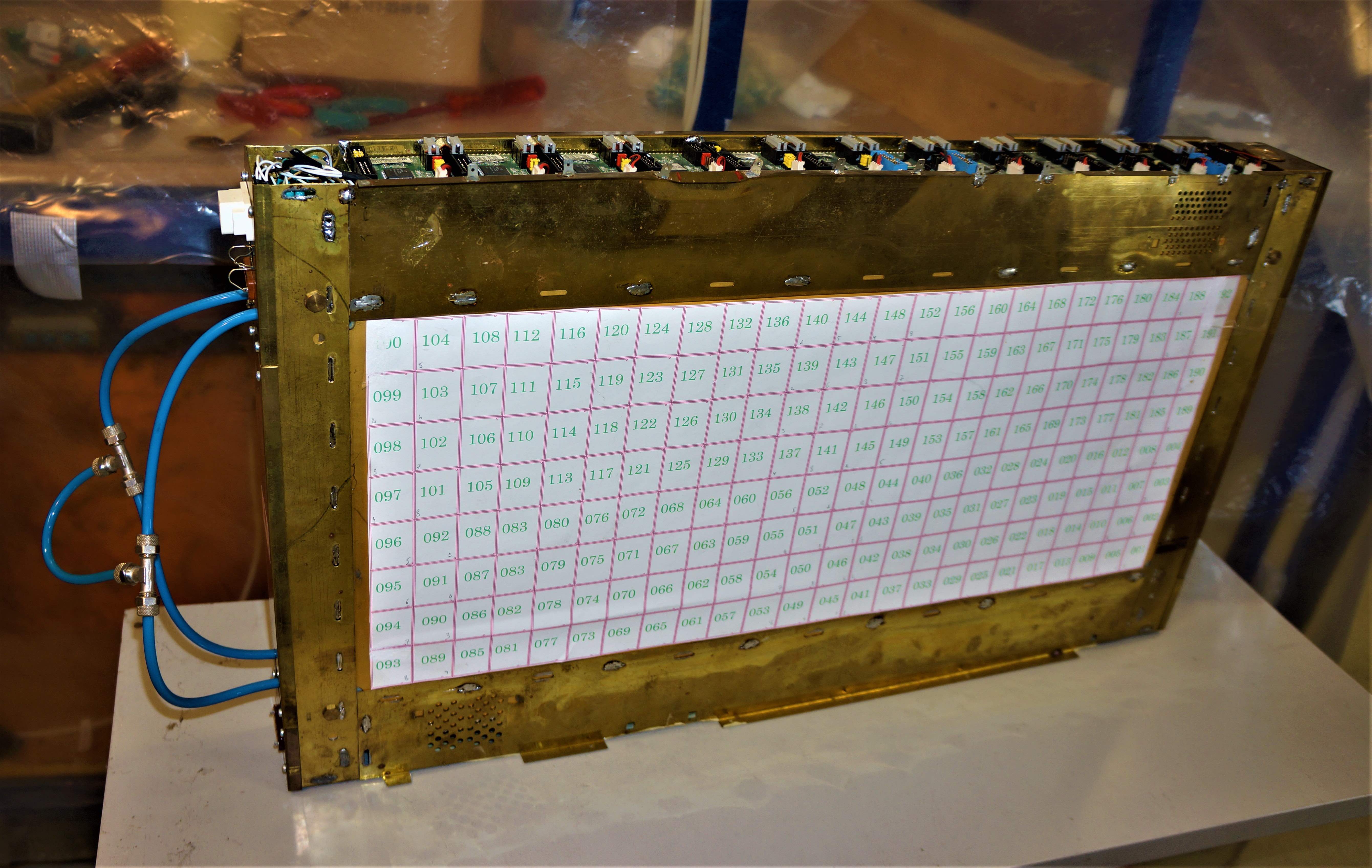} \hfill
  \includegraphics[height=0.30\linewidth]{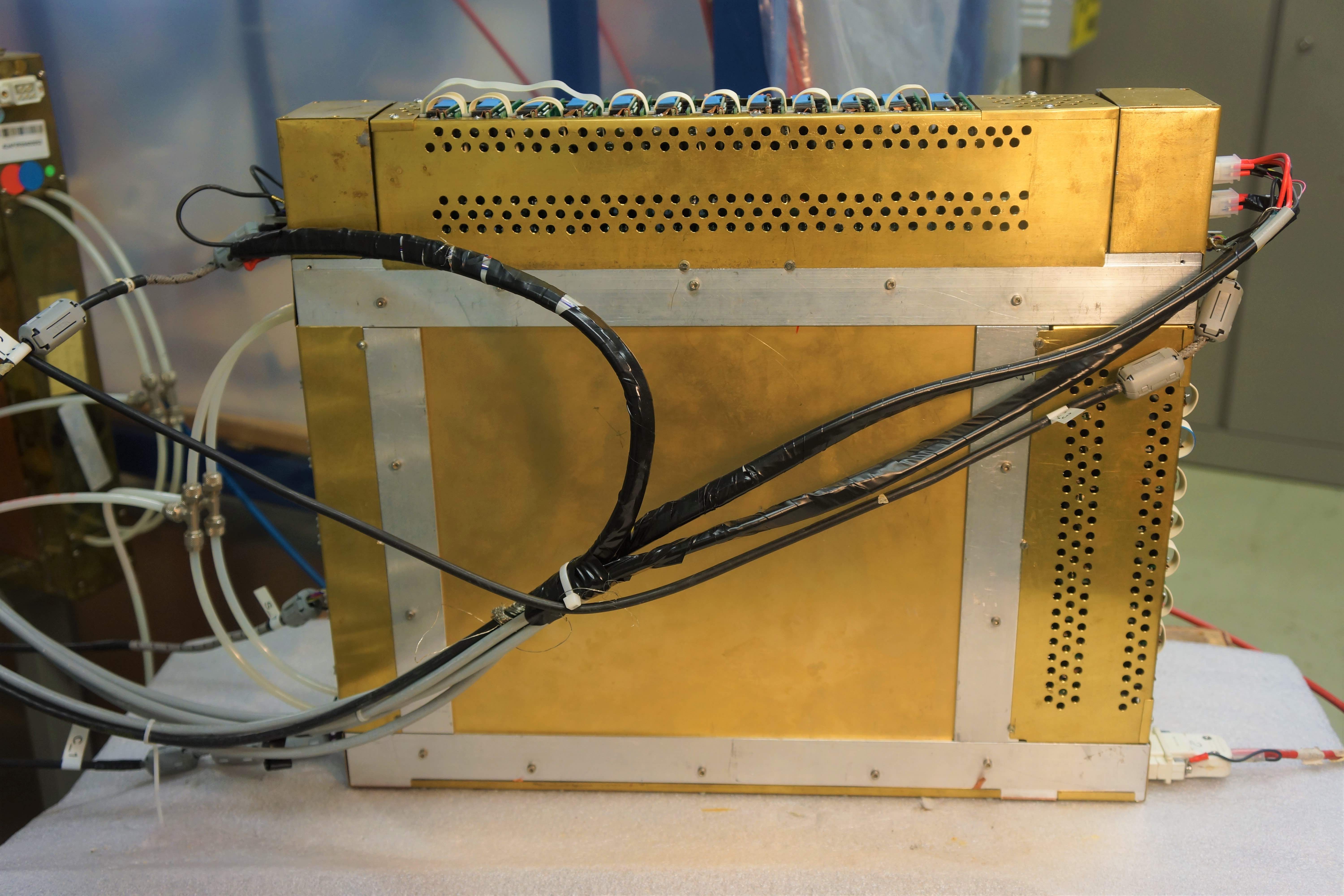}
  \caption{Left: M2R2 new pad chamber prototype. Right: M2R1 new pad
    chamber prototype.}
  \label{fig:M2R12_proto}
\end{figure}

New pad detectors with increased granularity have been
proposed~\cite{Bondar_2020} for these innermost regions to mitigate
inefficiency from dead time. A first prototype for an M2R2 chamber,
see figure~\ref{fig:M2R12_proto} (left), has been designed and
constructed in 2016. A prototype for region R1 of M2 and M3 stations,
see figure~\ref{fig:M2R12_proto} (right), was also constructed in
2020. Both prototypes successfully passed the necessary tests and were
accepted for mass production.  The new \Mwpc{s} with full pad readout
could be installed during \runthree or \Lsthree, depending on the
readiness of the chambers.  To mitigate the effect of DIALOG dead
time, the granularity of the logical channels has been increased by
replacing some Intermediate Boards~\cite{LHCb-TDR-004} with \Node
boards in regions of particularly high rates, namely regions R2, R3
and R4 of station M2 (right behind the \Hcal) and R4 of station M5
(where the rates are dominated by interactions with \Lhc materials
placed behind the LHCb detector). The replacement of Intermediate
Boards is expected to reduce the inefficiency by about $20\%$, which
will be reduced by an additional $\sim40\%$ when the three innermost
regions will be equipped with new pad chambers.

\subsection[Long-term operation of the MWPCs]{Long-term operation of the \Acr[s]{mwpc}s}
\label{ssec:muonLTO}

It is foreseen that the muon system \Mwpc{s} will be operated over the
whole LHCb life time, for a total integrated luminosity of about
50\invfb. Special care was therefore taken in limiting as much as
possible severe damage to the wire chambers and strategies were
developed to fix and recover \Mwpc{s} showing operational problems, in
particular due to sparks or high currents.

In nine years of operation in a high radiation environment, the LHCb
muon detectors did not show visible gain reduction or any other
apparent performance deterioration.  However, during this period,
about 19\% of the chambers were affected by sudden appearance of high
currents in at least one of their wire layers, with a total of about
100 wire planes per year suffering \Hv trips during \Lhc operations.
In order to prevent future severe damage and to ensure long-term
operability of the \Mwpc{s}, a detailed review of the status of the
chambers has been conducted during \Lstwo. The findings of these
systematic studies, described in details in
ref.~\cite{muonbib:Albicocco:2019jvz}, are summarised in this
section.

\begin{figure}[t]
  \centering
  \includegraphics[width=0.9\linewidth]{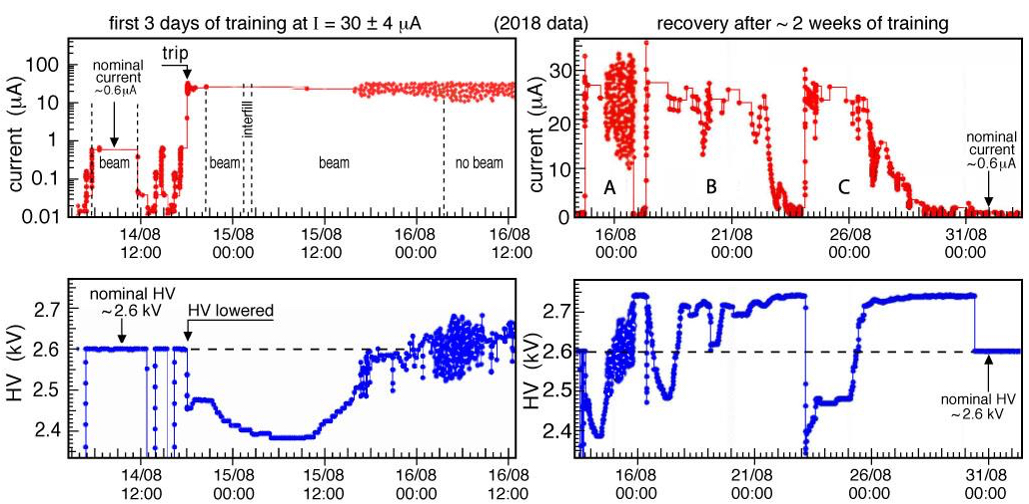}
  \caption{Typical recovery procedure for a gap (data are from a
    chamber in region M5R3). The plots on the left show (top) the
    current and (bottom) the \Hv setting during a period of about
    three days around the first appearance of the \Hv trip and the
    subsequent start of \Hv training. The plots on the right show
    (top) the current and (bottom) the \Hv setting during the full
    recovery procedure, which lasted about two weeks.  The nominal \Hv
    setting for this gap is 2600\aunit{V}. in normal conditions, the
    average current in presence of colliding beams is about
    0.6\aunit{$\upmu$A}. Reproduced from~\cite{muonbib:Albicocco:2019jvz}. \textcopyright\ 2019 CERN. CC BY 3.0.}
  \label{fig:longTermMWPC}
\end{figure}

High currents observed in some of the chambers during operation were
found to be originating from localised areas usually located on the
cathodes. This has been verified by direct inspection of broken
\Mwpc{s}, where carbonised or stained spots where observed on the
cathodes, in some cases in correspondence of broken or damaged
wires. These currents were found to be due to sustained discharges
that, in addition to locally damaging the wires and the cathodes,
generate noise and may lead to \Hv trips when exceeding the set
threshold, causing temporary detection inefficiency.  Inspection of
the behaviour of these currents suggests that they are likely due to a
Malter-like effect~\cite{PhysRev.50.48}. Even if the latter is often
associated to ageing, the analysis of the \Hv trip distribution in
time and in different regions of the detector, as well as direct
inspection of damaged chambers, indicates that ageing due to prolonged
irradiation is not the underlying cause and that most of the \Hv trips
are connected with imperfections existing since the chamber
construction.

A method for a noninvasive recovery on site has been developed and
applied individually to all problematic gaps, consisting in \Hv
training cycles carried out in presence of colliding beams and with
\Mwpc{s} working with the standard gas mixture. The training procedure
lasts typically many weeks (two months on average), before the
affected wire plane is recovered and restored back to normal
operation.  A typical example of the appearance of a self-sustained
current and of the recovery procedure during LHCb operation with beams
is shown in figure~\ref{fig:longTermMWPC}.  The four-fold redundancy
of the muon system \Mwpc{s} and the \Hv training on site allowed to
keep the whole muon detector continuously close to 100\% efficiency
for almost a decade.  Moreover the recovery procedure developed and
refined over the years has been shown to be effective, and less than
1\% of the chambers had to be replaced because of \Hv trips in nine
years of operation. The percentage of gaps treated with this method in
the past and showing repeated high current problems decreased with
time and was measured to be about 10\% during the last two years of
\Lhc operation.

\begin{figure}[t]
  \centering
  \includegraphics[width=0.6\linewidth]{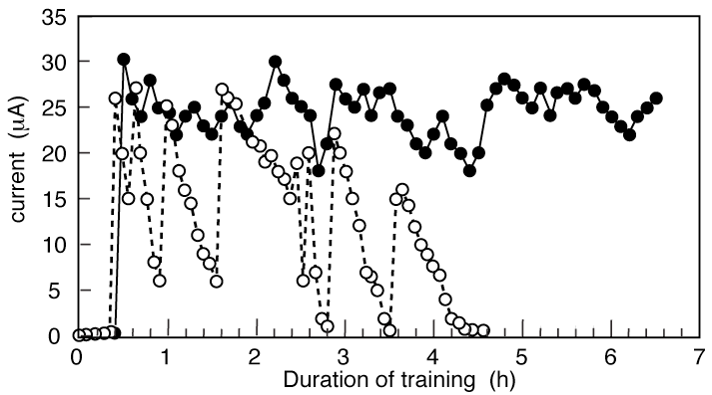}
  \caption{Current in the \Mwpc during the Malter-effect recovery
    training: default mixture (full circles) is compared with a
    mixture containing $\sim2$\% of oxygen (open circles). Reproduced from~\cite{muonbib:Albicocco:2019jvz}. \textcopyright\ 2019 CERN. CC BY 3.0.}
  \label{fig:longTermMWPC_O2}
\end{figure}

In order to make the training procedure faster and even more
efficient, a method for accelerated recovery has also been
investigated during \Lsone.  A set of four \Mwpc{s} removed from the
apparatus because of persistent high currents, underwent the standard
\Hv training procedure, but $\sim2$\% oxygen was added to the default
gas mixture.  Figure~\ref{fig:longTermMWPC_O2} shows the results of
this procedure obtained for one of the wire planes.  While no current
decrease is seen after more than six hours of \Hv training with the
standard gas mixture, consistently with observations described above,
\Hv training in presence of oxygen is much faster and the discharge
current is observed to drop down to zero in about four hours after a
few \Hv cycles.  After the training with oxygen, all of the four
\Mwpc{s} were fully recovered and installed back on the apparatus
being still operational today.  The noninvasive character of the
recovery techniques discussed in this section makes them an important
ingredient for the long-term operation of the muon system.  The
results obtained through the fast recovery suggest that in the future
a small amount of oxygen could be added to the working gas mixture
during the \Lhc winter stops, either for targeted recovery
interventions, or for conditioning while exposing the chambers to a
high intensity radioactive source.  A permanent use of a small amount
of oxygen during detector operation is also being considered.

\section{Online system}
\label{sec:online}
\subsection{System architecture}

The upgraded \online system consists of a continuation and evolution
of the successful \Acr[m]{ecs} from \runone and
\runtwo~\cite{LHCb-TDR-007}, a new \Acr[m]{tfc} for clock, synchronous
and asynchronous commands distribution for the trigger-free readout,
and a significantly increased \Acr[m]{daq}. Further hardware and
software subsystems are alignment and calibration frameworks, the
\online monitoring, storage and the infrastructure to operate all
these systems and the event-filter farm. The system hardware consists
of a single type of a powerful custom-made, flexible \Fpga board and
commercial off-the-shelf hardware.

\begin{figure}[t]
  \centering
  \includegraphics[width=0.9\textwidth]{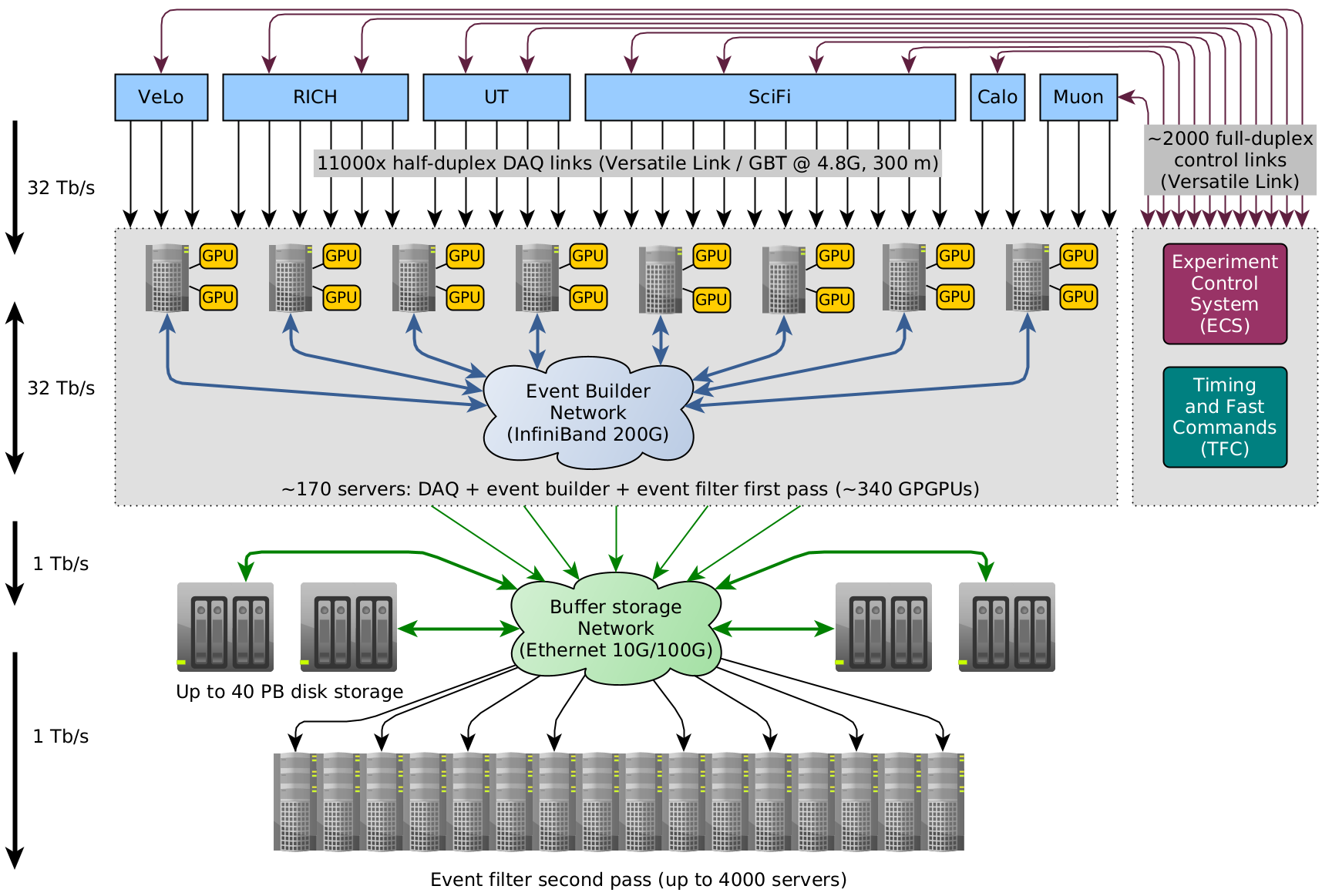}
  \caption{Upgraded LHCb \online system. All system components are
    connected to the \Ecs shown on the right, although these
    connections are not shown in the figure for clarity. Reproduced from ~\cite{10.3389/fdata.2022.1008737}. CC BY 4.0.}
  \label{fig:online-system}
\end{figure}

Most of the system naming scheme is derived from the previous LHCb
online system~\cite{LHCb-TDR-007}, often adapted to recall the 40\mhz
master readout clock.

\subsection{Data acquisition system}
\label{ssec:LHCbDAQ}

LHCb uses a synchronous readout where for \emph{every bunch-crossing}
all \Acr[m]{fend} elements participating in the data taking send data,
possibly after zero-suppression.  Therefore, from the point of view of
the \Fend electronics, the LHCb online system can be justifiably
described as ``trigger-free'', despite the fact that subsequent event
selection mechanisms implemented in software are called
``\Acr[m]{hlt}s''.  In the LHCb online scheme, the readout elements
are grouped in \emph{partitions}.  A partition is a set of \online
resources, which can be read out, monitored and controlled
independently, for example a part of a subdetector, a full subdetector
or a group of subdetectors. Multiple partitions can run
simultaneously, which is a very powerful tool for commissioning and
testing. Partitioning is implemented by both the \Tfc and the \Ecs.

The LHCb \Acr[m]{daq} is shown in figure~\ref{fig:online-system}. It
consists of a farm of \Acr[f]{eb} servers hosting the back-end
receiver boards (\Tellfourty boards) and the \Acr[p]{gpu} running the
\Hltone application. Data processed by the \Eb and the \Hltone are
then sent to the \Hlttwo for further processing and final storage.

Data are transported over half-duplex multimode optical fibres from
the detector underground level through a service shaft up to a
data-centre on the LHCb site surface. The radiation hard \Acr{vl}
protocol is used, with most subdetectors using the \Acr[f]{gbt}
protocol at the OSI-layer~2~\cite{osimodel}.

Specific to LHCb is that the links dedicated to data transmission are
used in half-duplex mode; there is only a single fibre carrying data
from the front-end to \Tellfourty boards, while control, configuration
and monitoring are out of band on dedicated connections. The links
dedicated to the control of the readout electronics, either to \Fend
or \Bend, are in full-duplex mode; \Tfc and \Ecs are transmitted to
the \Fend electronics sharing the payload on the links whereas only
\Ecs are received back, still utilising the same optical duplex links.

The \Daq links are received by the \Tellfourty boards described below,
which then push the data into the memory of the \Eb servers. After
event-building the completed events are passed to the \Hltone running
on \Gpu{s} installed in the \Eb servers. Accepted events are stored on
the \Hltone buffer storage and then read by the \Hlttwo processes for
final selection. Accepted events are consolidated into files and sent
to permanent storage.

The various components of the \Daq system are described in more detail
in the following~sections.\looseness=-1

\subsubsection[Common readout and control PCIe40 board]{Common readout and control \Acr[s]{pciefty} board}
\label{sssec:pcie40}
 
LHCb has chosen to use a single, custom-made board for data
acquisition, slow control and fast, synchronous and asynchronous
control. The board concatenates data, transforming input streams based
on a custom protocol into an output stream based on a standard
protocol used in the data centre. The input stream is the \Acr{vl}
running the \Gbt protocol on top for all detectors except the \Velo,
which uses the \Gwt protocol. The output stream is \Pcie Gen 3 with a
bandwidth of 120\gbps.  This board has a \Pcie form-factor
(three-quarter length, standard height, dual slot) and it is designed
according to \Pcie specifications.  Hinting at the board interface and
at the 40\mhz experiment's driving clock, the board has been called
\Pciefourty.

\begin{figure}[t]
  \centering
  \includegraphics[width=0.9\textwidth]{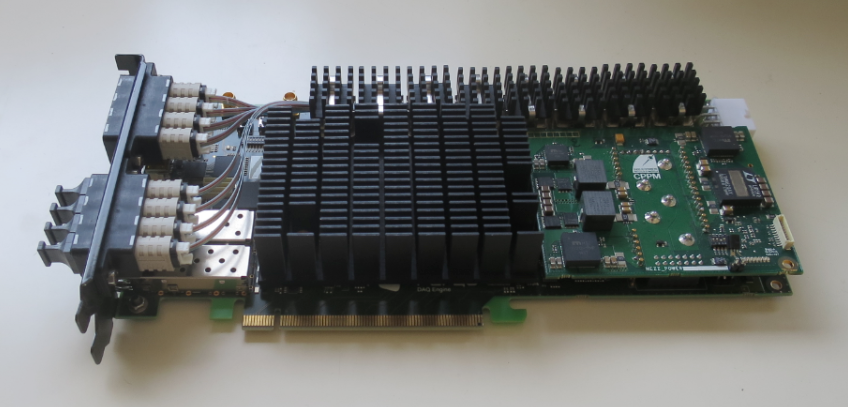}
  \caption{View of the \Pciefourty board.}
  \label{fig:online-pcie40}
\end{figure}

This board, shown in figure~\ref{fig:online-pcie40}, is based on an
\Fpga\footnote{\Trmk{Intel} Arria10.} featuring 72
serialiser/deserialiser (SerDes) out of which 16 are used for the
host-interface to the PC. The front-end electronics can be connected
via up to 48 bidirectional optical links.  The bandwidth of each link
is up to 10\gbps in both directions. The \Fend facing links are
directly connected to optical transceiver modules (MiniPods) to
achieve high density.\footnote{\Trmk{Broadcom} MiniPod.} Physically,
these links appear as eight MPO-12 connectors on the board. In
addition, two serial links are reserved for the timing and fast
control with a pluggable optical module. The cooling of the \Fpga and
MiniPods is obtained through a custom heat sink and relies on the air
flow of the PC server. Components of the board require six different
voltages which are provided by a daughter card connected to the main
12\volt of the PC server. The card supplies up to 45\amp on 0.9\volt
to the core of the \Fpga, in a sustained regime.  Finally, a series of
phase-locked loops are implemented to distribute clocks at 40 and
240\mhz with low jitter. Their phases are controlled at the level of
200\ps with respect to the \Lhc main clock.

When used for data acquisition, the \Pciefourty interfaces the \Fend
electronics with the \Eb. In that configuration, only the receiver
direction is used and no transmitter MiniPods are fitted in order to
save cost. In this configuration, the board is called
\Tellfourty. When used for slow and synchronous controls, the board is
called \Solfourty and all bidirectional links are active. Typically, 3
(8) \Tellfourty (\Solfourty) are installed per PC server respectively.

Different functionalities are obtained by charging different firmware
versions in the \Fpga, with at least one per subdetector (see
section~\ref{ssec:firmware}).  A special firmware version is also
implemented for the master-module of the synchronous readout, the
\sodin board, which acts as readout supervisor.

\subsubsection{Firmware}
\label{ssec:firmware}
By reconfiguring the onboard \Fpga with dedicated firmware, the
\Pciefourty can be used to serve very different roles within the
upgraded LHCb experiment: fast control, clock and slow control
distribution and data acquisition. The firmware of the board contains
an interface layer code, common to all the boards, whose aim is to
interface the hardware with the user firmware using common blocks, for
example \Gbt decoder/encoder and \Pcie IP cores. The actual user
firmware defines the flavour of the board and its functionality within
the upgraded readout architecture. This considerably reduces the
number of components to be developed and optimises personpower as well
as effort on firmware design and maintenance. The environment to
develop the firmware for each configuration of the boards is common
across the entire LHCb experiment, with only the user code being
exclusive.  The different board flavours are:
\begin{itemize}
\item \sodin (readout supervisor),
\item \Solfourty (main interface),
\item \Tellfourty (data acquisition).
\end{itemize}

\begin{figure}[h]
  \centering
  \includegraphics[width=14cm]{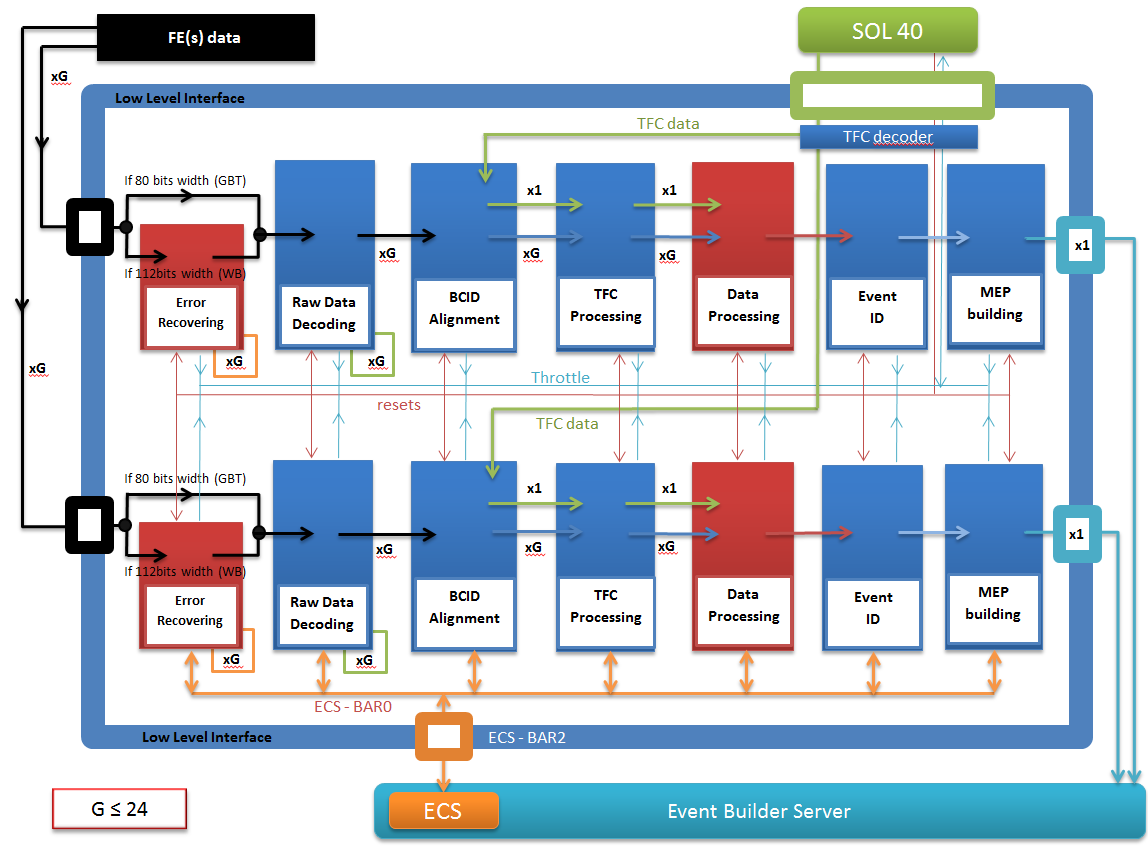}
  \caption{\Tellfourty firmware architecture. Common blocks in blue;
    specific subdetector blocks in red.}
  \label{tell40arch}
\end{figure}

The \Solfourty board serves three main purposes:
\begin{itemize}
\item interface all the readout boards to the \sodin by fanning-out
  the synchronous timing and trigger information and fanning-in
  throttle information~\cite{tfcupgrade};
\item interface all the \Fend electronics to the \sodin by relaying
  the clock, timing and command information onto fibres towards the
  \Fend electronics~\cite{LHCb-PUB-2012-017};
\item relay the \Ecs information by interfacing the slow control
  protocols at the \Fend electronics~\mbox{\cite{tfcscacore,tfcscacore2}.}\looseness=-1
\end{itemize}
The user code dedicated to the functionality of the readout of events
from the trigger-free architecture faces considerable
challenges. Events arrive from the \Fend to the \Tellfourty boards
asynchronously across all input links due to the variable latency in
compression/zero-suppression mechanisms. Thus the code must be able to
handle a large spread in time between fragments of the same event. The
readout code of the board must be able to decode the data frames from
the \Fend, realign them according to their \Bxid, build an event
packet and send it to the \Daq network.  Figure~\ref{tell40arch}
illustrates the architecture of the \Tellfourty firmware. Low level
interfaces and features which can be common to all the subdetectors,
like decoding or data time alignment, are developed centrally for the
collaboration (coloured in blue in figure~\ref{tell40arch}). Each
subdetector's user will only develop a few specific blocks which will
be fitted within the common architecture (coloured in red in
figure~\ref{tell40arch}). This optimises personpower and reduces the
complexity of the development allowing for faster integration,
commissioning and maintenance.

The LHCb \online group provides engineering support and development to
the different LHCb subdetector groups, mainly in relation to readout
board firmware, low-level software for front-end configuration and
data acquisition, and control system components. A proper
firmware/software framework based around continuous integration has
been put in place to mitigate the very heterogeneous nature of the
different subsystems.  This approach reduces the time and effort
required to detect, reproduce, and correct integration issues during
the development cycle of the firmware.\looseness=-1

\subsubsection{Firmware and software frameworks and continuous integration}

In addition to the \online developer team, other members of the LHCb
collaboration from several institutes participate actively in the
firmware and software development. The benefits of automated
integration testing are evident in such a geographically distributed
structure, since issues like failing builds or failing tests can be
flagged and reported automatically to the participants regardless of
their location or timezone. Source control is organised using \git and
\git repositories managed through the \gitlab infrastructure provided
by the Information Technology department at CERN.

The readout \Fpga firmware is organised into logically separate \git
submodules, allowing maintainers of the corresponding pieces of
functionality to version their code independently. A dedicated
top-level repository tracks the state of the firmware submodules, and
provides a unified location from which different \Fpga firmware codes
can be built. When developers are ready to integrate their changes,
they submit a merge request which automatically triggers a dedicated
\gitlab continuous integration pipeline. This pipeline executes Questa
RTL simulations according to several predefined firmware
configurations and, if successful, executes the Quartus \Fpga
synthesis.

Given the complexity of modern high-capacity \Fpga{s}, producing all
required permutations requires of the order of 100 hours of
computation. A series of optimisation steps are thus
applied. Distributing synthesis jobs across a small computing cluster
reduces this turnaround to a single day.  As a further optimisation,
the process keeps the result of each synthesis job in an internal
cache and tracks changes in all firmware components across successive
invocations of the same job configurations. This automated dependency
tracking is used to selectively trigger only pipeline jobs whose
dependencies have changed. For example, changes to a specific
subdetector implementation, or to board-specific logic, will result in
repeating only the jobs associated to that specific subdetector, or
that specific board, respectively.

Both software and firmware are automatically packaged in RPM format
and published to dedicated RPM repositories for distribution, as is
customary on all Linux installations deployed at~CERN.

\subsubsection{Event-building}

To perform the event selection, the full-software LHCb trigger
requires the complete event information from all the
subdetectors. Consequently, event-building, the assembly of all pieces
of data belonging to the same bunch-crossing, is done for every
collision of non-empty bunches at a 40\mhz rate.\footnote{The
  information for each bunch filling is transmitted synchronously to
  the \Tellfourty{s} by the \Tfc system; data recorded during empty
  crossings are normally dropped.}

The \Eb system consists of a farm of servers hosting the \Tellfourty
boards receiving data from the subdetectors and the \Gpu{s} running
the \Hltone application. As each server receives only the data from
the subdetectors connected to the corresponding \Tellfourty{s}, all
the \Eb nodes must be interconnected with a high performance network
able to transmit the full information to the node which is in charge
for the full event assembly.  The architecture of this high-speed
network is the main driver for the overall system design, which has
been optimised such that the costs for handling the expected input
event rate are minimised.

For any high-speed network the number and reach of the links are the
main cost-drivers. Therefore, the network links are used
bidirectionally so that their number is kept comparatively small. In
addition, the system has been designed to be very compact and
physically installed in only two modules of LHCb's data centre
(described in section~\ref{sec:datacentre}), keeping the physical
length of the network links at a minimum.

The 200\gbps \Acr{hdr} InfiniBand technology has been chosen for the
\Eb network implementation, which not only offers comfortable data
throughput but allows to fully exploit modern \Pcie interfaces.

The \Pciefourty cards needed to read out all the LHCb subdetectors are
hosted in 162 servers with up to three cards per server and avoiding
server sharing among different subdetectors. Each server is connected
through two \Acr{hdr} ports to the \Eb network. Each server acts in
turn as data-source and data-sink in the event-building process, where
cyclically every node acts as full event builder (sink) and receives
data from all other servers (sources). In order to achieve optimal
network performance, data from several thousand bunch-crossings are
packed together and these multievent fragment packets are treated as
unit data blocks in the event building.

When a builder unit has received data from all sources, the sources
are reordered to facilitate subsequent processing. Completed events
are stored in a shared memory buffer and are handed over to the
\Hltone selection process. The \Hltone application pushes the required
event data to a \Gpu installed in each event-builder server. The
events selected by \Hltone are then sent via a separate 10G/100G
Ethernet network to temporary storage, from where they are accessed by
the alignment and calibration processes and by the second-stage
selection (\Hlttwo).

\subsection{Timing and fast control}
\label{sec:tfc}
 
The \Tfc is responsible for controlling and distributing clock, timing
and trigger information, synchronous and asynchronous commands to the
entire readout system as described in~\cite{LHCb-PUB-2011-011}. The
system must maintain synchronisation across the readout architecture,
provide the mechanisms for special monitoring triggers and manage the
dispatching of the events to the trigger farm. It regulates the
transmission of events through the entire readout chain taking into
account throttles from the readout boards, the \Lhc filling scheme,
calibration procedures and physics decisions if any, while ensuring a
coherent data taking acquisition across all elements in the readout
architecture. The specifications, functionalities and the full details
of the system are described in ref.~\cite{tfcupgrade}.

Generally, the signals generated and propagated by the \Tfc system to
the entire readout system~are:
\begin{itemize}
\item the \Lhc reference clock at 40\mhz, that is the master clock of
  all the electronics synchronised to the master clock of the \Lhc
  accelerator;
\item commands to synchronously control the processing of events in
  the readout board~\cite{tfctimingupgrade} or front-end electronics;
\item calibration and specific subdetector commands for the detector
  electronics.
\end{itemize}
In addition, \Fend electronics configuration is generated by the \Ecs
and relayed by the \Tfc system to the \Fend boards. At the hardware
level the \Tfc system is implemented using the common \Pciefourty
card. Details of the \Tfc aspects in this card are discussed in
ref.~\cite{Alessio:2018xzo}.

\subsubsection[Functionalities of the TFC system]{Functionalities of the \Acr[s]{tfc} system}

The main functionalities of the \Tfc system are:
\begin{itemize}
\item \emph{Readout control}: control of the entire readout system is
  made by one of the \Tfc Masters in the pool. The control of the
  readout implies controlling the trigger rate, balancing the load of
  events at the processing farm and balancing the occupancy of buffers
  in the electronics. The \Tfc system auto-generates internal triggers
  for calibration and monitoring purposes in a programmable way, as
  well as a full set of commands in order to keep the system
  synchronous. The details of the specifications for the \Fend and
  \Bend are described in detail in ref.~\cite{LHCb-PUB-2012-017}.
\item \emph{Event description}: a data bank, containing information
  about the identity of an event as well as the trigger source, is
  transmitted by the central \Tfc Master to the farm for each event as
  part of the event data.
\item \emph{Partitioning}: this is achieved by instantiating a set of
  independent \Tfc Masters in SODIN, each of which may be invoked for
  local subdetector activities or used to run the whole of LHCb in a
  global data taking. An internal programmable switch fabric allows
  routing of the information to the desired destination.
\item \emph{Coarse and fine time alignment}: the \Tfc distribution
  network~\cite{tfctiming} transmits a clock to the readout
  electronics with a known phase, kept stable at the level of about
  50\ps, and a very low jitter ($< 5$\ps). The latency of the
  distributed information is fully controlled and kept constant. Local
  alignment at the front-end of the individual \Tfc links is required
  to ensure synchronisation of the experiment. This alignment relies
  on the synchronous reset commands together with Bunch Identifier and
  Event Identifier checks.
\item \emph{Run statistics}: information about the trigger rates, run
  dead-time, number of events accepted, type of events accepted, bunch
  currents, luminosity and load of buffers is stored in a database to
  allow retrieving run statistics and information per run or per \Lhc
  fill.
\end{itemize}

\subsubsection[TFC architecture, timing and control distribution]{\Acr[s]{tfc} architecture, timing and control distribution}

\begin{figure}[b]
  \centering
  \includegraphics[width=10cm]{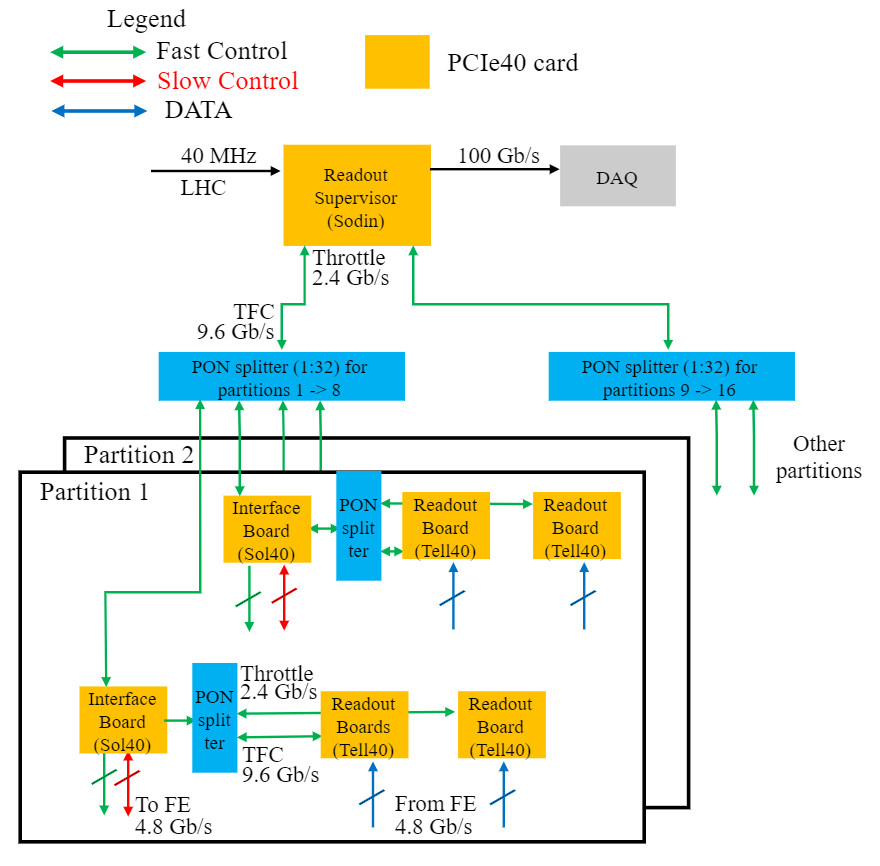}
  \caption{Logical architecture of the upgrade \Tfc system.}
  \label{TFCarchitecture}
\end{figure}

The upgraded \Tfc architecture and data flow are represented in
figure~\ref{TFCarchitecture}.
The readout supervisor \sodin is the \Tfc Master, responsible for
generating the necessary information and commands to be interfaced to
the \Lhc clock distribution system. The subdetector readout
electronics comprised of \Fend and \Bend boards are connected to the
\sodin via a network of bi-directional optical links via multiple
\Solfourty interface boards and passive optical splitters. These
connections define the partition granularity and their topology
defines a partition, controlled by the \Tfc to run any ensemble of
subdetectors simultaneously.

These connections utilise different technology, protocol and bandwidth
according to their destination and purpose:
\begin{itemize}
\item the connections between the \Solfourty and the \Fend boards use
  the \Gbt protocol with Forward-Error correction enabled, at 4.8\gbps
  (including the bits needed for error correction), for fast and slow
  control distribution to the \Fend and for slow control back from the
  \Fend;
\item the connections between the \sodin and the \Solfourty as well as
  between the \Solfourty and the \Tellfourty{s} utilise Passive
  Optical Network technology (PON) using PON splitters to reach
  multiple destinations, running at 9.6\gbps in the downstream
  direction for clock and command distribution and at 2.4\gbps in the
  upstream direction for throttle information;
\item the connection to the \Lhc clock is via an electrical interface
  (LVDS) from the LHCb clock reception and distribution
  system~\cite{tfctimingupgrade}. The \sodin receives the 40\mhz \Lhc
  clock as well as the 11.245\khz revolution clock (commonly referred
  to as the orbit pulse) and uses this to synchronise the event
  information to the \Lhc collisions;
\item lastly, \sodin is interfaced to the rest of the \Daq by its
  \Pcie interface, transmitting roughly 100\gbps of event bank
  information.
\end{itemize}

\begin{figure}[t]
  \centering
  \includegraphics[width=\textwidth]{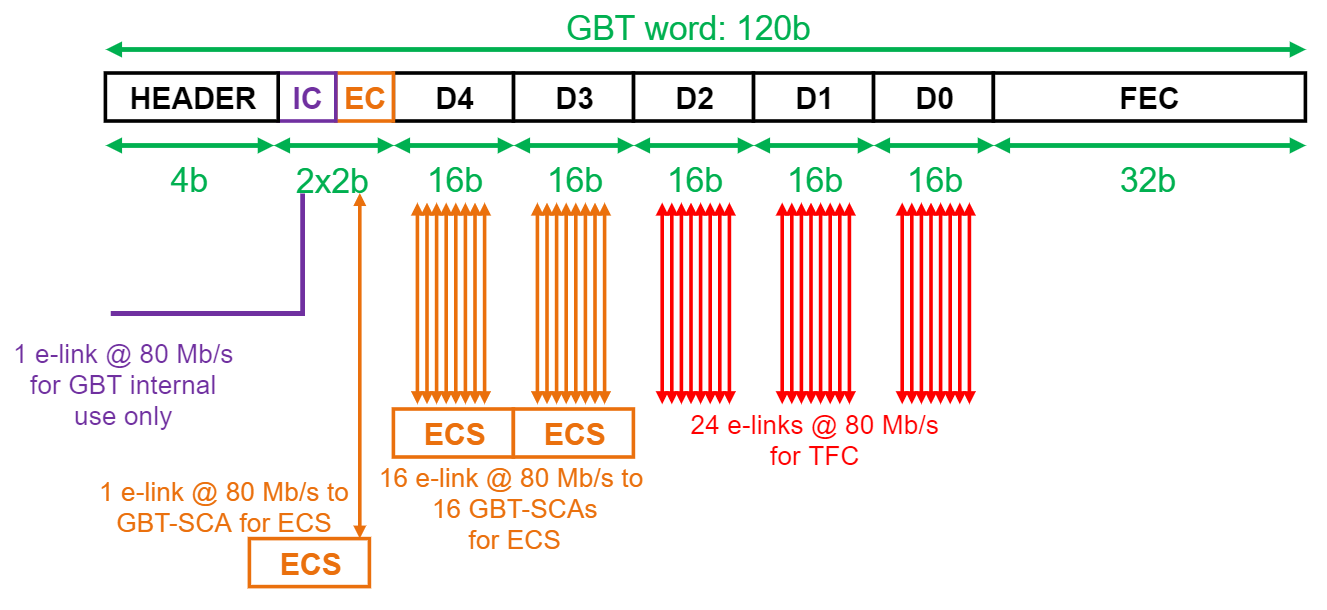}
  \caption{Schematic view of the packing mechanism to merge \Tfc and
    \Ecs information on the same \Gbt links towards the \Fend
    electronics. \Gbt words are subdivided into small e-links.}
  \label{TFCECStoFE}
\end{figure}

LHCb employs a unique mechanism to merge the \Tfc and \Ecs information
in the same duplex link to the front-end, via their interface to the
\Solfourty boards. The \Tfc information is packed into the \Gbt
verbatim at 40\mhz, while the \Ecs information can span multiple words
and so make use of the available bandwidth as needed.  At the
front-end each subdetector developed its own architecture in order to
decode this information and use it locally.  The logical scheme of the
merging is shown in figure~\ref{TFCECStoFE}, as a generic example. In
fact, the \Solfourty boards may be cascaded and configured differently
to support different requirements in terms of number of links and
bandwidth as well as supporting different architectures at the \Fend
boards. This is entirely done via a mixture of configuration registers
in the firmware as well as configuration parameters at compile time,
while keeping the firmware development common to all subdetectors.

\subsection{Experiment control system}

\label{sec:experiment-control-system}
\begin{figure}[b]
  \centering
  \includegraphics[width=\textwidth]{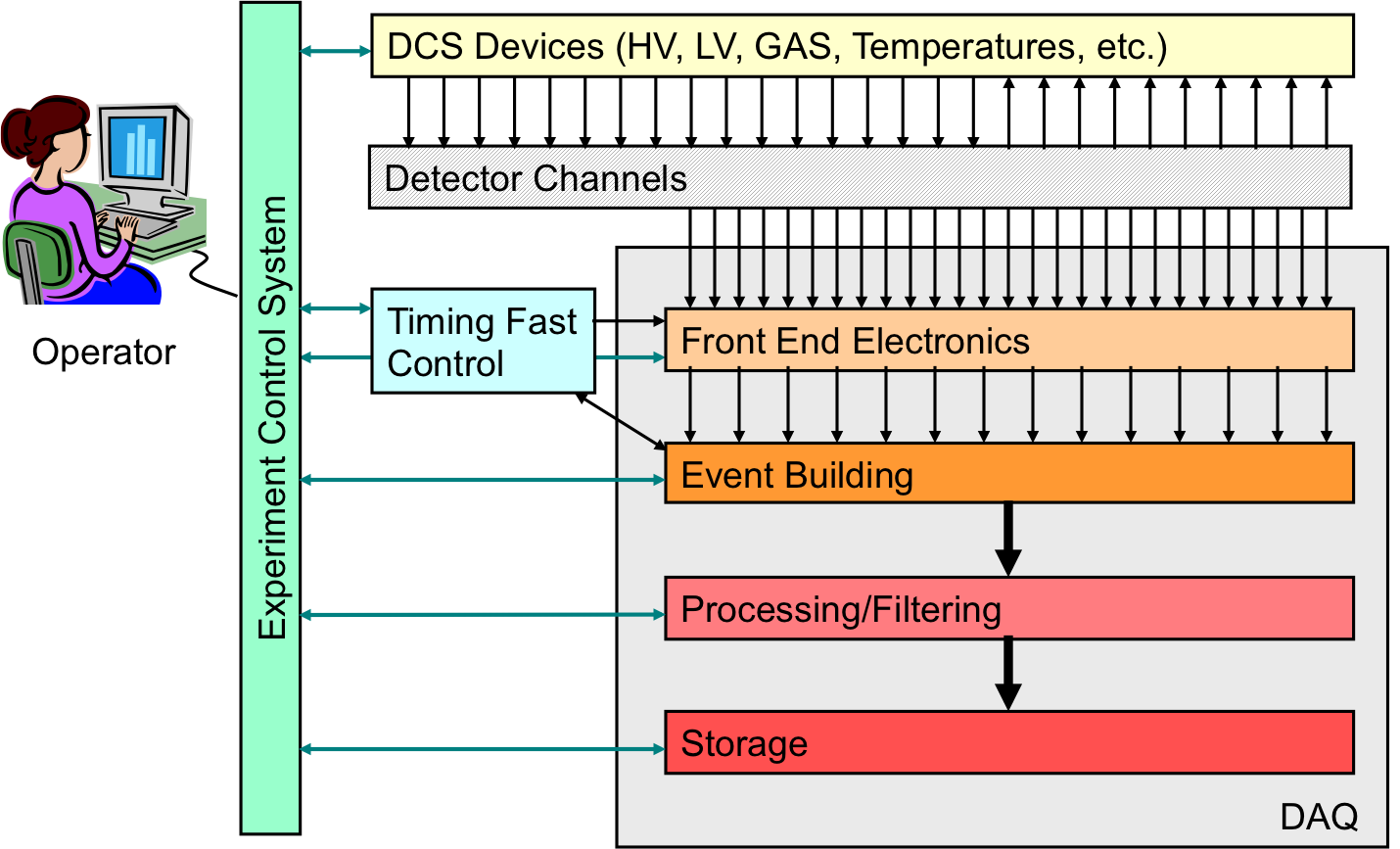}
  \caption{Scope of the \Ecs.}
  \label{fig:ecs-overview}
\end{figure}

The \Ecs is in charge of the configuration, monitoring and control of
all areas of the experiment; this comprises classical slow controls of
high and low voltages, fluid-systems, various sensors as well as
monitoring and control of the \Daq and \Hlt systems. It provides an
homogeneous and coherent interface between the operators and all
experimental equipment, as shown in figure~\ref{fig:ecs-overview}.

The \Ecs for the upgraded detector is an evolution of the current
system, described in~\cite{LHCb-TDR-007}.  It is still developed in
the context of the \Jcop~\cite{myersjcop}, a common development
between the four \Lhc experiments and CERN. The project defined a
common architecture and a framework to be used by the experiments in
order to build their detector control systems.

\subsubsection{Architecture}

\Jcop adopts a hierarchical, highly distributed, tree-like structure
to represent the structure of subdetectors, subsystems and hardware
components. This hierarchy allows a high degree of independence
between components, for concurrent use during integration, test or
calibration phases. It also allows for integrated control, both
automated and user-driven, during physics data-taking.  LHCb adopted
this architecture and extended it to cover all areas of the
experiment.

\begin{figure}[h]
  \centering
  \includegraphics[width=\textwidth]{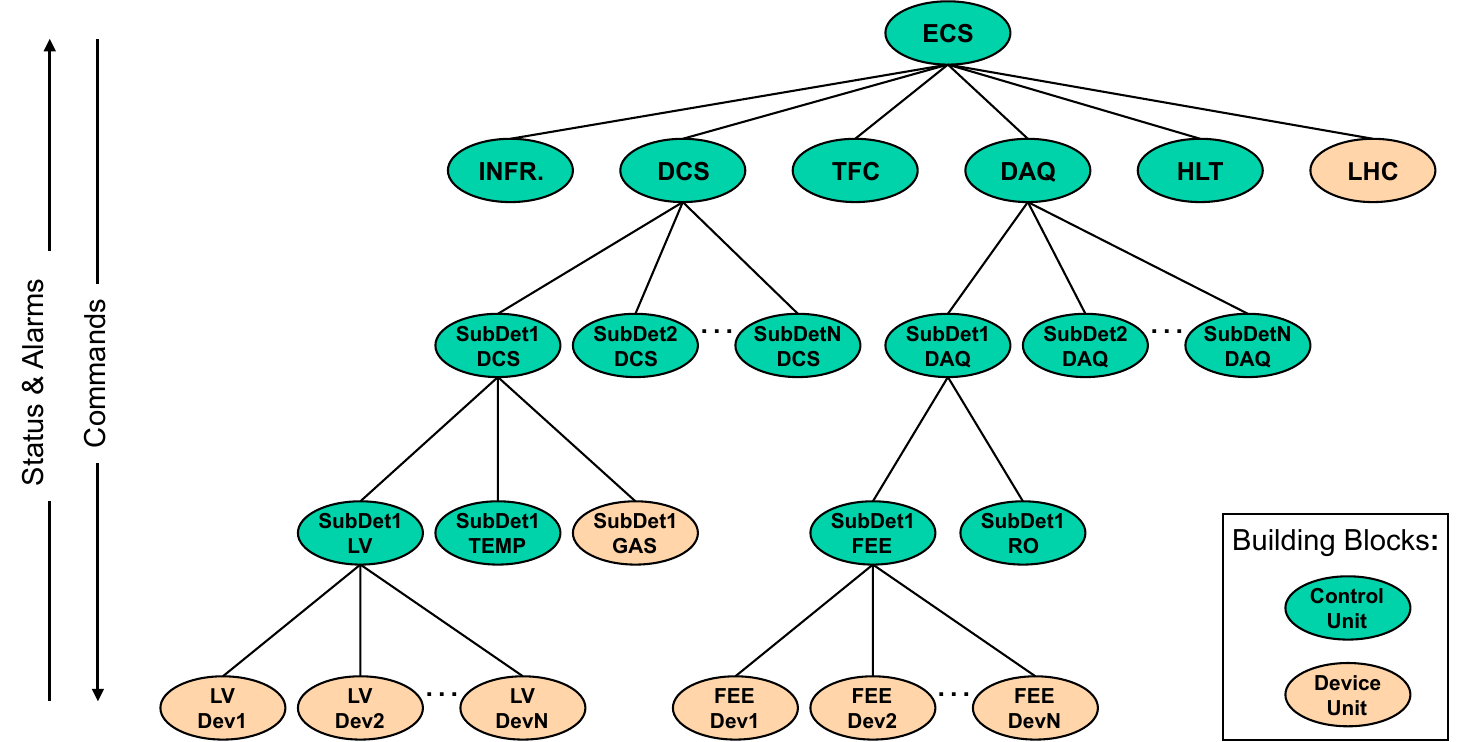}
  \caption{Simplified \Ecs architecture. Reproduced from~\cite{claragrenoble}. CC BY 4.0.}
  \label{fig:ecs-arch}
\end{figure}

Figure~\ref{fig:ecs-arch} shows a simplified version of LHCb's control
system architecture.  The building blocks of this tree can be of two
types: Device Units (DU), the tree leaves, which are capable of
driving the equipment to which they correspond, and Control Units (CU)
which correspond to logical subsystems and can monitor and control the
subtree below them.

\subsubsection{Framework}

The \Jcop framework provides tools for the integration of the various
components in a coherent and uniform manner.  It is built upon a
\Scada system.\footnote{Formerly called PVSS~II, now WinCC-OA,
  \url{http://www.etm.at}.}

While the \Scada system offers most of the needed features to
implement a large control system, the CUs described above are abstract
objects which are better implemented using a modelling tool. For this
purpose \smipp~\cite{smixx} was integrated into the framework.  \smipp
is a toolkit for designing and implementing distributed control
systems. Its methodology combines three concepts: object orientation,
finite-state machines and rule-based reasoning.  The \Jcop framework
is also complemented with LHCb specific components, providing for the
control and monitoring of LHCb equipment or components such as \Daq
electronics boards, power supplies or \Hlt algorithms.

\subsubsection[DAQ and electronics control]{\Acr[s]{daq} and electronics control}

The upgraded electronics are integrated into the control system
following the philosophy described above. Standard LHCb components
have been developed which allow users to configure, monitor and
interact with their electronics. The upgrade electronics
specifications document~\cite{LHCb-PUB-2011-011} contains requirements
and guidelines for electronics developers, so that common software can
be implemented.

\begin{figure}[h]
  \centering
  \includegraphics[width=\textwidth]{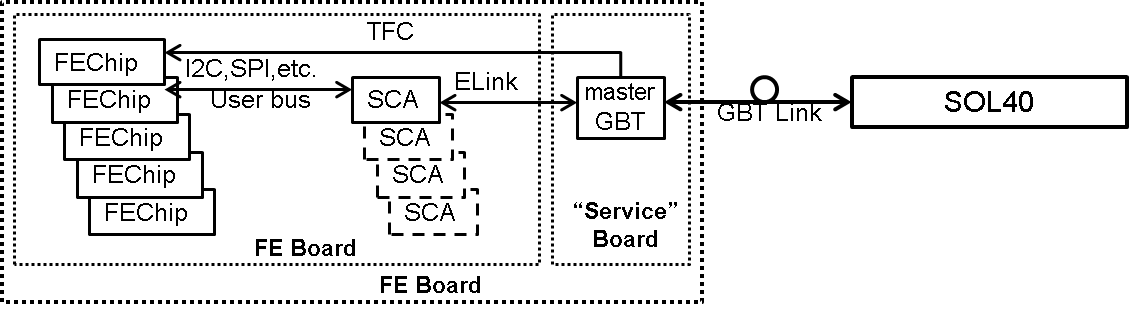}
  \caption{{\Fend}-\Ecs interface. Reproduced from~\cite{LHCb-PUB-2011-011}. CC BY 4.0.}
  \label{fig:fe-ecs}
\end{figure}

As described in section~\ref{sec:tfc}, the \Ecs interface to the \Fend
electronics is implemented via \Solfourty interface boards, using the
\Gbt system. This bi-directional link allows writing and reading of
configuration and monitoring data. The \Gbtsca chip provides an
interface between the \Gbt and standard protocols such as \I2c,
\Acr{spi} or \Acr{jtag} and can be mounted on the \Fend modules, as
shown in figure~\ref{fig:fe-ecs}.

A generic server process running inside the PC hosting the \Solfourty
cards provides the interface between the \Fend electronics and the
\Scada system.  Similarly to the \Fend electronics, the software for
the configuration and monitoring of \Bend boards is maintained
centrally in the form of \Jcop components providing for the high-level
description and access to all electronics components.

\subsubsection{Guidelines and templates}

Configurable framework components are distributed to the subdetector
and subsystem teams in order to build their specific control
systems. In order to ensure the coherence and homogeneity of the
system, detailed guidelines specifying naming and colour conventions
have been prepared and distributed.  Whenever possible, the code
necessary to implement the guidelines and conventions or the code to
implement the finite-state machine behaviour specified for the
different LHCb domains is also centrally provided in the form of
templates.

\subsubsection{Operations and automation}

\begin{figure}[t]
  \centering
  \includegraphics[width=0.9\textwidth]{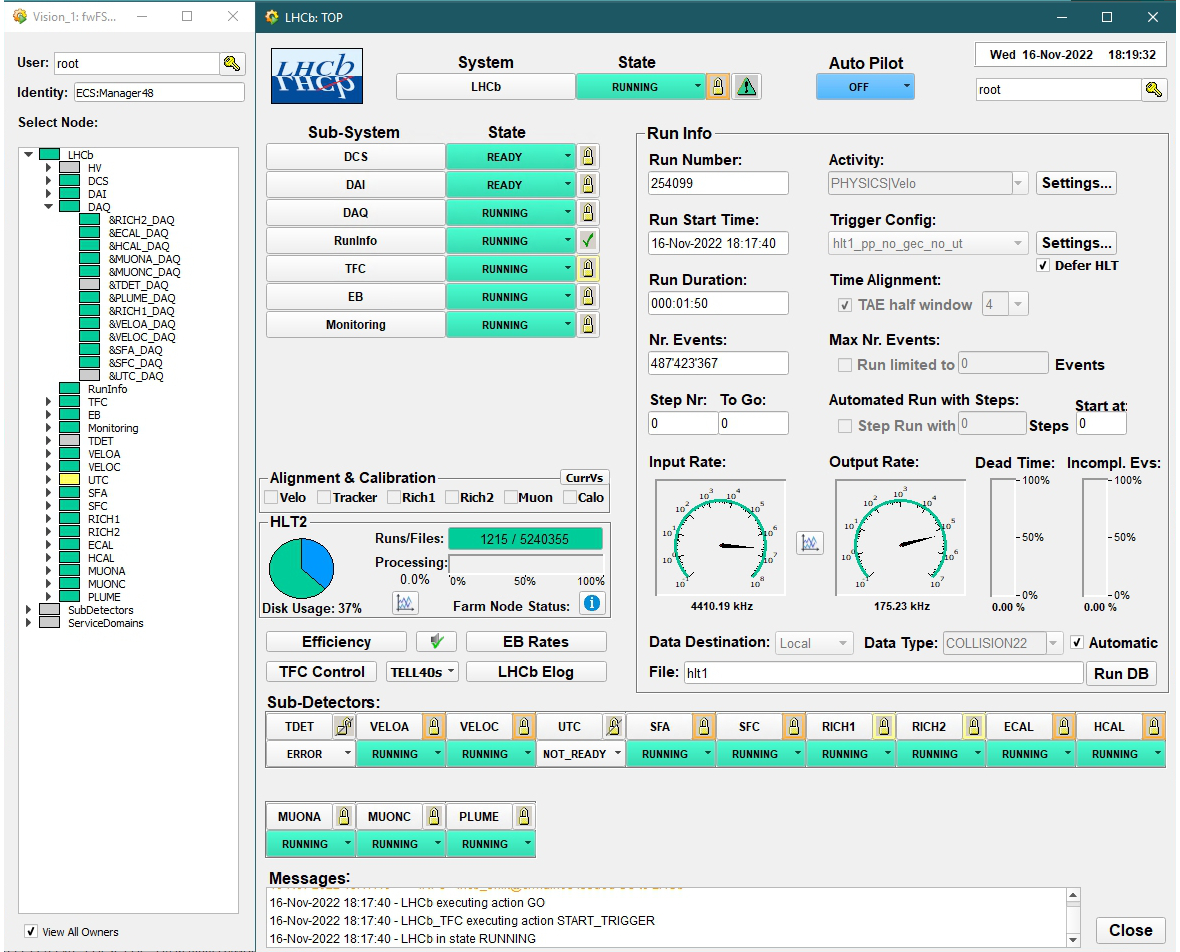}
  \caption{LHCb Run Control panel.}
  \label{fig:ecs-run-control}
\end{figure}

Like in the previous system, all standard procedures and, whenever
possible, error recovery procedures are automated using the \Jcop
framework finite-state machine tools~\cite{claragrenoble}.  The
experiment's operation, in terms of user interfaces, is again based on
the \Jcop framework and \Scada system, providing a global Run Control,
control panels for detector systems, and alarm screens.  As an
example, the Run Control panel is shown in
figure~\ref{fig:ecs-run-control}.

\subsection{Monitoring, alignment and calibration}
\label{sec:onlinecalib}

The software trigger applications move event data through the
selection stages of \Hltone and \Hlttwo. In between and at the end of
these stages, events are stored in large global buffers. This
mechanism makes data available also for monitoring and intermediate
calibrations performed by dedicated applications and processing
mechanisms, which are described in this subsection.

\subsubsection{Monitoring}

Automatic and interactive monitoring are essential tools to ensure the
quality of the recorded data. Broadly speaking, monitoring works with
two categories of input data.  The first category is represented by
histograms and counters produced by selection algorithms and other
processes throughout the system. These are collected by so-called
adder-processes, which accumulate quantities and histograms specific
for partitions (see section~\ref{ssec:LHCbDAQ}), and are identified by
run and fill number.  Interactive analysis of these quantities is done
using the \monet~\cite{Monet} web-based application. Data are acquired
and published via LHCb's standard publish-subscribe framework, the
distributed information management (DIM) system
protocol~\cite{GASPAR2001102}. Automated algorithms can access data
with the same mechanism to generate alarms in the experiment control
system.  The second category consists of fully assembled events
obtained after event building which are designated for monitoring by
random selection or by the trigger processes. These events are tagged
for monitoring when they are made available in the shared memory of
the individual processing nodes. Monitoring-tagged events are
duplicated and picked up by the standard data processing, but also
sent to the monitoring farm, via a separate network. In the monitoring
farm they are first received by a distribution software layer
implemented using the same shared memory architecture as the rest of
the system. This layer implements various access patterns to
monitoring data:
\begin{itemize}
\item every event is guaranteed to be delivered to a single monitoring
  application (\emph{consumer});
\item every event is guaranteed to be delivered to one of a group of
  similar consumers for load balancing;
\item events are offered to all interested consumers.
\end{itemize}
Guaranteed delivery is ensured by back-pressure. This mechanism can
ultimately block the data flow, but an excess of input data is
normally detected by the \Ecs before a complete stall. The
distribution layer duplicates event data as needed. Consumers need to
decode the raw-data, but otherwise have no limitations other than
available computing resources. They are implemented using the same
software frameworks as used by the trigger algorithms. They produce
counters and histograms as output, which are monitored by the
shift-crew and automated analysis systems.

\subsubsection{Calibration and alignment}

Calculating calibration and alignment constants is crucial for the
optimal selection of events. As explained earlier, events selected by
\Hltone are stored in a buffer storage. From this storage the events
are accessed by sampling processes running on the event-filter farm
nodes normally used for the \Hlttwo selection algorithms. Because the
calculations are distributed, alignment and calibration algorithms can
be run very quickly, even if large samples are required to obtain the
required precision. The constants are typically available in a few
minutes for the parameters which need to be updated as soon as
possible in \Hltone. The update is immediately propagated to \Hltone
when the new parameters are available via a run change. Later, \Hlttwo
consistently uses the same parameters. The other parameters used only
in \Hlttwo processing are usually available less than one hour after
data taking and can be immediately fed into the \Hlttwo algorithms,
which can start event processing, organised by \Lhc run and fill, only
after these parameters are available.  This process is described in
more detail in section~\ref{sec:RTA:AlignAndCalib}.

\subsection{Computing farm}

While the first level selection (\Hltone) is done on \Gpu cards
installed in the event-builder nodes, the second level selection
(\Hlttwo), as well as calibration and alignment processes, are running
on a large farm of general purpose CPU servers called the event-filter
farm or computing farm. These servers are typically dense computing
nodes, requiring half a rack-unit per server. They are connected via
an Ethernet network, with a speed which tries to match approximately
their relative processing power. The servers are quite heterogeneous
because they come from many different procurement procedures. More
than 3000 servers are available. Up to 80 can be packed into a single
rack. When new batches of servers are acquired, the oldest servers are
retired, which maximises reliability and optimises electrical power
usage. Although the servers are currently not yet fully operated as a
computing cloud, many characteristics of cloud computing apply. All
nodes run the same operating system (currently, CentOS Linux 7), do
not store anything relevant locally, and can be easily replaced, which
makes it easy to use a wide variety of different machines of widely
varying age and performance. From the point of view of the \Ecs, their
key characteristic is their processing power, in which they
differ. Local storage is only used for scratch space and the OS
installation. Management is done using industry standard
tools.\footnote{Among them Foreman~\cite{foreman} and
  Puppet$^{\text{\textregistered}}$~\cite{puppet}.} Physically they
are installed in the LHCb data centre described in this paper in
section~\ref{sec:datacentre}.

\subsection{Infrastructure}

The event-builder and event-filter farms are implemented directly on
dedicated servers to simplify management and have more flexibility,
although it can be expected that their deployment over a cloud system
will grow over time. In contrast, the monitoring, the \Ecs, various
infrastructure services such a logging, low-level system monitoring
and others run on a typical enterprise infrastructure.

To ensure a uniform environment, a virtualisation cluster has been
implemented, running on top of RedHat Enterprise Virtualisation
hypervisors. The event-builder and event-filter applications, except
as mentioned, are run on virtual machines. Many of these applications
are determined by peak-load and many are very memory intensive for
comparatively little CPU use. Logically, the virtualisation is divided
into two clusters. The first cluster provides the core control system
function, hosting all services needed for safe and controlled data
taking. These services are provided with guaranteed resources which
are fully redundant, ensuring that if a hypervisor or group of
hypervisors fail, there are enough resources on the remaining
hypervisors to completely take over after a restart. The second
cluster hosts all services which are not critical, such as development
or general purpose machines and certain control functions not required
during data taking.  Enterprise virtualisation requires a shared
storage system, which is provided by a commercial
filer.\footnote{NetApp$^{\text{\textregistered}}$~\cite{netapp}
  filer.} An independent filer provides all the shared file-systems
for the entire cluster.

To accommodate more appropriately the growing number of services
originating in the cloud computing world, a
Kubernetes~\cite{kubernetes} cluster is also available. Services
running on the cluster must provide redundancy at the application
level, because in this setup no attempt is made, beyond basic best
practices, to provide hardware redundancy.

The computing infrastructure is comprised of almost 5000 physical
servers and more than 10000 network ports. In addition to physics
data-taking, the infrastructure is used to serve other computing needs
of LHCb.

\subsection{Performance}

\begin{figure}[t]
  \centering
  \includegraphics[width=\textwidth]{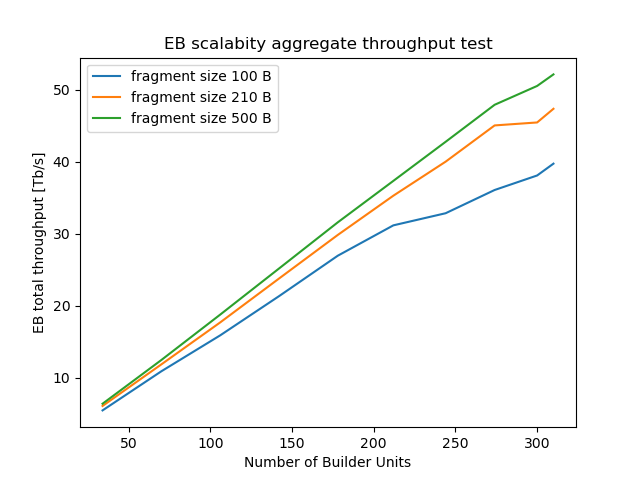}
  \caption{Aggregated data-rate in the LHCb event-builder network as a
    function of the number of builder units for various nominal
    event-fragment sizes. }
  \label{fig:eb_scaling}
\end{figure}

The performance of the event-building network has been studied using
simulated data generators to emulate the \Tellfourty
boards. Figure~\ref{fig:eb_scaling} shows the scaling of the total
event-builder data throughput as a function of the number of builder
units per event-builder server (each server implements two independent
builder-units). The deviations from ideal (i.e.\ linear) scaling
behaviour are believed to be related to remaining hardware and
low-level tuning issues, in particular the underlying fragment
(message) size, which will be addressed during the commissioning
period. However, already at this level the performance requirements
are met with a sufficient margin.

\section{Trigger and real-time analysis}
\label{sec:trigger}
The objective of the trigger system is to reduce the data volume,
which reaches 4\tbyps at the nominal instantaneous luminosity in \Prpr
collisions, to around 10\gbyps which can be recorded to permanent
offline storage. As discussed in section~\ref{sec:introduction}, this
reduction of a factor 400 is complicated by the high rate of
potentially interesting signals which can be at least partially
reconstructed in the detector acceptance. Over 300\khz of bunch
crossings contain a partially reconstructed beauty hadron and almost
1\mhz of bunch crossings contain a partially reconstructed charm
hadron~\cite{LHCb-TDR-016}.\footnote{A particle is considered
  partially reconstructed when only part of the decay products are
  reconstructed and identified.} It is therefore not possible to fully
pursue a traditional inclusive trigger strategy in which events
containing the signals of interest are identified using a small set of
generic signatures.

Instead, the majority of LHCb physics channels must be selected by the
trigger by fully reconstructing and identifying the specific signals
of interest, while saving only a limited subset of information about
the rest of the event~\cite{LHCb-DP-2019-002}.  This \emph{real-time
  analysis} approach, pioneered by LHCb in \runtwo data
taking~\cite{LHCb-DP-2016-001}, necessarily requires that the trigger
performs a full offline-quality reconstruction, enabled by an
alignment and calibration of the detector performed in
quasi-real-time. The physics motivations behind real-time analysis
have been documented in detail
refs.~\cite{Fitzpatrick:1670985,Gligorov:2018fuk}. They result
in a two-stage trigger system: a first inclusive stage or
\Acr[m]{hlt},\footnote{This terminology is a legacy of the previous
  experiment where the trigger had a hardware level (L0) and a
  software level (\Hlt).} the \Hltone, based primarily on charged
particle reconstruction which reduces the data volume by roughly a
factor of 20, and a second stage, the \Hlttwo, which performs the full
offline-quality reconstruction and selection of physics signatures.  A
large disk buffer is placed between these stages to hold the data
while the real-time alignment and calibration is being performed, as
described in section~\ref{sec:onlinecalib}.

\subsection{Physics requirements}
\label{sec:RTA:requirements}

Because of the necessity to perform real-time analysis, the physics
requirements of the \Hlttwo trigger stage are easily expressed though
demanding: it must perform the full offline-quality detector
reconstruction on all events, using the offline-quality alignment and
calibration provided in real-time, and support on the order of 1000
independent selections for signals of interest without any restriction
on the topology of those physics channels (see
section~\ref{ssec:rta_hlt2} for more details).

The bulk of the non-trivial physics requirements concern \Hltone,
where a partial reconstruction is performed with the attendant
trade-offs between speed, efficiency, and output rate.  The objective
of \Hltone is to reduce the event rate to a level at which the data
can be buffered to disk for real-time alignment, calibration, and
further processing in \Hlttwo, while maintaining high efficiency
across the LHCb physics programme.  This rate depends on the
throughput of the \Hlttwo full detector reconstruction and on the
maximum write-speed possible to the LHCb disk
buffer.\footnote{Throughput is the number of events processed per unit
  time and is expressed in \hz.} It is therefore helpful to approach
the \Hltone physics requirements by looking at the signal rates
expected in the LHCb upgrade, which have been documented in
ref.~\cite{Fitzpatrick:1670985} and served as the initial
motivation for the development of LHCb real-time analysis processing
model.

As discussed before, the signal rates are dominated by charm and
beauty hadron decays which are at least partially reconstructible
within the LHCb acceptance. Signal rates for other areas of the LHCb
physics programme such as electroweak physics or studies of quarkonia
are significantly lower. While some short-lived hadronic resonances
are produced at higher rates, measurements of these states are limited
to the percent-level by luminosity knowledge or data-driven
determination of reconstruction efficiencies. Therefore the full data
set is only needed in sparsely populated regions of the kinematic
parameter space where the event rate is negligible.  Strange hadrons
have signal rates which are more than an order of magnitude higher
than charm and beauty. However, accepting strange hadrons with high
efficiency would conflict with the main physics aims of the
experiment, which concern charm and beauty hadrons. Therefore
preference is given to rare strange hadron decays, whose signal rates
are negligible compared to the singly-Cabibbo suppressed charm hadron
decays at the centre of LHCb charm studies.

Since the signal rates are so high, it is of paramount importance that
the \Hltone selects events due to the presence of signal as opposed to
fake (``ghost'') tracks or random combinations of
tracks.\footnote{Fake tracks are the result of the pattern recognition
  algorithms that may reconstruct false trajectories from random
  combinations of hits.} If this requirement is met, at a luminosity
of $2\times 10^{33}$\invcma\invsec a maximal \Hltone output rate of
2\mhz is from first principles sufficient to cover the full LHCb
physics programme.

Having defined the maximal \Hltone output rate which the trigger
system should be able to support, it is important also to define the
smallest \Hltone output rate at which it remains efficient for LHCb
signals of interest. This is important because at times of very high
collider efficiency the \Hltone output rate might need to be reduced
in order to avoid overflowing the disk
buffer. Studies~\cite{Fitzpatrick:2017als} show that it is possible
to reduce the \Hltone output rate to around 500\khz while remaining
relatively efficient for LHCb charm physics channels.  This rate is
thus considered as a lower limit for the \Hltone output rate.

All these requirements imply that the \Hltone must be able to
reconstruct and select at least the following physics signatures:
\begin{itemize}
\item Tracks or two-track vertices displaced from the primary \Prpr
  interaction, or \Pv. This signature can be used to select any event
  containing a long-lived hadron or \tauon lepton, which covers the
  vast majority of LHCb analyses;
\item Leptons, particularly muons, regardless of their displacement
  from the \Pv. Displaced leptons can be selected as any other tracks,
  although the efficiency can be kept higher for the same output rate
  by using lepton identification criteria to allow displacement- or
  {\pt}-based criteria to be loosened. Non-displaced (di)leptons are
  particularly important for spectroscopy studies, exotic searches,
  and electroweak physics.
\end{itemize}
Based on these constraints, requirements on the track reconstruction
itself can be defined:
\begin{itemize}
\item \Hltone must reconstruct all tracks in the \Velo detector
  acceptance in order to accurately determine the position of primary
  vertices and calculate the displacement of tracks and secondary
  vertices produced in the decay of long-lived particles;
\item \Hltone must be able to reconstruct tracks regardless of whether
  they are displaced from the \Pv or not;
\item \Hltone must be able to reconstruct track momenta at the percent
  level, in order to ensure a fast rise of the turn-on curve of the
  \Hltone efficiency versus the \Hlttwo reconstructed particle
  momentum (where the resolution is around $0.5\%$);
\item the \Hltone reconstruction must provide an accurate and precise
  covariance matrix of the track measurement closest to the beam line,
  in order that the turn-on curve of \Hltone efficiency versus the
  \Hlttwo reconstructed particle displacement remains sharp;
\item \Hltone must be able to identify tracks as muons or non-muons.
\end{itemize}
It is crucial that \Hltone is able to reconstruct tracks above a
certain kinematic threshold which depends on the physics in
question. Experience from \runone and \runtwo shows that a \pt
threshold of 500\mev is adequate for most of the beauty and charm
physics programme, although being able to reconstruct tracks down to
$\pt = 200$\mev would be highly beneficial for many areas of charm
physics studies. For rare strange hadron decays it is important to
reconstruct tracks with a \pt as low as possible, and a more natural
cutoff is to require 3\gev of momentum, which corresponds to the
lowest momentum at which it is possible to identify muons in LHCb.

While there are studies which could benefit from the reconstruction of
calorimeter quantities (like e.g.\ energy deposits by photons) in
\Hltone, this is not considered as a mandatory requirement because a
good portion of the signal efficiency can be achieved using tracks
alone. A similar argument applies to electron identification or the
reconstruction of downstream tracks, specifically $\KS\to \pip\pim$
and $\Lz\to\proton\pim$ decaying outside the \Velo. Both would
significantly benefit certain areas of the physics programme but they
do not represent a strict requirement for a functioning \Hltone.\looseness=-1

\subsection{Design overview}

\subsubsection{Hardware design}

\begin{figure}[t]
  \centering
  \includegraphics[width=.9\textwidth]{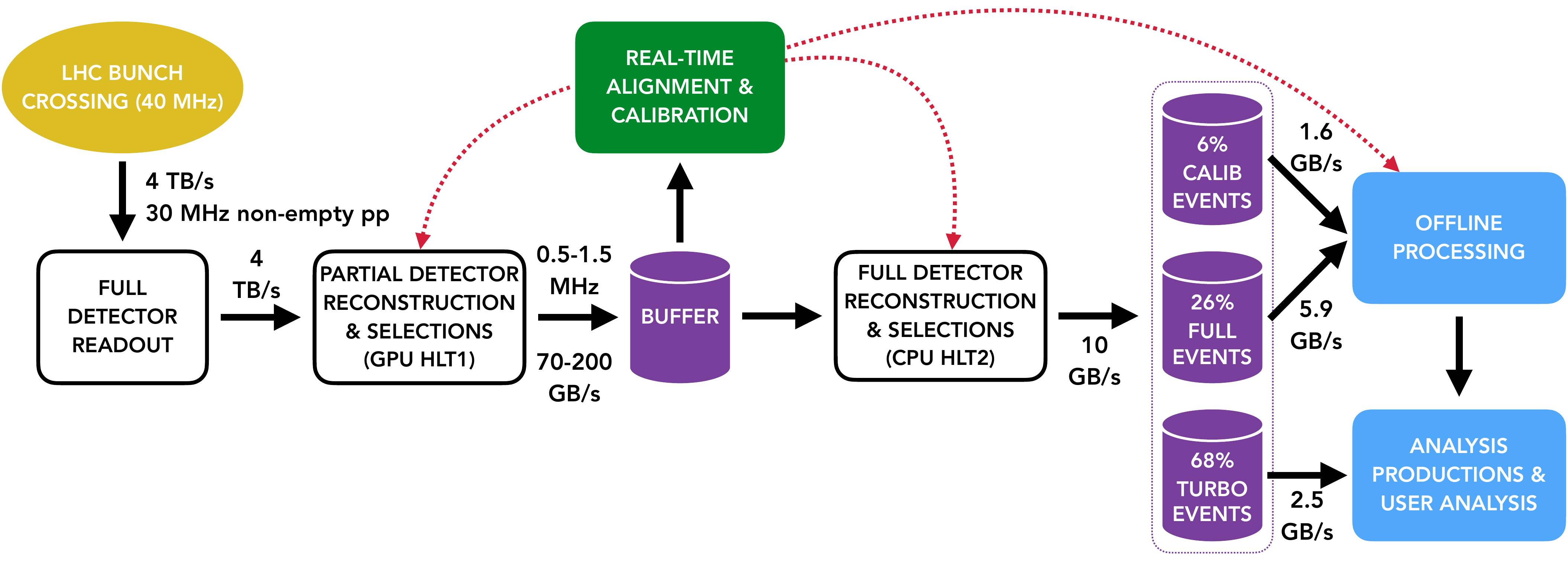}
  \caption{Online data flow. Reproduced with permission from~\cite{LHCb-FIGURE-2020-016}.}
  \label{fig:RTA:dataflow}
\end{figure}

As described in section~\ref{sec:online} the centrepiece of LHCb
trigger is the full detector readout and event building at
30\mhz.\footnote{Although the \Lhc bunch crossing frequency is 40\mhz,
  the rate of events with some signal visible inside the LHCb
  acceptance (\emph{reconstructible} events) is $\sim 30\mhz$.} In
addition to its inherent flexibility, this approach also matches the
described physics requirements for \Hltone, most notably the
requirement that all the tracking algorithms must reconstruct the
events at 30\mhz. Since the tracking detectors constitute
approximately two-thirds of LHCb overall data bandwidth, the cost of a
hardware trigger allowing the other one-third (principally \rich and
calorimeters) to be read out at a lower rate does not justify the
added complexity and reduced flexibility of such a system. The trigger
data flow from detector to offline storage is illustrated in
figure~\ref{fig:RTA:dataflow}.\looseness=-1

The online disk buffer serves two purposes: it stores events selected
by \Hltone while real-time alignment and calibrations are being
performed, and it allows events selected by \Hltone to be buffered for
processing between LHC fills. This in turn effectively increases the
processing power of the \Hlttwo computing resources. Simulations of
the LHC fill structure~\cite{LHCb-DP-2019-001} result in an optimal
buffer size of around 30\aunit{PB}, which allows around
80\aunit{hours} of LHC collisions to be buffered at an \Hltone output
rate of 1\mhz.  The disk buffer architecture is constrained not only
by the total size but also by I/O limitations of individual disks,
which impose a minimum number of disks in the system.

Because the detector is fully read out upfront, there is in principle
total freedom to choose the computing architecture with which to
process the data; the only constraints are the available budget and
the capacity (racks and cooling) of the data centre. LHCb has sought
to take advantage of this freedom since about the end of \runone,
pursuing the development of high-throughput reconstruction algorithms
on both CPU and \Gpu architectures. Eventually, a full cost-benefit
analysis~\cite{LHCb:2021kxm} led to the choice to implement \Hltone
on \Gpu{s} while remaining with the same CPU architecture used during
\runone and \runtwo for \Hlttwo. Because the \Gpu{s} are hosted in the
event-building servers as described in section~\ref{sec:online}, there
is a limit of around 500 \Gpu{s} which can be installed for running
\Hltone. As the baseline \Hltone described in the remainder of this
paper requires only around 200 latest generation \Gpu{s}, this limit
does not introduce a significant constraint into the system.

\subsubsection{Software design}
\label{Sec:RTAsoft}

The software design of the LHCb upgrade trigger is guided by the
principle that LHCb's real-time and offline processing should be as
similar as possible. More concretely, the upgrade code base is set up
so that any offline reconstruction and selection is a specific
reconfiguration of the same underlying algorithms as the real-time
version. Much of the quality assurance and validation machinery is
also shared, easing the maintenance burden. This convergence is
partially a result of the physics requirements, in particular the need
to run the full offline-quality reconstruction within the trigger. In
fact, the LHCb code base has been evolving in this direction
throughout \runone and \runtwo, so the almost total convergence
achieved for the upgrade can be seen as a natural completion of this
process.\looseness=-1

Most of the code base is a mixture of \cpp used for data processing
and algorithms, \textsc{python} for job configuration and
\textsc{CUDA} in the {\Gpu}-based \Hltone. The CPU code base is built
on top of the \gaudi~\cite{Barrand:2001ny,MatoGaudiDesign:1998}
framework.

\textls[-10]{Both are optimised for multithreaded execution.  The GPU code base is
implemented within the cross-architecture \allen
framework~\cite{Aaij:2019zbu}. \allen is based on a single source
code, that can be compiled for both CPUs and \Gpu{s}, supporting
different \Gpu architectures through \textsc{CUDA} and \textsc{HIP}
languages. The framework also provides a custom memory manager to
optimise \Gpu memory usage, and a multievent scheduler to handle the
data and control flow of events processed in batches during
data-taking In the context of event simulation \Hltone \allen
algorithms are converted to \gaudi algorithms automatically and event
processing is controlled by \gaudi to fit into the offline processing
chain (section~\ref{subsec:DPA}).}\looseness=-1

In order to ensure thread-safety, the design of algorithms ensures
that there are no race-conditions and that results are not altered by
the execution order of the threads. This in turn allows the
configuration to largely define the data flow, with the control flow
automatically deduced by the scheduler; manual overriding remains an
option to resolve ambiguities or enforce a preferred execution
order. The detector geometry and conditions are read in from dedicated
databases, as described further in
sections~\ref{sec:conddb}~and~\ref{sec:detgeodesc}. Although the
scheduler is fully flexible, in what follows, for simplicity of
exposition, the description of the reconstruction and selections will
be separated as if the trigger executed a monolithic reconstruction
followed by multiple selection algorithms.

The code base is maintained by a common effort between the developers
of LHCb's various software projects, with continuous integration,
testing, and code review before any changes are made. A rotating team
of (less experienced) shift takers supported by (more experienced)
maintainers supervises this process and carries out the task of
merging and releasing specific versions of the code for use in
production.

\subsection{First trigger stage}

The \Hltone first level trigger design follows the requirements
described in section~\ref{sec:RTA:requirements}. The baseline
implementation in terms of reconstruction and selections is described
here. Key performance figures are documented in detail in
ref.~\cite{LHCb-TDR-021} and are not repeated here.

\subsubsection{Reconstruction}

The baseline \Hltone reconstruction focuses on finding the
trajectories of charged particles which originate within the LHCb
vertex detector and traverse the rest of the LHCb tracking system
(specifically the \Ut, \Scifi, and muon chambers). The objective is to
measure their momenta with percent-level precision, associate each
particle to the \Prpr collision where it was produced and measure its
displacement from that \Prpr collision, and identify the particle as a
muon or non-muon.

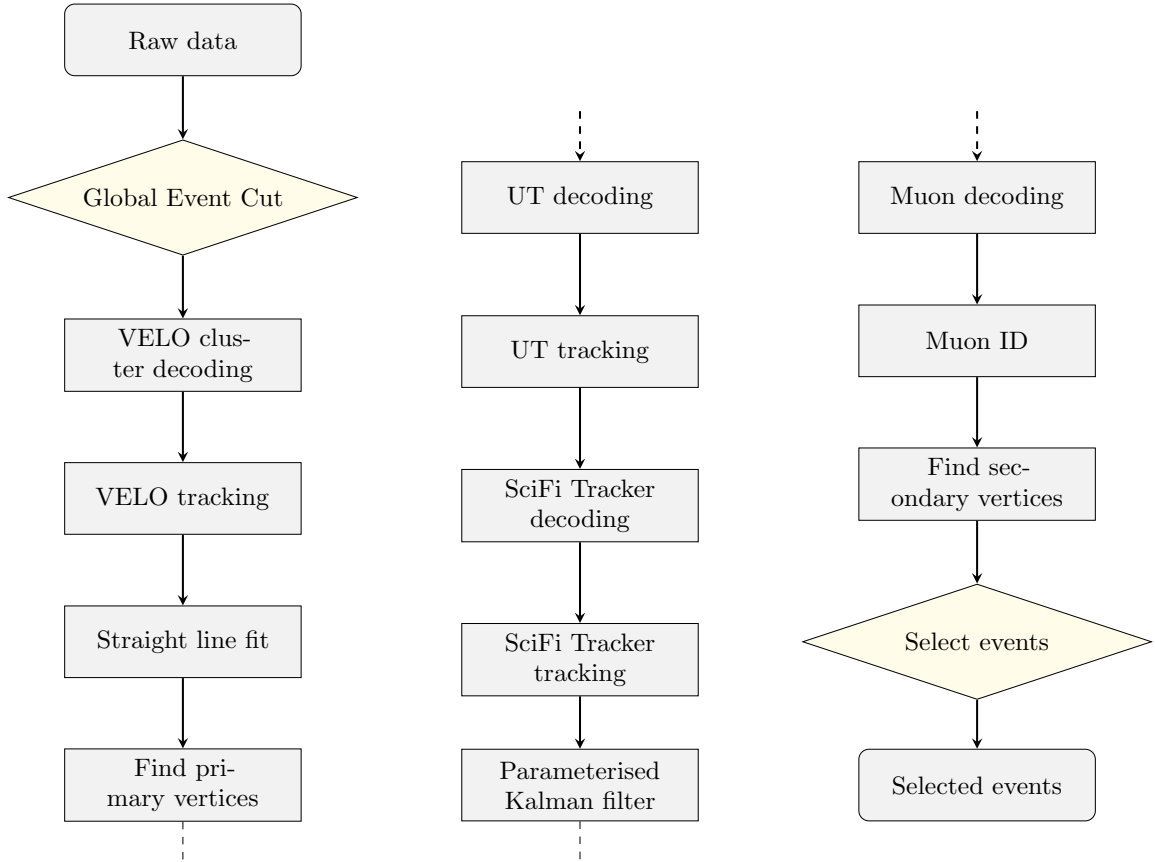
\begin{figure}[t]
  \centering
  \scalebox{0.95} {
    \begin{tikzpicture}[node distance=2cm]
      \node (initialise) [startstop] {Raw data};
      \node (gec) [decision, below of=initialise] {Global Event Cut};
      \node (velo_decoding) [process, below of=gec, yshift=-0.2cm] {\Velo cluster decoding};
      \node (velo_tracking) [process, below of=velo_decoding] {\Velo tracking};
      \node (simple_kalman_filter) [process, below of=velo_tracking] {Straight line fit};
      \node (find_primary_vertices) [process, below of=simple_kalman_filter] {Find primary vertices};
      \node (find_primary_vertices_arrow) [below of=find_primary_vertices, yshift=0.8cm] {};
      
      \node (ut_decoding) [process, right of=gec, xshift=3.5cm] {\Ut decoding};
      \node (ut_decoding_arrow) [above of=ut_decoding, yshift=-0.65cm] {};
      \node (ut_tracking) [process, below of=ut_decoding, yshift=-0.15cm] {\Ut tracking};
      \node (scifi_decoding) [process, below of=ut_tracking, yshift=-0.15cm] {\Scifi decoding};
      \node (scifi_tracking) [process, below of=scifi_decoding, yshift=-0.15cm] {\Scifi tracking};
      \node (kalman_filter) [process, right of=find_primary_vertices, xshift=3.5cm] {Parameterised Kalman filter};
      \node (kalman_filter_arrow) [below of=kalman_filter, yshift=0.8cm] {};
      
      \node (muon_decoding) [process, right of=ut_decoding, xshift=3.5cm] {Muon decoding};
      \node (muon_decoding_arrow) [above of=muon_decoding, yshift=-0.65cm] {};
      \node (muon_id) [process, below of=muon_decoding] {Muon ID};
      \node (find_secondary_vertices) [process, below of=muon_id] {Find secondary vertices};
      \node (select_events) [decision, below of=find_secondary_vertices] {Select events};
      \node (output) [startstop, right of=kalman_filter, xshift=3.5cm] {Selected events};
      
      \draw [arrow] (initialise) -- (gec);
      \draw [arrow] (gec) -- (velo_decoding);
      \draw [arrow] (velo_decoding) -- (velo_tracking);
      \draw [arrow] (velo_tracking) -- (simple_kalman_filter);
      \draw [arrow] (simple_kalman_filter) -- (find_primary_vertices);
      
      \draw [arrow] (ut_decoding) -- (ut_tracking);
      \draw [arrow] (ut_tracking) -- (scifi_decoding);
      \draw [arrow] (scifi_decoding) -- (scifi_tracking);
      \draw [arrow] (scifi_tracking) -- (kalman_filter);
      
      \draw [arrow] (muon_decoding) -- (muon_id);
      \draw [arrow] (muon_id) -- (find_secondary_vertices);
      \draw [arrow] (find_secondary_vertices) -- (select_events);
      \draw [arrow] (select_events) -- (output);
      
      \draw [dashed_arrow] (ut_decoding_arrow) -- (ut_decoding);
      \draw [dashed_arrow] (muon_decoding_arrow) -- (muon_decoding);
      \draw [dashed] (kalman_filter) -- (kalman_filter_arrow);
      \draw [dashed] (find_primary_vertices) -- (find_primary_vertices_arrow);
    \end{tikzpicture}
  }
  \caption{Baseline \Hltone sequence. Reproduced with permission from~\cite{LHCb-TDR-021}. Rhombi represent algorithms reducing the
    event rate, while rectangles represent algorithms processing
    data.}
  \label{fig:hlt1_sequence}
\end{figure}

The reconstruction sequence is shown schematically in
figure~\ref{fig:hlt1_sequence}. It is preceded by a \Gec which removes
a fraction of the events with a very large number of tracks, which
cost a disproportionate amount of computing time to reconstruct while
having a worse detector performance. The cut can in principle be
applied based on the occupancy of any combination of subdetectors. The
baseline criterion is to reject the $7\%$ of busiest minimum bias
events based on \Ut and \Scifi occupancies. This will be revisited
during detector commissioning to make sure that the cut uses
information from those subdetectors whose data occupancies best match
simulation. As necessary, special reconstruction and selection
sequences can be deployed which bypass this \Gec, e.g.\ for
electroweak physics or very forward high transverse momentum exotic
signatures. These are not discussed further here for brevity.

From the algorithmic point of view, the reconstruction sequence is
straightforward:
\begin{enumerate}
\item tracks are reconstructed in the \Velo and are used to locate the
  positions of the primary vertices where the beam collisions
  occurred;
\item tracks are extrapolated to the \Ut, and subsequently to the
  \Scifi detector, based on a minimum allowed momentum and/or
  transverse momentum. A magnetic field parametrisation is used to
  predict the track position and speed up the reconstruction. In the
  baseline configuration, the minimum \pt threshold is set to 500\mev,
  but this value will be updated once the performance is evaluated
  with real data. Reducing this threshold requires improved
  computational performance but increases the number of tracks
  available to physics selections and consequently the signal
  efficiency. This is particularly important for signatures typical of
  strange and charm quark physics;
\item above the \pt threshold, the track momentum is known with a
  precision better than 1\% across the full momentum range of
  interest. The momentum is thereafter used as input to a
  parameterised Kalman filter~\cite{Kalman_1960,FRUHWIRTH1987444}
  which estimates the position and covariance matrix of the particle
  at the beam line. This information is in turn used to calculate the
  particle's displacement from all primary vertices with a resolution
  similar to the one achievable in \Hlttwo with a fully aligned and
  calibrated detector. Then, selections typically apply requirements
  based on a minimal displacement from the primary vertex which is
  equivalent to at least two or three standard deviations;
\item tracks are identified as muons or non-muons and fitted to a
  common origin to form two-body displaced vertex candidates, which
  are then input to the selections.
\end{enumerate}
This reconstruction sequence largely matches that of LHCb \runtwo \Hlt
in conceptual terms. The main difference is that only a simplified
Kalman filter is used in order to reduce processing time. Further
detailed documentation can be found in refs.~\cite{LHCb-TDR-021,
  LHCB-FIGURE-2020-014}.

The sequence processing time is further reduced by exploiting
prereconstructed hit clusters from the \Velo, built at an early stage
into the \Velo \Tellfourty \Fpga firmware by a 2D cluster finding
algorithm.  The \Fpga algorithm identifies VELO clusters in a fully
parallel way by pattern matching, and computes their topology and
position centroid in real time providing then the results to
\Hltone~\cite{Bassi:2825279}.\looseness=-1

\subsubsection{Selections}
\label{ssec:hlt1:selections}

The \Hltone selections are broadly divided into four categories: the
primary inclusive selections for the bulk of LHCb physics programme;
selections for calibration samples essential to a data-driven
evaluation of the reconstruction performance; selections for specific
physics signatures not covered by the inclusive triggers; and
technical triggers for luminosity determination, monitoring,
calibration and alignment. Here, a brief description of the physics
logic of these selections is given, while their performance is
described in section~\ref{sec:RTA:SelPerf}.

The primary inclusive selections are:
\begin{itemize}
\item a \emph{two-track vertex trigger}, requiring large transverse
  momentum and significant displacement from all primary vertices in
  the event; this trigger provides the bulk of the efficiency for
  nonmuonic signatures;
\item a \emph{displaced single-track trigger}, requiring large
  transverse momentum and significant displacement from all primary
  vertices in the event~\cite{Gligorov:2011zya}. This is the most
  inclusive of the \Hltone triggers as well as the least biasing
  (because it fires on a single particle) of the hadronic
  triggers. However, because of the abundance of genuine displaced
  hadrons in upgrade conditions, particularly from charm decays, its
  main role is to provide redundancy for the two-track vertex trigger;\looseness=-1
\item a \emph{displaced single muon trigger}, which operates similarly
  to the single-track trigger, but requires that the track be
  identified as a muon; this in turn allows the transverse momentum
  and displacement requirements to be significantly relaxed;
\item a \emph{displaced dimuon trigger}, which operates similarly to
  the two-track vertex trigger, but requires that both tracks be
  identified as muons; the transverse momentum and displacement
  requirements can be relaxed even further compared to the single muon
  trigger;
\item a \emph{high-mass dimuon trigger}, which does not make any track
  displacement requirements, but requires that the dimuon invariant
  mass be above 2900\mev; this trigger is particularly important for
  the study of charmonia, whether in prompt production or from the
  decays of beauty hadrons;
\item a \emph{very high transverse momentum muon trigger}, which does
  not make any \Gec requirement and is used for electroweak physics
  and searches for exotic signatures;
\end{itemize}

When selecting calibration samples, it is particularly important to
ensure a good coverage across the kinematic spectrum. Therefore, while
charmonia are already selected by the high-mass dimuon trigger, the
calibration selections focus on the decays of charm hadrons such as
$\Dz\to\Km\pip$, which are critical for a data driven evaluation of
particle identification performance. Analogous selections are foreseen
for other charm hadron species. Additional selections are used to
create samples enriched in tracks which traverse regions of the
detector particularly critical for the tracker alignment or \rich
mirror alignment. \Hltone throughput scales to ${\cal O}(100)$
selections. It is therefore likely that numerous triggers for specific
hard-to-select signals will be implemented as data taking
progresses. Examples from previous LHCb data-taking periods included
diproton triggers, diphoton triggers, decay-time unbiased charm
triggers, and others.\footnote{Although there is no full calorimeter
  reconstruction in the \Hltone, a rudimentary calorimeter clustering
  has been implemented in \allen, allowing to reconstruct electrons
  and photons. This was possible due to computing resources freed by
  the calorimeter preprocessing (\Acr[m]{llt}) discussed in
  section~\ref{sssec:calo_electronics_LLT}.}

\subsection{Second trigger stage}
\label{ssec:rta_hlt2}

The second level trigger (\Hlttwo) uses the information provided by
the real-time alignment and calibration of the detector to perform an
offline-quality reconstruction, followed by ${\cal O}(1000)$ selection
algorithms which decide whether or not to retain any given event. If
an event is retained, the selection algorithms specify which portions
of the full event are recorded to permanent storage following the
real-time analysis paradigm developed during
\runtwo~\cite{LHCb-DP-2016-001,LHCb-DP-2019-002}.

\subsubsection{Reconstruction}

The LHCb reconstruction is divided into four main components: charged
particle pattern recognition, calorimeter reconstruction, particle
identification, and the Kalman fit of reconstructed tracks which
allows their parameters to be measured with the best possible
precision and accuracy (see
section~\ref{ssec:rta_hlt2}).\footnote{Within LHCb, \emph{Kalman fit}
  identifies a fit of track parameters based on a Kalman filter which
  combines the coordinates of an ensemble of selected hits and
  additional information like magnetic field and material distribution
  maps.} As the upgrade detector design is not fundamentally different
from that of the previous LHCb detector, its reconstruction is also
conceptually similar to what was used during \runone and
\runtwo. Nevertheless some changes and improvements have been possible
in specific areas, as described here.

\begin{figure}[t]
  \centering
  \includegraphics[width=0.7\textwidth]{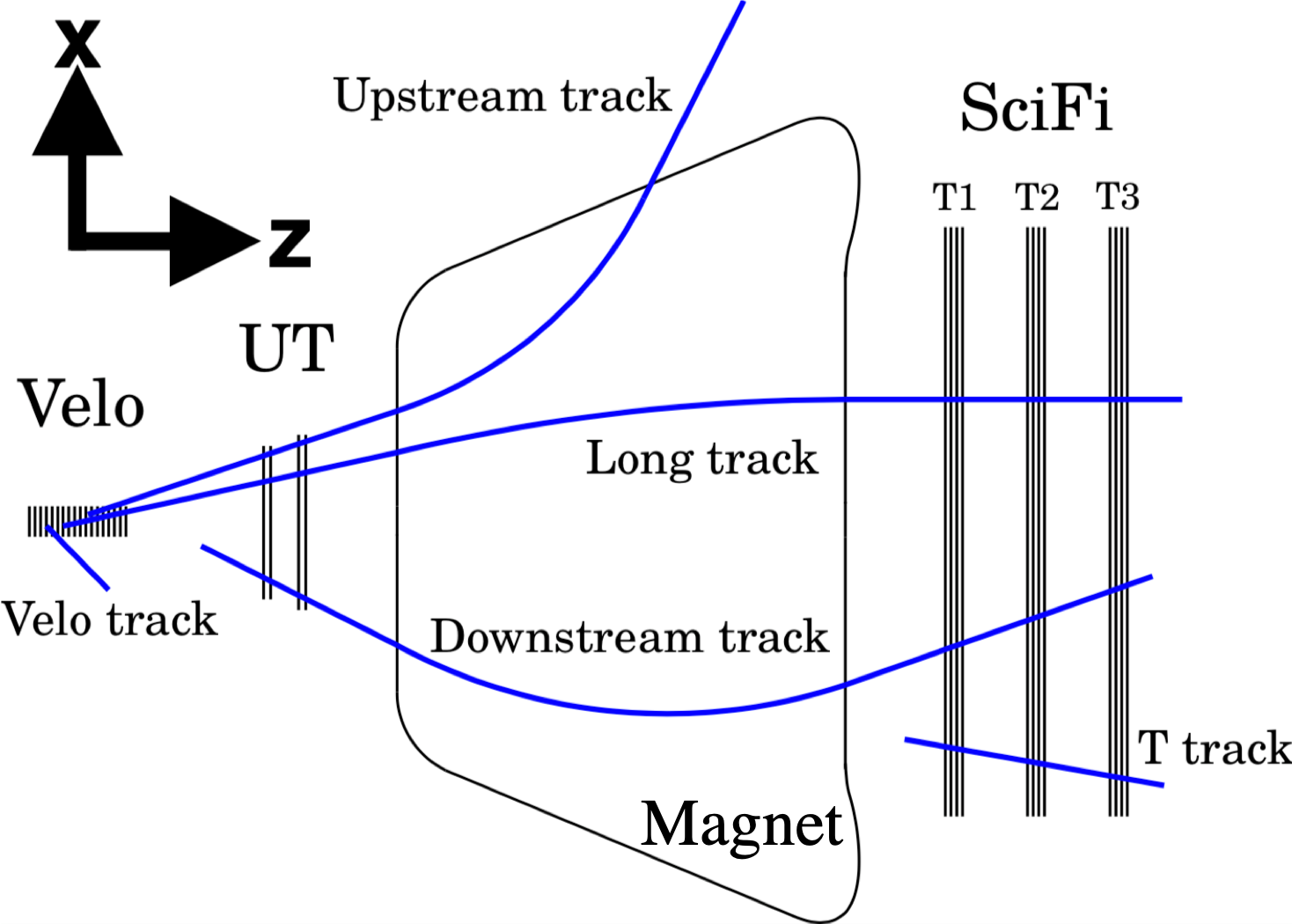}
  \caption{Track types in the LHCb detector bending plane. Reproduced from~\cite{LHCb-DP-2008-001}. \textcopyright\ 2008 IOP Publishing
Ltd and Sissa Medialab. All rights reserved.}
  \label{fig:RTA:track_types}
\end{figure}

\paragraph{Charged particle pattern recognition.}
Different tracking algorithms exist to reconstruct different track
types, illustrated in figure~\ref{fig:RTA:track_types}. Tracks which
originate in the vertex detector (\emph{\Velo tracks}) are used to
determine the positions of the primary \Prpr collisions, a process
known as primary vertex finding. The combination of \Pv positions and
track trajectories, in turn, allows tracks which originate from the
decays of long-lived particles and are therefore displaced from the
\Pv to be precisely identified. As there is effectively zero magnetic
field inside the \Velo, these tracks must be extrapolated into the
region covered by the \Ut (\emph{upstream tracks}) and \Scifi
(\emph{long tracks}) in order to measure their momentum. Long tracks
have the most precise and most accurate momentum determination and are
used in nearly all LHCb analyses. In addition to the \emph{forward}
algorithm which extrapolates \Velo tracks to the \Scifi, a second
redundant reconstruction path (seeding) performs a standalone
reconstruction of track segments in the \Scifi (\emph{T tracks})
before matching them to \Velo tracks and optionally \Ut hits. In
addition, \Scifi seeds are extrapolated to the \Ut and used to form
\emph{downstream tracks} in order to reconstruct particles which
originate outside the \Velo but before the \Ut. Downstream tracks
provide the bulk of LHCb statistical power for the study of decays
involving strange hadrons. The track extrapolations used in all of
these pattern recognition algorithms are, for reasons of speed, based
on parametric models of trajectories in the LHCb magnetic
field. Duplicated tracks (\emph{clones}) can be formed when different
algorithms reconstruct the same track segment in one of the
subdetectors, for example when a long track and a downstream track
share a T-station seed. These are filtered by removing duplicates
within individual pattern recognition algorithms. Following the Kalman
fit, a global clone-killing algorithm uses the fit quality to perform
a final arbitration between overlapping \Velo, long, and downstream
tracks and removes the remaining duplicates.

The charged pattern recognition algorithms have undergone significant
evolution from the \runone and \runtwo code in order to make them
better able to efficiently use modern multicore CPU architectures.  An
example of such optimised algorithms is described in detail in
ref.~\cite{Hennequin:2019itm}, while their performance is
documented in section~\ref{sec:performance}.

\paragraph{Kalman fit.}
\label{ssec:kalmanfit}
While the pattern recognition is used to accurately group detector
hits into collections corresponding to individual charged particles, a
separate step based on a Kalman filter (referred to as the
\emph{Kalman fit}) is required in order to determine the properties of
charged particle trajectories with maximum accuracy and precision.
Several approaches exist: a detailed Kalman fit which uses lookup
tables to describe the magnetic field and material distribution and
Runge-Kutta methods to propagate the particle trajectories, an
intermediate solution where the propagation is performed as in the
detailed Kalman fit but the interactions with the detector material
are parametrised, and a fully parametric Kalman fit also parametrising
the particle propagation through the LHCb magnetic field. The last
option is described in more detail in ref.~\cite{Billoir:2021srr}.
Despite a significant investment of time and effort, a full
parallelisation of the detailed Kalman fit has remained
difficult. However, significant gains in speed have been achieved by
simplifying the data structures used inside the algorithm and reducing
the amount of output data to what is strictly necessary for physics
analysis. In addition, the software framework is sufficiently flexible
to allow the use of a mixture of parametric and detailed Kalman
processing, depending on the specific application. While the
parametric Kalman fit achieves a performance suitable for nearly all
physics applications, the detailed Kalman fit remains necessary for
use in the detector alignment procedure, as well as for certain
analyses which require the ultimate precision on track states.  The
detailed Kalman fit as well as the solution with the parametrised
material description will be therefore both available in \Hlttwo and
in offline data processing and will be used depending on the specific
needs.

\paragraph{Calorimeter reconstruction.}
\begin{figure}[t]
  \centering
  \includegraphics[width=\textwidth]{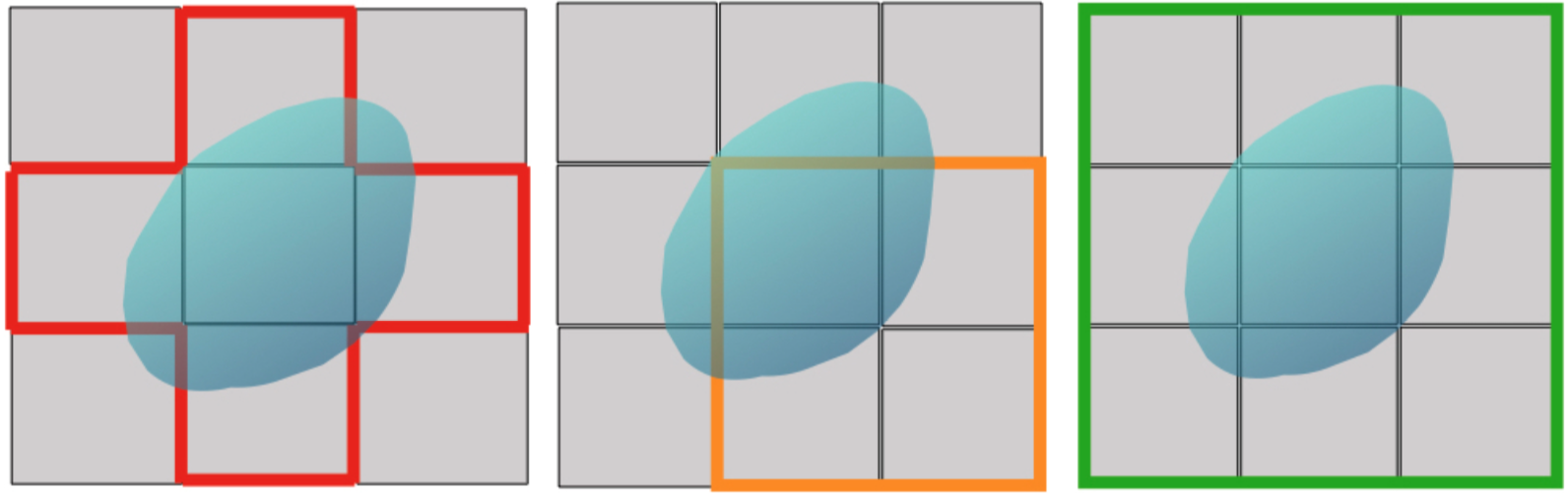}
  \caption{Shapes used for the upgrade calorimeter reconstruction,
    referred to as \emph{cross} (left), $\mathit{2\times 2}$~(centre)
    and $\mathit{3\times 3}$~(right).}
  \label{fig:RTA:ShowerShapes}
\end{figure}
Although the LHCb calorimeter system comprises a hadronic and an
electromagnetic calorimeter, during \runonetwo the \Hcal was used
almost exclusively for the first level hardware trigger which has now
been removed for the upgrade. At present there are no upgrade analyses
which foresee using \Hcal information in the real-time processing and
therefore only the \Ecal is reconstructed.

\Ecal clusters are formed from $3\times 3$ cells, as shown in
figure~\ref{fig:RTA:ShowerShapes}. Because of the higher pileup
expected in upgrade conditions and the limited spatial granularity of
the calorimeter, a combination of cross and $2\times2$ shapes is
needed to obtain the most accurate position and energy resolution for
each individual cluster.  Multivariate algorithms based on the shower
shapes and individual \Ecal cell energies are used to distinguish
single-photon clusters from those formed by the merger of multiple
photons, most notably in highly boosted $\piz\to\g\g$ decays. Electron
clusters are identified by extrapolating tracks to the \Ecal region
and subsequently matching them to \Ecal clusters. Dedicated algorithms
for calorimeter reconstruction have been prepared for the upgrade,
with completely new structure and inherent algorithmic logic to
improve the computational and physics performance; the calorimeter
reconstruction performance is documented in
section~\ref{sec:performance}.

\paragraph{Particle identification.}
LHCb uses a combination of its two \rich detectors, the \Ecal, and the
muon system in order to identify the five basic long-lived charged
particle species --- electron, muon, pion, kaon, and proton. Unlike in
central detectors, tau leptons are considered as any other composite
particle in LHCb and there are no centralised identification
algorithms for them; the same holds for neutral particles decaying in
the detector. The efficiency to correctly identify the charged
particle corresponding to any given track is heavily dependent on the
charged particle species and on the dominant backgrounds. In general,
as shown quantitatively in section~\ref{sec:performance}, performance
depends most strongly on the particle momentum, then on its
pseudorapidity, and then on the detector occupancy. The different
subdetectors dominate particle identification performance in different
momentum regimes, except for muons where the muon system plays a
dominant role in all cases. The particle identification performance
depends critically on an accurate knowledge of the track trajectory
within each particle identification subdetector, and therefore
requires the use of Kalman fitted tracks in order to achieve the best
results.

The standalone \rich reconstruction is described in detail
elsewhere~\cite{LHCb-DP-2012-003}. It has undergone few conceptual
changes but the software has been rewritten and reoptimised for
computational efficiency and speed.

The standalone muon reconstruction has been gradually improved
throughout \runonetwo, and for the upgrade, new variables have been
introduced which further improve the performance in the high-occupancy
regime of the upgrade by considering correlations between the hits in
the different layers of the muon detector~\cite{LHCb-DP-2020-002}.

While optimal absolute performance is achieved by combining the
information provided by each subdetector into global multivariate
classifiers, it is equally important to have a particle identification
performance which can be accurately calibrated using data
tag-and-probe samples, as described in
section~\ref{sec:PIDCalib}.\footnote{In the tag-and-probe method
  typically a two-body decay of a resonance (e.g.\ a \jpsi) is used to
  identify a track independently of \Pid algorithms. One of the two
  tracks is used to \emph{tag} the identity of the other
  (\emph{probe}) which can be used to calibrate the \Pid algorithm.}
This is particularly true of the precision measurements which make up
the bulk of LHCb physics programme and which require permille-level
control of particle identification efficiencies and misidentification
rates in order not to be limited by systematic errors. For this
reason, multiple multivariate classifiers, trained on simulation and
tuned to have a better and more stable performance in different
kinematic regions, exist. The choice of which classifier is optimal
for any given analysis and where particle identification information
is used at the trigger level is left to the analysts which have to
ensure the relevant samples exist which can calibrate these
classifiers in the kinematic regions of interest.

\subsubsection{Selections}
\label{ssec:hlt2:selections}

Unlike \Hltone, where a limited number of largely inclusive selections
are sufficient to select the most interesting events, \Hlttwo relies
on about one thousand different selection algorithms, each tuned for a
particular signal topology and/or physics analysis. Although the
software framework provides full flexibility to schedule interleaved
sequences of reconstruction and selection steps, in practice almost
all selection algorithms are executed once the complete
offline-quality reconstruction has been performed.

In order for these selections to achieve the necessary computing
throughput, tracks and neutral objects are zipped together with
particle identification information into Structure-of-Array data
structures which can be efficiently processed in parallel. Both
rectangular-cut-based and multivariate or artificial
intelligence-based selections can be deployed.

In addition to identifying which events should be recorded to
permanent storage, each selection algorithm identifies which subset of
event data to record. If multiple selection algorithms decide to
record an event, the superset of information requested by them is
recorded. This \emph{real-time analysis} paradigm (also named \turbo
analysis) allows the rate of recorded events to be increased by
decreasing the volume of information recorded for each event. The
\turbo concept was already deployed during \runtwo and a detailed
description can be found in
refs.~\cite{LHCb-DP-2016-001,LHCb-DP-2019-002}.  The \turbo mechanism
allows full flexibility on the amount of event information that is
stored (\emph{selective persistence}), from the bare minimum of two
tracks and vertex coordinates for a two-body decay, up to the full
event information, depending on the specific physics channel under
study, as described in ref.~\cite{LHCb-TDR-018}. As also shown in
figure~\ref{fig:RTA:dataflow}, while a majority of triggered events
will be saved in the reduced \turbo format, the majority of the data
volume will consist of the calibration and traditionally triggered
(\emph{full}) events. In order to minimise the overall data volume,
\Hlttwo selections will be grouped into \emph{streams}, with all
selections belonging to a stream sharing an underlying physics logic
and recording similar sets of event information. Streams can be
configured according to broad physics channels like e.g.\ charm
physics, hadronic beauty decays, leptonic decays, electroweak physics,
etc.\ and will evolve during the experiment's lifespan, as needed.

\subsection{Alignment and calibration}
\label{sec:RTA:AlignAndCalib}

The fact that \Hlttwo performs a full offline-quality event
reconstruction and selects the majority of events based on the
real-time analysis paradigm also necessitates an offline-quality
alignment and calibration of the detector in real-time. This serves
two separate purposes. The first is to provide the most accurate
alignment and calibration parameters to the real-time reconstruction
and selections. This ensures that the physics parameters of interest,
such as particle masses or decay-times, are computed with the best
possible resolution, maximising the selection efficiency. In addition,
calibration parameters can be stored and made available for offline
physics analysis without (in most cases) the need for further
calibrations, simplifying the analysis workflow. The second is to
provide large tag-and-probe samples which can be used offline to
calibrate the difference between the detector performance in real data
taking and its performance in simulation. This is particularly
critical for physics studies which require a permille-level accuracy
of single-particle reconstruction and identification efficiencies for
each particle type.

Alignment and calibration procedures have been designed to maximise
the physics reach and analysis flexibility rather than imposing
centralised calibrations. For example, analyses which can benefit from
offline recalibration can be stored in the \FULL stream. A notable
example is electroweak physics in which the very high transverse
momentum of the signal decay products are very sensitive to any
residual misalignment. The optimal alignment needed for electroweak
physics can be obtained using large samples of very high momentum
tracks (typical of \W and \Z boson decays) which can be collected only
after a few months of data taking, thus requiring an offline
calibration which can be performed for example once per year. Even
these analyses, however, benefit from having the best possible
alignment and calibration available in real time, as it minimises the
difference in measured quantities used in online and offline
selections and therefore simplifies the modelling of corrections for
such effects.

\begin{figure}[t]
  \centering
  \includegraphics[width=\textwidth]{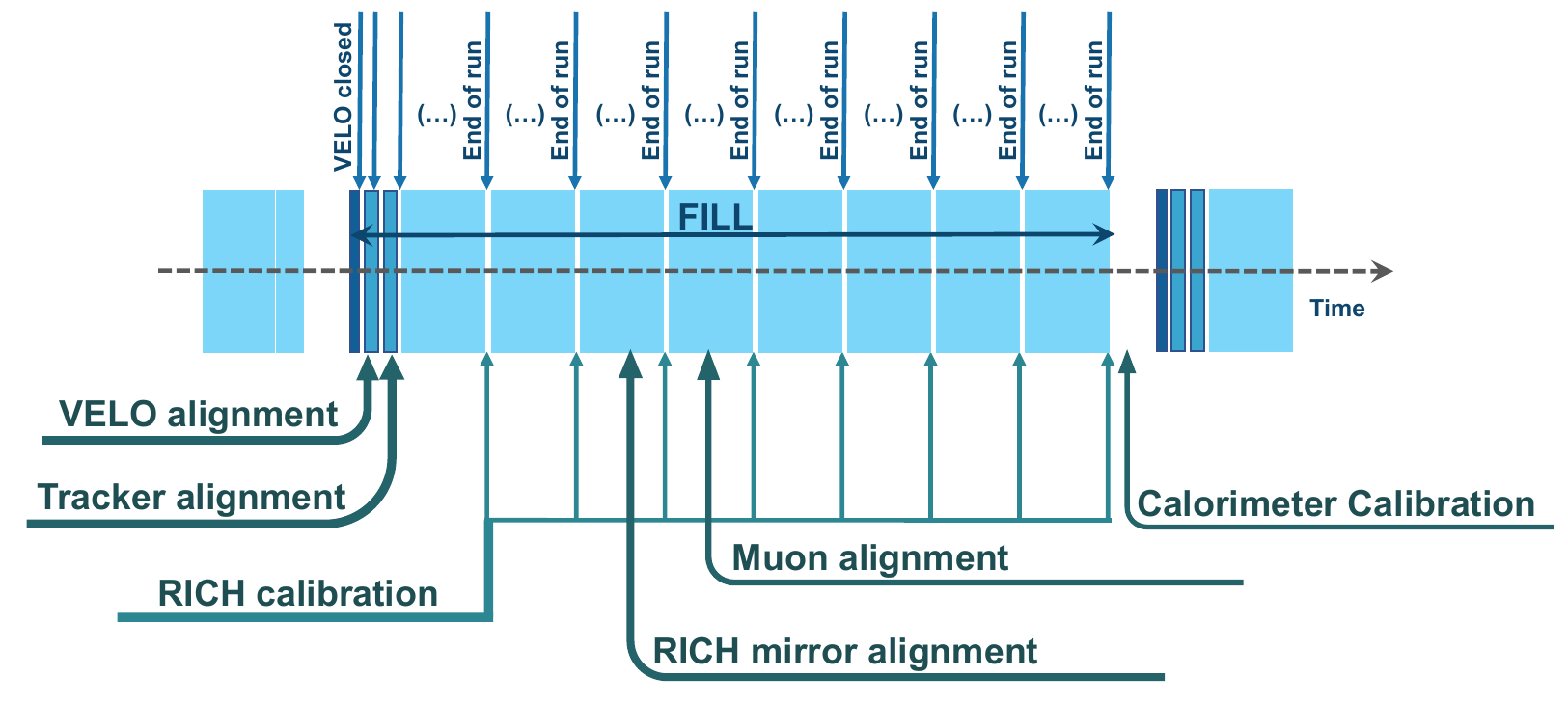}
  \caption{Schematic view of the real-time alignment and calibration
    procedure starting at the beginning of each fill.}
  \label{fig:RTA:global_alignment}
\end{figure}

Each step of the real-time alignment and calibration procedure uses
different input samples and is performed at a different
frequency. This is illustrated in
figure~\ref{fig:RTA:global_alignment} based on \runtwo
operations. While the strategy will remain the same for the upgraded
detector, the details will naturally evolve with commissioning
experience.

\subsubsection{Global alignment}

A detailed description of LHCb global alignment procedure and strategy
can be found in refs.~\cite{Hulsbergen:2008yv,Amoraal:2012qn}
and~\cite{LHCb-DP-2012-002,BORGHI2017560}. The alignment of the LHCb
detectors proceeds in a sequence, with the \Velo aligned first,
followed by the \Ut and \Scifi detectors, the \rich mirror alignment,
and finally the muon detector alignment.  Extensive studies of
alignment stability~\cite{LHCB-FIGURE-2019-015} have been performed
during \runtwo in order to make sure that all possible lessons from
the \runonetwo LHCb detector are applied to the upgrade.

For the upgrade, the optimisation of the alignment configuration has
been studied extensively using simulated samples, taking into account
the survey measurements of each subdetector and its mechanical and
thermal behaviour. It will be further tuned on the first data and its
performance followed over time to identify and correct any trends.

The \Velo alignment requires a track sample traversing all the modules
in any azimuthal and radial position. Since the residual magnetic
field in the \Velo region is negligible, a large sample of minimum
bias events can be used for this purpose. This sample is enriched with
a sample of beam collision on the residual gas upstream of the \Velo
region, to select tracks originating far from the collision region and
crossing several \Velo modules. This data sample is collected in a few
seconds at the beginning of the fill. The \Velo alignment is performed
using an additional constraint that tracks should come from their
associated \Pv~\cite{Amoraal:2012qn}. This alignment is performed
at the start of each fill to account for the opening and closing of
the \Velo detector. In \runtwo, only the alignment parameters related
to the relative displacement of the \Velo halves were varying after
the closing, and similar behaviour is expected also in the upgrade.

The alignment parameters used by the reconstruction are only updated
if the observed changes exceed a certain tolerance; in this case the
new alignment parameters become the reference values for the next
fill, and so on. During \runtwo, it was observed that updates were
required on average every few fills. When parameters are updated, they
are picked up in \Hltone via a run change, and the alignment used in
\Hltone and \Hlttwo is always kept consistent to minimise systematic
uncertainties associated with the reconstruction and selection
efficiencies. Since the \Velo alignment takes a few minutes to run,
while a fill lasts around 8-10 hours, the fraction of data which is
treated with an imperfect alignment in each fill is rather small.

For the alignment of the \Ut and \Scifi detectors, a selection of
signal tracks from the decays of well known resonances (notably \Dz
and \jpsi) are used. The constraint of the daughter kinematics to the
mother particle mass~\cite{Amoraal:2012qn} is a powerful tool to
significantly improve the alignment in the tracker system. Around
$2\times 10^4$ reconstructed resonance decays are required, which are
collected by dedicated \Hltone selection algorithms and streamed to
the alignment tasks. As the rate of fully reconstructed \Dz and \jpsi
signals will be larger than 10\khz during the upgrade, such samples
can be collected in a matter of seconds and the tracker alignment can
therefore be executed at the start of every fill immediately following
the \Velo alignment. Once again, actual updates to the alignment
parameters only occur when tolerances are exceeded.  In \runtwo this
typically occurred only for some elements of the detectors after
magnet polarity changes or interventions on the detector during
technical stops.

The \Rich mirror alignment requires a sample of tracks selected such
that their Cherenkov photons are distributed equally among the
different \richone and \richtwo mirrors; this effectively means
down-weighting tracks in higher occupancy areas of the detector and
preferentially selecting high-purity tracks in the more peripheral
areas of the \rich system. It consequently takes significantly longer
to select the alignment samples, which may take up to a few
hours. These corrections are not used in \Hltone and thanks to the
disk buffer there is enough time to evaluate \Rich alignment before
running the \Hlttwo reconstruction.

Finally the muon alignment is performed using \jpsi decays. Although
very important for the performance of the L0 hardware trigger during
\runone and \runtwo, the muon detector alignment has a small impact on
the overall tracking system performance. As the upgraded detector no
longer uses a hardware trigger, the muon detector alignment should
only need to be updated at the beginning of the data taking and in
case of an opening of the muon stations due to a hardware
intervention.

\subsubsection{Calorimeter calibration}

The performance and calibration of the calorimeter system is described
in detail in ref.~\cite{LHCb-DP-2020-001}. As the \Ecal and \Hcal are
unchanged for the LHCb upgrade apart from their readout, no
significant changes to the calibration procedure are foreseen. Due to
the removal of the hardware trigger stage, only the \Ecal requires
accurate calibrations to ensure efficient electron and photon
identification.  The gain of \Pmt{s} is first calibrated in-situ using
the LED monitoring system. Subsequently this LED system is used to
monitor changes in the gain due to ageing. After each fill, the \Pmt
high voltages are adjusted so that the LED signal of the ongoing fill
matches the LED reference value. This procedure, introduced during
\runtwo data taking, is able to control the ageing-induced \Pmt gain
variations at the level of 1--2\%. Secondly, a fine-grained
calibration is performed for each \Ecal cell based upon the
reconstructed \piz mass in that cell, achieving the ultimate possible
accuracy. This calibration is performed once per month using
$\sim 3\times 10^8$ minimum bias events.

\subsubsection{RICH gas refractive index calibration}
\label{sec:PIDCalib}

In addition to the \Rich and muon alignment and \Ecal calibration,
optimal particle identification performance also requires the
calibration of the \rich gas radiator refractive index. As the
refractive index is in particular highly sensitive to temperature and
pressure variations, these calibrations are performed on a per-run
basis using dedicated calibration samples selected by
\Hltone~\cite{LHCb-DP-2012-003}. Since the \Rich is not used in
\Hltone, calibrations can be applied for each run while the data is
stored on the disk buffer and made available to \Hlttwo.

\subsubsection{Tag-and-probe samples for offline data-driven corrections}

The calibrations described so far ensure that LHCb takes data with an
optimally performing detector at all times. However they do not
address differences between the detector performance in data and
simulation, which are of central interest to physics analyses. These
differences are studied by collecting large tag-and-probe samples
which allow for single particle reconstruction and identification
efficiencies to be measured in the same way using data and
simulation. The tag-and-probe methods are applied to pion, kaon,
proton, electron, and muon samples within the LHCb acceptance; they
also enable the study of charge asymmetries in reconstruction and
particle identification.\looseness=-1

The samples collected to measure particle identification efficiencies
are described in ref.~\cite{LHCb-DP-2018-001}. Extensive studies have
been undertaken during \Lstwo in order to further optimise the
kinematic coverage of these samples, particularly taking advantage of
the removal of the hardware trigger to enhance coverage at the edges
of the kinematic and geometric acceptance of the LHCb detector. During
\runone and \runtwo, these methods were able to measure particle
identification efficiencies with an accuracy of a few permille, and it
is expected that the same performance can be maintained.

For the study of track reconstruction efficiencies, large samples of
displaced $\jpsi\to\mumu$ decays, in which one muon is fully
reconstructed while the other is reconstructed in only part of the
tracking system are used~\cite{LHCb-DP-2013-002}. Tracking
efficiencies measured with muons have historically been used also as a
proxy for hadronic reconstruction efficiencies. However, alternative
dedicated methods to directly measure hadron track efficiencies have
been proposed and the relevant trigger selections implemented. All
these methods require a dedicated reconstruction for the probe track,
which must be carefully optimised to fit within the real-time timing
budget. Because electrons lose a great deal of their energy to
bremsstrahlung radiation, it is not possible to measure their
reconstruction efficiencies with displaced $\jpsi\to\epem$ decays
alone. Thus, samples of $\B\to X\jpsi(\to\epem)$ decays, where
$X\in \Kpm, \Kstarz, \phiz$, are reconstructed using only the \Velo
segment for the probe electron. The additional kinematic constraints
of the \B mass and the distance of flight between the production \Prpr
collision vertex and \B decay vertex give sufficient information to
suppress backgrounds and measure the per-electron reconstruction
efficiency with subpercent level systematic
uncertainties~\cite{LHCb-DP-2019-003}.

\section{Software and computing}
\label{sec:software}
As discussed in the previous section, the software trigger inherits
the two-stage model already implemented in \runone and \runtwo and
relies heavily on the mechanism of selective persistence to optimise
the amount of data stored. Moving the event reconstruction and
selection to the online domain implies that centralised offline data
processing involves only the final data preparation for physics
analysis. This process consists of data \emph{skimming} and
\emph{slimming}, if required, and data streaming according to physics
content in order to optimise the amount of data that users need to
access for their analysis.\footnote{\emph{Skimming} is the process of
  a further event rejection before the analysis, while \emph{slimming}
  is the reduction of the amount of information stored for a certain
  signal event. Skimming reduces the number of events but not their
  size on storage, while slimming reduces the event size but does not
  reject events.} This offline data processing is globally known as
\sprucing (see section~\ref{subsec:DPA}). The last step of the data
processing flow is represented by the physics analysis which may or
may not be centralised.  The \sprucing utilises the same selection
framework employed in \Hlttwo, while physics analyses proceed through
\emph{analysis productions}, a centralised way to produce artefacts
(e.g.\ ntuples) that are subsequently utilised for physics
measurements.

The demanding processing rate of the trigger applications implies a
redesign of the core software framework \gaudi, with the goal of
optimising it for speed on current computing architectures, without
compromising physics performance.  The main lines of development in
this major redesign consist of the usage of multithreading and vector
registers, the optimisation of the data model, the extensive
modernisation of code and algorithmic improvements, the adoption of
modern technologies for the detector description and conditions data.

The offline computing model follows the distributed computing paradigm
already exploited in \runone and \runtwo.  The amount of storage
needed for the recorded data is driven by the trigger output bandwidth
of 10\gbytes per second of LHC collisions.  The offline computing work
is dominated by the production of simulated events. Several avenues
are exploited in order to mitigate the resource requirements.  The
following sections describe in more detail the concepts outlined
above.

\subsection{Core software}
\label{subsec:coreSW}

The \gaudi core software
framework~\cite{Barrand:2001ny,LHCb-TDR-011} is the common
infrastructure and environment for the software applications of
LHCb. It was designed and implemented before the start of the LHC and
the LHCb experiment, and it has been in production without major
modifications ever since. Its main design principles remain still
valid, however a review and modernisation were needed to meet the
challenges posed by the LHCb upgrade and make it flexible to adapt to
forthcoming challenges.

Previously, the major limitations of \gaudi were its weak scalability
in RAM usage and its inefficient handling of CPU-blocking
operations. These limitations have been addressed by introducing
multithreading, a task-based programming model, and a concurrent data
processing model where both interevent and intraevent concurrencies
were considered.  The introduction of a multithreaded approach changes
the programming paradigm and introduces new guiding design principles,
such as thread safety and reentrance, the declaration of data
dependencies (needed for scheduling and concurrency control) and the
immutability of data (that simplifies concurrency control and allows
to focus on control- and data-flow rules).

The implementation of a concurrent, task-based framework according to
the above design principles required a deep revision of the LHCb code
base and a refactoring of many existing components. In particular, the
declaration of data dependencies between \gaudi algorithms implied the
explicit declaration of the input and output data requirements of
algorithms. Categorising algorithms accordingly, and providing common
interfaces and implementation rules for each category allowed
code-developers to integrate their algorithms in the new task-based
framework with minimal effort.\looseness=-1

\begin{figure}[t]
  \centering
  \begin{minipage}{.5\textwidth}
    \centering
    \includegraphics[width=\linewidth]{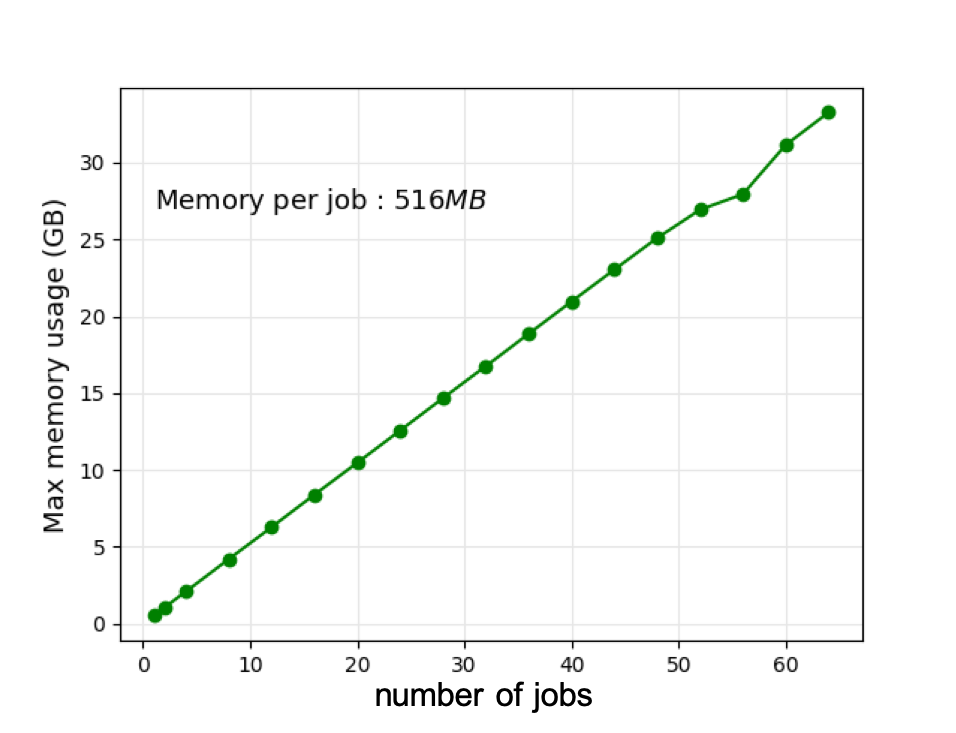}
  \end{minipage}\hfill
  \begin{minipage}{.5\textwidth}
    \centering
    \includegraphics[width=\linewidth]{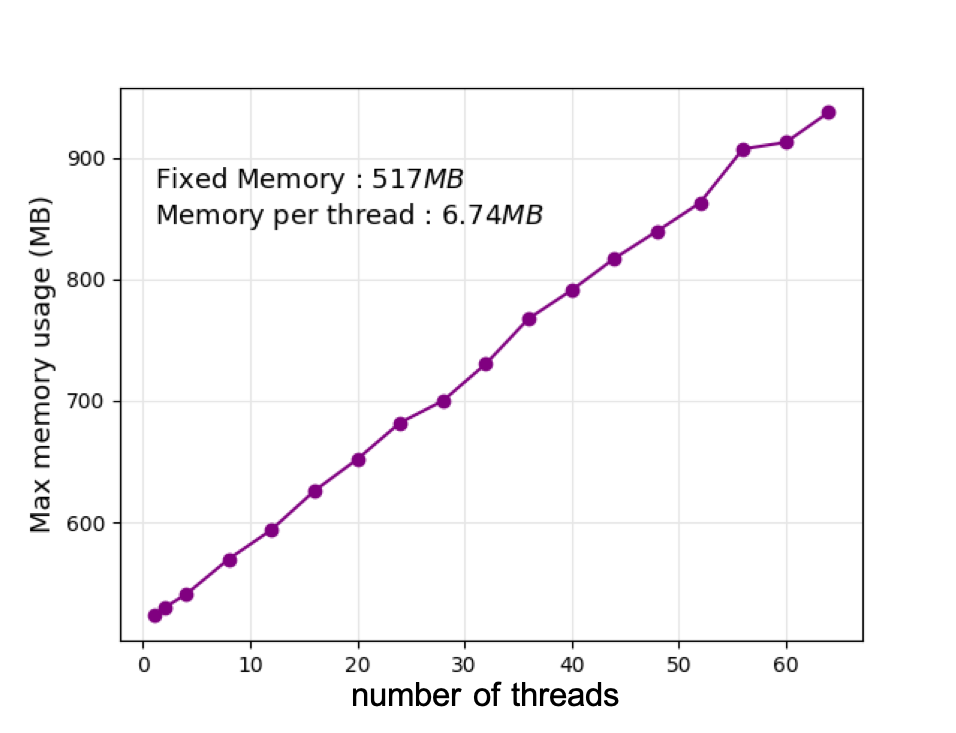}
  \end{minipage}%
  \caption{\looseness=-1Memory consumption of a prototype \Hltone
    application when run in (left) multijob and (right)~multithread
    modes.  The application was run on 3000 events per thread on a
    reference server node with 20 physical cores and a factor~2
    hyper-threading. Note that the $y$-axis scale of the right plot does
    not start at zero. }
  \label{fig:multithread}
\end{figure}

The introduction of multithreading in \gaudi had significant effects
on the memory footprint of applications. As an example,
figure~\ref{fig:multithread} shows the memory utilisation of a
prototype \Hltone application when executing many single-threaded jobs
or one multithreaded job. Tests were run on a reference server
node.\footnote{Equipped with two \Trmk{Intel} Xeon E5-2630 CPUs.}  The
memory increase is about 0.5\gbytes per job in the former case, while
it is about two orders of magnitude smaller (6.7\mbytes/thread) in the
latter case. At large number of processes or threads, the memory usage
in the multithread approach was found to be a factor of about 40
smaller than the usage in the multiprocess approach.

\begin{figure}[t]
  \centering
  \includegraphics[width=\linewidth]{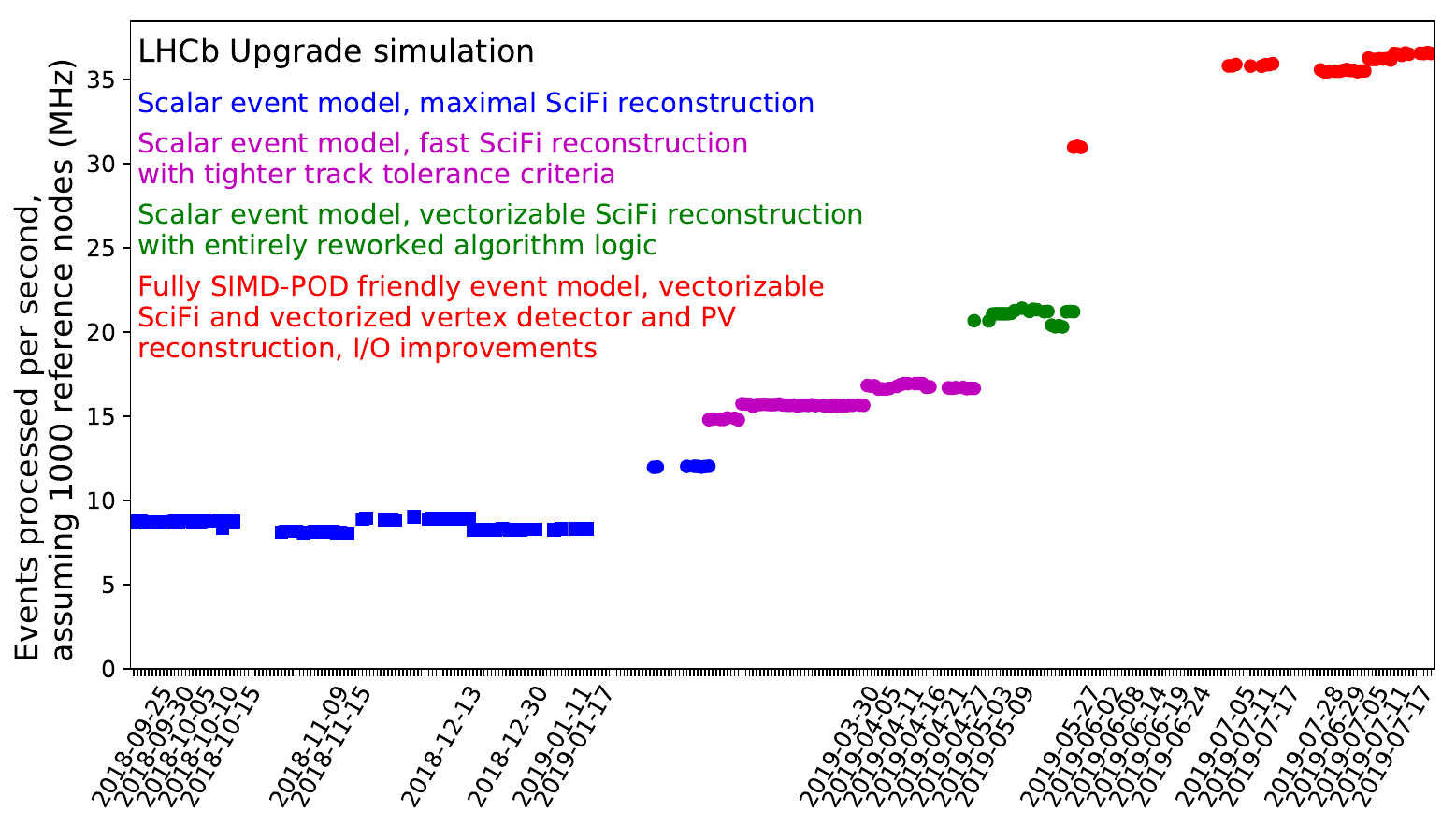}
  \caption{Evolution of the throughput of the CPU-based \Hltone
    prototype application between autumn 2018 and summer 2019, as
    measured on a reference server.}
  \label{fig:HLT1_CPU_per}
\end{figure}

In addition to the introduction of multithreading and to the
improvement of the core software framework, the performance of the
LHCb software has been optimised by following complementary approaches
and techniques. The software stack has been modernised by using the
latest \texttt{C++} versions, introducing a full code review and
suppressing code that was no longer used. The computationally
intensive parts of the code have been adapted, and data structures
have been reviewed, to exploit architectural features such as vector
registers and nonuniform memory access (NUMA) domains, and to
effectively use memory caches. In addition to these purely
computing-related aspects, algorithmic improvements have been made, in
some cases by completely changing the strategy of the most
time-consuming algorithms.  In these cases, it has been carefully
verified that the performance of the relevant data analyses was not
affected.  An example, which summarises the improvements on software
performance due to the points mentioned above, is shown in
figure~\ref{fig:HLT1_CPU_per}.  In this figure, the evolution of the
throughput of the \Hltone application between the autumn of 2018 and
summer of 2019 is shown, as measured on a reference server by using
simulated minimum bias events in nominal upgrade data-taking
conditions. The key changes in the reconstruction algorithms during
this period are colour-coded and described in the legend. The
throughput improvement due to the introduction of single-instruction
multiple-data (SIMD) instructions and data structures suitable to be
used on vector registers are clearly visible.\footnote{Although
  eventually \Hltone has been implemented on \Gpu{s}, the example is
  illustrative of the improvements introduced by the software upgrade
  concepts that are implemented in CPU-based \Hlttwo and offline
  analysis applications as well as on GPU applications.}

\subsection{Conditions database}
\label{sec:conddb}

Conditions data describe the information about the detector that is
required for data processing (e.g.\ detector configuration, alignment
or calibration constants, environmental parameters). Conditions data
may have a fine granularity, down to the level of individual detector
elements. The space of conditions data has a three-dimensional
structure defined by a geographical location, the time evolution, and
a versioning applied to the entire condition data set.  LHCb uses a
Conditions Database to keep track of the time-dependent, nonevent data
required to process the data collected by the experiment. For many
years, conditions have been stored in a set of SQLite files that were
accessed by means of \textsc{COOL}~\cite{Valassi2014}, until a
Git-based solution was developed and commissioned in 2017. The
Conditions Database is stored in a \git repository hosted on the CERN
\gitlab instance. Conditions data are accessed by a library, which
also adds the time dimension (i.e.\ the interval of validity for a
given condition) to the filename and versioning dimensions provided by
a \git repository. The \texttt{YAML} standard is used to persist
condition data.  Conditions values in the Conditions Database are used
in different contexts and have to support different use cases, in
particular simulation and real data processing. In simulation, the
response of the detector for fixed sets of conditions is exploited.
In real data processing, the time evolution of conditions is followed,
with the \git versioning of data used to track the evolution of the
response of subdetectors, or improved alignment algorithms.
Conditions data can be generated automatically in the online
environment, for example calibration and alignment parameters or
temperature and pressure values retrieved from probes, or produced
manually for example in offline workflows that require expert analysis
before inclusion in the database. In the former case, conditions are
automatically pushed to the \gitlab project and published to the
\textsc{CERNVM} file system
(\textsc{CVMFS})~\cite{bib:cvmfs,Buncic_2010,Blomer_2011}, while
the offline conditions have to be proposed in the form of \gitlab
merge requests, to undergo a review before being integrated and
published.

\subsection{Detector geometry description}
\label{sec:detgeodesc}

The description of the geometry of the \lhcb detector serves a wide
range of applications, from simulation to detector alignment,
visualisation, and computation of material budget. It must fulfil the
\lhcb needs in terms of flexibility, management, integration with the
conditions database, and speed of navigation between detector
elements.  The Detector Description for HEP (or \ddforhep)
toolkit~\cite{frank_markus_2018_1464634} is used for the \lhcb
detector description. It replaces an XML-based
framework~\cite{Ponce:2003pr}, designed and implemented within the
collaboration, but not optimised for modern computing architectures
and no longer maintained. The \ddforhep toolkit builds on the
experience gathered by LHC experiments and aims to bind the existing
tools for detector description, simulation and visualisation to
produce a consistent toolkit. Its structure is modular, with a generic
detector description model as its core, and a set of nonmandatory
extensions or plugins that can be used when needed.  The object model
and visualisation are provided by the ROOT framework. For simulations,
the \geant toolkit is used.

\subsection{Software infrastructure}
\label{subsec:SWinfra}

The \lhcb software stack depends on many software packages developed
at CERN, in other scientific institutions, or in industry.  Many of
these packages are open source. External packages must be tracked,
versioned, compiled and distributed.  \lhcb uses the \emph{LCG
  releases}~\cite{bib:LCG}, which are prepared by the EP-SFT groups
at CERN, for packages that are common among the LHC experiments. These
releases are prepared using the
\textsc{LCGCmake}~\cite{bib:LCGCmake} tool.

As many developers are involved in the development of the millions of
lines needed for the experiment framework, a strong version control
system and good practices are crucial to ensure the quality of the
software.  The \lhcb code base is split into several projects,
versioned and managed independently, each having a distinct goal.
Each of these projects is managed using the \git~\cite{bib:git}
version control system. This allows keeping the full history of all
code changes.  Code reviews are prevalent in the software industry, as
a way to improve code quality and harmonise development practices
across organisations. \lhcb \git projects are hosted on the CERN
\gitlab server, which also provides features allowing better
collaboration between the members of the teams. New features to the
code are peer-reviewed before being merged into the main code
base. The projects are organised around the \jira~\cite{bib:jira}
task management and the \gitlab issue management systems, as deployed
by the CERN IT department. This allows the developers within the LHCb
collaboration to follow the evolution of the projects and collaborate
in a constructive manner.

A Software Configuration Database has been built to track all projects
and their dependencies~\cite{1742-6596-898-10-102010} using the
\textsc{Neo4j} graph database~\cite{bib:neo}.  This information is
crucial to the management of the software in the short term, but is
also necessary to identify the software needed to continue analysing
the LHCb experiment data in the long term.

In order to ensure the quality of the software produced by the
developers, automatic builds of the software are performed, as
described in ref.~\cite{1742-6596-664-6-062008}. This
infrastructure relies on the industry-tested
\jenkins~\cite{bib:jenkins} automation server as deployed in the
CERN IT \textsc{OpenShift} service.  The build nodes are provided by
the CERN IT department in the form of
\textsc{OpenStack}~\cite{bib:openstack} virtual machines.  Results
are gathered and displayed on a custom made web
application~\cite{Clemencic:2134602}.

Unit tests are run straight after the builds and the results are
published to the LHCb Dashboard.  Integration tests requiring more
resources are run using \lhcbpr, the LHCb performance and regression
testing service~\cite{1742-6596-898-7-072037}. \lhcbpr is
responsible for systematically running regression tests, collecting
and comparing results of these tests so that any changes between
different setups can be easily observed.  The framework is based on a
microservice architecture where a project is broken into loosely
coupled modules communicating with each other through APIs. The test
service requests from \lhcbpr information on how to run tests, then
runs them and finally saves the results back to \lhcbpr. Users have
the ability to retrieve and analyse these test results.

The LHCb software is distributed using the CERN software deployment
service \textsc{CVMFS}.  Private installations of the software stack,
as well as full management of the installed packages, are also
possible.  In this case, the applications are packaged in the
\textsc{RPM} package manager~\cite{bib:rpm} format, which allows
specifying dependencies for applications relying on external packages
(e.g.\ an installation of the analysis package relies on more than a
hundred different packages).

The LHCb software stack is supported on 64-bit Intel
architecture. Other architectures (ARM, IBM Power, \Gpu{s}) are
handled by the LHCb build and release infrastructure, providing that
the operating system used by LHCb can be installed, that packages
using \textsc{LCGCMake} are ported and released, and that nodes for
that architecture can be managed by the experiment's instance of
\jenkins.

The \lhcb software is preserved in the long term by the same tools
used in the LHCb development process, more specifically the version
control system as well as the \textsc{CVMFS} repositories, virtual
machines and containers. Databases containing information about the
software artefacts, their dependencies and their use in physics
analysis ensure that applications can be rebuilt and rerun if
necessary.\looseness=-1

Several collaborative tools are used to coordinate the efforts and
increase the proficiency of software developers. Rather than
developing ad hoc solutions, the general strategy is to monitor the
available tools and trends and adapt them to the specific use cases of
LHCb.

\subsection{Simulation}
\label{subsec:simulation}

\begin{figure}[t]
  \centering
  \includegraphics[width=\textwidth]{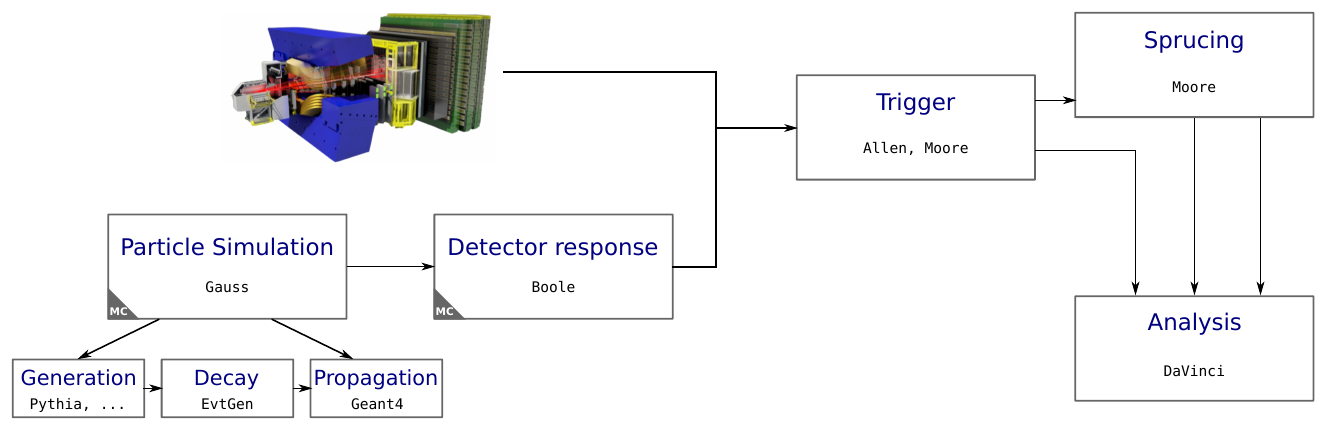}
  \caption{\label{fig:sim:dataflow}Schematic representation of the
    LHCb upgrade data flowand the related LHCb application, with an
    emphasis on simulation.}
\end{figure}

Monte Carlo samples have been essential for the detector design and
preparation of the data processing and will be instrumental to the
physics analysis of the \lhcb upgrade.  In order to define, tune and
validate the reconstruction and selection algorithms, Monte~Carlo
samples are processed through a data flow identical to that of real
data, as shown in figure~\ref{fig:sim:dataflow}. Simulation is
followed by the online HLT and offline data processing described in
sections~\ref{sec:trigger} and~\ref{subsec:DPA} exactly as for real
data.

The software to generate simulated events in \lhcb is conveniently
encapsulated in two separate applications.  The first is the \gauss
package, responsible for the event generation and simulation of
particle interactions in the detector volumes which result in energy
deposits or \emph{hits} in the subdetector sensitive elements.  The
second is the \boole package, in charge of modelling the detector and
readout electronics response by converting the hits into specific
subdetector signals.  This broad modularity is essential to profit
from the experience continuously gained during the commissioning and
operation of the subdetectors.

The generator-level information is transparently propagated through
all the data processing steps to allow detector or physics performance
studies comparing reconstructed and \emph{true} (i.e.\ as produced by
the physics generator) quantities.\footnote{Physics generators are
  dedicated software packages that statistically generate particle
  four-momenta and decays based on first-principle physics quantities
  such as cross sections and branching ratios, using Monte Carlo
  techniques.}

Like all other \lhcb applications, \gauss and \boole are built on the
\gaudi core software framework and were heavily modernised with the
adoption of parallelisation and multithreading and by implementing a
new geometry description.

In addition, a new \gaudi-based experiment-independent simulation
framework, named \gaussino~\cite{Muller:Gaussino,Siddi:2704513},
was introduced to decouple widely used simulation software packages
like \geant or physics generators from LHCb-specific developments.

Simulation is the main consumer of LHCb computing resources as
demonstrated by the fact that about 80\% of the CPU resources made
available to the collaboration during \runtwo were employed to
generate, simulate and process Monte~Carlo events.  Therefore, to be
ready to cope with the rapidly growing demand for simulated events due
to the increase in LHCb integrated luminosity, code optimisation and
modernisation has been complemented with the adoption of approximated
simulation methods, referred to as \emph{fast} and \emph{ultra-fast
  simulations} (see also section~\ref{subsec:compresources}).  On a
similar line, developments aiming at reducing the storage resources
for simulated events were also pursued, as discussed in
section~\ref{subsubsec:sim:fast}

The software and physics performance of the simulation suite is
monitored exploiting the software infrastructure described in
section~\ref{subsec:SWinfra}.

\begin{figure}[h]
  \centering
  \includegraphics[width=0.9\textwidth]{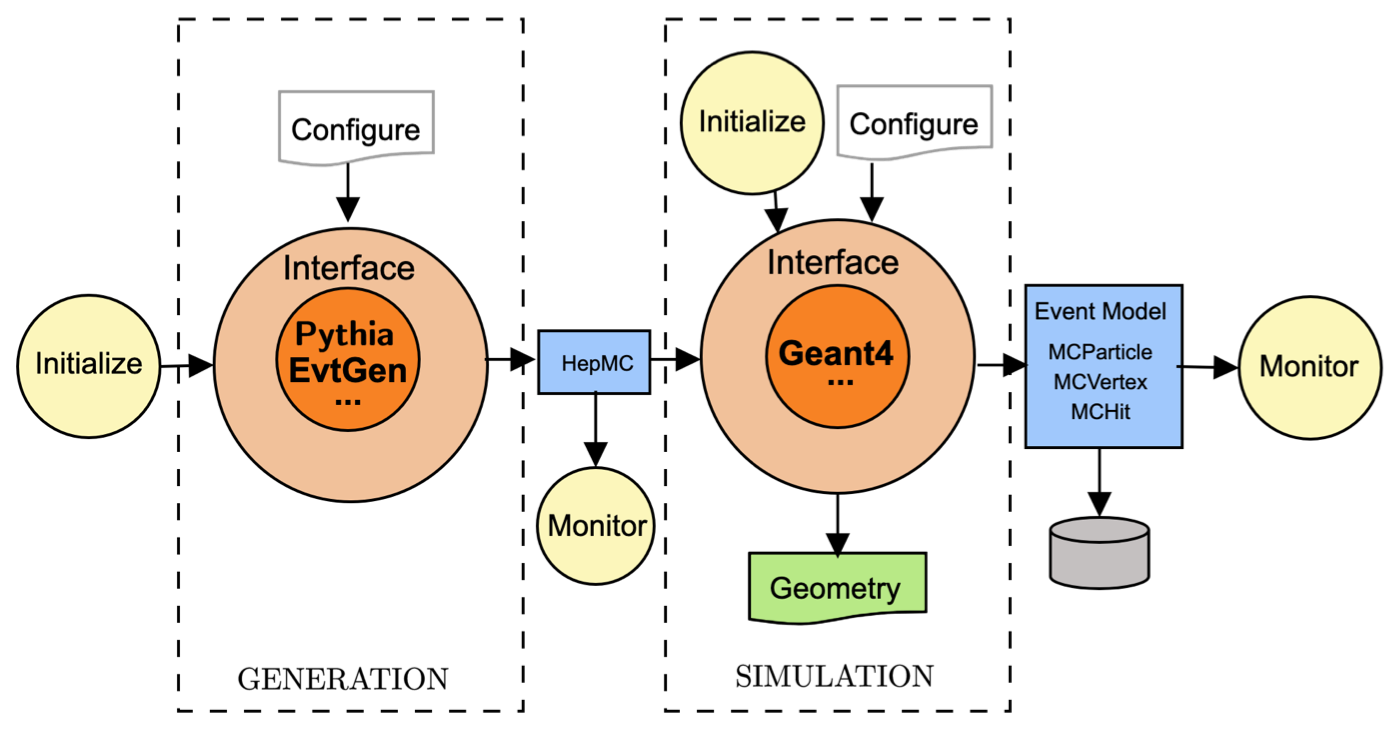}
  \caption{Schematic structure of the \gauss application. Reproduced from~\cite{LHCb-PROC-2010-056}. \textcopyright\ 2022 IOP Publishing
Ltd. CC BY 3.0.}
  \label{fig:sim:gauss}
\end{figure}

\subsubsection{Gauss}
\label{subsubsec:sim:gauss}

The \gauss application facilitates modelling of the physics processes
occurring in the \Prpr collisions and takes care of the transport of
the resulting particles through the experimental apparatus, including
their interactions with the magnetic field and the detector and
infrastructure material.

\gauss has been extensively used by LHCb since the early development
phases of the experiment and has undergone various changes to support
evolving needs~\cite{Clemencic_2011}.

Figure~\ref{fig:sim:gauss} presents an overview of the structure of
the \gauss application.  The simulation process consists of two
subsequent phases.  In the \emph{generation} phase the physics process
is obtained by dedicated generators, such as
\pythiaeight~\cite{Sjostrand:2007gs,LHCb-PROC-2010-056} and
\evtgen~\cite{Lange:2001uf}.  In the \emph{simulation} phase the
generated particles are transported through the experimental
apparatus, relying on the \geant toolkit, on resampling techniques or
on custom parametrisations, finally providing particles, vertices and
energy deposits (\emph{hits}) in the LHCb event data format.  The
\texttt{HepMC}~\cite{Dobbs:684090} format is used to transfer
information between different tasks within the generation phase and to
pass the generated particles to the simulation step.

The design of \gauss is based on \gaudi.  Selection of components and
steering of processing phases is achieved through a unique and
coherent configuration system which represents a fundamental building
block of the \gauss application.  The modular architecture of \gauss
makes it an excellent candidate from which to model a \gaudi-based
core simulation framework for future experiments.

\begin{figure}[t]
  \centering
  \includegraphics[width=0.40\textwidth]{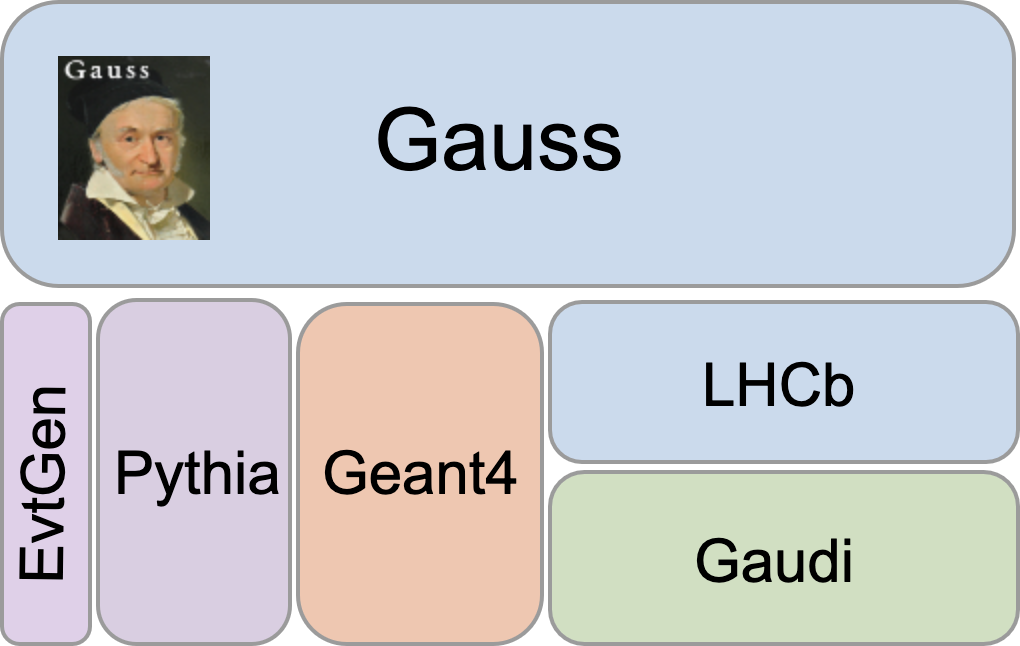}
  \hspace{1cm}
  \includegraphics[width=0.40\textwidth]{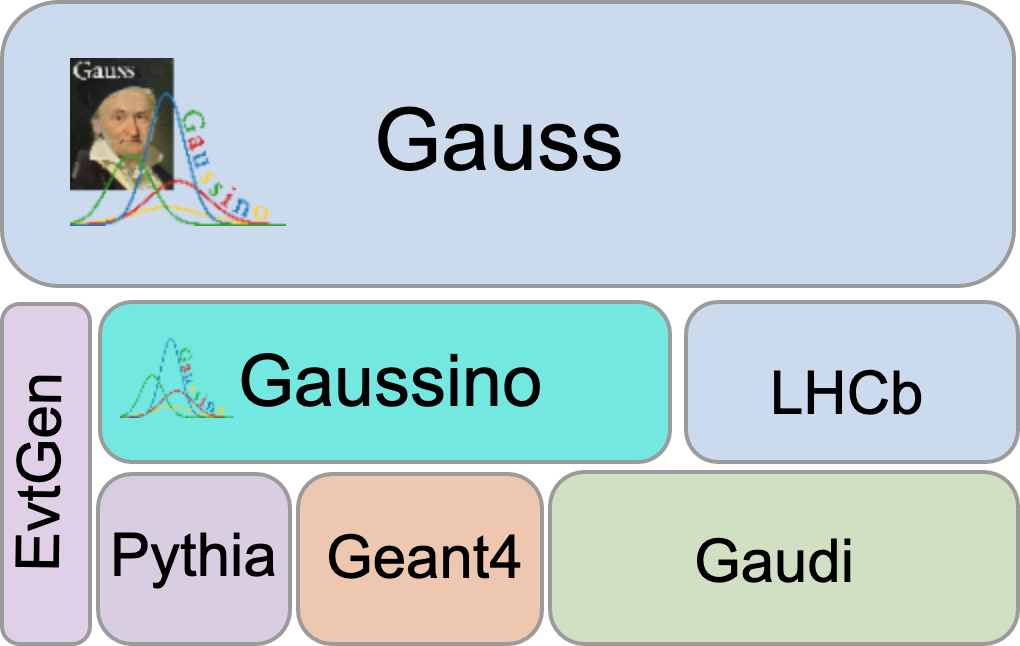}
  \caption{\label{fig:sim:gauss:GonG}Graphical representation of the
    dependencies in the simulation software stack in (left) \runonetwo
    and (right) in the upgrade, where the experiment agnostic package
    \gaussino decouples \gauss from \pythia and \geant. Reproduced with permission from~\cite{Mazurek:2022MX}.}
\end{figure}

The experiment-agnostic simulation core features were packaged in
\gaussino, providing interfaces to widely used packages such as
\pythia or \geant. \gauss is built on \gaussino and adds the
LHCb-specific simulation functionalities as depicted in
figure~\ref{fig:sim:gauss:GonG}.

To be compatible with the multithreading model adopted for the LHCb
upgrade, \gauss must be compatible with a multithreaded scheduling.
This required the migration from the in-house geometry description
software package used for \runone and \runtwo to the \ddforhep
package~\cite{frank_markus_2018_1464634}.  Multithreading is handled
by \gaussino which is also equipped with a general interface to steer
the transfer of geometry information to \geant.

The excellent modularity introduced through \gaussino and the upgraded
\gauss allows also to easily simulate \runonetwo events thanks to its
support for the legacy geometry and data persistence.

\subsubsection{The new Gaussino experiment-agnostic core simulation framework}
\label{subsubsec:sim:gaussino}

The migration of the LHCb simulation to a multithreaded computing
model implied the development of a general thread-safe interface with
external simulation packages, such as \geant or physics generators, a
task which was one of the main drivers of the development of
\gaussino, explicitly designed for a much wider use than for LHCb
only.

\gauss and \gaussino have an identical structure, shown in
figure~\ref{fig:sim:gauss}.  The same architectural design was also
retained, but using more modern \gaudi features in the implementation.

The \gaussino generator phase is essentially derived from that of
\gauss~\cite{LHCb-PROC-2010-056}, which was repackaged extracting the
parts not specific to LHCb.  \gaussino provides a thread-locking
infrastructure to encapsulate and protect the execution of external
tools not designed for multithreading, such as those including
\textsc{FORTRAN} dependencies.  Thread-safe libraries, such as
\pythiaeight, can also be executed in multiple independent
thread-local instances achieving faster execution at the expenses of a
larger memory footprint.

The simulation phase, closely tied to \geant, required a full redesign
of the interfaces to various components to make them more
experiment-independent and compatible with \geant multithreading.  The
same design choice taken by ATLAS was adopted to make the different
\gaudi and \geant concurrent models work together.

\begin{figure}[t]
  \centering
  \includegraphics[width=0.48\textwidth]{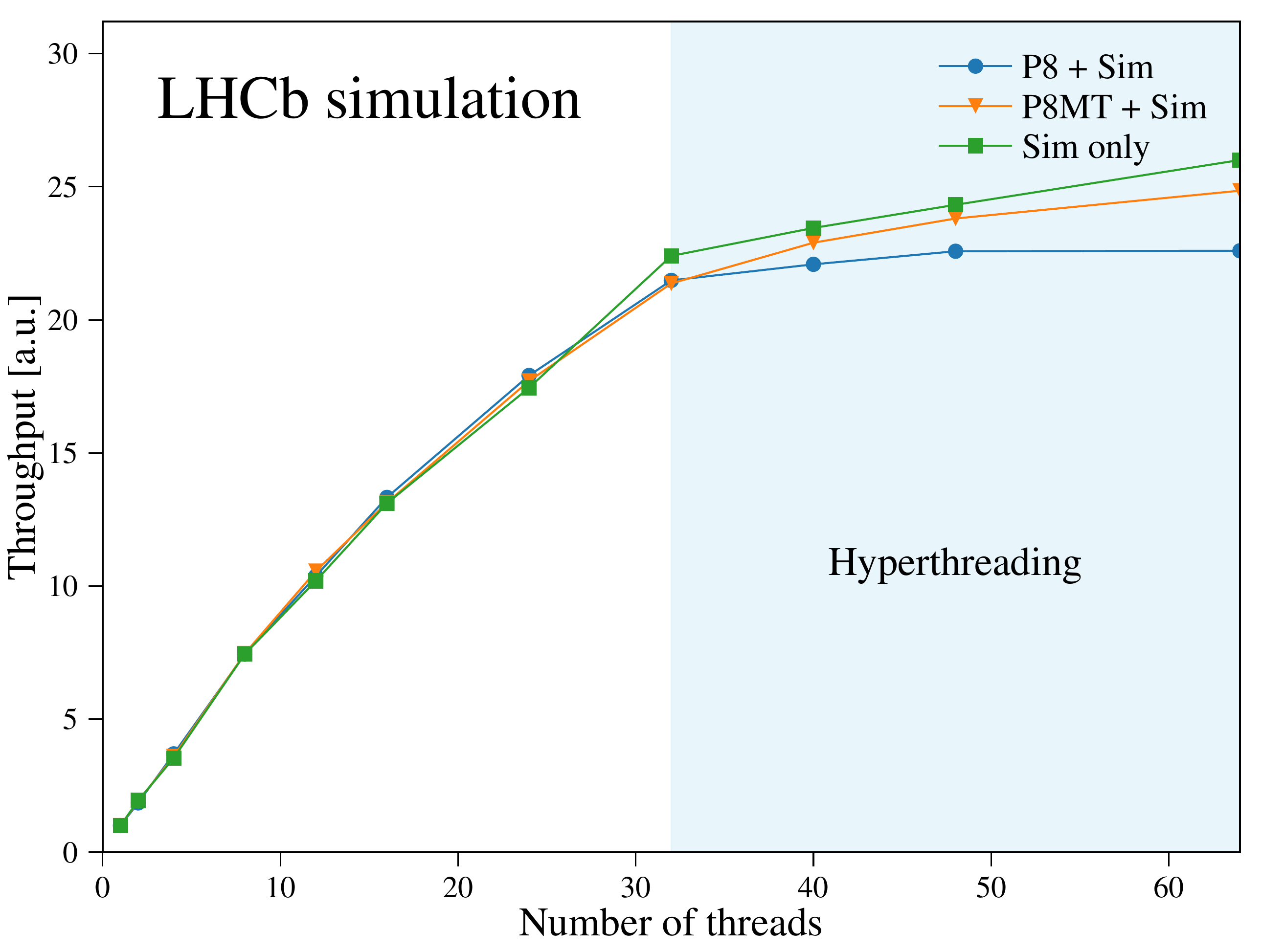}
  \includegraphics[width=0.48\textwidth]{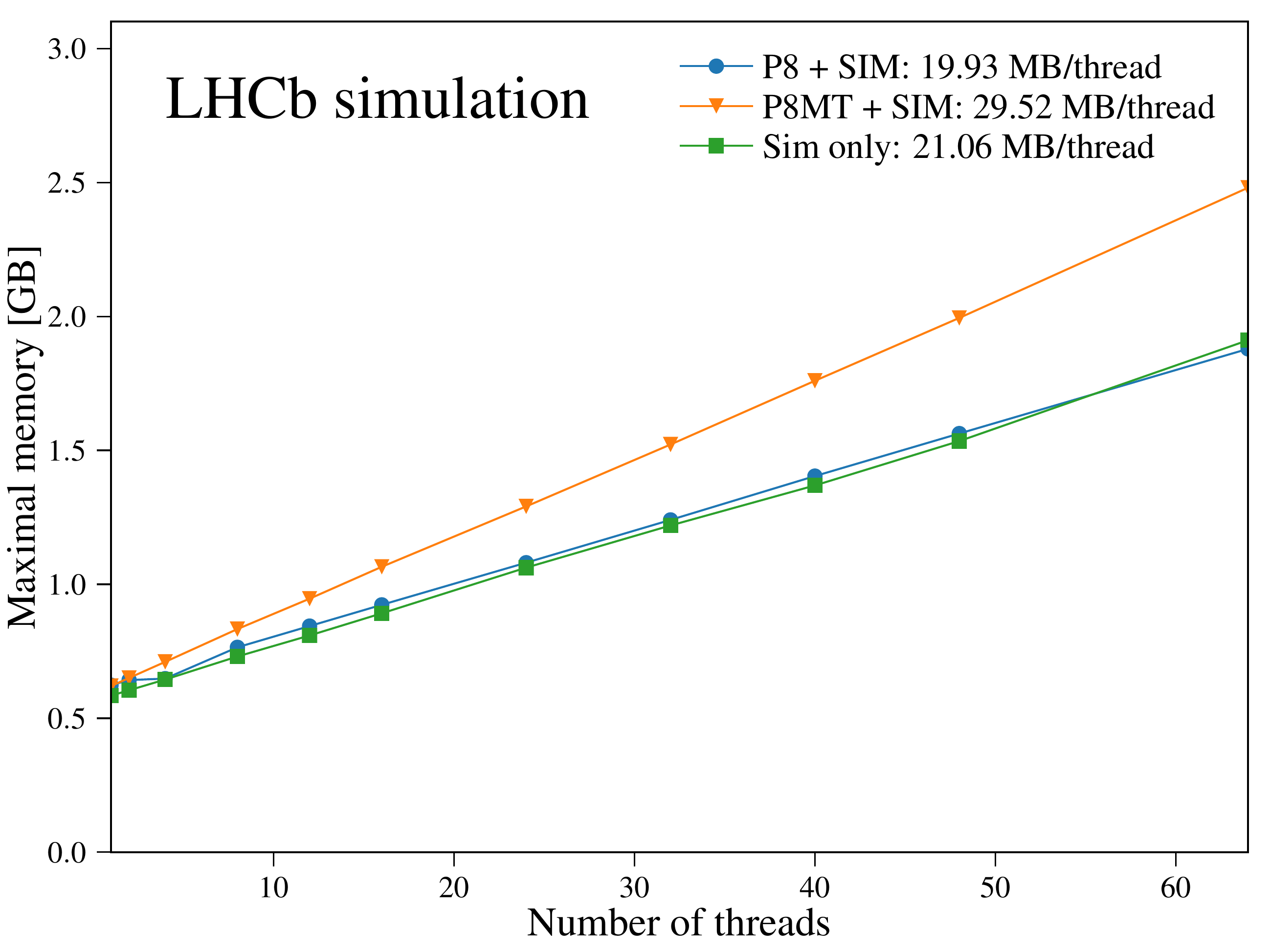}
  \caption{\label{fig:sim:gaussino:sim} Throughput and memory scaling
    for the generation of \Prpr collisions producing at least a \Dz
    meson with beam conditions as found in the 2016 data-taking period
    in LHCb.  Shown are the curves for a shared (P8) and a
    thread-local (P8MT) interface to \pythiaeight, followed by the
    \geant-based simulation.  The contribution of the simulation
    phase, as obtained by reading the generated events from file
    (``Sim only''), is also shown. Reproduced with permission from~\cite{LHCb-FIGURE-2019-012}.}
\end{figure}

As an example of the performance of this approach, the evolution of
the memory occupation and event throughput as a function of the number
of threads involved in the simulation of \Prpr collision producing at
least a \Dz meson~\cite{LHCb-FIGURE-2019-012} is reported in
figure~\ref{fig:sim:gaussino:sim}.  While the throughput increases
almost proportionally with the number of threads, a large fraction of
memory is shared, enabling the adoption of high-performance computing
resources with several tens of cores to generate Monte Carlo samples.

\gaussino provides an additional interface to steer the interaction
with \geant giving the possibility of replacing its detailed
description of physics processes with fast simulation models for a
specific detector.  This interface was developed with the aim to
minimise the work needed to implement fast simulation models for \lhcb
and facilitate their integration in \gauss. A facility to ease the
production of training data sets for fast simulation models is also
available in \gaussino.

\subsubsection{Filtered, fast and ultra-fast simulations}
\label{subsubsec:sim:fast}

While crucial to optimally adapt the LHCb simulation to modern
computing architectures, the software modernisation effort described
above is not sufficient alone to match the amount of necessary
simulated events with the computing resources pledged to LHCb.
Therefore, considerable effort has been devoted to the development of
simulation techniques and technologies with a lesser impact on the
computing and storage resources.  Some of the developed strategies
have already been deployed and widely adopted during \runtwo, enabling
a 70\% reduction of the average computing power and a 60\% reduction
of the average storage occupation per simulated
event~\cite{LHCb-PUB-2022-011}.

In particular, \emph{filtered} Monte~Carlo simulation and a new output
format were introduced during \runtwo.  In a filtered simulation,
produced events are rejected by analysis-specific criteria before
storing them. By the end of \runtwo, only 13.9\% of the simulated
events was preserved, drastically reducing the impact of simulation on
the storage resources~\cite{Corti:2019805}. Filtered Monte~Carlo
productions are used also for the upgrade, where the analysis-specific
criteria are defined by the \Hlt or by the \sprucing selections
(section~\ref{subsec:DPA}).  The new output format, with selectively
persisted event information, complemented by the corresponding Monte
Carlo generator information, well matches the \turbo mechanism (see
section~\ref{ssec:hlt2:selections}) and is expected to be used for the
majority of simulated samples in most cases in combination with Monte
Carlo filtering.

Once a careful software optimisation is achieved, to further reduce
the computing resources needed for event simulation it is necessary to
compromise on the accuracy of simulated event features by replacing
the detailed simulation based on \geant physics models with fast
simulation techniques based on either resampling methods or
parametrisations of the detector response.  The choice of the features
on which a degradation of the accuracy is acceptable, the level of
reliability expected from a simulated sample, as well as the
statistical precision needed or, correspondingly, the number of events
to simulate, depend on the specific needs of the data analysis and no
unique solution can make optimal usage of the computing resources.  A
palette of simulation options spanning from \geant-based
\emph{detailed simulation} to fully parametric options has thus been
developed to efficiently cover as many use cases as
possible~\cite{Whitehead:2702933}.

The most widely adopted fast-simulation option (see for example
refs.~\cite{LHCb-PAPER-2022-002,LHCb-PAPER-2022-001,LHCb-PAPER-2022-008,LHCb-PAPER-2021-017})
is implemented in the \redecay package~\cite{Muller:2644956}.  Once
a signal process is identified, for example the production of a heavy
hadron, multiple random instances of its decay products are propagated
through the detector while the detector response to the rest of the
event is computed only once and superposed to the decay products of
the signal particle. An overall decrease of the computing power per
simulated event by a factor of 10 to 20 is achieved using the \redecay
technique, introducing an effect of event-to-event correlation which
is only relevant for a minor fraction of analyses.

To study detector-induced effects that can be considered independent
of the overall event particle multiplicity, it is often useful to
simulate large samples of a specific process, e.g.\ the decay of a
heavy-flavoured hadron.  In these cases, the full simulation of the
\Prpr collision can be avoided by generating the heavy hadrons
according to parametrised spectra of their kinematic variables. These
simulations, referred to as \emph{particle guns}, are widely adopted
for example to study background contributions and trigger effects in
charm physics, or to analyse specific detector effects such as charge
detection asymmetry~\cite{LHCb-PAPER-2021-016}.  Using particle guns,
an overall increase in speed by a factor of 50 with respect to a
detailed simulation is achieved~\cite{LHCb-PUB-2021-003}.

A third fast-simulation option already deployed in production relies
on a simplified geometry where entire subdetectors are removed,
possibly complementing the simulation by sampling the nonsimulated
features from parametrised distributions at analysis level.  For
example, \emph{tracker-only} simulations can save up to 90\% of the
computing power per event by skipping the simulation of the optical
photons in the \rich detectors and of the electromagnetic and hadronic
showers in the calorimeters.  The \Pid information is then added
\emph{a posteriori} using data-driven techniques originally developed
to calibrate the simulation on unbiased data
sets~\cite{LHCb-DP-2018-001}.

However, in some cases, partial detector simulations may require ad
hoc solutions, for example in hardware trigger emulation.  A more
general solution, with similar CPU speedup, is obtained by replacing
the simulation of time consuming processes with specific
parametrisations obtained from dedicated simulated samples and
deployed as part of the simulation software stack. In particular,
techniques based on resampling from hit
libraries~\cite{Rama:2728527} and querying generative
models~\cite{Chekalina:2759239} have been developed to speed up the
simulation of the energy deposits in the calorimeter, taking advantage
of novel features of the \geant package designed to replace part of
the propagation with custom statistical models and available via
\gaussino as described earlier in this section.

\begin{figure}[t]
  \centering
  \includegraphics[width=0.8\textwidth]{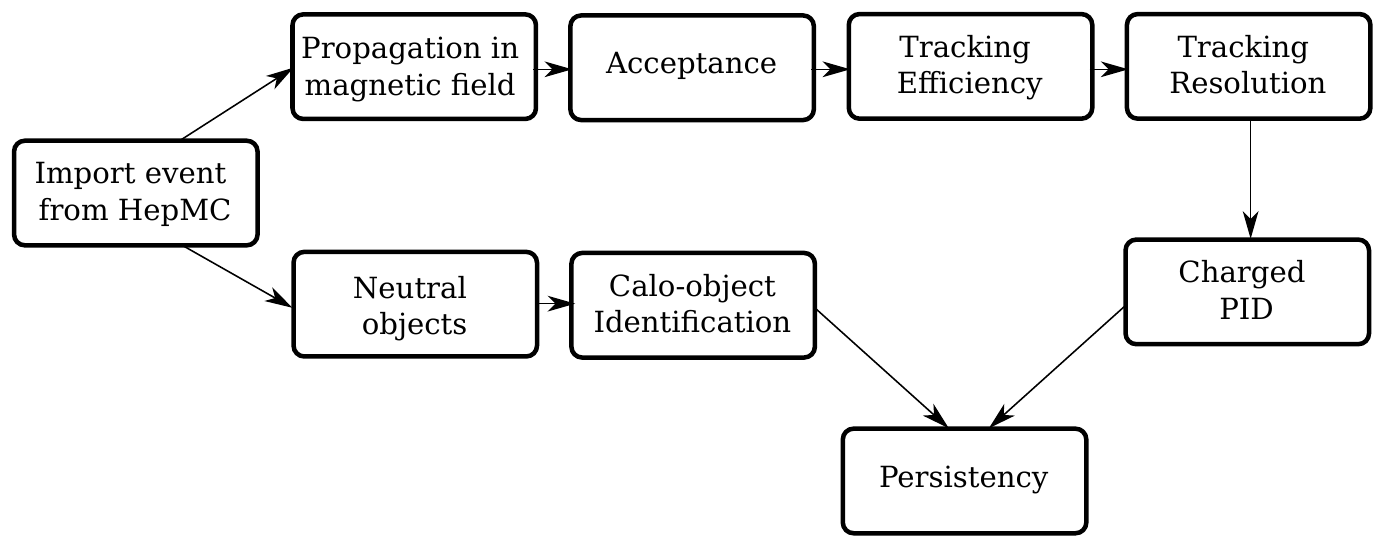}
  \caption{\label{fig:sim:lamarr:flow}A flow-chart representing the
    \lamarr project as a pipeline of parametrisations.}
\end{figure}

Machine learning techniques are also widely adopted to build
statistical models of the detector
response~\cite{Anderlini:2789455} and make it feasible to define
parametrisations of extreme complexity to reproduce the whole detector
simulation and reconstruction procedure by combining simple random
generators and deterministic formulas.
Figure~\ref{fig:sim:lamarr:flow} depicts the extension of \gauss
devoted to such an ultra-fast simulation option, named \lamarr,
designed as a pipeline of parametrisations, taking in input
generator-level quantities and producing reconstructed, analysis-level
variables on output.  The \lamarr framework aims at easing the
deployment of machine learning models in \gauss, providing a common
infrastructure to data preparation and persistence
configuration. Models are developed in \root
\textsc{TMVA}~\cite{Hocker:2007ht,TMVA4},
\textsc{scikit-learn}~\cite{Scikit-learn-paper} or
\textsc{Keras}~\cite{chollet2015keras} and deployed as compiled shared
objects, possibly relying on the \textsc{scikinC}
package~\cite{Anderlini:2022ltm}.

Original machine learning models were developed to enable training on
real data, introducing statistical background subtractions in the
training procedure~\cite{Maevskiy:2678418}, and adopted for the
parametric simulation of the \Pid features obtained from the \rich,
calorimeters and muon detectors, and their
combinations~\cite{Barbetti:2826210}.

\subsubsection{Digitisation}

Except for \lamarr and its ultra-fast simulation approach, the output
of simulation applications must be processed by \boole to simulate the
detector response (\emph{digitisation}).

\boole converts the simulated hits into the same format as obtained
from the \Daq of each subdetector, so that they can be processed by
the common reconstruction and selection applications.  Firstly, \boole
simulates the subdetector technology and electronics response,
including imperfections such as noise, cross-talk and dead channels.
Then, the output of this stage is packed in \emph{banks} which are
passed to the event builder emulator to produce a raw data buffer
identical to the output of the LHCb \Daq chain.  Information of each
hit digitisation history is preserved in the \boole output to allow
detector and physics performance studies.  The \boole application has
been completely overhauled to match the upgraded \Daq and detector
technologies.

\subsubsection{Monte Carlo production}

A correct and efficient production of Monte Carlo samples requires a
well developed strategy and accurate prioritisation to match the
physics analysis needs~\cite{Corti:2019805}.  To maximise the CPU
usage efficiency, the Monte Carlo production system has been upgraded
to enable continuous integration tests in order to verify the
consistency of simulation requests before submitting them to the
distributed computing system.  Simulation quality assurance tools were
also improved and include the monitoring of the event simulation
process by checking their physics output.

\subsection{Offline data processing and analysis}
\label{subsec:DPA}

\begin{figure}[t]
  \centering
  \includegraphics[width=\textwidth]{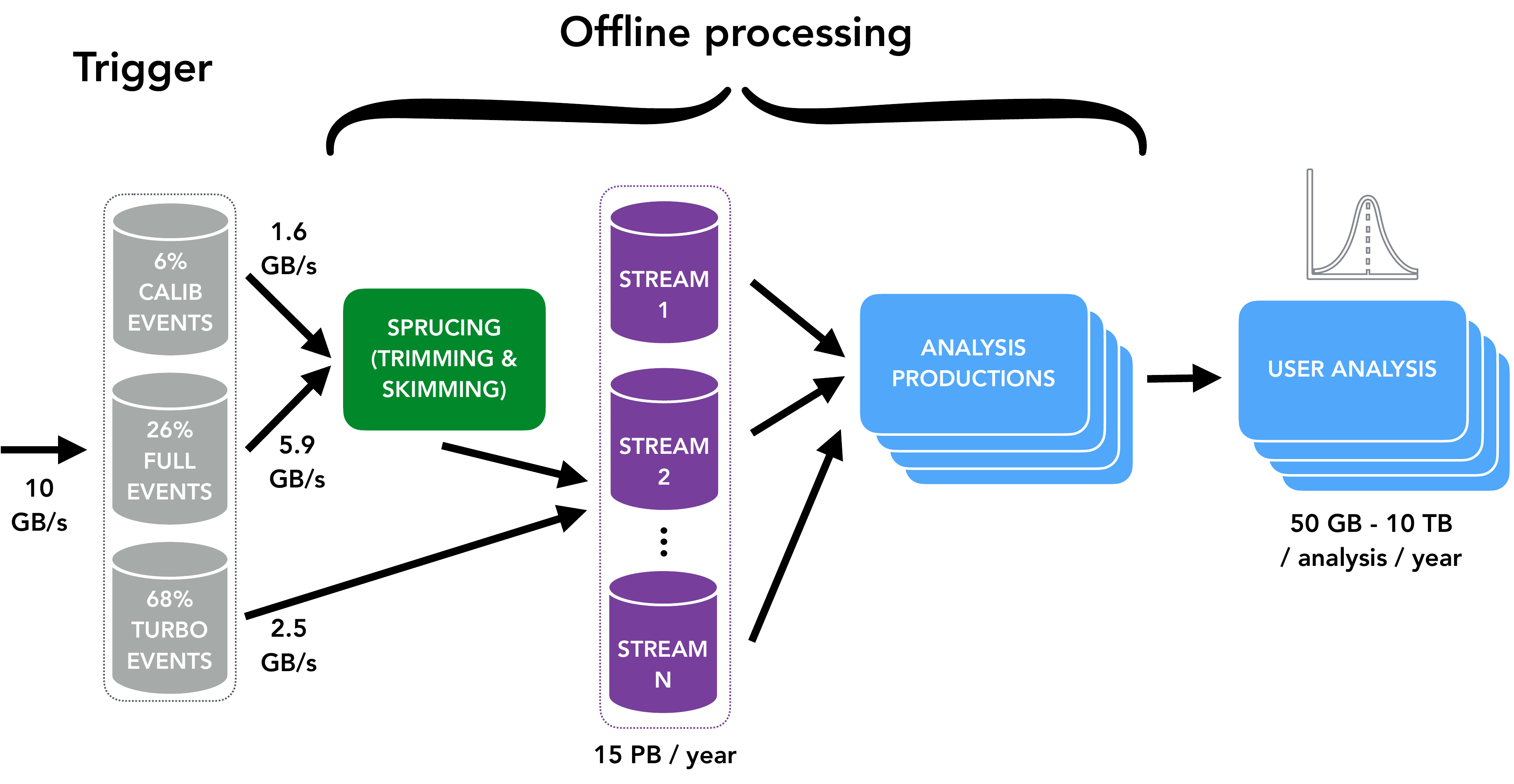}
  \caption{Offline data flow. Reproduced with permission from~\cite{LHCb-FIGURE-2020-016}.}
  \label{fig:DPA:dataflow}
\end{figure}

The offline data processing flow can be seen in
figure~\ref{fig:DPA:dataflow} and is detailed in this subsection.

\subsubsection{Sprucing}

The offline data processing is globally known as \sprucing.  The
\sprucing code base uses the same application as \Hlttwo, namely the
\moore application. Furthermore, the same algorithms and tools are
shared between \hlttwo, \sprucing and the offline analysis software
project \davinci~\cite{davinci} (see section~\ref{sec:offlineana}),
namely the functor based selection and combinatorial algorithms.

The \sprucing performs three main functions: it applies further
selections to data saved into the \FULL stream (data skimming); it
enables the tuning of the amount of event information to be persisted
in the final output files through the \turbo mechanism (data
slimming); it streams the data into a number of different physics
stream files and create file summary records (FSR) that store metadata
about the file content in the output \root files (see
section~\ref{subsec:dataprocessing_flow}. Data skimming and slimming
can also be combined as needed.  ` A typical example of data skimming
is represented by events selected by inclusive topological \hlttwo
trigger selections~\cite{LHCb-DP-2019-001}, which were instrumental in
\runone and \runtwo. These \hlttwo selections persist the full event
information into the \FULL stream. These events are further processed
by exclusive \sprucing algorithms that perform again the particle
reconstruction and apply further selections to reduce the data volume
to be saved to disk (data skimming).  The same procedure can be
applied to specific samples, such as data sets used to derive data
driven calibrations in particle identification algorithms

Data in the \FULL stream can also be used to tune the amount of event
information that the selective persistence of \Hlt selections saves on
output, in view of a future implementation in \Hlttwo, reducing the
size of the events saved on disk for further physics analysis (data
slimming).

Normally, data which are saved in the \turbo stream by \Hlttwo are not
further processed except for possible conversion to the final offline
data format, creation of FSRs and streaming (\emph{pass-through}
\sprucing). These steps correspond to the tasks performed in \runtwo
by the \tesla application~\cite{LHCb-DP-2016-001}.

Irrespective of the intermediate processing (skimming, slimming or
pass-through), data from both \FULL and \turbo streams are eventually
distributed into physics \emph{streams} optimised to allow analysts to
access reduced data sets categorised by physics topic. This optimises
data processing time for a specific analysis as well as the use of
computing resources such as disk access.

Compared to \Hlttwo, offline \sprucing benefits from less strict
limits on CPU time for selection algorithms. While no recalibration
and no rerunning of the pattern recognition is planned, the \sprucing
selections will have more time for example for the analysis of complex
cascade or many-particle final state decays where the number of track
combinations to be tested increases very rapidly. In addition, offline
\sprucing allows detailed analysis of physics topics where the full
event information has to be taken into account such as electroweak
physics or jet reconstruction.

\textls[-10]{Both exclusive and pass-through \sprucing will run concurrently with
data taking while global re\sprucing campaigns will take place in
end-of-year shutdowns whereby data will be staged~from~tape.}\looseness=-1

\subsubsection{Distributed analysis productions}

The output of the \turbo stream and of the \sprucing is split into
multiple streams which are directly accessible to analysts.  In the
\runonetwo (known as \emph{legacy}) data model these data sets would
be processed by submitting user jobs to LHCb\dirac that filter one of
these streams to select physics-quantities of candidates for a
specific analysis, typically resulting in a reduction in data volume
of $\mathcal{O}\left(10^{3}\right)$ (see section~\ref{subsec:DIRAC}
for all details on distributed analysis).  While this model works well
for smaller data sets, scaling has been problematic with legacy
analyses requiring many thousands of jobs.  This causes the majority
of analysts to be affected by site downtime, infrastructure
instabilities and other distributed computing issues.  These problems
are compounded by the imperative nature of user jobs where each one
has exactly specified input data and cannot be adjusted to adapt to
current grid conditions.  To deal with these issues and with the
foreseen much larger data volume to be processed by analysts in
\runthree and \runfour, a new strategy for analysis data processing
has been developed, based on centralised \emph{analysis productions}.

Analysis productions are an extension of the LHCb{\dirac}
transformation system, which has been primarily used for the
centralised processing of LHCb data and simulation.  Productions are
submitted declaratively by providing the \gaudi configuration and
bookkeeping query for the input data, which enables LHCb{\dirac} to
automatically handle failures and adjust the way in which files are
grouped.  Information about productions and the provenance of files is
permanently stored in the LHCb\dirac bookkeeping, enabling high
quality analysis preservation and additional safety checks to be
performed.  Interactive analysis work can directly interact with
LHCb{\dirac} to obtain the location of the analysis production output
data, thereby reducing the need to copy data manually and further
supporting analysis preservation efforts.

Good testing of productions is essential as invalid productions have
the potential to waste computing resources and cause instability in
LHCb\dirac itself. To provide assurance that \gaudi configurations
provided by analysts are correct, extensive tests are run in \gitlab
continuous integration prior to submitting the productions and the
results of these tests are summarised on a dedicated website.

\subsubsection{Offline analysis}
\label{sec:offlineana}

While in the previous LHCb data processing flow, the trigger software
was largely based on the previously designed offline analysis
framework, the opposite is true for the upgraded LHCb data
taking~\cite{Skidmore:2022rza}. As much as possible, the new
offline analysis framework is based on the software developments made
for the trigger (see section~\ref{Sec:RTAsoft}). This approach avoids
duplication of effort, profits from the major developments outlined in
section~\ref{sec:trigger} and guarantees a similar look-and-feel of
the software to be used by analysts. Most importantly, all basic
analysis building blocks are shared between the trigger, the \sprucing
and the offline analysis, thus guaranteeing that the same software is
used to compute the same quantities online and offline.

Following this approach, the \davinci analysis software is descoped
(compared to the versions designed to process the legacy data) to be
only used for producing output tuples from input data (including
simulation). Its core is an algorithm to produce \emph{ntuples} that
stores measured quantities using \emph{functors} provided by the
\textsc{ThOr} framework~\cite{Ferrillo:2806414} within
\moore.\footnote{Ntuples are collections of variables related to an
  event, typically stored in a format suitable for \root or
  \textsc{python} packages. Functors are \cpp objects used in the LHCb
  code to return calculated kinematic variables.} These are the same
functors that are used in the trigger or \sprucing algorithms, thus
ensuring a one-to-one correspondence between applied selection
requirements and observables used offline. It should be emphasised
here that since no full offline reconstruction is foreseen in the
computing model, the input data objects (tracks, clusters, etc.) are
identical online and offline~\cite{LHCb-TDR-018,LHCb-TDR-016}.

\subsubsection{Data and analysis preservation}
\label{subsec:APOD}

The centrally produced samples are preserved on the distributed
computing infrastructure and catalogued in a dedicated LHCb
bookkeeping system (see section~\ref{subsec:DIRAC}). In 2020 LHCb
ratified the CERN Open Data policy~\cite{CERN-OPEN-2020-013}. In
accordance with the access policies outlined in this document, LHCb
will make the output of the \turbo selections as well as the output of
the \sprucing available to the public through the CERN Open Data
portal. The software necessary to read these files will be preserved
as \textsc{CVMFS} releases. In order to enable secure access for third
parties to the replicas of the data stored on the grid, a web-based
interface is under development, which will allow users to configure
analysis production jobs with minimal knowledge of the LHCb software.

To facilitate flexibility and creativity in the end-user analysis,
implementing a statistical interpretation of the filtered data, only
minimal constraints are placed on the necessary analysis code. While
commonly used libraries, such as \textsc{RooFit},
\textsc{ScikitHEP}~\cite{ScikitHEP}, etc.\ are available through
\textsc{CVMFS} and can be managed using the \textsc{Conda} package
manager, analysts can write custom scripts and custom routines to best
answer their specific analysis tasks. For each publication the
respective code is preserved in the respective physics working group
\gitlab repositories. Intermediate files, in particular filtered
ntuples, are usually stored on the \textsc{EOS} storage system at
CERN.

\subsection{Distributed computing}
\label{subsec:DIRAC}

The distributed computing of \lhcb is based on the \dirac
interware~\cite{DIRAC-communities-2017}, and its LHCb\dirac extension.
The \dirac project is developing interware to build and operate
distributed computing systems. It provides a development framework and
a rich set of services for both workload and data management tasks of
large scientific communities.  The LHCb{\dirac} infrastructure relies
on database backends and services. The databases are provided by the
CERN/IT database infrastructure (\emph{database on
  demand}~\cite{Aparicio:2134564}).\footnote{The CERN/IT database
  infrastructure is provided by \Trmk{Oracle}.} The services run on a
dedicated computing infrastructure also provided by CERN/IT.

\dirac was started by \lhcb as a project for accessing computing grid
resources.  In 2009, following interest of other communities, the
project has been open sourced.  Now, it is a truly open source project
hosted on \github and released under the \texttt{GPLv3} license.
Within the following years, \dirac has been adopted by several
experimental communities both inside and outside high energy physics,
with different goals, intents, resources and workflows.

\begin{figure}[t]
  \centering
  \includegraphics[width=0.68\textwidth]{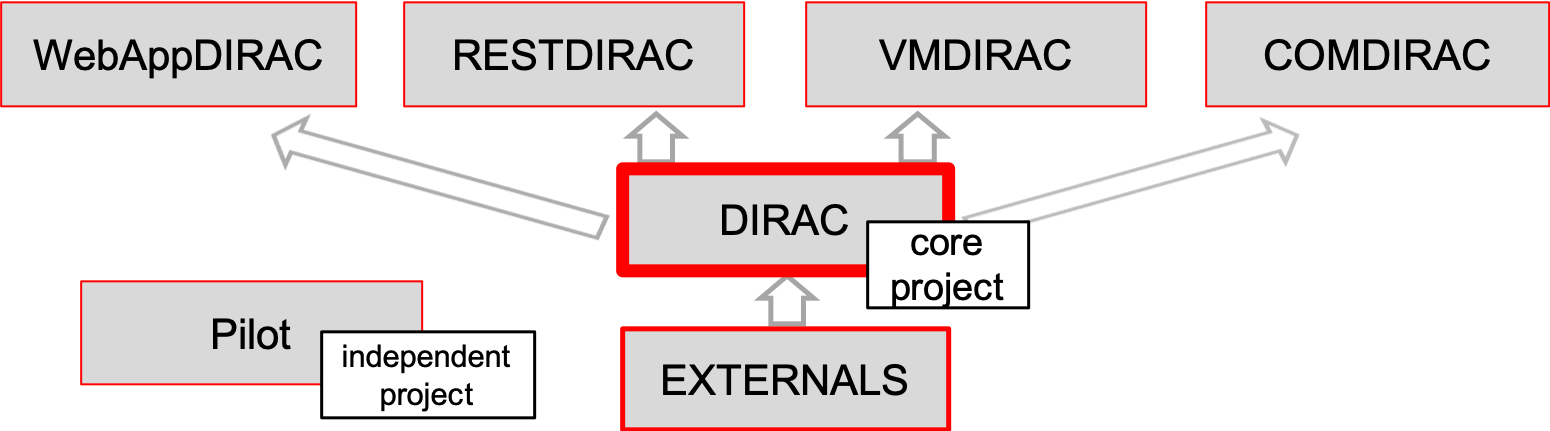}
  \vspace{1cm}
  \includegraphics[width=0.68\textwidth]{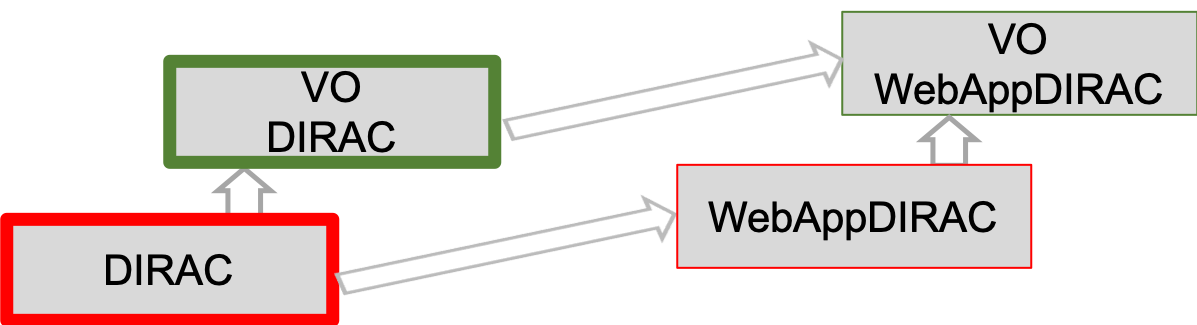}
  \caption{Diagram of a DIRAC release components. The concepts of
    (top) \emph{horizontal} and (bottom) \emph{vertical} extensibility
    are illustrated. Reproduced from~\cite{DIRAC-communities-2017}. \textcopyright\ 2022 IOP Publishing Ltd. CC BY 3.0.}
  \label{fig:dirac-horiz-vert}
\end{figure}

In order to accommodate different requirements, \dirac has been
designed with extensibility in mind. The core project (\dirac) can be
\emph{horizontally} extended by adding projects that are independently
versioned but nevertheless strongly interdependent, as they concur to
form a \dirac release.  All projects are hosted together on \github,
share the same license, and are maintained by the same set of users.
The \dirac core project is the glue that keeps the satellite projects
together. It also depends on software not maintained by the \dirac
consortium, which is collected in \emph{externals}. Horizontal
extensions include: \textsc{WebAppDirac}, the \dirac web
portal~\cite{WebAppDIRAC-2015};
\textsc{RESTDirac}~\cite{RESTDIRAC-2012}, which extends some \dirac
services by providing a representational state transfer (REST)
interface; \textsc{VMDirac}~\cite{VMDIRAC-CHEP2012}, which allows to
create virtual machines on clouds; \textsc{COMDirac}, which extends
the \dirac user interface. The \textsc{Pilot}~\cite{DIRAC-pilots-2017}
project is instead independent from all the other ones. This scheme is
illustrated in figure~\ref{fig:dirac-horiz-vert} The \emph{vertical}
extensibility of \dirac enables users and \Acr[p]{vo} to extend the
functionalities of the basic projects, in order to provide specific
functionalities.\looseness=-1

\lhcb uses fully the functionalities provided by \dirac, but has
customised some of the \dirac systems by implementing the
LHCb{\dirac}~\cite{LHCbDIRAC-CHEP2012} extensions.  New
\texttt{Bookkeeping}~\cite{BKK-2014} and \texttt{Production
  Management}~\cite{LHCbDIRAC-Production-2012} systems have been
created, both also providing GUI extensions within the
LHCb\textsc{WebAppDirac}, (the \lhcb extension of
\textsc{WebAppDirac}).  The \dirac \textsc{Pilot} project is extended
within the LHCb\textsc{Pilot} project.

It is essential that \dirac provides a transparent and uniform
interface for \Vo{s} to access resources that are more and more
heterogeneous which implies that IaaS (Infrastructure as a Service)
and IaaC (Infrastructure as a Client) models must also be supported.
This is realised in \dirac by a generic pilot
model~\cite{DIRAC-pilots-2015}, where a plugin mechanism enables easy
adaptation on a wide range of computing
resources~\cite{ARC-CT-2015,HTCondorCE-2015,VAC-2017}, including cloud
resources, \Hpc centres and the servers of the \lhcb online farm.

The \dirac system scales in terms of traffic and data set growth, and
maintainability.  In terms of traffic growth, \dirac closely follows
the most modern architectural directives. Nevertheless technological
updates such as the usage of message queues, \textsc{python 3},
multiprocessing and centralised logging systems add the required
robustness to the system.  In addition, the relational databases used
in \dirac are adequate to ensure full scalability for the data set
growth.

System and software maintainability has also been taken in due
consideration in the constant evolution of \dirac, by implementing a
proper monitoring, easily maintainable code with increasing
functionality tests, the use of continuous integration tools and
performance tests, better documentation and user support.

The \dirac jobs are handled by the workload management system (WMS)
and by the \dirac data management system (DMS).  Data sets are
retrieved through a bookkeeping tool. The \lhcb bookkeeping is a
metadata and provenance catalogue used to record information about the
data sets.  In order to cope with the rapidly increasing data size,
the main bookkeeping tables and indexes are partitioned. This allows
to run more efficient queries using, for example, partition-wise
joins.

\subsection{Computing model}
\label{subsec:compmodel}

The new paradigm for the trigger selection process, described in the
previous section, implies necessary changes in the offline computing
model.  Owing to the five-times higher instantaneous luminosity and
higher foreseen trigger efficiency, the LHCb upgrade has a signal
yield per time unit approximately ten times higher than that of the
\runonetwo \lhcb experiment. The pileup also increases at upgrade
instantaneous luminosity, resulting in an average event size increase
by a factor of three. As a consequence a data volume larger by more
than a factor of 30 is expected. A corresponding necessity to generate
significantly larger samples of simulated events arises, as the number
of events to simulate is proportional to the integrated luminosity.
The computing resource requirements are substantially mitigated by the
novel real-time data processing model and by the massive use of fast
simulation techniques, as discussed in the previous sections.

\subsubsection{Data processing flow} 
\label{subsec:dataprocessing_flow}

Building on the experience developed during LHC \runtwo, in the \lhcb
upgrade most of the activities related to data processing, such as
event reconstruction and calibration and alignment of subdetectors,
are performed online.  The output produced by the \Hlt is stored on
tape through three streams: \FULL, \turbo and calibration (\turcal).
The \turbo stream undergoes only minimal offline processing before
being stored to disk. The \FULL and \turcal streams instead requires
further offline filtering.

The online reconstruction and trigger selection process execute the
order of thousand trigger lines, each of which is associated to one of
the three (\turbo, \turcal or \FULL) streams that are subsequently
stored offline.  The offline processing of the \turbo stream, which
comprises the bulk of the events, is performed by the {\textsc{Tesla}}
application that converts the information from the raw format into
\root I/O objects such as tracks, calorimeter clusters and particles,
ready to be used for physics analysis, adds the luminosity
information, and persists them on disk in the appropriate
format. \turbo events are also classified into streams for an easier
access at analysis stage.

Events in the \FULL and \turcal streams are further processed offline
by the \sprucing application, which reduces the event size and
performs a further event selection before storing the events on disk
(data slimming and skimming, see section~\ref{subsec:DPA}).  Each of
the \sprucing selections is associated to a specific analysis, or
group of closely related analyses, and the output information can be
persisted at the appropriate level.  An average event retention of
80\% is obtained.

\begin{figure}[t]
  \centering
  \includegraphics[width=\linewidth]{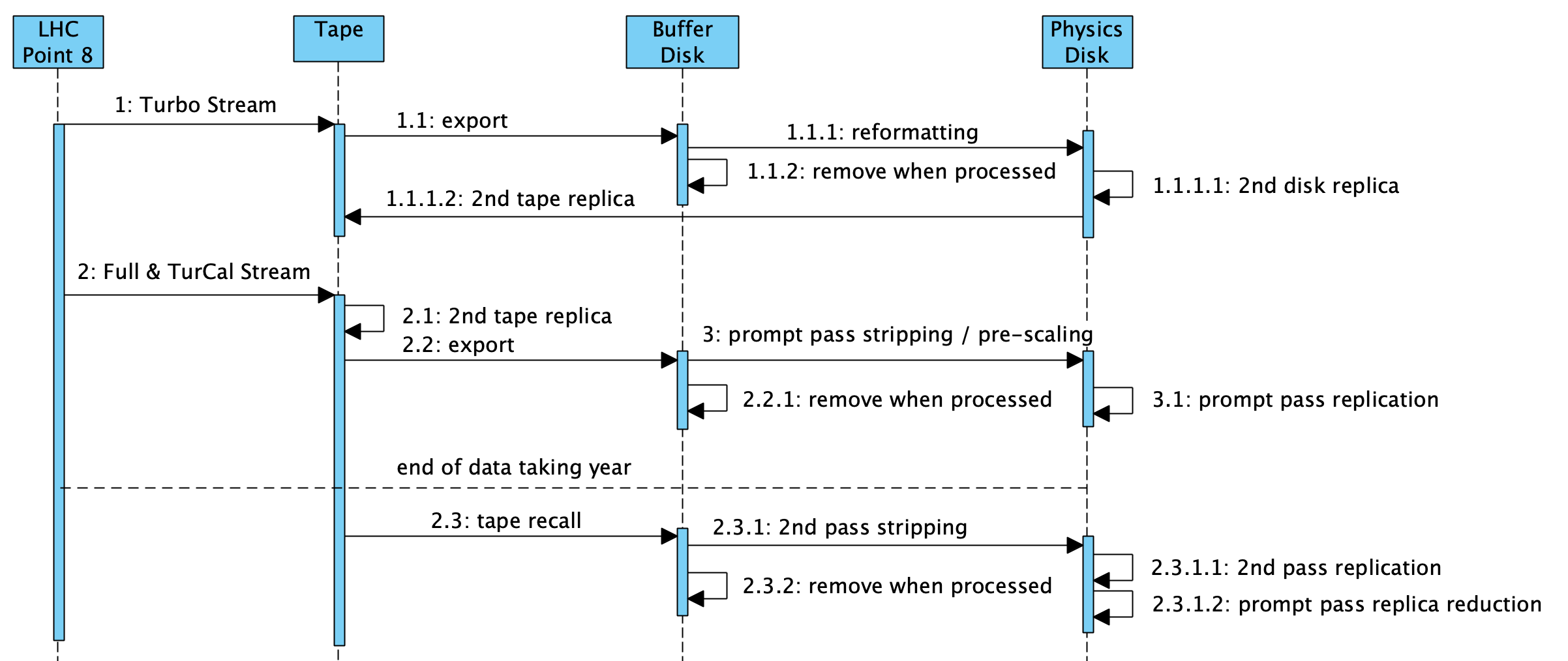}
  \caption{\small The LHCb offline data processing workflow. Reproduced from~\cite{LHCb-TDR-018}. CC BY 4.0.}
  \label{fig:offlinelogical}
\end{figure}

The offline data and processing flow is described in
figure~\ref{fig:offlinelogical}. The \turbo, \FULL and \turcal streams
are exported from the LHCb data centre. One copy of all data is stored
at the CERN tape system. One additional copy of the \FULL and \turcal
raw data is stored at another Tier 1 tape system. All data are also
copied to intermediate buffer disk storage. Data are then immediately
processed by the appropriate stream dependent applications, as
previously explained, and saved on disk. The \turbo data are simply
streamed and put onto disk storage and as a second copy on a Tier 1
tape system.  The first \sprucing pass happens synchronously with data
taking and can be prescaled if needed.  Two replicas of the data will
be kept on disk after this first processing pass.  A second processing
pass (re\sprucing, implementing updated selections and calibrations)
is typically performed after the data taking period, usually during
the LHC winter shutdown. When the second processing pass has been
performed, the number of copies of the previous processing saved on
disk is reduced.  One copy of each \sprucing pass is kept also on tape
archive.

In the streaming scheme described above each user typically analyses a
small fraction of the whole data set.  In order to avoid bottlenecks
due to each user chaotically running jobs on individual streams as
desired, the data processing for user analysis is organised in
centrally-managed productions (analysis productions), further
described in section~\ref{subsec:DPA}.  In addition, users are allowed
to submit jobs to offline resources, using the \textsc{Ganga}
framework~\cite{Ganga-2015, Ganga-2017}, for analysis prototyping and
testing purposes and other cases, such as running parametrised
pseudo-experiment simulations, and performing fits or other further
stages of the analysis.

The production of Monte Carlo events is described in detail in
section~\ref{subsec:simulation}.  The simulated events are produced by
two applications, \gauss~\cite{Clemencic_2011} and
\boole~\cite{sw:boole}, taking care of the event generation and
propagation through the detector, and of the digitisation,
respectively.  Various Monte Carlo simulation techniques that are
faster than the standard, detailed {\geant}-based simulation, are
available, as described in section~\ref{subsec:simulation}.

The simulation workflow in the LHCb Upgrade is very similar to the one
used for \runone and \runtwo. A number of steps are run in
sequence. The intermediate files created at the end of each step are
transient and deleted when no longer necessary. The only notable
exception to this workflow is the fully parametric simulation where
the data are saved directly in the form of high-level objects that are
ready to be used in physics analysis. In this case, the digitisation,
trigger emulation and event reconstruction steps are skipped.

\subsubsection{Resource provisioning}
\label{subsec:provisioning}

The provisioning of resources for the computing infrastructure
provided by the Worldwide LHC Computing Grid (WLCG) follows a pledging
scheme where computing sites provide a dedicated amount of resources
to the experiments. The pledged resources are based on requests
submitted by the experiments for the forthcoming years and
accompanying resource usage reports. Both documents are provided twice
a year to the relevant funding bodies.  The WLCG infrastructure is
setup in tier levels. The Tiers used by LHCb are the Tier~0 at CERN,
major Tier~1 sites in several countries and approximately ninety
additional Tier~2 sites both in countries with Tier~1 centres and in
other countries.  CPU resources are provided on all Tier levels. Tape
storage is only provided at Tier~0 and Tier~1 sites. Disk storage is
provided at Tier~0, Tier~1 and a limited number of Tier~2
sites. Limiting the storage resources (tape and disk) to a restricted
number of sites has proven to be a successful operational model.

Using the \emph{mesh processing} paradigm~\cite{Arrabito:1565912},
the so called Tier~2 \emph{helper sites} are attached to one or more
storage sites. Each helper site receives a payload, downloads the
input data from the remote storage site, processes the files locally
and subsequently uploads the output data again to the same storage
site from which the input data were downloaded. This concept is used
for data reprocessing campaigns to increase the throughput but also
during prompt processing in case the Tier~0 and Tier~1 sites do not
have enough resources to cope with the load.  The \emph{\gaudi
  federation} concept~\cite{Haen:2134567} is used to read input
data from other than the initially foreseen storage sites.  Within
this scheme, an application is deployed with additional information on
the location of all replicas of all needed input data files.  If the
first priority copy of a file is not readable (for example because the
file is corrupted or the disk storage is not available), the
application searches over the network for remote replicas of the input
file across the federation, and reads data from there.  This concept
is especially useful for user analysis files where multiple replicas
can be available.

Data handling and data replication follow a \emph{democratic}
principle where data are replicated over all possible storage sites
depending on the available space and the capacity of the corresponding
site. For raw detector data, the smallest block to be replicated is
represented by the files corresponding to one detector run. Derived
data sets are also kept on the same storage site. In case of data
replication, all descendant files from one run are also replicated.
Intermediate files within a simulation job (typically executed on a
Tier~2 site) are stored on a topologically close Tier~1 site and
deleted after they have been processed by the corresponding
application. The final simulated event files, ready for user analysis,
are also uploaded to a topologically close Tier~1 site and then
replicated following the democratic data replication policy. This
principle has proven to be successful, as it has made the data set
handling operations easier and allowed to optimise the load of
applications using the distributed data.

In addition to the CPU resources pledged via the WLCG, LHCb also uses
several additional computing resources in an opportunistic and/or ad
hoc way. They come from two different sources: the LHCb online farm
and opportunistic resources not owned by and not under the control of
the experiment.  The online farm is used by \lhcb for offline data
processes outside data-taking periods. It is possible in principle to
run concurrently online and offline applications, thanks to the fact
that the same hardware infrastructure is used in both cases, without
any intermediate virtualisation layer, and to the implementation of a
\emph{fast stop} mechanism that allows to switch between offline and
online usage within a time of about 1-2 hours.  Opportunistic
resources not owned by \lhcb include \Hpc centres and resources hosted
on WLCG sites that are not pledged to \lhcb. In both cases, the main
usage is for simulation, as it does not require input data.  The
amount of tasks that can be performed with these CPU resources is
unpredictable as the priority given to the LHCb applications compared
to other users is lower, and LHCb jobs are essentially used to fill
otherwise unexploited CPU time.

\subsubsection{Resource requirements}
\label{subsec:compresources}

The production of simulated events dominates the offline CPU computing
needs. The number of events to be simulated is estimated to be
$\sim 5\times 10^9\invfb$ of integrated luminosity per calendar
year. The production of events simulated according to a given
data-taking year extends normally up to six years afterwards. The
amount of corresponding resource requirement is mitigated by
exploiting faster simulation options. The storage needs are instead
dominated by data and crucially depend on the \Hlt output bandwidth of
10\gbytes per live second of LHC\@. While the associated tape needs
are not compressible, a mitigation is achieved for disk. A fraction of
about 70\% of triggered events are saved in the \turbo
format. However, the majority (6.5\gbyps out of 10\gbyps) of the
bandwidth is taken by the remaining 30\% of events in the \FULL and
\turcal streams, where the entire event is persisted. The events in
these two streams are therefore slimmed and filtered offline by the
\sprucing process, such that the total (logical) bandwidth to be saved
on disk is only $\sim 30\%$ of the original.  The extrapolated
throughput to tape and disk for the three data streams are reported in
table~\ref{tab:run3-rates}.

\begin{table}[t]
  \centering
  \caption{Extrapolated throughput to tape and disk for the \FULL,
    \turbo and \turcal streams. }
  \label{tab:run3-rates}
  \begin{tabular}{ |lccccc|}
    \hline
    Stream & Rate fraction & TAPE & TAPE & DISK & DISK\\
           &               & throughput & bandwidth & throughput & bandwidth \\
           &               & (GB/s) & fraction & (GB/s) & fraction\\
    \hline
    \FULL &  26\% & 5.9 & 59\% & 0.8 & 22\% \\
    \turbo &  68\% & 2.5 & 25\% & 2.5 & 72\% \\
    \turcal &  \phz6\% & 1.6 & 16\% & 0.2 & \phz6\% \\
    \hline
    Total &  100\% & 10.0 & 100\% & 3.5 & 100\%\\
    \hline
  \end{tabular}
\end{table}

The impact of simulated events on storage requests is small, since all
data produced during the intermediate production steps are not saved
and the simulation output is persisted in a compressed format, thus
achieving a size reduction per event of a factor up to twenty.  In
addition, analysis-dependent filtering criteria are also applied to
reduce the number of events written on storage.  The offline
reprocessing of the \FULL and \turcal streams requires a significant
reading throughput from tape. The needed throughput is estimated by
considering that reprocessing is done over a two-months period, with a
provisional buffering space corresponding to two weeks of data
staging. Half of data is staged at Tier~1 sites, the other half at
CERN, 50\% of which is then transferred at the Tier~1 sites for
processing.
\begin{table}[t]
  \centering
  \caption{Summary of the LHCb upgrade computing model
    requirements. Top section: main assumptions of the model. Bottom
    section: indicative resource requirements.}
  \label{tab:executive}
  \begin{tabular}{ |lccc|}
    \hline
    \multicolumn{4}{|c|}{Model assumptions}\\
    \hline
    ${\cal L}$ [cm$^{-2}$s$^{-1}$] & \multicolumn{3}{c|}{$2\times10^{33}$}\\
    Pileup  & \multicolumn{3}{c|}{6}\\
    Running time [s]  & \multicolumn{3}{c|}{$5\times 10^6$ ($2.5\times 10^6$ in 2021)}\\
    Output bandwidth (GB/s) & \multicolumn{3}{c|}{10}\\
    Fraction of \turbo events & \multicolumn{3}{c|}{73\%}\\
    Ratio \turbo /\FULL event size & \multicolumn{3}{c|}{16.7\%}\\
    Ratio full/fast/param. simulations & \multicolumn{3}{c|}{40:40:20}\\
    Data replicas on tape & \multicolumn{3}{c|}{2}\\
    Data replicas on disk & \multicolumn{3}{c|}{2 (\turbo); 3 ({\FULL}, \turcal)}\\
    \hline
    \multicolumn{4}{|c|}{Resource requirements}\\
    \hline
    WLCG Year& Disk (PB)& Tape (PB)& CPU (kHS06)\\ 
    \hline
    2021 &  66  & 142 & 863  \\
    2022 & 111  & 243 & 1.579  \\
    2023 & 159  & 345 & 2.753  \\
    2024 & 165  & 348 & 3.467  \\
    2025 & 171  & 351 & 3.267  \\
    \hline
  \end{tabular}
\end{table}
For safety reasons, in general two copies of all data that are
impossible to be regenerated are saved on tape.  Therefore, two copies
of the primary data sets, i.e.\ those originating from the online
system, are stored on tape.  An archive copy of the offline processed
data for each of the three streams is also saved on tape.  After the
offline processing, two copies of the \turbo stream and up to three
replicas of the \FULL and \turcal streams will be saved on disk.  A
single copy of all the simulated events is kept on tape, while two
copies of the most used simulated data sets ($\sim 30\%$ of the total)
is stored on disk.

A summary of the computing model parameters and an indicative
estimation of computing resource requirements is given in
table~\ref{tab:executive}.

\section{Performance}
\label{sec:performance}
The expected performance of the upgraded LHCb detector has been
extensively studied with detailed simulations. Dedicated samples of
specific physics channels have been generated and processed through
the full LHCb detector simulation to study in particular the \Hltone
reconstruction and selection efficiency. In the LHCb upgrade
full-software trigger scheme, the physics selections are largely
performed at \Hlttwo level; their efficiency will only be available
when the specific analyses will be carried out and finalised during
data taking.

\subsection{Computational performance}

\begin{figure}[h]
  \centering
  \includegraphics[width=\linewidth]{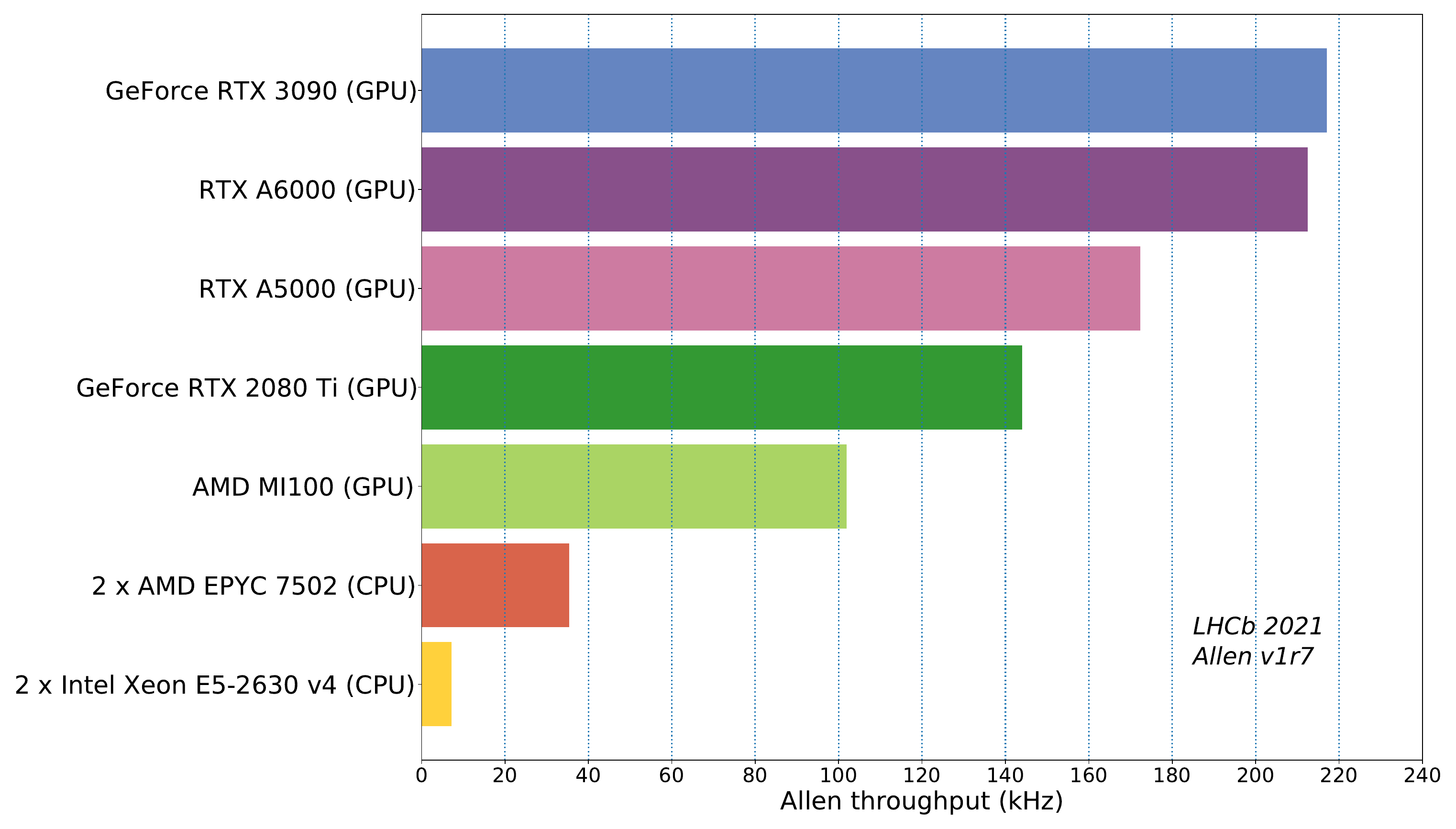}
  \caption{Throughput of the \hltone application on a selected
    subset of current generation \Gpu cards and a representative
    modern CPU server. Reproduced with permission from~\cite{LHCB-FIGURE-2020-014}.}
  \label{fig:hlt1_throughput}
\end{figure}

\begin{figure}[t]
  \centering
  \includegraphics[width=.95\linewidth]{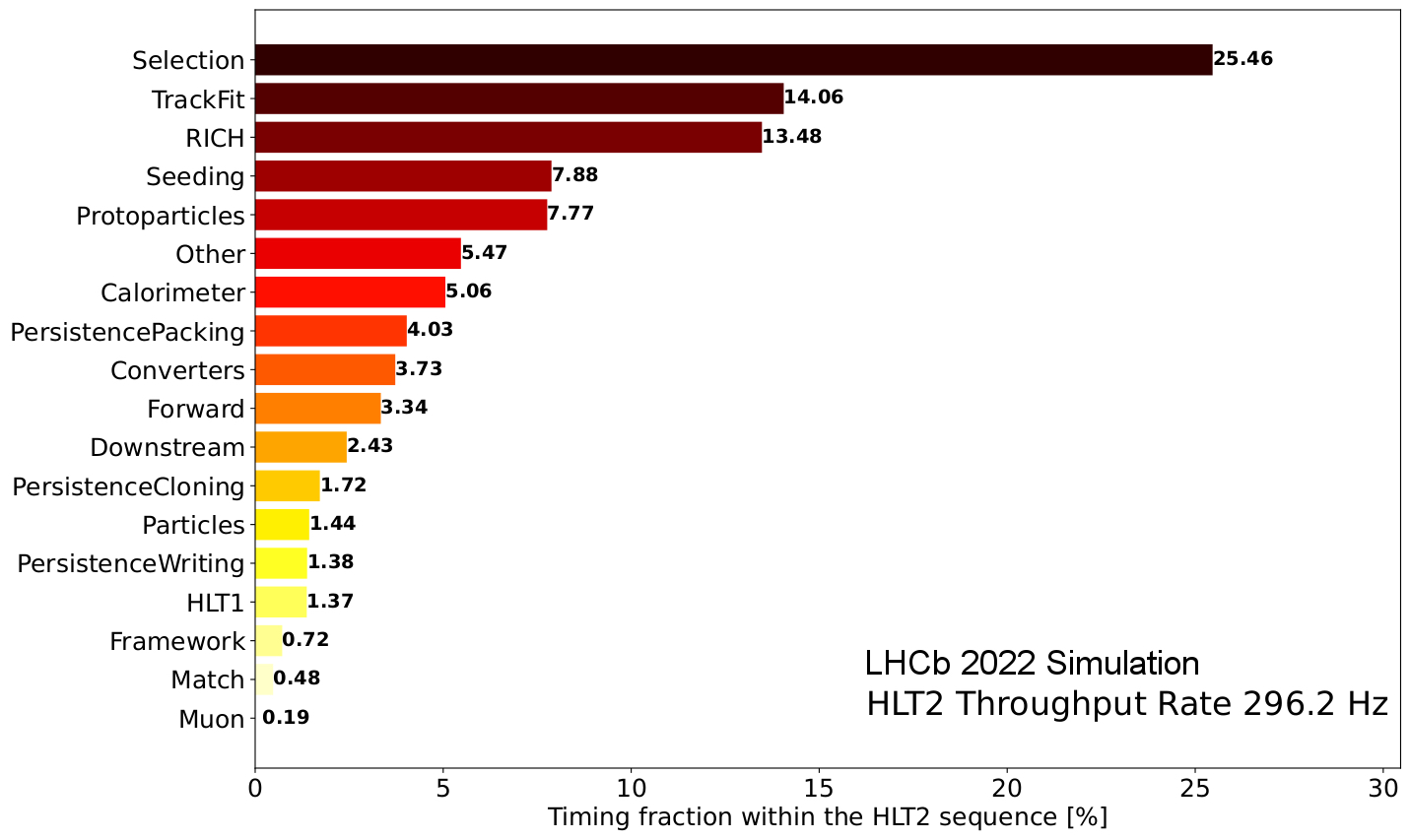}
  \caption{Throughput of the \Hlttwo application and fraction of
    \Hlttwo resources used by different parts of the reconstruction
    and selections, measured on a representative \Hlttwo server. A
    total of 1111 selection algorithms were executed as part of this
    test. Reproduced with permission from~\cite{LHCB-FIGURE-2021-003}.}
  \label{fig:hlt2_throughput}
\end{figure}

The physics performance of the upgraded LHCb detector critically
depends on the computational performance of its real-time software,
which enables the relevant algorithms to be deployed in \Hltone and
\Hlttwo as described in section~\ref{sec:trigger}. The performance is
measured in Hertz, giving the number of events which can be processed
each second on a representative processing unit. The results obtained
with the \Hltone application running on a range of currently available
\Gpu cards are shown in figure~\ref{fig:hlt1_throughput}. The \Gpu
card used by LHCb for the first data taking run is the
A5000.\footnote{NVIDIA RTX A5000 graphics card.} The figure also shows
the performance of the \Hltone code on a modern CPU server which is
relevant for running \Hltone in simulation on the grid where \Gpu
resources are not necessarily available. The computational performance
of \hlttwo is shown in figure~\ref{fig:hlt2_throughput} for a
reference CPU server which will be used for \runthree data
taking.\footnote{About 3500 \Hlttwo nodes are installed in the LHCb
  data centre.} In both cases the computational performance, albeit
evaluated on simulation, is comfortably adequate for the foreseen
nominal \runthree data taking conditions.

\subsection{Reconstruction performance}

In the upgraded LHCb the event reconstruction is performed in
real-time at the trigger level. The reconstruction efficiencies in
\Hltone and \Hlttwo have been studied in great detail while designing
the trigger algorithms. The main results are shown in the next
sections.

\subsubsection{Tracking performance}

Performance figures are produced with simulated event samples of
$\Bs\to\jpsi\phiz$, $\Bs\to\phiz\phiz$, $\Bz\to\Kstarz\epem$ and
$\Dp\to\KS\pip$ (5000 events per magnet polarity), including the
effect of \Velo and \Scifi radiation damage after 5\invfb, as well as
a sample of $\Dp\to\KS\pip$ decays enriched in tracks originating
outside the vertex detector.

\begin{figure}[p]
  \centering
  \includegraphics[scale=0.32]{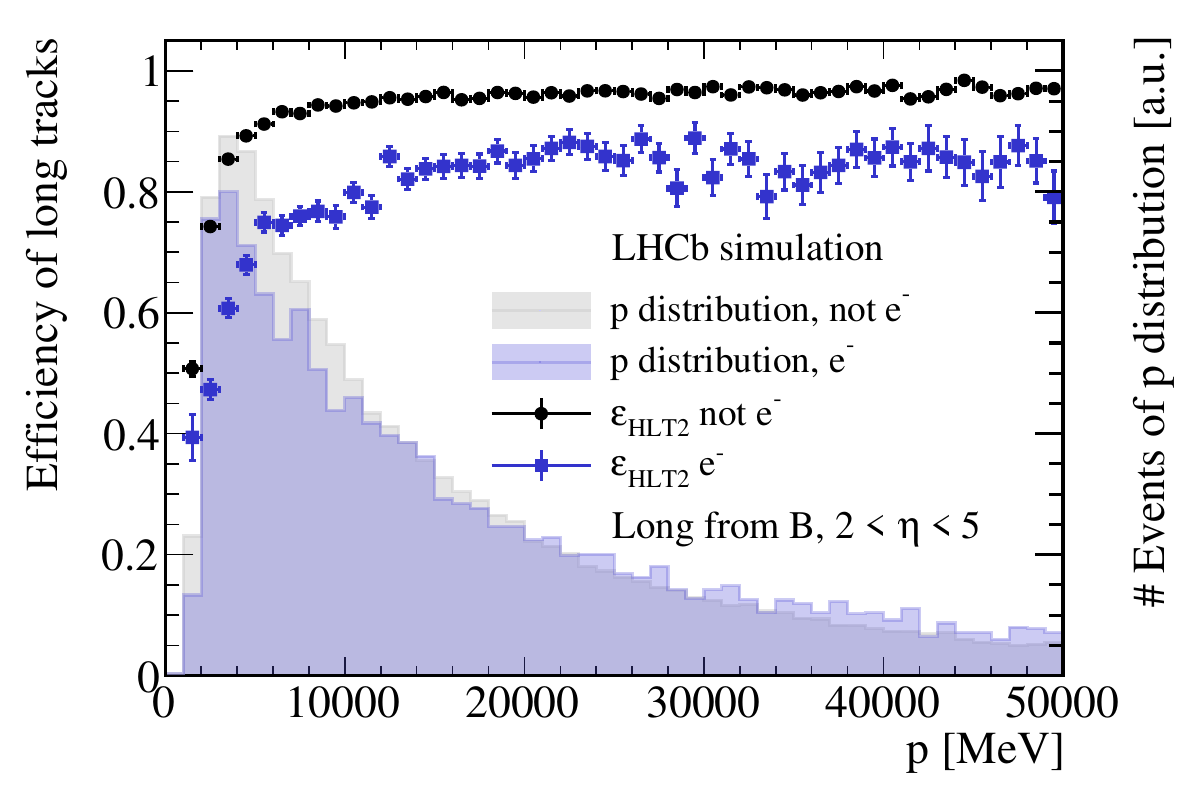}
  \qquad
  \includegraphics[scale=0.32]{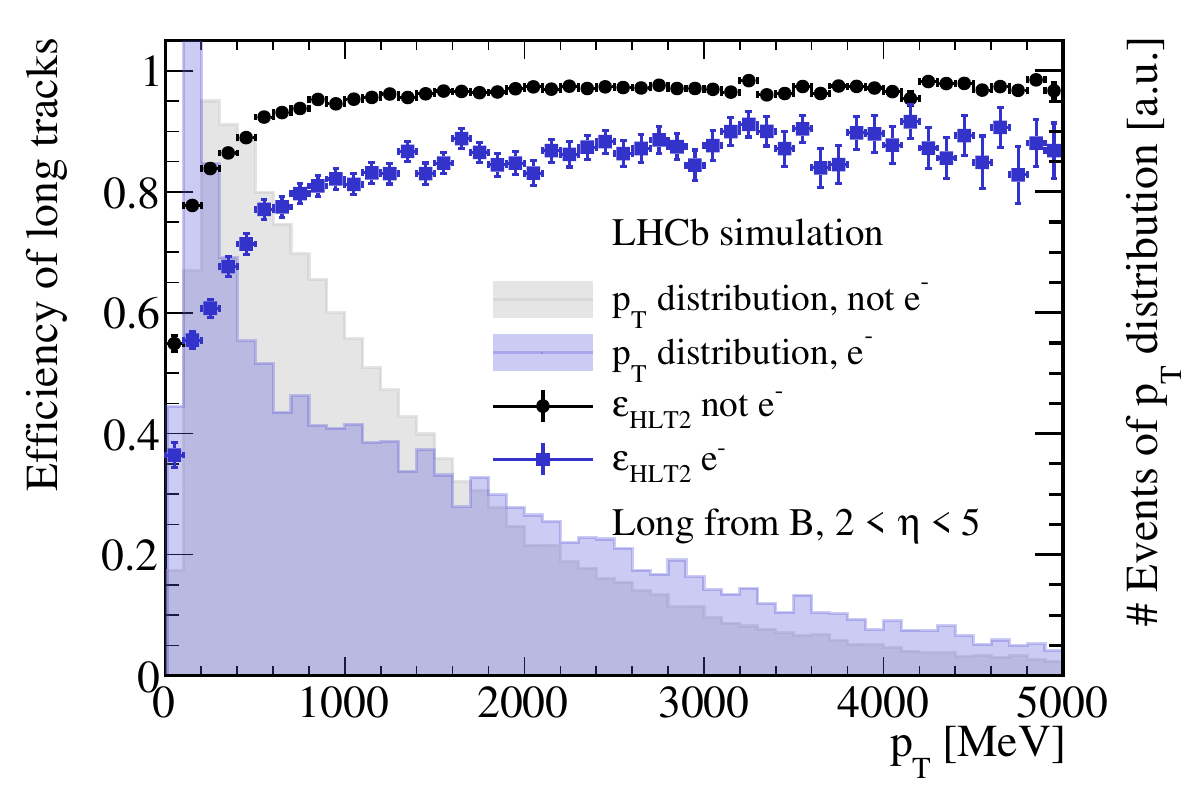}
  \\
  \includegraphics[scale=0.32]{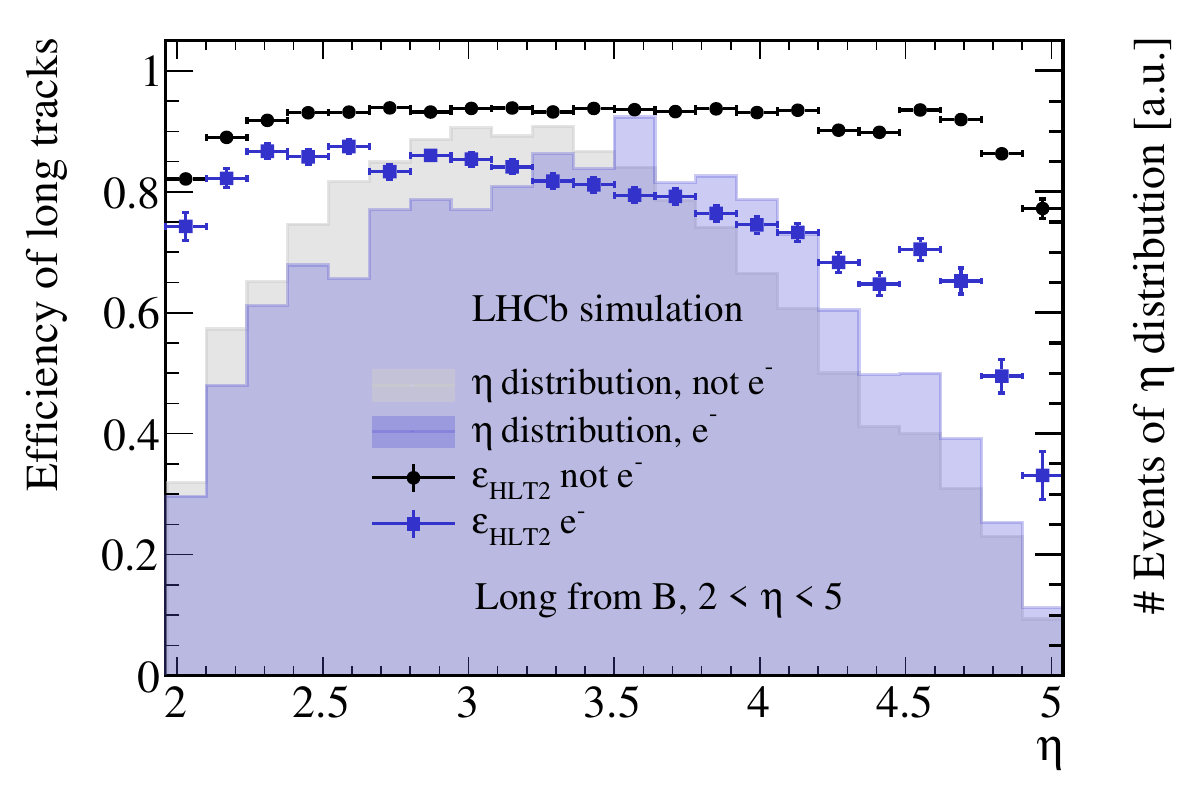}
  \qquad
  \includegraphics[scale=0.32]{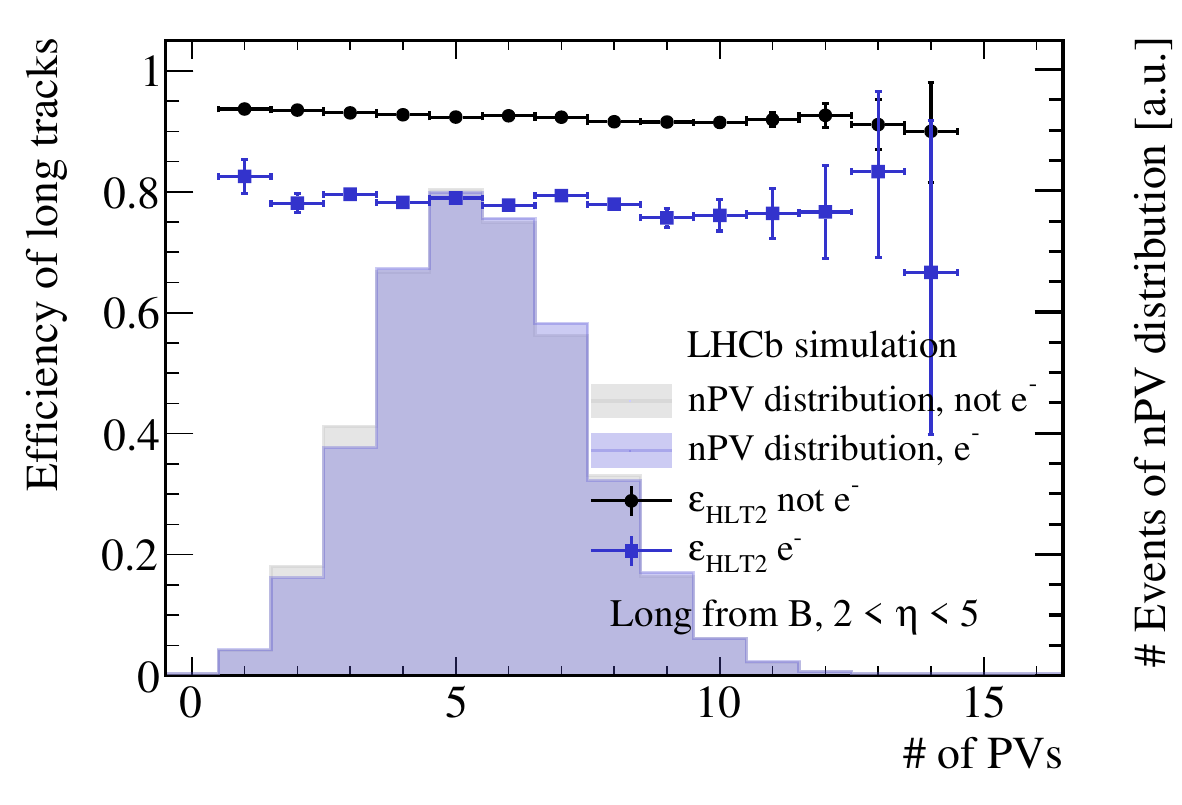}
  \caption{Long track reconstruction efficiency versus momentum \ptot,
    transverse momentum \pt, pseudo-rapidity \Peta, and number of
    primary vertices for long reconstructible electrons (blue squares)
    and non-electron (black dots) particles from \B decays within
    $2<\Peta<5$. Shaded histograms show the distributions of
    reconstructible particles. Reproduced with permission from~\cite{LHCB-FIGURE-2021-003}.}
  \label{fig:bestlong}
\end{figure}

\begin{figure}[p]
  \centering
  \includegraphics[scale=0.32]{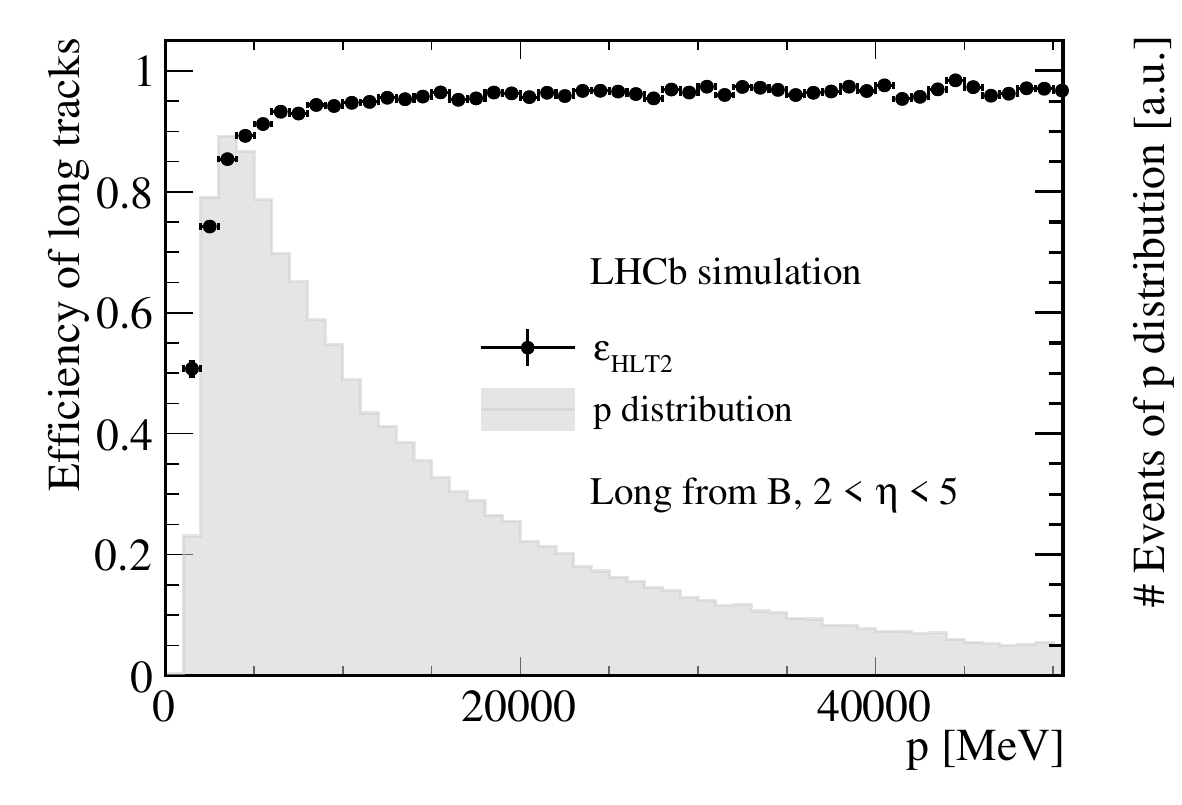}
  \qquad
  \includegraphics[scale=0.32]{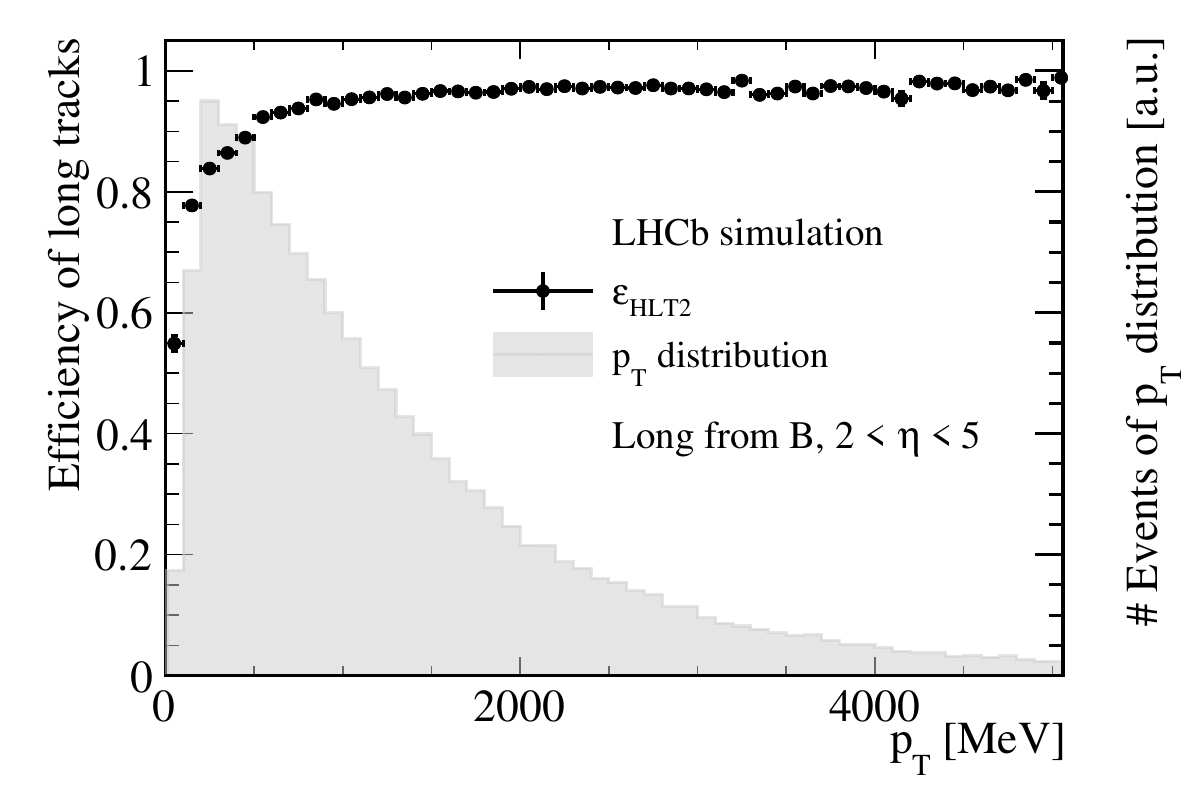}
  \\
  \includegraphics[scale=0.32]{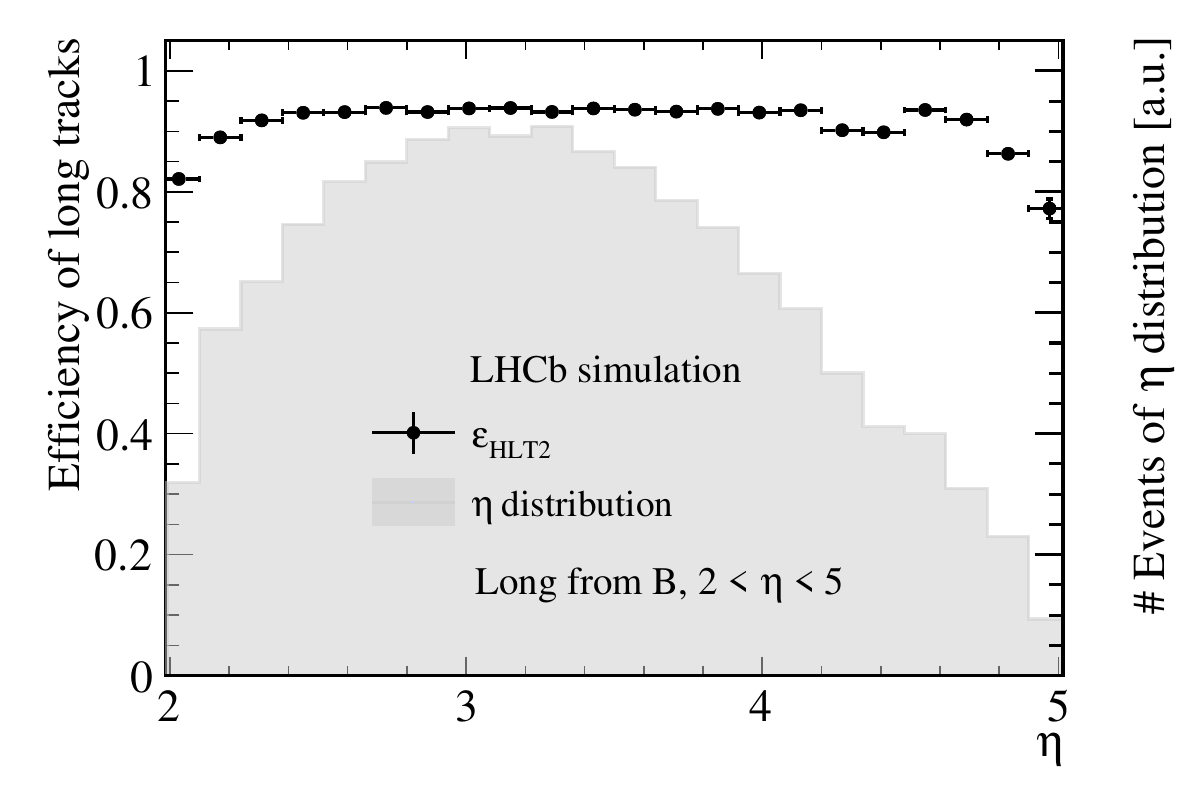}
  \qquad
  \includegraphics[scale=0.32]{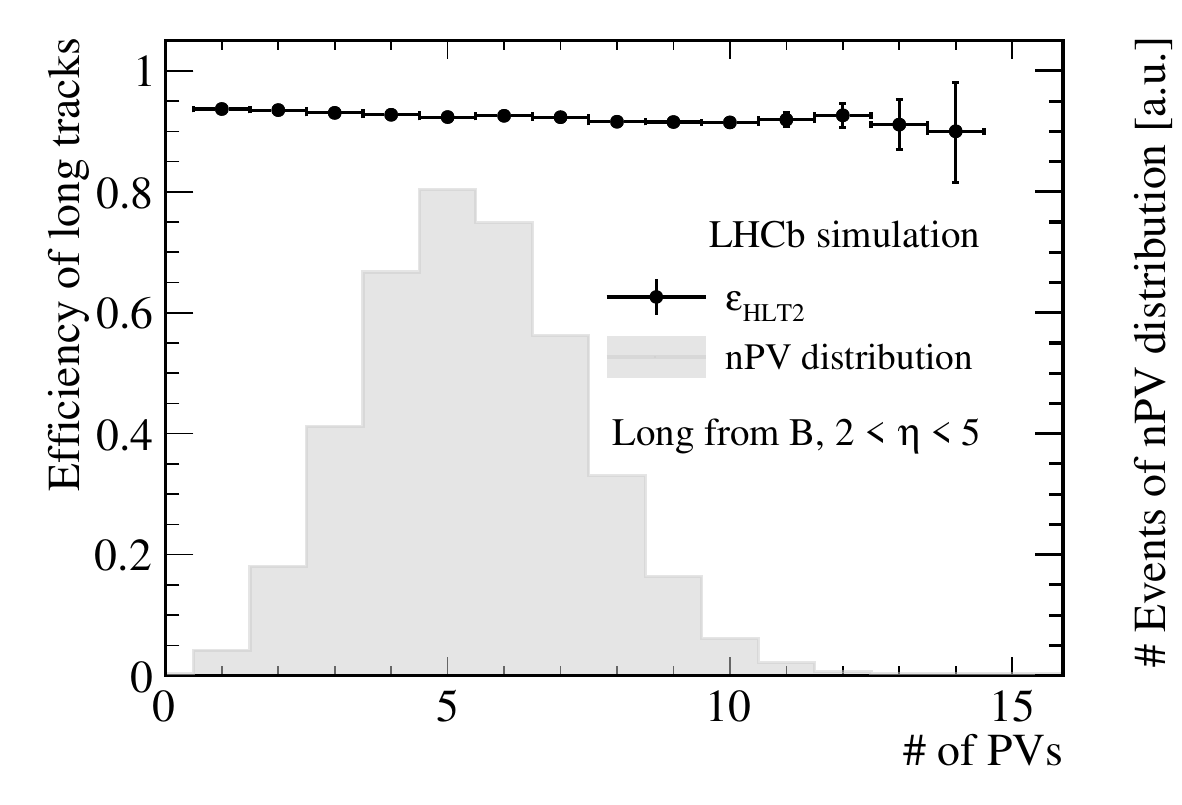}
  \caption{Long track reconstruction efficiency versus momentum \ptot,
    transverse momentum \pt, pseudo-rapidity \Peta, and number of
    primary vertices for long reconstructible particles from \B decays
    within $2<\Peta<5$. Shaded histograms show the distributions of
    reconstructible particles. Reproduced with permission from~\cite{LHCB-FIGURE-2021-003}.}
  \label{fig:bestlong_all}
\end{figure}

\begin{figure}[p]
  \centering
  \includegraphics[scale=0.32]{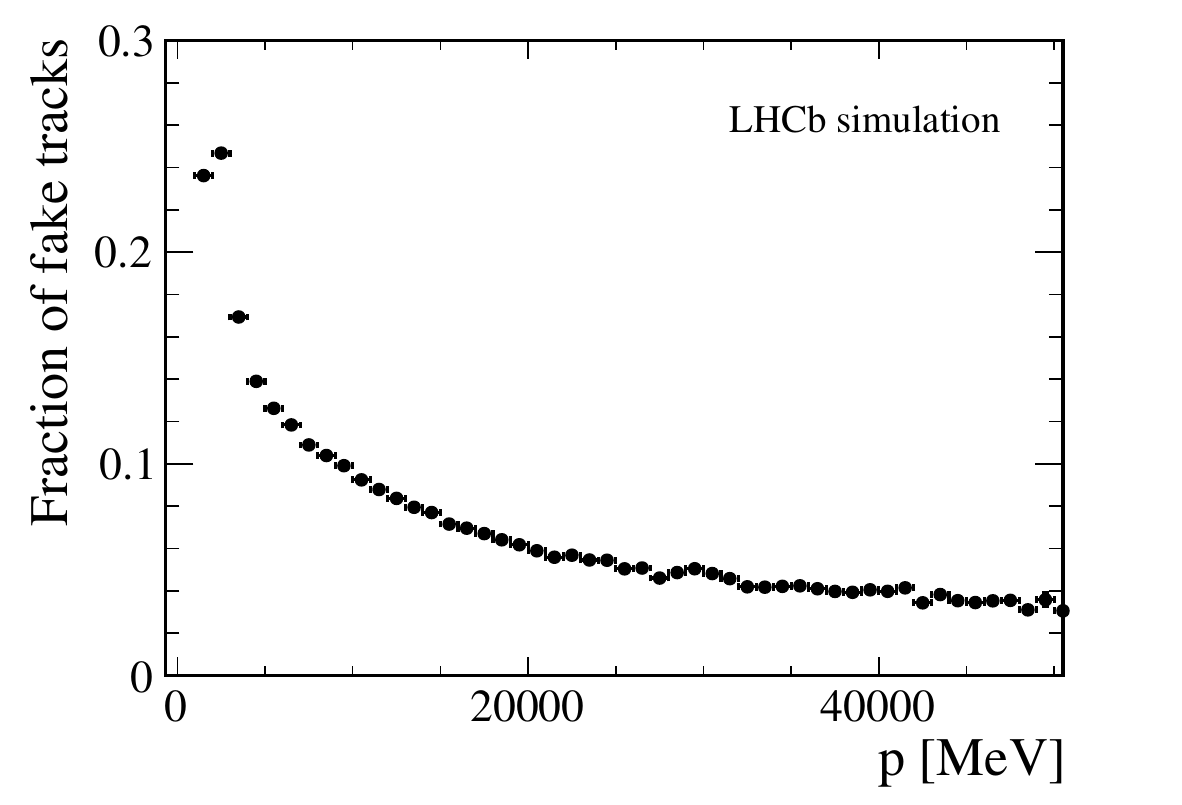}
  \qquad
  \includegraphics[scale=0.32]{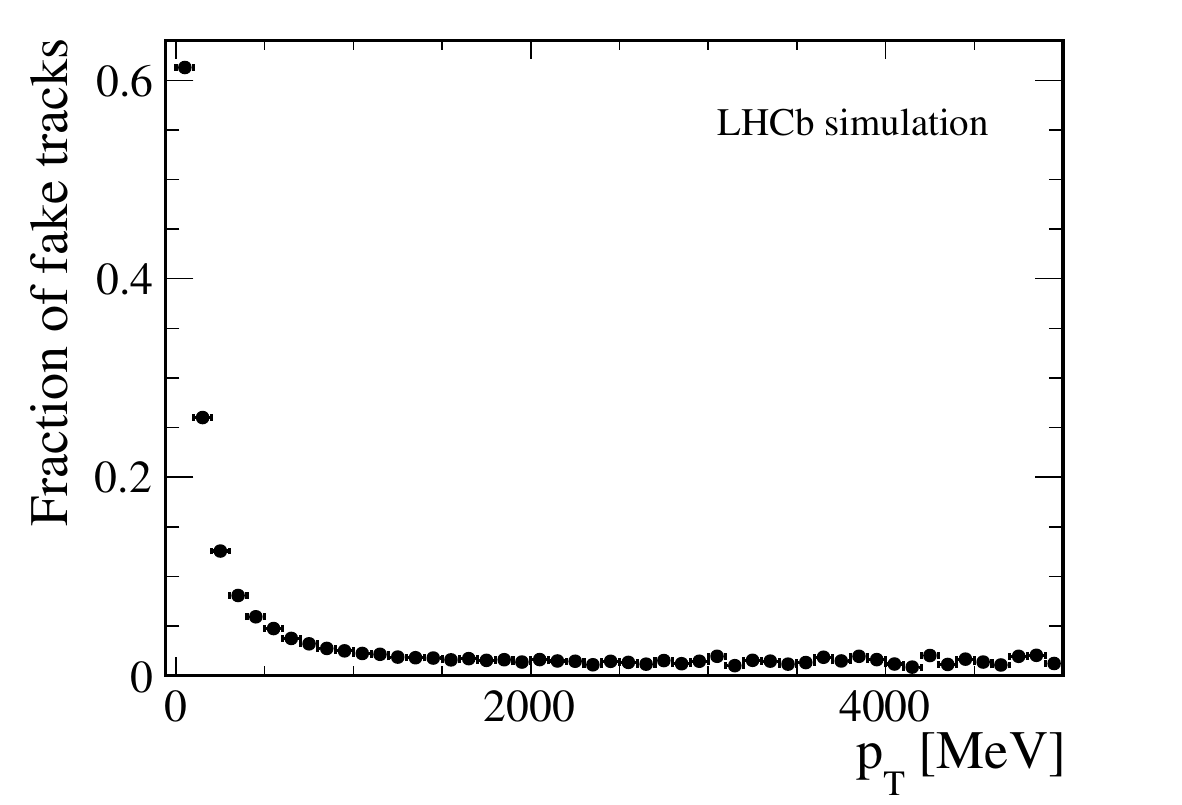}
  \\
  \includegraphics[scale=0.32]{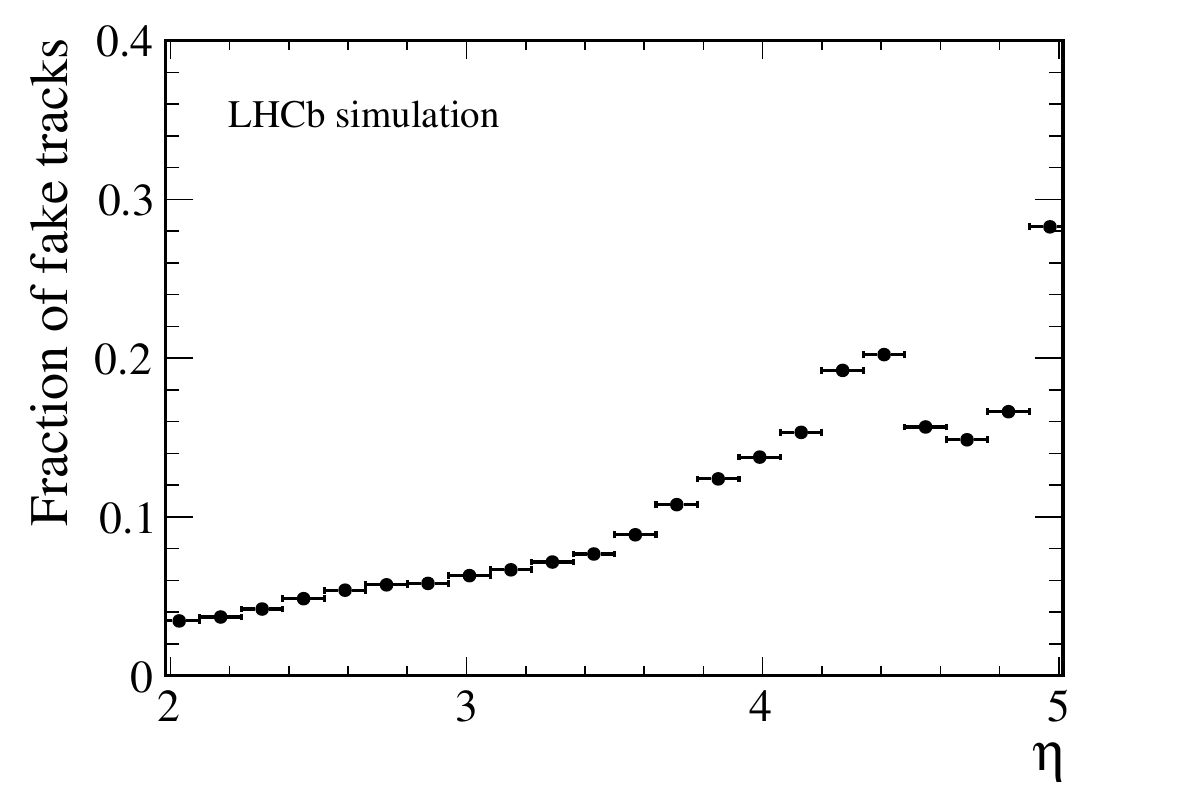}
  \qquad
  \includegraphics[scale=0.32]{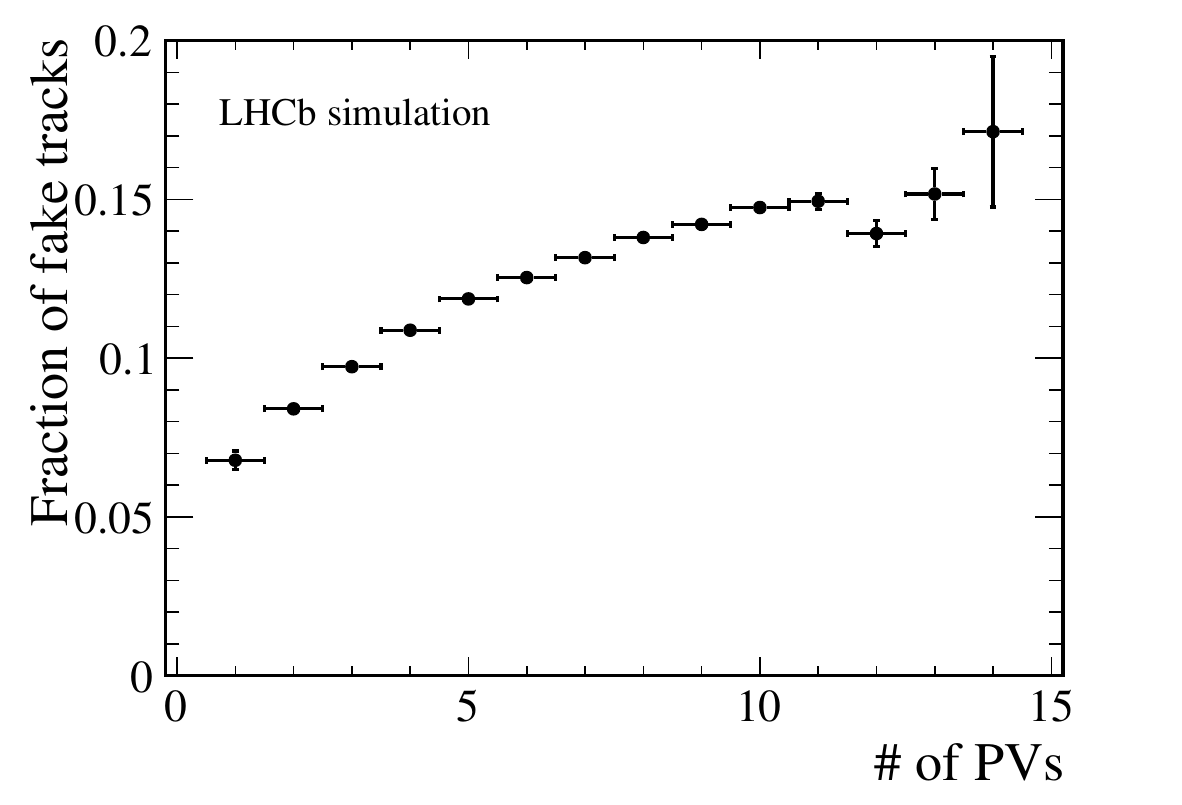}
  \caption{Ghost rate of long tracks reconstructed by the forward and
    match tracking algorithms as a function of momentum \ptot,
    transverse momentum \pt, pseudo-rapidity \Peta, and number of
    primary vertices. Reproduced with permission from~\cite{LHCB-FIGURE-2021-003}.}
  \label{fig:bestlong_ghostrate}
\end{figure}

\begin{figure}[p]
  \centering
  \includegraphics[scale=0.3]{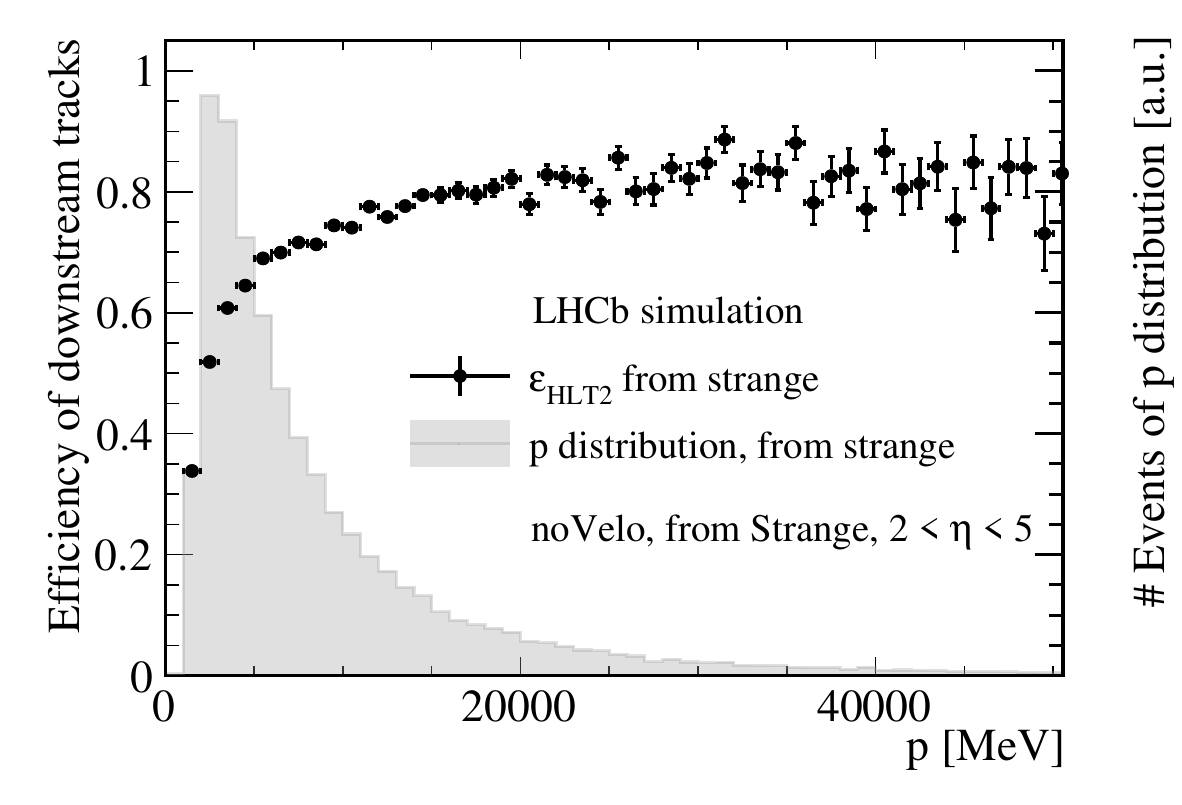}
  \qquad
  \includegraphics[scale=0.3]{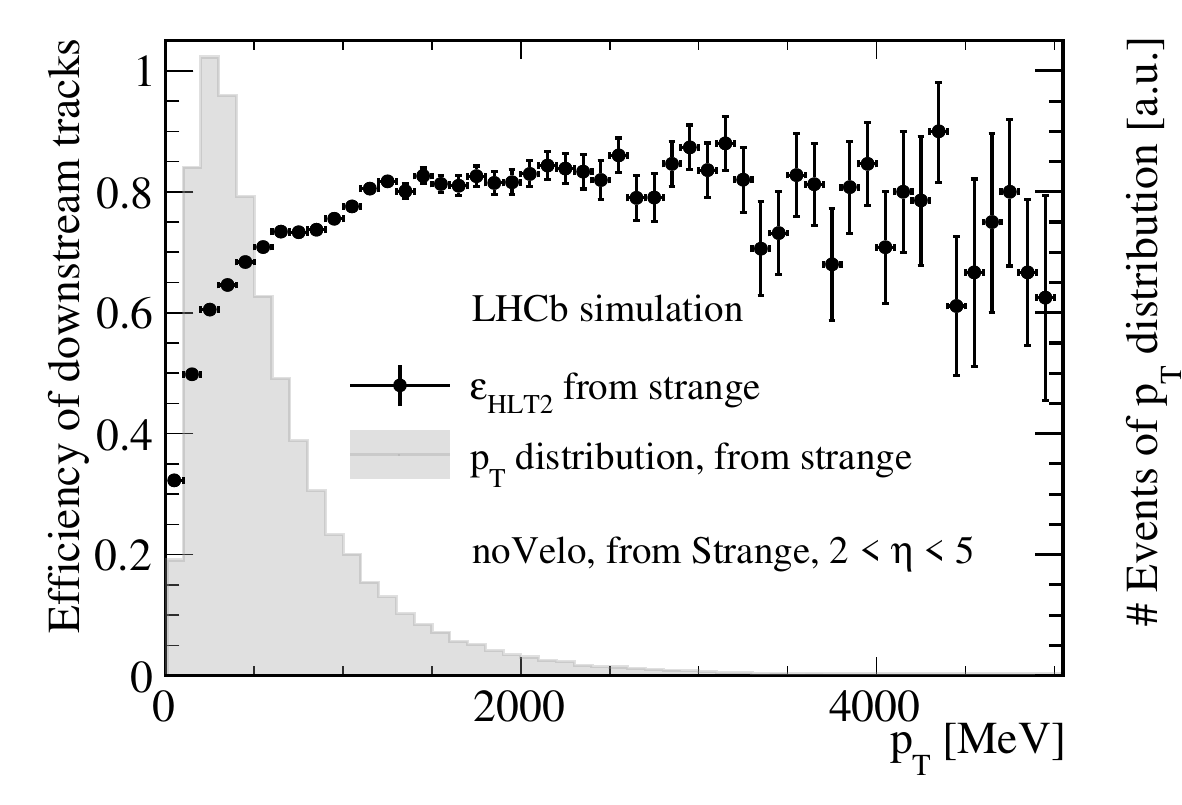}
  \\
  \includegraphics[scale=0.3]{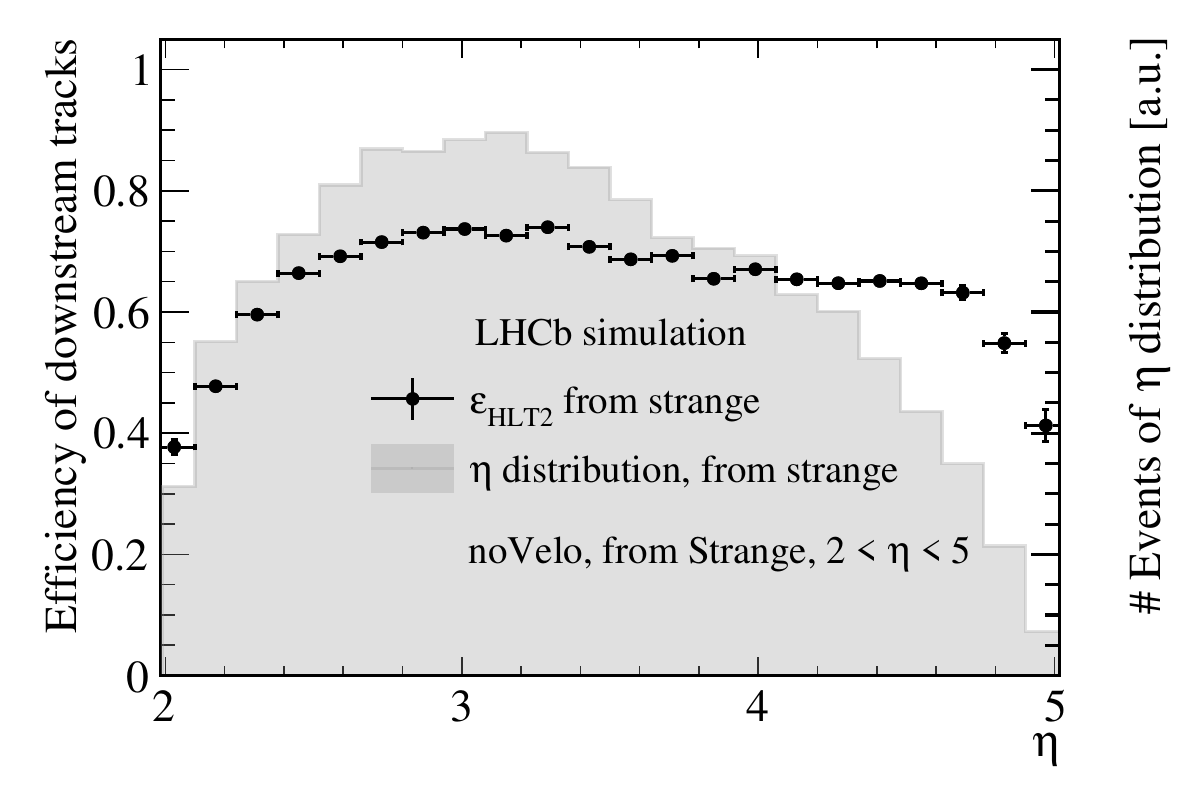}
  \qquad
  \includegraphics[scale=0.3]{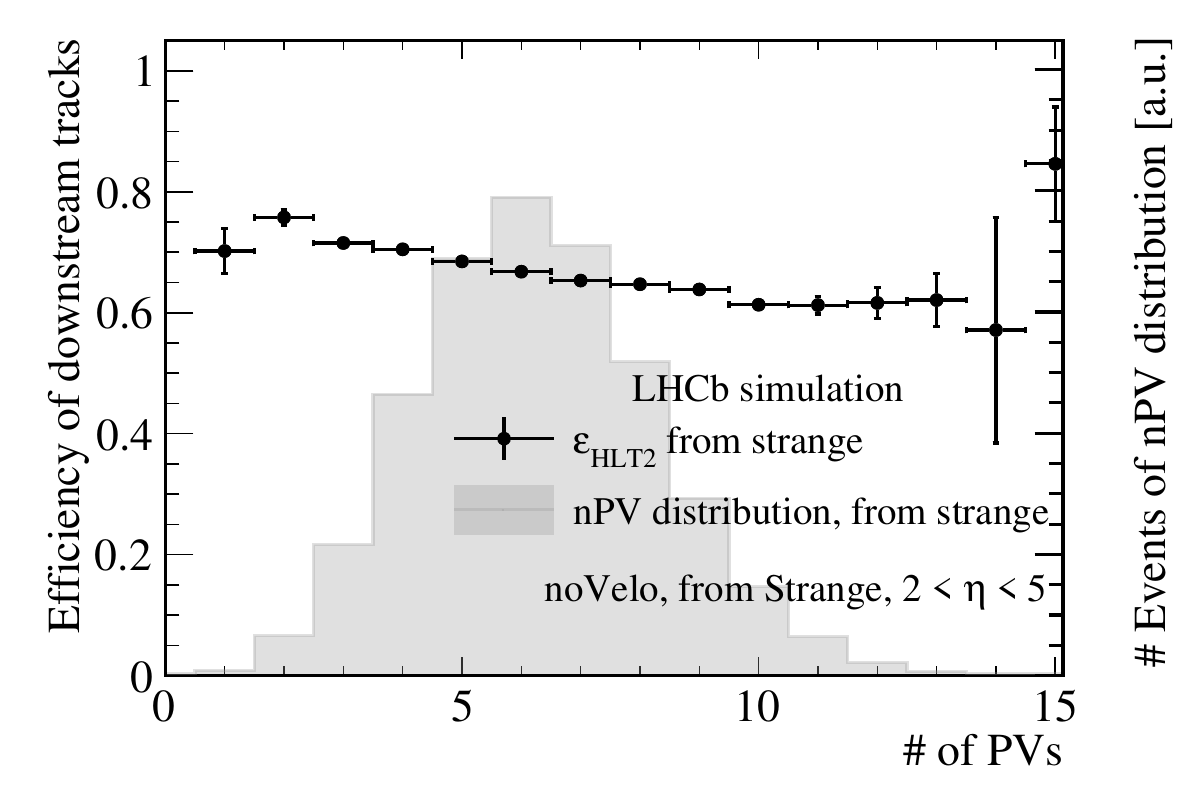}
  \caption{Downstream track reconstruction efficiency versus momentum
    \ptot, transverse momentum \pt, pseudo-rapidity \Peta, and number
    of primary vertices for reconstructible particles from long-lived
    particle (marked as \emph{strange} in the legend) decays within
    $2<\Peta<5$ that have no hits in the \Velo. Shaded histograms show
    the distributions of reconstructible particles. Reproduced with permission from~\cite{LHCB-FIGURE-2021-003}.}
  \label{fig:bestdown}
\end{figure}

\begin{figure}[p]
  \centering
  \includegraphics[scale=0.3]{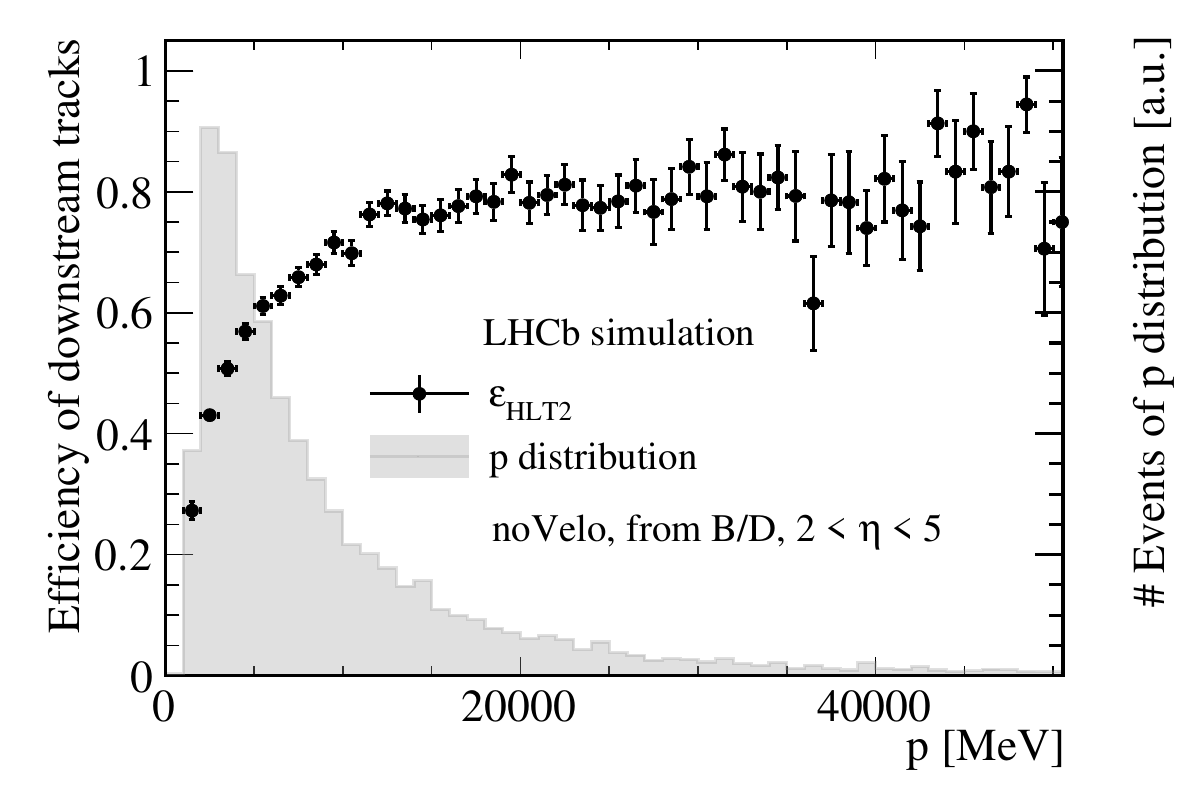}
  \qquad
  \includegraphics[scale=0.3]{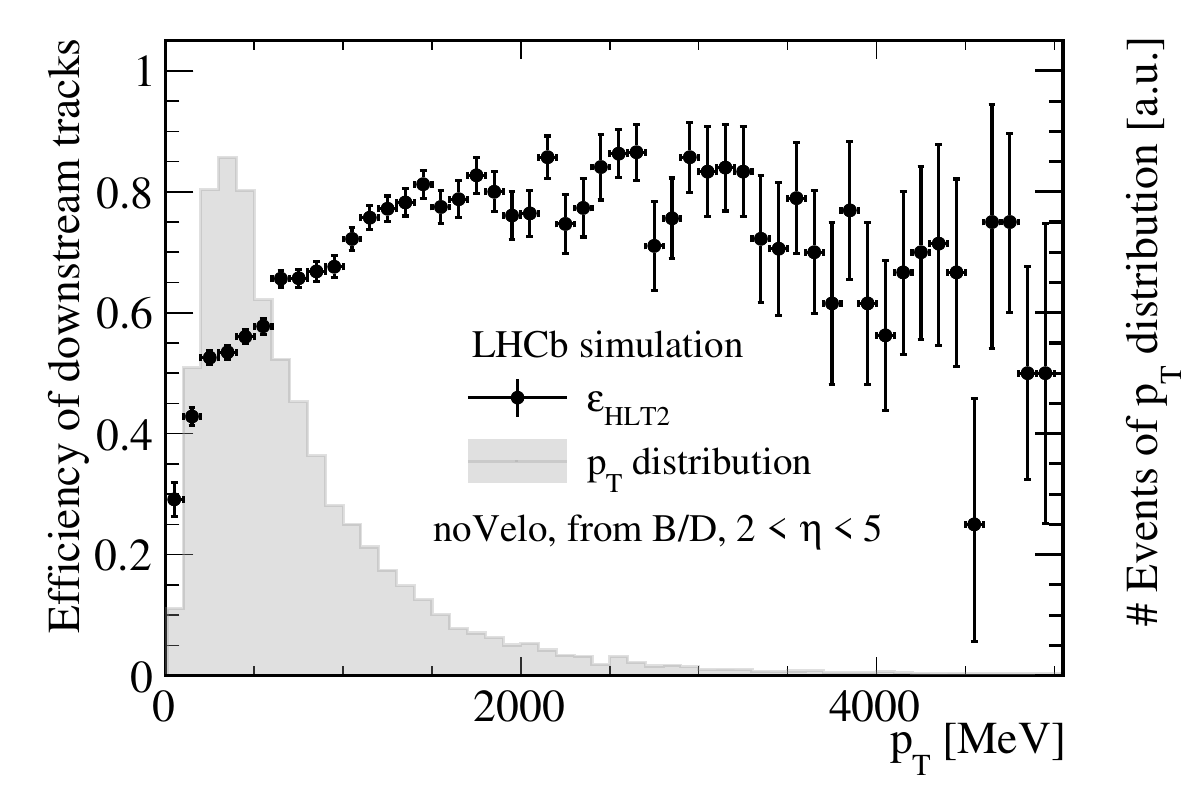}
  \\
  \includegraphics[scale=0.3]{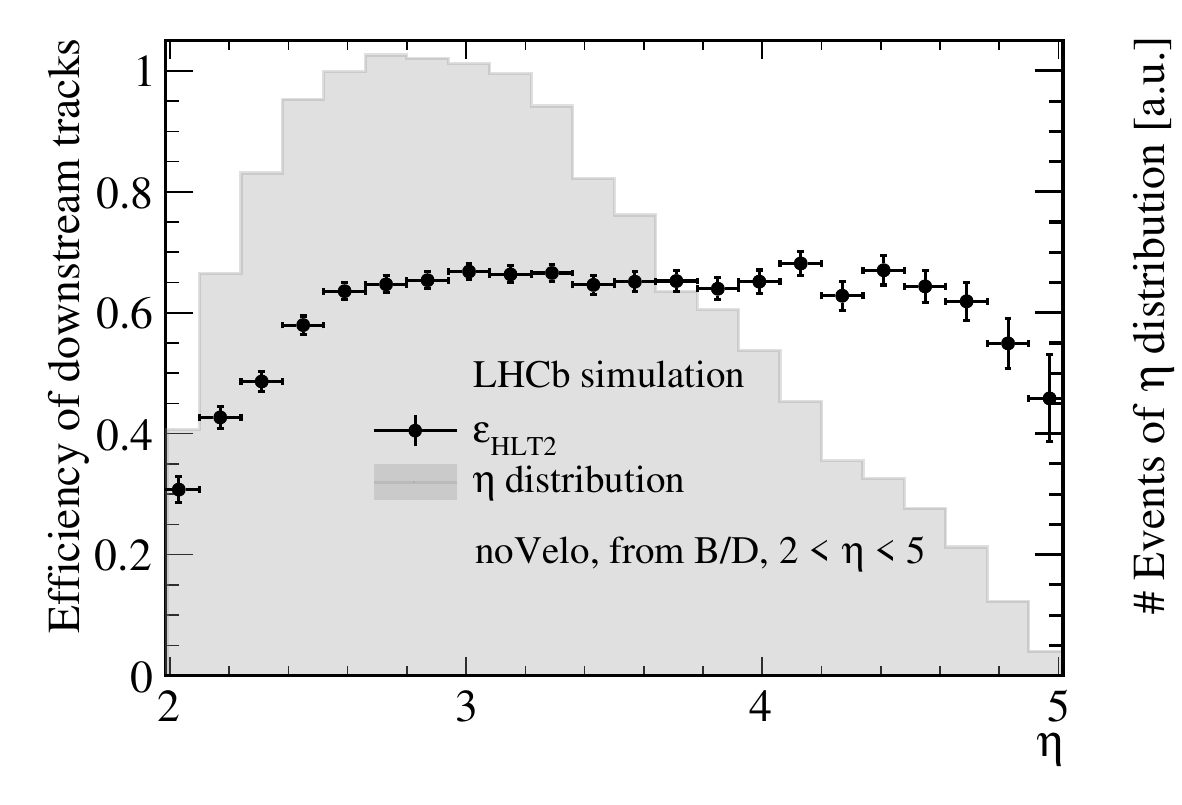}
  \qquad
  \includegraphics[scale=0.3]{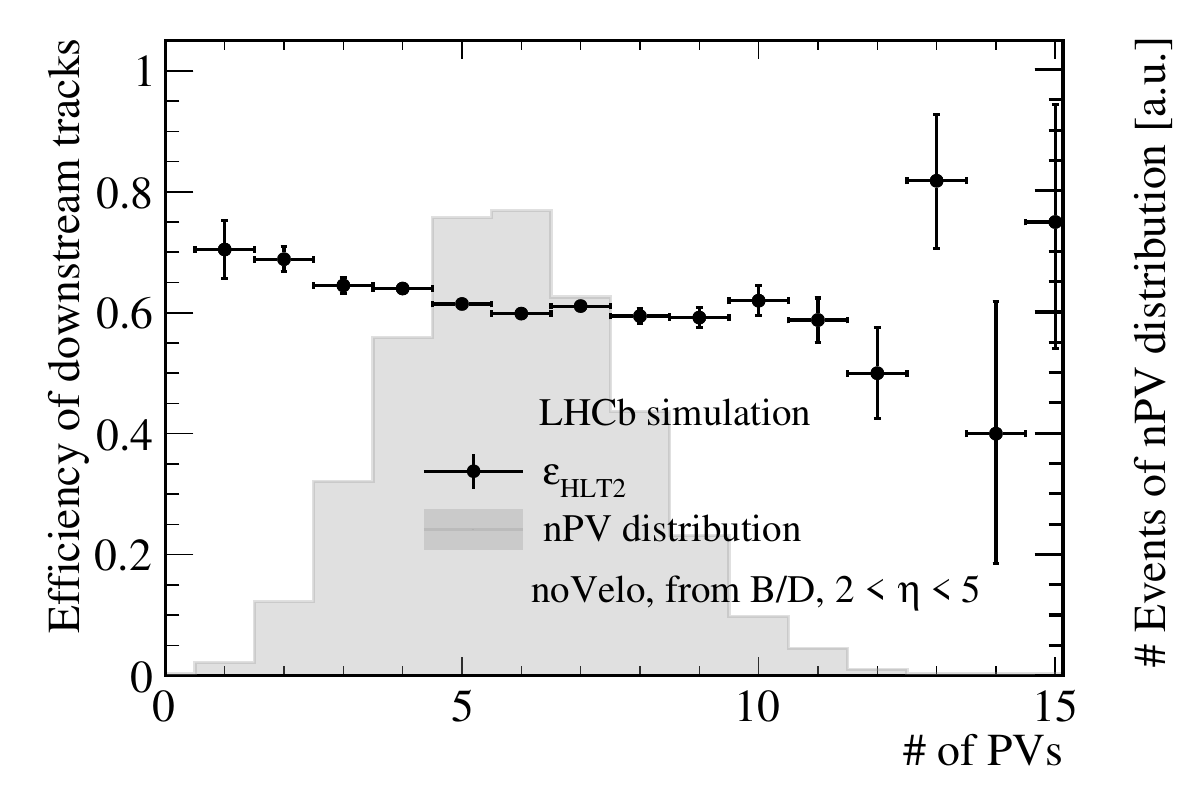}
  \caption{Downstream track reconstruction efficiency versus momentum
    \ptot, transverse momentum \pt, pseudo-rapidity \Peta, and number
    of primary vertices for reconstructible particles from \B/\D
    decays within $2<\Peta<5$ that have no hits in \Velo. Shaded
    histograms show the distributions of reconstructible particles. Reproduced with permission from~\cite{LHCB-FIGURE-2021-003}.}
  \label{fig:bestdown_fromBD}
\end{figure}

\begin{figure}[p]
  \centering
  \includegraphics[scale=0.32]{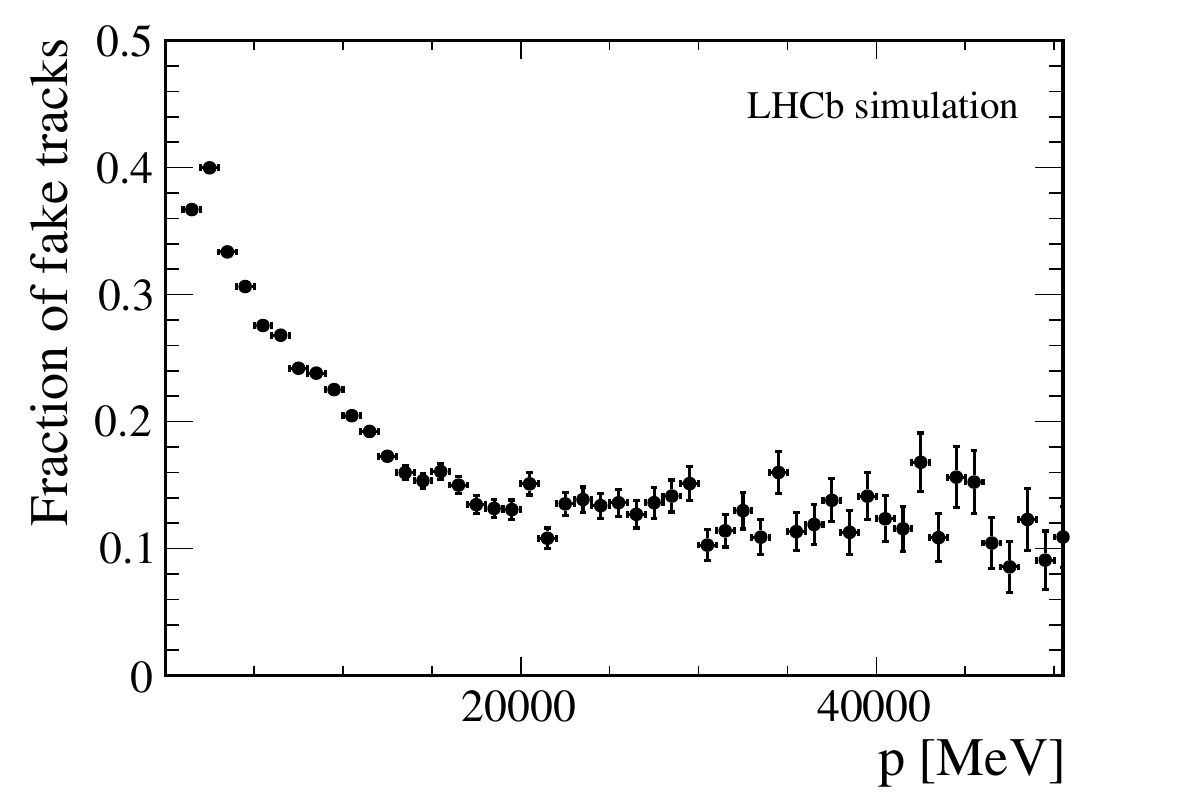}
  \qquad
  \includegraphics[scale=0.32]{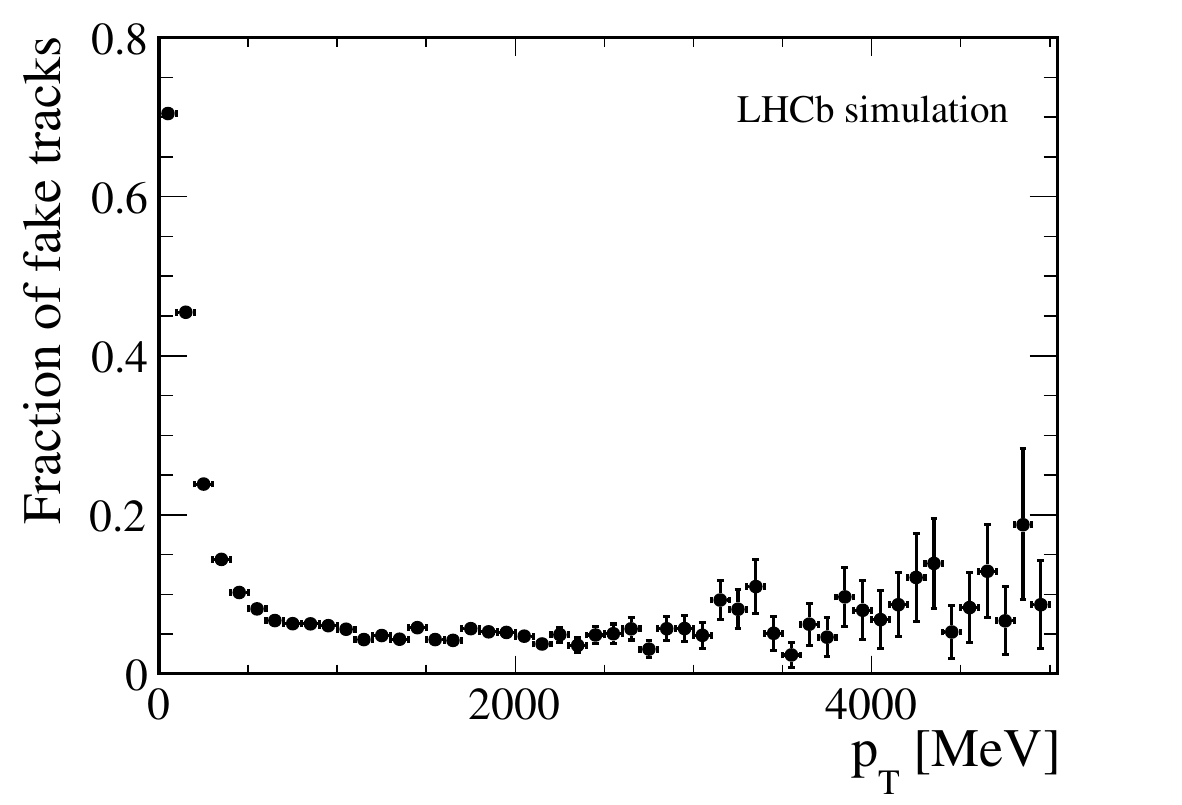}
  \\
  \includegraphics[scale=0.32]{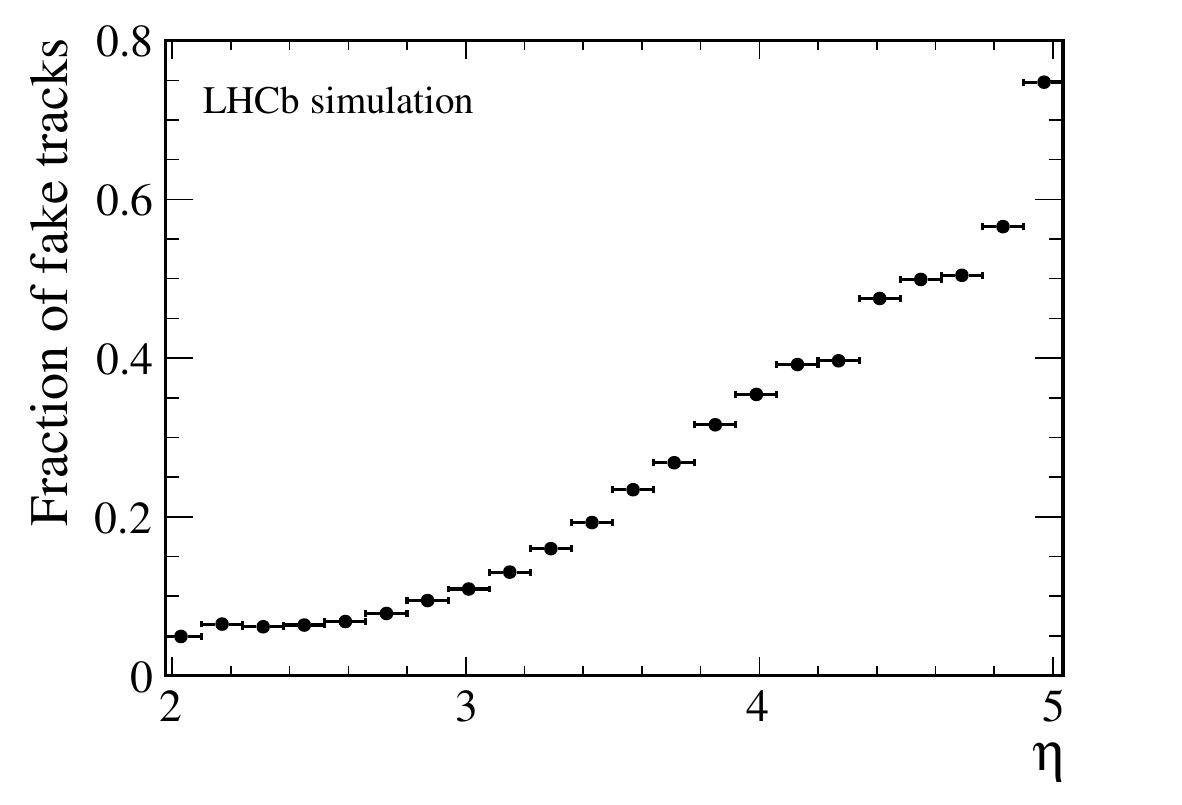}
  \qquad
  \includegraphics[scale=0.32]{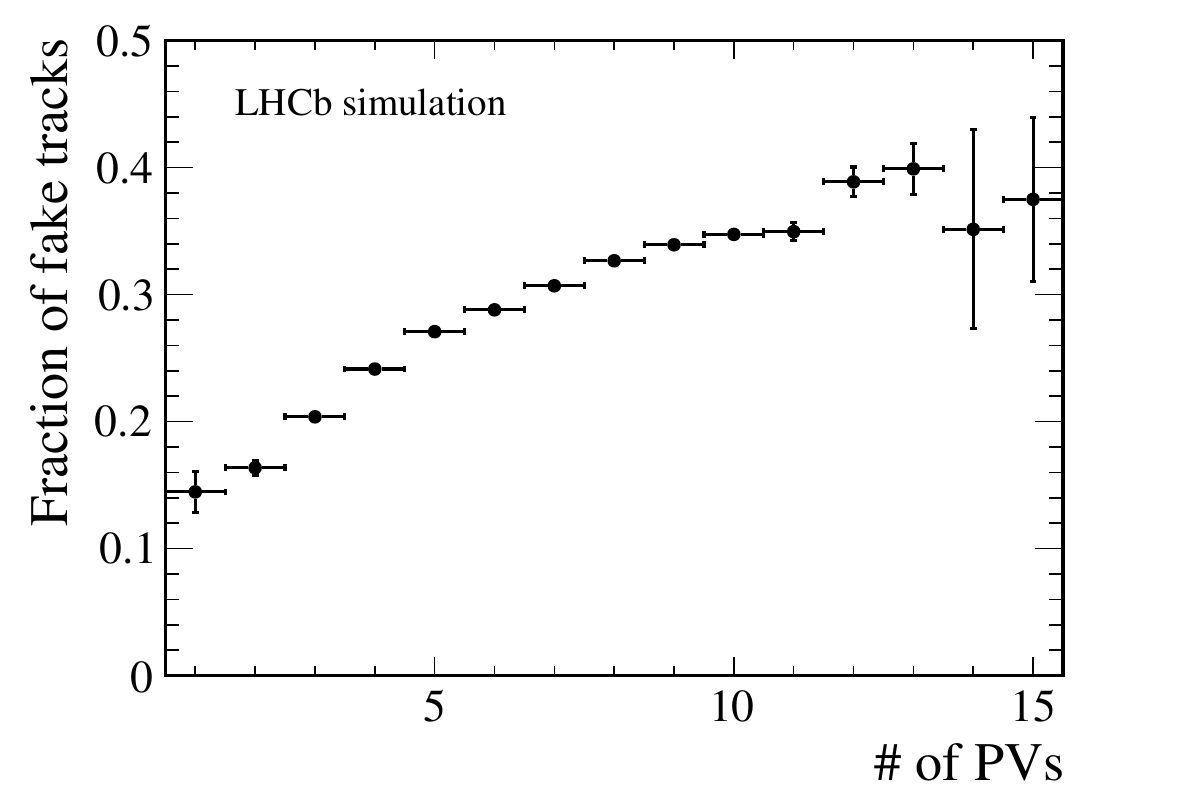}
  \caption{Ghost rate of downstream tracks reconstructed by the
    forward- and match-tracking algorithms as a function of momentum
    \ptot, transverse momentum \pt, pseudo-rapidity \Peta and number
    of primary vertices. Reproduced with permission from~\cite{LHCB-FIGURE-2021-003}.}
  \label{fig:bestdown_ghostrate}
\end{figure}

\begin{figure}[p]
  \centering
  \includegraphics[scale=0.32]{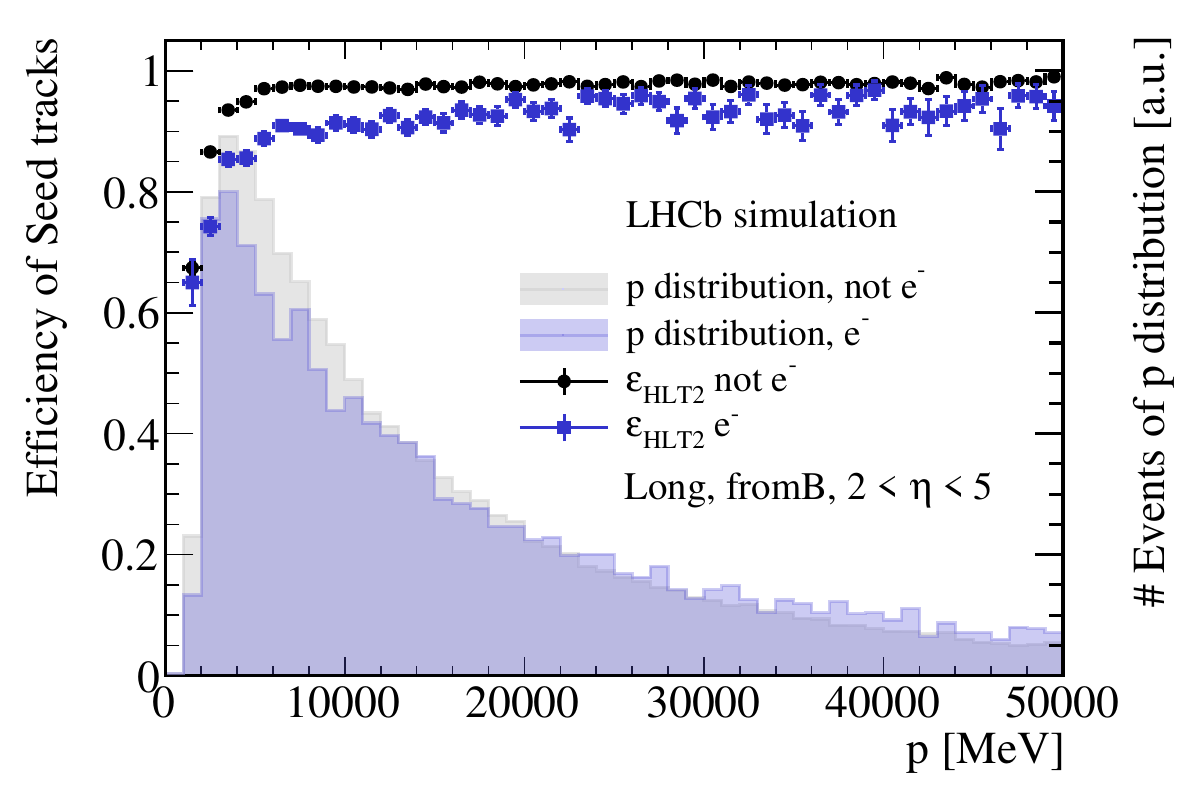}
  \qquad
  \includegraphics[scale=0.32]{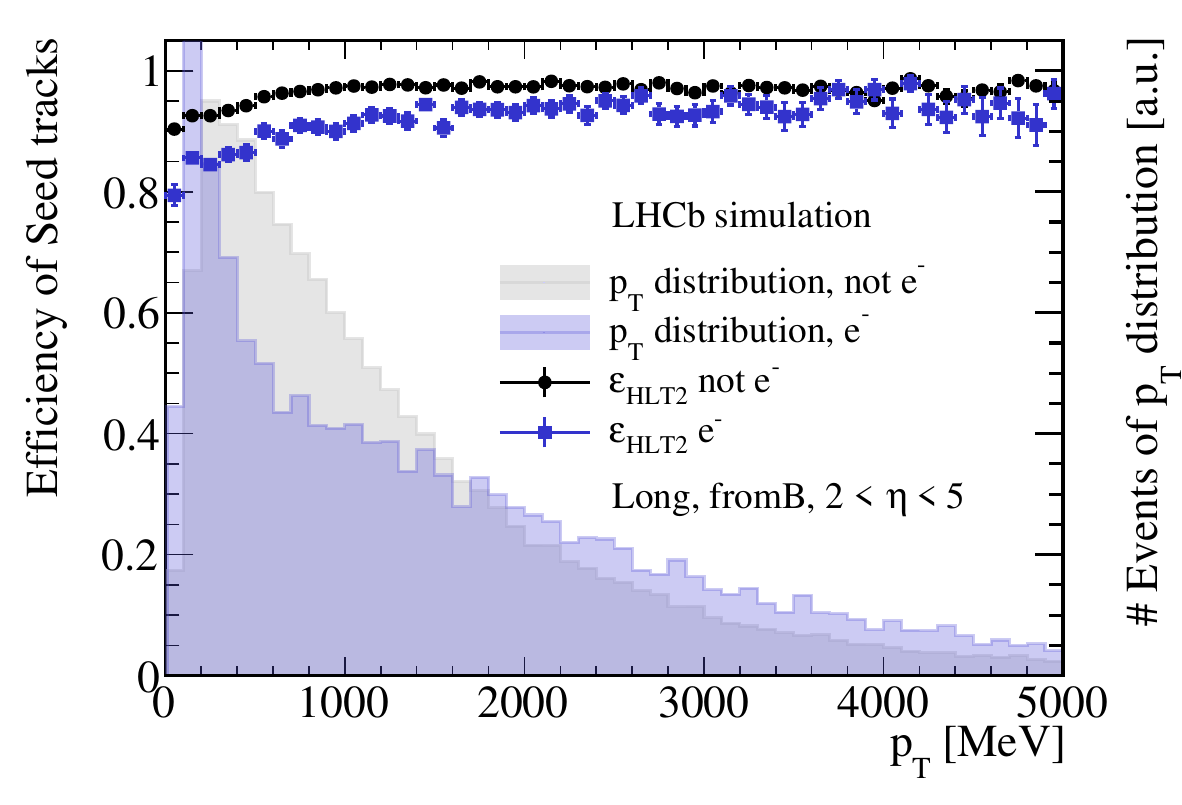}
  \\
  \includegraphics[scale=0.32]{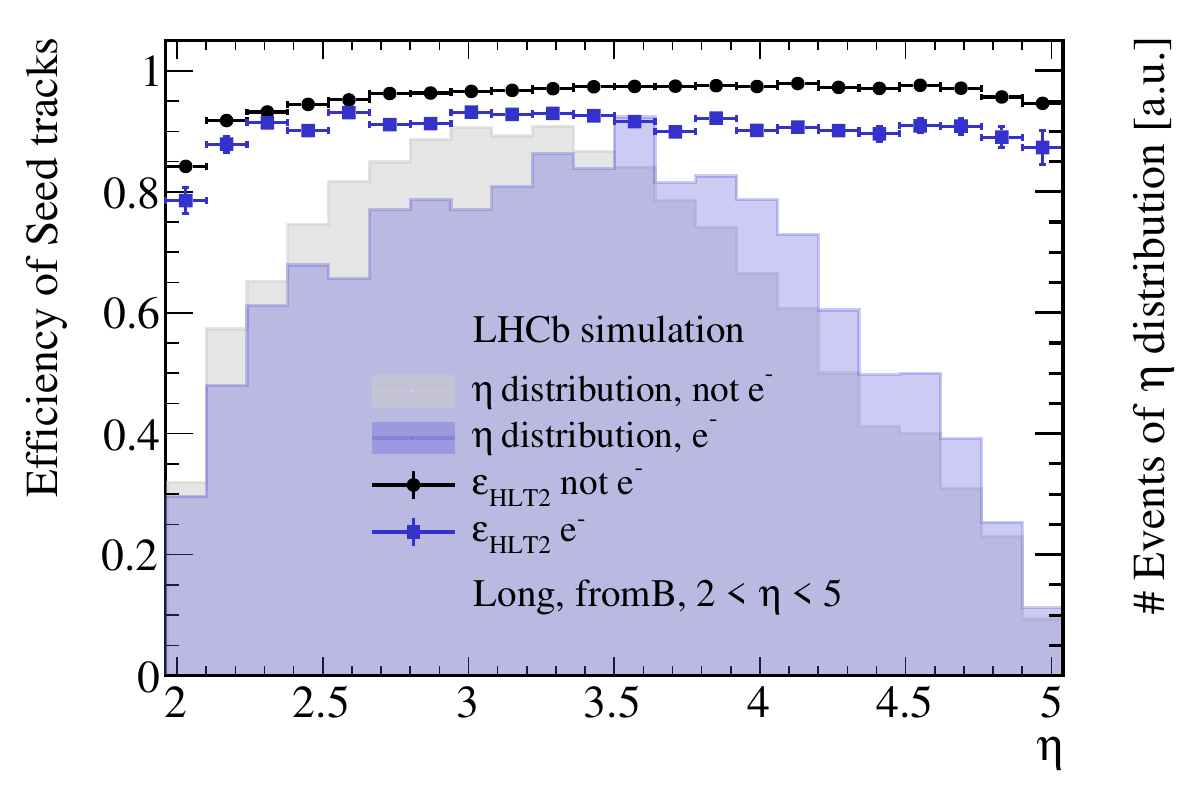}
  \qquad
  \includegraphics[scale=0.32]{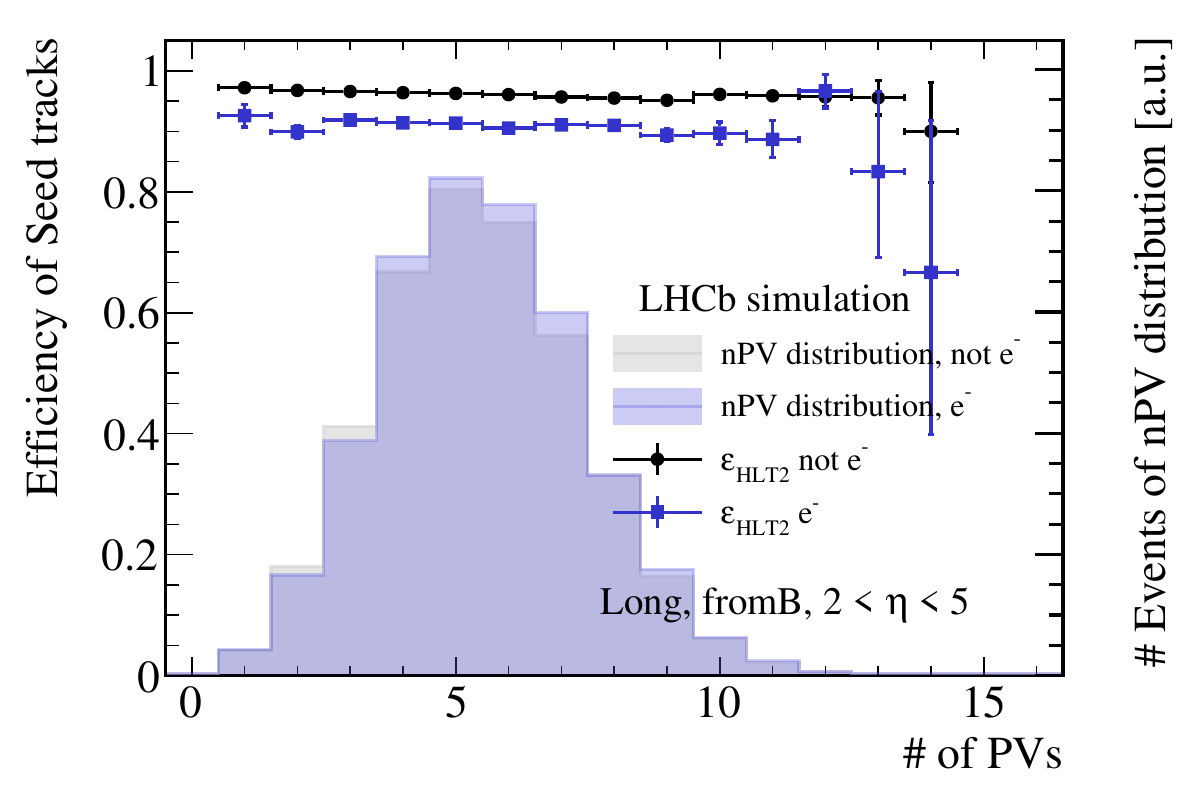}
  \caption{Seeding track-reconstruction efficiency versus momentum
    \ptot, transverse momentum \pt, pseudo-rapidity \Peta, and number
    of primary vertices for long reconstructible electrons (blue
    squares) and non-electron (black dots) particles within
    $2<\Peta<5$. Shaded histograms show the distributions of
    reconstructible particles. Reproduced with permission from~\cite{LHCB-FIGURE-2021-003}.}
  \label{fig:seed_eff}
\end{figure}

\begin{figure}[p]
  \centering
  \includegraphics[scale=0.32]{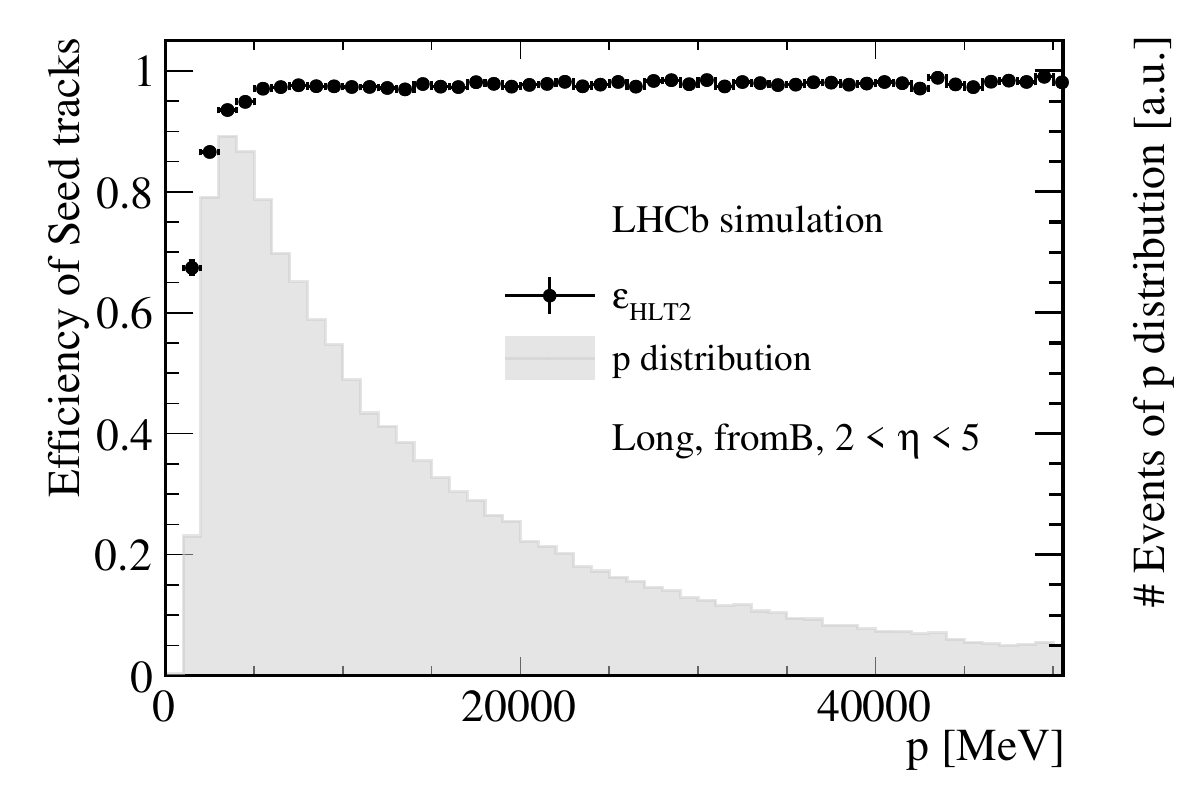}
  \qquad
  \includegraphics[scale=0.32]{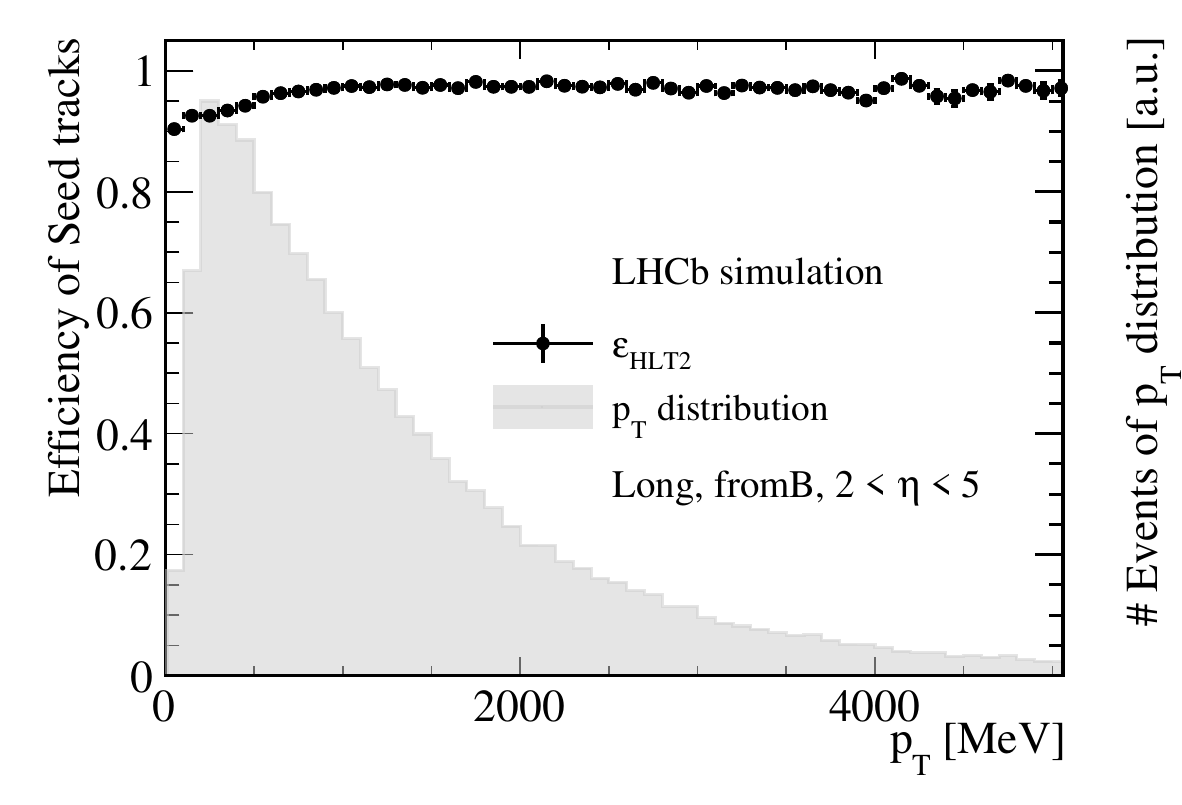}
  \\
  \includegraphics[scale=0.32]{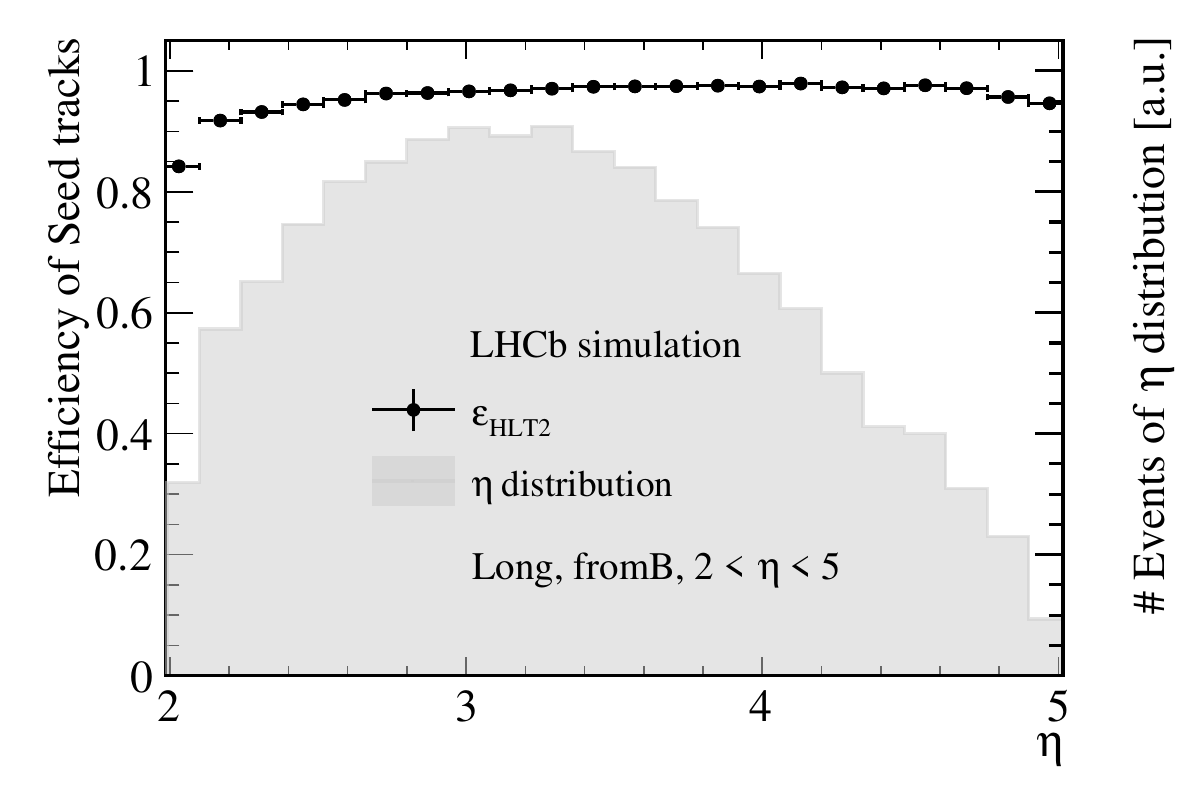}
  \qquad
  \includegraphics[scale=0.32]{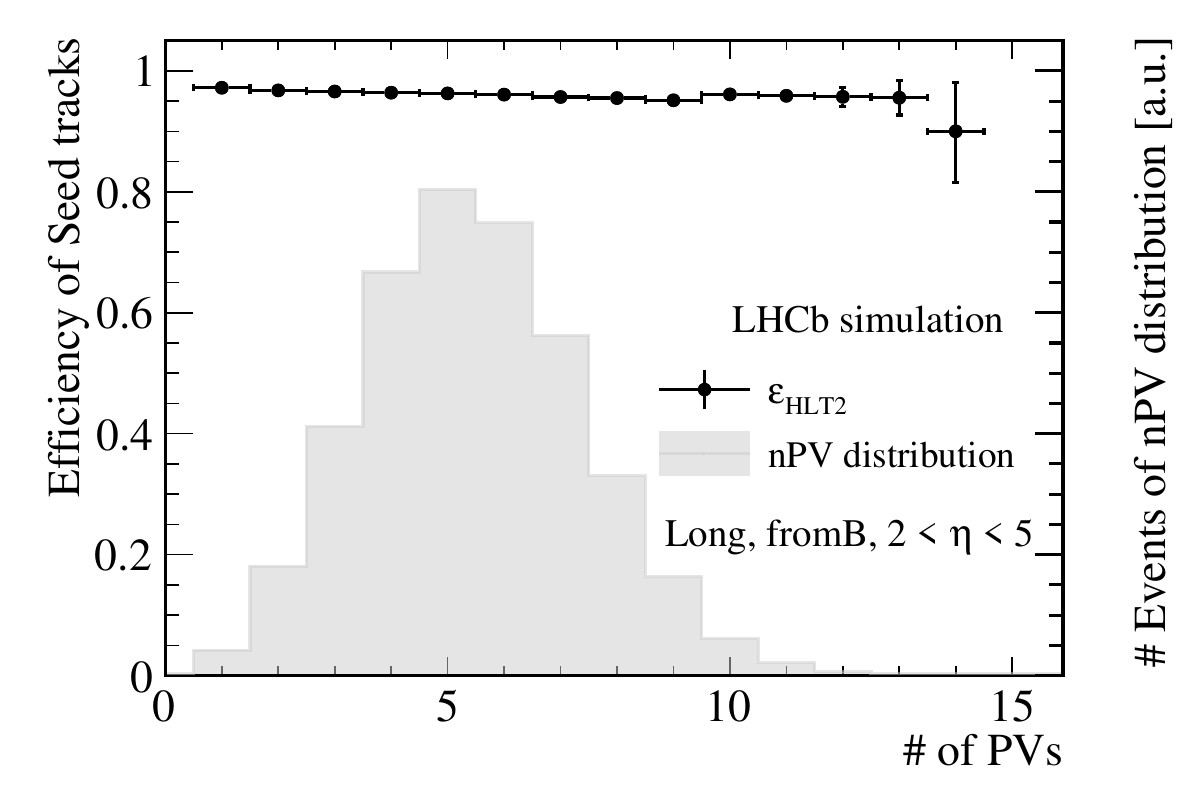}
  \caption{Seeding track reconstruction efficiency versus momentum
    \ptot, transverse momentum \pt, pseudo-rapidity \Peta, and number
    of primary vertices for long reconstructible particles from \B
    decays within $2<\Peta<5$. Shaded histograms show the
    distributions of reconstructible particles. Reproduced with permission from~\cite{LHCB-FIGURE-2021-003}.}
  \label{fig:seed_effall}
\end{figure}

\begin{figure}[p]
  \centering
  \includegraphics[scale=0.32]{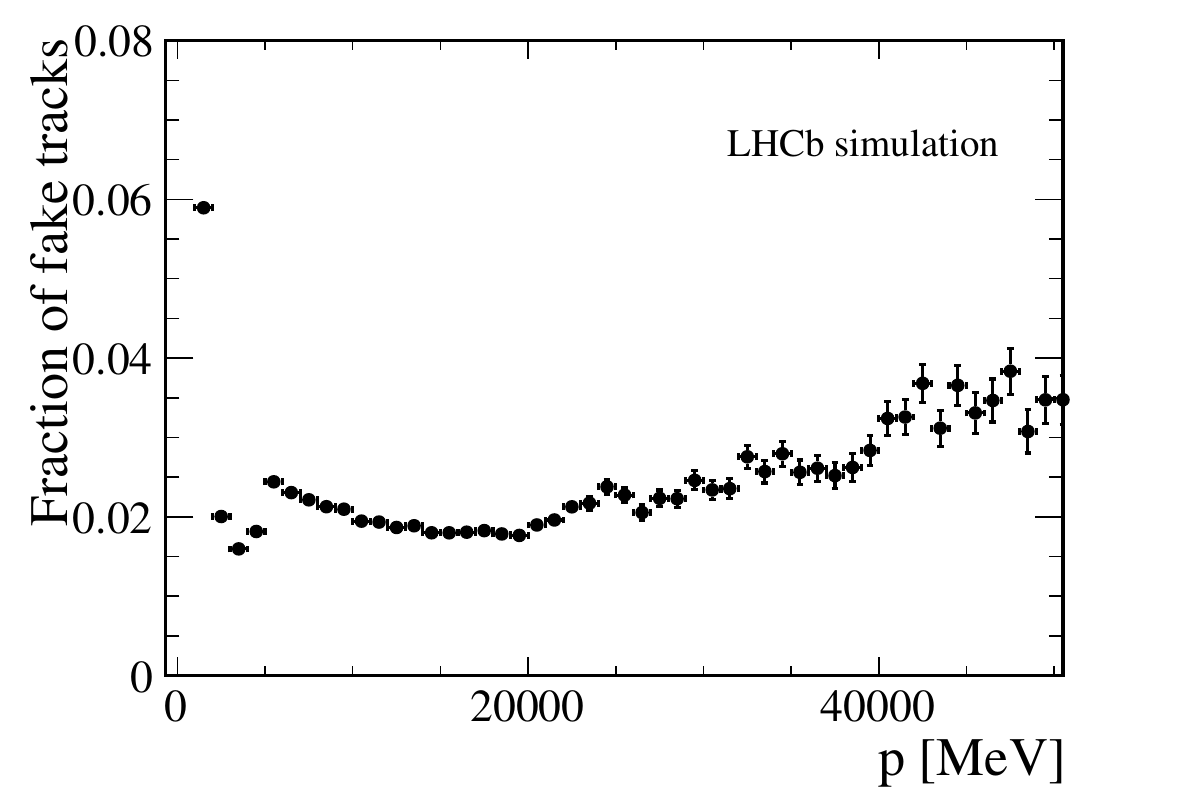}
  \qquad
  \includegraphics[scale=0.32]{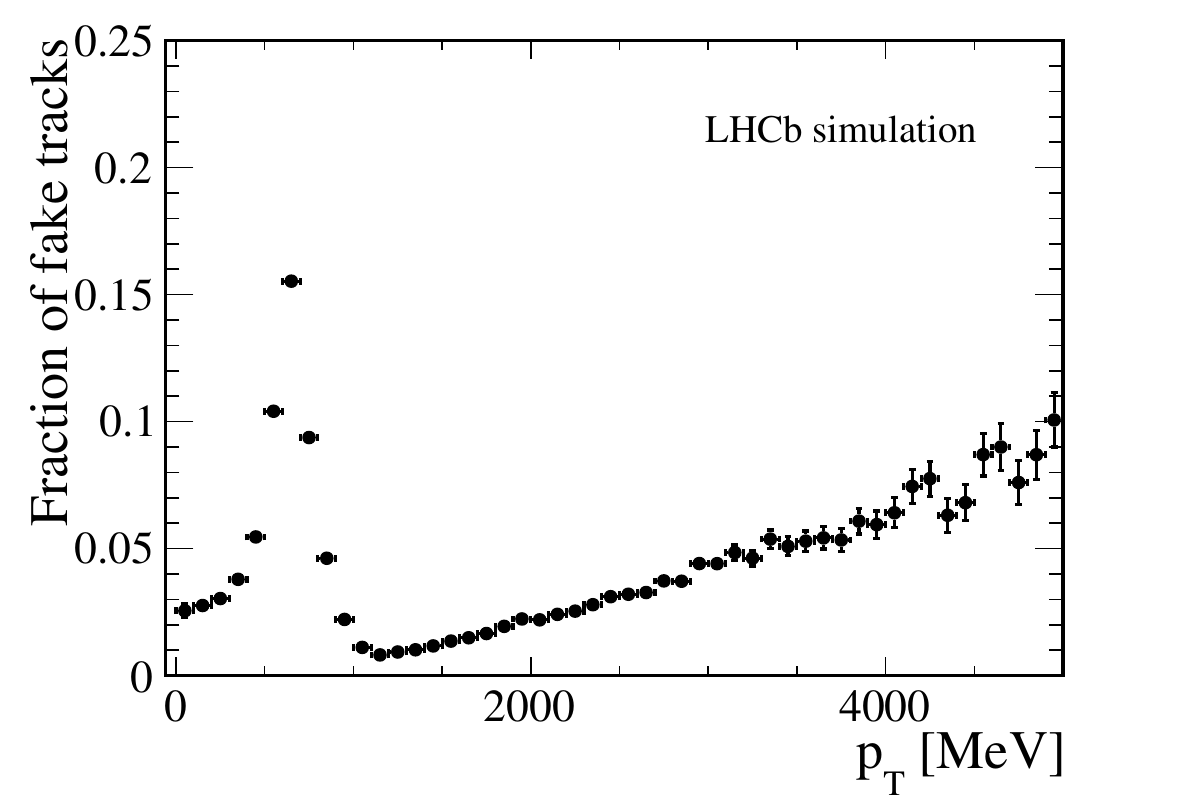}
  \\
  \includegraphics[scale=0.32]{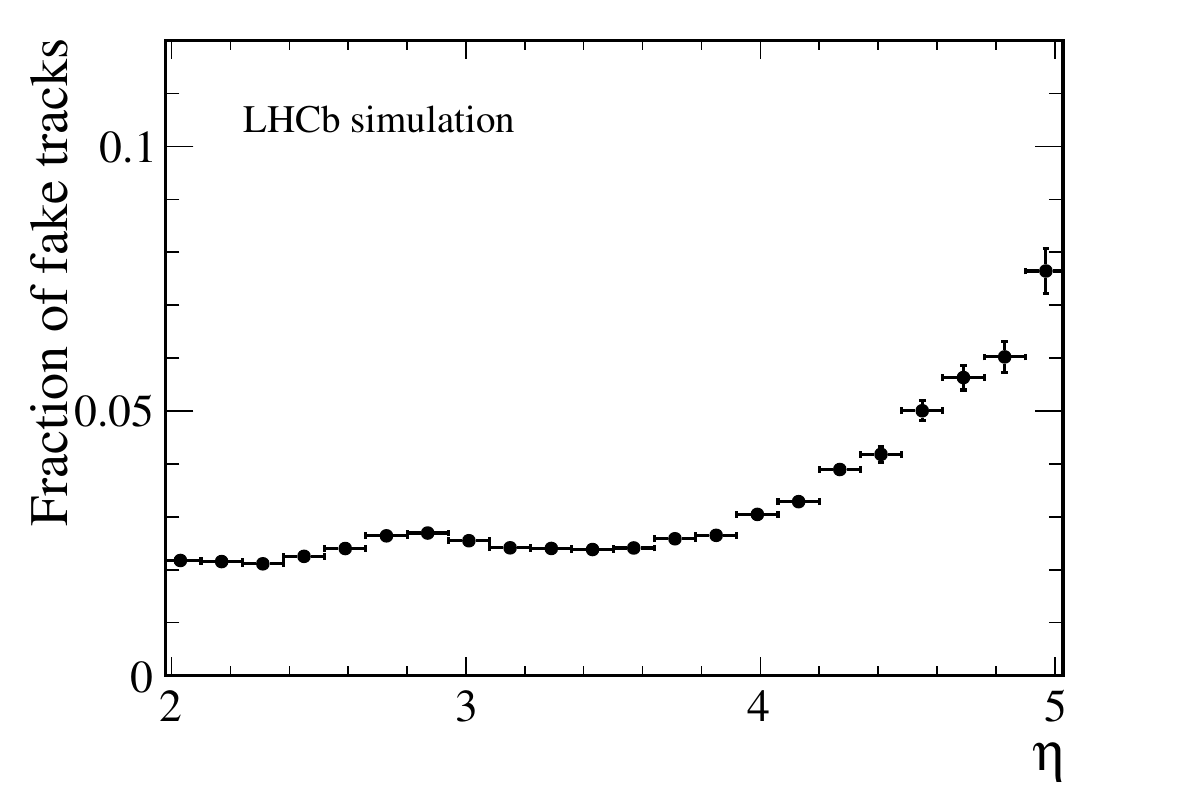}
  \qquad
  \includegraphics[scale=0.32]{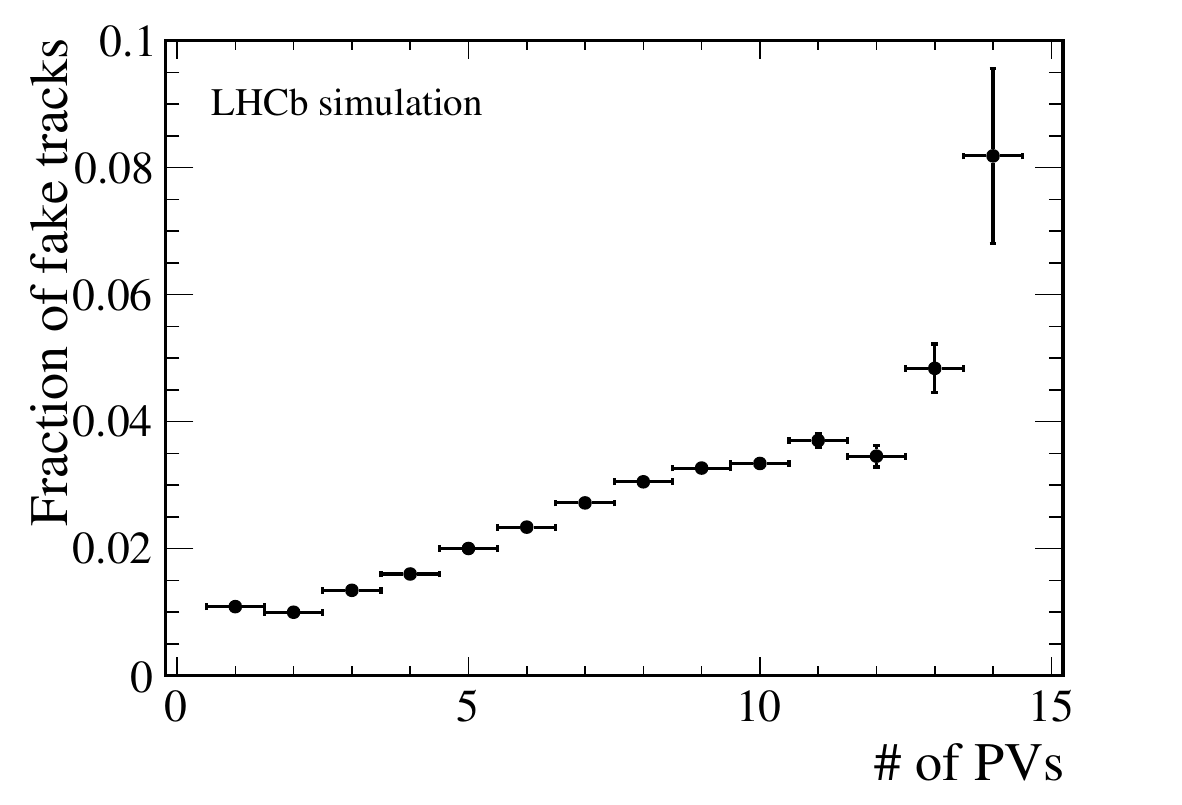}
  \caption{Ghost rate of standalone seeding tracks as a function of
    momentum \ptot, transverse momentum \pt, pseudo-rapidity \Peta,
    and number of primary vertices. The prominent peak in the ghost
    rate at low transverse momentum (top-right panel) results from a
    combination of geometric and kinematic effects. Reproduced with permission from~\cite{LHCB-FIGURE-2021-003}.}
  \label{fig:seed_ghostrate}
\end{figure}

\begin{figure}[p]
\centering
\includegraphics[scale=0.32]{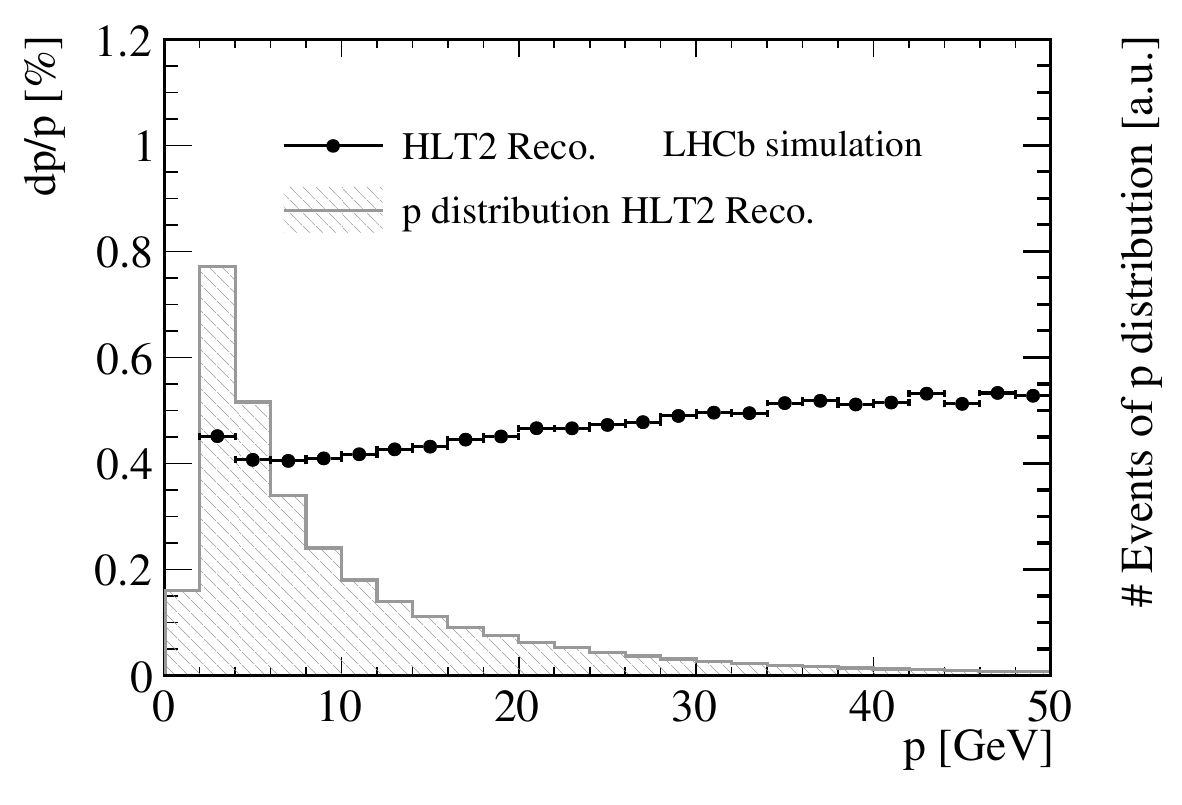}
\qquad
\includegraphics[scale=0.32]{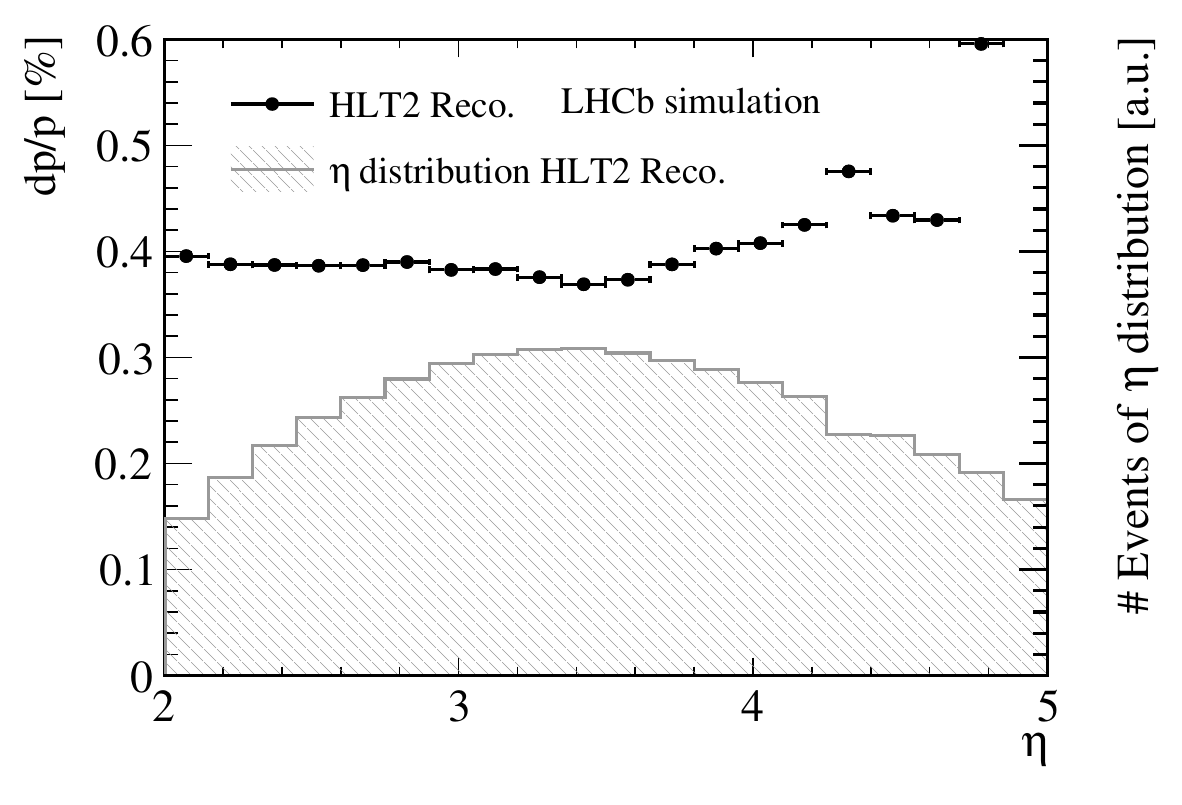}
\caption{Relative resolution of the momentum of reconstructed
  tracks as a function of momentum \ptot, and  pseudo-rapidity $\Peta$. Reproduced with permission from~\cite{LHCB-FIGURE-2021-003}.} 
\label{fig:trackres}
\end{figure}

\begin{figure}[p]
  \centering
  \includegraphics[scale=0.32]{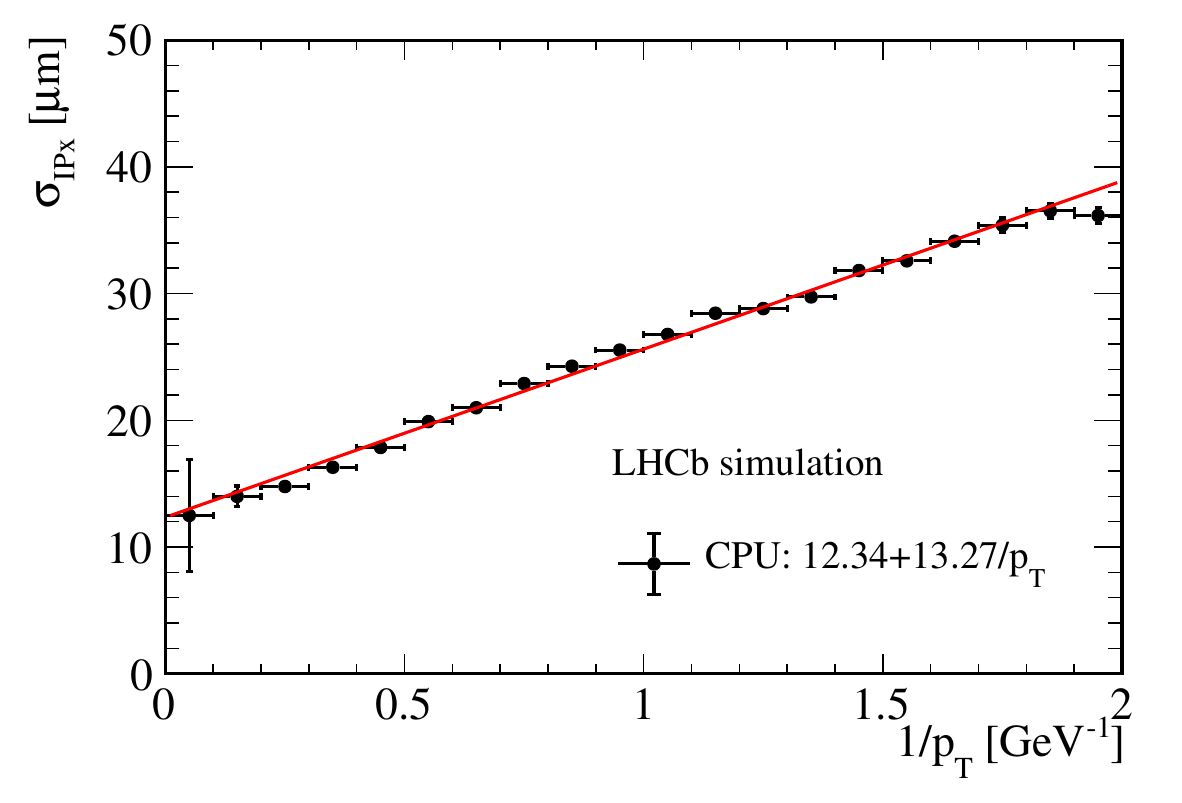}
  \qquad
  \includegraphics[scale=0.32]{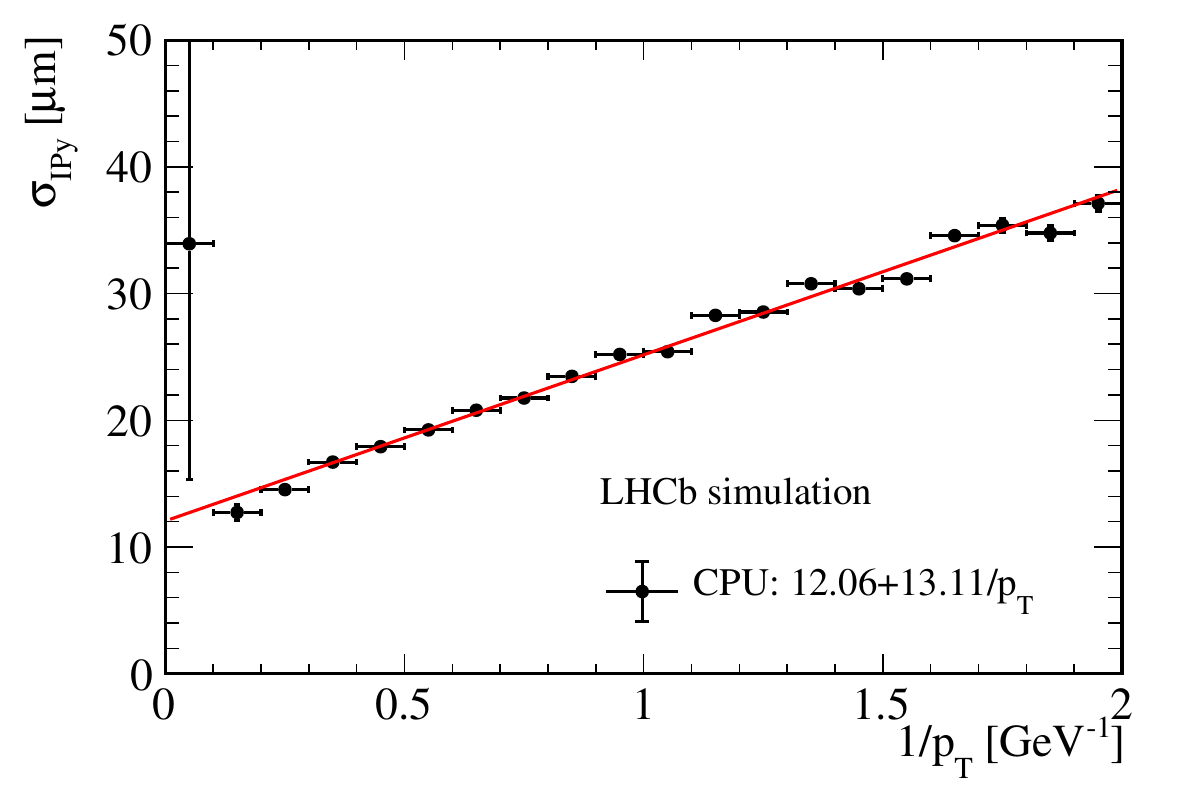}
  \caption{Resolution of the $x$ projection of the impact parameter,
    $\sigma_{\mathrm{IPx}}$ (left) and $\sigma_{\mathrm{IPy}}$ (right)
    as a function of the inverse of transverse momentum $1/\pt$. A
    minimum bias sample is used for the \Ip resolution study. Reproduced with permission from~\cite{LHCB-FIGURE-2021-003}.}
  \label{fig:IPres}
\end{figure}

\begin{figure}[t]
  \centering
  \includegraphics[width=0.46\textwidth]{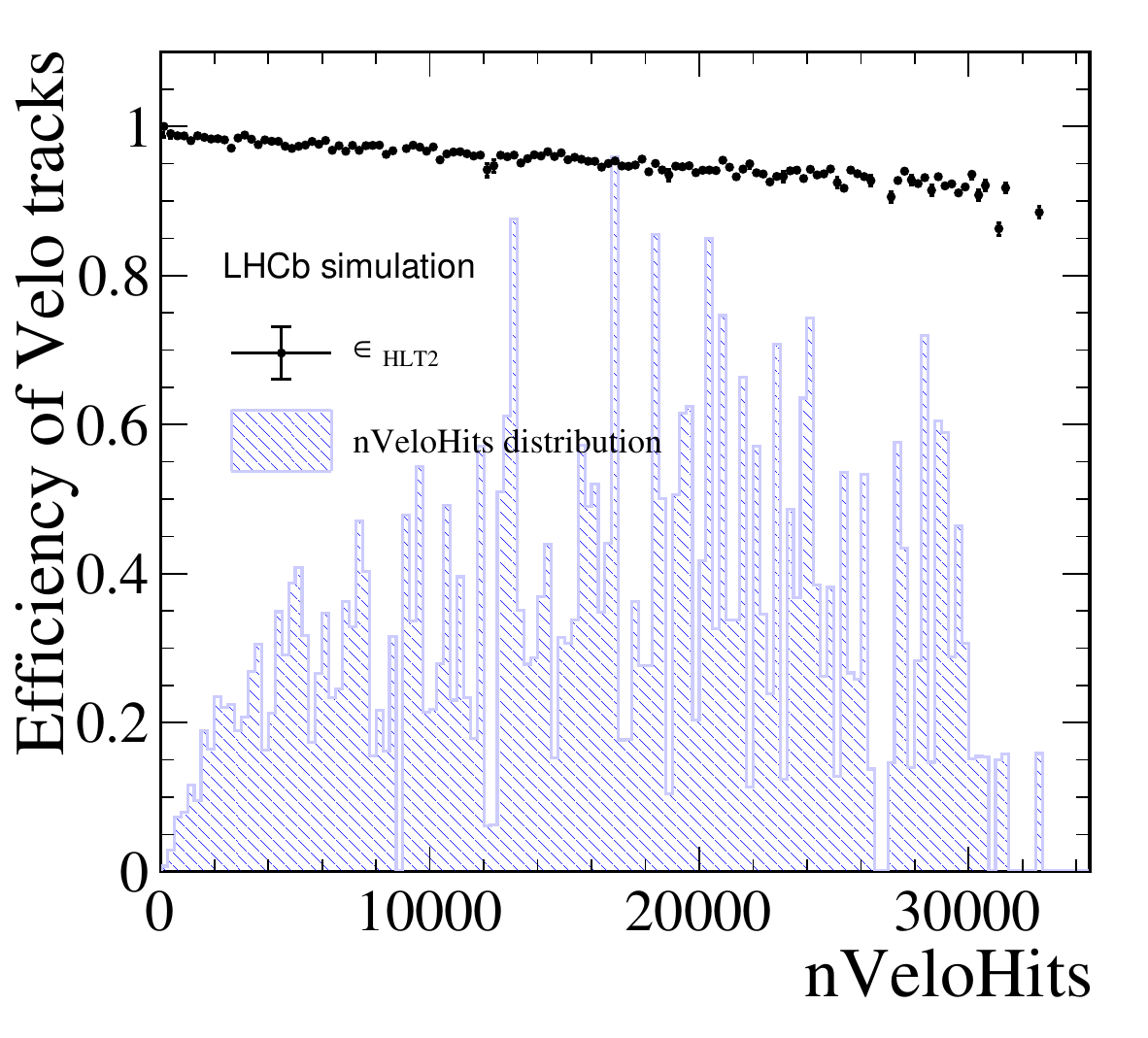}
  \qquad
  \includegraphics[width=0.46\textwidth]{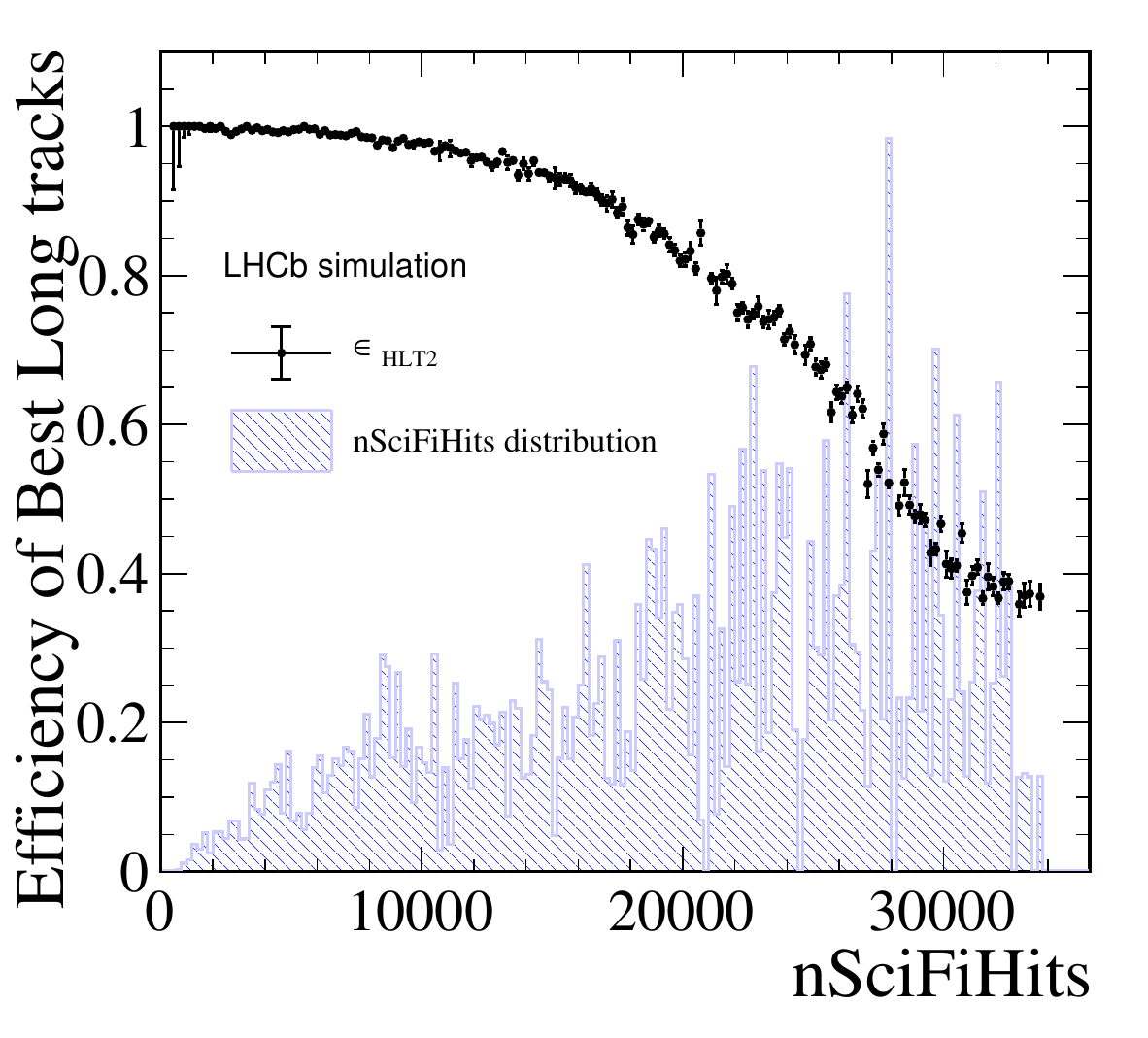}
  \caption{Reconstruction efficiency of (left) \Velo and (right) long
    tracks as a function of the occupancy in the vertex detector and
    \Scifi, respectively. Reproduced with permission from~\cite{LHCB-FIGURE-2021-003}.}
  \label{fig:perf_ions}
\end{figure}

The tracking performance in the baseline \hlttwo reconstruction is
shown in terms of reconstruction efficiency and corresponding
fake-track reconstruction rate (\emph{ghost rate}) for long and
downstream tracks as a function of momentum \ptot, transverse momentum
\pt, pseudo-rapidity \Peta, and number of primary vertices.
Figures~\ref{fig:bestlong} and~\ref{fig:bestlong_all} show the track
reconstruction efficiency for long tracks originating from \B-meson
decays while the corresponding fake-track rate is shown in
figure~\ref{fig:bestlong_ghostrate}.  The downstream-track
reconstruction efficiency for particles originating in strange and \B
or \D decays is shown in Figs~\ref{fig:bestdown}
and~\ref{fig:bestdown_fromBD}, respectively; the corresponding
fake-track rate is reported in
figure~\ref{fig:bestdown_ghostrate}. Finally the seeding
reconstruction efficiency for particles originating in \B decays and
reconstructible\footnote{In the context of this section
  \emph{reconstructible} means an object (track, vertex) or an event
  with visible activity in the detector that satisfies some minimum
  requirements to allow its reconstruction by software algorithms.} as
long tracks is shown in figures~\ref{fig:seed_eff}
and~\ref{fig:seed_effall}, and the corresponding fake-track rate is
displayed in figure~\ref{fig:seed_ghostrate}.
Figures~\ref{fig:trackres} and~\ref{fig:IPres} show the momentum and
\Ip resolution for tracks after the Kalman fit. Finally, the tracking
efficiency is measured as a function of detector occupancy in
lead-lead collisions, in order to understand the suitability of the
upgraded LHCb detector for heavy ion data taking. This is shown in
figure~\ref{fig:perf_ions}. A gradual degradation of performance is
observed with occupancy but, encouragingly, there is no sharp edge
where performance collapses.\looseness=-1

\begin{figure}[t]
  \centering
  \includegraphics[width=0.8\textwidth]{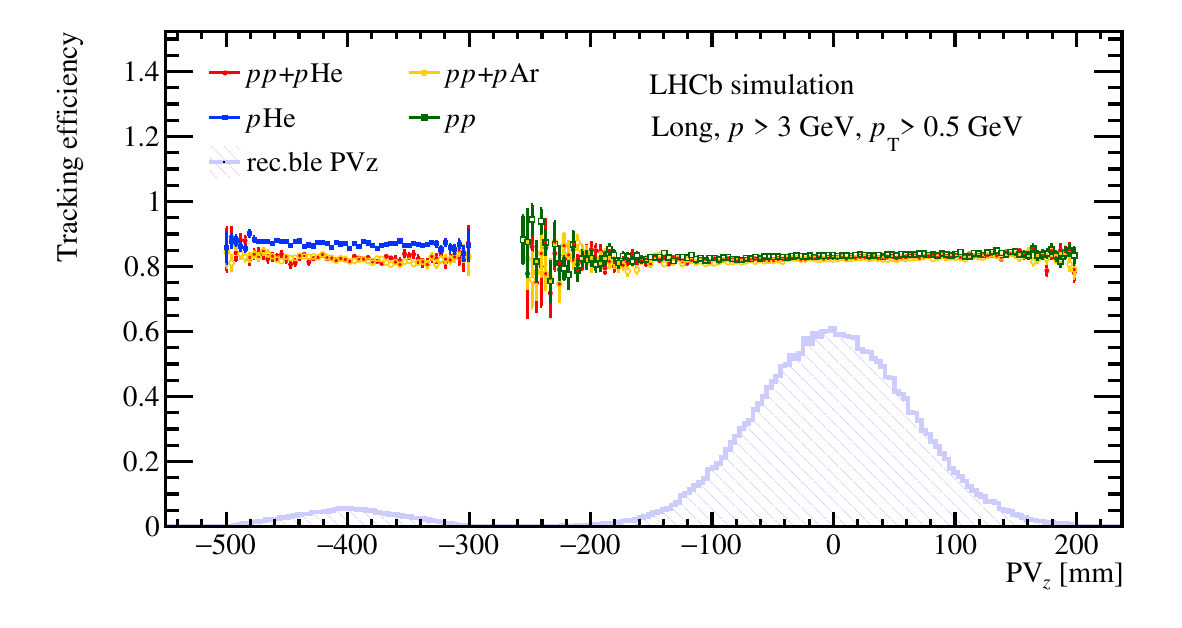}
  \includegraphics[width=0.8\textwidth]{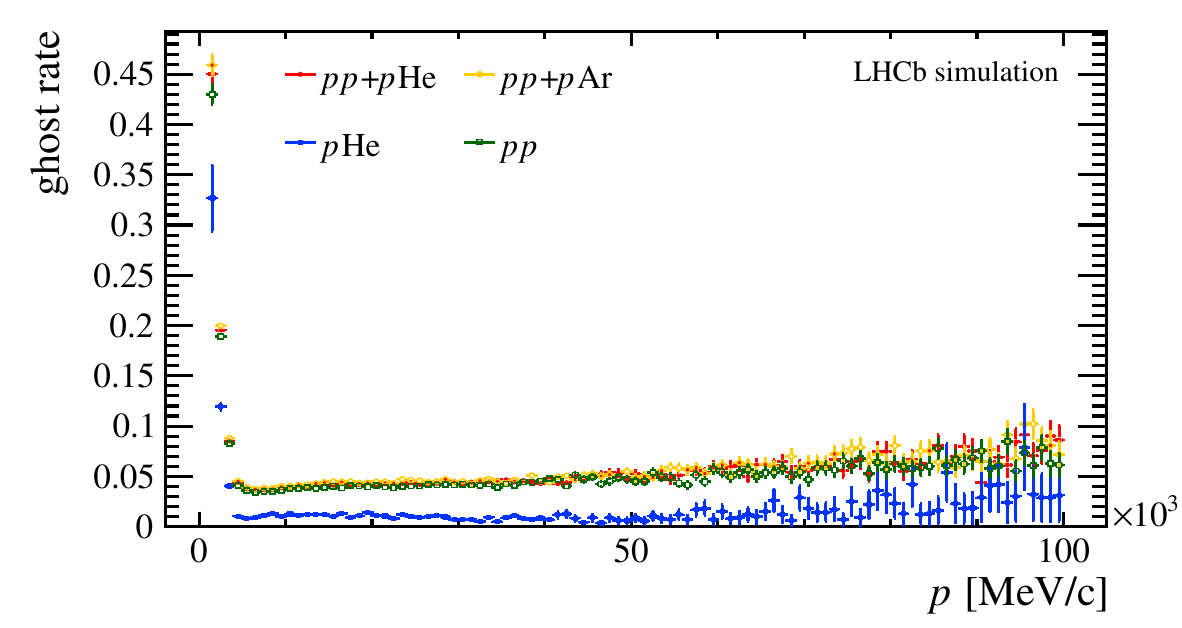}
  \caption{Top: track reconstruction efficiency as a function of the
    $z$ position of the primary vertex ($PV_z$), for simulated samples
    with stand-alone (blue) \pHe and (green) \Prpr, and overlapping
    (red) \Prpr $+$ \pHe and (orange) \Prpr $+$ \pAr collisions. The
    distribution of $PV_z$ for reconstructible \Pv{s} is also shown
    (shaded histogram, arbitrary units). Bottom: corresponding rate of
    fake reconstructed tracks as a function of track momentum \ptot. Reproduced with permission from~\cite{LHCB-FIGURE-2021-003}.}
  \label{fig_SMOG2:TrackReco}
\end{figure}

\begin{figure}[t]
  \centering
  \includegraphics[width=0.79\textwidth]{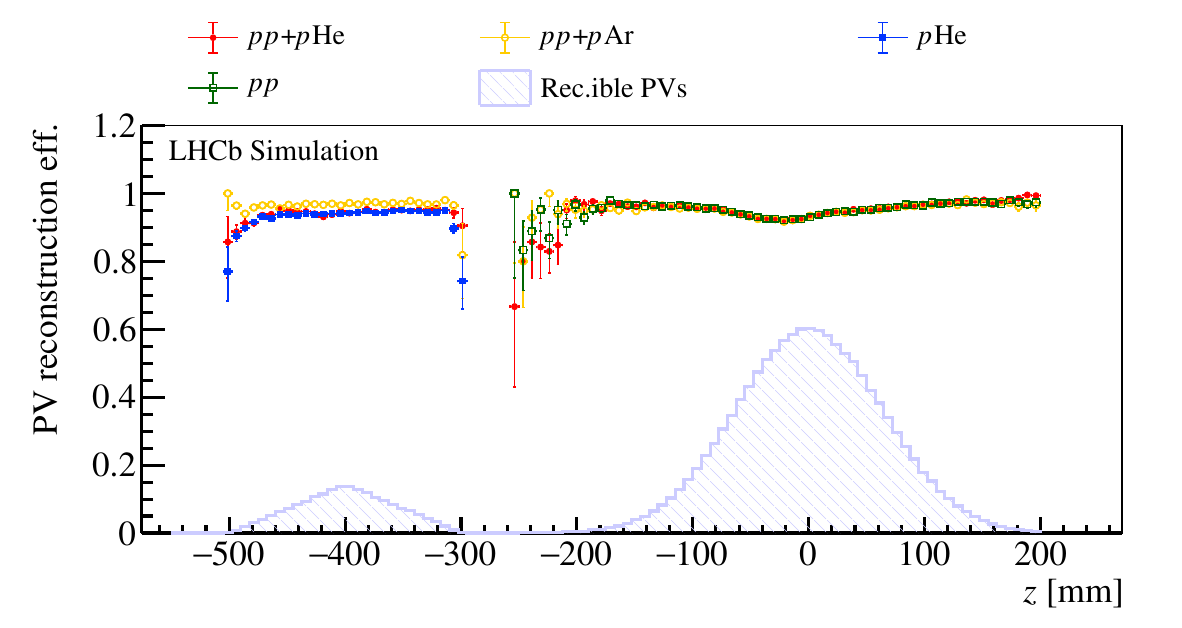}
  \includegraphics[width=0.79\textwidth]{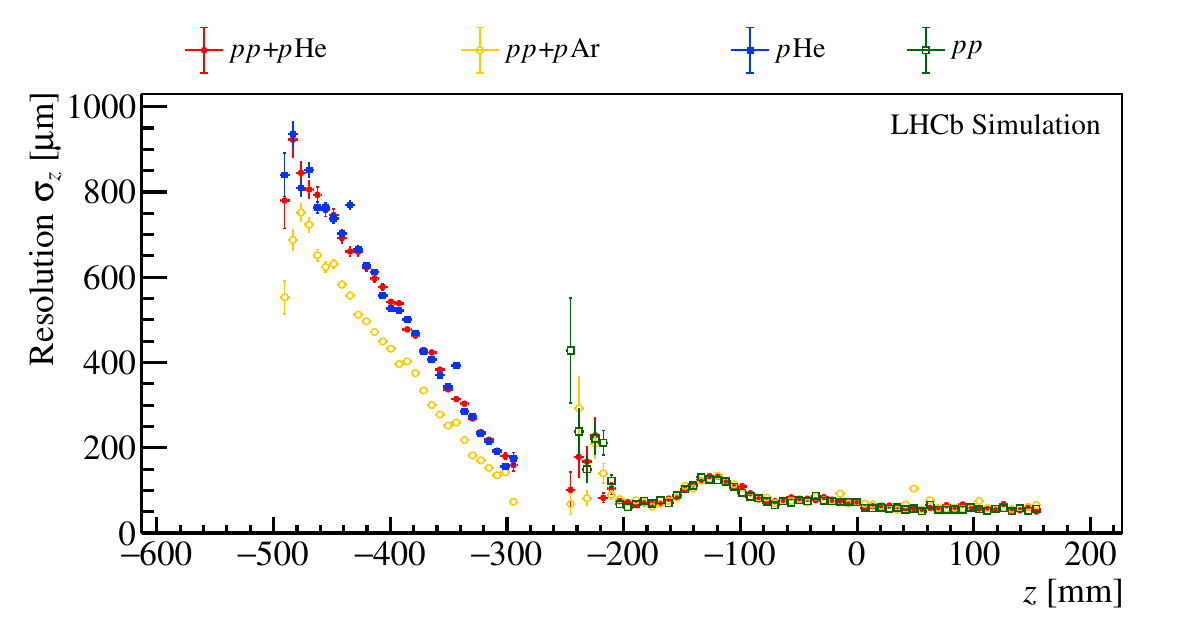}
  \caption{Primary vertex reconstruction (top) efficiency and (bottom)
    resolution as a function of the $z$ coordinate measured on
    simulated samples with stand-alone (green) \Prpr, (blue) \pHe and
    overlapping (red) \Prpr $+$ \pHe and (orange) \Prpr $+$ \pAr
    collisions. Reproduced with permission from~\cite{LHCB-FIGURE-2021-003}.}
  \label{fig_SMOG2:PVReco}
\end{figure}

A simultaneous data taking of beam-beam and beam-gas collisions can be
envisaged due to the new SMOG design, with well separated gas-target
and beam-crossing regions, and since the expected rate of beam-gas
collisions is one order of magnitude smaller than the beam-beam
rate. Due to the tight constraints of the online reconstruction
framework discussed in section~\ref{sec:trigger}, the simultaneous
data-taking must not spoil the track reconstruction performance in
\Prpr collisions while ensuring good tracking efficiency for beam-gas
events.  Results of the studies carried out on simulated samples of
stand-alone proton-helium (\pHe) collisions and with overlapping \Prpr
and \pHe or \pAr are shown here. The upgrade luminosity conditions and
one proton-gas collision per event confined in the SMOG cell region
$z \in [-500, -300] \mm$ according to the expected gas pressure into
the storage cell are assumed.  The track reconstruction efficiency for
stand-alone \pHe and \Prpr collisions, and for overlapping \Prpr and
\pHe or \pAr are shown in figure~\ref{fig_SMOG2:TrackReco} as a
function of the $z$ position of the primary vertex ($PV_z$), along
with the corresponding fake track rate as a function of the track
momentum \ptot.  The results are obtained with long tracks with
$\ptot > 3\gev$ and $\pt > 0.5\gev$.
The primary vertex reconstruction efficiency is also investigated with
the same simulated event samples.  The results are shown in
figure~\ref{fig_SMOG2:PVReco}. In the top plot the \Pv reconstruction
efficiency is reported as a function of the $z$ coordinate. The
distribution of the reconstructible \Pv{s} (in arbitrary units) is
also shown. In the bottom plot, the \Pv $z$ resolution is shown as a
function of $z$.
The time needed by the \Hltone application to process overlapping
beam-gas and \Prpr events increases by about 1--3\% with respect to
\Prpr collision processing time.  These studies demonstrate the
feasibility of simultaneous running in \Prpr and fixed-target mode.\looseness=-1

\subsubsection{Calorimeter performance}

\begin{figure}[p]
  \centering
  \includegraphics[scale=0.32]{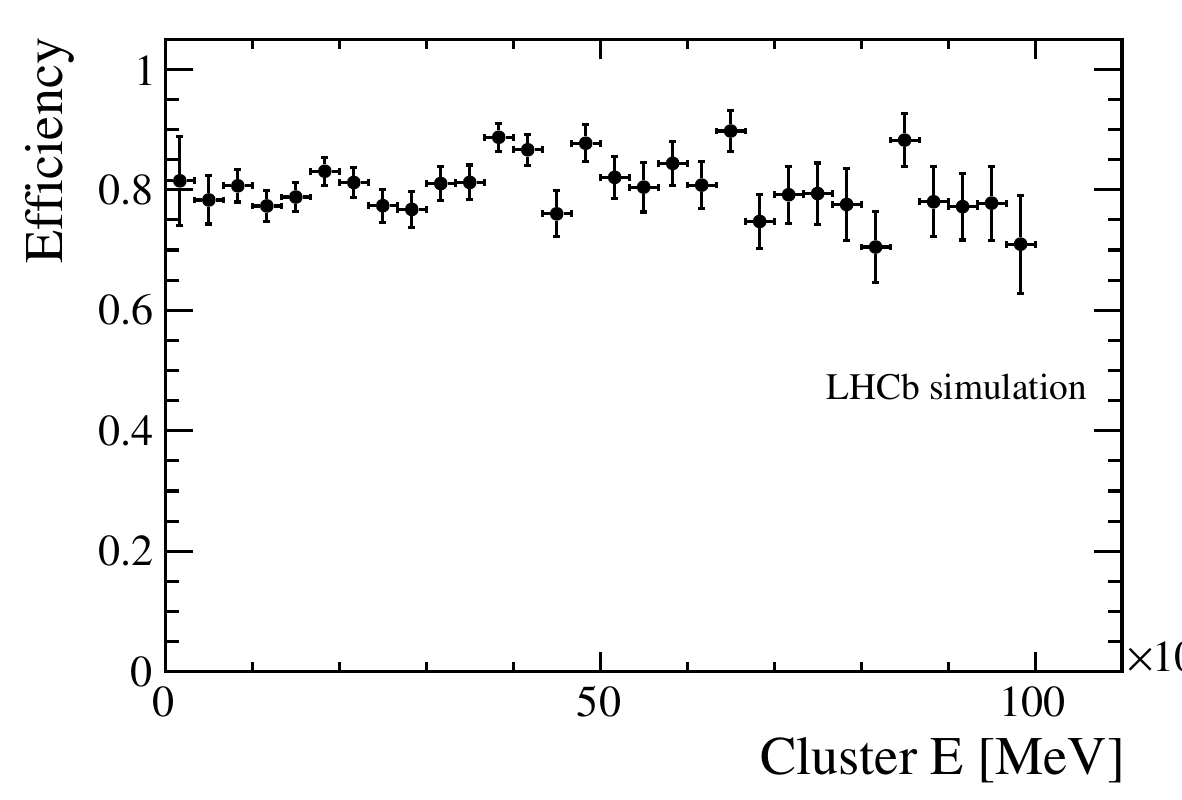}
  \qquad
  \includegraphics[scale=0.32]{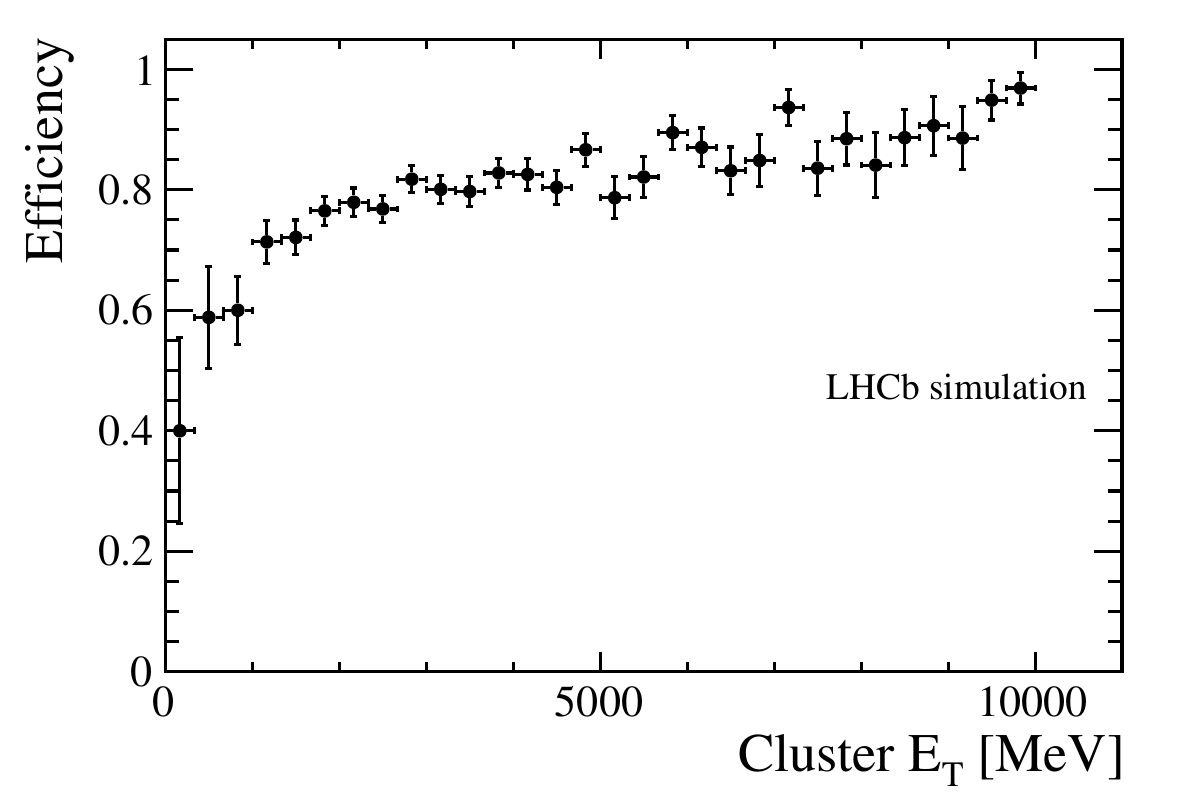}
  \\
  \includegraphics[scale=0.32]{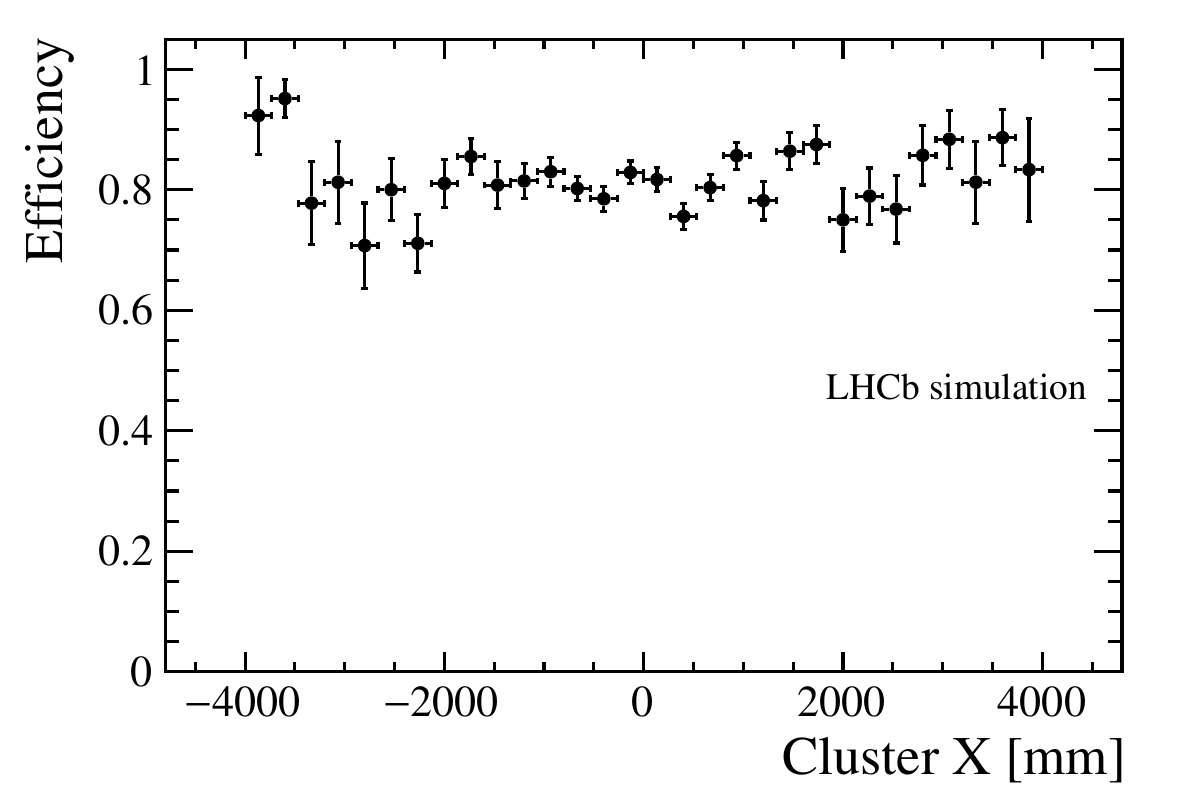}
  \qquad
  \includegraphics[scale=0.32]{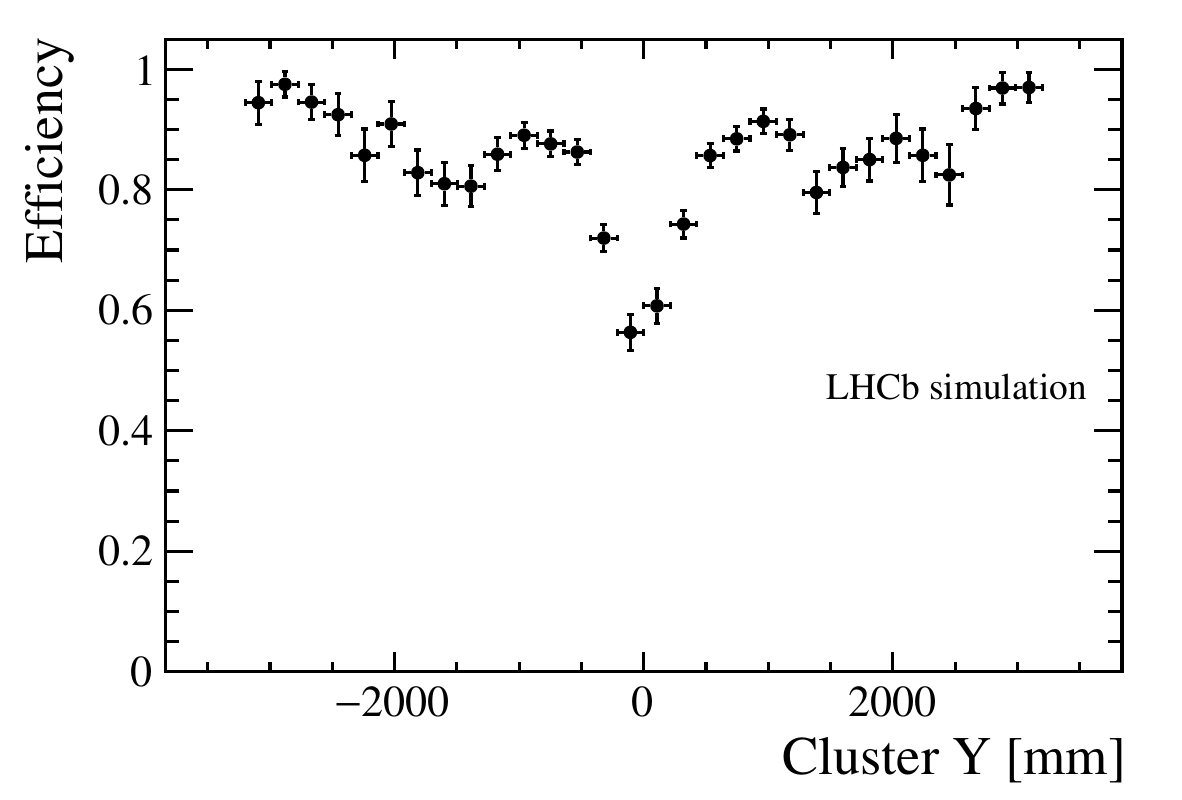}
  \caption{\Ecal cluster reconstruction efficiency versus energy $E$,
    transverse energy \et and $x$ and $y$ position in the \Ecal for
    reconstructible photons from \BdKstGam decays. Reproduced with permission from~\cite{LHCB-FIGURE-2021-003}.}
  \label{fig:calo_eff}
\end{figure}

\begin{figure}[p]
  \centering
  \includegraphics[scale=0.32]{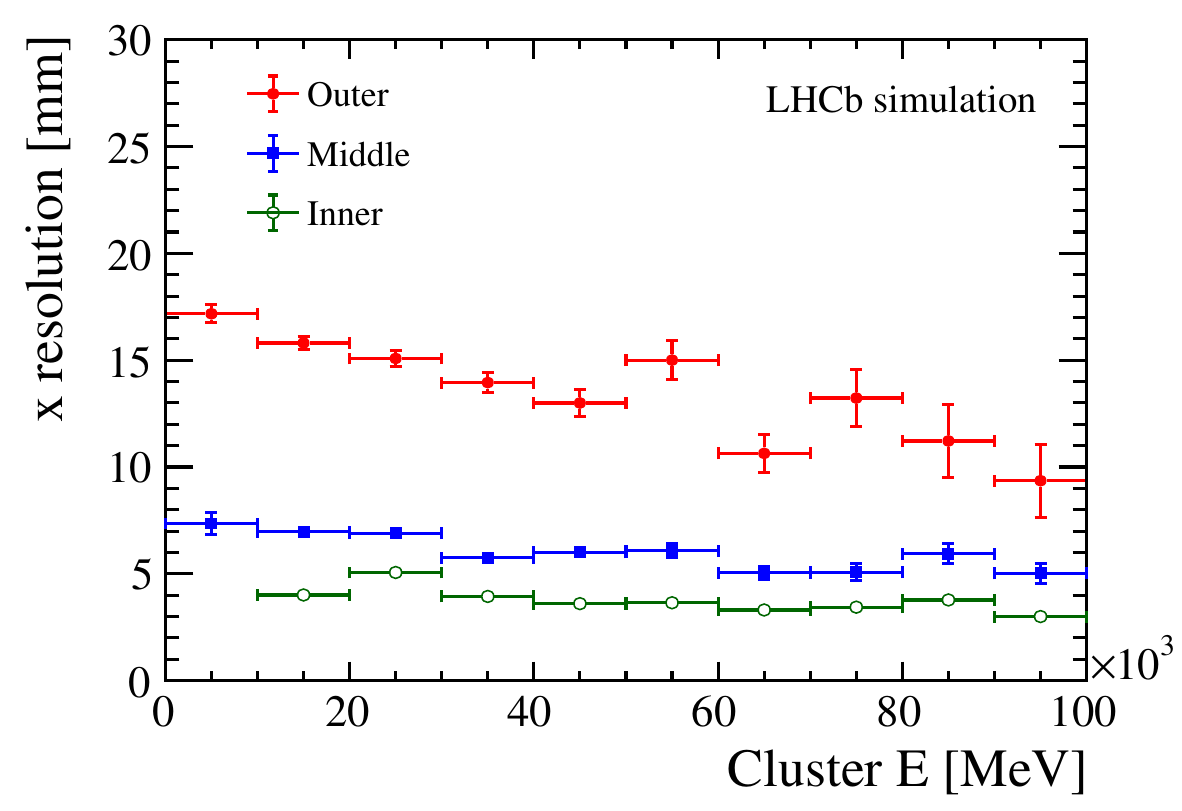}
  \qquad
  \includegraphics[scale=0.32]{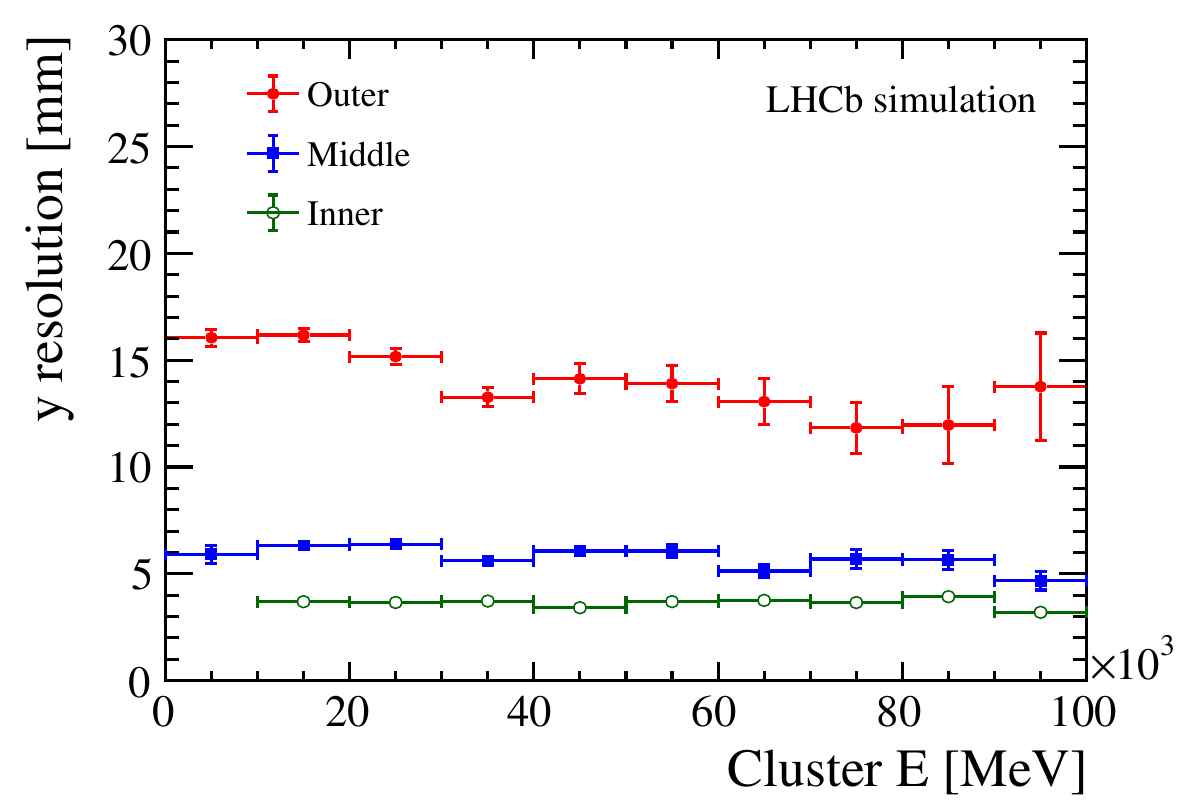}
  \caption{\Ecal-cluster (left) $x$ position and (right) $y$ position
    resolution versus energy for reconstructible photons from
    \BdKstGam decays. Reproduced with permission from~\cite{LHCB-FIGURE-2021-003}.}
  \label{fig:calo_res}
\end{figure}

\begin{figure}[p]
  \centering
  \includegraphics[scale=0.32]{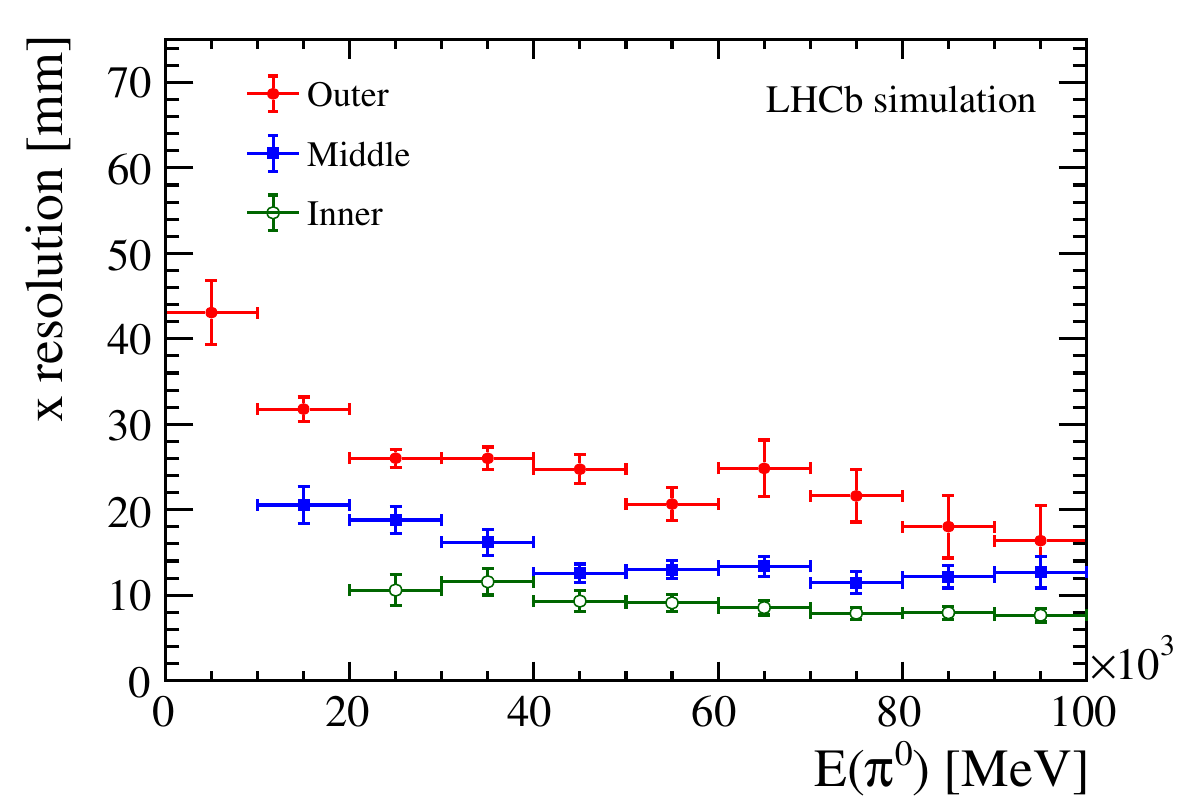}
  \qquad
  \includegraphics[scale=0.32]{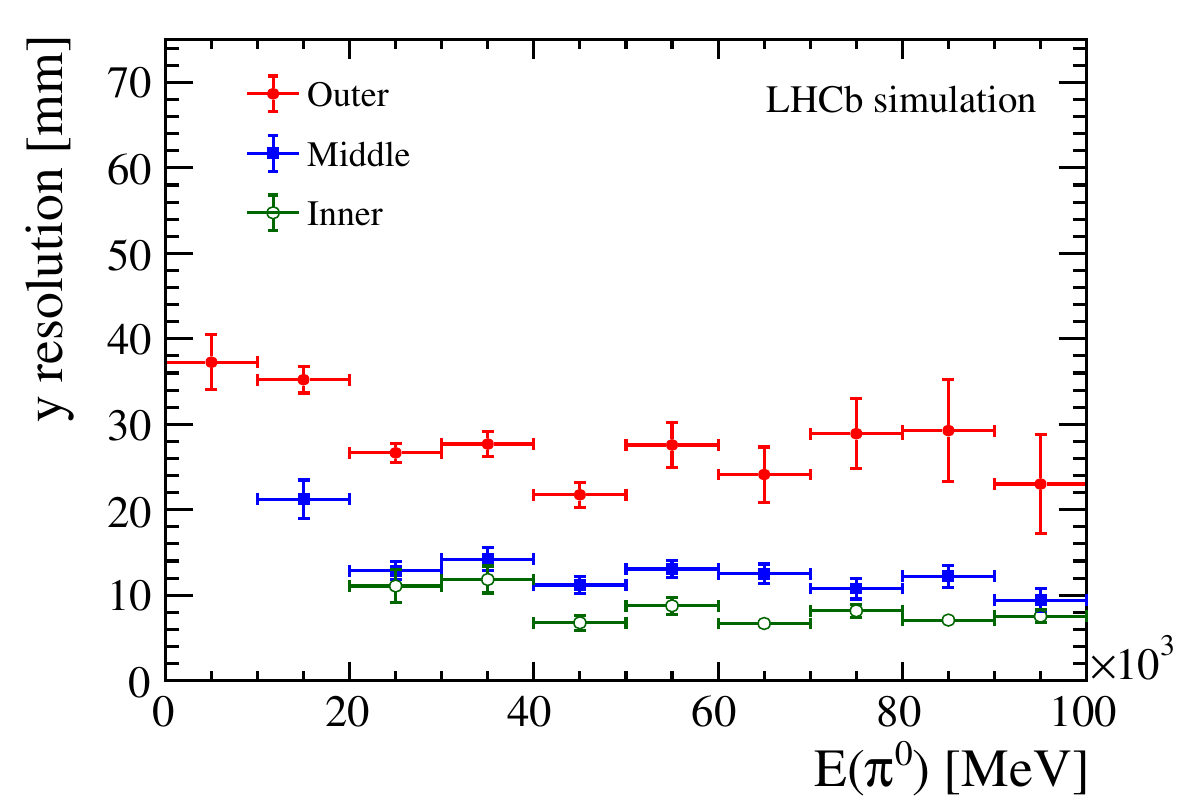}
  \caption{Merged \piz (left) $x$ position and (right) $y$ position
    resolution versus energy for ${\piz\to\g\g}$ from \B~decays. Reproduced with permission from~\cite{LHCB-FIGURE-2021-003}.}
  \label{fig:calo_res_pi0}
\end{figure}

In figure~\ref{fig:calo_eff} the efficiency of reconstructing \Ecal
clusters from the energy deposited by photons is shown, while the
position resolution of the reconstructed clusters is given in
figure~\ref{fig:calo_res}. These figures are produced with \BdKstGam
simulation samples.  Figure~\ref{fig:calo_res_pi0} shows the \Ecal
cluster position resolution for merged $\piz\to\g\g$ decays from \B
decays.  The resolutions are shown separately for the three \Ecal
regions, which are characterised by differing sizes of the \Ecal cells
which affect in particular the position resolution.

\begin{figure}[t]
  \centering
  \includegraphics[scale=0.32]{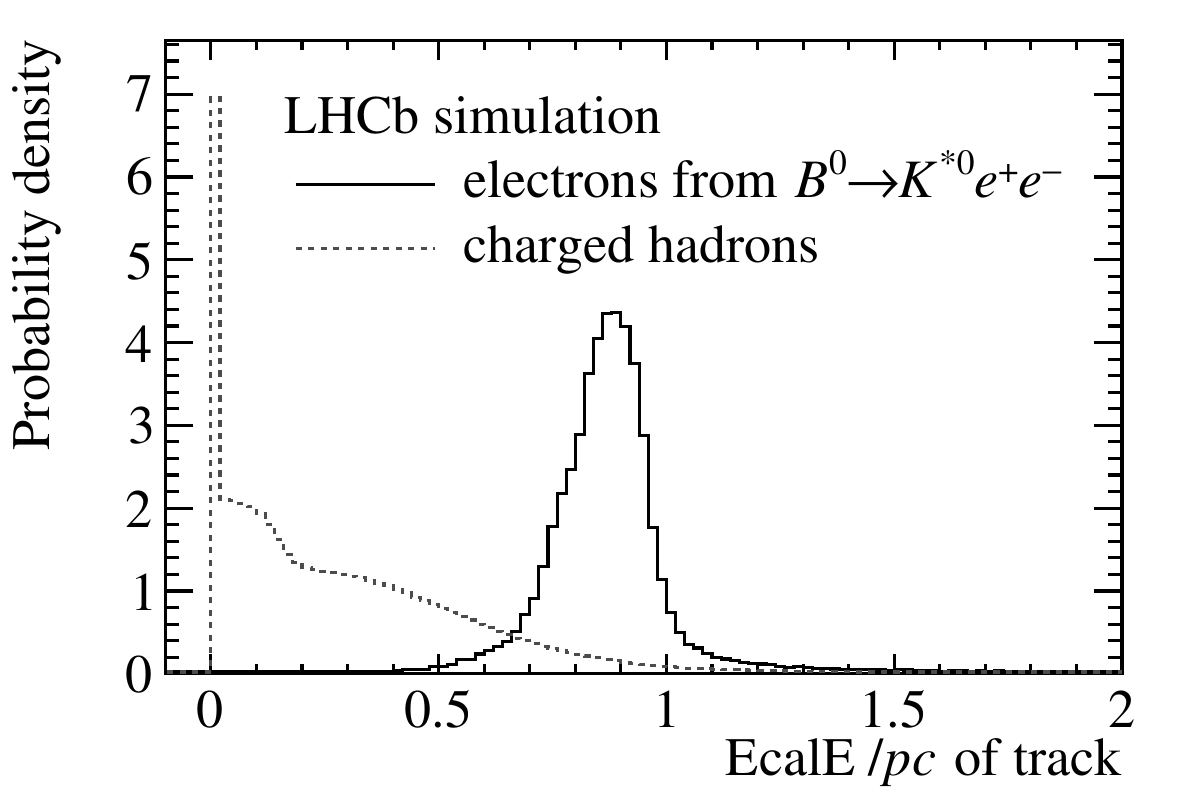}
  \qquad
  \includegraphics[scale=0.32]{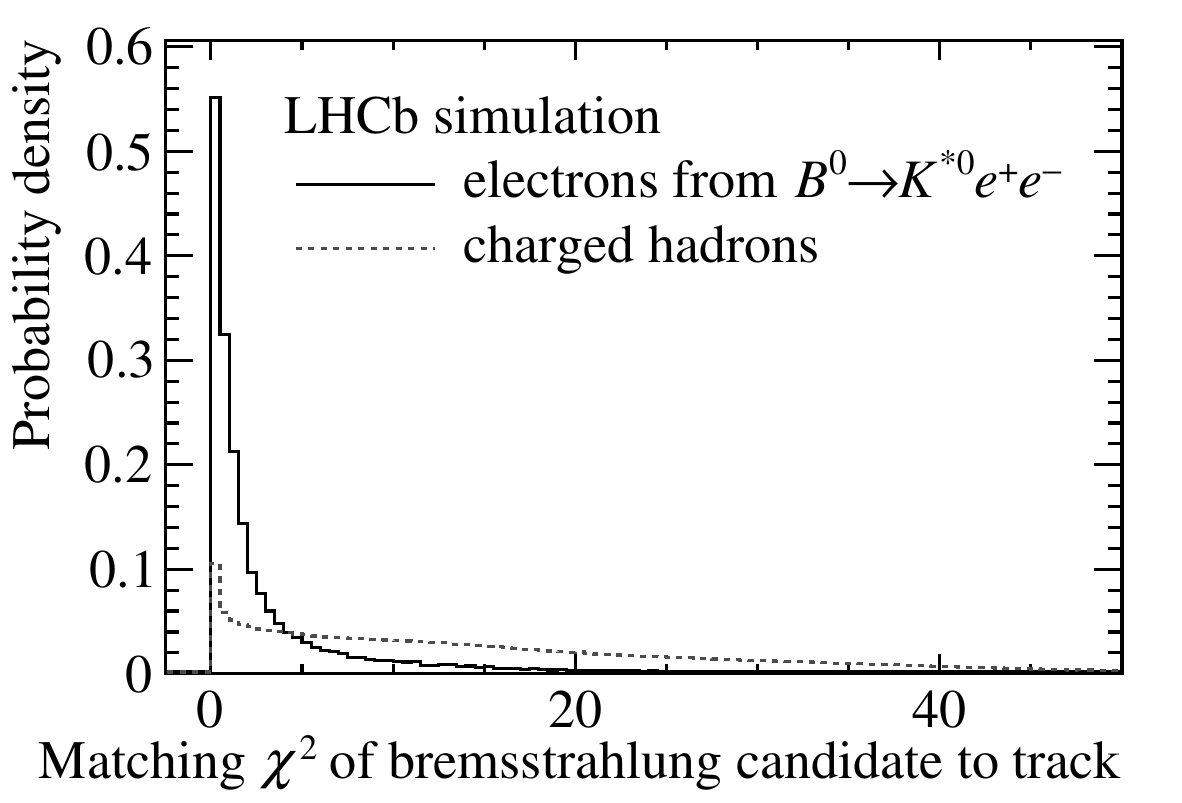}
  \caption{Main electron \Pid variables for the \Ecal: distributions
    for signal and background separately for the variables (left)
    EcalE/\ptot and (right) matching \chisq of a bremsstrahlung
    cluster candidate to a track. The distributions of the
    bremsstrahlung matching \chisq are conditional on having a cluster
    candidate in a $3\times 3$ cell grid around the bremsstrahlung
    track extrapolation. Reproduced with permission from~\cite{LHCB-FIGURE-2021-003}.}
  \label{fig:calo_epid}
\end{figure}

The power of the \Ecal variables to separate electrons from other
charged particles is shown in figure~\ref{fig:calo_epid} by the
distribution of the ratio between \Ecal energy EcalE and track
momentum \ptot, where the variable EcalE is the sum of energies of the
\Ecal cells intersecting the track extrapolation and those compatible
with potential bremsstrahlung emissions. These bremsstrahlung
emissions are determined by projecting the track direction before
bending in the magnetic field to energy deposited in the \Ecal.  These
plots are produced with \BdKstee simulation samples.

\subsection{Selection performance}
\label{sec:RTA:SelPerf}

To cover the current LHCb physics programme, ${\mathcal O}(100)$
selections are deployed in \Hltone and ${\mathcal O}(1000)$ in
\Hlttwo, with their number expected to increase significantly with the
evolution of physics studies. The precise balance between efficiency
and rate for each physics signature will only be established once the
detector is commissioned and the backgrounds observed in data during
this new high-pileup regime are understood. Nevertheless, the general
selection performance is evaluated for certain archetypal \Hltone and
\Hlttwo selections on a subset of representative signal topologies. In
all cases the efficiency is calculated on events in which the signal
topology of interest can be fully reconstructed inside the LHCb
detector acceptance (i.e.\ factorising out the geometrical
acceptance), but no other offline selection criteria are applied. This
rather loose normalisation is used to illustrate the work which the
collaboration will have to do in order to fully benefit from the
removal of the first-level hardware trigger, which biased all signals
to have large transverse momentum. In the case of \Hlttwo selections,
the efficiencies are calculated relative to signal events passing
\Hltone requirements, in order to factorise the performance of the
first- and second-level triggers.

\begin{figure}[h]
  \centering
  \includegraphics[scale=0.32]{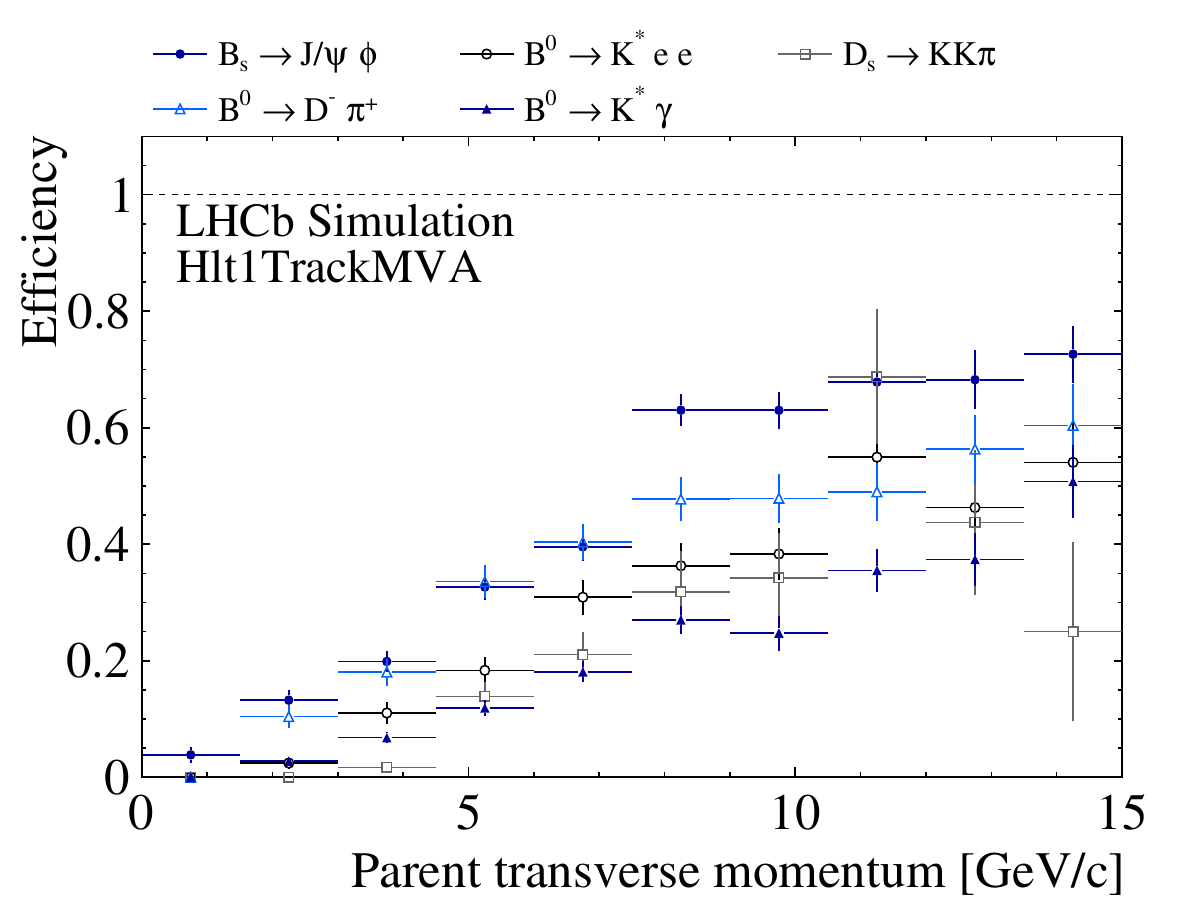}
  \qquad
  \includegraphics[scale=0.32]{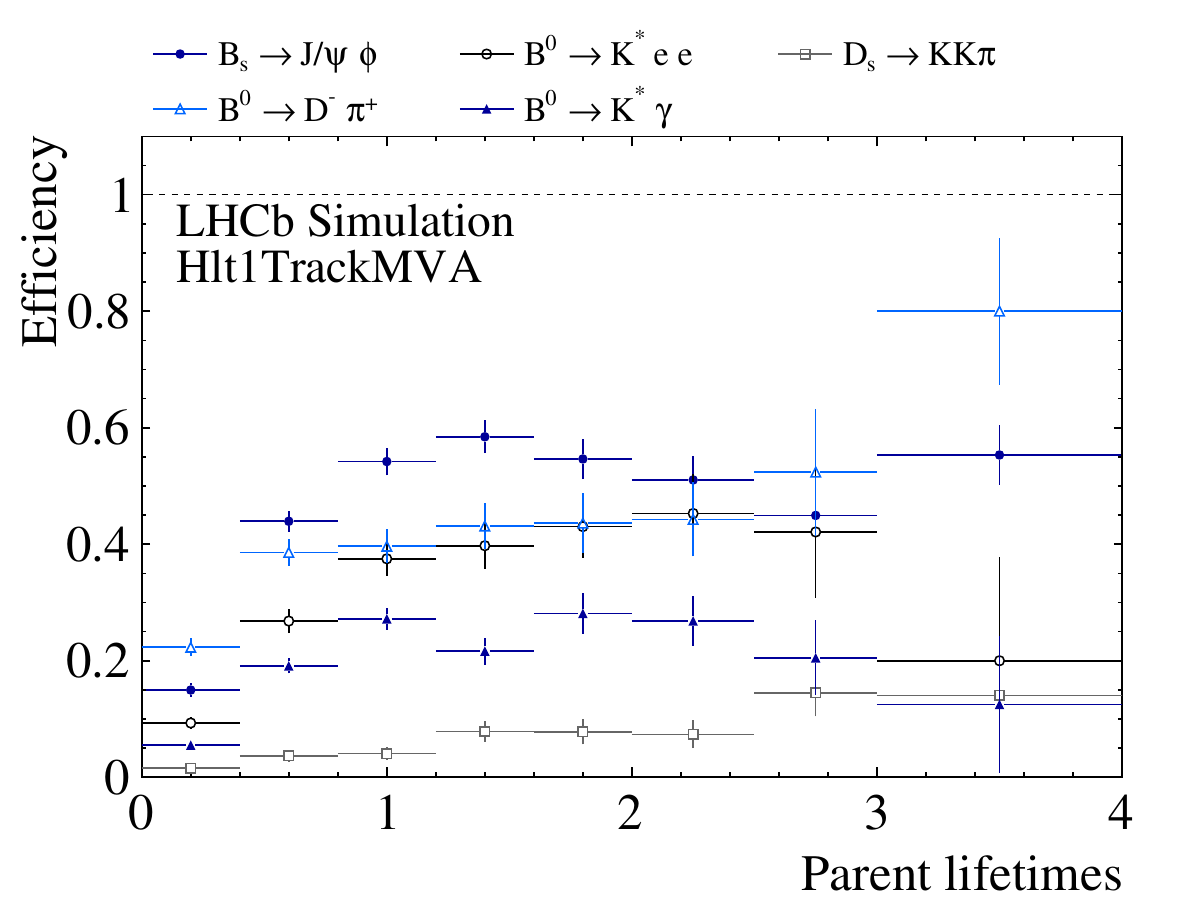}
  \\
  \includegraphics[scale=0.32]{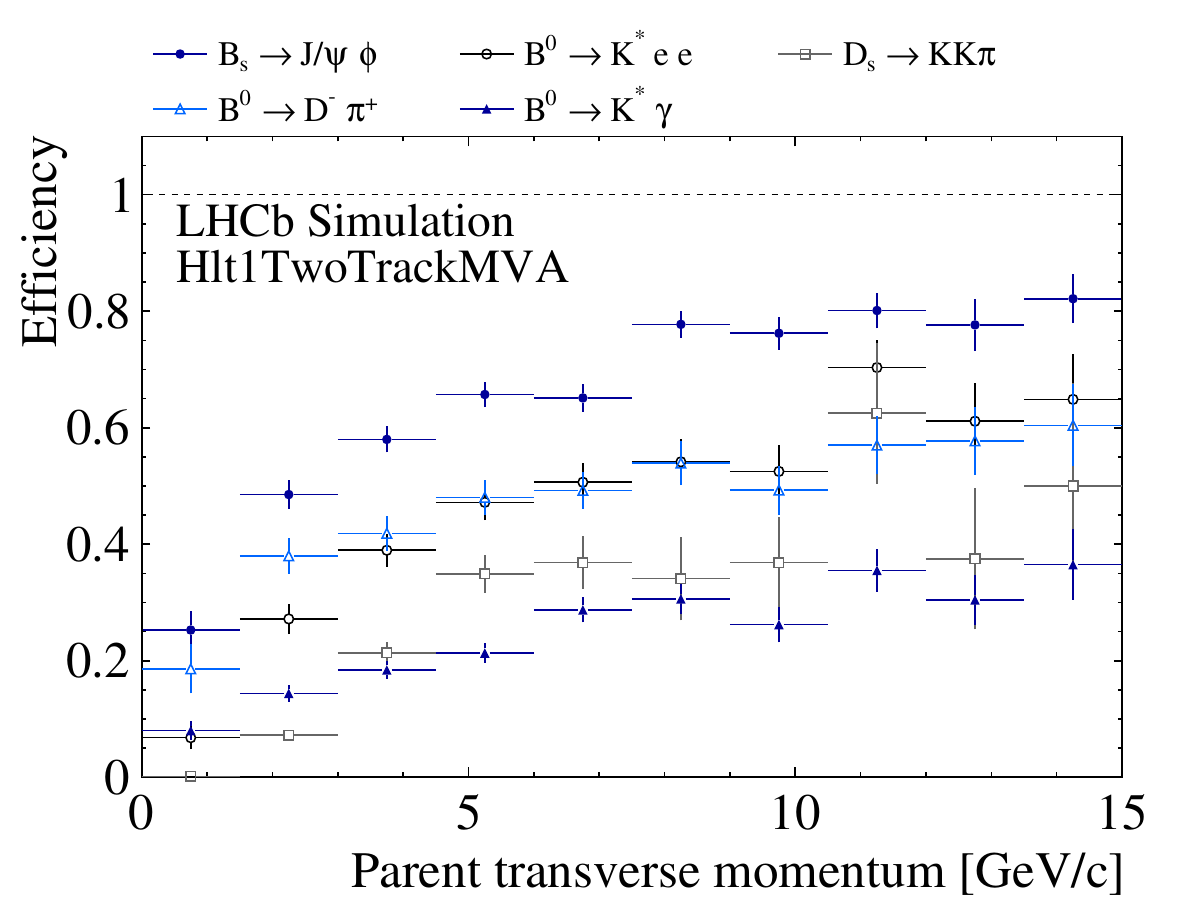}
  \qquad
  \includegraphics[scale=0.32]{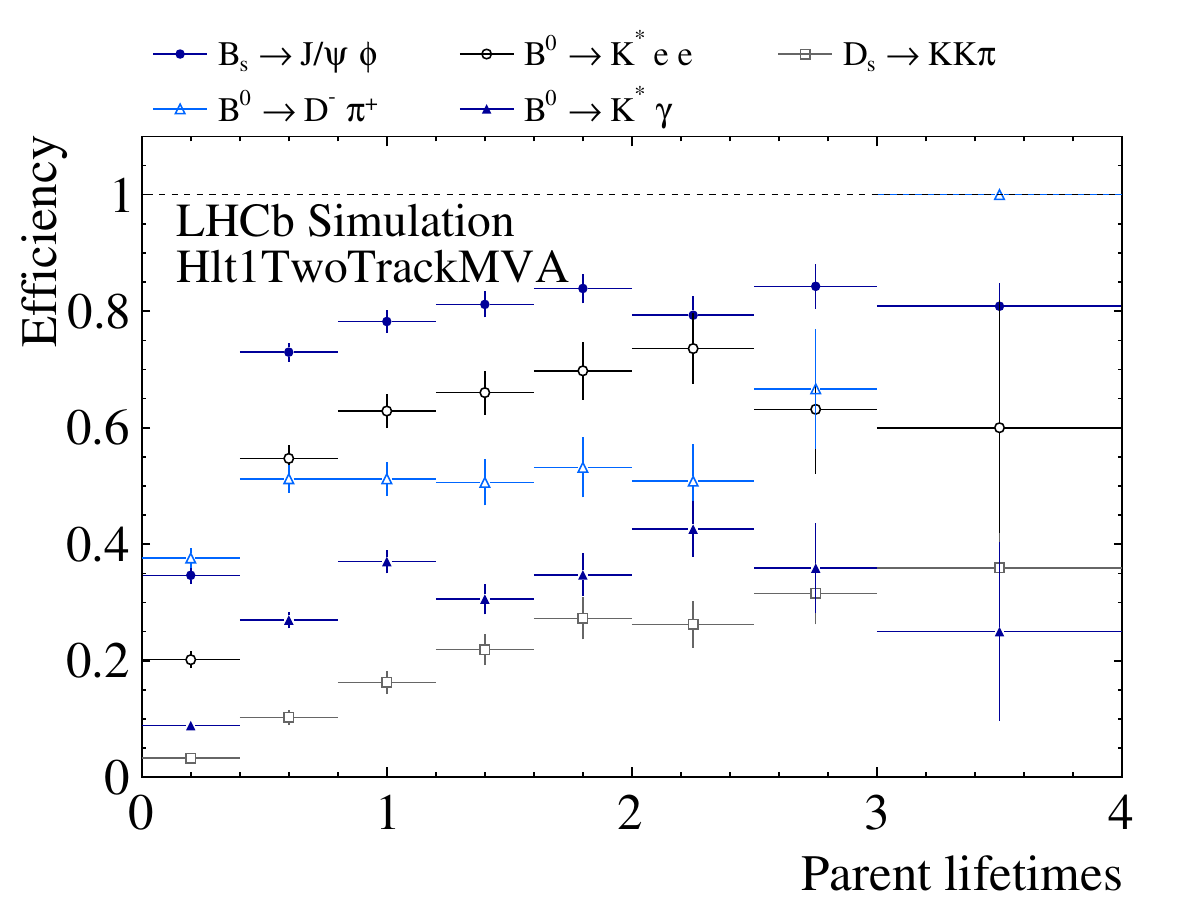}
  \caption{Performance of the \Hltone inclusive selections as a
    function of (left) parent-particle transverse momentum and (right)
    parent-particle decay time. The top row plots are the single-track
    selections, while the bottom row plots are the two-track displaced
    vertex selections. The signal topologies are indicated in the
    legend above each plot. The decay time plots are drawn such that
    the $x$ axis is binned in units of the lifetime for each hadron in
    its rest frame. Reproduced with permission from~\cite{LHCB-FIGURE-2021-003}. }
  \label{fig:selections_perf_track_mva_hlt1}
\end{figure}

\begin{figure}[p]
  \centering
  \includegraphics[scale=0.23]{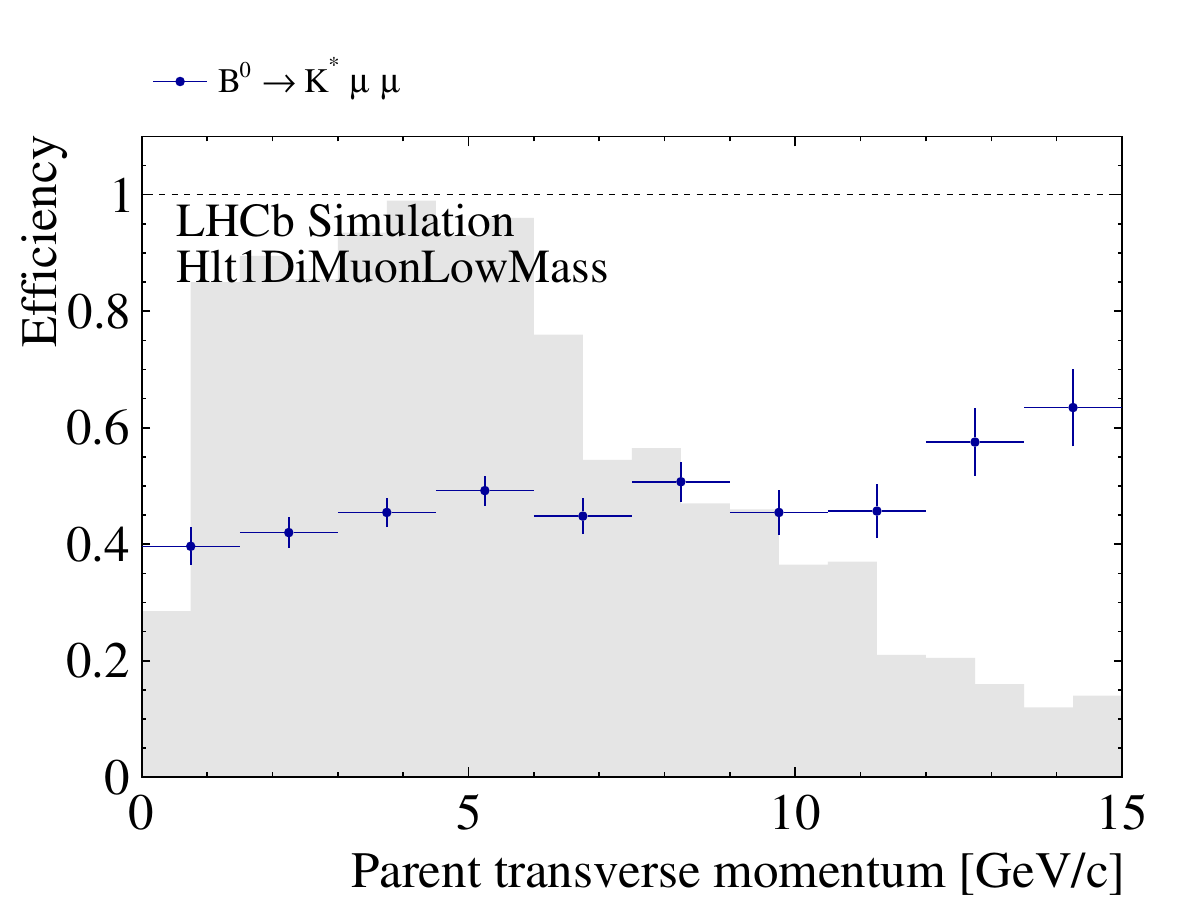}
  \includegraphics[scale=0.23]{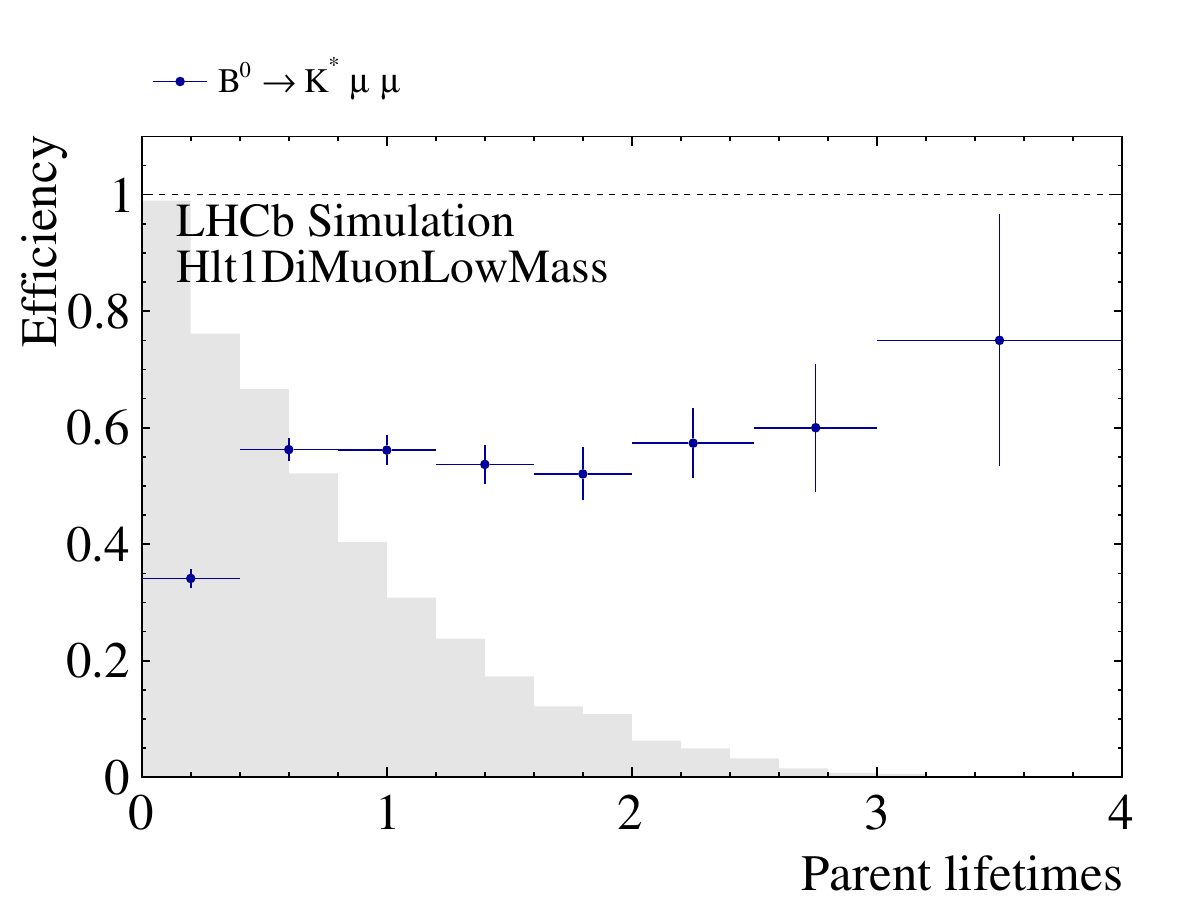}\\
  \includegraphics[scale=0.23]{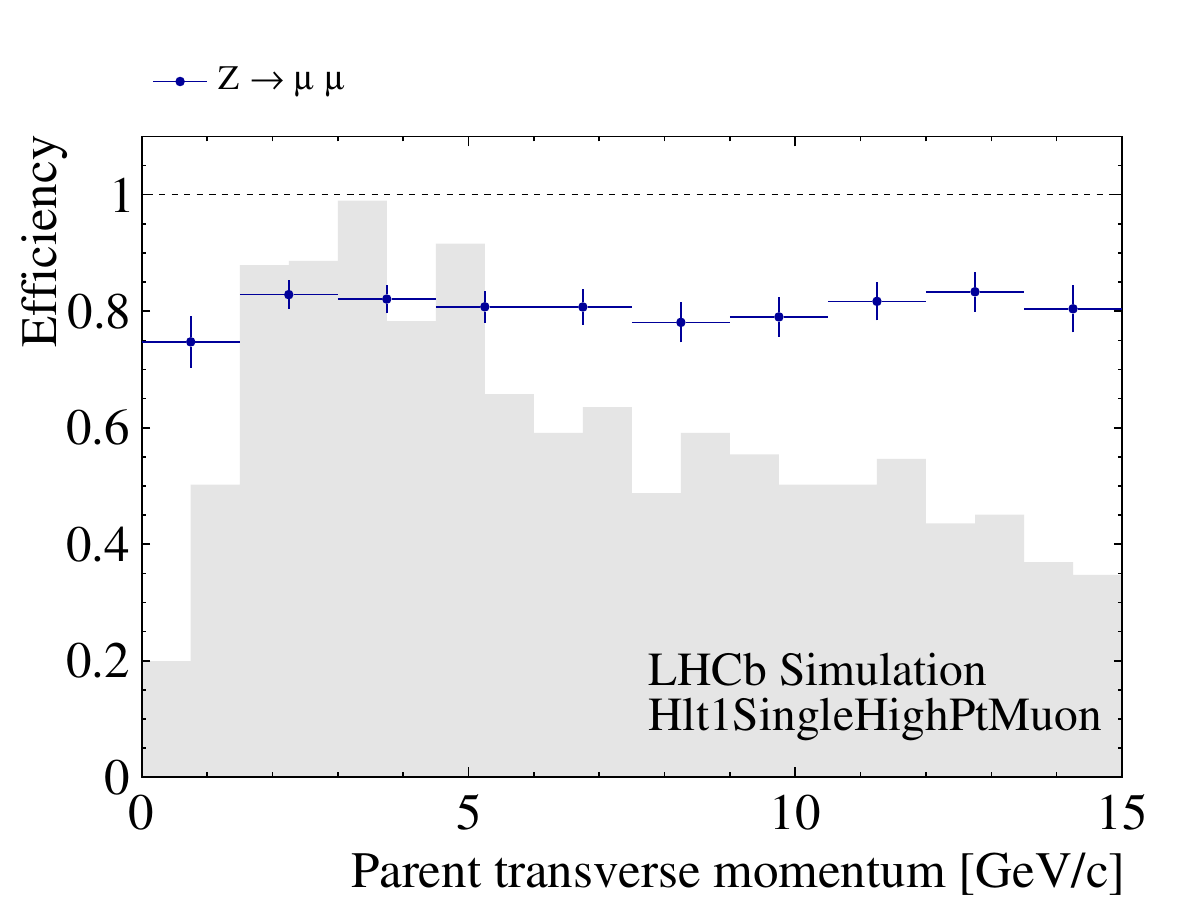}
  \caption{Performance of the \Hltone muon selections. The signal
    topologies are indicated in the legend above each plot. In the top
    row the performance of the dimuon selections is plotted as a
    function of (left) parent-particle transverse momentum and (right)
    parent-particle decay time. The decay time plot is drawn such that
    the $x$ axis is binned in units of the lifetime for each hadron in
    its rest frame. In the bottom row the performance of the single
    high-\pt muon selection is plotted as a function of parent
    transverse momentum. The shaded histograms indicate the
    distribution of the parent particle prior to any trigger
    selection. Reproduced with permission from~\cite{LHCB-FIGURE-2021-003}.}
  \label{fig:selections_perf_muon_hlt1}
\end{figure}

\begin{figure}[p]
  \centering
  \includegraphics[scale=0.23]{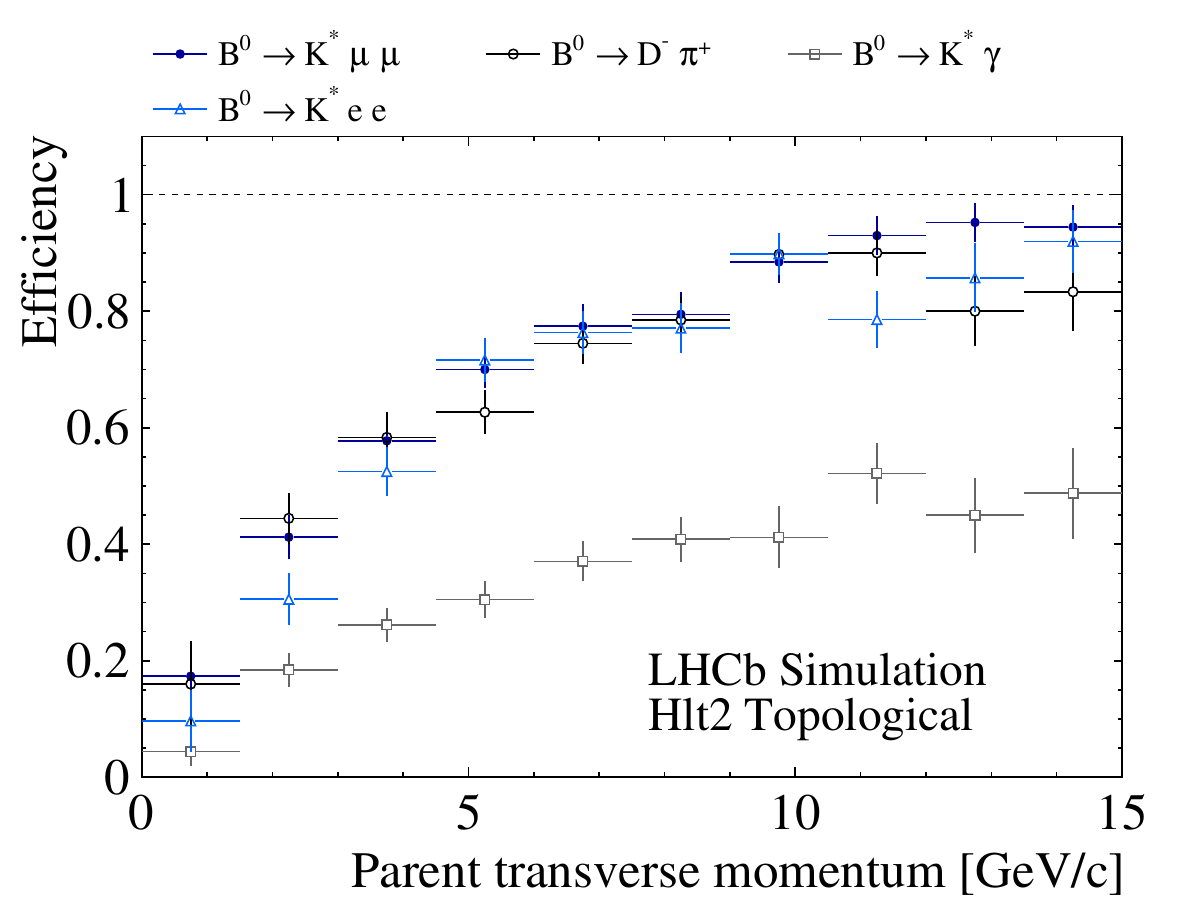}
  \includegraphics[scale=0.23]{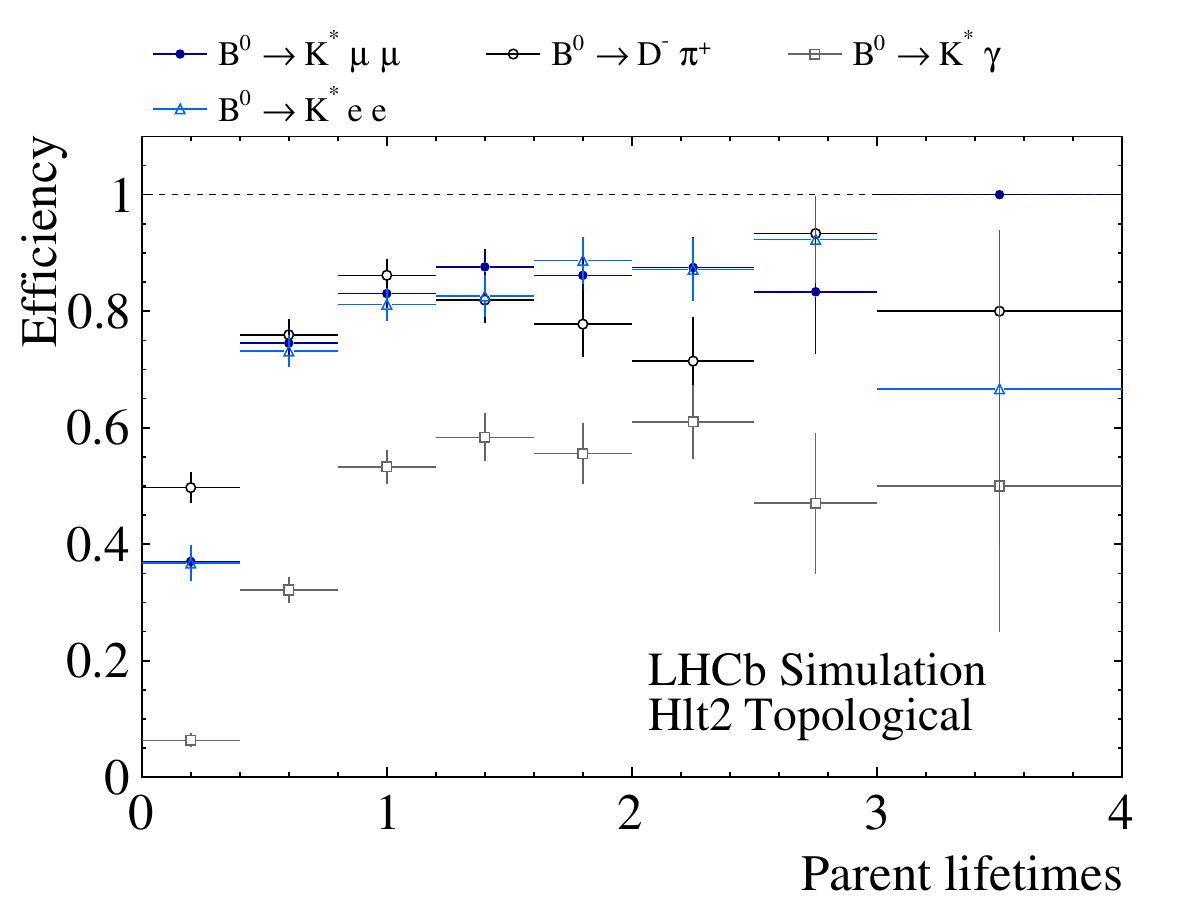}\\
  \includegraphics[scale=0.23]{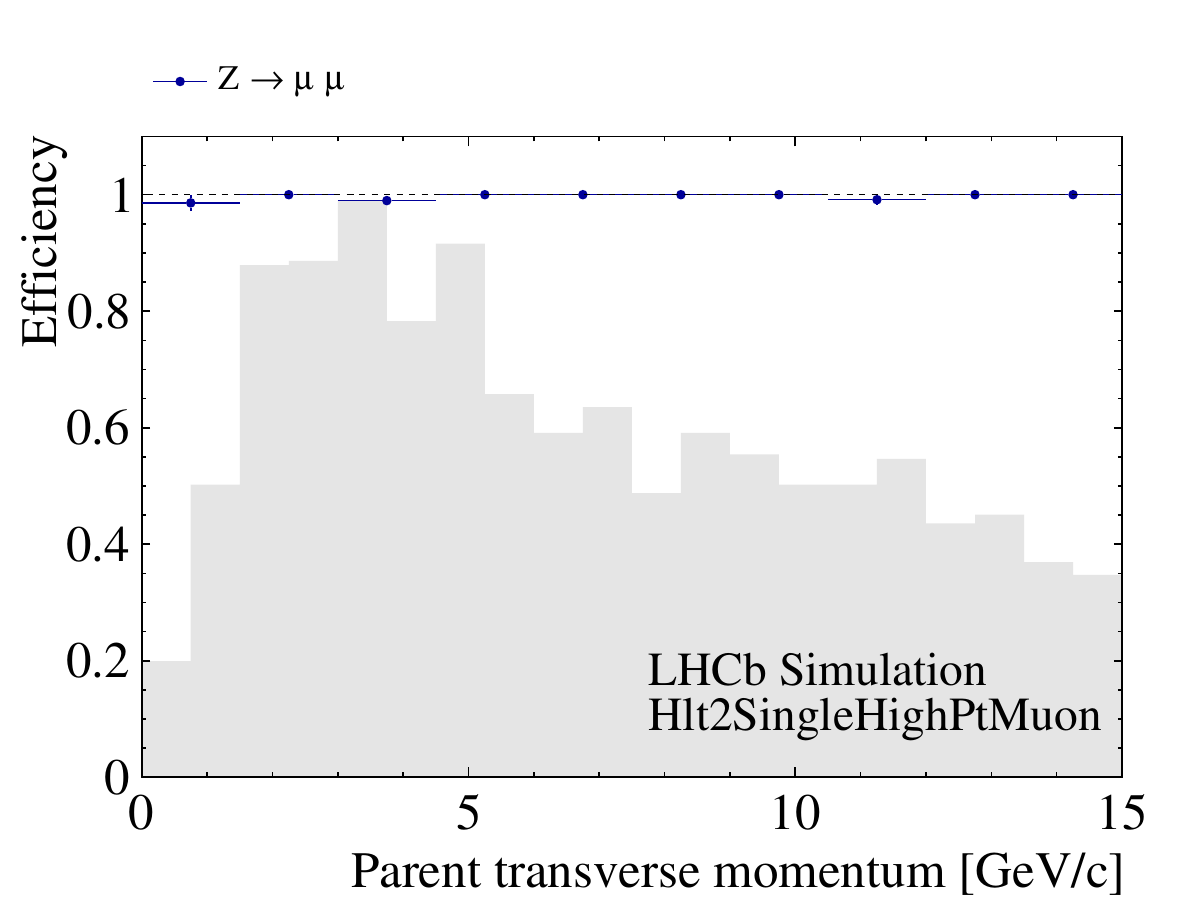}
  \caption{Performance of the \Hlttwo inclusive selections. The signal
    topologies are indicated in the legend above each plot. In the top
    row the performance of the inclusive displaced-vertex selections
    is plotted as a function of (left) parent-particle transverse
    momentum and (right) parent-particle decay time. The decay-time
    plot is drawn such that the $x$ axis is binned in units of the
    lifetime for each hadron in its rest frame. In the bottom row the
    performance of the single high-\pt muon selection is plotted as a
    function of parent transverse momentum. The shaded histograms
    indicate the distribution of the parent particle prior to any
    trigger selection. Reproduced with permission from~\cite{LHCB-FIGURE-2021-003}.}
  \label{fig:selections_perf_inclusive_hlt2}
\end{figure}

\begin{figure}[htbp]
  \centering
  \includegraphics[scale=0.32]{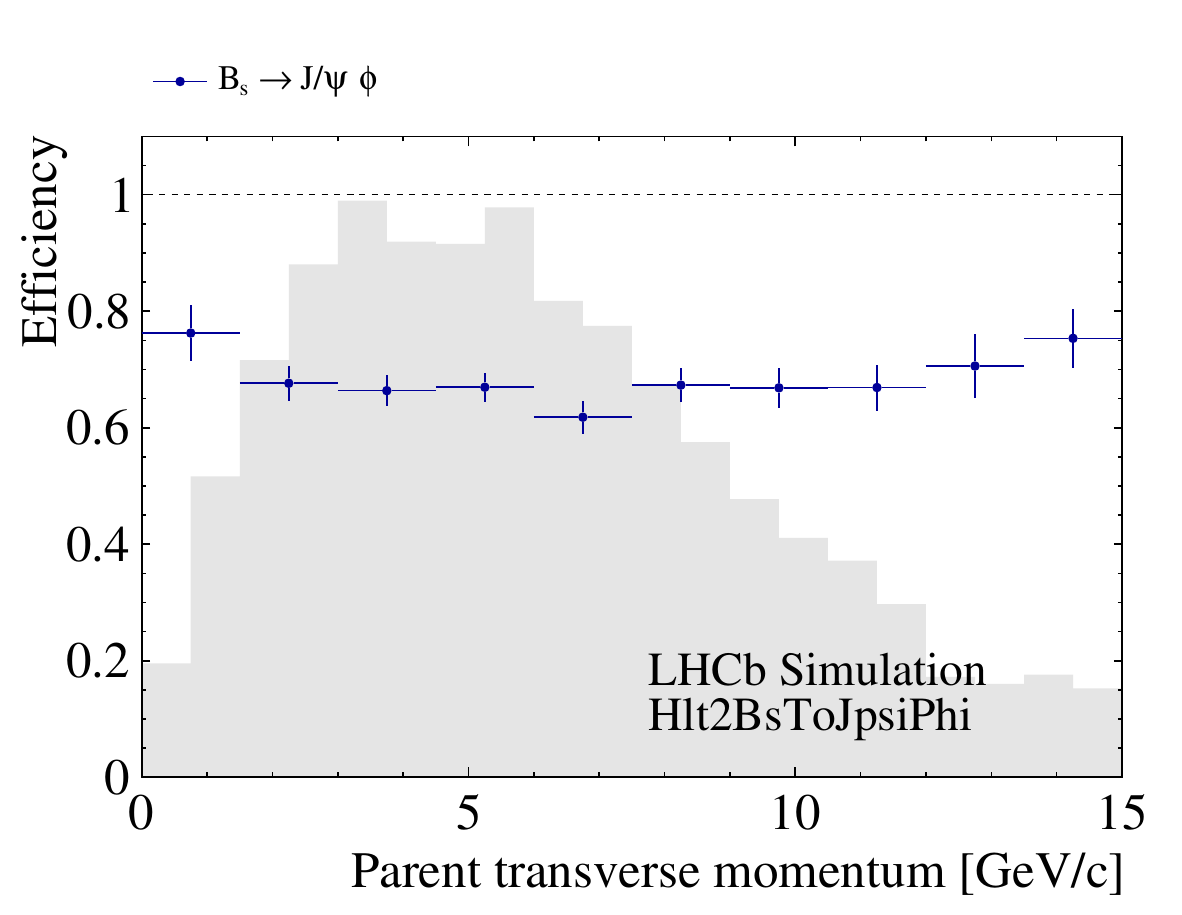}
  \qquad
  \includegraphics[scale=0.32]{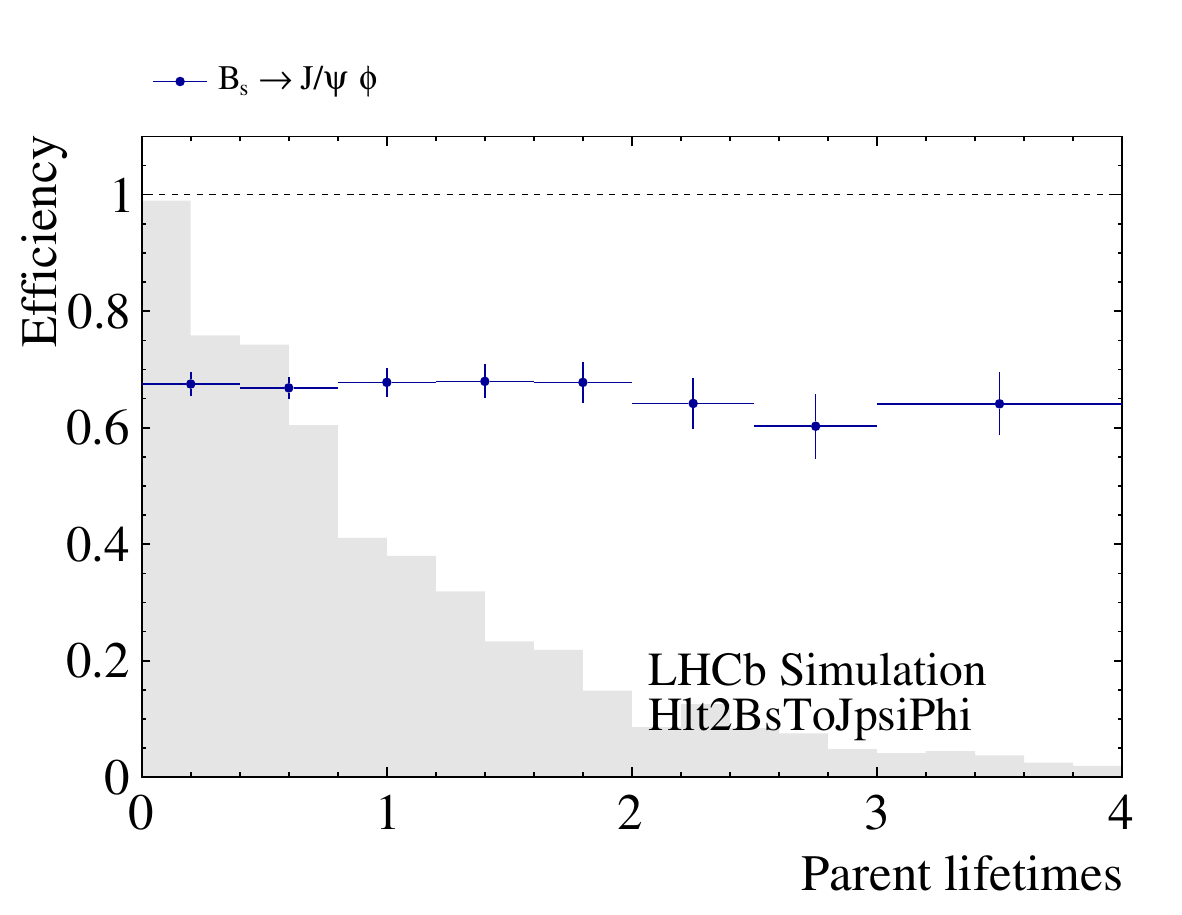}
  \\
  \includegraphics[scale=0.32]{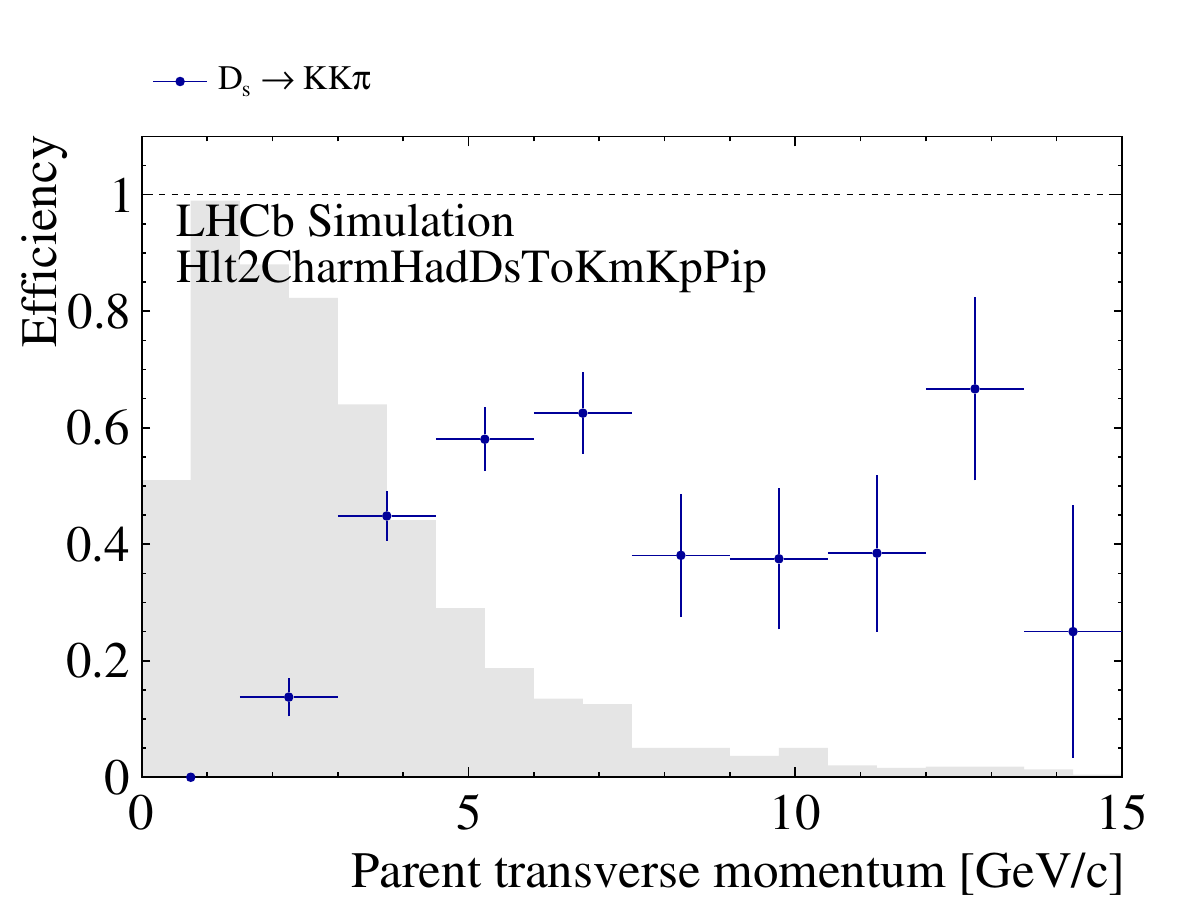}
  \qquad
  \includegraphics[scale=0.32]{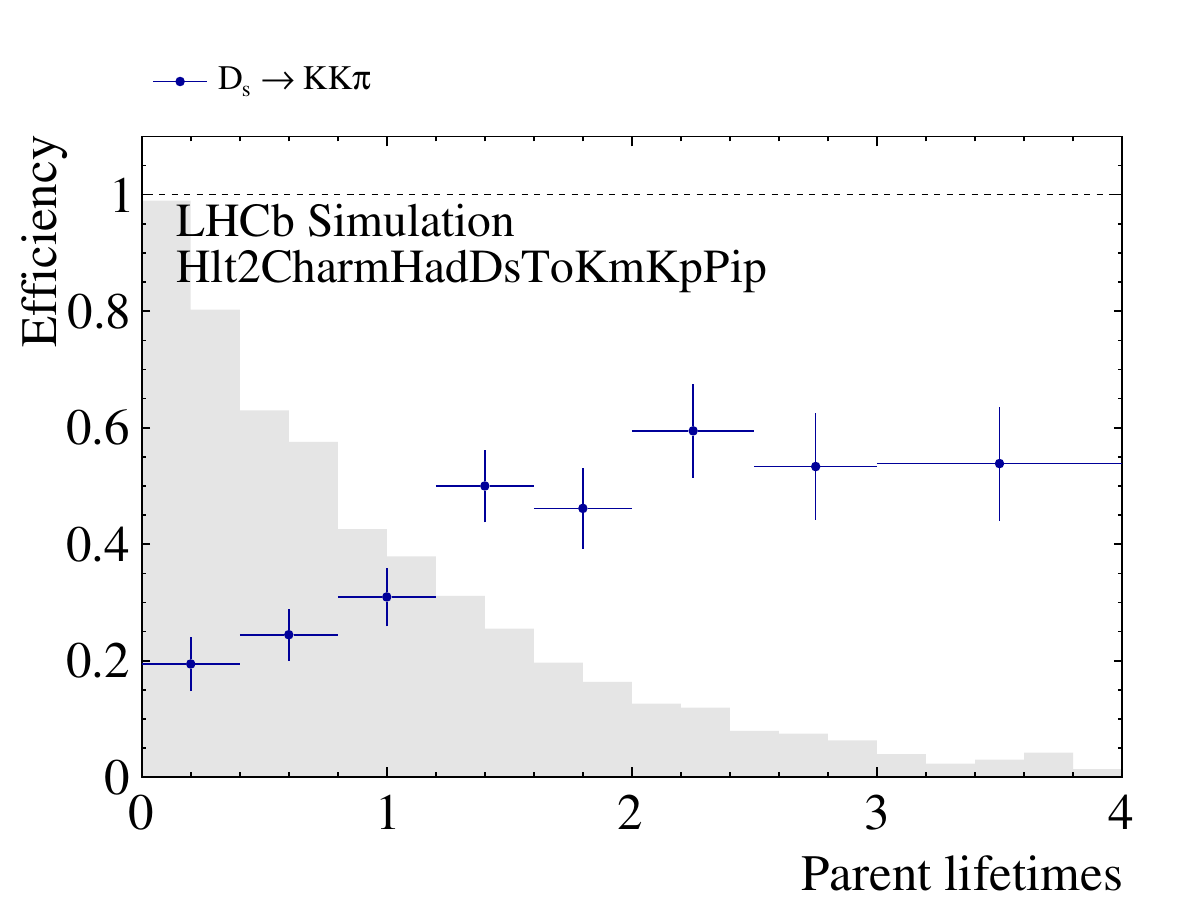}
  \caption{Performance of example \Hlttwo exclusive selections as a
    function of (left) parent-particle transverse momentum and (right)
    parent-particle decay time. The signal topologies are indicated in
    the legend above each plot. The decay-time plots are drawn such
    that the $x$ axis is binned in units of the lifetime for each
    hadron in its rest frame. The shaded histograms indicate the
    distribution of the parent particle prior to any trigger
    selection. Reproduced with permission from~\cite{LHCB-FIGURE-2021-003}.}
  \label{fig:selections_perf_exclusive_hlt2}
\end{figure}

The performance of \Hltone inclusive selections --- a single displaced
track trigger and a displaced vertex trigger --- is shown in
figure~\ref{fig:selections_perf_track_mva_hlt1}.  The performance of
\Hltone inclusive muon selections is shown in
figure~\ref{fig:selections_perf_muon_hlt1}.  The performance of the
\Hlttwo inclusive triggers is shown in
figure~\ref{fig:selections_perf_inclusive_hlt2}. Finally, the
performance of two representative exclusive \Hlttwo triggers is shown
in figure~\ref{fig:selections_perf_exclusive_hlt2}. While the plotted
efficiencies may appear low in many cases, this is because of the lack
of offline criteria in the denominator. For the same reason the
\Hlttwo efficiencies, where the denominator are events passing \Hltone
conditions, are higher. Nevertheless, the examples of \Hlttwo single
high-\pt muon triggers and the exclusive $\Bs\to\jpsi\phiz$ triggers
show that exclusive triggers targeting specific well-defined decay
chains can achieve excellent efficiencies even with respect to
reconstructible decays.

\clearpage

\section{Summary}
\label{sec:summary}
The LHCb upgraded experiment has been described including the
detector, the online system, the all-software trigger and the software
and computing infrastructure.

The upgrade of LHCb consists of: a tracking system including a new
silicon-pixel vertex detector, a new silicon-strip tracker upstream of
the dipole magnet and a new scintillating-fibre tracker downstream of
the dipole magnet; a complete rebuild of the photon detection system
of the Cherenkov detectors using multianode photomultipliers tubes;
and redesigned and updated electronics of the calorimeters and the
muon detector. A novel all-software trigger running on GPUs and on a
dedicated computing farm has been deployed, and a completely renewed
online system installed. To match the new trigger scheme, the software
code base and computing model have been fully redesigned and
reimplemented.

The performance of the new detector systems, as studied in the
laboratory and with test-beam measurements, has been discussed, along
with the expected overall experiment performance, estimated through
Monte Carlo simulations, and is found to be as good as the previous
experiment, if not better, while facing much higher luminosity and
pileup running conditions.

The upgraded experiment will significantly extend the physics
programme of LHCb, providing substantially larger statistics for
precision studies and new physics searches, and opening new fields of
investigation not only in the flavour physics domain but also, as a
general purpose detector, in heavy ion, electroweak and fixed target
physics.

\section*{Acknowledgements}
%
%
\noindent We express our gratitude to our colleagues in the CERN
accelerator departments for the excellent performance of the LHC. We
thank the technical and administrative staff at the LHCb
institutes.
We acknowledge support from CERN and from the national agencies:
CAPES, CNPq, FAPERJ and FINEP (Brazil); 
MOST and NSFC (China); 
CNRS/IN2P3 (France); 
BMBF, DFG and MPG (Germany); 
INFN (Italy); 
NWO (Netherlands); 
MNiSW and NCN (Poland); 
MEN/IFA (Romania); 
MICINN (Spain); 
SNSF and SER (Switzerland); 
NASU (Ukraine); 
STFC (United Kingdom); 
DOE NP and NSF (USA).
We acknowledge the computing resources that are provided by CERN, IN2P3
(France), KIT and DESY (Germany), INFN (Italy), SURF (Netherlands),
PIC (Spain), GridPP (United Kingdom), 
CSCS (Switzerland), IFIN-HH (Romania), CBPF (Brazil),
Polish WLCG  (Poland) and NERSC (USA).
We are indebted to the communities behind the multiple open-source
software packages on which we depend.
Individual groups or members have received support from
ARC and ARDC (Australia);
Minciencias (Colombia);
AvH Foundation (Germany);
EPLANET, Marie Sk\l{}odowska-Curie Actions and ERC (European Union);
A*MIDEX, ANR, IPhU and Labex P2IO, and R\'{e}gion Auvergne-Rh\^{o}ne-Alpes (France);
Key Research Program of Frontier Sciences of CAS, CAS PIFI, CAS CCEPP, 
Fundamental Research Funds for the Central Universities, 
and Sci. \& Tech. Program of Guangzhou (China);
GVA, XuntaGal and GENCAT (Spain);
SRC (Sweden);
the Leverhulme Trust, the Royal Society
 and UKRI (United Kingdom).

\clearpage

\printglossary[type=\acronymtype]
\clearpage





\addcontentsline{toc}{section}{References}
\setboolean{inbibliography}{true}
\bibliographystyle{LHCb}
\bibliography{upgrade_paper}

\newpage
\centerline
{\large\bf LHCb collaboration}
\begin
{flushleft}
\small
R.~Aaij$^{32}$\lhcborcid{0000-0003-0533-1952},
A.S.W.~Abdelmotteleb$^{50}$\lhcborcid{0000-0001-7905-0542},
C.~Abellan~Beteta$^{44}$,
F.~Abudin{\'e}n$^{50}$\lhcborcid{0000-0002-6737-3528},
C.~Achard$^{9}$,
T.~Ackernley$^{54}$\lhcborcid{0000-0002-5951-3498},
B.~Adeva$^{40}$\lhcborcid{0000-0001-9756-3712},
M.~Adinolfi$^{48}$\lhcborcid{0000-0002-1326-1264},
P.~Adlarson$^{77}$\lhcborcid{0000-0001-6280-3851},
H.~Afsharnia$^{9}$,
C.~Agapopoulou$^{13}$\lhcborcid{0000-0002-2368-0147}
% ...

{\footnotesize \it

% ...
$^{9}$Universit{\'e} Clermont Auvergne, CNRS/IN2P3, LPC, Clermont-Ferrand, France\\
% ...
$^{13}$LPNHE, Sorbonne Universit{\'e}, Paris Diderot Sorbonne Paris Cit{\'e}, CNRS/IN2P3, Paris, France\\
% ...
$^{32}$Nikhef National Institute for Subatomic Physics, Amsterdam, Netherlands\\
% ...
$^{40}$Instituto Galego de F{\'\i}sica de Altas Enerx{\'\i}as (IGFAE), Universidade de Santiago de Compostela, Santiago de Compostela, Spain\\
% ...
$^{44}$Physik-Institut, Universit{\"a}t Z{\"u}rich, Z{\"u}rich, Switzerland\\
% ...
$^{48}$H.H. Wills Physics Laboratory, University of Bristol, Bristol, United Kingdom\\
% ...
$^{50}$Department of Physics, University of Warwick, Coventry, United Kingdom\\
% ...
$^{53}$School of Physics and Astronomy, University of Glasgow, Glasgow, United Kingdom\\
$^{54}$Oliver Lodge Laboratory, University of Liverpool, Liverpool, United Kingdom\\
% ...
$^{77}$Department of Physics and Astronomy, Uppsala University, Uppsala, Sweden, associated to $^{53}$\\
% ...
}
\end{flushleft}

\end{document}